# Main Concepts and Principles of Political Economy

## Production and Values, Distribution and Prices, Reproduction and Profits

## Prelude to a Reconstruction of Economic Theory

by

**Christian Flamant**

**Main Concepts and Principles of Political Economy**
By Christian Flamant

To my wife Anne,

And my children Nicolas, Aurélie, Alexandre and Camille

# Preface

The intention to write this book arose from the double ascertainment that economic science is in deep trouble, to such a point that it can hardly be considered as a science, but that at the same time no prominent economist has so far published a book which could be widely considered as a true reconstruction of this science, based on solid foundations. When I use the terms “economic science”, I refer to the theory that is socially considered as constitutive of this science, because it is widely taught in all universities over the world and developed in numerous articles published in the most prestigious economic reviews, i.e. the neo-classical theory, based on the paradigm of equilibrium on the market as the best way to allocate scarce resources.

This theory has nevertheless failed, on two different but related grounds. First, at a purely theoretical level the theory has been criticized and refuted over the years by a number of heterodox economists. The best synthesis of all these criticisms might well be the book of Steve Keen: “Debunking Economics”, published in 2001 for the first edition, and in 2011 for a revised and expanded edition. Second, at a more concrete and empirical level, neo-classical theory has been the doctrinal basis of neo-liberal economic policies implemented in most developed economies by almost all governments, whatever their location on the political spectrum, since 1980. These policies have in the vast majority of rich countries encouraged changes in favor of liberalization, financialization and globalization, which have resulted in disastrous consequences worldwide, i.e. a global reduction in the rate of growth, a rise in unemployment and finally the world financial crisis of 2008, which the neo-classical theory had been unable to foresee, and is still unable to explain correctly.

To be sure a small number of economists had predicted the occurrence of the crisis, and Steve Keen was one of them. His own explanations can be found in a series of articles which have been regrouped in a new book, published in 2015: “Developing an economics for the post-crisis world”. These explanations can also be found in chapter 14 of “Debunking Economics” (second edition), which develops a monetary model of capitalism. This model relies on a Keynesian theory of money, as interpreted in particular in the works of Minsky. On this basis Keen develops a theory of debt which is consistent and has great heuristic value. However none of his two books can be considered as a full-fledged economic theory, encompassing an integrated theory of production, distribution, values and prices.

Indeed, after examining various theoretical alternatives to neo-classical theory, in the last two chapters of Debunking Economics, Keen himself recognizes that “at present, however, the various non neo-classical schools of thought have no coherent theory of value as an alternative to the neoclassical school’s flawed subjective theory of value” (page 443). This is undoubtedly a big problem, because it is a “central organizing concept” (ibid.). Economic theory still needs therefore to be reconstructed, and as for any construction, this has to start from the very foundations of economic analysis. These foundations are the essential characteristics which constitute the logical fundaments of a theory, and therefore its main and primary assumptions, allowing to define its basic concepts. This is the reason why they are called principles. To try and define such fundamentals principles was therefore the main motivation for writing this book.

As for its epistemological bases, they rely on the idea that the world of experience exists independently of the theories that try to explain it, and that each theory is nothing else than a picture of the world. I refer here to what Wittgenstein wrote in his Tractatus Logico-Philosophicus: “a picture is a model of reality” (Wittgenstein, 1921, p. 9). Each theory has its own set of concepts, which constitute a paradigm, in the sense that Kuhn gave to this word, on second thoughts, i.e. “ ‘a disciplinary matrix’ - ‘disciplinary’ because it is the common possession of the practitioners of a professional discipline and ‘matrix’ because it is composed of ordered elements of various sorts” (Kuhn, 1974, p. 3). These elements are symbolic generalizations, models, and exemplars, the latter being defined as “concrete problem solutions, accepted by the group as, in a quite usual sense, paradigmatic” (Kuhn, 1974, p. 4).

A paradigm can be analyzed as a system of concepts, and in such a system a special emphasis must be placed on these particular concepts which are the basic and simple assumptions underlying the other elements of the system. These assumptions constitute by themselves a set of concepts which are explained and defined on the basis of the experience that we can derive from our perception of the real world. A concept is always an element which can be defined only through its relations within the system of concepts to which it belongs. But what makes the specificity of these assumptions that can be considered as principles is that they must relate to a context which is outside the theory per se, because it is the world of experience as we perceive it.

The role of these principles is indeed a fundamental one, since they define the nature and give its signification to the object which is built by a theory. They constitute indeed the semantics of a theory. A problem can therefore arise when some of these assumptions are implicit, and when they happen to be contradictory to other implicit or explicit assumptions or to the process of construction of some concepts. The view on which this book is based is that it is indeed what has befallen to the neo-classical theory: despite the fact that the syntax of this theory, under the form of an impressive mathematical apparatus, is generally correct, the loose definition of its principles has resulted in a semantics which is highly flawed, and made it unable to give a coherent description of the world as it is. This inherent shift from reality will ultimately lead to the loss of its scientific statute.

It is to contribute to this process that this book has been written and it is why it will attempt to rebuild economic theory from its main concepts and principles. If it will do so by trying to stick to the common perception that ordinary human beings can have of the real world, it will also try to use the teachings of various theories, considered as scientific ones, which have been developed in other unrelated fields of knowledge, such as the mathematical theory of measurement or the physical theory of dimensional analysis.

"Pure logical thinking cannot yield us any knowledge of the empirical world; all knowledge of reality starts from experience and ends in it. Propositions arrived at by purely logical means are completely empty as regards reality."

Albert Einstein, *Ideas and Opinions* (1933)

# CONTENTS

## About the author

Christian R. Flamant is a retired Economist. He graduated in 1970 from Science Po - Paris, and obtained a PhD in Economics in 1977 from the University of Paris 1 - Panthéon-Sorbonne. He was a Research and Teaching Assistant in Economics at the University of Paris 1 from 1972 until 1979, and a lecturer in National Accounting and Economic Analysis, for the last two years (1977-1979). In 1979 he joined the French Development Agency (AFD), where he worked for 32 years, until his retirement at the beginning of 2012. He held various positions in several countries, in particular as a country or regional director in Vanuatu, Angola, Ghana, Horn of Africa (Djibouti, Ethiopia, etc.) and Yemen. He was seconded to the IMF from 1982 until 1985, and to the OECD as a Principal Administrator in the Development Cooperation Directorate, from 1998 until 2001.

After retiring he went back to what had been his first professional interest: pure economic theory. He wrote two articles published in the World Economics Association journals:

- the first one on Piketty: "Again on Piketty's *Capital in the Twenty-First Century* or Why National Accounts and Three Simple Laws Should not be a Substitute for Economic Theory" in the *World Economic Review,* N°5 (July 2015);
- the second one on Sraffa: "Commodities do not produce commodities: A critical view of Sraffa's theory of production and prices" in the *Real World Economics Review*, issue N° 72 (30 September 2015).

He devoted the last three years to the continuation of a reflection that was initiated with his thesis on "Hypotheses and concepts in the theories of value, capital and distribution". More than forty years after his thesis had left some questions unanswered, he feels that he has at last succeeded in redefining the most basic concepts of production, values, distribution and prices, allowing to establish in this book a series of fundamental principles of political economy. His conviction is indeed that Economic Theory, even though it will always remain an incomplete science, needs to be rebuilt starting from its foundations, which these concepts and principles might now help to provide.

# LIST OF ABBREVIATIONS

*out of matrix equations

$a$ : share in value of fixed capital obtained by section I, producing fixed capital, over total fixed capital produced during a defined period of time

$\alpha = \frac{a}{1-a}$ : ratio of the share in value of fixed capital obtained by section I, and the share in value of fixed capital obtained by section II, producing consumption goods

$b$ : share in value of fixed capital obtained by section II, producing consumption goods, over total fixed capital ($b = 1-a$)

$c$ : share in value of capitalist consumption over total consumption

$c_I$ : share in value of consumption of capitalists of section I over total consumption

$c_{II}$ : share in value of consumption of capitalists of section II over total consumption
$c_I + c_{II} = c$

$C$ : total value of all consumption goods produced during a period

$C_c$ : total value of consumption goods obtained by capitalists during a period

$C_w$ : total value of consumption goods obtained by workers during a period ($C = C_c + C_w$)

$\gamma = \frac{c}{1-c}$ : ratio of the share in value of capitalist consumption over the share in value of workers consumption

$c*$ : a parameter setting the consumption of all capitalists of section I which can be obtained through selling fixed capital to capitalists of section II at a price without capitalist consumption multiplied by $\frac{1}{1-c*}$

$\gamma* = \frac{c*}{1-c*}$ : the corresponding markup (because $\frac{1}{1-c*} = 1 + \frac{c*}{1-c*}$)

$I$ : price of total investment during a period of time, equal to the price of fixed capital produced during the same period ($I = I_I + I_{II}$)

$I_I$ : price of investment in section I producing fixed capital, during a period

$I_{II}$ : price of investment in section II producing consumption goods, during a period

$k$ : primary rate of surplus-value in a model without capitalist consumption ( $k = \frac{L_I}{L_{II}}$ i.e. the ratio of value produced in section 1 to value produced in section 2)

$k_c$ : rate of surplus-value in the general case, with capitalist consumption ( $k_c = \frac{k+c}{1-c}$ )

$K$ : total value of fixed capital previously accumulated and in operation during a given period of time

$L$ : total value of all commodities produced in a given period of time (L = $L_I + L_{II}$ )

$L_I$ : value of all commodities produced in section 1, producing fixed capital

$L_{II}$ : value of all commodities produced in section 2, producing consumption goods

$p_i$ : price of commodity $i$

$P$ : total price of the product of a given period of time ( $P = P_I + P_{II}$ ), equivalent to $Y$

$P_I$ : price of the product of section I, equivalent to $Y_I$

$P_{II}$ : price of the product of section II, equivalent to $Y_{II}$

$\Pi$ : overall amount of monetary profits

$\Pi_I$ : amount of monetary profits in section I

$\Pi_{II}$ : amount of monetary profits in section II

$\pi$ : share of monetary profits in the price of total product ( $\pi = \frac{\Pi}{Y}$ )

$\pi_I$ : share of profits of section I in the price of product of section I ( $\pi_I = \frac{\Pi_I}{Y_I}$ )

$\pi_{II}$ : share of profits of section II in the price of product of section II ( $\pi_{II} = \frac{\Pi_{II}}{Y_{II}}$ )

$q$ : organic composition of capital, originally a marxist concept equal to $\frac{K}{V}$, the ratio of the value of total accumulated fixed capital over the value of labor-power, in a simple model without capitalist consumption

$q_c$ : organic composition of capital in a model with capitalist consumption

$r$ : average rate of profit:

- can be calculated in terms of prices or values,
- can be calculated either the Marxist way $\left( r = \frac{S}{K+V} = \frac{k}{q+1} \right)$, or according to the current definition, i.e. $r = \frac{S}{K}$ (in terms of values) or $r = \frac{\Pi}{It}$ (in terms of prices)

$r_c$ : average rate of profit in the general case with capitalist consumption

$r_{Ic}$ : average rate of profit in section I, in the general case with capitalist consumption

$r_{IIc}$ : average rate of profit in section II, in the general case with capitalist consumption

$S$ : total amount of surplus-value (in value terms)

$t$ : average lifespan of accumulated fixed capital

$u_i$ : fraction of available fixed capital obtained by capitalists of section I at the i$^{th}$ stage of purchases of fixed capital within this section

$v_i$ : value of commodity $i$

$V$ : total value of labor-power ($V = V_I + V_{II}$)

$V_I$ : value of labor-power in section I

$V_{II}$ : value of labor-power in section II

$\bar{w}$ : average monetary wage per unit of time, equivalent to total wages divided by the total number of (physical) hours of work

$W$ : total price of labor-power, equivalent to total money wages

$W_I$ : total price of labor-power in section I, equivalent to money wages paid in this section

$W_{II}$ : total price of labor-power in section II, equivalent to money wages paid in this section

$Y$ : total incomes distributed during a given period of time, equivalent to $P$, the price of the product ($Y = Y_I + Y_{II}$)

$Y_I$ : amount of incomes distributed in section I during a given period, equivalent to $P_I$

$Y_{II}$ : amount of incomes distributed in section II during a given period, equivalent to $P_{II}$

$\Psi$ : a parameter which is by definition the sum $1+\alpha+\gamma+\gamma^*+\alpha\gamma$

# LIST OF FIGURES AND TABLES

## LIST OF FIGURES

## LIST OF TABLES

# Chapter 1. Introduction: the basic concepts and principles. Commodities, Exchange and Money

This introductory chapter recalls what are the main steps of the approach pursued in this book, and how the theory that it develops is linked to its social background. It then shows why economic variables are dimensional variables, and explains what is dimensional analysis. At this preliminary stage, only the most basic variables are examined. These variables are commodities, exchange and money, and their dimensional nature has a bearing on their conceptual definition. This is in particular the case for the definition of money, which is closely related to the theory of measurement.

## 1. Main steps of the approach

In order to try and contribute to a reconstruction of economic theory, starting with its foundations, the book starts with this first introductory chapter devoted to the definition of the most basic economic variables: commodities, exchange and money. These are indeed the three most basic concepts corresponding to unavoidable realities with which most human beings have to deal with on a daily basis to survive in the real world. Defining these concepts will involve to state some basic principles, closely related to the world of experience.

To do so, we will have to recall first that these concepts cannot be defined in a vacuum, but against the background of this real world. We will have also to acknowledge that the corresponding variables are used in quantitative reasoning, which implies that they must be measured. But such a measurement is possible only to the extent that these variables have a dimension, in the scientific sense of the term. This will lead us to recall briefly the main elements of dimensional analysis. Similarly, we will show that the definition of money implies to rely on the theory of measurement.

The main concepts being defined, and since commodities have to be produced before being obtained and used by people who need them, the first and second part of the book will deal with the production process and the different roles played by different kinds of commodities in this process. The first part will show that this process is still wrongly analyzed by economic theory, and why theories of production based on the idea of production as surplus are necessarily a theoretical stalemate.

This will allow to show in the second part that once production is correctly defined as a transformation process, it becomes possible to address the question of the value of commodities, basing their equivalence at this theoretical stage.

The third part will deal with the question of exchange, considered as an exchange between commodities and money. This will be done from a macroeconomic point of view, simultaneously shedding light on the question of distribution and prices, considered as monetary prices.

Finally the findings achieved in the preceding parts will give us some tools to deal in the fourth part with the controversial question of the rate of profit and its evolution, which is closely related to the reproduction of the economic system.

## 2. The background

At the outset let it be clear that the theory exposed here is not background independent. This notion is well known in physics, where one can refer for instance to a 2005 article by Smolin: "the case for background independence", and in particular in general relativity, which is essentially background independent. However it is generally left aside in economic theory, which is part of social sciences, and where there can be no such thing as background independence, which simply means that the defining equations of the theory cannot be independent from the social context, i.e. the social and political organization of the society in which individuals enter into economic relations.

This social and political background is to some extent the framework of the disciplinary matrix that Kuhn, as mentioned above, considers as constitutive of a paradigm. This means that it is impossible not to refer to this specific social and political background because it plays a role in the definition and determination of some important variables of the theory. But the reverse is also true: the solutions of the equations of a theory may have an influence on social relationships, if they become a constituent of the ideology of a society.

This background against which a theory must be built and principles established is therefore in this book the social structure of the capitalist mode of production, since it is the dominant mode of production in the real world. Apart from a country like North Korea, or from isolated tribes in the Amazon forest or in Papua New Guinea, this capitalist mode of production prevails indeed everywhere in the world, even to a great extent in a so-called Communist country like China.

It is not the objective of this book to enter into a full and detailed description of this mode of production. Suffice it to say that it is characterized by the separation of individuals into two broad categories: those who own the means of production, who for convenience will be called capitalists, and the other ones, with no other property than their capacity to work, who will be called workers.

To be sure, in the real world a number of workers own some means of production, through the ownership of stocks of various companies or enterprises, and could be considered at the same time as both workers and capitalists, but this is not relevant for the present analysis. This only shows that we are here not in the real world, but in the theoretical world of language and concepts, aiming at providing a picture of the world, distinct from the real world and giving only an approximate vision of it. Obviously the objective pursued here is that these concepts make a picture which is as close as possible of what is empirically perceived as the real world.

## 3. Economic variables have a dimension

At this stage, and without going as far as claiming that there is no science if there is no measurement, one has to recognize that economic science deals with quantities and must therefore be able to perform quantitative analyses. This makes the question of measurement a central one for economic theory, and brings to the fore a theory which is rarely heard of in economic circles, because of its origins in Physics, and which is dimensional analysis. This analysis has to be used because economists, like physicists, deal with quantities, and because it is this analysis which has to be applied when dealing with quantities in a scientific way. Dimensional analysis is well explained in several books of Physics, including a short but excellent course which was taught by Ain N. Sonin at the MIT, under the title: "The physical basis of dimensional analysis" (Sonin, 2001, pp.1-57). Although it does not seem necessary to expose here the full theory in all its details, it seems appropriate just to recall a few of its findings that can be useful also for economists, through a less than one page recapitulation which can be found in section 2.8 of Sonin's course:

1. A *base quantity* is a property that is defined in physical terms by two operations: a comparison operation, and an addition operation. The comparison operation is a physical procedure for establishing whether two samples of the quantity are equal or unequal; the addition operation defines what is meant by the sum of two samples of that property.

2. Base quantities are properties for which the following concepts are defined in terms of physical operations: equality, addition, subtraction, multiplication by a pure number, and division by a pure number. Not defined in terms of physical operations are: product, ratio, power, and logarithmic, exponential, trigonometric and other special functions of physical quantities.

3. A base quantity can be measured in terms of an arbitrarily chosen *unit* of its own kind and a *numerical value*.

4. A *derived quantity* of the first kind is a product of various powers of numerical values of base quantities. A derived quantity is defined in terms of numerical value (which depends on base unit size) and does not necessarily have a tangible physical representation.

5. The *dimension* of any physical quantity, whether base or derived, is a formula that defines how the numerical value of the quantity changes when the base unit sizes are changed. The dimension of a quantity does not by itself provide any information on the quantity's intrinsic nature. The same quantity (e.g. force) may have different dimensions in different systems of units, and quantities that are clearly physically different (e.g. work and torque) may have the same dimension.

6. Relationships between physical quantities may be represented by mathematical relationships between their numerical values. A mathematical equation that correctly describes a physical relationship between quantities is *dimensionally homogeneous*. Such equations remain valid when base unit sizes are changed arbitrarily.

7. The categorization of physical quantities as either base or derived is to some extent arbitrary. If a particular base quantity turns out to be uniquely related to some other base quantities via some universal law, then we can, if we so desire, use the law to redefine that quantity as a derived quantity of the second kind whose magnitude depends on the units chosen for the others. All base quantities that are transformed into derived quantities in this way retain their original physical identities (i.e. their comparison and addition operations), but their numerical values are measured in terms of the remaining base quantities, either directly via a defining equation or indirectly by using a unit that is derivable from the remaining base units.

8. A system of units is defined by (a) the base quantities, (b) their units, and (c) the derived quantities, each with either its defining equation or the form of the physical law that has been used to cast the quantity into the derived category. Both the type and the number of base quantities are open to choice (Sonin, 2001, pp. 32-33).

Recalling these main elements of dimensional analysis provides us with a useful tool to now address the definition of our main concepts, starting with the concept of commodities.

## 4. Commodities

The first sentence of Capital by Karl Marx has not lost its actuality. It states: “The wealth of societies in which the capitalist mode of production prevails appears as an ‘immense collection of commodities’; the individual commodity appears as its elementary form” (Marx, 1867, p. 27). In fact this starting point is not truly different from that of Walras, who names his main book, published in 1874**: “**Elements of Pure Economics or the Theory of *Social Wealth”,* and for whom commodities are likewise the components of this wealth.

Over the whole spectrum of economic schools, it seems therefore that a consensus exists to consider commodities as a central concept of economic theory. This provides us with a good reason similarly to start this analysis from commodities, the question being then to build up a concept of commodity which meets the criteria for being considered as a scientific one.

A first observation that can be made about commodities is that commodities appear in vast quantities of various types and dimensions: some commodities can have their quantities expressed in terms of mass (in tons or kilograms), some others in terms of length (in meters or kilometers, or in terms of hours (in hours, days, or years). This brings us back to the relevance of dimensional analysis to address the question of this diversity of dimensions. It is therefore useful to recall one important consequence of the main rules of dimensional analysis exposed in the introduction, again on the basis of Sonin’s course:

“Suppose that, in a specified physical event, the numerical value $Q_o$ of a physical quantity is determined by the numerical values of a set $Q_1...Q_n$ of other physical quantities, that is,

$Q_o = f(Q_1, Q_2,..., Q_n),$

A physical equation must be dimensionally homogeneous. Dimensional homogeneity imposes the following constraints on any mathematical representation of a relationship like this equation:

(1) Both sides of the equation must have the same dimension;

(2) Wherever a sum of quantities appears in *f*, all the terms in the sum must have the same dimension;

(3) All arguments of any exponential, logarithmic, trigonometric or other special functions that appear in f must be dimensionless" (Sonin, 2001, p. 21).

It is my deep conviction that not only physical equations, but also economical equations must be homogeneous, meaning that economical quantities must have the same dimension, in order for these economical equations to be as logically meaningful as physical ones, and thus for economic theory to be eligible to a scientific status.

For economists commodities must indeed have similar characteristics which make it possible to consider them as measurable quantities and as such to make them comparable. The problem which cannot be ignored is that there are millions of different commodities, with very different properties. This fact of observation is of paramount importance but at the same time so obvious that it cannot in itself be considered as a primary and overarching principal. It is more an element of the background underlying all others. It seems indeed more important to focus on the elementary form of commodity, which makes it indispensable to define, as a starting point, what we intend as a commodity, i.e. what is the signification of this word.

It is a fact of experience: commodities are objects which have numerous properties, some of them being measurable, which allows to treat them as quantities, with each type of quantity having by definition its own dimension. For instance some types of objects can be characterized inter alia by their length, a spatial dimension, which means that the corresponding numerical value has the dimension of a length. But for other types the mass of the objects can be the main element (in so far as it is indispensable in their case to set their price), and the corresponding numerical value has the dimension of a mass. These objects may also have derived quantities, like an area, a volume, a density, which have no physical meaning, but are also dimensions. Commodities have simultaneously a number of other unmeasurable properties, like their color, or their shape, etc...From this simple observation we can therefore derive a first principle:

**Principle 1**: An elementary commodity is a multidimensional object

As stated above by Marx there are vast numbers of commodities, and it is obvious that the dimensions of commodities differ widely from one commodity to another, which implies as a corollary to this first principle that:

**Corollary to Principle 1**: Commodities are heterogeneous

It is because of this heterogeneity from one commodity to another that economic theory has been obsessed by the quest for a particular property of commodities which could provide the basis for describing the economic world in quantitative terms. This implies to find a property allowing to define otherwise heterogeneous commodities in terms of a common quantity. This would make it possible to decide whether two quantities of commodities $A$ and $B$ having this property are equivalent ($A = B$), or non-equivalent ($A \neq B$). Dimensional analysis tells us that there must also be an addition operation that defines what is meant by the sum $C = A + B$ of two quantities of commodities sharing this property. If indeed it is the case, dimensional analysis ensures us that "base quantities with the same comparison and addition operations are of the same kind (that is, different examples of the same quantity). The addition operation A + B defines a physical quantity C of the same kind as the quantities being added" (Sonin, 2001, p. 10).

One of the reasons for this quest for a common property is that economic theory has founded the equivalence of commodities in the exchange of these commodities, by considering that exchange is an equivalence relation. But before addressing the question of exchange, one must first raise the preliminary question of the existence of commodities. Indeed, and this is an obvious truth, commodities cannot be created ex nihilo, by spontaneous generation. Whatever the way they come into existence, commodities must pre-exist to exchange. The recognition of this fact leads to a kind of overarching principle:

**Principle 2**: Commodities must have been produced before being exchanged

## 5. Exchange

Economists have always viewed the exchange of two commodities as an exchange between two equivalent objects, which has led some of them to write an equation such as for instance:

2 m of fabric = 5 kg of wheat

To be sure, these economists were aware that such an equation is a pure nonsense, because of the different dimensions of the two objects on each side of the equation. They tried therefore to find out what would be this hidden common dimension, shared by both commodities, which would make the equation homogeneous and give it a logical sense. In other words they intended to replace this equation by another one, where on each side of the equation there would be two common and comparable magnitudes having exactly the same measurement. Economists call value this magnitude, so that instead of the previous equation, we would get:

value of 2 m of fabric = value of 5 kg of wheat

Such an equation has now a logical sense, provided that this particular magnitude is properly defined. The question which has therefore to be answered is: what is the common dimension of this magnitude named value, which makes different values of various heterogeneous commodities comparable, and therefore measurable. This question was relevant for classical economists as well as neo-classical economists, despite the different answers that they gave to it.

## 5.1. The neo-classical approach

At the beginning of the neo-classical theory, a first attempt had been made by Jevons, in his "theory of political economy" published three years before Walras's book, to identify the exchange relation, perceived as an exchange between quantities, to a ratio of utilities: "the ratio of exchange of any two commodities will be inversely as the final degrees of utilities of the quantities of commodities available for consumption after the exchange is effected" (Jevons, 1871, pp. 95-96). This is written in a symbolic form:

$\frac{\Phi(1)^a}{\Psi(1)^b} = \frac{y}{x} = m,$ with $a$ and $b$ two commodities, $x$ and $y$ their quantities and $\Phi$ and $\Psi$ their utilities.

But the utility given to one commodity is a subjective feeling which has no material existence, and exists only in the minds of individuals. Therefore this homogeneity concerns only one individual. Even if we suppose that a single individual can establish a kind of "utility scale" between various commodities, which is quite debatable, such an homogeneity cannot be demonstrated when there are many of them, each one with his own subjectivity and his own scale of utility. Among the infinite space of all individual subjectivities, no homogenization is possible. Therefore homogeneity is not demonstrated, but only postulated, which does not allow correctly to produce the concept of value, as the common dimension of commodities associated to the relation of equivalence of exchange and to its measurement.

A more sophisticated approach in the neo-classical tradition started with Léon Walras (op. cit.) and was based on the well-known parable according to which, at the beginning of the story, all individuals have some endowments of diverse quantities of various commodities, and meet on a marketplace. There, after a number of bargaining rounds, and with the assistance of some kind of "market commissioner", equilibrium of overall supplies and demands for all commodities is reached. This means that the exchange ratios between all commodities are determined in terms of one of them, which is taken as a standard of value and provides a unit of account. Then real exchanges can take place in accordance with these exchange ratios, and the market is cleared. These are the roots of the theory of general equilibrium.

This position has been proven wrong on purely logical grounds by a number of economists. A very good synthesis of all these criticisms is given in "Debunking Economics" the already mentioned book by Steve Keen. Let us just mention the circularity behind this kind of reasoning: prices are defined when equilibrium is reached, but in order to reach such an equilibrium prices must have been announced out of equilibrium at the beginning of the process.

## 5.2. The classical approach

The classical approach is somewhat different, because it does not seek in exchange as such the common characteristic which makes commodities equivalent, but goes upward to the production process of commodities to emphasize the fact that all of them have been produced by human labor, which later on at the stage of the exchange process appears on both sides of the equivalence relation. This is so because for classical economists in the production process labor has been embodied, or for Marx "crystalized" into commodities.

Such an approach deserves nevertheless the same kind of criticism as the neo-classical one, on several grounds. First, labor in itself has no material content. It is not a substance but an activity, which is as such immaterial. There is no real thing in labor which can be transferred or embodied, or crystalized into a commodity. Second, there are many types of labor, from manual labor, implying physical activities of many kinds, to intellectual labor, which requires almost no physical activity. This is what Marx tried to conceptualize through the distinction he made between simple labor and complex labor. Moreover labor can differ greatly from one trade to another.

To be sure, Marx was fully conscious of this problem of heterogeneity of labor and tried to overcome it by developing the concept of abstract labor, supposed to be a homogeneous type of labor. But the pages that he devoted to that are much more philosophical and metaphysical than theoretically satisfactory. They do not provide the key which would allow to transform two commodities into comparable quantities having a common dimension of abstract labor. At this stage, abstract labor remains therefore an abstraction with no tangible meaning.

Moreover, both schools of thought remain in fact theoretically stuck by a conception of exchange which does not allow for a correct introduction of the concept of money.

### 5.3. Exchange and money

Both schools have a conception of exchange as an equivalence relation between two commodities. But such a conception is also factually wrong, because it has no empirical foundation: exchanges as they take place in the real world are not exchanges between commodities, since they have nothing to do with barter. In the world as it is commodities do not exchange against commodities, but against money.

Furthermore many empirical researches show that, despite the fact that money has long been represented through some kinds of commodities like gold or silver coins, and although it can still be represented through some objects of a material nature, like notes or coins, money in the world in which we live is essentially of an immaterial nature, and has lost most of its material representations. If money is not a commodity, an exchange between a commodity and money cannot be thought of as an exchange between two commodities.

If classical as well as neoclassical economists have overlooked this fact and made the same mistake on this question, although they were aware of the existence of money as a medium of exchange, it is simply because they could not abandon the view, which appeared to them well founded empirically, that money was a commodity. This was indeed the case for Marx as well as for all of the neoclassical economists. This means for instance that to transform the previous "equation" supposed to represent an exchange between two commodities into an equation reflecting an exchange between one commodity and money, they would write:

2 m of fabric = 3 g of gold (if gold is the commodity-money)

Or to write it more logically:

Value of 2 m of fabric = value of 3 g of gold

In such a formulation, it is clear that an exchange with money is still equivalent to an exchange between two commodities, with the only difference that one of these two commodities has a special status as a standard of value, allowing it to play the role of a universal medium of exchange. But this does not prevent this standard of value from keeping the dimension of a commodity.

One of the reasons behind the fact that money is not a commodity lies in the nature of money, which has more to do with the existence of debt than with the weight of a gold or silver coin. The recent and fascinating book by anthropologist David Graeber: "Debt, the first 5000 years", provides us with a detailed historical account of this reality, showing in particular the very ancient linkages between debt and money. In his book Graeber sheds light on the way this relationship has emerged, much earlier than was previously thought, and has evolved over centuries in very different societies. He shows also that the nature of money as a general equivalent has always been disconnected from its material representation in various old or silver or bronze coins.

On a more theoretical ground, Keynes's "Treatise on Money", published six years before his "General Theory" and a little forgotten, has shown that a fundamental characteristic of money is liquidity. Finally, it is now widely recognized that in the real world money is created ex nihilo by commercial banks when they make new loans.

The question of the conceptual nature of money will be discussed later. Nevertheless all these observations, based on experience, are sufficient to enunciate two new principles:

**Principle 3**: Money is not a commodity

**Principle 4**: Commodities exchange against money and money exchanges against commodities: commodities do not exchange against commodities

At this juncture it is not enough to observe that the vast majority of commodities do not have their price set on a "market", on which, through the intervention of an auctioneer, and at the end of a kind of iterative convergence process, relative prices as well as monetary prices would be established. It has moreover to be recognized that for an exchange relation to be considered as a relation of equivalence, both quantities on each side of the equality have to be of the same dimension.

Therefore, if exchange - as it is the case - is a monetary one, this means that both quantities have to be expressed in money. In other words, when exchanges take place, commodities exchanged against money must already be measured in money, i.e. expressed in terms of their money value. This logical necessity is indeed what can be observed in the real world: commodities which are on sale have already a monetary price which has been established before commodities are made available to buyers, and buyers already possess money. This deserves to be formulated in terms of a new principle:

**Principle 5**: Commodities have a monetary price before exchanges take place

Where these monetary prices and money come from will be dealt with later in the course of this book, because the nature of money needs to be specified first in more details.

# 6. Money

The links between money and debt and money and liquidity have already been mentioned, but money is more often defined in the economic literature through its three main functions: money serves as a medium of exchange, as a store of value, and as a unit of account.

Of these three functions, it is the third one which is by far the most important at a conceptual level, because it is the logical precondition for the existence of the two other ones. Indeed money could not be used as a medium of exchange if it were not a unit of measurement, and could neither be used as a store of value if one could not measure the quantity of money which is stored.

The nature of a unit of measurement and of measurement is nevertheless a question which is rarely dealt with by economic theory. It is indeed traditionally the mathematical theory of measurement which addresses this issue, and it is more often associated to physics. Physicists, more than economists, are familiar with this theory. But this question is of such an importance, even for economic theory, that it seems quite useful to give now, if not an exhaustive presentation of the mathematical theory of measurement, at least some rather substantial explanations about its content.

## 6.1. The mathematical theory of measurement

On this technical question we rely on an old but rather simple presentation made by Doneddu, in a classic book on the basics of modern mathematical analysis, where chapter 3 deals with the "Measure of magnitudes" and is broadly used in the following pages (Doneddu, 1963, pp. 224-244). It is indeed a reliable presentation of this theory of measurement.

By definition a real measure (i.e. a measure by a real number) of a set of objects, named *E*, comes down to the establishment of an isomorphism between all the equivalence classes that can be defined on this set through a given equivalence relation and the additive semi-group of $R^+$, the sub-group of positive real numbers.

For such an isomorphism to exist, a certain number of assumptions have to be verified.

1. If one admits the existence of an operation of real measurement, it is necessary, within the set of all equivalence classes resulting from a given equivalence relation, to define an operation of addition, which to each couple of equivalence classes establishes a correspondence with a class named sum of the two first ones. It two elements belong to classes $p$ and $q$, the sum of these two elements represents class $p + q$, under the condition that we can define on set $E$ an addition which is associative, commutative and has a neutral element (the zero class).

It is considered that the set of equivalence classes is moreover totally ordered by the following relation: if element $p$ is smaller than element $q$, then $p \leq q$.

Introducing an addition which is associative, commutative and has a neutral element makes the set $E$ of equivalence classes a totally ordered commutative semi-group with a neutral element.

2. The existence of an isomorphism with $R+$ also implies the introduction of the axiom of Archimedes.

Let us indeed give the following definition of measurement:

*Definition 1*: To measure the elements of a set E consists in establishing for any element $x$ of $E$ a correspondence with a real number and only one, named "real measure of $x$", written $\mu(x)$, and such that for a non-zero element $u$ of $E$, chosen once and for all and named "measurement unit", the corresponding real number is 1, and also such that to the sum of two elements of $E$ corresponds the sum of their measures:

$$\mu (u) = 1$$
$$\mu (x + y) = \mu (x) + \mu (y)$$

This means that function $x \to \mu(x)$ defined on $E$ and with images in $R^+$ must be a homomorphy for addition.

The problem of measurement comes down to building this function $x \to \mu(x)$. It is easily solved for all the elements which are a multiple of the unit $u$. Solving the problem for all the elements of $E$ which are not a multiple of unit $u$ needs the introduction of the axiom of Archimedes: "whatever the non-zero elements $x$ and $y$ of $E$, there is a natural number $n$ such that $n.x > y$".

If E verifies this axiom, its elements are classes of Archimedean magnitudes. The set $E$ is then an Archimedean semi-group. If $E$ is an Archimedean semi-group then it can be shown that to any couple of elements $(x, y)$ $(x \neq 0)$ of set $E$ corresponds a natural number $q$, representing an equivalence class, and only one, such that :

$q\,x \leq y < (q + 1)\,x$, $q$ being the Euclidean quotient of $y$ by $x$

To find $q$ means to realize the Euclidean division of $y$ by $x$, with this division having a rest $r$ such that:

$[q\,x \leq y < (q + 1)\,x] \Longleftrightarrow [y = q\,x + r \text{ and } r \leq x]$

$q$ is only the approximated measurement of $x$ at the nearest whole.

The next step consists in going from the semi-Archimedean group isomorph to $Q^+$ (the positive rational numbers) to a set isomorph to $R^+$. Such an operation does not pose any particular problem, and therefore its demonstration will not be presented, in order not to unduly lengthen this section. It must just be recalled that it involves the introduction of the axiom of bisection and of the axiom of absence of gaps (this is Dedekind's method for the construction of real numbers).

It is now possible to draw from this construction of the operation of measurement, as it has been proposed, two important consequences:

*First consequence*: change of unit and stability domain of the equivalence relation of measurement.

The real measurement of a magnitude has been defined by assuming that the measurement of some particular magnitude *u* was 1. Any other magnitude is then measured in relation to *u*, by comparing it to a multiple or sub-multiple of *u*. Let us take v as a non-zero element of *E*. If we name $\mu_u(x)$ the measurement of *x* when *u* is taken as the unit, let us find the relation between $\mu_u(x)$ and $\mu_v(x)$.

We just have to consider the function $f(x) = \dfrac{\mu_u(x)}{\mu_u(v)}$ of *E* into $R^+$ (with $\mu_u(v) \neq 0$)

By operating in $R^+$, we have:

1) $f(v) = \dfrac{\mu_u(v)}{\mu_u(v)} = 1$

2) $f(x+y) = \dfrac{\mu_u(x+y)}{\mu_u(v)} = \dfrac{\mu_u(x)}{\mu_u(v)} + \dfrac{\mu_u(y)}{\mu_u(v)} = f(x) + f(y)$

$f(x)$ is therefore the measurement of *x* when *v* is taken as the unit : $f(x) = \mu_v(x)$,

Which implies: $\mu_v(x) = \dfrac{\mu_u(x)}{\mu_u(v)}$

We have therefore the following property: whatever the non-zero elements *u* and *v* of an Archimedean semi-group: $\mu_u(x) = \mu_u(v).\ \mu_v(x)$

This property is a characteristic of the measurement operation. It allows to define what is called the domain of stability of the equivalence relation. In the present case, i.e. the equivalence relation of exchange, the domain of stability is made of all of the systems where exchange takes place and which leave this relation invariant while maintaining the property which has just been defined : $\mu_u(x) = \mu_u(v).\ \mu_v(x)$

Since a measurement relates to a domain of stability, the definition of the conditions according to which a system constitutes a domain of stability is an important one in the context of a research focusing on the dimension associated to an equivalence relation. This definition will indeed imply to deal with the question of the standard of value corresponding to the measurement unit.

*Second consequence:* Influence of the construction of real measurement on the structure of the corresponding set of magnitudes.

Let us recall first that the construction of the set of magnitudes and the construction of the set of measurements of these magnitudes are simultaneous. The two sets have therefore the same mathematical structure, by construction. Let us recall also that to make the real measurement of a magnitude comes down to considering it as belonging to an additive semi-group isomorph to $R^+$. It derives from this that the set of magnitudes is complete, because it has been admitted, by definition, that every infinite suite of nested intervals included in E, and whose length tends to zero, never defines a gap.

If E is complete, then to each real number α corresponds one element y of E having α as its measurement when x is taken as unit.

$x \neq 0, \exists y \mid \{y = \alpha . x\} \Longleftrightarrow \mu_x(y) = \alpha$

It must be noted that this formula does not define the outer product of a number by an element of *E*, whatever the way it could be done. In particular the outer product of α by *x* only exists if the measurement of y when *x* is taken as unit is α. This outer product is only another way to say that α measures *y* in terms of *x*. In particular such a product only exists within the domain of stability of the equivalence relation.

This remark having been made, we can introduce a second definition:

*Definition 2:* To any couple α, *x* of a number $\alpha \in R^+$ and an element $x \in E$ ($x \neq 0$) it is possible to associate one element of *E*, named product of α by *x*, noted α *x*, and defined as the element of *E* having α as its measurement when *x* is chosen as the unit.

If $x = 0$, then $\alpha.0 = 0 \ \forall \alpha \in R^+$

From this definition three properties can be deducted:

*1st Property*:
$\forall\alpha, \forall\beta, \forall x \in E$ Assuming that $x \neq 0$ and taking *x* as the measurement unit:
$y = \alpha x \ \Leftrightarrow \ \alpha = \mu(y)$
$z = \beta x \ \Leftrightarrow \ \beta = \mu(z)$
Then: $\alpha + \beta = \mu(y) + \mu(z) = \mu(y + z)$
But $y + z = (\alpha + \beta)x$

Therefore: $\boxed{(\alpha + \beta).x = \alpha x + \beta x}$

*2nd Property*:

$\forall\alpha, \forall\beta, \forall x \in E$

Assuming that $x \neq 0$ and β ≠ 0, and that $y = \beta.x \neq 0$

Taking *x* or *y* as the unit and using the first property:

$\mu_x(z) = \mu_x(y).\mu_y(z) = \beta.\mu_y(z)$

With $z = \alpha.y$,

We obtain $\mu_x(z) = \alpha.\beta$

Therefore $z = (\alpha.\beta)x$

But $z = \alpha.y = \alpha(\beta.x)$

Therefore: $\boxed{\alpha.(\beta.x) = (\alpha.\beta).x}$

*3rd Property:*

$\forall\alpha \in R^+, \forall x, \forall y \in E$

Let us take any unit u:

$$\mu\left[\alpha\left(x+y\right)\right] = \alpha.\mu\left(x+y\right)$$
$$= \alpha\left[\mu(x)+\mu(y)\right]$$
$$= \alpha.\mu(x)+\alpha.\mu(y)$$
$$= \mu(\alpha.x)+\mu(\alpha.y)$$

Therefore: $\boxed{\alpha.(x + y) = \alpha x + \alpha y}$

It must be emphasized that these theorems are valid only for the domain of stability of the equivalence relation. But they are then fully valid.

Definition 2 introduces only a quite specific outer operation and does not come down, in general, to accept that one can multiply any scalar number by any magnitude. By definition the operation $y = \alpha.x$ is a notation for $\mu_x(y) = \alpha$, which is allowed because the objects to be measured are "infinitesimally adjustable". Besides, the demonstrations of the three properties show well that you always have to go back to the definition to grasp the sense of outer product $\alpha.y$. The application that defines real measurement: $x \to \mu(x)$ from *E* into $R^+$, has the following properties:

$\mu\ (x + y) = \mu\ (x) + \mu\ (y)$
$\mu\ (\alpha.y) = \alpha.\ \mu\ (y)$

So we can say that real measurement is pseudo-linear.

It follows from this whole demonstration that all the objects belonging to set *E* must have the same dimension, and that the theory of measurement is thus closely related to dimensional analysis. The mathematical theory of measurement having thus been fully exposed, it is also possible now to go further in the understanding of the nature of money.

## 6.2. The nature of money

From the definition of money as a unit of measurement most economists go one step further to define money as a measure of value, usually without providing any preliminary definition of the dimension of this value. However we know from principle 4 that money appears in exchange, and from principle 5 that exchange takes place when a quantity of money is equivalent, not to a value, but to the monetary price of a commodity. From this it follows that money is the unit of measurement of prices, and not of values. This also implies that the question of the nature of money must be coherent with the mathematical theory of measurement, which in turn necessarily constrains the answer that can be given to this question. Indeed, as a measurement, money cannot be anything else than a set of numbers belonging to the $R^+$ set. This provides us with the theoretical justification of the next principle:

**Principle 6**: As the measurement of a dimension of commodities which appears when it is exchanged against them and is called their price, money is a pure set of numbers belonging to the set of positive real numbers $R^+$. As such money has itself no dimension, and is composed of scalar numbers

Corollary to Principle 6: Monetary prices are therefore also dimensionless, scalar numbers

It results from this principle that money has no material content, which is perfectly coherent with the way it is created. In the world as it is, money is indeed created ex nihilo by banks. These economic agents are of such importance in present day capitalists economies, as the institutions responsible for money creation, that banks have to be added to the two categories of economic agents that were defined previously, those of workers and capitalists.

It is widely agreed as an empirical fact that money is created by banks, and this has a bearing on some other features of money which give more precisions about its nature, because money is created by banks when they make loans. This allows us to state that money is an "asset-liability", to use a term coined by Bernard Schmitt, a French theorist who developed the circuit theory of money on the basis of Keynes's theories, in his book "Money, wages and profits", published in 1966. He shows that money is more precisely a double asset and a double debt.

It is a double asset because as soon as it is created through the registration of a loan in the books of a bank, the bank bears a claim on the borrower to whom it has extended a loan. It appears in the bank books on the assets side of its balance sheet. This claim is redeemable at the time which has been agreed for repayment, and it will stay in the books up to this time. Simultaneously the amount of money corresponding to the loan is credited to the account of the borrower, where it appears as a liquid monetary asset for this borrower. At the same time, it is also a double debt, because it is registered as a demand deposit on the liabilities side of the bank balance sheet, and it is a debt for the borrower, which will stay as such up to the time when it has to be repaid. When this repayment occurs, all these four accounts are cleared, and money disappears. This brings us to the next principle:

**Principle 7**: Money is created ex nihilo by banks through credits that they provide to other economic agents, taking the form of a double asset and a double debt: money is cancelled by the repayment of these credits

The fact that money is created ex nihilo does not imply that its creation happens by chance. Money is closely linked to production, because one of the essential features of the capitalist mode of production is that wages are paid in money. If we assume that there is no money, and situate ourselves at the beginning of a new production process or period, then there are capitalists owning the means of production, and workers which need to be hired in order to start a new production cycle. These workers have to be paid in money, and capitalists have to turn to banks and to borrow from them in order to get this amount of money.

This observation constitutes the starting point of what is usually considered as the "monetary theory of production", which as recalled by Dillard in his 1980 article "A Monetary Theory of Production: Keynes and the Institutionalists", goes back to Keynes, and to the title of his course of lectures at Cambridge University at the autumn of 1932. It is important enough to be considered as a principle, being understood also that wages paid to a worker are not a pure amount of money, but an amount of money paid for a defined period of time:

**Principle 8**: Wages are paid in money and have the dimension of dimensionless numbers per unit of time

It must be noted that because of the nature of money as a measurement, the corresponding quantities that money measures cannot be themselves dimensionless: there is logically no such thing as the measurement of a measurement. The only operations that can be performed on measures are changes of unit, which are of no help to answer the question of the dimension of the magnitudes measured by money, i.e. the values of commodities. So far this question is still left unanswered. This is the reason why we must now turn to production, which might bring about a solution to this problem.

Production will be dealt with in two parts. The first one will explore theories of production based upon the notion of surplus, and will show that they constitute a theoretical dead-end. The second one will try and shed some light on the nature of production as a transformation process.

# PART I - PRODUCTION WITH A SURPLUS: A THEORETICAL DEAD-END

“Nothing is created, nor in the operations of art, nor in those of nature, and we may assume as a principle that in any operation there is an equal amount of matter before and after operation ; that the quality and quantity of the ingredients is the same, and that there are only changes, modifications". Antoine Lavoisier (1789), in his Elementary Treatise of Chemistry.

# Chapter 2. The Ricardian theory of production, distribution and value

To understand conceptual problems still left unresolved today, it is often very useful to go back to the great economists of the past, who grappled with these same problems. This is the reason why the Ricardian theory of production and distribution, closely linked to Ricardo's theory of value and prices, will be exposed and criticized in this chapter.

## 1. The Ricardian problematics and its basic data

### 1.1. The Ricardian problematics

David Ricardo was concerned by the future of the capitalist mode of production, which was just starting its prodigious development during his lifetime. In the first words of the preface to his major book "Principles of Political Economy and Taxation" published in 1817, he considered that: "the produce of the earth, all that is derived from its surface by the united application of labor, machinery and capital, is divided among three classes of the community; namely, the proprietor of the land, the owner of the stock or capital necessary for its cultivation, and the laborer by whose industry it is cultivated" (Ricardo, 2001 ed., p. 5).

A little further, he continued: "But in different stages of societies, the proportion of the whole produce of the earth which will be allotted to each of these classes, under the names of rent, profit, and wages, will be essentially different; depending mainly on the actual fertility of the soil, on the accumulation of capital and population, and on the skill, ingenuity and instruments that will be employed in agriculture" (Ricardo, 2001, p.5).

Ricardo was interested in the reproduction of the system, and hence by the share going to the owners of capital, because he had well perceived that the accumulation of capital realized from the profits would play a major role in this reproduction. This is why he added: "To determine the laws which regulate this distribution, is the principal problem in Political Economy" (Ricardo, 2001, p. 5).

### 1.2. The exogenous data

The exogenous data of the Ricardian theory are first the quantities of the various commodities which are produced and the techniques available, meaning the quantities of labor and means of production used in the production of each commodity. But it is also the level of real wages, i.e. the quantities of goods consumed by workers, which are fixed in the long run exogenously at the minimum of subsistence, by the law of Malthus. This law was indeed accepted as perfectly valid by Ricardo, himself a friend of Malthus.

Ricardo also considers fixed capital as a wages fund: a machine is but the materialization of the past spending of the wages fund, which has gone one step further in the production process. Profit is thus the difference between what is produced, on the one hand, and the current wages plus the wages fund necessary for this production, on the other hand: it is thus a residue.

Finally Ricardo uses the important assumption that the share of production going to each capitalist, divided by the amount of capital employed, must be equal for all of them, which implies that the rate of profit must be uniform in all branches.

It is on the basis of this problematics and of these data that Ricardo has devised successive solutions to the question of distribution that led him to view as the principal difficulty the question of value. These solutions will now be examined.

## 2. A first try to solve the distribution problem: "the Essay on Profits"

### 2.1. A mathematical presentation of the Essay on Profits

It is a well-accepted opinion in the economic literature since the Introduction of Sraffa to the Principles of Ricardo that within the framework of Ricardo's simple model of the "Essay on Profits" published in 1815, i.e. two years before the "Principles", he had succeeded in solving the problem of the determination of profits while avoiding the problem of valuation (Sraffa, 1951, pp. xiii- lxii).

In this model Ricardo assumes that the workers' wages consist only of corn, which is a generic term for wheat, since Ricardo makes all his calculations in terms of wheat. Therefore the only wage-good is wheat, and the production of wheat is realized with only labor and wheat, because the other means of production can also be represented by wheat, as a wage-good.

In the branch which produces wheat, the share of profits in the net product and its rate on the invested capital are consequently determined as a ratio of physical quantities as soon as the level of wages is set and as the conditions of production on the marginal land are given, which allows to get rid of the rent. With the assumption of the uniformity of the profit rate across the branches, through the moves of capital from one branch to another, we get the principle put forward by Ricardo that: "the general profits of stock being regulated by the profits made on the least profitable employment of capital in agriculture" (Ricardo, 1815, p. 13)

On the basis of these rather simplifying assumptions, which will be discussed below, the model allows for the determination, not only of the profit rate, but also of the exchange ratios of all the other commodities in terms of wheat, since the means of production of the other branches can always be expressed in terms of a quantity of wheat.

This can be easily demonstrated, taking the following notations:

$r$ is the profit rate
$L_i$ is the quantity of labor required for the production of the quantity $A_i$ of commodity $i$.

We designate by index $i$ = 1 the branch which produces wheat, the wage-good consumed by workers. The amount of wages advanced in this branch is equal to $L_i$ multiplied by the quantity $w$ of wheat paid to workers per unit of labor time, so that $L_i w = A_{i1}$.

We designate by $A_{ij}$ the quantity of commodity j used for the production of commodity $i$.

Then we note $a_{ij}$ the ratio $\frac{Aij}{Aj}$, with $\sum_{i=1}^{n} a_{ij} \leq 1 \ \ \forall j$

If we resize to one the quantities of the various commodities, and if we name $p_1, p_2, ...p_n,$ the price of the $n$ commodities, the system can be presented as follows :

$$(1) \quad (a_{11}p_1)(1+r) = 1$$

$$(2) \quad \begin{vmatrix} a_{21}p_1 + a_{22}p_2 + ... + a_{2n}p_n(1+r) = p_2 \\ ...................................................... \\ a_{n1}p_1 + a_{n2}p_2 + ... + a_{nn}p_n(1+r) = p_n \end{vmatrix}$$

In equation 1, the price of wheat is taken as the unit of measurement, equal to 1, corresponding to the overall quantity of wheat produced, which is the standard of value. Equation (1) of the system determines the profit rate and the system (2) of $n$-1 equations determines the $n$-1 prices of the commodities other than wheat in terms of the overall quantity of wheat produced, which corresponds to the measurement unit. In this system it is not possible for any commodity or basket of commodities other than wheat to play the role of standard of value, because otherwise the rate of profit could not be determined, and the system of equations (2) could not be solved. The prices, expressed in terms of the measurement unit (the number 1) are numbers which can only correspond to values which have the dimension of quantities of wheat, and nothing else. To demonstrate this assertion, let us transform the system into a matrix format.

Let $A$ be the matrix of technical coefficients and P the vector of prices. The system can then be written:

$$(3) \ (1 + r) A P = P$$

Which gives $[I - (1 + r) A] P = 0$

It is not possible to say that this system has a solution (positive prices and one unique positive profit rate) up to a scalar multiple. For the price vector obtained as a solution to be positive (i.e. $p_i \geq 0, \forall i$) it is indeed needed that matrix A be irreducible, according to the Perron-Frobenius theorem, which is demonstrated by Gantmacher, in his 1959 book "Applications of the Theory of Matrices" (pp. 64-65).

But matrix $A$ is reducible, because one of its lines has only one non-zero element. One could then get negative prices, which would be devoid of economic significance.

The system cannot therefore be written under this format. It is necessary first to eliminate from matrix *A* the first line, which gives the conditions of production of wheat. Let us call *A'* the matrix obtained by doing so and *P** the price vector ($p_2, p_3, ...p_n$).

The system can then be written:

$$(4)\quad \left| \begin{array}{l} a_{11} p_1 (1+r) = p_1 \\ (1+r) A' P = P^* \end{array} \right. .$$

But such a system has no solutions, because matrix *A'* is not a square matrix and vector *P* has not the same dimension as vector *P**. It is therefore necessary to eliminate one column of matrix *A'* by eliminating one price. The only price that can play this role is price $p_1$ because otherwise the two price vectors on each side of the equation would not be identical and the system would have no solution, since it would be written under the format:

$(1 + r) A'' P' + (1 + r) U = P^*$

Taking $p_1 = 1$, naming *C* (for corn!) the column vector ($a_{21}, a_{31}, ...a_{n1}$), and *A** the matrix resulting from the elimination of vector *C*, the system cannot but be written under the following format:

$$(5)\quad \left| \begin{array}{l} a_{11}(1+r) = 1 \\ (1+r) A^* P^* + (1+r) C = P^* \end{array} \right.$$

Therefore $P^* = \left[ I - (1+r) A^* \right]^{-1} \cdot (1+r) C$ (6)

Or: $P^* = \left[ I + (1+r) A^* + (1+r)^2 A^{*2} + ... + (1+r) A^{*n} \right] . (1+r) C$ (7)

If we develop this expression for any price, we get:

$$p_i = (1+r) a_{i1} + \sum_{k=2}^{n} (1+r)^k A_i^{*k} C \qquad (8)$$

Where $A_i^{*k}$ is the $i^{th}$ line of matrix $A^{*k}$

From equation (8) we can observe that a price $p_i$, although being a scalar number, can be analyzed as corresponding to the sum of the quantity of wheat employed directly in the production of commodity *i*, to which is added the profit of branch *i*, on the one hand, and the quantities of wheat employed indirectly in the production of intermediate goods used in the production of commodity *i*, to which is added the profit corresponding to the different stages, on the other hand. To choose the price of wheat as the measurement unit, since this measurement corresponds to a system with the dimension of wheat, comes down to choosing as a standard of value a pure quantity of wheat.

Although this simple model allows for the determination of the profit rate *r* without any need to determine the exchange ratios (the prices), it is clear that it allows nevertheless to calculate these exchange ratios in terms of wheat, the standard of value for the model. It is striking that the method used by Ricardo to determine the profit rate while avoiding the valuation problem allows however to determine the value of commodities, which has the dimension of a quantity of wheat! It is equally stunning to realize the correctness of Ricardo's findings, however within the theoretical boundaries corresponding to his peculiar and rather restrictive assumptions, without the assistance of the mathematical apparatus that we used.

A first conclusion is that the profit rate is determined exclusively in the sphere of production of the wage-good, wheat. And it is because the price of wheat is chosen as the measurement unit that a variation in the price of wheat cannot have any influence on the price of wheat, which is by definition invariable.

A variation in the profit rate can therefore come from only two causes: either a variation in the technical conditions of the production of wheat, i.e. in the number of workers necessary for the production of a given quantity of wheat on the marginal land; or a variation in the level of real-wage, i.e. the quantity of wheat advanced by worker. These two causes of variation in the profit rate can combine, even though it is always possible to determine precisely the exact influence of each of the two causes in this variation.

When the profit rate changes in the sector producing wheat, capital moves are supposed to ensure the transmission of this variation to the other sectors, where prices in terms of wheat adjust consequently.

When the only origin of this variation is a change in the technical conditions of wheat production, the only cause of the variation in the other commodities price is the change in the profit rate. When the modification of the profit rate comes from a change in the level of real wages, the variation in the other commodities price results not only from a change in the profit rate, but also from the change in the real wage of the workers employed in each of the corresponding branches (in other word vector C varies, in a homothetic way if real wage is the same in all of the branches). But it is always possible to isolate within the variation in prices the shares resulting from the modification in the profit rate and from the change in real-wage, respectively.

Finally the price of commodities may change only because of a modification in the technical conditions of production of anyone of these commodities (if it enters directly or indirectly into the production of all the other ones). The influence on prices of such a modification can also be isolated.

But the most important conclusion that one can derive from this model is certainly that it is always possible to compare the price of commodities when they change, whatever the cause of this change: a change in either the profit rate, or the level of the real-wage, or the conditions of production. Values are indeed nothing else than quantities of wheat, and it is always possible to compare a quantity of wheat to another one. It is similarly always possible to measure the evolution of the amount and share of wages or the amount and share of profits in terms of wheat in the product when the system itself changes, or/and to compare the whole product in terms of wheat during a period to the product in terms of wheat in another period. A given quantity of wheat is indeed in this model an invariable standard of value, corresponding to the measurement unit (the number 1) of the prices, which are scalar numbers.

One might nevertheless consider as superfluous to explore the properties of such an old and simplified model, dating back two hundred years ago. In fact it is not so, because this model gives us a good hint on what an economic system with money as a commodity would look like. It suffices indeed to replace wheat by gold, to say that gold is the money-commodity and that wages are paid in money (gold) to be transported into a world looking to some extent like the world of the eighteenth century!

## 2.2. The underlying conception of production and distribution

At the time of Ricardo, agriculture was still the most important sector, employing the majority of population, and for Ricardo production in agriculture depended on the fertility of the land. This fertility resulted itself from the laws of nature, giving more or less "productive power" to different types of land. Agricultural production resulted in the creation of a surplus, the size of which depended on the level of fertility of the different sorts of lands under cultivation.

In the "Essay on Profits" this fertility can be measured unequivocally, as a ratio of two quantities of wheat: wheat produced over wheat used in the process of production. This can be done independently of the price system, which is not needed to define fertility. Production in other branches is seen by analogy.

If we consider now that wages are part of the surplus, in each of the *j* branches there is at least the surplus needed to cover the cost of the wage-good (wheat) paid to their workers. This is the case when $\sum_{j=1}^{n} a_{1j} = 1$. There is also the possibility of an additional surplus for each of the commodities *i* for which $\sum_{j=1}^{n} a_{ij} < 1$.

In this model the distribution of the product among the three classes identified by Ricardo has nothing to do with the social or political background, or with the will of men. Indeed the rent of land comes from a purely natural phenomenon, the difference in the fertility of various types of land, and from the growth of population, which obliges to put under cultivation new lands of decreasing fertility, with the least fertile land paying no rent. As Ricardo writes: "Rent then is in all cases a portion of the profits previously obtained on the land. It is never a new creation of revenue, but always part of a revenue already created" (Ricardo, 1815, p.18).

The level of real wages is also depending on the growth of population, which keeps it at a level that is just sufficient to allow for the subsistence of laborers, in accordance with Malthus's law.

Finally the profit rate is determined in agriculture. Profit is the difference between the product of the least fertile land and the amount of wages in terms of wheat paid for the labor required, directly or indirectly, by the existing technique of production. The profit rate is the ratio of this last quantity and this same amount of wages. Once this rate is known, it is easy to compute the amount of profits in agriculture, given the fertility of the lands, first, and finally in all the other branches. Distribution is therefore a kind of natural phenomenon, which regulates the shares of each class in the output of the system, i.e. its net product.

### 2.3. A highly questionable model

A first and obvious remark can easily be made regarding the assumptions of this model. It is that they violate two of the principles that we already established in the previous chapter. First, they contradict principle 3, according to which money is not a commodity, because the standard of value here is a commodity, in this case wheat. And second, as a consequence, they also contradict principle 8, according to which wages are paid in money, as we defined it, because wages in this model are real wages, not paid in money, but in wheat.

This model relies indeed on very simplifying assumptions, according to which the only wage-good employed in the production of corn is corn itself, and that corn is produced only from itself and from labor. This is indeed a matter for criticism, which will now be discussed.

The fact is that Ricardo isolates the sector producing wheat, which is the only sector whose production enters in the production process not only of the wheat sector itself, but also of all the other sectors, without any exception, because wages in all of these other sectors are paid in wheat. Conversely the production of these other sectors does not enter at all into the production process of wheat.

It is this very peculiar feature which allows for the determination of the profit rate in the wheat sector, where the surplus is the difference between the amount of wheat produced and the amount of wheat employed in the production process of wheat, among others to pay for wages. It is the reason why the profit rate is determined independently of prices.

However, outside of the wheat sector and at the level of the system as a whole, there is absolutely no reason why there should necessarily be a surplus of wheat, which must be used as an intermediate good in some sectors, and not only for the payment of wages. There will be an overall surplus of wheat only when $\sum_{j=1}^{n} a_{1j} < 1$.

But then, if it is the case, such a surplus implies that wheat will be also a consumption good, which is quite a logical problem, because it is difficult to understand why the same good, wheat, can be an intermediate good when it is consumed by workers, and a consumption good when it is consumed by the other classes of society: the owners of the lands and the owners of capital.

This difficulty can be resolved only if we admit that, as a pure consumption good, wheat is not produced by the agricultural sector as such, in the raw form it has when it has just been harvested, but rather by the agribusiness sector, where it is processed to be made edible for consumers. In this case there is no longer a surplus of raw wheat, because all the raw wheat produced is used either to pay wages or as an intermediate good transformed in other sectors.

Moreover edible wheat is then produced in a separate production process and is necessarily different from unprocessed wheat: it is another good, different from raw wheat. And since no one can imagine that workers should consume unprocessed, non-edible wheat, real wages paid in a genuine consumption good, i.e. processed wheat, imply that other goods than raw wheat must appear on the left side of the equation in the sector producing wheat. This would completely remove the possibility to determine the profit rate in the wheat sector alone, and in fact would invalidate the whole model!

But the same analysis can be made as soon as we consider that there are obviously other inputs than wheat (be it processed or unprocessed) in the sector producing wheat. Then it is no longer possible to determine the profit rate as a pure ratio of quantities. The price system must be determined otherwise, even though, as we shall see in the next section, it can still be possible to determine the profit rate before the price system. All this shows thus that this first Ricardian model is too simple to be considered as an appropriate image of the real world.

Another kind of criticism can be made as regards the conception of production as a technical or physical process, in particular in the wheat sector as such, where there is necessarily a big surplus, because a limited quantity of wheat is used to produce a certainly much bigger quantity, which will be used not only in the wheat sector itself, but also in all of the other sectors, if only to pay for wages. It is also this surplus, once realized through the sale of wheat, which in the wheat sector will take the form of profits alone, on the marginal land, and of profits and rents on the other, more fertile lands.

This conception of production as the creation of a surplus, inherited from the Physiocrats, does not seem to be a problem for Ricardo, because he perceives this phenomenon like a farmer of his time who sows one quintal of wheat seeds per hectare and harvests 8 quintals of wheat. The surplus of 7 quintals is attributed to the productive power of land, its fertility. Production can therefore be conceived, quite naively, as a kind of spontaneous generation process. However, in the physical world where cultivation takes place, there is no such thing as spontaneous generation. At the time when Ricardo wrote his Essay on Profits, Lavoisier had already discovered the principle of mass conservation, which implies that mass can neither be created nor destroyed, although it may be rearranged in time and space, or the elements associated with it may be changed in form.

In the case of wheat production Lavoisier's principle implies that the mass of the wheat harvested, and its difference with the mass of wheat sowed, has to come from physical elements other than seeds, brought into the system by other sources. Agronomists now know what these elements are and also from which sources they come from.

One of these elements is water, in the form of rain or irrigation, which provides hydrogen and oxygen. Another one is air, which provides carbon dioxide and through the role of chlorophyll allows the synthesis of sugars. Legumes directly absorb nitrogen from the air in the soil, with the help of bacteria in their roots. For all of the other plants, like wheat, nitrogen is removed from the ground almost entirely in the form of nitrates. This brings us to this other source which is the ground itself, which acts as a substratum, and also provides multiple elements: at the height of its growth, one hectare of wheat absorbs 2 kg daily of nitrogen, 6 kg of potassium, and 1 kg of phosphorus, as well as sulfur, calcium, magnesium and trace-elements.

All this shows well that from a physical point of view there is therefore no such thing as the spontaneous generation of a surplus. One might think nevertheless that there can be an economical surplus, owing to the fact that a number of these elements are free and not appropriated. Therefore they have no price (or a price equal to zero), and do not have to enter in the economical production process. This is indeed true, but we will show later that for other theoretical reasons the notion of surplus has to be rejected.

# 3. The problem of value and prices for Ricardo

## 3.1. The problem in terms of labor values

For Ricardo, profit is determined as a ratio between production and the consumption necessary for such a production. In the model examined in the last section there was a branch where this ratio had the form $\frac{A - A_{11}}{A_{11}}$ and was thus perfectly determined.

Let us abandon now the assumption that workers' wages are constituted by only one commodity. Let us assume for instance that workers receive in exchange for their labor not only a given quantity of wheat (produced by branch 1) but also a given quantity of woolen cloth (produced by branch 2). Even if we imagine a branch *i* where production is realized only with labor, to the exclusion of any other means of production, the ratio between the surplus obtained in this branch and the consumption (under the form of real wages) needed for this production now takes the form $\frac{A_i - (A_{i1} + A_{i2})}{A_{i1} + A_{i2}}$. The quantities in the numerator and denominator are not homogeneous (they do not have the same dimension). The profit rate therefore cannot be determined in such a way.

The situation remains obviously the same if we examine the whole system with n branches, where the profit rate is then: $r = \frac{\sum_{i=1}^{n} A_{ij} - \sum_{ij=1}^{n} A_{ij}}{\sum_{ij=1}^{n} A_{ij}}$.

The determination of the profit rate implies that the numerator and denominator of the ratio be made homogeneous, i.e. that the quantities of commodities that appear here be expressed in a common measurement unit. Since profit is realized through the exchange of commodities, this measurement unit must be for Ricardo that of the exchange ratios of commodities, and must also be such that these ratios ensure the equalization of profit rates among the different branches. The determination of the profit rate requires therefore a theory of measurement, which in turn implies a theory of value, as the dimension to be measured.

It is not simple to explain how Ricardo attempted to solve this problem, since he did not succeed in doing so in a satisfactory way. One can indeed find, inextricably mingled in his works, a theory of measurement of the exchange ratios of commodities through the labor time, direct or indirect, necessary to produce them, and a theory of measurement through the prices of production, defined as the prices that ensure the uniformity of profit rates.

Two texts, apart from the first chapter of Ricardo's "Principles" and from some letters in the correspondence, allow nevertheless to shed some light on his attitude when confronted to this question. These texts are the two successive versions of the study entitled "Absolute value and exchangeable value", written in the very last weeks preceding Ricardo's death, on September 11, 1823 (Ricardo, 1823, pp. 361-397 and pp. 397-412).

What appears fundamental to Ricardo from these texts is what he names the absolute value of commodities, i.e. the measurement of the values of commodities through a measurement unit which has to be invariable in space and even more in time, or whose variations can be referred to an invariable standard. The determination of such an absolute value seems to him necessary to understand the variation of the exchange ratio of a commodity with another one, which he names the exchange value. One has indeed to know in which one of these two commodities lies the cause of this variation.

It is on this theoretical basis that Ricardo considered since the "Principles" that the absolute value of a commodity had to reflect its production cost, which in turn had to be understood as the quantity of labor directly or indirectly necessary to produce it. This position seemed indeed to give him an invariable measurement of value:

"It has been said that we are not without a standard in nature to which we may refer for the correction of errors and deviations in our measure of value, in the same way as in the other measures which I have noticed, and that such standard is to be found in the labor of men. The average strength of a thousand or of ten thousand men it is asserted is always nearly the same, why then not make the labor of man the unit or standard measure of value? If we are in possession of any commodity which requires always the same quantity of labor to produce it that commodity must be of uniform value, and is eminently well qualified to measure the value of all other things. And if we are not in possession of any such commodity, we are still not destitute of the means of accurately measuring the absolute value of other things, because by correcting our measure, and making allowance for the greater or less quantity of labor necessary to produce it we have always the means of referring every commodity whose value we wish to measure to an unerring and invariable standard" (Ricardo, later version, unfinished, 1823, pp. 401- 402).

Let us forget, at least temporarily, what has been said in chapter one on the heterogeneity of labor, and suppose that this problem has been solved. The insoluble problem which Ricardo nevertheless had to face was to attempt to introduce the profit rate as such within the system of absolute values determined by quantities of incorporated labor.

To show it let us use again the notation of the preceding sections. Moreover let us call $v_i (i = 1,...,n)$ the absolute value of commodities $i$ in terms of labor quantities. Commodities designated by $i = 1,...,k$ are wage goods.

Ricardo then defines the profit rate as the ratio between the value of the surplus and the value of wage goods:

$$r = \frac{\sum_{i=1}^{n} v_i A_i - \sum_{i=1}^{k} v_i A_i}{\sum_{i=1}^{k} v_i A_i}$$

If we introduce the profit rate such as just defined in the calculation of values, the system becomes the following, using matrix notation:

With $V = (V_i)$ $i = 1,..., n$
$L = (L_i)$ $i = 1,..., n$

$$(1+r)Av + (1+r)L = V$$
$$\Rightarrow [I - (1+r)A]V = (1+r)L$$
$$\Rightarrow V = [I - (1+r)A]^{-1}(1+r)L$$

The value of a commodity *i* is then:

$$v_i = (1+r)L_i + \sum_{k=2}^{n} (1+r)^k A_i^k L \qquad \text{where } A_i^k = i^{th} \text{ line of matrix } A^k$$

If we develop this last expression, attributing the *k* index to the quantity of labor used during the $k^{th}$ "period" ( $L_{ik} = A_i^k L$ ), we get:

$$v_i = (1+r)L_{i1} + (1+r)^2 L_{i2} + .... + (1+r)^n L_{in}$$

This expression allows us to reach the same conclusions as those that Ricardo himself drew from his assumptions. So defined, the value of commodities depends on the profit perceived on the means of productions employed, whose amount varies with the relative importance of these means of production: "every alteration in the permanent rate of profits would have some effect on the value of all these goods, independently of any alteration in the quantity of labor employed on their production" (Ricardo, 1817, p. 33).

Then, when the profit rate varies, the value of commodities varies whereas the quantity of labor has not varied. Consequently, the absolute value of commodities is equal to the quantity of labor required to produce them only when this rate is equal to zero. As Ricardo also writes: "If all commodities were produced by labor employed only for one day there could be no such thing as profits for there would be no capital employed, beyond that of which every laborer is in possession before he commences to work - there could as we have seen be no variation in the value of labor, but commodities would vary as labor was more or less productive" (Ricardo, 1823 – rough draft, p. 365).

Let us consider the exchange ratio between two commodities, *i* and *j*. It is expressed as follows:

$$\frac{v_i}{v_j} = \frac{L_{i1}(1+r) + L_{i2}(1+r)^2 + .... + L_{in}(1+r)^n}{L_{j1}(1+r) + L_{j2}(1+r)^2 + .... + L_{jn}(1+r)^n}$$

This exchange ratio will not be equal to the ratio of the quantities of labor incorporated in each commodity, except in two cases:

- Firstly when the profit rate is zero, whatever the conditions of production of the two commodities,
- Secondly, when the conditions of production of the two commodities are identical, whatever the profit rate, which implies that we have:

$$\frac{L_{i1}}{L_{j1}} = \frac{L_{i2}}{L_{j2}} = \cdots = \frac{L_{in}}{L_{jn}}$$

When this condition is not met, the exchange ratio will be different from the ratio of the quantities of labor incorporated in the compared commodities. Ricardo was aware of that, since he wrote: "The difficulty then under which we labor in finding a measure of value applicable to all commodities proceeds from the variety of circumstances under which commodities are actually produced" (Ricardo, 1823 – rough draft, p. 368).

The exchange ratio might be equal to the ratio of quantities of incorporated labor only if profit rates in branches *i* and *j* were different and in a ratio that would authorize this equality, with $r_i$ and $r_j$ such as:

$$\frac{L_{i1}(1+r_i) + L_{i2}(1+r_i)^2 + .... + L_{in}(1+r_i)^n}{L_{j1}(1+r_j) + L_{j2}(1+r_j)^2 + .... + L_{jn}(1+r_j)^n} = \frac{L_i}{L_j}$$

With $L_i = L_{i1} + L_{i2} + .... + L_{in}$

And $L_j = L_{j1} + L_{j2} + .... + L_{jn}$

But such a case obviously contradicts completely the basic assumption of the uniformity of profit rates, which is for Ricardo a fundamental feature of capitalist competition. The conclusion that Ricardo drew from his analysis was therefore that labor did not provide an invariable standard for the measurement of value:

"It must then be confessed that there is no such thing in nature as a perfect measure of value, and that all that is left to the Political Economist is to admit that the great cause of the variation of commodities is the greater or less quantity of labor that may be necessary to produce them, but that there is also another though much less powerful cause of their variation which arises from the different proportions in which finished commodities may be distributed between master and workman in consequence of either the amended or deteriorated condition of the laborer, or of the greater difficulty or facility of producing the necessaries essential to his subsistence" (Ricardo, 1823 – later version, p. 404).

In fact, at the root of Ricardo's problems there is a definition of the profit rate which is uncoherent, as soon as it is introduced in the calculation of values in terms of incorporated labor.

Indeed the formula $r = \frac{\sum_{i=1}^{n} V_i A_i}{\sum_{i=1}^{k} V_i A_i} - 1$ does not allow for the determination of the profit rate, as soon as these values $v_i$ are themselves depending on the profit rate. But in the opposite case where values are defined exclusively in terms of incorporated quantities of labor the calculation of the profit rate becomes contradictory with the hypothesis of uniformity of the profit rate and has no significance, because as we just shew there are as many different profit rates as there are commodities produced in different conditions.

### 3.2. The same problem in terms of production prices

What happens now if we reject the measurement of values in terms of incorporated labor? Is it possible to determine the profit rate in a coherent way, if we interpret the Ricardian analysis in terms of production prices, defined as the exchange ratios which ensure the equalization of the profit rates among branches? In other words, if we define now the profit rate by the following expression:

$$r = \frac{\sum_{i=1}^{n} p_i A_i}{\sum_{i=1}^{k} p_i A_i} - 1, \text{ with } p_i = \text{production price of commodity } i$$

Will we not meet the same contradiction as when we interpret it in terms of labor values?

A whole school of economic theory and in particular Von Thunen, Bohm-Bawerk and Walras, answered yes to this question. For more detailed comments on this point, one can refer to Dobb's article "The SRAFFA system and critique of the neo-classical theory of distribution" (Dobb, 1980, pp. 347-362).

Among others, Walras thus accused Ricardo of having one equation to determine two unknowns in affirming that the price is determined by the sum of profits and wages and that profits are determined by the difference between the price of commodities produced and the amount of wages.

However, as soon as 1898, an early Russian mathematician economist named Dmitriev, in his "Economic essays on value, competition and utility" did justice to these criticisms and provided the demonstration that the profit rate can be determined within the framework of the Ricardian model interpreted in terms of production prices (Dmitriev, 1974).

To show it let us take the preceding notation. Moreover, since we abandon the measurement through the quantities of incorporated labor, and given the conception of wages adopted by Ricardo, labor must be replaced by the quantities of wage-goods paid to the workers.

The equations which determine the production prices are therefore the following:

$$\begin{aligned}
&(a_{11}p_1 + a_{12}p_2 + .... + a_{1n}p_n)(1+r) = p_1 \\
&(a_{21}p_1 + a_{22}p_2 + .... + a_{2n}p_n)(1+r) = p_2 \\
&\text{.........................................................} \\
&(a_{n1}p_1 + a_{n2}p_2 + .... + a_{nn}p_n)(1+r) = p_n
\end{aligned}$$

This system, once transcribed in matrix form, is represented by the equation:

$$(1+r)AP = P$$

For mathematical reasons already exposed this system only provides significant solutions, from an economical point of view, when matrix A is irreducible. This requirement translates at an economical level in the fact that each commodity has to enter directly or indirectly in the production of all of the other ones. But for Ricardo only wage-goods meet this condition, since we can always resolve the production of a good to an expense of labor during successive periods. It must be noted however that in fact the expression "wage-goods" has in this model a broader meaning than the one admitted by Ricardo, because it encompasses not only the goods consumed by workers, but also the goods that enter directly or indirectly in their production.

Let us continue to consider that commodities designated by $i = 1,...,k$ are wage goods (in this broader sense). From a mathematical point of view, this means that matrix A can always be reduced in the following way:

$$(1+r)\begin{bmatrix}
a_{11} & a_{12} & \cdots & a_{1k} & 0 & 0 & \cdots & 0 \\
a_{21} & a_{22} & \cdots & a_{2k} & 0 & 0 & \cdots & 0 \\
\cdots & \cdots & \cdots & \cdots & \cdots & \cdots & \cdots & 0 \\
a_{k1} & a_{k2} & \cdots & a_{kk} & 0 & 0 & \cdots & 0 \\
a_{k+1,1} & a_{k+1,2} & \cdots & a_{k+1,k} & a_{k+1,k+1} & a_{k+1,k+2} & \cdots & a_{k+1,n} \\
a_{k+2,1} & a_{k+2,2} & \cdots & a_{k+2,k} & a_{k+2,k+1} & a_{k+2,k+2} & \cdots & a_{k+2,n} \\
\cdots & \cdots & \cdots & \cdots & \cdots & \cdots & \cdots & \cdots \\
a_{n,1} & a_{n,2} & \cdots & a_{n,k} & a_{n,k+1} & a_{n,k+2} & \cdots & a_{n,n}
\end{bmatrix}
\begin{bmatrix} p_1 \\ p_2 \\ \cdots \\ p_k \\ p_{k+1} \\ p_{k+2} \\ \cdots \\ p_n \end{bmatrix}
=
\begin{bmatrix} p_1 \\ p_2 \\ \cdots \\ p_k \\ p_{k+1} \\ p_{k+2} \\ \cdots \\ p_n \end{bmatrix}$$

In matrix notation, this translates into the following system:

$$(1+r)\begin{bmatrix} A_1 & 0 \\ W & A_2 \end{bmatrix}\begin{bmatrix} P_1 \\ P_2 \end{bmatrix} = \begin{bmatrix} P_1 \\ P_2 \end{bmatrix}$$

We have called 0, $A_1$, $W$ and $A_2$ each of the different submatrices. Submatrix $A_1$ means that the $k$ wage-goods enter in their own production. Since wage-goods constitute some kind of basket of goods which are all needed for the subsistence of workers, no element of this submatrix should be equal to zero. The existence of submatrix 0 reflects the fact that the production of the $k$ wage-goods requires the use of wage-goods, but do not require any input from the $n$ - $k$ sectors producing other goods. These other goods are not consumed by the workers, and are therefore considered by Ricardo as "luxury goods". Submatrix $W$ means that the production of luxury goods obviously requires wage-goods. Submatrix $A_2$ means that luxury goods can enter in their own production. $P_1$ is the price vector of wage goods and $P_2$ the price vector of "luxury goods".

Under this format, such a system has no economically significant solutions, because matrix A is reducible, and it must therefore be reduced into two sub-systems:

(1) $(1+r)A_1P_1 = P_1$

(2) $(1+r)WP_1 + (1+r)A_2P_2 = P_2$

Sub-matrix $A_1$ is an irreducible square matrix (of order k). Matrix $A_2$ is also a square matrix of dimension $(n-k, n-k)$. The two other matrices have no reason to be square matrices, and are rectangular ones: matrix $W$ is of dimension $(n-k, k)$, and matrix 0 is of dimension $(k, n-k)$.

For this reason only the first matrix equation (1), which contains the conditions of production of the wage-goods and is equivalent to $A_1P_1 = \frac{1}{1+r}P_1$, allows the calculation of the profit rate.

The Perron-Frobenius theorem indeed ensures us that the only eigenvalue of an irreducible matrix with non–negative terms to which corresponds a positive eigenvector is the dominant eigenvalue of this matrix, i.e. the highest one.

If we call $\alpha$ the dominant eigenvalue of matrix $A_1$, equation (1) is equivalent to:

$A_1P_1 = \alpha P_1$, with $\alpha = \frac{1}{1+r}$, and this implies that $r = \frac{1-\alpha}{\alpha}$

The profit rate is thus directly deducted from the dominant eigenvalue $\alpha$ of matrix $A_1$, and vector $P_1$ is itself the eigenvector associated to this eigenvalue. Moreover we know that the eigenvalue of a matrix is determined independently of the eigenvector, as a solution to the equation $(A_1 - \alpha I) = 0$ [1].

[1] Indeed $A_1P_1 = \alpha P_1 \Rightarrow A_1P_1 = \alpha IP_1 \Rightarrow (A_1 - \alpha I)P_1 = 0$

We arrive again to the same conclusion as that which Ricardo never ceased to uphold since the “Essay on Profits”, according to which the profit rate is determined exclusively by the conditions of production of wage-goods, independently of their price. This means that in such a system the production prices and the profit rate are not determined simultaneously: the profit rate is determined first.

Once the profit rate is determined the eigenvector $P_1$ of matrix $A_1$ can be determined, up to a scalar multiple, and the price of wage-goods is therefore also determined, which comes down to adopting as a measurement unit the price of anyone of these goods.

Once we know the profit rate and the price of wage-goods, we can also determine the price of the other goods, as a solution to equation (2), such as:

$$P_2 = \left[I - (1+r)A_2\right]^{-1}(1+r)WP_1$$

For a good *i*, this price is expressed as:

$$P_i = (1+r)W_iP_1 + \sum_{k=2}^{n}(1+r)^k A_{2,i}^k W_i P_1 \qquad \text{with } A_{2i}^k = i^{th} \text{ line of matrix } A_2^k$$

It should be no surprise that we obtain again the result of the “Essay on Profits” according to which the price of “luxury goods” can be expressed solely in terms of the price of wage-goods and the profit rate. When the whole system of production prices is known it is then possible to determine the amount and the share of wages, as well as the amount and the share of profits.

Moreover, we can see that a production price can be identified to a pure production cost only for luxury goods. This cannot be the case for wage-goods, because the price of anyone of these wage-goods depends not only on the proportion in which the other ones enter into its own production, but also on the proportion in which it enters itself in the production of the other goods. In passing we also see that, contrary to what we could think at first glance, the level of wages as such is not exogenous: these are indeed the quantities of wage-goods which are exogenous. Their price is endogenous.

It has therefore be demonstrated that the Ricardian theory, at least from a mathematical point of view, and when interpreted in terms of production prices, allows to determine in a coherent way the distribution of the net product between the workers and the capitalists.

### 3.3. The contradictions and limitations of the Ricardian analysis

First we must ask ourselves whether this second interpretation in terms of production prices is able to answer the fundamental concern of Ricardo: to understand the evolution of distribution over time. This comes down to asking if it is possible to compare the share of wages and profits from one period to another, when these shares change, and therefore to measure the variation of income distribution between wages and profits.

Let us begin by noting that for given conditions of production the profit rate is determined independently of the price which is chosen as a measurement unit. On the other hand, the price of all the commodities, and consequently the level of wages and the amount (and not the rate) of profits, depend on the price chosen as a measurement unit. As Ricardo indicates: “We are possessed then of plenty of measures of value and either might be arbitrarily selected for the purpose of ascertaining the relative value of commodities at the time they are measured” (Ricardo, 1823 – rough draft, p.381).

Let us note however, that the share of wages and profits in the income is always the same, whatever the measurement unit. Indeed the price vectors are collinear.

The problem is that in the Ricardian system such as the one that we just exposed every variation in wages has to be analyzed as a variation in the conditions of production, because wages are made of goods. When such wages change, matrix A of technical coefficients is modified and especially the conditions of production of the commodity whose price is adopted as the measurement unit. Obviously, it is still possible to calculate the profit rate, the price system and the distribution resulting from a variation in the wage level. But it is not possible to compare this new system to the previous one, and consequently the new distribution to the previous one, because the measurement unit has been modified and is no longer the same.

We cannot but conclude that the evolution as such of prices and distribution is undetermined, because of the impossibility to find out an invariable measurement unit as soon as we are at the level of exchange ratios implying the introduction of a uniform profit rate into the analysis. We can nevertheless affirm that if the conditions of production of wage-goods deteriorate, or if the quantities of wage-goods increase, these circumstances will translate into a decrease in the profit rate. The dominant eigenvalue $\alpha$ of matrix $A_1$ is indeed an increasing function of the elements of this matrix, and the profit rate is consequently a decreasing function, since $r = \frac{1-\alpha}{\alpha}$.[2]

This is exactly the definitive conclusion which Ricardo had reached: “We are without any (measurement unit) by which to ascertain the variations in the values of commodities for one year, for two years or for any distant portions of time” (Ricardo, 1823, rough draft, p. 381). And: “there can be no perfect measure of the variations in the value of commodities arising from an alteration in these proportions (into which commodities are divided for wages and profits) as the proportions will themselves differ according as the commodity employed for the measure may be produced in a shorter or longer time” (Ricardo, 1823, later version, p. 404).

At a first level, we can conclude from this analysis of Ricardo’s Political Economy that if he did not succeed in producing a correct concept of value, his approach, as it appears particularly in his last text, “Absolute value and exchangeable value”, written shortly before his death, provides nevertheless some interesting elements.

[2] On this point, see again Gantmacher. Op.cit.

One of his problems and in fact his main mistake was that Ricardo wanted to integrate production and distribution, and therefore absolute values in terms of incorporated labor with exchangeable values as exchange ratios resulting in a uniform rate of profits, i.e. production prices. He could not succeed, because values and prices do not have the same dimension, which he almost understood, but not to the point of finding out a way to pass from one to the other. In any case his treatment of wages as real wages, or wage-goods, would have prevented him from progressing towards a possible solution.

This leads us to a second level, which is the conception of production and distribution. In fact it had not changed much since the Essay on Profits, since the essential change was the replacement of a single wage-good, wheat, by a number of different wage-goods, which can come from any sector, and not only from agriculture. To be sure, it is a little more realistic to consider that workers survive not only by consuming wheat, but also a variety of other goods.

But this does not allow to get rid of the criticism that his system continues to violate two principles already established. First, it contradicts principle 3, according to which money is not a commodity, because the standard of value here is still a commodity, even if it is no more wheat, but can be any one of the various wage-goods. And second, as a consequence, it also contradicts principle 8, according to which wages are not established as real wages, but in money, in which they are paid.

Instead of wheat alone, these are now all of the wage-goods, in the broad sense that have been defined, which are singled out as the only goods that enter, directly or indirectly, in the production of all the other ones. This derives from the assumptions made by Ricardo, because the payment of wages in wage-goods makes them the perfect substitute to labor, i.e. the only element in the system to enter in the production of every good.

It is the reason why the determination of the profit rate takes place in the wage-goods sector, where there is a surplus, as the difference between the amount of wage-goods produced and the amount of wage-goods employed in the same production process. It has to be so, because on top of the payment of wages in the wage-goods sector there must remain an amount of wage-goods available to pay for wages in all of the other sectors, those producing “luxury goods” (in fact goods that, once produced, do not re-enter into the production process of any wage-goods). It is also the reason why the profit rate is determined independently of prices, through the eigenvalue $\alpha$ of matrix $A_1$.

However, outside of the wage-goods sector and at the level of the system as a whole, there is absolutely no reason why there should necessarily be a surplus of wage-goods. It is all the more the case that, due to the broad definition of wage-goods, they also include all the goods entering in their production (otherwise they could not be produced). Therefore wage-goods must be used not only to pay for wages, but also as intermediate goods in some sectors, and not only for the payment of wages. There might be an overall surplus of a given wage-good $j$ only when for this particular wage-good $\sum_{i=1}^{n} a_{ij} < 1$.

But then, if it were the case, such a surplus would imply that this wage-good, for the quantity not entering into the production process, would be also a “luxury good”, i.e. in fact a consumption good, which is quite a logical problem, because it is impossible to understand why the same good, whatever it is, could be an intermediate good when it is consumed by workers, and a consumption good when it is consumed by the other classes of society: the owners of the lands and the owners of capital. In fact, even though it is certainly true that these other classes consume some specific goods that are not necessary for their subsistence (the reason why they are called “luxury goods”) they also have to consume for their subsistence a number of goods, like food etc., that are similar to the goods consumed by workers.

Such a contradiction can be resolved only if we admit that wage-goods are true consumption goods, which have to be treated as such, meaning as goods that do not re-enter as intermediate goods into the production process. In passing it is of great importance to be very clear about what that means to enter in the production process, since it has to do with the very definition of the concept of production. Let us precise therefore that to enter (or re-enter) in the production process implies first to be bought by a producer from another producer, and second and furthermore to be transformed one way or another into another good through the production process. This is the definition of intermediate goods, which are part of what is often called circulating capital. In this sense wage-goods, which are not bought on the market by producers to be given in kind to the workers whom they employ, for payment of their labor, and are not transformed in the production process, do not meet the criteria to be considered as intermediate goods. They are pure consumption goods, which once produced are bought, not by producers, but by individuals, be they workers or capitalists, and never go back to re-enter for transformation into the production process.

If therefore wage-goods are considered as what they truly are, i.e. consumption goods that do not enter in the production process, this implies that it is no longer possible to make them appear on the left side of the equations which were used to represent this process. They will only appear on the right side or the equations. This would make them quite similar to luxury goods, which would be quite inappropriate to designate usual consumption goods. But this should also be the case for all the goods so far considered as “luxury” goods, which - once produced, have no reason to appear on the left side of the equations, as if they went back into their own production process. In other words, submatrix $A_2$ should also be a zero submatrix. As for the most appropriate common term for both kinds of consumption goods, it should be “final goods”.

Does it mean that the presentation used so far must be completely abandoned? No, because there still remain true intermediate goods, which are finished products, but not final products. Indeed all of these goods enter directly or indirectly in the production of all other goods, but are never used, neither as consumption goods, nor as capital goods.

This leads us to state first a corollary to the last principle, stated at the end of chapter one:

**Corollary to Principle 8**: Wages are not paid to workers in wage-goods.

We can also state two new principles, to be added to the first eight ones:

- First, the following principle:

**Principle 9**: Workers buy consumption goods out of their wages previously paid in money.

- With the following corollary:

**Corollary to Principle 9**: there are no specific consumption goods bought by the workers that could be isolated as pure wage-goods.

- And second, as an additional principle:

**Principle 10**: Intermediate goods are bought, not by individuals, but by producers, to enter the production process, where they are physically transformed into other intermediate goods or final goods.

- With this last corollary:

**Corollary to Principle 10**: Goods bought by workers as consumption goods, as well as all the other consumption goods, are not intermediate goods, but final goods.

Moreover, one must observe that the conception of distribution underlying the main and the last works of Ricardo (the "Principles" and the text on "Absolute and Exchangeable Value", respectively) has not really changed from that of the "Essay on Profits". Indeed, if the model that we analyzed in this chapter is actually more complex than the previous one, distribution is still completely disconnected from its social background, let alone from the balance of power between the various classes of society. The level of wages is given and considered as a natural phenomenon, because it is still Malthus's law of population which maintains it at the minimum level of subsistence. Since the technical conditions of production, and in particular the number of workers in each branch, are also given, this determines the overall amount of wages. Then the amount of profits is obtained by difference, as a residue. In this sense, distribution still depends on the conditions of production.

In any case and finally, the problem that also stays unresolved is that prices in this last Ricardian model still have the dimension of a commodity, whose price is chosen as the measurement unit, whatever it may be (as long as it is an intermediate good). It follows, as Ricardo himself recognized it, that no comparison in time between two price systems is possible as soon as there is the slightest variation in the conditions of production, which would cause necessarily a variation in the conditions of production of the good chosen as the standard of value.

The Ricardian system remains however quite a rich one, owing to all the questions it raises, and despite or maybe because of its flaws. It provides us with a number of insights on the problems that need to be overcome to answer actual economic questions. But however instructive it may be, it has to be seriously modified to make some progress toward the resolution of these questions.

Is it still possible nevertheless, on a basis where there would be no wage-goods, and therefore where distribution would have to be defined differently, to design a coherent system to determine both distribution and prices, as well as their variation over time? This is what is going to be examined in the next chapter, devoted to the theory of Sraffa, such as exposed in his well-known book “Production of commodities by means of commodities”.

# Chapter 3. Distribution and prices in Piero Sraffa's theory

This chapter will expose first Sraffa's theory of production as production of a surplus, and of distribution as a sharing of this surplus between workers and capitalists. It will show also that Sraffa's treatment of the problem of reproduction is quite different from Ricardo's. As for its construction of an invariable standard of prices, it is closer to Ricardo's "Essay on Profits" than to Ricardo's "Principles". Moreover, the way according to which this standard is built makes it impossible to analyze changes in the system of production. The main utility of Sraffa's system has been in fact to destroy theoretically the neo-classical theory of capital and its view of the rate of profit as the marginal productivity of capital.

To begin with, let us recall that the neo-Ricardian or neo-Cambridgian theory, which has developed after the publication in 1960 of Sraffa's only book, "Production of commodities by means of commodities", has been widely perceived as a return to the classical and more particularly Ricardian political economy.

Such a conception seems however somewhat questionable, to the extent that it relies on a superficial interpretation of Ricardian analyses. It starts from the idea, which is right, that the problem of finding an invariable measurement of values has been an essential concern for Ricardo. Since Sraffa manages to build up such an invariable standard of value, this conception then deducts ipso facto that he solves the problem which had been left unsolved by Ricardo, and is therefore following his steps in the same footprints, with more success.

However, the analysis of the preceding chapter has shown that Ricardo, far from leaving the problem unsolved, has given it an unequivocal answer, but a negative one, by showing that there cannot be such an invariable measurement of values as soon as conditions of production change. Therefore the solution proposed by Sraffa can only be developed in another type of problematic, which is not exactly the Ricardian one.

This will be shown at several levels: first the level of reproduction, second the level corresponding to the link between the conception of wages and the construction of a standard of measurement. Finally it will be shown that the nature of Sraffa's assumptions have a bearing on the meaning of production prices.

## 1. Sraffa's problematic of reproduction: quite different from Ricardo's

### 1.1. Reproduction and accumulation of capital

The divergence in problematics between Sraffa and Ricardo appears as soon as the first paragraph in the preface of "Production of commodities..", where Sraffa underlines that: "the investigation is concerned exclusively with such properties of an economic system as do not depend on changes in the scale of production or in the proportions of 'factors'."(Sraffa, 1960, Preface, p. 5)

If we could consider that Ricardo's problem consists in determining the distribution of the product at a given point in time, Sraffa would be perfectly right in adding that: "this standpoint (…) is that of old classical economists from Adam Smith to Ricardo" (Sraffa, 1960, p. 5). It is true indeed that for Ricardo the problem is not simultaneously to determine equilibrium prices and quantities, but to determine prices and distribution for produced quantities of commodities and production techniques which are both given.

But for Ricardo this constitutes only a logical necessity, and as such only a first step for an analysis which is embedded in a larger framework, i.e. the reproduction of the whole economic system. The determination of profits is of such an importance for Ricardo because it represents a logical precondition of a second question, which is the accumulation of capital. This will allow in turn the reproduction of the economic system on an enlarged scale, when the additional means of production newly accumulated start producing.

When Sraffa warns us at the very beginning of his book that the change of the economic system stays outside of his analysis, the implicit message is that he does not intend to address the reproduction of the system, unlike the Ricardian analysis. The reason must be sought after in the nature of the invariable standard that Sraffa wants to build up. This standard, as we shall see, is indeed invariable only to changes in distribution, whereas for Ricardo the problem which is unsolvable is that of the invariability of the standard in case of changes in production. Indeed for Ricardo any variation in distribution must be analyzed as a variation in the production system, as it was shown in the preceding chapter.

### 1.2. The question of returns

These previous remarks allow us to better understand a second precision provided by Sraffa in his preface, which is that "no changes in output and (at any rate in Parts I and II) no changes in the proportions in which different means of production are used by an industry are considered, so that no question arises as to the variation or constancy of returns" (Sraffa, 1960, p. 5). It is because the concept of return implies a comparison between the price systems of two distinct production systems, and because such comparison would need a standard that would be invariable to changes in the system, which is not the case. Sraffa's theory leads therefore to the absolute rejection of any assumption about returns.

This impossibility of formulating an assumption about returns is quite significant of the difference in problematics between Sraffa's and Ricardo's analyses. For the last one, an analysis of reproduction cannot indeed be separated from an assumption on returns. This does not mean that the Ricardian concept of returns is identical to the neo-classical one. On the contrary Ricardo does not intend to make the quantity of one factor vary while the other ones remain constant, in order to determine simultaneously distribution, quantities and prices, since for Ricardo the quantities which are produced and the technique are given, as well as the quantities of wage-goods.

The problem is to determine the evolution in the value of wage-goods in order to determine the amount available for accumulation, when all the data are changing. And the evolution in the rent of land is not the only change to be concerned, as the whole chapter on wages in the "Principles" clearly shows. The Ricardian concept of returns does not refers to "factors of production", but to the productivity of labor, the measurement of which requires a theory of value based on a measurement unit invariable with respect to changes in the system of production. This explains the use by Ricardo of the labor theory of value.

It is indeed by assuming a decreasing productivity of labor in the sector producing wage-goods that Ricardo concludes to the possibility of a deadlock in the accumulation of capital, because the profits rate will fall below the level needed for capitalists to continue investing. To be sure, this analysis is not fully rigorous to the extent that the link between the fall in returns in terms of labor value and the fall in profits rate is not correct, at least in the "Principles" (if not in the "Essay on Profits"). It remains that the concept of returns with the meaning that has just been defined plays an essential role in the Ricardian analysis of reproduction, whereas it is completely absent from Sraffa's analysis.

This particular point concerning returns seems thus to constitute a good test of the change in problematics, since it shows that the accumulation of capital, which is central to Ricardo, is outside of the ambit of the theory of production prices. One can certainly agree that a system that does not register any change, neither in the quantities that are produced, nor in the technique of production, is reproducing identically. But such a reproduction has nothing to do with the conception of reproduction for Ricardo, because it belongs to a logic which is exterior to the accumulation of capital.

However, it is by examining the way through which Sraffa's constructs his invariable standard that one can best perceive the effects of this difference in problematics.

## 2. Wages and the construction of an invariable standard

### 2.1. Two Ricardian models with wages as wage-goods

One could think that Sraffa deals with the problem of the invariable standard only in chapter 4, entitled "the Standard commodity" of "Production of commodities", thus starting to address it from paragraph 23 of his book. In fact the reasoning which will lead him to the construction of the standard commodity is touched upon as soon as paragraph 8, in which Sraffa abandons the Ricardian conception of wages.

Up to this point Sraffa develops two models. The first one, that Sraffa names "production for subsistence" can be represented in matrix format by the following equation:

$$AP = P \text{ with} \sum_i a_{ij} = 1, \ \forall j$$

This system has a solution other than $P$ = 0, as soon as one of the prices is taken as a measurement unit, since the determinant of matrix ($I - A$) is zero, because of the condition $\sum_i a_{ij} = 1$, which means the absence of a surplus. In this model no problem of distribution arises, although one can only interpret it as a situation in which the totality of output is used, either as means of production or for the consumption of workers.

The second model, entitled "production with a surplus" is based on the implicit assumption that quantities produced are higher than quantities used in the production process or as means of subsistence for the workers. It is the condition for a surplus to appear, and this condition is not a uniquely technical one, since it implies that workers receive only the quantities of goods necessary for their subsistence. As Sraffa writes: "wages (are) consisting of the necessary subsistence of the workers and thus entering the system on the same footing as the fuel for the engines or the feed for the cattle" (Sraffa, 1960, § 8, p. 10).

This model is therefore nothing else than the Ricardian model exposed in the previous chapter, and which takes the form $(1+r)A.P = P$. It is important to note that the apparition of a profit rate depends on the existence of a surplus.

In this model and for reasons already exposed, Sraffa shows that only the commodities which enter directly or indirectly in the production of all the others play a role in the determination of the system, and he decides for this reason to name them "basic products". The profit rate is determined only by the conditions of production of these basic products. As for the equations of production of non-basic products or "luxury goods", they can determine only the price of these goods themselves, once the price of basic products is known (anyone of these prices being chosen as a measurement unit).

In order to clarify in advance some ideas that will be exposed later, it is important to note that in this "Ricardian" system the only data relating to wages are the quantities of wage-goods by labor unit, and not the price of these quantities of wage-goods. In particular the fact to add to the price system an additional equation like: $S_1 p_1 + S_2 p_2 + ... + S_k p_k = w$ (where $S_1$, $S_2$, … , $S_k$ correspond to given quantities of the commodities that are part of the wage-unit), does not help in solving the system. Indeed the information conveyed by this equation is already embodied in the system itself through the production coefficients of wage-goods.

As a matter of fact, one could think to use such an equation to present the system of production prices under the format:

$(1+r)(A^1P + wL) = P$, where $A^1$ is the column block of the production coefficients of basic goods other than wage-goods.

But this would have no sense, since $A^1$ is not a square matrix, and therefore has no inverse matrix.

It follows than in a "Ricardian" system of production prices, one has to determine first the profit rate, then the production prices, including the price of the wage-goods, which allows finally, and only then, to determine the level of *w*, which is not given, but is an unknown variable of the system, in terms of any price that has been chosen as a measurement unit.

As we showed in the previous chapter, the conception of wages which is at the basis of this model does not allow to take into account a change in distribution, because of the impossibility to define an invariable measurement unit within this framework. It is therefore no surprise that the first step made by Sraffa in the construction of his invariable standard consists in abandoning this conception of wages. But then, what conception does he adopt?

### 2.2. A new conception of wages as part of a surplus

Sraffa starts by expressing the idea that: "besides the ever-present element of subsistence, they (wages) may include a share of the surplus product. In view of this double character of the wage it would be appropriate, when we come to consider the division of the surplus between capitalists and workers, to separate the two component parts of the wage and regard only the "surplus" part as variable; whereas the goods necessary for the subsistence of the workers would continue to appear, with the fuel, etc., among the means of production" (Sraffa, 1960, § 8, p. 10).

Sraffa nevertheless abandons this conception to adopt that which consists in considering the totality of wages as variable.

We can however note that this distinction between wages partially or totally taken out of the surplus has no consequence from a formal point of view. This was demonstrated in 1974 by Maurisson in his PhD thesis, devoted to the works of Piero Sraffa (pp. 153-154). Indeed, in the case where only part of the wages are a part of the surplus, the matrix of production methods A contains two column blocks: the first one is $A^W$ reflecting the conditions of production of subsistence goods, and the second one is $A^N$, for the other goods entering the production process.

Thus $A = (A^N, A^W)$.

Let us call $L$ the vector of direct quantities of labor, and $P$ the column vector of production prices, which subdivides into a vector $P_W$ for the price of subsistence goods, and a second one $P_N$ for the price of the other goods:

$$P = \begin{pmatrix} P_N \\ P_W \end{pmatrix}$$

The system of production prices then can be written under a matrix format:

$(1+r)AP + w'L = P$, with w' the wage corresponding only to the part payed out of the surplus.

This is equivalent to: $(1+r)\left(A^N P_N + A^W P_W\right) + w'L = P$

And gives us: $(1+r)A^N P_N + rA^W P_W + A^W P_W + w'L = P$

This last equation allows us to express the total amount of wages in nominal terms (meaning in terms of the measurement unit) $w$. We can indeed write:

$$wL = A^W P_W + w' L$$

By pre-multiplying by vector $L^{-1}$, inverse to the left of column vector $L$, we get:

$$w = L^{-1} A^W P_W + w'$$

We define $W = L^{-1} A_W$ as the line vector of quantities of subsistence goods by labor unit, which gives:

$$w = W P_W + w'$$

The whole system of production prices can therefore be written:

$$\left(1+r\right) A^N P_N + r A^W P_W + wL = P$$

When all of the wages are considered as a part of the surplus, it means that there are no wage-goods and that all the elements of vector W are zero, which implies that the column block $A^W$ is zero. In such a case w = w' and the system of production prices can be written:

$$\left(1+r\right) A^N P_N + wL = P$$

It is therefore under a mathematical format which is identical to the previous system [which was $\left(1+r\right) AP + w' L = P$], of which it constitutes therefore a particular case. The essential difference between both systems is that wage-goods appear no more in the second system as basic goods and therefore do not participate any more in solving this system. Solving it will now be considered within the framework of this last model, which is the one kept by Sraffa.

### 2.3. The consequences of wages not being advanced

Before exploring the functioning of this model, it seems useful to open a parenthesis in order to understand that when wages are no longer considered as advanced (whether it is a fraction or the totality of these wages), which means that no profit is perceived on them, their expression in terms of wage-goods becomes ipso facto undetermined. It means that wages cannot in any case be identified to a quantity of wage-goods. To show it the simplest thing to do is to argue by contradiction, precisely by imagining that wages are identified to a quantity of wage-goods, and that this quantity is not advanced, but taken from the surplus.

If therefore we imagine a hypothetical system of production prices where wage-goods exist as such but are not advanced, this system will be of the form:

$$\left(a_{11}p_1+...+a_{1k}p_k\right)\left(1+r\right)+a_{1,k+1}p_{k+1}+...+a_{1n}p_n=p_1$$

…………………………………………………….. .

$$\left(a_{k1}p_1+...+a_{kk}p_k\right)\left(1+r\right)+a_{k,k+1}p_{k+1}+...+a_{kn}p_n=p_k$$

$$\left(a_{k+1,1}p_1+...+a_{k+1,k}p_k\right)\left(1+r\right)+a_{k+1,k+1}p_{k+1}+...+a_{k+1,n}p_n=p_{k+1}$$

.........................................................................................

$$\left(a_{n,1}p_1+...+a_{n,k}p_k\right)\left(1+r\right)+a_{n,k+1}p_{k+1}+...+a_{n,n}p_n=p_n$$

With $P_1 = (p_1, p_2, ..., p_k)$ = the price vector of basic goods other than wage goods

And $P_2 = (p_{k+1}, p_{k+2}, ..., p_n)$ = the price vector of wage-goods

Is it possible to solve this system? This comes down to asking whether it is possible to calculate the unknown variables which are the profit rate and the production prices.

This can be done by presenting the system under the following matrix format:

$$A'P=P$$

With $A'=\left[\left(1+r\right)A^1 \vdots A^2\right]$

$A^1$ and $A^2$ are the column blocks corresponding respectively to the technical coefficients of the basic goods other than wage-goods, on the one hand, and the wage-goods, on the other hand.

We get then: $\left(I-A'\right)P=0$

For the system to have a solution other than the trivial one $P = 0$, it is necessary and sufficient that we have: determinant of $\left(I-A'\right)=0$.

If it were the case, solving the equation canceling the determinant would allow to determine the profit rate and the prices.

But it cannot be the case. Let us show it, naming $\overset{t}{A'}$ the transposed matrix of matrix *A'*.

Indeed: det. $\left(I-A'\right)=0 \Rightarrow \det.\left(I-\overset{t}{A'}\right)=0 \Rightarrow \exists X\neq 0 \mid \overset{t}{A'}X=X$

$$\Rightarrow \overset{t}{X}\,{}^{t}\!\left(\overset{t}{A'}\right)=\overset{t}{X}\Rightarrow \overset{t}{X}A'=\overset{t}{X}$$

$$\Rightarrow \exists \sigma \neq 0 \middle| \sigma A' = \sigma \text{, avec } \sigma = \overset{t}{X}$$

The fact that the determinant is zero entails therefore that we must have $\sigma A' = \sigma$, which means that the n equations are non-independent, and implies that anyone of them can be deducted from all the others.

However the system does not verify the condition $\sigma A' = \sigma$, although it is necessary for solving it, because it would imply:

$\sigma(1+r)A^1 = \sigma_1$, with $\sigma_1$ the vector made of the k first elements of $\sigma$,

$\sigma A^2 = \sigma_2$, with $\sigma_2$ the vector made of the n-k last elements of $\sigma$.

But only the condition $\sigma(1+r)A^1 = \sigma_1$ can be verified, whereas we have $\sigma A_2 \neq \sigma_2$ in the general case. There is indeed no reason why we should have $\sigma A_2 = \sigma_2$ because this condition, translated in economic terms, would mean that the calculation of the profit rate and prices would need the assumption that there is no surplus of wage-goods. Such an assumption is not acceptable, because the consumption of capitalists must at least partly be made of the same goods as the consumption of the workers.

It has thus been demonstrated that the representation of the system of production prices with wage-goods as part of a surplus (and consequently non-advanced), and such as *A'P = P*, cannot be accepted. But is the representation under the form $(1+r)A^1P_1 + A^2P_2 = P$ acceptable? The answer is again a negative one, because matrices $A^1$ and $A^2$ do not have an inverse (they are not square matrices).

The only remaining possibility for a representation of the production system consists in distinguishing two sub-systems, the first one which gives the price of basic goods other than wage-goods, the second one which gives the price of wage-goods. These sub-systems are:

$$(1+r)A_1^1P_1 + A_1^2P_2 = P_1$$

$$(1+r)A_2^1P_1 + A_2^2P_2 = P_2$$

We can see that for solving one sub-system only, one has to consider as given either $P_2$, in order to determine $r$ and $P_1$, or $P_1$, in order to determine $r$ and $P_2$. But then one of the two matrix equations has no more reason to exist, because either $P_2$ or $P_1$ are already determined. We can note that if it is necessary to give oneself vector $P_1$ in order to determine vector $P_2$, the converse is not true. It is indeed sufficient to give oneself the level of wages: $w = WP_2$, with $W$ the vector of the quantities of wage-goods by unit of labor, because we have:

$$A_1^2P_2 = WP_2L_1$$

It remains that when we give ourselves w to determine $r$ and $P_1$, the sub-system giving the price vector $P_2$ of wage-goods cannot be calculated. The system is indeed over-determined because it contains $n - k + 1$ equations, which are:

$$(1+r)A_2^1P_1 + A_2^2P_2 = P_2 \quad \text{(n - k equations)}$$

$$w = WP_2 \quad \text{(1 equation)}$$

This for $n$ - $k$ unknowns (the $n$ - $k$ prices of wage-goods).

We arrive thus to this conclusion, at first sight rather curious, that as soon as we give ourselves $w$, i.e. the level of wages as a share of the surplus, while wanting at the same time to continue to consider wages as a quantity of wage-goods, the nature of these wage-goods is not classifiable, their quantity is undetermined, and the price of wage-goods stops having any significance. In fact they cannot even be calculated as long as we try to define wages simultaneously as the price of labor and as a quantity of wage-goods.

One other important conclusion is that the conception of wages as quantities of wage-goods is compatible only with the classical conception of wages as advanced wages. As soon as wages are not advanced their economic expression in terms of wage-goods is undetermined. The classical notion of real wages disappears with the assumption that wages are a part of the surplus. This does not mean that we cannot in this case calculate the price of "wage-goods", but that this calculation is made ex-post, with the level of wages w being given as the price of labor, which cannot be identified beforehand as a quantity of wage-goods.

The converse of the proposition that we just demonstrated that any wages considered as quantities of wage-goods are necessarily advanced is however not true. It is possible to consider wages as advanced without necessarily considering them as quantities of wage-goods.

It must finally be noted that this demonstration has a corollary, which is that it is not sufficient, for a commodity to be a basic commodity, that it enters directly or indirectly in the production of all the other ones, but it is also necessary that its utilization gives rise to the perception of a profit rate. In fact a commodity cannot be perceived as entering directly or indirectly in the production of all the other ones unless it is designed beforehand to give rise to this perception.

# 3. The construction of the invariable standard

## 3.1. A new system with n+2 unknown variables

After the explanations provided in the last section on the theoretical status of wages, we can now understand that since the search for an invariable standard requires from Sraffa to renounce to the classical conception of wages as a quantity of wage-goods, an essential difference appears in the solving of the system, with regard to the second model, i.e. the "Ricardian" one, analyzed by Sraffa.

This difference lies in the fact that the system of production prices now presents itself with $n+2$ unknowns, i.e. the *n* prices, the profit rate $r$ and the wage *w*, and this for *n* equations. To give oneself one of the prices, chosen as a unit of measurement, no longer suffices for solving the system, which has still one unknown, and remains undetermined. The only means to remove this indetermination, in order to be able to solve the system, is to give oneself also one of the distribution variables, chosen as an independent variable.

If we focus on the equations of production for basic goods only, in which the wage-goods corresponding to the wage *w* obviously have no longer their place, the system of production prices presents itself under the following format:

$$(1+r)AP + wL = P$$

Matrix *A* is by assumption irreducible and supposed to be "productive", which implies that $\sum_i a_{ij} \leq 1$ for every j, and $\sum_i a_{ij} < 1$ for at least one j. Such a system has therefore always a solution in the form of a positive price vector, provided that one of the prices is taken as a measurement unit and one of the two distribution variables is given.

However it must be acknowledged that solving the system cannot be done directly, through the simultaneous determination of *n* unknowns, i.e. $n-1$ prices and the dependent distribution variable. The calculation of the second distribution variable is indeed logically prior to calculation of prices.

The system can indeed be represented in the following matrix format:

$$P = w\left[I-(1+r)A\right]^{-1}L$$

Such a matrix equation can be solved only after *w* and *r* are known. On this basis, how is a standard supposed to be invariable with respect to the variations of distribution going to be built? We will try to understand that in the next sub-section.

### 3.2. The determination of the second distribution variable

What is now at stake is to define the conditions that have to be met for a price chosen as a measurement unit to become an invariable standard. To understand it, the simplest solution consists in examining the way in which the system of production prices is solved when the price of any one of the *n* commodities is chosen as a measurement unit.

Let us therefore take the price $p_i$ of commodity i as this unit, such as $p_i = 1$. The calculation of the second distribution variable, the first one being supposed known, can be made from the equation giving the price of this commodity, by developing this equation. We get then:

$$p_i = w\left[l_{i0} + (1+r)A_iL + (1+r)^2 A_i^2L + \ldots + (1+r)^n A_i^nL\right]$$

$$1 = w\left[ l_{i0} + (1+r) l_{i1} + (1+r)^2 l_{i2} + ... + (1+r)^n l_{in} \right]$$

This equation defines a relation between the wage and the profit rate. This means that for a given level of the profit rate *r*, between 0 and a maximum reached when $w = 0$, this equation allows to calculate the corresponding level of wage. Reciprocally, for a given level of wage, between 0 and a maximum reached when $r = 0$, this equation allows to calculate the corresponding profit rate. For reasons of positivity of prices, it must be noted that the profit rate *r* calculated this way is the highest of the roots of this equation of degree *n* in *r*. It must also be noted that if the profit rate is a pure number, as regards the wage it is necessarily expressed in terms of the price chosen as the measurement unit.

Moreover any price, of a commodity or an aggregate of commodities – provided that they are basic commodities, can be chosen as a measurement unit. It entails that there can be an infinite number of relations between w and r, as many as the number of prices that can be chosen as measurement units.

Any price having been chosen as the measurement unit, and once the distribution variables have been determined (one or the other being given) by the equation of this price, the totality of prices can then be determined by the matrix equation:

$$P = \left[ I - (1+r) A \right]^{-1} wL$$

### 3.3. Changes in distribution and the measurement unit

What happens now if distribution changes, with production techniques remaining the same?

It can be directly deducted from the preceding equation that the whole price system is going to be modified. But contrary to what one might think at first sight, the new price system corresponding to the new state of distribution is not comparable to the previous one, although it is still the same commodity whose price is taken as the measurement unit. Indeed the fact that this price stays equal to 1 does not prevent the change in distribution from influencing it. This appears clearly if we develop the equation giving the price $p_j$ of commodity *j* in terms of the price $p_i$ of commodity *i*, which is equal to 1:

$$\frac{p_j}{1} = \frac{l_{j0} + (1+r) l_{j1} + (1+r)^2 l_{j2} + ... + (1+r)^n l_{jn}}{l_{i0} + (1+r) l_{i1} + (1+r)^2 l_{i2} + ... + (1+r)^n l_{in}}$$

It is clear from this equation that the modification in prices following a change in distribution is the result of two separate effects. First an obvious effect which results from the modification of the equations of these prices themselves (i.e. a change in the numerator of the preceding ratio). Let us call intrinsic effect the effect on the prices. Second a less obvious effect at first sight which is due to the repercussion of the variation in distribution on the price taken as measurement unit itself (i.e. a change in the denominator of this same ratio). We will call extrinsic effect this effect produced on the prices by this variation in the “numéraire”, which results from the intrinsic effect produced by a change in distribution on the "numéraire" itself.

Thus the new price system resulting from a variation in distribution is not comparable to the old one. It follows necessarily that a comparison is no more possible for the net product of the system $\sum(I-A)P$ as well as for the amount of profits $\sum rAP$ and the amount of wages $\sum\left[I-(1+r)A\right]P$. It is of course the same for the share of profits $\frac{\sum rAP}{\sum(I-A)P}$ and the share of wages $\frac{\sum\left[I-(1+r)A\right]P}{\sum(I-A)P}$.

The evolution of distribution is therefore undetermined.

### 3.4. The existence of an invariable measurement unit

To suppress this indetermination comes down to building up an invariable measurement unit. It is now possible to affirm that for a commodity to be able to play this role, it is sufficient for its price not to be submitted to the intrinsic effect produced by a change in distribution.

Since the price of such a commodity is $p_i = (1+r)A_iP + wl_i$ (1)

It means that we must have:

$$\forall(w,r),\ dp_i = (1+r)a_i dP + drA_iP + dwl_i = 0 \quad (2)$$

But the absence of any intrinsic effect when distribution changes implies that an increase in profits must be exactly compensated by a decrease in wages, and reciprocally, a necessary condition which writes as:

$$drA_iP + dwl_i = 0 \quad (3)$$

The conjunction of equations (2) and (3) entails necessarily that we have also:

$(1+r)A_i dP = 0$, and therefore $dP = 0$ (4)

As a consequence a necessary condition for the price of a commodity to be free from any intrinsic effect is that the price of its means of production must be equally invariable as regards the effects of changes in distribution. The same condition applies to the means of production of these means of production and so on. We can deduct immediately from this demonstration that the standard commodity cannot consist in a unique good and that it necessarily includes all the commodities entering directly or indirectly in the production of all the others, i.e. all the basic goods.

The price of each of these commodities must necessarily verify equation (2), which can be translated in synthetic terms by summing up all these equations. It must be noted that, since the invariable standard must necessarily include all the basic goods, the wage rate *w* is necessarily expressed in terms of the price of all these commodities. And since the maximum amount that *w* can reach when r is zero is the price of the net product or $\sum(I-A)P$, the simple requirement of simplifying the calculation leads to express *w* in terms of this price, which makes *w* vary from zero to 1. If moreover we assume that $\sum l_i = 1$, the sum of equations (2) can be written under the form:

$$\sum(I-A)dP = r\sum AdP + w\sum(I-A)dP + dr\sum AP + dw\sum(I-A)P = 0,\ \forall(w,r) \qquad (5)$$

From the definition of the standard we have $dP = 0$, and therefore equation (5) is verified only if we have:

$$dr\sum AP + dw\sum(I-A)P = 0 \qquad (6)$$

Or: $$-\frac{dr}{dw} = \frac{\sum(I-A)P}{\sum AP} = \text{constant} \qquad (7)$$

Indeed, because *dP* equals zero by definition, then *P* = constant.

In this case, we have also:

$$\sum(I-A)P = r\sum AP + w\sum(I-A)P \qquad (8)$$

Or: $$\sum(I-A)P = r\sum Ap + w\sum(I-A)P \qquad (9)$$

Therefore by combining equations (7) and (9), we get:

$$-\frac{dr}{dw} = \frac{\sum(I-A)P}{\sum AP} = \frac{r}{1-w} = \text{constant} \qquad (10)$$

And since this relation must be verified for every value of the couple (*w*, *r*), it is necessarily verified for w = 0. In this case we get $\frac{\sum(I-A)P}{\sum AP} = R$, i.e. the maximum profit rate of the system under review. And therefore we have also: $-\frac{dr}{dw} = \frac{r}{1-w} = R \quad \forall(w,r)$

It results from this demonstration that a necessary condition for the existence of an invariable measurement unit is the existence of a linear relationship between *w* and *r*, such as $r = R(1-w)$. It is worth remarking that this whole demonstration has been conducted without even asking the question of the exact nature of this measurement unit!

Indeed it was sufficient to make the assumption that such a unit existed. However, since the demonstration was done under the definitional constraint that *dP* = 0, it follows that the wage *w* is necessarily expressed in this invariable measurement unit, without ours knowing what it is made of. We thus find back, but in a much quicker way, Sraffa's ascertainment according to which: "It is curious that we should thus be enabled to use a standard without knowing what it consists of" (Sraffa, 1960, p.37).

We must note in passing that the necessity to have a linear relationship between *w* and *r* for a standard commodity to exist implies conversely that it is impossible to define a standard commodity when wages are advanced, even though no reference is made to quantities of wage-goods, and wages are expressed in terms of net product. In this case indeed the system of production prices should be written as: $(I - AP) = rAP + (1+r)wL$

And it is obvious that the relationship between *w* and *r* would be no more linear.

### 3.5. The nature of the standard commodity and its consequences

The uniqueness of the standard commodity has so far not been demonstrated, but this can be proven quite simply. Indeed we just need to observe that in order to build this commodity it is necessary to find a set of multipliers $Q = (q_1, q_2, ..., q_n)$ such as:

$$\frac{Q(I-A)P}{QAP} = R \qquad \forall P \qquad \text{[see equation (5) above]}$$

Or $QAP = \frac{1}{1+R} QP$

Vector *Q* is the left eigenvector of matrix *A*, corresponding to the eigenvalue $\frac{1}{1+R}$, which is the dominant eigenvalue of matrix A. This matrix being irreducible, the elements of vector *Q* are positive, and this vector is unique within a homothety. For it to be perfectly defined it is sufficient to add to the system the relation $\sum QL = 1$, which has the advantage of identifying the total quantity of labor of the initial system of production prices to the total quantity of labor of the Standard system. The standard commodity, whose essential function is to express wages in the standard system, is thus the net product of the standard system, such as $Q(I-A)P = 1$.

We can remark that the invariable nature of the price of the standard commodity comes from the invariability of the ratio between its price and that of its means of production, always equal to *R*, the standard ratio of the system, whatever the situation of distribution. This property results from the fact that this ratio can be assimilated to a simple ratio of quantities, with the same commodities that appear in the same proportions in the net product and in the means of production. The standard commodity is thus an homothetic commodity, which is produced with only itself. This constitutes an analogy to the Ricardian system where wheat is equally a homothetic commodity. It is both the only basic commodity and an invariable measurement unit. This was emphasized by Benetti and Cartelier in an article on Production prices and Standard of prices, published in 1975.

This analogy is nevertheless not total, because wheat in this last system remains an invariable standard even when its standard ratio is modified (on this point we refer to the previous chapter), which is not the case here, because any modification of $R$ leads to a modification of $Q$ and of the standard commodity (as well as of the linear relation between $R$ and $w$).

We have seen so far that the existence of a linear relation between $r$ and $w$, with $w$ being expressed in terms of the standard commodity, is a necessary condition for the existence of this standard. The converse of this proposition can now be demonstrated, which will make this necessary condition appear as a sufficient condition. For this purpose all that we need is to show that the linear relation between $r$ and $w$ is not confined to the standard system, but remains verified in the initial system of production prices, provided that wages are paid in terms of the standard commodity. This point was highlighted as soon as 1962 by Newman, in an article on "Production of commodities by means of commodities", published in the Swiss Journal of Economics and Statistics (Newman, 1962).

Let us name $r^*$ the profit rate in the initial system. This profit rate $r^*$ is the ratio between the amount of the initial net product remaining available after payment to the workers of a fraction $w$ of the net standard product, on the one hand, and the price of the means of production of the initial system; on the other hand.

Thus we have:

$$r^* = \frac{\sum_j p_j \left(1 - \sum_i a_{ij}\right) - w \sum_j p_j \left(q_j - \sum_i q_i a_{ij}\right)}{\sum_j p_j \sum_i a_{ij}} \qquad (1)$$

Since $r^*$ is the profit rate in each branch, we have also:

$$r^* = \frac{p_i - \sum_j a_{ij} p_j - l_i w \left(\sum_j p_j \left(q_j - \sum_i q_i a_{ij}\right)\right)}{\sum_j a_{ij} p_j} \qquad (2)$$

From equation (2) and the fact that $q_i > 0$ for every $i$, we obtain:

$$r^* = \frac{q_i p_i - q_i \sum_j a_{ij} p_j - q_i l_i w \left[ \sum_j p_j \left( q_j - \sum_i q_i a_{ij} \right) \right]}{qi \sum_j a_{ij} p_j} \tag{3}$$

And by summing up for each good j:

$$r^* = \frac{\sum_i q_i p_i - \sum_j p_j \sum_i q_i a_{ij} - w \sum_i q_i l_i \left[ \sum_j p_j \left( q_j \sum_i ai_j \right) \right]}{\sum_j p_j \sum_i q_i a_{ij}} \tag{4}$$

Since by construction of the $q_i$ multipliers, we have $\sum_i q_i a_{ij} = \frac{1}{1+R} q_j$, and that, again by construction, vector $q$ being defined within a homothety, we have $\sum_i q_i l_i = 1$; since finally R has the same value in the initial system and in the standard system, equation (4) can be put under the form:

$$r^* = \frac{\sum_i q_i p_i - \frac{1}{1+R} \sum_j p_j q_j - w \left( \sum_i p_i q_i - \frac{1}{1+R} \sum_j p_j q_j \right)}{\frac{1}{1+R} \sum_j p_j q_j} \tag{5}$$

Since each $p_i$ and each $q_i$ are positive, equation (5) reduces itself to:

$$r^* = \frac{\left( \sum_i q_i p_i - \frac{1}{1+R} \sum_j p_j q_j \right)(1-w)}{\frac{1}{1+r} \sum_j p_j q_j} \tag{6}$$

And since $\sum_i q_i p_i = \sum_j p_j q_j$, we finally get:

$$r^* = \frac{\left( 1 - \frac{1}{1+R} \right)(1-w)}{\frac{1}{1+R}} = \frac{\frac{R}{1+R}(1-w)}{\frac{1}{1+R}} \tag{7}$$

$\Rightarrow r^* = R(1-w)$ in the initial system (8)

We showed thus that the rate of profit $r^*$ of the initial system is identical to the rate of profit of the standard system, when w is expressed in terms of the standard commodity. On the other hand, the actual share of wages in the initial system *w** is not identical to the share of wages w in the standard system.

Indeed $w^* = \frac{(I-A)P - rAP}{I-AP}$, (9)

Whereas $w = \frac{Q(I-A)P - rQAP}{Q(I-A)P}$ (10)

It is obvious that the two ratios are identical only when $w = 0$, $r = R$.

From this property we can draw an important conclusion: it is that wages can no more be considered as an independent variable, so that only the profit rate is able to play this role.

Indeed, like equation (9) shows well, the share of wages *w** in the net product of the initial system depends on the price system and the profit rate: *w** can therefore be calculated only when the profit rate and the prices are known. As a consequence, it is this share *w** which constitutes a residue. If we took thus the share of wages in the standard system *w* as the independent variable, we would arrive to the contradictory phenomenon that the same profit rate *r* would be a dependent variable and the share of profit a residue in the standard system, but an independent variable in the initial system, because – being as such different from *w**, it does not depend on prices.

The unique means to get rid of this contradiction consists in considering that only the profit rate can be an independent distribution variable in the production price system, as soon as we define an invariable measurement unit. We must note indeed that this proposition is verified only when wages are expressed as a proportion of the net product, which implies first the construction of the standard commodity.

Sraffa himself was well aware of that, because he notes: “the choice of the wage as the independent variable in the preliminary stages was due to its being regarded as consisting of specified necessaries determined by physiological or social conditions which are independent of prices or the rate of profits. But as soon as the possibility of variation in the division of the product is admitted, this consideration loses much of its force. And when the wage is to be regarded as “given” in terms of a more or less abstract standard, and does not acquire a definite meaning until the prices of commodities are determined, the position is reversed. The rate of profits, as a ratio, has a significance which is independent of any prices, and can well be given, before the prices are fixed. It is accordingly susceptible of being determined from outside the system of production, in particular by the level of the money rates of interest” (Sraffa, 1960, § 44, p.39).

# 4. Nature of the assumptions and significance of production prices

## 4.1. The invariable measurement unit and the conception of distribution

As we demonstrated in the previous section Sraffa succeeded in constructing a measurement unit invariable as regards changes in distribution, something that Ricardo precisely considered impossible. But the theoretical change to achieve such a result has been important, first because it involves a radical change in the problematic of production prices, linked to the abandonment of the Ricardian assumption concerning the nature of wages. This cuts the links, which to Ricardo was essential, between production and distribution. This rupture between production and distribution is certainly an essential ingredient of the post-Ricardian economics.

In Ricardo's paradigm, this is indeed the conception of wages as a quantity of wage-goods that provides the link between production and distribution. Wage-goods are produced and enter directly in the production of all commodities: in Sraffa's system, they would be basic goods. Therefore any variation in wages implies a variation of the whole production system. The reproduction of the capitalist mode of production, which is the main problem for Ricardo, does not take place in a homothetic way, but through a deformation in the structure of production and distribution. Moreover, once the quantity of wage-goods by unit of labor is given (through Malthus' law of population), the rate of profits and the level of wages in terms of prices are endogenous variables of the Ricardian system.

These features disappear in Sraffa's theory. The process to construct the standard commodity indeed forbids to consider the wage, not only as a quantity of wage-goods[3], but also as advanced, and this even if the wage is nothing else than a fraction of the net product. Because the construction of the standard implies an invariant structure of the production system, it is clear that the reproduction of this system completely defies intelligibility, and on this question Sraffa, despite the fact that he has been able to construct an invariable standard, albeit from another theoretical ground, joins Ricardo.

More important still is the fact that distribution itself cannot be explained within the theoretical boundaries of Sraffa's system. Indeed, in the Ricardian system of production prices, distribution is determined within the system once the quantity of wage-goods (which is not a distribution variable) is known, and it is the evolution of distribution which is left undetermined, because of the lack of a numéraire which would be invariable to the changes in distribution. In other words, if the concept of real wage is meaningful in the Ricardian theory, the notion of a change in real wage is meaningless.

In Sraffa's theory, and precisely because of the conception of the wage which is implemented for purely logical reasons - that have been analyzed, distribution has to be completely imported from outside in the production price system, through the need to give oneself the profit rate. Distribution thus goes out of the ambit of this analysis, it is undetermined and therefore unexplained. In this connection, Sraffa's remark, previously cited, and stating that the rate of profits "is accordingly susceptible of being determined from outside the system of production, in particular by the level of the money rates of interest" cannot be taken seriously. As Kregel clearly showed in a paper presented to the Sraffa symposium of 1973 in Amiens Kregel, 1973), there is no conception of money that would be compatible with Sraffa's system of production prices.

[3] The measurement of the wage in terms of the price of a given good is however formally equivalent to the conception of the wage as a quantity of wage-goods with only one wage-good, i.e. the good whose price is chosen as the measurement unit.

For these various reasons, Sraffa's theory cannot be fully interpreted as constituting a return to the Ricardian theory, from which it clearly departs because some of its basic assumptions, and therefore its problematic, are different. If we want nevertheless to reconcile Sraffa's model and Ricardo's analysis it is more with Ricardo's first model, that of the "Essay on Profits" , that an analogy exists, because Sraffa's standard commodity plays the same role, as wheat in the "Essay on Profits", and for the same reasons: it is an homothetic commodity. It seems therefore important not to confuse both theories. However we will not go as far as saying that Sraffa's system is a regression from this last system, to the extent that wheat offered at least the advantage of being an invariable numéraire with regard to changes in the production system.

We must nevertheless be conscious of the very peculiar character of the underlying assumptions that allow for the construction of the standard commodity. These assumptions are always implicit in "Production of commodities", because, as Joan Robinson rightly pointed out in the review of Sraffa's book that she wrote in 1961, shortly after its publication, there is no preliminary discussion of assumptions in this book.

Going back to the assumptions regarding wages, Sraffa first adopts, for just a while, the assumption that part of wages consist in subsistence goods, the remaining part being taken from the surplus. We already showed that it is equivalent, on a formal level, to the conception adopted later, of wages totally paid out from the surplus. However, in case of a variation of the part of wages coming from the surplus, it requires a supplementary assumption according to which the share of wages made of subsistence goods is not modified, because otherwise it would become impossible to build up the standard commodity. But this in turn implies that any variation in $w$ must necessarily translate into a change in the budget coefficients of subsistence goods such as the quantity of these goods consumed by workers remains unchanged, in spite of the change in the wage level. In other word the income elasticity of subsistence goods must be zero, whatever the variation in w. One must admit the eminently restrictive character of such assumptions and can understand therefore why Sraffa, although he did not discuss this kind of assumptions, did not pursue this "hybrid" case.

What to think now of the assumption according to which the wage is defined in terms of the standard commodity, this one being made of basic goods (means of production), owing to the conception of wages finally adopted by Sraffa ? If we took it literally, it might mean that workers are paid with the means of production corresponding to their share of the surplus, and hence they would become capitalists, which is absurd. In order not to face such a nonsense, one must get back in the analysis the specificity of the goods consumed by workers, although they are rejected by Sraffa into the limbo of non-basic goods. We arrive therefore to this paradoxical observation, even though it is an implicit assumption of the system: although labor enters directly into the production of every commodity, those commodities that are bought by workers are not basic goods, otherwise they would become capitalists.

We are thus facing the following alternative:

- Either we take into consideration only the sub-system of basic commodities, which are in fact the means of production of this system. Then wages are nothing else than a fraction of the means of production and workers become ipso facto capitalists, which is absurd;

- Or we admit this last assumption, according to which commodities consumed by workers are non-basic ones, but we need to re-introduce in the analysis the sub-system of non-basic commodities. What happens then when the wage is modified?

We could think first that the quantities of non-basic commodities consumed by workers will be modified, but such a modification can be defined only if the quantities of these goods are determined. However, as we saw previously, this determination implies to have an additional equation identifying $w$ and the price of these goods, which makes the system over-determined. As a consequence, the variation in the quantities of consumption goods remains undetermined. In any case this variation (undetermined) must necessarily cause a variation in the use of basic commodities needed for the production of these non-basic goods, and then a variation (undetermined) of the standard system. In fact all this should not be a surprise, since Sraffa warned us that changes in the economic system were outside the scope of his analysis. We must therefore admit that a change in distribution leaves the produced quantities of commodities and the economic system unchanged, a very peculiar and restrictive hypothesis indeed.

If the produced quantities are thus given, how can we imagine a variation in distribution? Some authors have considered that it would be sufficient to add the assumption that workers can save and that capitalists can consume the same goods as workers. But this would imply that any change in distribution would be equivalent to a change in workers' savings and capitalists' consumption, in identical and opposite proportions, and furthermore such an hypothesis does not hold, since it comes down to adding to the production price system additional equations: these equations would identify total savings (from workers and capitalists) to the global price of basic commodities, and total consumption to the global price of non-basic commodities. As we already saw, such a system, and in particular the distribution variables, would be overdetermined. The production price system is indeed a close system[4], in which it is irrelevant to inject assumptions regarding the realization of the net product, realization being external to the logic of the system. Therefore any variation in distribution, especially concerning the wage, is illusory in such a system. And the invariable standard of production prices, invariable to illusory distribution changes, is itself but a dummy construction.

[4] As soon as the profit rate has been given - and it alone can be given, the system has as many equations as unknowns and is completely determined.

### 4.2. The blurred significance of production prices

First, we must now emphasize something which should not have escaped the reader familiar with Sraffa's theory, which is that this whole chapter has addressed only part I, entitled "Single product and circulating capital", of Sraffa's book. On this somewhat limited basis, what can be the significance of production prices?

On the negative side, they are characterized mainly by a series of impossibilities:

- Impossibility to provide an instantaneous determination of distribution, precisely because the link between production and distribution that existed in the Ricardian system is severed, albeit rightly so: indeed the Ricardian conception of wages does not constitute an economic theory of distribution, even though it provides a measurement for a given state of distribution;

- Impossibility to reflect otherwise than in an illusory way the variations in distribution, the origins of which remain in any case unexplained;

- Impossibility to analyze the accumulation of capital, i.e. the reproduction of the capitalist mode of production and therefore to provide any analysis of growth, as well as technical progress;

- Impossibility to reflect, if not the existence of a stable structure of differentiated profit rates, at least the process of differentiation of profits rates.

The only object for which Sraffa's theory of production prices is adequate is the measurement of production prices, but it refers back to a field which is external to the analysis, with the indetermination of the given variable that is the rate of profits.

On the positive side, the sub-title of Sraffa's book: "Prelude to a Critique of Economic Theory" seems nevertheless to be fully justified. Maybe not so much because the book has allowed on certain points to demolish the dominant neo-classical theory, but because it demonstrates, by reduction to the absurd, the vanity of attempting to build up an economic theory based on exchange ratios between commodities taking into account the existence of profit, without asking the question of its nature. It is thus clearly a prelude.

Moreover this prelude is not without bearing some fruits, because it allows to get rid of the neo-classical misconception of capital as a factor of production paid for at its marginal productivity. The system of production prices again clearly demonstrates the vanity of this attempt, as soon as we are outside of an economic system with only one good, because the measurement of capital made of heterogeneous goods depends on the profit rate, and is therefore unable to define this variable. But this analysis, which forbids us to consider capital as a factor of production, does not allow in itself to define clearly the concept of capital. If we know indeed that capital is made of means of production fetching a profit, i.e. of basic commodities, we do not understand what is this profit which is indispensable to provide a definition of capital as such.

These problems come obviously from the initial assumptions of Sraffa's analysis, and from the fact that they do not respect some of the basic principles that were highlighted at the end of the previous chapter. To be sure, and this is some progress compared to Ricardo's paradigm, workers are no longer paid in wage-goods. But they are still not paid in money, because there is no money in this theory. Workers are paid by sharing a part of the standard net product, made itself of the standard commodity, and this standard commodity cannot play the role of money because it is not invariable to changes in the system. In other words the standard commodity is not a scalar, but has the dimension of a basket of basic commodities. Therefore it cannot play the role of a bridge between past, present and future. If there is progress compared to Ricardo, it is therefore small progress.

And this all the more so that the rate of profits is no more a residue, but on the contrary has to be given before prices are determined, which implies that exchange, i.e. the realization of the commodities at such prices, has taken place. One cannot see indeed how any producer could make any profit without having sold first his commodities on the market. It follows that production and exchange are not clearly distinguished and therefore "collide" theoretically: production prices are supposed to be known before exchange takes place, but can be calculated only when the rate of profits is known, i.e. after exchange has taken place.

If we are brought to such problems and contradictions, it might be the sign that there are in fact deeper flaws in Sraffa's conceptual foundations. That it is indeed the case, in particular as regards the very conception of production, will be shown in the next chapter.

# Chapter 4. Commodities do not produce commodities: the biased concept of production in Sraffa's theory

In this chapter we will go further in exploring Sraffa's theory of production as a circular process and as production of a surplus, and show why it entails a number of theoretical contradictions, in particular with the introduction of fixed capital. The introduction of land and non-produced resources also creates an intractable logical problem, which ends up by invalidating Sraffa's whole theoretical system.

The previous chapter discussed the links between distribution, prices, and the invariable standard in Sraffa's theory, and highlighted a few problems concerning these questions. But it confined itself to the first part of *Production of commodities by means of commodities*, where there is only circulating capital, without questioning the underlying conception of production. It did so because Sraffa himself does not clarify his conception of production at any time neither within this first part, not in the two other parts of the book.

Our interpretation of this strange attitude (not defining production at first while dealing with *production* prices) is that Sraffa's theoretical target was the neo-classical theory of distribution, seen as a mere extension of the neo-classical theory of prices to this particular kind of goods that are "factors of production". In fact in 1960, when Sraffa published *Production of commodities by means of commodities*, it was perceived by a number of economists as a kind of theoretical bomb. Indeed it hit a heavy blow to the neo-classical theory of distribution, by showing that in a system with multiple heterogeneous goods it was impossible to define capital as a factor of production of which the marginal productivity was determining the rate of profit. However, doing so implied keeping some assumptions corresponding to the neo-classical paradigm, as regards the very nature of production and of the goods involved in this process, in particular fixed capital. To what extent this was done unknowingly is difficult to say, but it is this preservation that brought about a number of flaws.

In the present chapter we will continue to revisit Sraffa's theory, trying now to go further by digging up to its very foundations, i.e. its most basic assumptions regarding the nature of production. This will lead to analyzing the definitions of the different categories of goods appearing in Sraffa's book, and playing a role in the production process as described by him: basic goods, subsistence goods, luxury goods, consumption goods and fixed capital goods. We will show why this classification is wrong, explaining in particular why subsistence goods should not be considered as intermediate goods, that consumption goods are final goods which are not part of a surplus, and that there is no surplus of intermediate goods.

We will demonstrate then that the method chosen by Sraffa to introduce fixed capital goods through joint production does not allow for the existence of a surplus for this kind of goods, and leads to several contradictions. These contradictions entail some serious consequences: neither the definition of basic goods nor the notions of surplus and of Standard commodity can be upheld at the level of the whole economic system. It comes therefore as no surprise that the problems created by the introduction of land or other non-produced means of production cannot be resolved. We will end up by showing how all these limitations explain why Sraffa's system has remained over the years such a puzzle, on which nothing has been built: in particular the system is incompatible with a Keynesian theory of money.

## 1. Sraffa's approach: production as a circular process

It is only in Appendix D of *Production of commodities by means of commodities,* i.e. in the very last pages of the book, that Sraffa finally discusses the question of production! He connects his conception of production to the old classical economists Quesnay and Ricardo who regarded the system of production and consumption as a circular process. He even adds that this conception, which he develops in his book, is "in striking contrast to the view presented by modern theory, of a one-way avenue that leads from factors of production to consumption goods" (Sraffa, 1960, Appendix D, p. 111-113).

As we indicated in the previous chapter, Sraffa says that his system is a generalization of that of Ricardo wherein wheat is both a factor of production and consumer good, which allows one to define a surplus, irrespective of the values or prices, and to determine the rate of profit regardless of them. In his system, what he calls the basic goods are generally playing the role of wheat, when, through the construction of the standard commodity, they appear in the same proportions in the means of production and the net product, which thus seems to validate his reasoning by generalizing the Ricardo's case to a system with multiple heterogeneous goods.

It was important to start by recalling this, because, as far as science is concerned, the field of validity of a scientific theory is generally constituted by the field delimited by its own assumptions, the first of them consisting in the definition of its concepts. It is also generally admitted that, for a theory to be considered as scientific, there is a need for consistency, both internally and externally. By this we mean that the theory's assumptions must not contradict themselves, which is the internal consistency, and that the assumptions must have a coherent link with the reality which the theory wants to describe, which is the external consistency.

Everybody will certainly acknowledge that the definition of production is a very important assumption, a fundamental one indeed, and in this regard it should normally have appeared, not in an appendix at the very end of *Production of commodities…*, but on the contrary in an introduction, at the very beginning of the book. However, and quite strangely, this is not what Sraffa does. Before starting to analyze his theses regarding production, let us state again that for Sraffa production means production of a surplus, and that to be able to compare this surplus to the means of production, there is an absolute logical need for the commodities to appear both in the means of production and in the surplus, which is therefore a forced corollary to his circular conception of production.

## 2. Production for subsistence

It is difficult to understand why Sraffa put back to the end of his book something as fundamental as his definition of production, but maybe the reason was that he did not want to put it too close to the first chapter of the book, entitled "Production for subsistence", because in this three-page chapter, there is no surplus, which seems completely contradictory to this definition of production. Indeed what Sraffa wants to show is that there can be a price system, even when there is no surplus. But can there be any production? Certainly not by Sraffa's own definition, unless we realize that the title of the chapter means that there is some kind of "individuals" or "producers" behind the scene, which barely survive (subsist), by the consumption of some of the commodities that are produced.

Saying that, however, which Sraffa implies when he writes that the system includes "the necessaries for the workers", means that on the left-hand side of the equations (like also on the right-hand side) some of the means of production are in fact consumption goods, or more precisely subsistence consumption goods, i.e. some kind of goods that are just sufficient to allow the "producers" to survive.

However this raises another difficulty, which has to do with the definition of consumption, something which, quite strangely also, does not appear in Sraffa's book. Let us therefore give a definition, and then discuss it. This definition is borrowed from Wikipedia (French version), and indicates that:

> "…consumption characterizes the act of an economic agent (consumer) that uses (final consumption) or transforms (intermediate consumption) goods and services. This use or transformation causes the immediate (non-durable goods) or progressive (durable goods) destruction of the items consumed. From a general point of view, consumption (value-destroying) opposes production (value-creating)."

This definition is both interesting and misleading. Interesting, because it establishes a distinction between the production process as an intermediate process, involving intermediate goods and services, and final consumption as a final process. But this definition is also misleading, because it tends to consider nevertheless both processes as consumption, since both would cause the destruction (whether productive or final) of the goods involved. The definition introduces however another distinction between final consumption, which "uses", and intermediate consumption, which "transforms".

What we would like to argue is that this distinction is not a subtle one, but a fundamental one, which should lead economists to reserve the use of the word "destruction" for only the goods which are used for final consumption. A good reason for that is in fact given by the very last sentence of this definition, pointing out that consumption, which is value-destroying, opposes production, which is value-creating. It is indeed logical that a process, like final consumption, which causes the immediate or progressive destruction, and in fact the disappearance of the consumption goods involved, be value-destroying, whereas conversely it is not at all logical that a process which is value-creating, like production, would cause the destruction of the intermediate goods involved. On the contrary, and if we want to stay connected to reality, it is easy to recognize that these goods used in the production process are not really destroyed in the sense that they would definitively disappear, because they are only transformed in the course of this process.

These explanations are not esoteric ones, and we do not want to split hairs on this question, but they are necessary in order for the concepts used to have a connection with the real world. In the real world indeed, final consumption is definitely a destruction. This is obvious for all the non-durable goods, like for instance food, for which there is simultaneity between consumption and destruction, this destruction being also a complete disappearance. But this is also the case for durable goods, even if this destruction can be a very long process for certain goods: in the end there is a complete impossibility for such a durable good to continue to be used as it was initially and during its whole lifespan.

On the contrary, in the case of intermediate goods, like all the raw materials, they most often disappear in the production process in their initial form, but only to reappear under another form at the end of this process, which means literally that they are not destroyed, but only transformed. This is so because their substance is still there, incorporated in the final goods of which they have become an element or a part. Furthermore, in a value-creating process, it is logical for the value of these intermediate goods as a dimension attached to them, not to disappear, but to be transferred to the value of the goods in which they become incorporated.

Evidence of this can be derived from the fact that at the end of the lifespan of a durable consumption good it is more and more frequent that it can be dismantled and recycled, which allows for many of the raw materials which they were made of, i.e. the intermediate goods, to reappear and be transformed again in new production processes, where they are reincorporated into new goods. As for the consumption goods they have been definitively destroyed by this recycling, which confirms their theoretical status of final goods. It must be emphasized that this recycling happens obviously in a completely new production cycle.

Within the production process itself, there are often some tailings, scraps or residues from these intermediate goods, and most of the time, in particular each time that these intermediate goods are valuable, they do not disappear but are recovered to be put back in the process, rather than being wasted. Therefore the word “destruction" is clearly inappropriate to name what happens to intermediate goods in the course of the production process, since they do not disappear completely, but on the contrary can reappear either during this process or at the end of the lifespan of the final consumption goods in which they are incorporated. It is only their original shape or form which are partially modified (rather than destroyed), to be transformed into other ones.

All this explains that final consumption goods cannot at all be treated conceptually like intermediate goods, since these last ones are not consumed but only transformed in the production process, and as such are not the object of an “intermediate” or “productive” consumption, but of a productive transformation. Therefore final consumption goods should not in any case be put on the left-hand side of the equations of the production system, which represents what enters into the production process, even if these goods are "means of subsistence", which can constitute as we have seen a part or the totality of "wage-goods".

Let us conclude that the analysis performed in this very first chapter of *Production of commodities by means of commodities*, devoted to "production for subsistence", is seriously flawed. Indeed there is no surplus, and therefore there should be no production, according to Sraffa's own definition of production. Even if one thinks that there is a kind of notional surplus, in the form of subsistence goods which would be both on the left and on the right-hand side of the equations, this would be contradictory to the demonstration, which has just been carried out, that consumption goods by their very nature are final goods which come out of the production process and as such can never enter into this process. This should be kept in mind when we now go further into the analysis of the concept of production in *Production of commodities by means of commodities*.

## 3. Production with a surplus

It is a little ironic in this context to observe that it is in the following chapter of *Production of commodities by means of commodities*, entitled "Production with a surplus" that Sraffa writes that the introduction of a surplus makes "the system becomes self-contradictory" (Sraffa, 1960, p. 6). But this is because the allotment of the surplus, which is supposed to be made by a uniform rate of profit, cannot be determined before the prices, and *vice-versa*. However Sraffa quickly resolves the difficulty by explaining that in fact both are determined simultaneously by the same mechanism.

The interesting thing in this chapter is rather that it shows that the emergence of the surplus has one effect which is the appearance, as part of this surplus, of a new class of products, "which are not used, whether as instruments of production or as articles of subsistence, in the production of others". As we mentioned previously, Sraffa names them "luxury goods".

Building upon the demonstration performed in the previous section, it can immediately be observed that, strictly speaking, this definition of luxury goods applies in fact to all consumption goods. This follows because we have shown that the fact that a consumption good be a subsistence good does not in itself transform it into an intermediate good, which belongs to a totally different category of goods, those that can be put on both sides of the equations representing a production process. It follows from this fact that as far as final consumption goods are concerned, no surplus can be determined as a difference between the quantities of these goods which are on the right-hand side and the quantities supposed – wrongly, to be on left-hand side of the equations.

To be sure, there are some particular goods or services which can be in the same physical form, both means of production in the form of intermediate goods (circulating capital), and final consumer goods. They are mostly fluids (water, energy), or some services. But electricity itself does not produce electricity: we know that this is not its use in the production process that makes it reappear for a larger amount at the end of this process. Even in this not so usual case, the total quantity of this particular kind of goods or services produced during a period totally disappears during the same period either as intermediate consumption or as final consumption, without therefore the apparition of a surplus or net product.

In this respect, and to be more precise, even the example of wheat is misleading, because in the real world wheat in the form of seeds (as an input) is increasingly a processed product which has undergone some specific treatments, or even has been genetically modified etc., and therefore is different from wheat as a pure consumer product. This last one is itself eaten only after being transformed (ground, conditioned, etc.), and this in a way which makes it a different product from wheat which has just been harvested. Let us also point out that for most agricultural products other than cereals, grains or seeds are in any case very different from harvested products.

Having arrived at this stage, and putting aside fixed capital for the time being, to adhere to Sraffa's approach, we must understand that there is no surplus either for intermediate goods: certainly the quantities of intermediate goods used in the production of these intermediate goods themselves are smaller than the total quantities of these goods which are produced. But it is obviously because the rest of these intermediate goods are used in the production process of final consumption goods or fixed capital. Moreover, in a self-replacing state, where Sraffa repeatedly locates his theory, there cannot be any place for stockpiles of intermediate goods, which are produced in the exact quantities needed for both their own production and the production of final goods. Therefore in a given period the total quantity of each of the intermediate goods which enters the production process, and appears on the left-hand side of all the equations of the production system using this intermediate good, is exactly equal to the quantity which comes out of the system, and appears on the right-hand side of the equation of the industry producing this good in the same system.

## 4. The introduction of fixed capital

To begin with the definition of fixed capital, given at the beginning of Chapter X of *Production of commodities by means of commodities* (§73), Sraffa does not elaborate a lot, and limits itself to writing that:

> "…we shall regard durable instruments of production as part of the annual intake of a process, on the same footing as such means of production (e.g. raw materials) as are entirely used up in the course of the year; while what is left of them at the end of the year will be treated as a portion of the annual joint product of the industry." (Sraffa, 1960, § 73, p.75).

There is therefore no doubt that fixed capital is considered in this book as a particular kind, or a sub-species, so-to-say, of intermediate goods, with the only difference that the life of the components of fixed capital is longer than the life of intermediate goods, which disappear (are transformed) in the production process at the same time as they transfer their value to the goods produced. Fixed capital is indeed made of "durable instruments", or machines, which implies that their use in the production process is a progressive one which lasts up to the end of their lifespan, during which they progressively transfer their value, up to the end of the production process in which they enter. The only difference with circulating capital seems therefore for Sraffa only a question of time, having to do with the longer durability of their lifespan and therefore of their participation in a production process. If one wonders why time is so important, the only coherent explanation is that the rate of profits is defined not only as a percentage, but as a percentage per unit of time (in fact, the year).

At a conceptual level, it is however extremely confusing to restrict the differences between both types of capital to just this single element of time. Indeed, as we have already pointed out earlier, intermediate goods participate in the production process in such a way as they are effectively and entirely transformed in this process, where they can no longer be found under their initial form at the end of it, because their material substance itself has been incorporated in the final goods which they have contributed to produce. In other words they enter into the process, but do not come out of it.

One must immediately recognize that this is not at all the case for fixed capital goods, which in the real world do not participate in the same way in the production process, whatever their durability and independently of the duration of this participation. Indeed fixed capital goods never disappear in the production process, where their true role is not to be incorporated in the structure of final goods, but on the contrary to participate in the transformation of the intermediate goods, which is an extremely different thing. In the real world, the tangible and visible role of fixed capital is not to transfer its value to the products, or to deserve the payment of a profit rate, but to help increase the productivity of the main actor of the production process, which is human labor.

Since the whole treatment of fixed capital by Sraffa is based on the particular question of the age of machines, as we shall see below, it is also important to note that in the real world there is no such thing as a once and for all clearly defined age of a machine, which would remain the same during its whole participation in the production process. In *Production of commodities by means of commodities*, we are clearly at a technical level, but even if we stay at this level there is no such thing as a predefined lifespan or age of whatever machine or piece of equipment, since this age will depend on many factors, like the intensity of the use of the machine.

A machine working eight hours a day with a single shift obviously will not have the same lifespan as exactly the same one working 24 hours a day with several shifts. The quality of maintenance, which can vary from a firm to another due to multiple factors, as well as during its own lifetime, can also greatly alter the real life duration of an equipment. Furthermore most machines are not even used during the whole duration of their nominal lifetime, for the well-known reason that they quickly become obsolete. New and cheaper or more ‘productive’ machines are indeed produced each year and after a few years make production with older machines become no longer competitive. Hence the replacement of machines becomes indispensable, even a long time before the day when they would have been worn out. The result is that in the real economic world the composition of a collection of machines and the way they are used vary continuously.

From a theoretical point of view all this is all the more annoying that the whole treatment of fixed capital by Sraffa, in § 76 of his book, consists in establishing a sub-system based on as many equations as there are separate processes which correspond to the successive ages of a given machine. He thus uses a method first developed by Torrens, who had introduced the notion of "residue of capital" in his "Essay on the Production of Wealth" (Torrens, 1821, pp. 28-29). Sraffa also indicates that "the quantities of means of production, of labor and of the main product are equal in the several processes in accordance, with the assumption of constant efficiency during the life of the machine" (Sraffa, 1960, p.78). This is hardly compatible with the fact that neither the lifespan nor the efficiency of a machine can ever be determined at any given point of time.

Nevertheless this is not the main criticism that can be made of the treatment of capital by Sraffa. To consider it, let us recall that for each machine, there is a sub-system having as many equations as the successive ages of this machine, for age 0, 1, 2,...*n*, where *n* is the lifetime of each machine. Each of the *n* equations represents the joint production of good *G* and of a machine of age 1 to *n* (on the right-hand side) by a machine of age 0 to *n* – 1 (on the left-hand side). This sub-system covers a whole range of years, as many as the lifespan of the machine. With a proper treatment, Sraffa then removes *n* – 1 equations corresponding to the machines of intermediate ages to finally obtain a single equation containing only the newly-produced machine, of age 0. This equation is the following (see § 76 of the book):

$$M_0 p_{m0} \frac{r(1+r)^n}{(1+r)^n - 1} + \left(A_g p_a + ... + K_g p_k\right)(1+r) + L_g w = G_g p_g$$

The first term represents the annual depreciation of the machine, i.e. the value supposed to be transferred by the machine to the final good G for a given year *n*.

However, when we come back to a whole system of production for a single given year, we therefore turn, as Sraffa does in § 83:

"from the standpoint of the life-progress of a single machine to the stand point of a complete range of *n* similar machines each being one year older than the preceding one, and thus forming a group such as we might find in a self-replacing system. The requirement that the life-sum of the depreciation quotas should be constant and independent of the rate of profits is now embodied in the fact that under all circumstances such a group is maintained simply by bringing in a new machine each year" (Sraffa, 1960, § 83, p.83).

All this is quite coherent and means that in this self-replacing production system, we have *n* machines in operation (from age 0 to age *n* – 1), with on the left-hand side of the *n* equations like the one above *n* different depreciations. As Sraffa clearly demonstrates also, and this demonstration is right, the price of the machines of the intermediate ages can vary with the rate of profits, but for a given rate, as it appears from Figure 6 (in § 83), the sum of these different prices is always equal to the initial value of the machine $p_{m0.}$

Up to now, everything might seem perfectly correct, but a problem arises when we realize that, when we come to the calculation of production prices, we are no longer in the sub-system in which it was innocuous to make appear *n* different machines of age 0 to *n*–1 (the machine of age *n* being withdrawn from the production process). Indeed, in a self-replacing state there is one and only one machine, of age 0, which is produced in a given year, with its own equation representing the conditions of its production. In this equation the quantity of the machine $M_0$ of age 0 (since it is new) appears on the right-hand side. And for a good G, and/or any other good in the production process of which this machine is used, including its own, there are in total as many equations for each good as the number of years *n* corresponding to the lifespan of this machine $M_0$.

Turning now to the way this whole system works, on the left-hand side of each of these equations there is a value:

$$M_0 p_{m0} \frac{r\left(1+r\right)^n}{\left(1+r\right)^n - 1}$$

which represents the contribution to production or the "transfer of value" of these machines of age 0 to $n - 1$ to the price of the goods that they help produce. We also know that the sum of these depreciations is $M_0 p_{m0.}$ But we are now in the real system itself, where only new machines of age 0 are really produced each year, each of these different machines with its own equation of production, and no longer in the sub-system where there was joint production of the machines of various ages. This implies that there are no equations corresponding to the real production of machines from age 1 to *n*, because these machines have in fact been produced previously, in earlier periods. This means that on the right-hand side of all these equations these machines of age 1 to *n* cannot and do not appear.

The interesting thing is that, as a consequence of this situation, and for the whole production system, when we sum up all the equations in which a machine Mo appears, we have on the left hand-side a value $M_0p_{m0}$ corresponding to the sum of the depreciations, and on the right hand side exactly the same value $M_0p_{m0}$ of the newly produced machine $M_{0.}$ This signifies clearly that, whatever the lifespan of the machines, in a self-replacing state the value of the quantity of the machines which is produced in each period corresponds exactly to the quantity which is supposed to "disappear" in the production process, where this value is supposed to be transferred to the value of the goods that they contribute to produce, to the tune of the total depreciation affecting the same machines of various ages.

The inescapable conclusion of this analysis is that there is no such thing as a surplus for machine *M*, nor for any machine, since the demonstration performed so far can obviously be generalized. Although all this can easily be understood intuitively (since in a self-replacing state there is no net production of fixed capital, but only its replacement), we have therefore analytically established a very important result, which is that in Sraffa's system as long as fixed capital is supposed to transfer its value to the goods produced, *there is no surplus of fixed capital*. This means reciprocally that, supposing that there is a surplus, it cannot include any fixed capital good.

## 5. The contradictions of Sraffa's system with fixed capital

At the point where we have arrived, we cannot but observe that we face a double contradiction: indeed Sraffa defines production as a circular process (meaning that what comes out of the process also enters or rather re-enters into it) through which a surplus is created.

However, what is to some extent circular in the production process, as exposed in *Production of commodities…*, is the production of intermediate goods and fixed capital, this last one being assimilated to a particular type of intermediate good with a span of life longer than the production period, because both are supposed to enter into and come out of the same production process. This explains why they appear on both sides of the equations describing this process. But at the same time, in a self-replacing state, where the whole Sraffa system is located, and as we have already shown, there is no such thing as a surplus of either intermediate goods or fixed capital, because fixed capital is itself a kind of intermediate good, and therefore there should be no production according to Sraffa's own definition of this concept.

As for consumption goods, we have already shown that these goods do not enter into the production process, but are destroyed (either instantaneously or more or less quickly, if they are durable) in the consumption process. It is clear therefore that there is no circular production process concerning consumption goods. However, since these consumption goods cannot be found on the left-hand side of the equations describing the process, but appear necessarily on the right-hand side, the difference between both sides, for consumption goods only, is necessarily made of the whole production of these consumption goods. One might possibly say that it constitutes a "surplus", although it would be an artificial one, because for the process as a whole the quantities of intermediate goods should normally be deducted from this "surplus", in order to obtain the true surplus of the system. But this is obviously impossible, because of the heterogeneity of the goods on both sides of the equations.

This brings us to a paradox: it is the Standard net product which has to be divided between wages and profits, but because what might actually be called a surplus and constitutes this net product is made only of consumption goods, then these profits should be devoted entirely to buying consumption goods! More importantly, this obviously contradicts the statement made by Sraffa according to which "the rate of profits in the Standard system thus appears as a ratio between quantities of commodities irrespective of their price" (Sraffa, 1960, bottom of p. 24). Indeed there is no such thing as a ratio between consumption goods and intermediate goods!

Another contradiction comes from the fact that Sraffa's system is built in such a way that:

> "…the ratio of the net product to the means of production would remain the same whatever variation occurred in the division of the net product between wages and profits and whatever the consequent price changes" (Sraffa, 1960, § 28, p. 24).

However, even with all other parameters remaining unchanged, if the rate of profit changes from one period to another during one of the n periods corresponding to the lifespan of a machine, the depreciations in the value of this machine already registered during the preceding periods will naturally remain unchanged, but the amount of the depreciations that will take place in the following periods, after the change in the rate of profit, will also change necessarily It follows that the sum of the *n* depreciations for a given machine will no longer be equal to the value of the machine, but will be lower or higher.

The same phenomenon would take place each time that the actual lifespan of a given machine would become shorter or longer than the original or nominal one. In both cases, either because of a change in the rate of profit or in the real life span of a machine, the resulting change in the overall amount of the depreciations will change the proportions in which the machine enters in the production process, and therefore the nature of the Standard system and the whole price system! This is indeed not compatible with Sraffa's system.

Before going further, let us go back to the definition of fixed capital as a kind of "long life" variety of intermediate goods, since we can now better realize that this vision is incorrect. Indeed what appeared on the left hand side of the equations, and that we did not questioned, has to be revisited, since it is clear that it was not the machine itself, but a purely virtual element, i.e. its depreciation, which varies, as Sraffa explains, both with the age of the machine and with the rate of profit. The only thing which does not vary, as we already pointed out, is the sum of the depreciations for the $n$ machines of age 0 to $n - 1$ in operation for a given period, which is always equal to $p_{m0}$, but only for a given rate of profit. This virtual quantity introduces an irreducible element of heterogeneity with the other intermediate goods: although they disappear in the process in their initial form, but only to be transformed, intermediate goods enter into it as real goods, and not as virtual ones.

# 6. The consequences for the Standard system

## 6.1. The notion of basic goods

An important conclusion that we can draw from these observations is that most goods are not basic goods, whose main property, as defined by Sraffa in § 6 of *Production of commodities…* is that they "…enter (whether directly or indirectly), in the production of all commodities". Indeed, since these goods are also produced, it means that one has to find them both on the left and right-hand side of the equations defining the system of production. We have shown that consumption goods do not meet this criterion, and at this stage we must admit that it is also the case for fixed capital goods. Indeed it is not these goods themselves, but only their depreciation that enters into the production process, and this depreciation, as defined by Sraffa, is not a tangible element, which could be put on the left-hand side of the equations in a non-contradictory manner. From this we must derive that fixed capital goods have to be considered as final goods, which like consumption goods can be found on the right-hand side, but not on the left-hand side of the equations.

In fact this corresponds to the treatment applied to depreciation by Keynes, who devotes the whole appendix of Chapter 6 of the *General Theory of Employment, Interest and Money* to the question of what he calls "the user cost". He defines the user cost "…as the reduction in the value of the equipment due to using it as compared with not using it" (Keynes, 1936, p. 66), and indicates that aggregate income, equivalent to aggregate supply price, is equal to A – U, or "as being net of aggregate user cost". It is clear that this user cost is quite similar to what Sraffa calls "the annual charge to be paid for interest and depreciation" for a machine, and that as we have seen, unlike Keynes, he includes it wrongly in the production cost.

As a result, the only goods that can be found on both sides of the equations are clearly the intermediate goods, strictly defined as goods that disappear (in the sense only of being transformed) within the production process, where they can be used in the production of any other good (final or intermediate) as well as in their own production (directly or indirectly). They are thus the only goods that can be called basics in the sense that Sraffa gives to this word. There is however no reason why there should be any surplus of these intermediate goods, which are produced for each of them in the same total quantity as the sum of the quantities used in all the various industries, as Sraffa calls them.

However, as Sraffa himself shows it, the fact that only intermediate goods *can be* basics does not mean that all intermediate goods *are* actually basic goods. As soon as § 6 of his book, Sraffa indicates indeed rightly that apart from luxury goods per se, there are two kind of intermediate goods that do not enter in the production of *all* commodities, and therefore are not basics: these are luxury goods that enter only in their own production, or only in the production of other luxury goods. Things turn out to be a little more complicated in part II of the book ("Multiple-Products Industries and Fixed-Capital"), where joint production is introduced, before being used to deal with fixed capital.

In chapter VIII, "the Standard System with Joint Products", of "*Production of Commodities by Means of Commodities"*, Sraffa demonstrates that there is also a Standard system and a Standard commodity in case of joint production, even though the Standard commodity can include negative quantities for some individual commodities. He indicates also that there are still three types of non-basics:

1. "Products which do not enter the means of production of any industries" (…).

2. "Products each of which enters only in its own means of production" (…).
3. "Products which only enter the means of production of an interconnected group of non-basics" (Sraffa, 1960, §57, p. 57).

Category 1) corresponds for Sraffa to what he calls luxury goods, but that we prefer to call final goods, for reasons which should be now easily understandable. For the two other categories, and in order to reformulate this description in a somewhat simplified way, we must consider that if there are intermediate goods that do not enter in the production of other intermediate goods, and enter only in the production of final goods (consumption goods or fixed capital good), and that we could name for this reason "ultimate intermediate goods", then they cannot be basic goods, and as such cannot be part of the Standard commodity. This leads to recognize that all intermediate goods cannot be basic goods, that in other words there is no identity between the set of intermediate goods and the set of basic goods of a given economic system, and to further restrict the content of the Standard commodity.

### 6.2. The notion of surplus

To be sure, some particular goods might prima facie be considered as basic goods, with an amount produced even greater than the amount used as an input (ignoring the differences already mentioned between their nature as an input and as an output). This is the case of some agricultural products.

But, as we have already shown in the second chapter about Ricardo's conception of production (see above p. 33), it appears so because there is a biological mechanism of organic production which as such does not come obviously from spontaneous generation, but precisely from the transformation of elementary goods, free or not. Indeed, using Lavoisier's formula, who has no reason to be false in economy, "nothing is lost, nothing is created, everything is transformed." These elementary goods are the oxygen and carbon in the air, Hydrogen and oxygen from the rain, nitrogen, potassium and phosphorus in soils, among others, which thanks to the biological mechanism of photosynthesis (among other processes at work), allow to harvest some agricultural products in higher quantities than those which have been sown. But the difference in quantities is used in the production of other goods, and in terms of material balance, the overall process is balanced, and it cannot logically be otherwise.

The same thing is true for production in its economic sense. Even for agricultural products, apart from the part that is self-consumed by small individual producers, products sold on the market for final consumption are not formally those harvested, because they have passed through various successive stages of preparation, packaging and transportation, which make them different from those goods which have been harvested, so that for them the concept of the surplus as a difference is not appropriate. As for mineral raw materials, mineral deposits already exist underground prior to their extraction, and it is clear that there can no more be any surplus in the form of any directly consumable good for any ore whatsoever.

These observations have an important consequence on the very nature of what can be called a surplus. They do not prevent us from considering that production is a circular process, at least for the goods that are part of the rightly-named “circulating capital”. In their case, one could even imagine that there is a surplus, but that would only be true for the subsystem, necessarily incomplete, which produces the circulating capital with circulating capital, since only a portion of the circulating capital is used in its own production process. But this surplus is a purely artificial one, it is an artefact, deriving from a mathematical property of this sub-system.

Indeed, at the level of the production system as a whole, all the circulating capital is used: the portion that is not directly used for the production of the circulating capital itself obviously enters in the production of final goods: consumer goods and fixed capital. And this circulating capital is always fully utilized: even in the case of expanded reproduction, the circulating capital produced in excess during a period compared to that produced during the previous period is fully used in the increased production of various goods, be they intermediate or final, during this same period. Indeed production prices should be considered actually as reproduction prices, which concern only a single period, during which there cannot be any surplus of intermediate goods (assuming the absence of stockpiles).

As for final goods, since they do not enter as such in the production process, they exist only as an output of the system, and not as an input. Therefore, it is true that we might think of a difference at an individual level between the output and the input: since for each of them this input is zero, the total quantity produced might be considered as a kind of individual surplus. But if we did so we would forget to deduct all the intermediate goods that are on the left-hand side of the equation of the system, and that have actually entered, directly or indirectly, into the production process of this final good. Similarly, at the level of all of these final goods, this is not true either, because at this global level all of the intermediate goods which enter as inputs into the whole production process have to be deducted from the whole output, which is impossible to perform due to the heterogeneity of both types of goods. As a consequence, a so-called surplus cannot be quantified, and has therefore no logical meaning.

### 6.3. The notion of Standard commodity

We already explained in the last chapter how Sraffa builds up his Standard commodity, as an invariable Standard of prices. Let us recall that Sraffa devotes Chapter IV of his book to this Standard commodity, which he defines as a commodity that would not itself change in value when the distribution between wages and profit changes. He notes in § 24 that the perfect composite commodity that could play this role:

> “…is one which consists of the same commodities (combined in the same proportions) as does the aggregate of its own means of production.” (Sraffa, 1960, § 24, p. 21).

In § 26 he calls the set of equations taken in the proportions of the Standard commodity, the Standard system. He continues by saying that:

> "…in any actual economic system there is embedded a miniature system which can be brought to light by chipping off the unwanted parts" (Sraffa, 1960, §26, p. 22).

and he adds that: "…this applies as much to a system which is not in a self-replacing state as to one which is". He then: "…takes as unit of the Standard commodity the quantity of it that would form the net product of a Standard system employing the whole annual labor of the actual system", calling it "the Standard net product."

Finally in § 28 Sraffa defines as the Standard ratio:

> "…the rate by which the total product of the Standard system exceeds its aggregate means of production, or the ratio of the net product to the means of production of the system" (Sraffa, 1960, § 28, p. 23), underlining that:
>
> "… the possibility of speaking of a ratio between two collections of miscellaneous commodities without need of reducing them to a common measure of price arises of course from the circumstance that both collections are made up in the same proportions – from their being in fact quantities of the same composite commodities." (Sraffa, 1960, § 28, p. 23-24).

It is easy to see that all this demonstration can be carried out only because the very first assumptions made by Sraffa establish an *ad hoc* classification of the goods, where all the goods, apart from the luxury goods as he names them, can in effect play both roles of means of production and of final products, which entitles them to being basic goods (except for some particular kinds of intermediate goods which are non-basics). This is what leads him to his particular treatment of fixed capital goods. However, we have shown that this classification is wrong, and that the only goods which are simultaneously means of production and products of the system are the intermediate goods, as part of the circulating capital, but that there cannot be a surplus of them, because they are (all of them) transformed in the process of production of final goods, be they consumption goods or machines, i.e. fixed capital goods.

To be sure, and we cannot deny it, at a theoretical level it is possible to extract from an actual economic system a Standard system, with its Standard commodity, Standard net product and Standard ratio. But it is a theoretical artefact: the Standard ratio R, for instance, is nothing more than a kind of efficiency indicator or index reflecting the degree or intensity according to which basic commodities are used in their own production, knowing that the remaining basics are not a surplus, but are necessarily used, directly or indirectly, in the production of all the other commodities. Moreover, if we suppose that wages are set as zero, R is then the maximum share of profits when they equal the whole surplus, i.e. the Standard net product. But because it corresponds to a surplus of this particular part of intermediate goods that is made of basics, R cannot be identified to profits, which are supposed to represent at least the fixed capital part of final goods.

It follows therefore that in an actual economic system where no confusion is made between the different types of goods, it is impossible to define, at the scale of the whole economic system, either a Standard commodity such as Sraffa's, or a Standard system, or a Standard net product made of true final goods, or a Standard ratio. Moreover Sraffa tells us in § 43 that:

> "…the last remaining use of the Standard net product is as a medium in terms of which the wage is expressed – and in this case there seems to be no way of replacing it" (Sraffa, 1960, §43, p. 38).

Consequently the fact that there is no such thing as a Standard net product made of true final goods implies in particular that the wage cannot be expressed in terms of this medium. Therefore we cannot but conclude on this point that the famous relation of proportionality $r = R\left(1-w\right)$ between the wage and the rate of profits established by Sraffa, and at the heart of his theory, cannot reflect the true functioning of an actual economic system, where for instance the share of profits should be such as to allow for the purchase of fixed capital goods.

## 7. The intractable problem of land and natural resources

An additional problem arises with the introduction of non-produced means of production, such as land and mines, in Sraffa's system, because this makes it reach again its conceptual limits. Indeed, in the same way that goods appearing only as products but not as means of production – such as consumer goods or fixed capital goods – i.e. in fact all final goods, are not basics, and cannot be part of the standard commodity, conversely land, or more specifically in this case the different land qualities, are among the means of production used to produce agricultural products, but obviously are not part of the product. They are not basic goods, but their existence implies the payment to their owners of a rent which is part of the production price.

When an agricultural product, such as wheat, is produced by several lands of different qualities (fertilities), to determine a price system including the price of this product (wheat), implies that the rent be removed from the system of equations, which would include otherwise more unknowns than equations. This in turn implies that the equation used for the production of wheat in the system of equations is that which corresponds to the land without rent, i.e. the least "fertile" land, but which yields however the average rate of profits. The rent of other more "fertile" lands, with a lower production cost per unit, can then be obtained as a differential rent. It is also clear that the least fertile land is one for which the production cost per unit of output is the highest, or conversely that which produces the lowest amount for a given production cost.

The problem, long known (and recognized by Sraffa himself), is that the determination of the least fertile land depends on the cost of production, i.e. on the price system. Fertility is not an absolute, exogenously given and intrinsic quality of the land, but a relative parameter, which itself depends on the price system. The introduction of land and generally of non-produced means of production and of their income, that constitutes rent, drives the system into a circular reasoning: to determine the price system, we must know which one of the different lands is the land without rent, i.e. the least "fertile" one, which implies to know the price system.

At a second level, the price system varies depending on the variation of the rate of profit or of the wage level. As a result, for different levels of the rate of profit and wages, we necessarily get different price systems, which leads to changes in the production cost of all goods produced with non-produced means of production, such as land or natural resources. These modifications in the price system in turn change the order of land fertility or of the 'productivity' of these natural resources.

Again, we see that "fertility" cannot be defined as a parameter which would be independent of distribution. The equation defining the method of production for the "marginal" land or natural resource can vary with distribution, with different equations corresponding to different marginal lands, depending on these changes in distribution. Therefore the system of equations defining the methods of production when using the least fertile land or other non-produced 'marginal' resources does not remain the same when distribution changes, which implies that the corresponding Standard commodity also varies. Therefore, since the rate of profit or the level of wages are defined in terms of the Standard commodity, it is the very notion of a variation in this rate of profit or level of wages that becomes irrelevant, because it is by definition impossible to compare two different Standard commodities: it is clear that each Standard commodity consists of heterogeneous goods in different proportions, each set corresponding to a different system of equations.

It must therefore be acknowledged that Sraffa's theory fails to provide a coherent conceptual framework which would be able to deal in a non-contradictory or non-circular manner with non-produced means of production and rent, in a way that would maintain the internal consistency of this theory.

## 8. Conclusions on Sraffa's system and the Standard commodity

Sraffa's system is a brilliant construction, which had the great merit to dismantle the neoclassical theory of distribution for which wages and profits are determined by the marginal productivity of labor and capital at the equilibrium. It shows well that when multiple heterogeneous goods and production are introduced, wages and profits have nothing to do with any "marginal productivity", but have to be defined as a share of a net product. It nevertheless results from the above demonstration that the system faces insurmountable problems, in fact partially recognized by Sraffa himself, especially when fixed capital and non-produced means of production are introduced in the system.

Moreover, as soon as a price system is needed to express and determine a state of distribution, and to the extent that the price system itself reflects a state of distribution, we fall into a circular reasoning: we must know the price system to measure distribution, but one must know the distribution to determine this system. Sraffa resolves this dilemma by defining an invariable Standard of value, which is invariable with respect to distribution, and in which distribution can therefore be expressed independently of prices. This invariable Standard is the Standard commodity, which leads to the well-known formula $r = R(1-w)$. But this is achieved by paying a heavy theoretical price:

- The Standard commodity is in fact a composite aggregate of heterogeneous goods, which are by definition basic goods, thus appearing to the right and left of the equations giving the prices. This Standard commodity made of basic goods allows to some extent to reason, with all the limitations which have been reported, as if we were in the universe with a single good of the neoclassical theory or the Essay on Profits of Ricardo, of which 'Production of commodities…' constitutes an attempt of generalization. But whereas prices in this Ricardo's system had the dimension of a quantity of wheat, in the theory of production prices all prices have the dimension of the Standard commodity, which is at the root of the other problems encountered by the theory ;

- Any change in the composition of the Standard commodity is a change of system which is unintelligible, since it is impossible to compare two different Standard commodities, i.e. two sets of different heterogeneous goods, with thus two different dimensions. Each Standard commodity corresponds to a period during which the methods of production cannot change : in the real world this implies that such a period must be extremely short ;

- Since production is generally a continuous process, with a continuous change in the methods of production and therefore in the Standard commodity, this prevents from considering this Standard of value as a unit of measure for money, which has to link the present to the future. This makes it impossible to introduce into this system a money with Keynesian characteristics, meaning – as we already saw in chapter 1, that its elements are pure numbers without any dimension, or scalars. I had shown that Keynes's units of measurement were not compatible with a Standard of value made of heterogeneous goods in a paper included in a collective book published in 1975: "Controversies on the Keynesian System".

- Therefore a change in distribution becomes unintelligible as soon as the methods of production change by the slightest amount, since the composition of the Standard commodity changes simultaneously;

- While prices are expected to ensure the reproduction of the system, no mechanism is provided to connect the expense of wages and profits to this reproduction, unless we assume that workers like capitalists share the same set of basic goods (that are in fact a particular kind of intermediate goods) in the same proportions, which is absurd;

- One could even say that the system cannot be strictly speaking the capitalist system, since the difference between workers and capitalists is suppressed by the fact that what they share is in fact the same Standard commodity.

Let us conclude that on the basis of the above analysis commodities do not produce commodities. As we shall see in more details in the following chapters, labor, helped by capital as a catalyst, produces commodities. Maybe Sraffa was a little aware of that, which would help explain the enigmatic character of his subtitle to “Production of commodities…”: “Prelude to a critique of economic theory”.

# PART II - PRODUCTION AND VALUES

"The laws and conditions of the Production of wealth partake of the character of physical truths. There is nothing optional or arbitrary in them. Whatever mankind produce, must be produced in the modes, and under the conditions, imposed by the constitution of external things, and by the inherent properties of their own bodily and mental structure…It is not so with the Distribution of wealth. That is a matter of human institution solely. The things once there, mankind, individually or collectively, can do with them as they like. They can place them at the disposal of whomsoever they please, and on whatever terms. Further, in the social state, in every state except total solitude, any disposal whatever of them can only take place by the consent of society, or rather of those who dispose of its active force". John Stuart Mill, in Principles of Political Economy (1848), Book II, Chapter I, § 1.

# Chapter 5. Production and values

This chapter will introduce the concept of production as a transformation process, considered as the physical transformation of intermediate goods into final goods. We will show that this transformation comes from the intervention of labor considered as an activity, and that on the other hand, fixed capital as a commodity is not transformed in this process, and is not an activity. Its intervention consists therefore in increasing the productivity of human labor. This is the reason why human labor alone can be at the origin of value, and why fixed capital cannot transfer its value to the product. Values must therefore be defined at the level of production, and human labor can be a standard of value first because it is not a commodity, and secondly because through its payment in money per unit of time it can be reduced to abstract labor, and acquire as such a dimension, which is the dimension of time.

## 1. The analytical distinction between production and distribution

The previous chapters showed that Ricardo as well as Sraffa were unable to build up a coherent theory of production and prices. Ricardo wanted to resolve the question of distribution and its variations within the framework of his theory of production, but failed because of the impossibility of constructing an invariable standard. Sraffa succeeded to some extent in building up this standard, but he had to abandon the idea of determining distribution within his system, because he had to introduce the profit rate from outside, with wages still determined within the system as being the remaining share of a surplus. However this implied a number of theoretical flaws, as was explained in the previous chapter. It will be shown now that there is a way to overcome these difficulties, which necessitates to go further than Sraffa in delinking production and distribution, and to adopt another conception of production. This will allow to solve in a coherent way the problem of determination of values and prices.

Maybe the first economist to separate clearly production from distribution was John Stuart Mill, to whom we refer in the quotation from his “Principles of Political Economy”, published in 1848, and highlighted above at the beginning of part II of this book. Although this distinction is contrary to the whole neo-classical tradition and paradigm for which incomes come from the productivity of production factors, it should go nevertheless without saying since it is indeed both a logical and chronological requirement. For commodities to be distributed among individuals (be they workers or capitalists, this does not matter), they must exist first, which means logically that they must have been produced first. In turn this implies that production must have taken place in chronological time before distribution.

This point, which corresponds to our second basic principle, is so straightforward that there is no need for justifying it through lengthy explanations. Suffice it to say that various modes of production differ in the social organization of production, but that production as a physical and technical process of creation of goods remains basically the same whatever the type of social organization.

To give but a simple example supporting this assertion, let us remark that the way water mills function did not changed when the mode of production based on slavery of the Roman Empire was replaced by the feudal mode of production, or later on by the capitalist mode of production. However nobody will deny that the social rules governing the distribution of the product of these mills, for instance wheat ground by them to be turned into flour, were deeply modified.

As Mill had well understood, the distinction between production and distribution is therefore a methodological requirement. It is to conform to this unescapable requirement that we will deal first with production, and try to deal at this sole level with the question of values.

## 2. Production as a transformation process

### 2.1. The background

It has been established so far that the problems encountered by Sraffa's theory could not but come from its very roots, i.e. from his initial assumption that production is a circular process: this leads him to an erroneous definition and classification of the goods which are part of this process, either as entering into it or as coming out of it. While Sraffa rightly demonstrates that there is no such thing as a "productivity" of capital, this is also what leads him to consider fixed capital goods as if they were intermediate goods, and to put the 'depreciation' of capital on the left of his equations, which implies that fixed capital transfers its value to the products, an assumption rejected by Keynes.

Does this mean that production must therefore be regarded, as Sraffa also wrote, to characterize the alternative view, as "a one-way avenue that leads from factors of production to consumption goods" (Sraffa, 1960, Appendix D, p. 111)? The previous chapter seemed not entirely to validate this view either. It has been shown indeed that production is a transformation process only for intermediate goods, and that part of these goods are used collectively in their own production, which indeed can be considered to some extent as a circular process.

Even in such a process, which concerns in fact all of the primary commodities, and however circular it may appear, the apparent existence of a surplus comes from a kind of "optical effect". Indeed in the production of agricultural products, we do not "see" this kind of particular intermediate goods which are brought about by nature: oxygen taken from the air, water coming from the rain, nitrogen obtained from the soil, etc. In other words, all the components which have been transformed into agricultural products do not appear as inputs on the left side of the production equations, but they do not appear on the right side either as outputs, because they have not been economically produced. Therefore they cannot be part of a surplus. Similarly, the mineral products extracted from underground already exist, and as such are not physically "produced": they are only taken from the geological layers where they have been enclosed for millions of years. But this must nevertheless be considered as a production in its economic sense, because it involves a transformation, if only of location.

At the individual level of each primary commodity, none of them directly enters indeed in its own production: no crude oil is directly used in the production of crude oil, or no apple in the production of apples: these are the apple seeds which are sowed to give apple trees…These are the phenomena which seem to give rise at the individual level of each good to a physical surplus, because it appears as a product on the right hand side of its own production equation, without appearing on the left hand side of the same equation, whereas there is in fact no such thing at the scale of the whole economic system. At this level indeed every intermediate commodity which appears as an output on the right side of the equations describing the production process appears in identical quantities on the left side.

To be sure, a good part of intermediate goods does not enter totally in the production of other intermediate goods, but it is because the remaining part is transformed into final goods, which implies that globally there cannot be any surplus of intermediate goods. As for final goods, they are either consumption goods or fixed capital goods, and for this very reason never enter again to be transformed in the production process, once this transformation has occurred and is over. There is no longer any circularity here, and we stand truly within the framework of a one way avenue, from primary commodities and intermediate goods to final goods, i.e. within a conception of production as a pure transformation process. This is what Sraffa missed to understand, because for him all goods, and even fixed capital ones, are intermediate goods, with the only exception of luxury goods.

If we reduce a final good to all of the components assembled to make it, and these components to their own components, and so on up to the most elementary components, there is no doubt that we will go back to a finite number of primary commodities or raw materials, all of them being from natural origin, and coming from the soil or from underground. These basic elements have indeed all been produced by cultivation or harvesting, in the case of agricultural products, or have been mined and extracted from the ground, as far as minerals are concerned. As we pointed out above, some may have come from the air, like water from rain or various gaseous elements. This confirms that production is primarily a natural and physical process, even though it is simultaneously a technical one.

Another observation which is easy to make is that the production of primary commodities has become over the centuries a more and more complex and capitalistic activity: apart from fixed capital, it requires also more and more intermediate goods, as well as energy, which formally speaking is not a material good, but appears in any production process. Among others, Steve Keen in his blog in the Real World Economic Review, pointed out recently the paramount role played by energy into production: “The abiding weakness of all schools of economics, ever since the Classicals - including today’s Neoclassical and Post Keynesian schools, which are normally at pains to point out how superior one is to the other - is this failure to acknowledge the key role of energy in production” (Keen, September 5, 2016). Keen even goes as far as putting energy at the same level as labor or capital, in a production function derived from the Cobb-Douglas one.

Although the important role of energy is certainly undeniable, one must nevertheless underscore that there is a big difference between labor and energy, which is that energy is itself produced, whereas labor is not a good which would be produced: labor is not on the right side of the production equations. Moreover, as soon as we abandon the idea of identifying labor to a quantity of wage goods (and we showed why this conception is wrong), labor cannot be replaced by any good, be it energy, in these equations.

This leads us to a new basic principle, which is the following:

**Principle 11**: Production, or the activity which creates goods, is a transformation process, which creates new goods through the transformation of pre-existing or already produced goods. As such production does not create any physical surplus.

We know also that the human activity consisting in the production of goods is labor, and we are thus lead to go back to and endorse Adam Smith's view put forward in the first sentence of the introduction to the "Wealth of Nations", and according to which: "The annual labor of every nation is the fund which originally supplies it with all the necessaries and conveniences of life which it annually consumes, and which consist always either in the immediate produce of that labor, or in what is purchased with that produce from other nations." (Smith, 1776, volume I, p. 1).

Let us recall finally that it was also shown in chapter 4 that fixed capital, while it plays an indispensable role in the production process by helping to transform intermediate goods, is not itself transformed in this process: no part of any machine, be it a specific part or a fraction of the whole machine, has ever been incorporated into one of the final goods in the production of which it is involved. Fixed capital is not either a factor of production in the sense that it would have a measurable "productivity", independent of distribution and of the price system.

We can thus formulate our next principle, with its corollary:

**Principle 12**: Labor is the human activity which produces goods through the transformation of other goods, and as such is the only factor of production.

All the above observations therefore explain why capitalist production can only be defined as a labor process of transformation of intermediate goods into final goods, which takes place in the context of specific social relations, where wages are paid in money to workers who do not own fixed capital goods. Fixed capital plays an important role in this process, but as a catalyst which as such is present and physically unchanged from the beginning up to the end of the production process, and increases in considerable proportions the productivity of labor. Therefore we can add to principle 12, as a corollary:

**Corollary to Principle 12**: Fixed capital, as well as energy, is not an activity, but a produced good, and as such is not a factor of production. As a good its specificity is that it is not transformed in the production process, but intervenes into it through increasing the productivity of labor. Fixed capital goods are therefore final goods.

On the basis of these premises, how can we progress in the understanding of the production process? If we adopt a step by step approach, we can start by focusing first on its physical properties, before going one step further in addressing the production process as a social process.

## 2.2. Production as a physical process

Let us consider an economy with a number $s$ of different and heterogeneous goods. These goods are designated by index $i$, and the branches producing them by index *j*. Quantities produced or outputs are designated by $A_j$, with $i = 1,...s$, and $j = 1,...s$. From what we learned from the analysis of Sraffa's system and the preceding observations, we can classify these *s* goods in three main categories:

- The first one, with indices going from 1 to *k*, comprises intermediate goods which enter directly or indirectly into the production of all the other commodities. We will call them intermediate goods of the first type. This category includes all of the primary commodities, and also all the goods that enter into the production of these primary commodities, as well as the goods entering in the production of the latter, and so on. As we previously pointed out these goods in their initial form do not enter usually in their own production: for instance raw coal does not enter as such in the production of raw coal, or coarse wheat in the production of wheat (it has to be transformed first into a different good, i.e. wheat seeds). Moreover, and although it is not always the case, a part of these goods may enter directly in the production of final goods, and the remaining part not: for instance most agricultural food products are not consumed as harvested, because they need some processing, packaging and transportation before they can be consumed. But they still enter indirectly in the production of final goods. Therefore, and for the sake of simplicity, we will not consider these sub-categories, which at this stage are not theoretically important.

- The second category, with indices going from *k + 1* to *n*, comprises intermediate goods which do not enter into the production of the previous category, but enter only in the production of final goods, and can also enter in their own production. They are generally appearing downstream in the production process. We will call them intermediate goods of the second type.

- The third category, with indices going from *n + 1 to s*, is made of final goods, which never enter in the production process, because they have reached their final form, and will not be transformed further. This category includes the two sub-categories of consumption goods and fixed capital goods.

We now designate by index *i* the branch of origin whose outputs are the means of production, i.e. the intermediate goods used as inputs in the production of goods $j$; and by $A_{ij}$ the quantity of good $i$ used for the production of good $j$. This means that the methods and techniques of production are given. They are based on science and technology, which is to a large extent independent of the economic system, and as such are external to the object of economic theory, although the choice of techniques among several available ones can obviously be an economic problem, which is as such relevant for economic theory and certainly needs to be addressed by it.

It must also be noted that at a given point in time, several different techniques can co-exist in a branch producing a particular good, or even in the case of a single producer, if a firm has for instance several different factories producing this same good. This implies that the theoretical level where we stand is neither a pure micro-economic level, nor a macro-economic one, but rather what we could name a "meso-economic" level. This implies also that all the quantities of the various goods used in the production of a given good correspond to the total quantities used in a given branch to product this given good, and that the average proportions in which they are combined may not correspond to the actual proportions of any specific individual producer of this given good.

We also introduce labor, designating by $L_i$, with $i = 1,...s$, the "quantities of labor" used in the production of commodity $i$. The question of the nature of a "quantity of labor" is going to be addressed below, but here too it must be emphasized that these quantities are total quantities.

Then the production system can be represented as follows, being understood that this scheme is not a matrix, and that in particular the indices *i* and *j* do not correspond to the usual matrix notation of rows and columns:

$$\begin{array}{cccccccccc}
A_{11} & ... & A_{k1} & 0 & 0 & 0 & and & L_1 & \to & A_1 \\
... & ... & ... & 0 & 0 & 0 & ... & ... & ... & ... \\
A_{1k} & ... & A_{kk} & 0 & 0 & 0 & and & L_k & \to & A_k \\
A_{1,k+1} & ... & A_{k,k+1} & A_{k+1,k+1} & ... & A_{n,k+1} & and & L_{k+1} & \to & A_{k+1} \\
... & ... & ... & ... & ... & ... & ... & ... & ... & ... \\
A_{1n} & ... & A_{kn} & A_{k+1,n} & ... & A_{nn} & and & L_n & \to & A_n \\
A_{1,n+1} & ... & A_{k,n+1} & A_{k+1,n+1} & ... & A_{n,n+1} & and & L_{n+1} & \to & A_{n+1} \\
... & ... & ... & ... & ... & ... & ... & ... & ... & ... \\
A_{1s} & ... & A_{ks} & A_{k+1,s} & ... & A_{ns} & and & L_s & \to & A_s
\end{array}$$

We must stress once more that fixed capital goods (the "machines") only appear on the right side of this representation, where they constitute a part of the goods numbered from *n + 1* to *s*. This is so because they are obviously produced and are an output of the system, but do not enter in this process to be transformed and therefore do not disappear in their original form during its implementation. They are present in the same condition at the beginning and at the end of the process, with the only difference of being older, by a time which represents the duration of this process. The situation is not the same, however, for the goods used for their maintenance, which are transformed, one way or another (they can even disappear), in the course of the production process.

It must also be stressed again that the above scheme is not a matrix, because the quantities $A_{ij}$ are not scalar but are quantities $A_{ij}$ of different heterogeneous goods with various dimensions. Because of this heterogeneity they cannot be added, which explains the absence of the addition symbol +, and at this stage we did not use either this sign + to note the intervention of labor in this transformation process.

### 2.3. Homogeneity and the nature of abstract labor

In order for the heterogeneous goods which are transformed and produced through the production process that we just described to be thought of as homogeneous commodities, we need to find a common property which can be assigned to them, and by definition this property will be named their value, in the sense that the existence of this common property makes them equivalent in terms of itself. What comes out of the above scheme is that the only common property of all these goods is to have been produced by labor, and this is precisely this property which had lead the first classical economists as well as Marx to develop the labor theory of value.

However, as we already indicated (see chapter 1), labor itself corresponds to a number of heterogeneous activities, with various skills differing broadly from one branch to another, and within a given branch the existence of various levels of skills, generally linked to various levels of education and responsibility. Marx himself was quite conscious of the problem, and addressed it by developing the concept of abstract labor, intended to overcome all these differences that result in the obvious heterogeneity of labor.

Marx introduces this concept of abstract labor in section 2 "The twofold character of the labor embodied in commodities", of chapter 1 of "Capital". There he first defines useful labor in relation to use-values: "in the use-value of each commodity there is contained useful labor, *i.e.*, productive activity of a definite kind and exercised with a definite aim" (Marx, 1887, p. 31). A few lines further he adds: "So far therefore as labor is a creator of use-value, is useful labor, it is a necessary condition, independent of all forms of society, for the existence of the human race; it is an eternal nature-imposed necessity, without which there can be no material exchanges between man and Nature, and therefore no life" (Marx, 1887, p. 31). And finally he arrives to abstract labor by indicating: "But the value of a commodity represents human labor in the abstract, the expenditure of human labor in general" (Marx, 1887, p. 32).

The developments that Marx devotes to this question are finally summarized in the very last paragraph of this section 2 of chapter 1 of "Capital", where he writes: "On the one hand all labor is, speaking physiologically, an expenditure of human labor-power, and in its character of identical abstract human labor, it creates and forms the value of commodities. On the other hand, all labor is the expenditure of human labor-power in a special form and with a definite aim, and in this, its character of concrete useful labor, it produces use-values" (Marx, 1887, p. 33).

On the basis of this section and of such citations, some commentators have interpreted abstract labor as being physiological labor, but it is not very helpful in a theory seeking a common dimension in order to measure values, because physiologically labor is an activity which differs from an individual to the other, each of them having his own physiology, with its own strength, etc.

Most commentators, however, have interpreted abstract labor sociologically, such as the great Hungarian philosopher and sociologist Georg Lukacs, a friend and associate to Max Weber. In his well-known book "History and Class Consciousness", he writes: "If we follow the path taken by labor in its development from the handicraft via co-operation and manufacture to machine industry we can see a continuous trend towards greater rationalization, the progressive elimination of the qualitative, human and individual attributes of the worker. On the one hand, the process of labor is progressively broken down into abstract, rational, specialized operations so that the worker loses contact with the finished product and his work is reduced to the mechanical repetition of a specialized set of actions. On the other hand, the period of time necessary for work to be accomplished (which forms the basis of rational calculation) is converted, as mechanization and rationalization are intensified, from a merely empirical average figure to an objectively calculable work-stint that confronts the worker as a fixed and established reality" (Lukacs, 1920, p. 88). To be sure, from a sociological point of view, this Marxist concept of abstract labor is certainly quite relevant, but as such, and within the framework of our quest for an economically relevant concept, it offers no particular basis for quantification, because abstract labor as defined has no particular dimension.

Nevertheless, Marx himself, although he certainly did not master the concept of dimension, was quite conscious of this necessity of reasoning in terms of quantities. This appears quite clearly if we refer to one of his works anterior to the publication in 1867 of his main book "Capital". This previous work is a second and rather long unpublished draft of chapter 6 of the later, dating back to 1864. Let us quote a particularly meaningful passage of this draft chapter: "Just as the owner of commodities is only interested in the use value of the commodity as the vehicle of its exchange value, so the capitalist is only interested in the labor process as the vehicle and instrument of the valorization process. Within the production process too - in so far as it is a valorization process - the means of production continue to be simply monetary values, for which the particular material shape, the particular use value in which this exchange value is expressed is a matter of indifference, just as labor itself does not count within the production process as productive activity of a particular useful character, but as the substance that creates value, as social labor in general which is being objectified, and of which the only interesting aspect is *its quantity"* (Marx, 1864, E-book).

To go further towards the understanding of abstract labor, one must in fact refer to a long forgotten economist, Isaak Illich Rubin. In his book "Essays on Marx's Theory of Value", originally published in 1924, he devotes one full chapter to discuss this particular concept of abstract labor, and shows well that the answer to the question of homogeneity can be found in Marx himself, despite all the complexities of his writings. Rubin emphasizes the fact that "Marx considers the quantity of working time expended by the worker the basic property which characterizes the quantitative determination of labor" (Rubin, 1924, chapter 14, p. 156). He refers in particular to Marx's book, "A Contribution to the Critique of Political Economy", published in 1859, and which is a kind of first draft of "Capital". He points out that "when we consider the distribution of total social labor among individuals and branches of production (…) different quantities of *labor* appear as different quantities of *labor-time.* Thus Marx frequently replaces labor with labor-time, and examines labor-time as the substance materialized in the product *(Critique,* pp. 23, 26)" (Rubin, 1924, p. 156).

Although Rubin, like Marx, is certainly wrong in identifying labor-time with "a substance materialized", he is completely right in highlighting time as the quantitative factor which transforms labor in abstract labor by stressing its dimension as labor-time, i.e. as time itself, because time, as a well-established physical concept, does not appear to have particular varieties. If we recognize that, we have therefore the solution to the homogeneity problem, because labor, despite of the many dimensions it can have as a physiological, social or productive activity, has only one dimension, which is not a substance materialized, and which is common to all its possible kinds and varieties, and it is time as a physical dimension, meaning a dimension of the physical and material world.

### 2.4. Values have the dimension of time

From what we just understood, it derives that various types of this particular activity which labor is  have no common dimension other than the time during which any particular type of activity is performed. Therefore the only way for abstract labor to be valid as a scientific concept underlying the concept of value is to be considered as labor reduced to its sole dimension of time, being aware that time as such is indeed a dimension of the physical world: time is a dimension of physics and one of the seven fundamental physical quantities in both the International System of Units and the International System of Quantities.

To acknowledge that is to understand that values are magnitudes that have the dimension of time, which is a perfectly homogeneous dimension, and that the values of commodities must be expressed in units of time. Moreover time presents the interest to be defined as a physical entity by the International Systems in charge of this task, and which also define the unit of time. It is equally important to note that this dimension of commodities, i.e. their value as time, can be defined at the stage of their production, seen as a purely physical and technical process, before they enter into exchanges and are the object of any distribution.

At this stage we can therefore formulate a new and fundamental principle:

**Principle 13**: The only way to think of commodities as homogeneous through a common property, which is by definition their value, is to consider that they are produced by abstract labor, i.e. labor reduced to its only dimension of physical time. Therefore values, and commodities through this common property, have the dimension of time.

**Corollary to Principle 13**: Since time in which values are expressed is the time of physics, measured in physical units like hours, days, weeks, months or years, values at the level of production of commodities must be considered as "physical values".

### 2.5. Fixed capital cannot transfer its value to the product

Although we already explained in sub-section 2.1. why fixed capital was not a factor of production, the last principle explains also why fixed capital cannot transfer its value to the products that it helps to produce through its catalytic role. Indeed from the last principle it derives that value is not a substance, but a property of commodities (among a number of other properties), which itself has the dimension of time. Each commodity has therefore the characteristic of having a value with the dimension of a time, and this characteristic is intrinsic to each of them.

It must be stressed that such a property, which as such is not a substance and has no material content, is not separable from the commodity to which it corresponds, and as such does not exist independently of this commodity. Since we know that fixed capital itself is not an intermediate commodity, and as such does not enter into the production process and is not transformed in this process, it follows that its value cannot be transferred either.

This gives us the following additional principle:

**Principle 14**: Since fixed capital is not itself transformed in the production process, its value as an intrinsic characteristic of commodities cannot be transmitted to the products in the production of which fixed capital participates as a catalyst.

A somewhat paradoxical consequence of this principle is that the intervention of fixed capital, through its increase in labor productivity, usually reduces the value of the products.

As for intermediate commodities, we have the corresponding corollary:

**Corollary to Principle 14**: The value of intermediate commodities is transferred to a product because these commodities themselves are transferred to this product through their transformation as a part or an element of it, in the course of its production process. In some cases (like energy) this transformation can be a disappearance.

This being said it remains to be seen whether values can be concretely determined, a question which is going to be addressed in the next chapter.

# Chapter 6. A first determination of values as physical values

At the stage where we now stand the question of distribution of the product is not yet a relevant one, including its distribution among various categories of workers performing various kinds of labor or more or less skilled labor, which is therefore not taken into account. We can thus consider that labor can and in fact must be represented through its dimension of time considered only as chronological time, without any consideration for the differences between various categories of labor. Now that we have linked values and time, and since time is a homogeneous and quantifiable dimension of the physical world, we can represent production in a mathematical way allowing for the determination of values, which will be calculated as quantities of labor time.

To this end, let us start first with a series of assumptions underlying the calculation of these values, which will therefore be pure physical values.

## 1. The assumptions

At the outset, let us indicate that we will from now on deal with objects that are made homogeneous by the establishment of a common property, their value, which has the dimension of time measured in physical time units: hours, or days, or weeks, of months, or years. It is the reason why we will no more speak of goods, which are heterogeneous, but of commodities. This being said, the assumptions behind this determination of values are the following:

1) There is only one method of production for each commodity, which is in fact the average method of production for the branch producing a particular commodity;
2) Each branch produces only one commodity, which avoids the problems of joint production (although these problems can be solved, which will be done later, because taking them into consideration now would introduce unnecessary complexity);
3) The only primary factor of production is labor, which is represented by the time during which it is performed, and as such is abstract labor devoid of all of its other possible dimensions. There is fixed capital (it is part of the product, as a particular type of final good), but it is not a factor of production in the usual sense;
4) The period of production is arbitrary, which may result in the fact that the production of some final goods which takes a particularly long time to be produced is not fully completed during a period (they can be registered as unfinished commodities);
5) All commodities have this same period of production, which therefore is not a purely technical data;

6) Although inputs are produced first during the period and outputs are produced when the necessary inputs are available, production is a continuous process, and therefore we will consider that all inputs and all outputs are produced during the same period.

In the sake of simplicity, the whole system is normalized by taking as a unit the overall produced quantity of each commodity.

Then we note $a_{ij}$ the ratio $\frac{A_{ij}}{A_i}$, with $\sum_{j=1}^{s} a_{ij} = 1$, meaning that, as we previously showed, there is no surplus of intermediate goods.

$a_{ij}$ is thus the share of total production or output of branch *i* ($A_i$) used as an input in the production of commodity *j* by branch *j* ($A_{ij}$). They are called technical coefficients, being understood that these coefficients are not defined like in National Accounting (i.e. $\frac{A_{ij}}{A_j}$).

$l_i$ is the labor-time it takes in branch *i* to produce one unit of commodity *i*, in time units.
On these bases, Morishima demonstrated that there are two different methods for the determination of values, in a book published in 1973: "Marx's Economics : a Dual Theory of Value and Growth". These two methods will now be developed, one after the other.

## 2. A first method for the determination of values

### 2.1. A first definition of values

On the above bases, let us first consider that there are s commodities, and that their values are the elements of a vector of values $V = [v_1 \quad ... \quad v_s]$. Since it is a row vector these elements are named $v_j$ (with *j = 1,..., s*). Let us define the value $v_j$ of one unit of commodity *j* as the overall labor time it takes to produce commodity *j*.

Since commodity *j* has been produced through the transformation of various other intermediate commodities, $v_j$ can be defined first as the sum of the direct labor time $l_j$ needed to produce commodity *j*, plus the indirect time corresponding to the value of the commodities that have been transformed to produce commodity *j*, which is the value of its means of production, i.e. the intermediate commodities *i* used in its production, be they of the first type, with *i = 1,...k,* or of the second type, with *i = k +1,...n.*

Thus $v_j = a_{1j}v_1 + a_{2j}v_2 + .... + a_{jj}v_j + ... + a_{nj}v_n + l_j$ or $v_j = \sum_{i}^{n} a_{ij}v_j + l_j$ (1)

In passing it must be stressed that $a_{jj}v_j$ is always equal to zero. Indeed in supposing that it would not be the case, then we could write:

$$v_j - a_{jj}v_j = a_{1j}v_1 + a_{2j}v_2 + .... + 0 + ... + a_{nj}v_n + l_j \quad (2)$$

And then we could change the dimension of the equation by dividing both members by $1-a_{jj}$ which in any case proves that $a_{jj}=0$ for any $i=j$, or that no commodity can enter directly in its own production, or in other words can be transformed into itself.

## 2.2. The calculation of values with the first method

On the basis of our first definition, we have therefore one equation for the determination of the value of each commodity, which means that there are s equations for s commodities. And since $a_{ij}$ coefficients are scalars (i.e. pure numbers) and all of the values have the same dimension as the $l_j$, which is time measured in time units, for instance one hour, or one month, or one year, the production system which has been exposed above (see on page 102) can now be made homogeneous. It is therefore possible to represent this production system mathematically, on the basis of the above equation extended to all commodities. In order to represent this system in matrix format we can try to use a square matrix which could be inverted, like in Sraffa's production system. But such a matrix taking into account all the s goods would be quite strange, since with its *s* rows and *s* columns it would appear as follows:

$$\begin{bmatrix} v_1 & \dots & v_s \end{bmatrix} \begin{bmatrix} a_{11} & \dots & a_{1k} & a_{1,k+1} & \dots & a_{1n} & a_{1,n+1} & \dots & a_{1s} \\ \dots & \dots & \dots & \dots & \dots & \dots & \dots & \dots & \dots \\ a_{k1} & \dots & a_{kk} & a_{k,k+1} & \dots & a_{kn} & a_{k,n+1} & \dots & a_{ks} \\ 0 & \dots & 0 & a_{k+1,k+1} & \dots & a_{k+1,n} & a_{k+1,n+1} & \dots & a_{k+1,s} \\ \dots & \dots & \dots & \dots & \dots & \dots & \dots & \dots & \dots \\ 0 & \dots & 0 & a_{n,k+1} & \dots & a_{nn} & a_{n,n+1} & \dots & a_{ns} \\ 0 & \dots & 0 & 0 & \dots & 0 & 0 & \dots & 0 \\ \dots & \dots & \dots & \dots & \dots & \dots & \dots & \dots & \dots \\ 0 & \dots & 0 & 0 & \dots & 0 & 0 & \dots & 0 \end{bmatrix} + \begin{bmatrix} l_1 & \dots & l_s \end{bmatrix} = \begin{bmatrix} v_1 & \dots & v_s \end{bmatrix}$$

In matrix notation this might be written as the following system: $VA+L=V$

However it is obvious that this notation leads us nowhere, because such a system cannot be solved. Indeed, despite the fact that matrix A is a square matrix, with its *s* rows and s columns, it is not indecomposable, because inter alia all of the elements of the last *n+1 to s* rows are zeros.

To understand how we could resolve the system, let us first use a simplified presentation with submatrices. The initial system can thus be transcribed into the following one:

$$\begin{bmatrix} V_1 & V_2 & V_3 \end{bmatrix} \begin{bmatrix} A_{11} & A_{12} & A_{13} \\ 0_{21} & A_{22} & A_{23} \\ 0_{31} & 0_{32} & 0_{33} \end{bmatrix} + \begin{bmatrix} L_1 & L_2 & L_3 \end{bmatrix} = \begin{bmatrix} V_1 & V_2 & V_3 \end{bmatrix}, \qquad \text{which we name system A}$$

The significance of the various submatrices and vectors appearing in this system deserves a few explanations. Let us begin with the row vector of values $V = [v_1 \quad ... \quad v_s]$: it can be decomposed into $V_1$, the vector of values of intermediate commodities of the first type, $V_2$ the vector of values of intermediate commodities of the second type, and $V_3$ the vector of values of final commodities:

$$V_1 = [v_1 \quad ... \quad v_k],\ V_2 = [v_{k+1} \quad ... \quad v_n], V_3 = [v_{n+1} \quad ... \quad v_s]$$

As for the different submatrices with positive elements $A_{11}$, and $A_{12}$ to $A_{23}$:

- Submatrix $A_{11}$ is a square matrix of dimension k, its elements reflect the fact that the first k intermediate goods enter collectively in their own production: they are primary commodities or commodities entering directly or indirectly in their production, and we now call them intermediate commodities of the first type. In passing, as we saw, at individual level no such commodity enters in its own production, which means that $a_{ij} = 0$ for every $i = j$ (this constitutes a difference with the situation of wage-goods in the Ricardian system) ;

- Submatrices $A_{12}$ and $A_{13}$ are rectangular matrices of dimensions (*k*, *n*-*k*) and (*k*, *s*-*n*) respectively. Their elements reflect the fact that these same k intermediate commodities of the first type enter into the production of what we now call intermediate commodities of the second type (this corresponds to submatrix $A_{12}$ and generally (but not always) into the production of final commodities (this corresponds to submatrix $A_{13}$) ;

- Submatrices $A_{22}$ and $A_{23}$ are a square matrix of dimensions (*n*-*k*, *n*–*k*) and a rectangular matrix of dimensions (*n*-*k*, *s*-*n*) respectively. Their elements reflect the fact that the *n* - *k* intermediate commodities of the second type enter into their own production (this corresponds to submatrix $A_{22}$) and generally (but not always) into the production of final commodities (this corresponds to submatrix $A_{33}$).

Similarly, the significance of the various 0 submatrices is the following:

- Submatrix $0_{21}$, of dimension *(n-k, k)*, reflects the fact that the production of the k intermediate commodities of the first type does not require any input from the *n* - *k* branches producing intermediate goods of the second type, because these last ones do not enter directly or indirectly in the production of the previous intermediate goods, and therefore in the production of all goods, but only enter in their own production and in the production of final goods ;

- Submatrices $0_{31}$ of dimension *(s-n, k)*, $0_{32}$ of dimension (*s*-*n*, *n*-*k*,) and $0_{33}$ of dimension (*s*-*n*, *s*-*n*) reflect the fact that the *s* - *n* final commodities by definition do not provide any inputs to the *k* branches producing intermediate goods of the first type (this corresponds to submatrix $0_{31}$), neither any inputs to the *n* – *k* branches producing intermediate goods of the second type (this corresponds to submatrix $0_{32}$), nor any inputs to their own production as final commodities (this corresponds to submatrix $0_{33}$).

As for the row vectors of labor quantities, each of them corresponds to one of the three types of commodities:

- $L_1 = (l_{1,}..., l_k)$ is the row vector of the quantities of direct labor used in the production of commodities $A_1$ *to* $A_k$, respectively, measured in physical time and in time units ;
- $L_2 = (l_{k+1,}..., l_n)$ is the row vector of the quantities of direct labor used in the production of commodities $A_{k+1}$ *to* $A_n$, respectively ;
- $L_3 = (l_{n+1,}..., l_s)$ is the row vector of quantities of direct labor used in the production of commodities $A_{n+1}$ *to* $A_s$, respectively.

As noted, system A cannot be resolved, because its matrix is not indecomposable, and has to be reduced. In order to do so, we can first try to suppress the three zero submatrices at the bottom of matrix *A*, called submatrices $0_{13}$, $0_{23}$ and $0_{33}$. This gives us a second system with a first decomposition of matrix *A*, now transformed in matrix *A'*, and which can be represented as follows:

$$\begin{bmatrix} V_1 & V_2 & V_3 \end{bmatrix} \begin{bmatrix} A_{11} & A_{12} & A_{13} \\ 0_{21} & A_{22} & A_{23} \end{bmatrix} + \begin{bmatrix} L_1 & L_2 & L_3 \end{bmatrix} = \begin{bmatrix} V_1 & V_2 & V_3 \end{bmatrix},$$ which we name system A'

But this system cannot be resolved either. Indeed matrix *A'* is not a square matrix, but a rectangular matrix of dimension (*n, s*), which again is not indecomposable, because of the presence of a submatrix $0_{21}$ on the lower left side of matrix *A'*.

In order to simplify this system one might think of merging submatrices $A_{12}$ and $A_{13}$ into one single submatrix $A_2$, and submatrices $A_{22}$ and $A_{23}$ into one single submatrix $A_3$. This would imply merging vectors $V_2$ and $V_3$ into one vector $V_2$, and vectors $L_2$ and $L_3$ into one single vector $L_2$. We would then obtain the following system:

$$\begin{bmatrix} V_1 & V_2 \end{bmatrix} \begin{bmatrix} A_1 & A_2 \\ 0 & A_3 \end{bmatrix} + \begin{bmatrix} L_1 & L_2 \end{bmatrix} = \begin{bmatrix} V_1 & V_2 \end{bmatrix}$$

One cannot but be struck by the resemblance of this system with the Ricardian system of chapter 2 (p. 45). By analogy to the latter, and in matrix format, we could therefore try similarly to write this system by splitting it in two sub-systems, in the following way:

$$V_1 = V_1 A_1 + L_1 \qquad (3)$$

$$V_2 = V_1 A_2 + V_2 A_3 + L_2 \qquad (4)$$

However it is obvious that matrix equation (2) above cannot be written because the expression $V_2A_3$ displays a big problem of dimensionality: indeed vector $V_2$ has $s - k$ elements corresponding to all the final commodities ($s - n$), plus the intermediate commodities of the second type ($n - k$), whereas submatrix $A_3$ has only n - $k$ rows, corresponding only to these last commodities. Multiplying $V_2$ by $A_3$ is therefore totally impossible.

The failure of this simplification attempt shows once more that it is impossible to assimilate wage-goods in the Ricardian sense, where they are viewed as intermediate goods, and final goods, these last one being absent on the left side of the equations.

We must therefore revert to system A', and in order to resolve it, rewrite it in a way which takes into account the existence of the three different types of goods introduced in sub-section 2.2. of the present chapter: intermediate goods of the first type, of the second type, and final goods. The system has therefore to be split into three subsystems, which allows to represent it in the following matrix format, where we have three matrix equations:

$$V_1 = V_1A_{11} + L_1 \qquad (5)$$

$$V_2 = V_1A_{12} + V_2A_{22} + L_2 \qquad (6)$$

$$V_3 = V_1A_{13} + V_2A_{23} + L_3 \qquad (7)$$

Only the first one is an independent matrix equation, from which we get:

$$V_1 = L_1\left(I - A_{11}\right)^{-1} \qquad (8)$$

Then, knowing $V_1$, the vector of values of intermediate commodities of the first type, from equation (8) we get $V_2$, the vector of values of intermediate commodities of the second type :

$$V_2 = \left(I - A_{22}\right)^{-1}\left(V_1A_{12} + L_2\right) \qquad (9)$$

And knowing $V_1$ and $V_2$, from equations (8) and (9) we obtain $V_3$, the vector of values of final commodities, without any other additional calculation than what appears in equation (7):

$$V_3 = V_1A_{13} + V_2A_{23} + L_3$$

In passing let us underline that it would not be relevant in this framework to consider the $k$ intermediate commodities of the first type as basics in the sense of Sraffa, because what distinguishes basics is not only the fact that they enter directly or indirectly in the production of all the other commodities, but also, as we showed it, the fact that a profit rate has to be levied on them. However there is no profit in this analysis, which remains strictly at the stage of production, i.e. before any distribution has taken place.

# 3. A second method for the determination of values

## 3.1. A second definition of values

The second method is based on a different way of defining values, which can also be found in Marx[5]. The value $v_j$ of a commodity *j* is now defined as the sum of the totality of direct labor time used to produce one unit of it. This includes the direct labor time used in its own production, i.e. in the transformation of the intermediate commodities used in this final process, plus the direct labor time used to produce these same intermediate commodities, and in turn the direct labor time used to produce their own means of production, and so on.

Since this time is proportional to the quantities of each intermediate good that enters ultimately, directly or indirectly, into the production of all the other intermediate or final goods, to make the calculation implies that we allocate this direct labor time to the quantity of each commodity which enters, directly or indirectly, in the production of all of the other ones.

Let us call $x_{ij}^*$ the quantity of commodity *i* used to produce one unit of commodity *j*. Then let $x_{ij}$ be the share of the total production of commodity *i* which is used directly and indirectly to produce one unit of commodity *j*. Thus we have:

$$x_{ij} = \frac{x_{ij}^*}{x_i}, \text{ where } x_{ij} \text{ is a scalar} \qquad (10)$$

Using the new definition of this second method, the values $v_j$ that we want to determine for the first *k* intermediate commodities are defined as:

$v_j = x_{1j}l_1 + x_{2j}l_2 + ... + x_{kj}l_k$ (we recall that there are no $x_{ij}$ for $i > k$, i.e. for final commodities)

$$\text{Or } v_i = \sum_{j=1}^{k} l_j x_{ji} \qquad (11)$$

## 3.2. The calculation of values with the second method

From our new definition, it is clear that we must first determine the quantities $x_{ij}$ appearing in the last equations. In order to do so without unnecessarily complicating the demonstration, we will assume that there are no intermediate commodities of the second type, and that therefore we have only *n* commodities, i.e. *k* intermediate commodities of the first type and *n* - *k* final commodities.

Then, with the first method, the system of three matrix equations (1), (2) and (3) above for determining values becomes a system with only two matrix equations:

[5] This is well explained by Morishima (op. cit.) where chapter 1 provides a demonstration.

$$V_1 = V_1 A_{11} + L_1 \tag{12}$$

$$V_2 = V_1 A_{12} + L_2 \tag{13}$$

Which gives us, as solutions to these equations:

$$V_1 = L_1 \left(I - A_{11}\right)^{-1} \tag{14}$$

$$V_2 = L_1 \left(I - A_{11}\right)^{-1} A_{12} + L_2 \tag{15}$$

Going back now to the second method, let us call $X_i$, with $i = 1,...n$, the column vector giving the quantities of the $n$ commodities that are produced by the system:

$$X_i = \begin{bmatrix} x_1 \\ x_2 \\ \cdots \\ \cdots \\ \cdots \\ x_n \end{bmatrix}, \text{ or } X_i = \begin{bmatrix} X_1 \\ - \\ X_2 \end{bmatrix}, \text{ with } X_1 = \begin{bmatrix} x_1 \\ x_2 \\ \cdots \\ x_k \end{bmatrix} \text{ and } X_2 = \begin{bmatrix} x_{k+1} \\ x_{k+2} \\ \cdots \\ x_n \end{bmatrix}$$

Like with the first method for calculating values, and in order to make this calculation possible, we need first to transform some of the quantities that we are dealing with into scalars, which is done by taking each of them as a unit.

Let us call $L_i$ the row vector of the direct quantities of labor time used in the production of the $n$ commodities $X_i$:

$L_i = l_1, ..., l_k, l_{k+1}, ..., l_n$, which in vector terms can be written:

$L_i = \left[L_1 \mid L_2\right]$, with $L_1 = l_1, l_2, ..., l_k$, and $L_2 = l_{k+1}, l_{k+2}, ..., l_n$

When we calculated the values of commodities with the first method, the quantities of each intermediate commodity that enter *directly* in the production of the other ones were known, since knowing them comes down to knowing the methods of production, which are among the data. These methods of production are given by the $a_{ij}$ coefficients, which as we recalled are named technical coefficients in standard input–output analysis. These $a_{ij}$ coefficients were indeed defined as the quantities of intermediate commodities $i$ (or outputs of branch $i$) used directly as inputs in the production of a commodity $j$ (output), with $i = 1, ..., k,$ and $j = 1, ..., n$.

In order to calculate these same values with the second method, we need now to go one step further and determine the quantities $x_{ij}$, which include the inputs of commodities $i$ entering *indirectly* in the production of commodities *j*, through their use in the production of other direct inputs, and so forth.

If we try to represent the system globally, as we did for the first method, the resolution of the system would therefore consist in the determination of all the elements of a matrix X which would look as follows:

$$\begin{bmatrix} x_{11} & \cdots & x_{1k} & x_{1,k+1} & \cdots & x_{1n} \\ \cdots & \cdots & \cdots & \cdots & \cdots & \cdots \\ x_{k1} & \cdots & x_{kk} & x_{k,k+1} & \cdots & x_{kn} \end{bmatrix} = \begin{bmatrix} X_{11} & | & X_{12} \end{bmatrix} \text{ of dimension } k \text{ x } n$$

Since each $x_{ij}$ represents the total amount of direct and indirect intermediate commodities *i* required as inputs to produce one unit of commodity *j*, in order to determine this amount, we can write for each intermediate commodity *i* a series of equations using the fact that technical coefficients $a_{ij}$ are a series of already known data (the methods of production), which are themselves elements of the following matrix:

$$\begin{bmatrix} a_{11} & \cdots & a_{1k} & a_{1,k+1} & \cdots & a_{1n} \\ \cdots & \cdots & \cdots & \cdots & \cdots & \cdots \\ a_{k1} & \cdots & a_{kk} & a_{k,k+1} & \cdots & a_{kn} \end{bmatrix} = \begin{bmatrix} A_{11} & | & A_{12} \end{bmatrix} \text{ of similar dimensions } k \text{ x } n$$

It is obvious that it is impossible to resolve a system using these matrices as such, because both are rectangular matrices and therefore are not invertible. It is indeed impossible to multiply matrix *A* by matrix *X*, because *A* has more columns than *X* has rows, although it is possible to multiply matrix $A_{11}$, which has only *k* columns, by matrix *X*, which has *k* rows.

Therefore, if we limit ourselves first to the *k* intermediate commodities of the first type, we can multiply matrix $A_{11}$ by matrix $X_{11}$ (the first *k* columns of matrix *X*). We also know that matrix $A_{11}$ is productive, because vector $X_1$ (with *i = 1, ..., k*), which gives the total production of the first k intermediate commodities, is non negative and such that $X_i \geq A_1 X_i \geq 0$. We are indeed in an open Leontief model, and this productivity entails some implications, which have been exposed among alia by Peterson and Olinick, in an article on Leontief models and Markov chains which they published in 1982 in “Mathematical modelling”. This implies first that:

$$\sum_{j=1}^{k} a_{ij} X_j \leq X_i \text{ for } i = 1, 2 \ldots k \tag{16}$$

This equation means that all the production of the first *k* intermediate commodities is not used in their own production, since a part of this production is used in the production of the *n - k* final commodities. This also implies that:

$$\sum_j \left( \sum_i a_{ij} \right) X_j = \sum_i \sum_j a_{ij} X_j < \sum_i X_i \qquad (17)$$

Therefore, since all intermediate commodities of the first type enter by definition in the production of all commodities other than themselves (i.e. in the production of final commodities), for any intermediate commodity *i* we must have:

$$\sum_j a_{ij} < 1 \qquad (18)$$

Inequality (16) also implies (by subtracting $a_{ii} X_i$ from both members), that:

$$0 \leq \sum_{j \neq i} a_{ij} X_j < X_i - a_{ii} X_i = (1 - a_{ii}) X_i \qquad (19)$$

Therefore $a_{ii} < 1$ for every *i* and matrix $I - A_{11}$ is non singular. On this basis a theorem guarantees that sub-stochastic matrix $A_{11}$ is productive. Then matrix $I - A_{11}$ is invertible and $(I - A_{11})^{-1} > 0$, with a solution which is the production vector $X_1 = (I - A_{11})^{-1} F$ (with $F$ as the vector of final demand).

Going back to this production vector $X_1$, i.e. the column vector giving the quantities of the *k* intermediate commodities of the first type that are produced by the system, and because matrix $X_{11}$ is also productive, we must write, for each of these *k* commodities, an equation which indicates that it is the sum of all its total uses, direct and indirect, entering as its outputs in the production of all the other *k* commodities, to which must be added the share of its output entering in the production of final commodities.

In an open Leontief model, this additional share is usually considered as final demand and for this reason it is generally called $f_i$ for commodity *i.* However it must be understood that it is not a demand for final commodities themselves, but only for the intermediate commodities entering in the production of these final commodities. Then we can write:

$$x_1 = x_{11} + ... + x_{1j} + x_{1k} + f_1$$
..........................................
$$x_i = x_{i1} + ... + x_{ij} + x_{ik} + f_i$$
..........................................
$$x_k = x_{k1} + ... + x_{kj} + x_{kk} + f_k$$

To continue in matrix notation, let us have:

$$X_1 = \begin{bmatrix} x_1 \\ .. \\ .. \\ x_k \end{bmatrix} \quad A_{11} = \begin{bmatrix} a_{11} & ... & ... & a_{1k} \\ ... & ... & ... & ... \\ ... & ... & ... & ... \\ a_{k1} & ... & ... & a_{kk} \end{bmatrix} \quad X_{11} = \begin{bmatrix} x_{11} & ... & ... & x_{1k} \\ ... & ... & ... & ... \\ ... & ... & ... & ... \\ x_{k1} & ... & ... & x_{kk} \end{bmatrix} \quad F_1 = \begin{bmatrix} f_1 \\ ... \\ ... \\ f_k \end{bmatrix} \text{ and } I_i = \begin{bmatrix} 1 \\ 1 \\ .. \\ 1 \end{bmatrix}$$

Where $I_1$ is used to represent what is known as the summation vector, i.e. a column vector of 1's, of appropriate dimension (here *k*). Then in matrix notation, Miller and Blair explained in their book, published in 2009: "Input-Output Analysis: Foundations and Extensions", that these equations that we just wrote above can be summarized as:

$$X_1 = A_{11} X_{11} I_1 + F_1 \tag{20}$$

It is well known that such an equation represents an open Leontief model, and can be easily resolved in the sense that the production vector $X_1$ as well as matrix $X_{11}$ can be determined for any given vector of "final demand" $F_1$. Let us show it by assuming that this final demand is represented by one unit for each of the *k* intermediate commodities. Then the above equations become:

$$x_1 = x_{11} + ... + x_{1j} + x_{1k} + 1$$
$$........................................$$
$$x_i = x_{i1} + .... + x_{ij} + x_{ik} + 1$$
$$........................................$$
$$x_k = x_{k1} + ... + x_{kj} + x_{kk} + 1$$

The $a_{ij}$ seem to have disappeared – and thus matrix $A_{11}$, but it is just because the elements $x_{ij}$ of these equations are obtained through a multiplication by matrix $A_{11}$. In order to determine the elements $x_{ij}$ appearing above, and before writing the detailed equations giving each of these elements, an example will help us to understand how they are obtained.

Let us take for instance element $x_{1i}$, i.e. the total quantity of commodity *1* (as a share of its total product) entering in the production of commodity *i*. It is the sum first of the share of commodity *1* going into its own production $a_{11}x_{1i}$ (which should be 0, because as we saw $a_{11}$ is normally equal to 0), plus the share of the production of commodity *1* going in the production of the share of the production of commodity 2 which goes itself in the production of commodity *i*, i.e. $a_{12}x_{2i}$ …, and so on, up to the share of the production of commodity *1* going in the production of the share of the production of commodity *k* which goes itself in the production of commodity 1, i.e. $a_{1k}x_{k1}$. On top of these intermediate uses of commodity *1*, we must also take into account the "final demand" for commodity *1*, which we have assumed to be equal to 1. Therefore the corresponding equations are:

$$
\begin{array}{lll}
x_{11} = a_{11}x_{11} + ... + a_{1k}x_{k1} + 1 & x_{1i} = a_{11}x_{1i} + ... + a_{1k}x_{ki} & x_{1k} = a_{11}x_{1k} + ... + a_{1k}x_{kk} \\
x_{21} = a_{21}x_{11} + ... + a_{2k}x_{k1} & x_{2i} = a_{21}x_{1i} + ... + a_{2k}x_{ki} & x_{2k} = a_{21}x_{1k} + ... + a_{2k}x_{kk} \\
\dots\dots\dots\dots & \dots\dots\dots\dots & \dots\dots\dots\dots \\
x_{i1} = a_{i1}x_{11} + ... + a_{ik}x_{k1} \quad \dots\dots & x_{ii} = a_{i1}x_{1i} + ... + a_{ik}x_{ki} + 1 \quad \dots\dots & x_{ik} = a_{i1}x_{ik} + ... + a_{ik}x_{kk} \\
\dots\dots\dots\dots & \dots\dots\dots\dots & \dots\dots\dots\dots \\
x_{k1} = a_{k1}x_{11} + ... + a_{kk}x_{k1} & x_{ki} = a_{k1}x_{1i} + ... + a_{kk}x_{ki} & x_{kk} = a_{k1}x_{1k} + ... + a_{kk}x_{kk} + 1
\end{array}
$$

In matrix notation this system becomes, first for the full system of equations:

$$X_{11} = A_1 X_{11} + I$$

$$\Rightarrow X_{11} - A_{11}X_{11} = I$$

$$\Rightarrow X_{11} = \left(I - A_{11}\right)^{-1} \qquad (21)$$

We recognize in this equation (21) the well-known solution to open Leontief models.

Whereas for a column vector, for instance $X^1$, i.e. the first column of matrix $A_{11}X_{11} + I$ (corresponding to the first column on the left of the above equations), we get:

$$X^1 = \begin{bmatrix} x_{11} \\ x_{21} \\ ... \\ x_{k1} \end{bmatrix} = A_{11}X_1 + I^1$$ , where $I^1$ is the first column of the identity matrix I: $I^1 = \begin{bmatrix} 1 \\ 0 \\ ... \\ 0 \end{bmatrix}$

Such a vector gives us the quantities of the various commodities used as inputs in order to produce the quantity $x_1$ of commodity *1* which needs to be produced by the system to satisfy the final demand of one unit of commodity *1*.

As for vector $X_1$, as defined above, giving the total quantities of all of the intermediate commodities to be produced by the system, in order to produce one unit of each intermediate commodity as “final demand”:

$X_1 = \begin{bmatrix} x_1 \\ .. \\ .. \\ x_k \end{bmatrix}$, we can determine it easily by using the summation vector $I_1$, such that:

$$X_1 = X_{11}I_1 + I_1$$

Once the outputs $x_{1i}, x_{2i}, ..., x_{ki}$ (the $i^{th}$ column of matrix $X_{11}$) used to produce one unit of intermediate commodity *i* are determined, it is easy to calculate the corresponding values, using their definition through this second method, given by equation (10) above:

$$v_i = \sum_{j=1}^{k} l_j x_{ji}$$

From this equation we get the values, i.e. in matrix notation the row vector $V_1$:

$$V_1 = L_1 X_{11} \qquad (22)$$

However it is clear that the *k* equations from which we started our demonstration above do not include all the equations in which the first *k* intermediate commodities of the first type intervene, and therefore do not include all the uses of these first *k* commodities, which also enter into the production of the other *n - k* final commodities.

Turning then to the production of these final commodities, we must add, for each intermediate commodity *i* entering in the production of a final commodity *j* (with *j = k+1 ,…, n)*:

- Firstly, a share $a_{1j}, ..., a_{kj}$ of the production of each intermediate commodity *1* to *k,* that will be directly required to produce one unit of each final commodity *j* (with *j = k +1* to *n*), for instance $a_{i,k+1}$ for the quantity of intermediate commodity *i* entering directly in the production of final commodity *k + 1*.
- Secondly, the indirect quantities of this intermediate commodity embodied into all of the other intermediate commodities entering in the production of this final commodity;

Therefore for each final commodity j we can write a series of equations, which for a final commodity *j*, is the following:

$$x_{1j} = a_{11}x_{1j} + .... + a_{1k}x_{kj} + a_{1j}$$
$$x_{2j} = a_{21}x_{1j} + ... + a_{2k}x_{kj} + a_{2j}$$
..............................................
$$x_{kj} = a_{k1}x_{1j} + ... + a_{kk}x_{kj} + a_{kj}$$

Similarly, for *j = k + 1* and for *j = n*, these sets of equations are:

$$x_{1,k+1} = a_{11}x_{1,k+1} + ... + a_{1k}x_{k,k+1} + a_{1,k+1}$$
$$x_{2,k+1} = a_{21}x_{1,k+1} + ... + a_{2k}x_{k,k+1} + a_{2,k+1}$$
..........................................................
$$x_{k,k+1} = a_{k1}x_{1,k+1} + ... + a_{kk}x_{k,k+1} + a_{k,k+1}$$

and

$$x_{1n} = a_{11}x_{1n} + ... + a_{1k}x_{kn} + a_{1n}$$
$$x_{2n} = a_{21}x_{1n} + ... + a_{2k}x_{kn} + a_{2n}$$
...............................................
$$x_{kn} = a_{k1}x1n + ... + a_{kk}x_{kn} + a_{kn}$$

With $\sum_{j=1}^{n} a_{ij} = 1, \forall i$, meaning that there is no surplus of any intermediate commodities *1* to *k*.

In matrix notation, these additional equations for all $n$ - $k$ commodities correspond to the multiplication of matrix $A_{11}$ by matrix $X_{12}$, this last one being a rectangular matrix of dimension *k* x *(n - k)*, to which is added a matrix $A_{12}$, obviously of the same dimension.

Going back to the values of the *n-k* final commodities, and to the second definition of these values, which says that they are the sum of the direct labor time used to produce the *k* intermediate commodities entering into their production, plus the direct labor time used in their own production, i.e. in the transformation of these same intermediate commodities used in this process into final commodities, we get therefore:

$$v_j = \sum_{i=1}^{k} l_j x_{ij} + l_j \tag{23}$$

Thus in matrix notation, we get the following equation for matrix $X_{12}$:

$$X_{12} = A_{11} X_{12} + A_{12} \tag{24}$$

This gives us the solution for matrix $X_{12}$:

$$X_{12} = \left(I - A_{11}\right)^{-1} A_{12} \tag{25}$$

Now that both matrices $X_{11}$ and $X_{12}$ have been determined, we can finally calculate also the values, for the *n* - *k* final commodities:

$$V_2 = L_1 X_{12} + L_2 \tag{26}$$

## 4. The equivalence of the two methods of value determination

To begin with, it is quite easy to show that the two methods that have been considered successively arrive at identical results for the calculation of values. To demonstrate it, it is sufficient to replace $X_{11}$ in equation (22) giving the values of intermediate commodities with the second method, by its expression in equation (21). Then we get:

$$V_1 = L_1 \left(I - A_{11}\right)^{-1} \tag{27}$$

We note that equation (27) is identical to equation (8) which gave the values of intermediate commodities of the first type with the first method.

Similarly, if we replace in equation (26) giving the values of final commodities with the second method $X_{12}$ by its expression in equation (25), we obtain:

$$V_2 = L_1 (I - A_{11})^{-1} A_{12} + L_2 \tag{28}$$

And we note again that equation (28) corresponds to what would happen of equation (7), which gave the values of final commodities with the first method, if there were no intermediate commodities of the second type, meaning that vector $V_2$ would not exist as a vector of values of intermediate commodities of the second type, but would be replaced by a new vector $V_2$ representing the values of final commodities.

## 5. The equivalence of total direct labor time and total product value

These same equations allow us to show also that the value of the product is equivalent to total direct labor time. To demonstrate it, let us call $y_1$ and $y_2$ the column vectors of the output of intermediate goods and final goods, respectively.

Then, from equations (8) and (9) we can write that the value of the total product is equal to:

$$V_1 y_1 + V_2 y_2 = L_1 (I - A_{11})^{-1} y_1 + L_1 (I - A_{11})^{-1} A_{12} y_2 + L_2 y_2 \tag{29}$$

This expression, from equations (21) and (26), is equivalent to:

$$V_1 y_1 + V_2 y_2 = L_1 X_{11} y_1 + L_1 X_{12} y_2 + L_2 y_2 \tag{30}$$

And this gives us indeed the overall quantity of direct labor time used in the production process, first for the production of intermediate commodities, i.e. $(L_1 X_{11} y_1 + L_1 X_{12} y_2)$, and second for the production of final commodities, i.e. $L_2 y_2$.

It is also possible to show that the value of the product is equivalent to the value of intermediate commodities used as inputs in the production process of both intermediate and final commodities, plus the quantity of direct labor time utilized in the production of the "surplus" of intermediate commodities and of final commodities.

Taking again into account the fact that we assume here that there are no intermediate commodities of the second type, equations (5) and (6) give us indeed directly:

$$V_1 y_1 + V_2 y_2 = (V_1 A_{11} + L_1) y_1 + (V_1 A_{12} + L_2) y_2 \tag{31}$$

$$\Rightarrow V_1 y_1 + V_2 y_2 = (V_1 A_{11} y_1 + V_1 A_{12} y_2) + (L_1 y_1 + L_2 y_2) \tag{32}$$

Bringing equations (30) and (32) together also allows us to see that the value of intermediate commodities, used as inputs, which is $V_1 A_{11} y_1 + V_1 A_{12} y_2$, is equal to the difference between the totality of direct labor time utilized in the production of all intermediate and final commodities, which is $L_1X_{11}y_1 + L_1X_{12}y_2 + L_2y_2$, i.e. total employment, on the one hand, minus the direct labor time utilized in the production of $y_1$ and $y_2$, i.e. $L_1y_1 + L_2y_2$, on the other hand.

From these findings we can derive a few teachings about values, which will be exposed in next chapter.

# Chapter 7. A few teachings about the concept of value

At this point there are some preliminary conclusions that we can draw in this chapter on the relevance of the concept of value and the conditions for this relevance, the first one having to do with the field of validity of this concept. We will also come back to the determination of values, to give a little more precisions on the conditions which make it possible. We will then be in a position to examine more closely the distinction between absolute values and relative values, which will help us to define the invariable standard of value. This will bring us to the mistakes made by Marx in the exposition of his labor theory of value, since they may partially explain some of the various criticisms still surrounding this theory. These criticisms, often based on misconceptions, will finally be discussed.

## 1. The field of validity of values

The first condition for the relevance of the concept of value has to do with its field of validity. What has indeed been shown so far is that this concept can be defined at the sole level of production, which means that values of commodities preexist, logically and chronologically, to the exchange of commodities on markets. Exchange of commodities as such plays therefore no role in the determination of values, which cannot be considered as exchange ratios, even though it is clear that these values will play a role in the determination of prices, defined as the exchange ratios of commodities against money, as we will see later.

Does it mean also that values can be determined as soon as there is production, whatever the underlying mode of production (which should be rather designated as a mode of distribution)? We should not go as far as that, since it has also be shown that the concept of value is very closely linked to and depends on another concept, that of abstract labor, which allows to consider labor purely in its dimension of time. Therefore it is only when abstract labor exists as such that the concept of value can remain a relevant one. This has to be related to one particular type of exchange whose conditions are determined before production takes place, and which is the exchange of labor against money, involving a definite quantity of labor time against a definite quantity of money.

To be sure, such an exchange is the case in the capitalist mode of production, where it is generalized, with most of the production made by wage-earners whose wages are indeed paid in money and defined as a number of monetary units for a specified quantity of time. The conditions of this exchange are determined before labor is executed, when workers are recruited. In turn this implies that values have a sense each time that there is paid labor and that this payment is determined and established as a specified amount of money for a specified quantity of time. To the extent that this could also take place, although to a more limited scale, in other modes of production, it would then certainly be relevant to also use the concept of value, when analyzing production relationships in these other modes of production.

## 2. The determination of values is possible and feasible

As we have seen, and on the bases of assumptions made in section 1 of chapter 6 (see p. 117), and which can be considered as empirically plausible, it has been possible to calculate values through two slightly different methods which were shown to be mathematically equivalent. These values are all positive (the reverse would be devoid of any significance) since the mathematical theory of matrices demonstrates that with any non-negative productive matrix such as matrix $A_{11}$ as defined (which means economically that each of the first k branches provides a surplus of intermediate commodities), we will always get positive solutions for the inverse matrix $\left(I - A_{11}\right)^{-1}$. On this point the reader can refer again to the book by Peterson and Olinick, which was cited previously.

A second condition for values to be positive, since they are given by $V_1 = L_1\left(I - A_{11}\right)^{-1}$, and $V_2 = L_1\left(I - A_{11}\right)^{-1} A_{12} + L_2$, is obviously that labor has to be indispensable to the production of each of the intermediate commodities, which implies that the labor vector $L_1$ be strictly positive ($L_1 > 0$).

These two conditions allow to deduct that means of production which are not produced, i.e. essentially land in its original state, have a zero value, which does not mean that they have a zero price, but loosens somewhat assumption 3 made previously (see p. 117), because primary factors other than labor time cannot appear in the equations.

Mathematically speaking, this assumption of a strictly positive vector $L_1$ can however be released to the extent that we would only have $L_1 \geq 0$, on the condition that matrix $A_{11}$ be not only productive, but also indecomposable, which means economically that each of the intermediate commodities must enter into the production of at least one intermediate commodity other than itself, or reciprocally that every intermediate commodity must be produced by at least one intermediate commodity other than itself. We know that it is the case for all intermediate commodities of the first type, by their very definition. In this case the value of an intermediate commodity will be positive, even though no direct labor is spent in its production. The indecomposable nature of matrix $A_{11}$ and the fact that $L_1 \geq 0$ (one of the $l_i$ is at least positive), ensure indeed that the production of each commodity requires at least some indirect labor.

As for final commodities, the positivity of their values requires that matrix $\begin{bmatrix} A_{12} \\ L_2 \end{bmatrix}$ formed by the junction of matrix $A_{12}$ and vector $L_2$ be non-negative and that none of its columns be empty. This means that any final commodity must be produced with either labor or an intermediate commodity.

From all this it derives that the calculation of values for both intermediate and final commodities is perfectly feasible, on the basis of data which are usually available, of assumptions which are empirically rather realistic ones and of conditions which are usually met. This theory is therefore economically significant and relevant, and will appear even more so if we address now the question of the distinction between absolute and relative values.

## 3. The distinction between absolute values and relative values and the invariable standard of value

It seems useful to devote a few lines to the distinction between absolute values and relative values, because some authors consider indeed that the system of values allows only for the calculation of *n* "absolute values", understood like quantities of labor time, whereas the system of prices would allow for the calculation of only *n – 1* prices. This would create some difficulty in the transformation of values into prices.

Let us recall first that as far as the system of values is a measurement system with the dimension of time, the values that we obtain are always relative to a standard of measurement which cannot be anything else than an arbitrary unit of time, such as one hour, one day, one week, one month, or one year, etc.). Then it is true that we can have *n* values for *n* commodities, each one being a certain amount of labor time measured in terms of this chosen unit of time. But it is also obvious that if the quantity of time corresponding to the value of one commodity is chosen as the measurement unit (or standard of value) - and we are free to do so, then there will be only *n - 1* values different from *1* which could still be considered as "absolute values", i.e. as specified amounts of labor time. Conversely, the price system can provide us with n prices different from 1, as soon as a basket of commodities (similar for instance to Sraffa's standard commodity) is taken as the measurement unit.

The lesson that we can draw from these considerations is that in fact every system of measurement always defines as many measurements as there are elements in the set to be measured, and therefore the fact that one of these measurement is or is not equal to *1* is a pure question of definition of this measurement unit, that can be internal or external to the set of elements to be measured (provided however that it has the same dimension as these elements). All this has no importance other than practical from an economical point of view.

Indeed this comes down to a question of vocabulary, being understood that usually when we talk about a certain magnitude, for instance a mass, we specify, if not the dimension of the measured element itself, at least implicitly the dimension of the unit of measurement: we will not usually say that such luggage has a mass of 15 kg, but that it weighs 15 kg, because we know that a kilogram is the measurement unit for the dimension of mass (and even the standard of measurement). We could say more accurately that its relative mass is 15, in terms of the measurement unit of mass, but it is not customary to say that.

These precisions being given, we must recall that there is nevertheless one fundamental divergence between the measurement of exchange ratios and the measurement of values. As we saw in the chapters on Ricardo and Sraffa the measurement of exchange ratios (i.e. for these authors the prices of production) can never be compared when exchange ratios are modified, because the measurement unit itself is defined in relation to a given state of these exchange ratios, depending itself on the state of techniques and on the state of distribution.

On the contrary the measurement of values can always be compared when they are modified, because they are defined either in terms of an invariable measurement unit (if it is a given quantity of labor time, chosen as the measurement unit), or in terms of measurement units that can themselves be compared (if the value of whatever commodity is chosen as the measurement unit, the modifications of this value can themselves be measured in terms of a given quantity of labor time). The measurement of values therefore refers to an absolute measurement unit, or a measurement unit which can always be expressed in terms of an absolute standard, for instance one hour of labor time. It is only in this well-defined sense of "measurement in terms of an absolute standard" that we can speak of "absolute values", if we think that it is necessary, although it is usually not the case. Let us recall that Marx himself does not use the expression "absolute value".

As for the expression "relative value", it is used by Marx in the restricted case where the measurement unit of values is not a quantity of labor time chosen arbitrarily and invariable, but the variable value of a given commodity, chosen as a standard commodity. Although values are as much relative when they are determined in terms of a given and fixed quantity of labor time, the expression "relative value" can indeed be considered as appropriate in this last case: indeed values are then twice relative, since the measurement unit itself can change, even though these variations may themselves be measured.

One has to be quite rigorous in trying to define laws of variation for this "relative value", and Marx himself makes some mistakes when he does so. In section 3 of chapter 1 of "Capital" (volume 1), he considers indeed (in point b) "quantitative determination of relative values", pp. 36-37):

1) That the "relative value" of a commodity rises and falls directly according to its own value (in terms of an invariable quantity of labor time), the value of the standard commodity B being constant.

2) That if the value of a commodity A is supposed to remain constant, its "relative value" expressed in the standard commodity B rises and falls inversely as the value of the standard commodity B.

3) That if the values of all commodities rise and fall simultaneously, and in the same proportion, their "relative values" would remain unaltered.

4) That when the values of a commodity A and of the standard commodity B vary simultaneously in the same direction but at unequal rates, or in opposite directions, the effect of all these possible different variations on the "relative value" of commodity A are diverse and may be deduced from a combination of rules 1) and 2).

In fact Marx seems to forget his own definition of relative value, which implies that the value of standard commodity B cannot be considered as a constant, whatever may be commodity A for which we contemplate a variation of its relative value. It is indeed enough for this commodity A to be an intermediate commodity to entail that a variation in its value (in terms of a quantity of labor time) results in a variation in the value of the standard commodity B, and this whatever this last one is an intermediate commodity or a final commodity (even though this final commodity is produced without using directly the first commodity). Then neither rules1) and 2) hold, no more than rule 4) which derives from them.

In fact, if we deal with this problem while taking into account this phenomenon, and in the light of what we learned from the determination of values, the conclusions that we can reach are more precise.

1) If we first consider the case where only the quantity of direct labor time used to produce a commodity *i* varies, while the direct labor time employed to produce all the other commodities remains constant as well as all the technical coefficients, we can only conclude that the relative value of commodity A in terms of the standard commodity is modified proportionally more than the relative value of any other commodity *j*. If commodity *i* is an intermediate commodity, the variation in its value will cause a variation in the value of all the other commodities, final commodities and intermediate commodities alike. If it is a final commodity, there will be no variation in the value of intermediate commodities, neither in the value of final commodities: it is the only case where rule 1) applies.

2) If however, the variation in the value of a commodity *i* comes from a technical improvement which reduces the quantity of one of its intermediate commodities, one can demonstrate equally that the value of commodity *i* diminishes proportionally more than that of any other commodity. Such a demonstration can be found in chapter 3 of Morishima’s book (op. cit.). The same findings can be made concerning the distinction between the case where *i* is an intermediate commodity and the case where it is a final commodity.

All this discussion may seem somewhat esoteric, but in fact it is not, because it has allowed us to discover that in the field of values we have found an invariable standard, which can be any specified quantity of labor time, in terms of a specified time unit. This is something that had been understood by Ricardo. And this invariable standard cannot be the value of a given commodity, for the reasons which were just explained. This is of fundamental importance for any economical reasoning, since it shows us at the same time what is a fundamental flaw for both the neo-classical theory and Sraffa’s theory of production prices: i.e. the contingent nature of their measurement unit.

It must indeed be understood that in any theory all of the measured elements have necessarily the same dimension as the measurement unit, which plays the role of the standard of prices in both of these last two theories. In the neo-classical theory, all commodities have thus the same dimension as the commodity that is chosen as money, and whose conditions of production change continuously with the change of techniques and the changes in distribution. In Sraffa’s theory all commodities have the same dimension as the standard commodity, and even wages and profits, as a share of the standard net product, are measured in terms of this standard commodity. The problem is that this standard commodity changes also continuously overtime, and is invariable only in a very narrow sense, to a change in the distribution which does not modify the techniques and structure of production.

On the contrary, we can consider that the labor theory of value, such as it has been exposed in the two previous chapters, has a truly invariable standard of value, which is an arbitrary quantity of physical time, and is a perfectly valid and relevant theory. As such this theory should be rehabilitated, despite the fact that it has been and is still so harshly criticized that it has fallen into desuetude. This rehabilitation would be all the more appropriate that these criticisms can be easily dismissed, as we shall see below. But we have to realize first that several mistakes made by Marx himself certainly facilitated the task of critics.

## 4. The labor theory of value and Marx's mistakes

In the light of everything that precedes, it is easy to show that Marx himself committed in the exposition of his labor theory of value a series of mistakes. To be sure, coming after the great classical economists Smith and Ricardo, he was in good company. But these mistakes, because of the confusion that they brought about in this complicated field, certainly helped in the rejection of the labor theory of value. This is why they will now be briefly recalled.

### 4.1. A first and main error: an inaccurate vision of exchange

The first and certainly the most fundamental error has to do with the very concept of value. Marx starts from exchange as an equivalence relation, but he conceives exchange as an equivalence relation directly between commodities. Then he considers as the only common property that allows to justify such an equivalence that commodities are the product of labor, and of abstract labor, and although he understands rightly that this equivalence between commodities can and must be established at the very level of production, he thinks that ipso facto this equivalence can be extended to the level of exchange, because he conceives it as an exchange between two commodities, one of them playing the role of money. Therefore he deducts in book one of Capital that exchange is an equivalence relation between values whose "substance" is labor whereas it is in fact labor time, i.e. time, which as such has no material content and is devoid of any substance.

He makes these mistakes despite the fact that he has correctly defined abstract labor as labor time, and because he does not from the start conceives exchange as a monetary exchange, i.e. as an exchange, not between two commodities, but between a commodity and money, this last entity having nothing to do with whatever commodity. This leads us to his second error.

### 4.2. A second error: money as a commodity

This can be considered in fact simply as a corollary of the previous error, since it derives in turn from the fact that when Marx introduces money, he immediately conceives money itself as a commodity, which entails that even monetary exchange is thought of as an exchange between two commodities, in the present case between any commodity and the commodity that plays the role of money. Then it becomes impossible for Marx to establish a clear conceptual distinction between an equivalence relation at the level of values and production, on the one hand, and an equivalence relation at the level of exchange and prices, on the other hand. For Marx this second equivalence relation at the level of exchange differs from the previous one only because of some specific properties of the money-commodity. This leads us to his third error.

### 4.3. A third error: the common dimension of values and prices

Because of the two last errors, values and prices belong for Marx to the same field of measurement, which implies that prices, even though they are transformed values, still remain comparable to values. What makes prices differ from values in Marx's theory is only that it is necessary to equalize profit rates on commodities whose conditions of production differ. But values and prices remain values and prices which are relative to the value or the price of a commodity chosen as a standard of value or prices (the standard commodity as money), with the common dimension of corresponding to a quantity of abstract labor in Marx's sense. It is what explains that for Marx prices are production prices, and not monetary prices.

What has been shown so far, is precisely - and unlike Marx, that value, as an equivalence relation between commodities, cannot be an equivalence relation at the level of exchange, because precisely commodities do not exchange according to their value, and because there is no exchange between commodities, but between commodities and money. Value can therefore correspond to an equivalence relation only at the level of production, as long as production is made with wage labor. As for monetary prices, they have the same nature as money, which means that they are pure scalars, which do not have therefore the same dimension as values, and do to not belong to the same space (in the mathematical sense of this last term). Consequently it would be senseless to try and directly compare values and prices.

### 4.4. A fourth error: the transmission of value by fixed capital

The last error concerns the assumption made by Marx that fixed capital transfers its value to the product. This point has already been developed previously, therefore we will not expand too much on it. Suffice it to say that Marx could not master the scientific concept of dimension, which was developed first by Maxwell, who published in 1863 an article on the question, and supervised after 1866 the experimental determination of electrical units for the British Association for the Advancement of Science. His work in measurement and standardization led to the establishment of the National Physical Laboratory.

Therefore Marx could not grasp the fact that value, as a property of commodities, has the conceptual status of a dimension, which is not a detachable property of the commodities that possess it, and cannot be transmitted as a material substance might be (even though he sees abstract labor as a kind of substance of value - which is wrong too…). Similarly, it would be senseless to say that an object with a given mass could transmit and add its mass to another one virtually, without undergoing the same process of being weighted together.

The fact is that fixed capital, and in this regard it differs from intermediate commodities, is not transformed into commodities during the production process. Whereas intermediate commodities disappear in their original form in the course of this process, fixed capital, and for instance the machines, remain unchanged during it, because they are essentially mere operators or catalysts of the transformation that intermediate commodities undergo. As such fixed capital is obviously indispensable to this process, where it plays a fundamental role, because production could not take place without it. But in itself it does not create anything more, and in particular it does not create any more surplus than a catalyst usually does: it facilitates and quickens the transformation of intermediate commodities into different and more elaborate commodities, be they other intermediate commodities or final ones.

# 5. Some misconceptions about the labor theory of value

The labor theory of value has been much criticized over one and a half century. However, as we shall see, most of these criticisms are innocuous, because they are based on some kind of misunderstandings: this justifies that we consider them as elementary misconceptions, which will be addressed first in the next sub-section. This being done, a more serious criticism, developed initially by Arun Bose and based on the theory of values as dated quantities of labor, will be examined in a second sub-section, where it will be shown that it can also be dismissed on theoretical grounds.

## 5.1. Elementary misconceptions

The usual misconceptions are based on examples of individual commodities or special circumstances showing that monetary prices are greatly divergent from labor values: it can concern for instance a diamond found by chance without any real work, differences in productivity of workers that have no influence on values, the high price of works of art for sometimes very little work, or on the contrary the large amount of labor time that can be spent on particular occasions to produce some particular kind of goods that are useless. Some comparisons concern old goods that had a high labor value in the past but have no more now.

These criticisms do not take into account the fact that commodities are produced in large numbers, by abstract labor, i.e. labor which is not only reduced to its only dimension of time, but is also socially necessary. Consequently values correspond to an average labor time performed with average skills and an average productivity. This point was emphasized by Marx, who wrote: “Some people might think that if the value of a commodity is determined by the quantity of labor spent on it, the more idle and un-skillful the laborer, the more valuable would his commodity be, because more time would be required in its production. The labor, however, that forms the substance of value, is homogeneous human labor, expenditure of one uniform labor power. The total labor power of society, which is embodied in the sum total of the values of all commodities produced by that society, counts here as one homogeneous mass of human labor power, composed though it be of innumerable individual units...The labor time socially necessary is that required to produce an article under the normal conditions of production, and with the average degree of skill and intensity prevalent at the time” (Marx, K., 1867, Chapter 1. Section 1, p. 29).

Another quotation of Marx on the question, much further in "Capital" deserves to be noted, because it brings about even more precision to the previous statement, and shows well that the notion of an individual value is irrelevant in his theory. This quotation is the following: "the real value of a commodity is, however, not its individual value, but its social value; that is to say, the real value is not measured by the labor-time that the article in each individual case costs the producer, but by the labor-time socially required for its production. If therefore, the capitalist who applies the new method, sells his commodity at its social value of one shilling, he sells it for three pence above its individual value, and thus realizes an extra surplus value of three pence" (Marx, K., 1867, Chapter 12, p. 223).

It follows that values (like in fact prices) are average values (or average prices), established over large quantities of similar or fungible products, which have been produced by a number of producers, possibly with different techniques or methods of production. These products must themselves be useful, or in Marxian vocabulary, must have a use value. Values are also meaningful over time, since comparisons at different periods help to understand that the use-value of some commodities has decreased due to innovations that occurred later, and have made them obsolete, or that technical progress and increased labor productivity has reduced their value.

## 5.2. Arun Bose's criticism: values as dated quantities of labor and the alleged existence of a commodity residue

### 5.2.1. A criticism based on Sraffa's theory

Arun Bose is an Indian economist whose criticism of Marx's theory of value was formulated in a book published in 1980, and entitled "Marx on Exploitation and Inequality". His criticism must be taken seriously, if only because it has been endorsed by Steve Keen, who in his previously cited book, indicates: "Bose argued…that if value was the essence of a commodity, then that essence could not be reduced solely to labor. Therefore, labor alone was not the essence of value: instead, both labor and commodities were the essence of value. His logic used a concept we saw earlier in chapter 6 (of Keen's book A.N.) the reduction of commodity inputs to dated labor" (Keen, 2011, pp. 431-432).

In passing let us recall that value is not, as we have shown, the "essence" of a commodity, but one of its many dimensions, in the sense that the word "dimension" has in physics. The word "essence" indeed is a kind of philosophical notion but not a scientific concept and has no theoretical meaning. This being said, Bose's criticism is indeed strictly based on chapter VI of Sraffa's book, a chapter which is entitled "Reduction to dated quantities of labor". This explains that we prefer to start our demonstration by first going back directly to "Production of Commodities…", and more precisely to this particular chapter, which we did not discuss specifically so far.

### 5.2.2. Reduction to dated quantities of labor in Sraffa's theory

In chapter VI of his book Sraffa defines "reduction to dated quantities of labor" as "an operation by which in the equation of a commodity the different means of production used are replaced with a series of quantities of labor, each with its appropriate 'date'" (Sraffa, 1960, § 46, p. 40). Sraffa's demonstration starts with the equation giving the production price of commodity "$a$", with the wage and prices expressed in terms of the Standard commodity:

$$\left(A_a p_a + B_a p_b + ... + K_a p_k\right)\left(1+r\right) + L_a w = A p_a \qquad (1)$$

Sraffa then replaces "the commodities forming the means of production of *A* with *their own* means of production and quantities of labor, i.e. with the commodities and labor which, as appears from their own respective equations, must be employed to produce those means of production; and they, having been expended one year earlier (§ 9), will be multiplied by a profit factor at a compound rate for the appropriate period, namely the means of production by $\left(1+r\right)^2$ and the labor by $\left(1+r\right)$" (Sraffa, 1960, p. 40).

It must here be recalled that in § 9 Sraffa "assumes that the wage is paid post factum as a share of the annual product, thus abandoning the classical economists' idea of a wage advanced from capital. We retain however the supposition of an annual cycle of production with an annual market." (Sraffa, 1960, p. 11).

Sraffa repeats the same operation, and next to the direct labor $L_a$ he places "the successive aggregate quantities of labor which we collect at each step and which we shall call respectively $L_{a1}, L_{a2}, ..., L_{an}, ...$" (Sraffa, 1960, p. 41). By summation for each period he thus obtains what he calls 'the reduction equation' for the product in the form of an infinite series:

$$L_a w + L_{a1} w\left(1+r\right) + ... + L_{an} w\left(1+r\right)^n + ... = A p_a \qquad (2)$$

Sraffa then continues as follows: "How far the reduction needs to be pushed in order to obtain a given degree of approximation depends on the level of the rate of profit: the nearer the latter is to its maximum, the further must the reduction be carried. *Besides the labor terms there will always be a 'commodity residue'*[6] consisting in minute fractions of every basic product; but it is always possible, by carrying the reduction sufficiently far, to render the residue so small as to have, at any prefixed rate of profits short of R, a negligible effect on prices. It is only at $r = R$ that the residue becomes all important as the sole determinant of the price of the product" (Sraffa, 1960, § 46, p. 41).

Let us state first that we are not at all here in the realm of production and values, because there are not only wages but also profits, which cannot be realized by capitalists before they have not only paid wages to the workers but also sold their products on the market. We are therefore in the realm of distribution and prices, and more precisely of production prices fixed in such manner that they allow these capitalists to earn an average rate of profits. It follows logically that a statement which could be true for prices might not be true for values. Let us recall also that we have already seriously criticized Sraffa's theory. However and as mentioned above, in doing so we had not touched directly upon this question of reduction to dated quantities of labor, which explains why it still deserves to be discussed.

[6] The italic characters are from the author.

A first remark concerns Sraffa's position that wages are paid *post factum*, as he says, i.e. once some labor has been accomplished. We have previously explained the theoretical reason why he is obliged to do so, and we cannot but endorse this position, all the more so that it corresponds to what is the case in the real world, to which any theory should stick as much as possible in its assumptions. Indeed workers are usually and almost systematically paid after labor has been carried out, be it after one day, one week, a fortnight or most often one month. Wage is therefore not advanced, in the sense that it is not paid before labor has been executed.

This creates a logical problem in Sraffa's system, where production takes place during one full year without being sold, i.e. without any revenues for the producer, because there is "an annual cycle of production and an annual market" (see again § 9). Indeed workers cannot stay one year without being paid, and simply wait until the market day arrives! In passing this point shows the influence of the Physiocrats on Sraffa's conception of production, which he acknowledges at the very end of his book, in appendix D "References to the literature" (Sraffa, 1960, p. 111): it is as if we were in a world of agricultural production only, with one harvest per year, and an annual market afterward.

All this means that if wages are paid in due time and are therefore not advanced from the point of view of the workers, they have to be advanced nevertheless from the point of view of the capitalists. Then, if workers start being paid from almost the beginning of the year (say after one week or one month of labor), why should wages be treated differently from the means of production bought at the beginning of the year and why should they not earn the producer the same profit rate, *prorata temporis,* as these means of production? Or conversely, if wages paid during almost the totality of the first year preceding the market do not earn this profit rate, why should the means of production employed simultaneously during this same period earn it? This is quite a logical problem, left unsolved by Sraffa, who does not even mentions it.

In fact these logical difficulties all come from Sraffa's erroneous conception of production, which he sees as a discontinuous and discrete process: even if he refers no more to agricultural production but to industrial production, it is as if all producers had to buy at the beginning of each year the full stock of means of production that will be processed all along the full year to come! Obviously this assumption has nothing to do with the real world, in which production is carried out more and more on a just-in-time basis with minimal stocks to avoid storage costs and constraints. Moreover in any case the only fact that production takes place means that the necessary intermediate commodities are available at any point in time.

Anyhow, and because of his wrong conception of production as the creation of a surplus which materializes once a year, what Sraffa misses completely is that production is fundamentally a continuous process, with extremely short production periods. These periods have recently been even more shortened by the progress of robotization: an automobile plant currently produces more than 1000 cars a day, or a steelworks more than 10 000 tons per day.

At this stage we can better understand that Sraffa's theory of dated quantities of labor boils down to a revival of the Austrian theory of production, such as formulated by Eugen von Böhm-Bawerk, in "The Positive Theory of Capital", a book that he published in 1891. For this economist, due to the *roundabout* nature of production, which necessarily involves the passage of time, the return on capital arises from this detour of production and the risk that it involves. Sraffa does not seem to be aware that the Russian economist Nikolai Bukharin had stricken a powerful blow at Böhm-Bawerk's theory of production in his 1927 book: "Economic Theory of the Leisure Class". Bukharin recalled rightly that production is a continuous process, which could not take place at any stage if the necessary intermediate commodities were not fully available to be transformed in this process.

We could not say it better than Bukharin does: "*But in reality we need neither to "wait" nor postpone consumption,* for the simple reason that the social product, whatever may be the section of production we are considering, is present *simultaneously in all the stages* of its manufacture, if we are dealing merely with a social production process. Marx already pointed out that the division of labor replaces the "succession in time" by a "succession in place." Karl Rodbertus describes the process as follows: "In all the 'enterprises' of all the branches of all stages of production, simultaneous uninterrupted labor is going on. While in the production establishments of the branches of raw production, new raw materials are being won from the earth, the production establishments in the branches of intermediate products are simultaneously transforming the raw materials of the preceding epoch into intermediate products; while the tool-producing factors are replacing tools that have been used up, and while, finally, at the last stage of production, new products are being turned out for immediate consumption (Karl Rodbertus: *Das Kapital,* p. 257. Berlin, 1884.)." (Bukharin, 1927, Chapter IV, section 3, § 5).

If we dig a little further into the arcana of Sraffa's theory, as we already did in chapter 4, it appears that all these theoretical difficulties that Sraffa encounters can be traced back to his wrong view of production as a circular process. This explains that we stand here in a system where we only have intermediate goods, which implies that wages are a share of a Standard net product made of intermediate commodities. Obviously this is absurd, because workers do not consume intermediate commodities, but consumption goods which are final commodities. Similarly it is absurd to imagine that the structure of this consumption does not change when wages change and when this share is modified.

A last reason why this notion of dated quantities of labor cannot be taken seriously is the following: however far we go back in the past when computing these quantities, we must assume that the Standard net product and the Standard commodity have not changed over the years, since otherwise the changes in distribution would be meaningless, being expressed in variable Standards. This implies that we must assume that the production system itself has remained the same over the years, keeping the same techniques and the same structure. This is absurd, but it is the theoretical ransom to be paid for having a Standard measurement unit which has is a basket of commodities, and not a truly invariable Standard. We could even be tempted to say that Sraffa's Standard measurement unit has the dimension of a basket of commodities, although this would be faulty: a basket of commodities as such cannot have a dimension, in the scientific sense of the term! Moreover, when the rate of profits changes, equation (2) above shows that it has to be changed also for the preceding years, so as to remain exactly the same for the whole period, whereas in the real world the profit rate changes continuously overtime.

### 5.2.3. The absence of a commodity residue

We can now go back to the original question: is there or not, as Sraffa claims, a commodity residue? To answer precisely the question, let us remark first that in equation (1), which is Sraffa's point of departure, there are no commodities in isolation, since we stand in the space of production prices, and all the quantities of commodities are multiplied by their price in terms of the Standard commodity. Therefore a commodity residue must be understood as equivalent to a price residue. If we reformulate that in mathematical terms, it means that equation (1) will never provide us with an absolute price, because there would always remain - to paraphrase Sraffa – "a (price) residue consisting of minute fractions of (the price) of every basic product".

To decide whether or not these production prices are approximate comes down to decide whether or not the equations giving the price of commodities converge. Hence we arrive to a manageable mathematical problem.

Indeed equations (1) and (2) in sub-section 5.2.1. correspond to Sraffa's system as described earlier in chapter 3, and can be synthetized in the following matrix equation (see above in section 3 of chapter 3), where P is the column vector of prices:

$$P = \left[ I - (1+r) A \right]^{-1} wL \tag{3}$$

Where L is a column vector of labor quantities in each branch, as a fraction of total labor.

This equation is equivalent to:

$$P = \left[ I + (1+r) A + (1+r)^2 A^2 + ... + (1+r)^n A^n \right] wL \text{ , with } n \to \infty \tag{4}$$

The reason why this function converges must be demonstrated. Let us recall first what was seen in chapter 2: from the Perron-Frobenius theorem, which we used in this chapter (we recall that a demonstration is given by Gantmacher, previously cited, we know that the only eigenvalue of an irreducible matrix with non–negative terms to which corresponds a positive eigenvector is the dominant eigenvalue of this matrix, i.e. the highest one. Therefore we can write:

$AP = \alpha P$ , where A is a non-negative, sub-stochastic and productive matrix and *P* is the price vector, which the Perron-Frobenius theorem guaranties is positive.

Secondly, from the construction of the invariable Standard commodity (see above in section 3 of chapter 3), we know that $\alpha = \frac{1}{1+R}$, with R being the maximum profit rate of this system.

It follows that $AP = \frac{1}{1+R} P$

Thirdly, since $w$ is expressed in terms of the Standard commodity, we know that $r$ must necessarily be the independent variable, and that we have the relation: $r = R(1-w)$, which means that for any value of $w > 0$ we also have $r < R$.

Therefore, for any such value of $r$, we have necessarily:

$AP < \frac{1}{1+r} P \Rightarrow (1+r)AP < P$, which implies that matrix $(1+r)A$ is also by definition a sub-stochastic and productive matrix. Moreover every positive power of a stochastic matrix is stochastic, a property which extends to sub-stochastic matrices, as is shown by Peterson and Olinick (op. cit.).

From this it follows that $(1+r)^n A^n \to 0$ when $n \to \infty$

It is now easy to show that there is no commodity residue in equation (4) above which gives the price vector of the system and that the limit of the sequence of powers of matrices corresponds to a price vector of which each element is itself a real number. In other words it can be shown that none of these numbers is an approximate number, because the approximation stays only in the notation. This assertion is supported by several demonstrations. Since I am not a true mathematician, I prefer to borrow these demonstrations from a recent article by the French mathematician Jean-Paul Delahaye, published in 2016 in the French edition of *Scientific American*, and titled “Est-il vrai que 0,999… = 1 ?” (“Is it true that 0.999…=1?”). This synthesis of well-known mathematical results definitively settles the question, which explains that these short demonstrations are reproduced hereafter.

1) We know that $\frac{1}{3} = 0.333...$, because indeed it is what we obtain if we realize the division.

But if we multiply both members of the equation by 3, we obtain: $1 = 0.999...$, an equality which shows that there is no residue between 1 and 0.999…!

2) Let us have $u = 0.999...$

Then let us multiply both sides of this equality by 10. We get: $10u = 9.999...$

Now let us subtract $u = 0.999...$ on both sides of this last equality. Then, because obviously $9.999... - 0.999... = 9$, we get: $10u - u = 9 \ \Rightarrow u = 1$

We are thus back to the equation: $1 = 0.999...$, which shows again that there is no residue!

In fact this is so because numbers are nowadays conceived as limits of numerical converging sequences, and the notation for expressing that conveys the false impression that there would be a residue. All numbers are indeed the limit of a series where each number has a finite decimal expansion: 0.333 is the limit of the series 3/10, 33/100, 333/1000, etc.

3) Let us assume that $0.999... < 1$, as if there were a residue, and consider the average *m* of both numbers $0.999...$ and 1: it is a number which is less than 1 and greater than $0.999...$, because the average of two numbers is always between these two numbers. The decimal writing of m starts therefore by 0.9. This average is also greater than 0.99 and less than 1: it is therefore a number of which the decimal writing starts by 0.99. By continuing like that, we establish that the decimal expansion of *m* is necessarily 0.999…

This implies that this average *m* is 0.999…, i.e. the smaller of the two numbers. But it is absurd, because the average of two different numbers cannot by definition be equal to any of them. The only explanation to this paradox is that the starting assumption is wrong. Since it is equally impossible to have 0.999…>1, we necessarily have 1 = 0.999… which means once more that there is no residue!

We could also have ended the demonstration by stating that $u = 0.999...$, and that by definition $m = \frac{(u+1)}{2} = u$, which implies that $u + 1 = 2u$, which gives $u = 1$, or $0.999... = 1$, as always!

4) A last demonstration starts by writing that *u* is the limit, when $n \to \infty$, of a series of which the general term is :

$$x_n = 0.9 + 0.09 + ... + 0.0...09$$

We can then make the factorization:

$$x_n = 0.9\left(1 + \frac{1}{10} + ... + \frac{1}{10^n}\right)$$

Then, knowing that $0.9 = 1 - \frac{1}{10}$,

We can use the well-known identity: $(1-a)\left(1 + a + a^2 + a^3 + ... + a^n\right) = 1 - a^{n+1}$, where we replace $a$ by $\frac{1}{10}$ to obtain the value of $x_n$ such that: $x_n = 1 - \left(\frac{1}{10}\right)^{n+1}$

The series of term $10^{n+1}$ tends towards infinity when n tends towards infinity, so that $\frac{1}{10^{n+1}}$ tends towards 0, and $x_n$ tends towards 1. Therefore we come again to the conclusion that $1 = 0.999...$, which shows once more that there is no residue !

To conclude on this question, all these demonstrations should have made it clear that for compelling mathematical reasons there is no such thing as a residue, when we are dealing with the convergence of numerical series, which is what this whole question comes down. It is hard to understand how Sraffa made such a mistake, and all the more so that, if it had not been a mistake, it is his whole theory of production prices that would have suffered from this approximate character of prices. Indeed prices as dated quantities of labor are only a modality of presentation of the fundamental equations of his system, and the criticism could have been extended to the whole system.

It is equally difficult to understand how Arun Bose, and unfortunately after him, Steve Keen, may have endorsed Sraffa's contention about a residue in § 46 of his book, without checking its validity. Maybe it could be so because both economists are Sraffians who adhere to and are convinced of the validity of the whole theory of production prices? Or more simply, not being true mathematicians, they are not familiar with the present mathematical definition of limits of numerical series, which is used with the theory of sets to construct the set of real numbers. Jean-Paul Delahaye also recalls that this conception of real numbers "allows to solve the problem of the comparison between 1 and 0.999…:

$$0.999... = \lim_{n\to\infty} 0.99...9 \text{ (with } n \text{ '9')} \Rightarrow 0.999... = \lim_{n\to\infty}\left(1-\frac{1}{10^n}\right)$$

$$\Rightarrow 0.999... = 1 - \lim_{n\to\infty}\frac{1}{10^n} = 1\text{" (QED)}$$ (Delahaye, 2016, p. 79)

The number 1 is thus the limit of a series where each element has a finite decimal expansion, and so it is with production prices, that are real numbers, representing shares of the Standard commodity, which means that there is no such thing as whatever commodity residue!

### 5.2.4. Dated quantities of labor as a wrong concept

The important teaching that we can draw from this sub-section is not only that there is no such thing as a residue of commodities in the definition of prices through their reduction to quantities of labor, but also that there are no such things as dated quantities of labor, because production is a continuous process, and the whole production of a period is produced during the same period, including all of the intermediate commodities used during this period. Any possible inventories of commodities coming from the preceding period(s) must be added to the production of the period because anyway they will be transformed in final commodities of the period and at the very beginning of it, and similarly intermediate commodities unused and stored at the end of the period must be deducted from its product, which is made of final commodities only. The only thing which intervenes in the production process and can be considered as dated are the machines produced in earlier periods and still in use during the period, but anyhow they do not transfer any value to the product of any period.

Now that we have dealt with this question of residue of commodities in the realm of production prices, by demonstrating the absence of such a residue, it is going to be very easy to address it in the realm of values. It is indeed immediate to reckon that if we start from equation (2) giving the production price of a commodity in sub-section **5.2.1.** in terms of "dated quantities of labor":

$$L_a w + L_{a1} w(1+r) + ... + L_{an} w(1+r)^n + ... = Ap_a$$

And if we remove the profits, and consequently the dates, then we obtain the following equation:

$$L_a w + L_{a1} w + ... + L_{an} w + ... = Ap_a$$

If we now remove the wages w from this last equation, it will be exactly similar to the equation giving the value of a commodity $v_j$ with the second definition of values that we exposed in chapter 6 where there is only direct labor time, which includes the direct labor time used in this commodity's own production, i.e. in the transformation of intermediate commodities used in this final process, plus the direct labor time used to produce these same intermediate commodities, and in turn the direct labor time used to produce their own means of production, and so on:

$$v_j = x_{1j} l_1 + x_{2j} l_2 + ... + x_{kj} l_k$$

With $L_a = x_{1j} l_1$, $L_{a1} = x_2 j l_2$, $L_{an} = x_{n+1,j} l_{n+1}$, both equations are similar, except for the number of intermediate commodities, which can easily be fixed and does not matter.

If we now go back to the set of all the equations for all intermediate commodities, in matrix format, we get what was equation (12) in chapter 6, giving the value for the second method of calculation of these values:

$$V_1 = L_1 X_{11}$$

But we have shown that this was equivalent to equation (17), i.e.:

$V_1 = L_1 (I - A_{11})^{-1}$, which can be replaced by its expansion in a power series:

$$V_1 = L_1 \left( I + A + A^2 + ... + A^n \right) \text{ with } n \to \infty$$

Since it has been already demonstrated that this sequence of power matrices converges for prices when the general term of the corresponding series is $(1+r)A$, it is immediate that it is also convergent when the general term is A, i.e. a sub-stochastic and productive matrix. And we know that A has to be so, otherwise there would be no surplus of intermediate commodities left to be used in the production of final commodities.

It follows that there is no residue either in the calculation of values, which are real numbers in terms of the chosen measurement unit, which is a unit of physical time.

This criticism of the labor theory of value on the basis of an alleged commodity residue has thus been definitively put aside. It is therefore time now to address another and last criticism, which deserves a full chapter, since it is a serious one, and has to do with the problem of joint production.

# Chapter 8. The labor theory of value and the problem of joint production

The criticism of the labor theory of value based on an alleged commodity residue has been definitively put aside in the previous chapter, and it is time now to address another and last criticism, based on the difficulties that the labor theory of value is supposed to encounter when it is confronted to the problem of joint production. In this chapter we will show that it is possible to solve this problem correctly.

The problem of joint production is a very tangled one, for a number of reasons which are going to be exposed in the present chapter. It is at the same time a serious one, judging by the fact that the alleged problems raised by joint production are the reason why Morishima, in the final chapter of his already cited book (Chapter 14: “The labor theory of value revisited”), put into question the validity of the labor theory of value under the form that he had exposed it so far, and proposed to revise it. He intended indeed to abandon the labor theory of value for a Von Neumann model of linear programming where equations are replaced by inequalities, and labor values are replaced by minimum labor requirements.

This problem of joint production is all the more complicated that there are several definitions of joint products, a point which we will examine first, noting that these definitions correspond partly to several levels of description of the real world and different theories of values and prices, which should not be unduly mixed.

## 1. Joint production and the various types of joint products

The whole difficulty of the matter comes from the fact that joint products cannot easily be distinguished from by-products. Things would be simple if everybody accepted a strict definition of a joint product as a product that results jointly with other products from processing a common input or several ones, with the proportions of these inputs going to each product impossible to distinguish. But in practice this is not the case, even if we turn to the *United Nations System of National Accounts (Studies in Methods*), updated in 2014. This system of accounts offers the possibility to use a matrix-form of presentation of the accounts, which implies to resort to input/output tables. The methodology to develop such I/O tables is explained in an associated *Handbook of Input-Output Table - Compilation and Analysis*, last updated in 1999. This handbook distinguishes, among the “secondary products” resulting from production technology, three distinct types of products:

a) “Exclusive by-products, or products that are not produced separately anywhere, e.g. molasses linked to the production of sugar, new scrap in metal industry ;

b) Ordinary by-products, or products that are technologically linked to the production of other products but are also produced separately elsewhere as main products. An example of this is hydrogen produced as a by-product in petroleum refining establishments, but also produced separately by other establishments in the chemical industry;

c) Joint products, or products that are more loosely linked technologically than ordinary by-products. The common costs shared by joint products are more significant in value than is the case for ordinary by-products. One example is milk and meat in the livestock industry which may be produced in a scale that depends on the demand for each product and the ratio of the two products may be varied in response to changing conditions of demand. One of the joint products may be produced separately elsewhere. Joint-products cannot be easily distinguished from by-products" (United Nations, Handbook, 1999, p. 77).

However, when we leave the area of statistics to turn to a purely theoretical definition, when Sraffa addresses the question of joint production in Chapter VII of "Production of commodities…" his definition is more straightforward, since he writes: "We shall now suppose two of the commodities to be jointly produced by a single industry (or rather *by a single process***,** as it will be more appropriate to call it in the present context)" (Sraffa, 1960, p. 51 - italic characters are from the author). The definition in this very last sentence is much more precise and in fact reduces the category of joint products to the above a) category, i.e. that of exclusive by-products. This means products which share not only one common input, but all of their inputs, and share them as already noted in proportions that are impossible to distinguish.

In the theoretical world this make sense, because if we suppose that in an otherwise common process of production for two different products there is only one input (either a commodity or a quantity of labor) for which the proportion going to each product can be distinguished, then the two corresponding production processes will be different, as little as this difference may be. Translated into mathematical language, it means that it will be possible to have two formally distinct and therefore independent equations: one for each process, and therefore that we will be no longer in the area where joint production creates a problem for the determination of individual values, i.e. when we have one equation corresponding to one single process for two unknowns values to be determined, but rather in an area where we can write two distinct equations.

In any case the linkage made by Sraffa between industry and process is quite appropriate, and also interesting because it draws our attention to the complexity of the real world.

## 2. The complexity of the real world as regards production

In the real world, or more precisely in the perception that we have of the real world, there are millions of producers. Some of them are small producers, among whom we can distinguish pure individuals who do not have a single employee and cannot be called capitalists in the usual sense of the word. There are also entrepreneurs who hire a few employees, small and medium enterprises, and bigger firms or companies, some of them, the largest ones, having more than a hundred thousand people on their payroll.

It is first in the agricultural sector, where there are sometimes millions of farmers in a single country - each of them producing but a few agricultural products, that there are quite certainly more producers than products, in spite of the many varieties which exist for each agricultural product. And although production processes for one particular agricultural product may not differ greatly, they certainly often differ from one producer to another, due to particular circumstances such as the differences in land quality.

Outside of agriculture anyway, although the number of producers for each particular product is usually smaller, most producers share the common characteristic of also producing several different products each, some of them being specific to only a few producers, and some being produced by many producers, although they may be sold under different brands.

Moreover, a number of producers may produce the same commodity, but usually each producer will have its own process of production, even though each process is only slightly different from the other producers', if only because the machines used by each producer are not exactly the same, or do not have the same age, or are not maintained similarly, which may affect the use of various inputs. A single producer may also use different processes, e.g. in different establishments, workshops or factories, for the same product. As a consequence, there are also millions of production processes, certainly more than the number of commodities themselves. The opposite case is in fact that of only one single producer for one product, which is the exact definition of a pure monopoly, and is not so frequent!

## 3. The statisticians view of production and joint production

All these various cases or circumstances exist in the real world and make a very complex picture, so that we could not build up a theory of production if we were to stay at this level of complexity. To reduce this complexity to a manageable theoretical level let us first turn to some degree of abstraction, and at the outset to statisticians. In order to describe the production process in an intelligible way, they use a number of simplifying techniques. First they compile data using as the statistical unit the establishment, which is a production unit consisting of either "an enterprise, or a part of an enterprise, that is situated in a single location and in which only a single, or (non-ancillary) productive activity … accounts for most of the value added" (SNA, para. 5.21). However an establishment may engage "in one or more secondary activities, [which] should be on a small scale compared with the principal activity" (SNA, para. 5.22).

Once statisticians get the data corresponding to all of the establishments, they compile them in order to get the production accounts of "industries". According to the SNA definition, an industry consists of a group of establishments engaged on the same, or similar, kinds of production activity. These activities are classified in categories, according to the UN International Standard Industrial Classification (or ISIC) which contains 17 major sections, 60 divisions, 169 groups and 291 industries (with four digits coding). More information on this question can be found in the *Handbook of Input-Output Table* (op. cit. p. 42).

Similarly, statisticians have to reduce the thousands of products of an actual productive system to a meaningful and manageable number: this is achieved through a statistical unit for products which is a unit of homogeneous goods and services, such as they appear in the "Central Product Classification" (or CPC). This classification is quite exhaustive, with all products being mutually exclusive. The detailed classification of products, which are either outputs of domestic production activities or imports from non-resident sources, consists of 10 sections, 69 divisions, 291 groups, 1036 classes, 1787 sub-classes and can accommodate up to 65,610 categories. The principles for classification used by CPC are the following:

a) "For transportable goods, categories of products should be based on the physical properties and the intrinsic nature of products, i.e. the raw materials of which they are made, their stage of production, the use they are intended for, the prices at which they are sold, whether or not they can be stored, etc.",

b) "Individual goods and services as far as possible should contain only goods and services which are produced by a single industry" (United Nations, 1999, p. 42).

As we can see, there can be therefore in these statistical data an important discrepancy between the number of industries and the much higher number of categories of commodities, which is contrary to our initial finding that in the real world there should be more industries, in the sense of production processes, than products. For statisticians, the matrix in which a column shows for a given industry the amount of each commodity it uses as an input is generally called the "use" matrix, and will have therefore more rows (the commodities) than columns (the industries). It will be a rectangular, commodity-by-industry matrix of dimension ($n$ x $m$) and of rank $n$. Similarly the matrix in which a column shows for a given commodity the amount of it produced by each industry, generally called the "make" matrix, will be a rectangular, industry-by-commodity matrix of dimension ($m$ x $n$) and of rank $n$.

Bearing in mind that statisticians as well as theoreticians have a common purpose of using the production accounts or equations to perform some statistical or mathematical operations, then their objective is to obtain symmetrical input-output tables, of either product-by-product or industry-by-industry tables. Indeed such tables will allow statisticians to work with square matrices, which are usually invertible, since only a square matrix can be inverted to obtain what is usually called the Leontief inverse matrix. For statisticians it means that a correspondence between ISIC and CPC systems of classification is needed: this is achieved, but at a price, which is the existence of "secondary products", since their classification contains more products than industries. Indeed in I-O accounting, the primary principle is that each industry is associated with a commodity that is considered as the primary product of that industry, and all other commodities produced by this same industry are considered as secondary products.

For instance in the United States system of national accounts, which is certainly one of the best systems currently available, since year 1997 I-O tables have incorporated a new classification structure known as the North American Industry Classification System (NAICS). NAICS was updated for 2002, and this corresponding NAICS has 20 major sectors with a total of 1,179 industries. The US system and I-O tables are described by Horowitz and Planting in a book published in 2006 and updated in 2009 by the Bureau of Economic Analysis, US Department of Commerce: *Concepts and Methods of the US Input-Output Accounts.*

In spite of the high number of industries in these accounts, there are nevertheless a number of secondary products, because with the first principle above as the main criterion, different products may come out of the same industry. To give but one example provided by the CPC, published in the Statistical Papers of the United Nations: “For example, meat and hides are both produced by slaughterhouses. These products are not listed together in one category or even in the same section of the CPC. Unprocessed hides are considered raw animal materials, and they are classified in section 0 (agriculture, forestry and fishery products), whereas meat is classified in section 2, among food products” (United Nations, 2015, p. 8). It must thus be clear that most of these secondary products of a given industry have nothing to do with true joint products or by-products.

The presence of these products in the make matrix cannot but have a perturbative influence on the calculations which can be made. The UN Handbook of Input-Output Table indicates indeed that to adhere to the classification of products it is necessary to separate secondary products or joint products from the main products of an industry. Several methods can be used for the treatment of these secondary products resulting from production technology: the negative transfer method, the aggregation or positive transfer method, and the transfer of outputs and inputs. But none of them is fully satisfactory, and sometimes they can involve negative values, which may arise for purely statistical reasons.

These explanations about the statistical field shed some light on the complexity of compiling data on production activities, and bring us finally to the theoretical world, which is Sraffa’s and Morishima’s world, and the world we are concerned with. Here there is no particular constraint on the maximum number of industries or commodities that can be dealt with. After showing in the next two sections why Sraffa’s as well as Morishima’ theories of joint production are wrong, we shall go back to the labor theory of value, demonstrating that values are not incompatible with joint production, and also that even when there is joint production, there is no need to start with square matrices.

## 4. Sraffa’s theory of joint production

There are two important things to note as regards the way Sraffa deals with joint production, which he starts to do in chapter VII of “Production of commodities”…: the first one regards his definition of joint production, which is right, and the second one concerns the way he tries to solve his system of equations, which is wrong and has nothing to do with values.

Let us underline first that in the universe of theory we can get rid of absolutely all of the secondary products, because there is no statistical limitation regarding the number of industries that we can accommodate. Therefore an “industry” can be defined much more precisely than in the field of statistics: instead of regrouping establishments where "only a single, or (non-ancillary) productive activity … accounts for **most** of the value added", which implies *a contrario* the existence of secondary activities and secondary products, nothing prevents us from deciding that an industry regroups all the establishments where a single productive activity accounts for *all* the value added. In fact it comes down to considering that at a first level of the theory there is a complete correspondence between each commodity and each industry, which is by the way the assumption that has prevailed so far, and exactly the same as Sraffa's and Morishima's assumption before they introduce joint production.

In passing these observations clearly show that the rather vague notion of industry, in the common sense that it has in the real world[7], is quite different from the statistical concept of industry, which is itself different from the theoretical concept of industry such as it can be defined by economists. It follows that the concept of industry used by statisticians should not be confused with the concept of industry used by theoreticians. It is an interesting epistemological finding.

As far as we are concerned, in adopting a quite restrictive concept of industry corresponding to a particular technical process we should also be aware that, since there are more establishments and therefore more producers than industries which by definition regroup them, we always deal (and any theory does so) with averages. Indeed the techniques and methods of production generally differ - even if it is sometimes very slightly, from one producer to another, for obvious reasons of differences in location, productivity, types of machines, etc.

Therefore the processes for producing the same commodity are never exactly the same for all producers. If we are at the level of the production of one commodity in the whole economy, it means that the conditions of production and the production process for a given commodity correspond necessarily to the average process for this commodity at a given time, with such an average spanning many establishments and many production processes. At this stage to each industry producing a particular commodity corresponds therefore a single (and average) production process.

Before introducing joint production in his theory, and as we already indicated, Sraffa is thus right in deciding that the elementary unit of production is an industry identified to a single process of production. The question of the validity of the labor theory of value in the case where there are several techniques to produce the same commodity has been raised, but Toker has shown quite convincingly, in a note on the "'negative' quantities of embodied labor", published in 1984 in the Economic Journal, that such a situation is not a problem, as long as it is understood that what he calls (following Marx) the "market value"[8] of a commodity "is the weighted average of its individual values, weights being the market shares of the respective techniques" (Toker, 1984, p. 152). From this observation it should be clear that each industry represents an average process of production, as long as joint production does not come into the picture.

[7]An industry is generally identified by broad categories of products, such as: construction industry, chemical industry, petroleum industry, automotive industry, electronic industry, meatpacking industry, and so on…
[8]Since we are here at the stage of production, commodities have not yet been put on the market. Therefore it would be preferable to refer to the output proportions of the commodity in question, even though these proportions should be based on the market shares of the previous production period.

After having rightly identified an industry with a single (and therefore an average) production process Sraffa then gets rid of the assumption that each commodity is produced by a separate industry, to suppose instead that a separate industry, and therefore a single process, can jointly produce two different commodities. From this it is clear that joint production as he defines it is identified to the production of exclusive by-products, which have nothing to do with secondary products that are produced by different processes. In such a case (a single process producing two different commodities) Sraffa explains rightly that there will be a single equation for the determination of two prices, and if the situation is generalized, more prices overall than equations, which is not sufficient to determine these prices.

Sraffa sees also clearly what constitutes a correct solution to overcome this difficulty: "In these circumstances there will be room for a second parallel process which will produce the two commodities by a different method and, as we shall suppose first, in different proportions. Such a parallel process will not only be possible, it will be necessary if the number of processes is to be brought to equality with the number of commodities so that the prices may be determined. We shall therefore go one step further and assume that in such cases a second process or industry does in fact exist" (Sraffa, 1960, p. 51).

Let us stress the fact that this assumption is a perfectly legitimate one, because in the real world, as we already noted, there are always a number of producers, and as already discussed there is no particular reason why it should not be the case for processes (or industries) producing joint products. There is also no reason for these processes to be strictly identical: on the contrary there is every reason to think that different producers will use production processes which are different, even though the differences are not very important. It is also quite probable that these producers will not produce the joint products (by-products) exactly in the same proportions.

In other words, since industries are identified with processes and since each one results from the aggregation of a number of producers and processes, there is no theoretical difficulty in doing things the other way around and disaggregating a given industry in as many processes (and industries) as the number of by-products that it produces.

Sraffa is therefore right when he generalizes his position in saying that the same result, i.e. the possibility to determine prices, would be achieved "provided that the number of independent processes in the system was equal to the number of commodities produced" (Sraffa, 1960, p.52), and by considering "a system of k distinct processes each of which turns out, in various proportions, the same k products" (Ibidem). It follows that "an industry or production-process is consequently characterized, no longer by the commodity which it produces, but by the proportions in which it uses and the proportions in which it produces, the various commodities" (Sraffa, 1960, P. 53).

On this basis, Sraffa's joint production equations present themselves as follows:

$$(A_1 p_a + B_1 p_b + ... + K_1 p_k)(1+r) + Lw = A_{(1)} p_a + B_{(1)} p_b + ... + K_{(1)} p_k$$
$$(A_2 p_a + B_2 p_b + ... + K_2 p_k)(1+r) + Lw = A_{(2)} p_a + B_{(2)} p_b + ... + K_{(2)} p_k$$
..............................................................................................................
$$(A_k p_a + Bkp_b + ... + K_k p_k)(1+r) + Lw = A_{(k)} p_a + B_{(k)} p_b + ... + K_{(k)} p_k$$

It is nevertheless at this juncture that problems start to arise. Indeed, to be able to deal with changes in the distribution (in *w* and *r*), prices and wages must be necessarily expressed in terms of the Standard commodity, the wage being defined as a share of the Standard net product, which Sraffa actually intends to construct in the case of joint production in chapter VIII of his book. In order to do so it is necessary for him to transform the above equations, through the definition of ad hoc multipliers, in such a way that products will appear in the same proportions on the left and right sides of these equations. This is a condition for the definition of R, the Standard ratio (which is equal to the maximum rate of profits when we have $w = 0$).

However, in this new context, some products may appear on the right side of the equations, as joint products, and not on the left side, as means of productions. Since these products cannot for this reason be part of the Standard commodity, they have to be eliminated, which implies the occurrence of negative multipliers and therefore of negative quantities in the Standard commodity. This does not seem to bother Sraffa, who writes that in the case of the Standard commodity "there is fortunately no insuperable difficulty in conceiving as real the negative quantities that are liable to occur among its components; these can be interpreted, by analogy with the accounting concept, as liabilities or debts, while the positive components will be regarded as assets" (Sraffa, 1960, pp. 56-57).

This quote shows that here we take another step further towards the realm of nonsense, because wages are defined in terms of the Standard commodity, like in fact all prices, which have necessarily the same dimension as the Standard, i.e. as their measurement unit, and wages are nothing else than a share of the net Standard product. It means therefore that with this new definition and construction of the Standard commodity, in Sraffa's theoretical universe workers would be paid in units of this strange thing, made not of final goods, i.e. consumption goods, but of intermediate commodities, and partly of debts and liabilities!

In the same vein, in the same chapter VIII of "Production of commodities", devoted to "the Standard System with Joint Products", Sraffa goes as far as explaining that: "Thus a Standard commodity which includes both positive and negative quantities can be adopted as money of account without too great a stretch of imagination provided that the unit is conceived as representing, like a share in a company, a fraction of each asset and of each liability, the latter as in the shape of an obligation to deliver without payment certain quantities of particular commodities" (Sraffa, 1960, p. 57). One must nevertheless acknowledge that such a view corresponds to a very strange conception of a money of account, which in fact borders to the grotesque.

When a theory thus leads to strange results, and in particular to totally unrealistic ones, like it is the case for Sraffa's theory of joint production, one could think that the theory would be abandoned or changed to produce a new or modified theory which would be able to generate more realistic results. But it is not the path followed by Sraffa, nor by his followers, who seem to think that since his theory is fine, the real world, and in particular the nature of wages and money, can be skewed is such a way that it would ultimately fit to the theory. This is however a strange epistemological attitude, to say the least.

But it is not Morishima's conception, since he decides on the contrary to modify the theory, but in so doing thinks that he can also throw the baby with the bath water, and therefore jettison the labor theory of value. This leads us to discuss now his conception, or rather his misconception, of the labor theory of value, because it is based on the failure of Sraffa's theory properly to integrate joint production.

## 5. Morishima's misconception of the labor theory of value

With the theory of joint production it is an important part of the theory of production prices which produces the bizarre outcomes that we just described, and which lead us quite far from the real world. One should normally consider that this theory is thus refuted and then conclude that there is something wrong at least with this part of the theory. On this background, the next step should be that this particular part of the theory should be abandoned, in order to be reworked or rethought on different bases. But for Sraffa this was impossible, because joint production is just an introduction and an indispensable element of his theory of fixed capital, which follows in the next chapter of his book. Owing to these shaky bases, it is no surprise that we already showed, in chapter 4, that this last one was also deeply flawed.

Quite differently from Sraffa, Morishima does recognize the absurdity of negative quantities, and he is convinced that he can remedy these difficulties by developing an alternative theory, based on the Von Neumann's model. But what is not understandable is why he considers at the same time that the theoretical flaws of the theory of production prices invalidate the labor theory of value, which has not at all the same conceptual background, in particular as regards its Standard, which as we demonstrated has nothing to do with any kind of composite commodity!

For Morishima this question of negative quantities of commodities appearing in Sraffa's theory of joint production is indeed considered as the crux of the matter, and as the basis for his strong criticism of the labor theory of value. Following Sraffa, he starts by considering that negative quantities of commodities appear in the theory of production prices. At the same time he believes that this theory can define the value of commodities as made of "embodied labor". For him it means ipso facto that there are negative labor values! But the values as he defines them, i.e. through the same kind of equations as the ones above, are nothing more than production prices when the rate of profits is equal to zero, which explains why he is wrong, and this on several grounds. Then we are no longer in a problematics of transformation of values into prices, but on the contrary in a reverse logic of transformation of prices into values, which is another kettle of fish!

First it must be recalled that labor values are defined at the level of production, before any sale has taken place, which implies that no profit has been obtained yet, whereas production prices are defined at the level of distribution, because they incorporate not only wages but also a uniform rate of profits, which implies (even if the rate of profit is equal to zero), that prices have been realized by the sale of commodities on the market.

Moreover, as we recalled above, the dimension of prices in this theory is the dimension of their measurement unit, i.e. the Standard commodity, which is a composite and heterogeneous commodity. It is indeed a necessary condition for the existence of a linear relationship between wages and the rate of profits. In passing, all these particularities show that to speak of production prices is quite inappropriate, and that it would be much better and accurate to name them distribution prices. It remains that these prices have nothing to do with time, which is the dimension of values.

Morishima wants nevertheless to demonstrate that the labor theory of value, like Sraffa’s theory of production prices (of which he considers that the labor theory of value is a particular case!) fails when it comes to joint production. This is the reason why, at the beginning of the final chapter of his book, he tries to show it by giving an example of negative labor values. But the big problem which arises then is that his example is completely absurd! (Morishima, 1973, Chapter 14, pp. 181-182).

To show it, let us present Morishima’s example with a system of two matrices, one use matrix of inputs and one make matrix of outputs. It replaces the simple Leontief model, in which one industry produces only one commodity, and each commodity is produced only by one industry. Leontief model is considered as a symmetric model, with therefore only one square matrix, which cannot handle joint production, because there is no distinction between commodities and industries.

The make-use model was introduced by the United Nations in the 1968 System of National Accounts, and is well explained in a paper published in 2002 by Guo, Lawson and Planting, from the US Bureau of Economic Analysis: “From Make-Use to Symmetric I-O Tables: an Assessment of Alternative Technology Assumptions”. We can therefore insert Morishima’s data into such a model, which is done in table 8.1.

**Table 8.1 - Morishima’s example of joint production presented in a make-use table**

| | Commodities Outputs | | | Industries (processes) | | | Total Output |
|---|---|---|---|---|---|---|---|
| | 1 | 2 | 3 | 1 | 2 | 3 | |
| Commodities Inputs | | | | | | | |
| 1 - Intermediate good | | | | 0.7 | 0.9 | 0.9 | 2.0 |
| 2 - New fixed capital good | | | | 0.5 | 0 | 0 | 1 |
| 3 - Old fixed capital good | | | | 0 | 0.5 | 0 | 0.5 |
| Industries (processes) | | | | | | | |
| 1 | 1.0 | 0 | 0.5 | | | | |
| 2 | 1.0 | 0 | 0 | | | | |
| 3 | 0 | 1 | 0 | | | | |
| | | | | Labor | | | |
| Value added | - 0.5 | 0.5 | 0 | 1 | 1 | 1 | |
| Total Input | 2.5 | 0.5 | 0.5 | | | | |

We can see from this table 8.1 that in Morishima’s example there is only one intermediate commodity, named good 1, and one capital good, which can serve for two periods. The new fixed capital good is designated as good 2 and the old fixed capital good is designated as good 3. The fixed capital goods are not used to product themselves but to produce the intermediate commodity, which can thus be produced by two industries (or processes), 1 and 2, utilizing the new and old fixed capital goods, respectively. Like in Sraffa’s treatment of fixed capital, the old capital good is a by-product of process 1 using the new capital good. The industry (or process) which produces the new fixed capital good is called process 3.

Input-output coefficients are given in the table, where Morishima assumes that process 2 requires a greater amount of the circulating capital good (i.e. 0.9) than process 1 (i.e. 0.7), because the former uses the old fixed capital good and the latter the new one. Then the so-called “values” are calculated from the following equations, each corresponding to an industry, where inputs are on the left side, and outputs on the right side:

$$0.7\lambda_1 + 0.5\lambda_2 + \ldots\ldots\ldots + 1 = \lambda_1 + \ldots\ldots + 0.5\lambda_3$$
$$0.9\lambda_1 + \ldots\ldots\ldots + 0.5\lambda_3 + 1 = \lambda_1 + \ldots\ldots + \ldots\ldots\ldots$$
$$0.9\lambda_1 + \ldots\ldots\ldots\ldots + \ldots\ldots\ldots + 1 = \ldots\ldots + \lambda_2 + \ldots\ldots\ldots$$

Solving these equations, Morishima obtains negative values: $\lambda_1 = -50$, $\lambda_2 = -44$, $\lambda_3 = -12$. But are this *ad hoc* system and its results significant? Obviously the answer is no, because its assumptions are incompatible with common sense, i.e. the sustainability and therefore the mere existence of the system, and this because of two errors:

- First the combined inputs of the intermediate good (i.e. 2.5) appearing in the last row of the table, are larger than the production of the same good (i.e. 2), which is incompatible with a self-replacing state ;

- Second, because the treatment of fixed capital does not correspond either to Sraffa's treatment of fixed capital – however flawed it may be (Sraffa, 1960, pp. 76-80), or to a self-replacing state. Since for Morishima as well as Sraffa, fixed capital transfers its value to the product, then if there is one unit of new fixed capital good 2 produced at each period by process 3 and serving for two periods, which is the case in Morishima's example, then at the beginning of each period and therefore during each period this same unit of new fixed capital good 2 should be used as an input, in whatever process, with half of the value of this new fixed capital good transferred to the product, and the other half transformed into the old fixed capital good 3, as a by-product of the processes where it has been used (in this example, process 1 only). Otherwise fixed capital does not transfer its value to the product, which is what happens in process 1 (even though it is wrong). It is the reason why there is a net input of 0.5 for the new fixed capital good in the above table, which is not compatible with a self-replacing state. However the treatment of the half-unit of old fixed capital good is correct, because at each period its full (residual) value of 0.5 is transferred to the product by industry 2, then without any by-product.

However Morishima acknowledges that "the case of all goods having negative values is obtained only when the system does not satisfy the conditions of productiveness." (Morishima, 1973, p. 182). In this example the corresponding input-coefficient matrix is not productive because the system uses more input of intermediate good 1 (2.5 units in total) than its produced quantity (only 2 units). One can therefore wonder why Morishima bothered to give an example which precisely did not meet these conditions, and has therefore nothing to do with the real world.

Then Morishima transforms his example by merging the two first processes producing good 1, and keeping only their average, which gives him the following system of equations, which he calls the neo-classical system:

$$\begin{aligned} 0.8\lambda_1 + 0.25\lambda_2 + 1 &= \lambda_1 \\ 0.9\lambda_1 + \ldots\ldots\ldots\ldots + 1 &= \lambda_2 \end{aligned}$$

He notes that we obtain again the same negative values of $\lambda_1 = -50$, $\lambda_2 = -44$, which is not surprising because the first equation is the average of the former two first ones: indeed this does not prevent the matrix from being non-productive, since there is 1.7 units of good 1 used as inputs whereas as it is easy to see only 1 unit is produced. He concludes from this second system that "the productiveness of the neo-classical system is necessary and sufficient for the positivity of the values of the circulating and new fixed capital good".

Finally from there he develops a last case where the input-coefficient matrix is productive, and which is the following:

$$\left.\begin{array}{l} 0.7\lambda_1 + 0.5\lambda_2 + ......... + 1 = \lambda_1 + ..... + 0.5\lambda_3 \\ 0.9\lambda_1 + ......... + 0.5\lambda_3 + 1 = \lambda_1 + ..... + ......... \\ 0.2\lambda_1 + ......... + ...... + 0.5 = .... + \lambda_2 + ......... \end{array}\right\} \text{ Morishima's system of equations}$$

He notes however that despite the fact that we obtain positive values for $\lambda_1$ and $\lambda_2$, at 7.5 and 2.0 respectively, negative values can still appear, which is the case for the old fixed capital good 3, its value being $\lambda_3 = -0.5$.

In fact it is so because Morishima does not fully realize the consequences of his own assumption that there is only one fixed capital good, which has a two period life and thus appears with two ages in the production processes: age 0, when it is produced by process 3 and used by process 1, and age 1 when it is used by process 2. To be sure, this does not prevent from naming the same goods of different ages goods 2 and 3, like he does, but it is not a pure convention, because there is a link between goods 2 and 3: indeed on the basis of Sraffa's and Morishima's assumptions regarding fixed capital and the transmission of its value to the product, good 3 is nothing more than good 2 which has lost, because it is one year old, a part of its value corresponding to its amortization in the process where it is used.

This forbids absolutely to write the first of the above three equations like Morishima does. In fact, to rewrite it properly, we must first take into account that fixed capital good 2 transfers in process 1 a value corresponding to its amortization, with its residual value, which Morishima seems to set at half its original value, appearing as a by-product in the form of good 3. We could be tempted to write it as in the following equation:

$$0.7\lambda_1 + 0.5\lambda_2 + ......... + 1 = \lambda_1 + ..... + (1 - 0.5)\lambda_2$$

But we can see that both terms in $\lambda_2$ cancel, which precludes any transfer of value to the product from the new fixed capital good 2. As a consequence, if there must be any transfer of value to $\lambda_1$ (the value of the intermediate good) the equation must be rather written as:

$0.7\lambda_1 + (1 - 0.5)\lambda_2 + ......... + 1 = \lambda_1$, which gives us:

$$0.7\lambda_1 + \lambda_2 + ......... + 1 = \lambda_1 + ..... + 0.5\lambda_2$$

If we then consider like our author that the residual value of good 2 appears as a by-product, under the form of a distinct good which is fixed capital good 3, it is not 0.5 units, but one full unit of this new good 3 which must necessarily appear on the right side of the equation:

$$0.7\lambda_1 + \lambda_2 + ......... + 1 = \lambda_1 + ..... + \lambda_3$$

The link between the two equations is obviously that $\lambda_3 = (1 - 0.5)\lambda_2$, assuming a 50% amortization of good 2 in the first period, which is a simple linear amortization rule (meaning that a fixed capital good is amortized in as many equal shares as the number of its periods of use), if we compare it to Sraffa's complicated demonstration, but corresponds to the fact that here there is no rate of profit to complicate the matter. But then in the second equation 0.5 $\lambda_3$ should in turn be replaced by $\lambda_3$. and we would therefore obtain quite a different - but coherent - system, which would be :

$$\left.\begin{array}{l} 0.7\lambda_1 + \lambda_2 + ........ + 1 = \lambda_1 + ..... + \lambda_3 \\ 0.9\lambda_1 + ..... + \lambda_3 ..... + 1 = \lambda_1 + ..... + ... \\ 0.2\lambda_1 + ..... + ..... + 0.5 = ..... + \lambda_2 + ... \end{array}\right\} \text{ Correct system of equations}$$

However the two first equations are not independent, because of the amortization rule which links $\lambda_3$ to $\lambda_2$ (here it is $\lambda_3 = (1-0.5)\lambda_2$), and would make the system overdetermined, with 4 equations and 3 unknowns. The only way to avoid that is to sum the first two equations, to obtain finally the following system, where we can see that $\lambda_3$ has been eliminated:

$$\left.\begin{array}{l} 1.6\lambda_1 + \lambda_2 + ....... + 2 = 2\lambda_1 \\ 0.2\lambda_1 + .... + ..... + 0.5 = \lambda_2 \end{array}\right| (1)$$

The elimination from the first equation of $\lambda_3$, i.e. the value of the by-product, confirms that it is not possible to define the value of a fixed capital good by treating it as a by-product, while treating it at the same time as an intermediate good which transfers its value through an amortization process, something which we demonstrated in a more general case in Section 4 of chapter 4 on the treatment of fixed capital in Sraffa's system.

In any case, this system (1) above has two positive solutions for values of goods 1 and 2, i.e. $\lambda_1 = 12.5$ and $\lambda_2 = 3$, with the amortization rule giving us the value of good 3, as $\lambda_3 = 1.5$.

From all the foregoing we cannot but infer that Morishima's ad hoc example is meaningless and quite flawed, and proves nothing. But our author does not realize it, and on the contrary uses this supposed inconsistency (coming in fact from his initial assumptions) to derive what he thinks are quite general rules dismissing the labor theory of value. Ultimately such an example justifies his attempt to build his case of replacing the labor value system by a Von Neumann model with inequations, from which are derived minimum labor requirements, rather than labor values. This transforms his system into a theory designed mainly for the choice of the most productive techniques, which has nothing to do with the real world. In this world indeed a number of techniques and processes exist at any point in time for the production of any commodity, and values as social values are necessarily an average of this multiple methods of production.

It seems nevertheless all the more useless to continue discussing at length Morishima's demonstration that it is in fact quite possible to show that joint production is perfectly compatible with the determination and calculation of non-negative labor values, which will be achieved in the next section.

# 6. Obtaining positive values in a realistic model of joint production

The difficulties that we encountered so far have shown clearly enough that joint production is for economists a true problem and at the same time a tricky phenomenon, which raises a number of question-marks about the best way to deal with it, in particular from a mathematical point of view. Statisticians are generally considered as more familiar with mathematics as many economists, and it should therefore be no surprise that the light at the end of the tunnel of joint production came from them. Indeed, as soon as 1968, they proposed two different methods for transferring secondary outputs and associated inputs by combining the use and supply matrices mathematically in order to build symmetric (and therefore invertible) input-output matrices. A quick presentation of these methods, and of the reason for selecting that which seems the most appropriate for the resolution of our problem, will allow us to propose a corresponding representation of a joint production system. This will in turn provide a background for a mathematical determination of values in such a system.

## 6.1. Statistical methods developed for dealing with joint production

These methods are exposed in the United Nations 1999 "Handbook of Input-Output Table", already cited. The first method was based on what was called an "industry technology assumption", which "assumes that inputs are consumed in the same proportions by every product produced by a given industry, which means that principal and secondary products are all produced using the same technology, i.e. the same input structure" (United Nations, 1999, p. 86). However this method was quickly found to break the fundamental economic rule that products with different prices at a given moment (i.e. secondary products, as opposed to the principal product of an industry) must reflect different costs or different technologies. Therefore the UN statisticians preferred to recommend the use of another assumption, i.e. the "commodity technology assumption", which assumes that the input structure of the technology that produces a given product is the same no matter where (by which industry) it is produced. Although this second method is indeed better adapted to the secondary products that statisticians have to deal with, "it tends to generate negative symmetric input-output tables and requires the make and intermediate matrices of the use table to be squared" (United Nations, 1999, p.87). This is the reason why this method is not widely used.

However, and as far as we are concerned, joint production is not defined in a way which applies to secondary products that appear in each industry (in the sense given to the word industry by statisticians), because these secondary products are produced with different costs, i.e. with different technologies. On the contrary our definition of joint production, given in section 1 of this chapter, precisely reduces the category of joint products to that of exclusive by-products, i.e. products that have exactly the same input structure. Therefore the criticism made to the industry technology assumption is not relevant in the case of exclusive by-products, and this assumption perfectly fits our definition of joint production.

This explains that we will adopt this industry technology assumption for our demonstration regarding the determination of values within a system of joint production, and all the more so that this assumption is also attractive for two reasons: first, it is applicable to the case of rectangular input-output tables, but - and this is the second and most important reason, the method always generates positive symmetric input-output tables, which can be represented by square matrices.

### 6.2. Joint production as a physical process

Before developing a mathematical solution, we can summarize the new view of production that emerges when we take into account joint production, by redrawing the scheme that we used (in section 2 of chapter 5) to describe production as a physical process. Let us stress that this scheme is not a matrix, and that each line represents an industry (as a distinct technical process). The *i* and *j* indices refer to commodities and industries, respectively. The situation prevailing previously, such as represented in the scheme on page 102, was rather simple, because to each commodity only one industry was corresponding, and thus the distinction made between industries was the same as the distinction between commodities, which were separated between three categories: intermediate commodities of the first type (listed from *1* to *k*), of the second type (listed from *k+1* to *n*), and final goods (listed from *n + 1* to *s*).

But now, in the more complicated world of joint production, the one-to-one correspondence between commodities and industries (or processes) has disappeared. To be sure the same distinction between various categories of commodities still exists, but we cannot use it in the same way to establish an equivalent distinction between industries. Indeed we can no more distinguish between those producing intermediate commodities of the first type, of the second type, and final commodities, because they can produce in their outputs several commodities belonging theoretically to anyone of these three categories. If we still can distinguish between industries, it is no more on the basis of their outputs, but only on the basis of the commodities that they use as their inputs: some will use only intermediate commodities of the first type, and others will use both categories of intermediate commodities. Thus there is no more any reason why the number of industries should coincide with the number of commodities. Therefore the industries (or processes) using only intermediate commodities of the first type will be listed from *1 to l,* the industries using both types of intermediate commodities will be listed from $l+1$ to $q$ and the industries producing final commodities from $q+1$ to *t*. Commodities will continue to be listed with the same indices as before.

As for the by-products of each industry (or process), most authors having worked on the question, starting with Sraffa, seem to consider that in the area of joint production any industry can produce any kind of commodity, belonging to any of the above categories, or at least such an assumption is implicit in their models, since the question is never really discussed. As far as we are concerned, and on the basis of the very definition of joint products as exclusive by-products having exactly the same structure of inputs, it seems nevertheless extremely doubtful that whatever industry should product jointly as by-products (distinct from secondary products) both intermediate commodities and final commodities, for a first and simple reason which has to do with the differentiation of final commodities.

This means that from one final commodity to the other, and even though two commodities are almost close substitutes, there are differences at the end of the supply chain which imply at the very least some differences in the structure of inputs that have been transformed to produce them. And we already noted that the slightest difference in the input structure of two commodities is sufficient to prevent them from being exclusive by-products. This is true for consumption goods as well as for these other final goods that are machines or more generally fixed capital goods. In fact by-products are more prone to be found in the upstream part of the production process, i.e. at the stage of production of primary commodities or commodities immediately derived from them, which are also by definition intermediate commodities.

Another and mutually reinforcing reason why by-products of these two different types seem quite improbable is that intermediate commodities are sold by producers only to other producers, and never to consumers. It implies that final goods sold to consumers reach them through distinct and separate distribution and marketing channels that increase or create their differentiation, since they change their cost and input structure. To be sure machines of fixed capital goods are also sold from producers to producers, but as we just noted it would be quite difficult to find out any plausible example of a fixed capital good as a by-product.

All these remarks obviously have a bearing on the possible representations of joint production with exclusive by-products. If we focus first on industries (as processes) whose only inputs are intermediate commodities of the first type (listed from *1* to *k*), we can reasonably assume that there will be processes (listed from 1 to *l*) that produce by-products of the same type (listed from *1* to *k*), and even simultaneously by-products that are intermediate commodities of the second type (listed from $k+1$ to *n*). But it seems highly improbable that such processes produce simultaneously final goods (listed from $n+1$ to *s*) as by-products, for the reason just developed, which explains that this possibility will therefore not be retained.

A second category of processes (listed from $l+1$ to *q*) includes those that use as inputs both types of intermediate commodities, and since by definition intermediate commodities of the second type do not enter in the production of intermediate commodities of the first type, the by-products resulting from these processes will only be intermediate commodities of the second type.

As for final commodities (listed from $n+1$ *to s*), even though it is difficult to imagine a process which would produce intermediate goods of the first type and simultaneously a consumption good or a fixed capital-good as by-products, one cannot *a contrario* preclude that some particular processes (listed from $q+1$ to *t*) might produce simultaneously intermediate goods of the second type and such final goods (as an example one can think of fruit-trees providing not only fruits – or final goods, but also some firewood – or intermediate goods). Then the corresponding production system would have to be represented as in the scheme appearing in table 8.2, where B*ij* corresponds to the quantity of product *i* used by industry (or process) *j*, and D*ij* to the quantity of product *i* produced by industry (or process) *j*:

**Table 8.2 - A schematic representation of a joint production system with by-products**

| | | | | | | | | | | | | | | | | | |
|---|---|---|---|---|---|---|---|---|---|---|---|---|---|---|---|---|---|
| $B_{11}$ | .. | $B_{k1}$ | 0 | .. | 0 | *and* | $L_1$ | $\rightarrow$ | $D_{11}$ | .. | $D_{k1}$ | $D_{k+1,1}$ | ... | $D_{n1}$ | 0 | .. | 0 |
| ... | .. | ... | .. | .. | .. | ... | ... | ... | ... | .. | ... | ... | ... | ... | .. | .. | .. |
| $B_{1l}$ | .. | $B_{kl}$ | 0 | .. | 0 | *and* | $L_l$ | $\rightarrow$ | $D_{1l}$ | .. | $D_{kl}$ | $D_{k+1,l}$ | ... | $D_{nl}$ | 0 | .. | 0 |
| $B_{1,l+1}$ | .. | $B_{k,l+1}$ | $B_{k+1,l+1}$ | .. | $B_{n,l+1}$ | *and* | $L_{l+1}$ | $\rightarrow$ | 0 | .. | 0 | $D_{k+1,l+1}$ | .. | $D_{n,l+1}$ | 0 | .. | 0 |
| ... | .. | ... | ... | .. | ... | ... | ... | ... | ... | .. | ... | ... | .. | ... | .. | .. | .. |
| $B_{1q}$ | .. | $B_{kq}$ | $B_{k+1,q}$ | .. | $B_{nq}$ | *and* | $L_q$ | $\rightarrow$ | 0 | .. | 0 | $D_{k+1,q}$ | .. | $D_{nq}$ | 0 | .. | 0 |
| $B_{1,q+1}$ | .. | $B_{k,q+1}$ | $B_{k+1,q+1}$ | .. | $B_{n,qD=V\hat{q}^{-1}+1}$ | *and* | $L_{q+1}$ | $\rightarrow$ | 0 | .. | 0 | $D_{k+1,q+1}$ | .. | $D_{n,q+1}$ | $D_{n+1,q+1}$ | .. | $D_{s,q+1}$ |
| ... | .. | ... | ... | .. | ... | ... | ... | ... | ... | .. | ... | ... | .. | ... | ... | .. | ... |
| $B_{1t}$ | .. | $B_{kt}$ | $B_{k+1,t}$ | .. | $B_{nt}$ | *and* | $L_t$ | $\rightarrow$ | 0 | .. | 0 | $D_{k+1,t}$ | .. | $D_{nt}$ | $D_{n+1,t}$ | .. | $D_{st}$ |

This table 8.2 is built on the assumption that despite the existence of exclusive by-products, which might potentially imply that any industry could simultaneously produce any kind of goods, be they intermediate or final, there are nevertheless various types of industries, depending on the nature of commodities that they produce. This seems indeed to correspond more closely to the real world. It means that industries noted from 1 to *l* use only basic goods of the first type to produce as exclusive by-products both basic goods of the first type, noted from 1 to *k*, and of the second type, noted from $k+1$ to *n*. Industries noted from $l+1$ to *q* use both types of basic commodities to produce as exclusive by-products basic goods of the second type only, noted from $k+1$ to *n*. Finally industries noted from $q+1$ to *t* use both types of basic commodities to produce as exclusive by-products both basic goods of the second type, noted from $k+1$ to *n* and final goods noted from $n+1$ to *s*.

It must also be emphasized that by-products, despite the attention devoted to them, are not so frequent in the real world, and that when they occur there are rarely more than two by-products produced by the same process. As a consequence, even in the areas (in the above table) where some outputs are deemed potentially to be found, most of the quantities in the cells could well be zero. But this should not jeopardize the mathematical representation of the system.

Coming back to the industry technology assumption, let us recall that its basic assumption is that inputs are consumed in the same proportions by every commodity produced by a given industry, and thus in fact by every process, since to each process corresponds one industry. Let us stress again the fact that if such an assumption cannot be considered acceptable in the case of secondary commodities which are produced with different technologies, i.e. with different actual costs, by the same industry, it is nevertheless perfectly appropriate and acceptable in the case of pure or exclusive by-products, which are produced with exactly the same technology, and therefore the same proportions in inputs and the same costs.

On the particular point of the definition of an industry in a system of joint production, since a process can produce several by-products, and in order to have as many equations as unknowns, we must adopt the same assumption as Sraffa, who on this particular question was quite right : in order for the number of processes to be brought to equality with the number of commodities we must assume that each time an industry produces several by-products it can be decomposed (or disaggregated) in as many industries (or processes) as the number of by-products that it produces. Alternatively, this same equality between the number of processes and by-products can also be achieved through another way, as Sraffa also rightly points out: "even if the two commodities were jointly produced by only *one* process, provided that they were *used* as means of production to produce a third commodity by two distinct processes; and, more generally, provided that the number of independent processes was equal to the number of commodities produced" (Sraffa, 1960, chapter 7, §50, p. 52).

From what we saw about processes and commodities, we can in any case consider this assumption as quite realistic, apart from the last corollary established by Sraffa, and according to which "*the number of independent processes was equal to the number of commodities produced".* This equality is indeed a very particular case, which has no reason to exist in the general case and above all in the real world. If we recall that an industry is an intellectual construction, built through the compilation of data from many individual producers, each of them most often having its own process, each industry represents an average process or method of production. Therefore it is an assumption which is clearly neither unreasonable nor unrealistic to make: in the real world there are certainly more processes (or industries) than commodities, which obviously has an important consequence on mathematical representation of production processes, since it means that matrices will not be square, but rectangular.

### 6.3. A mathematical presentation of joint production

The starting point of this mathematical presentation is based on the make-use input-output tables developed by statisticians from the US Bureau of Economic Analysis (Guo, Lawson and Planting, 2002, pp. 1-42) and derived from the Input-Output Tables of the SNA as established by UN statisticians (United Nations, 1999, p. 88). To begin with the notations, we will keep the indices used so far, and therefore:

*s* is the number of commodities (decomposed in *k*, $n-k$ and $s-n$)

*t* is the number of industries (decomposed in *l*, $q-l$ and $t-q$)

*U*, of dimensions (*s* x *t*) is the intermediate table of the use table (commodity by industry)

*B*, of dimensions (*s* x *t*) is the use coefficient matrix (commodity by industry)

*V*, of dimensions (*t* x *s*) is the make matrix (industry by commodity) transposed of the supply table *M* of dimensions (*s x t*) – commodity by industry, describing domestic production

*D*, of dimensions (*t* x *s*) is the commodity output proportions matrix (industry by commodity)

$g_t$ is the column vector of industry output

$q_s$ is the column vector of commodity output

$\hat{g}$ is the diagonal matrix of industry output

$\hat{q}$ is the diagonal matrix of commodity output

With these notations, the scheme of a joint production system as represented above in table 8.2 can be represented in table 8.3, where we have the two matrices U and V as defined above, and where the elements in capital letters represent quantities of the various commodities:

**Table 8.3 A Make-Use Input-Output table for joint production**

| | Commodities | | | | | | | | | Industries | | | | | | | | | Total |
|---|---|---|---|---|---|---|---|---|---|---|---|---|---|---|---|---|---|---|---|
| | *1………k* | | | *k+1………n* | | | *n+1………s* | | | *1………l* | | | *l+1………q* | | | q+1………t | | | Output |
| Com. | | | | | | | | | | Use Table: U | | | | | | | | | |
| *1* | | | | | | | | | | B11 | … | ... | B1,l+1 | … | B1q | B1,q+1 | … | B1t | |
| … | | | | | | | | | | … | … | … | … | … | … | … | … | … | |
| *k* | | | | | | | | | | B*k1* | … | … | Bk,l+1 | ... | Bkq | Bk,q+1 | ... | Bkt | |
| *k+1* | | | | | | | | | | 0 | ... | 0 | Bk+1,l+1 | ... | Bk+1,q | Bk+1,q+1 | ... | Bk+1,t | *q* |
| … | | | | | | | | | | ... | ... | ... | ... | ... | ... | ... | ... | ... | |
| *n* | | | | | | | | | | 0 | ... | 0 | Bn,l+1 | ... | Bnq | Bn,q+1 | ... | Bnt | |
| *n+1* | | | | | | | | | | 0 | ... | 0 | 0 | ... | 0 | 0 | ... | 0 | |
| … | | | | | | | | | | ... | ... | ... | ... | ... | ... | ... | ... | ... | |
| *s* | | | | | | | | | | 0 | ... | 0 | 0 | ... | 0 | 0 | ... | 0 | |
| Indus. | Make Table: V | | | | | | | | | | | | | | | | | | |
| *1* | D11 | … | … | Dk+1,1 | ... | Dn1 | 0 | ... | 0 | | | | | | | | | | |
| … | … | … | … | … | … | … | … | … | … | | | | | | | | | | |
| *l* | D1l | … | … | Dk+1,l | … | Dnl | 0 | … | 0 | | | | | | | | | | |
| *l+1* | 0 | … | 0 | Dk+1,l+1 | … | Dn,l+1 | 0 | … | 0 | | | | | | | | | | *g* |
| … | … | … | … | ... | … | … | … | … | … | | | | | | | | | | |
| *q* | 0 | … | 0 | Dk+1,q | … | Dnq | 0 | … | 0 | | | | | | | | | | |
| *q+1* | 0 | … | 0 | Dk+1,q+1 | … | Dn,q+1 | Dn+1,q+1 | | Ds,q+1 | | | | | | | | | | |
| … | ... | ... | ... | ... | ... | ... | ... | ... | ... | | | | | | | | | | |
| *t* | 0 | … | 0 | Dk+1,t | ... | Dnt | Dn+1,t | ... | Dst | | | | | | | | | | |
| Total | | | | | *q'* | | | | | | | | | *g'* | | | | | |
| Input | | | | | | | | | | | | | | | | | | | |

Regarding now the notations, so far in this book we used to name $a_{ij}$ the production (or input-output) coefficients, which correspond to the share of the total quantity of a product *i* used by an industry (or process) *j*.

Since one industry produced only one commodity, and reciprocally, these coefficients corresponded to a commodity-by-commodity square matrix, and we will keep them in this role. But the industry technology assumption, with which we are now working, allows us to start from a more general case of rectangular input-output tables, which are normally expected in the case where there are more industries (*l*, *q* or *t*) than commodities (*k*, *n* or *s*). As we shall see this will not prevent us from generating symmetric input-output tables.

Reserving thus the $a_{ij}$ notation to the particular case of a commodity-by-commodity square matrix, the data upon which we must rely are as usual the methods or processes of production, represented by the share of each commodity's total output that is used as an input by each industry. We will rename these coefficients as $b_{ij}$, in the case of the use coefficient matrix (commodity by industry) derived from the use matrix *U* in the above table, and which is therefore named *B*. This commodity-by-industry, direct requirements matrix *B* has *s* rows, corresponding to *s* commodities, and *t* columns, corresponding to *t* industries (or processes), and is therefore a rectangular matrix of dimensions (*s* x *t*). It shows (in rows) the share of each commodity's total output that is used as an input by each industry, which explains why the last rows from $n+1$ to s are made of zeros, because final goods are not used as inputs. It also shows (in columns) the commodity composition of each industry's total inputs.

Matrix *B* is derived by dividing the use matrix *U* of dimensions (*s* x *t*) by the vector of industry total output *g* of dimensions (*t* x *t*):

$$B = U\hat{g}^{-1} \qquad (1)$$

Commodities Industries

[ 1………*l* *l*+1………..q q+1………..t ]

$$\begin{bmatrix} 1 \\ .. \\ k \\ k+1 \\ .. \\ n \\ n+1 \\ .. \\ s \end{bmatrix} \quad \begin{bmatrix} b_{11} & b_{1l} & b_{1,l+1} & b_{1q} & b_{1,q+1} & b_{1t} \\ & & & & & \\ b_{k1} & b_{kl} & b_{k,l+1} & b_{kq} & b_{k,q+1} & b_{kt} \\ 0 & 0 & b_{k+1,l+1} & b_{k+1,q} & b_{k+1,q+1} & b_{k+1,t} \\ & & & & & \\ 0 & 0 & b_{n,l+1} & b_{nq} & b_{n,q+1} & b_{nt} \\ 0 & 0 & 0 & 0 & 0 & 0 \\ & & & & & \\ 0 & 0 & 0 & 0 & 0 & 0 \end{bmatrix} \qquad (2)$$

In the case of joint production, one particular process can produce several different products. Even though in the real world pure or exclusive by-products are quite rare, and there are not so often more than two exclusive by-products for a particular joint production process, from a theoretical point of view we must consider a general case in which any process is able to produce a variety of different joint-products, in the sense of exclusive by-products.

In order to give a mathematical representation of this reality, we will use below a "commodity-output-proportions" matrix *D*, which is an industry-by-commodity matrix. It is a matrix which has t rows (the industries) and s columns (the commodities), and is therefore also a rectangular matrix, of dimensions (*s* x *t*). It shows (in rows) the commodity composition of each industry's total output of various commodities, which are exclusive by-products. It also shows (in columns) the share of each commodity's total output that is produced by each industry.

This "commodity-output-proportions" matrix *D*, which shows the shares of each commodity's total output that are produced by each industry, is obtained by dividing the make matrix V of dimensions (*t* x *s*) by the vector of commodity output *q* of dimensions (*s* x *s*):

$$D = V\hat{q}^{-1} \qquad (3)$$

$$\begin{array}{c} \\ \text{Industries} \\ \begin{bmatrix} 1 \\ .. \\ l \\ l+1 \\ .. \\ q \\ q+1 \\ .. \\ t \end{bmatrix} \end{array} \quad \begin{array}{c} \text{Commodities} \\ [\, 1 \ldots\ldots\ldots k \quad k+1 \ldots\ldots\ldots\ldots n \quad n+1 \ldots\ldots\ldots\ldots s \,] \\ \begin{bmatrix} d_{11} & d_{k1} & d_{k+1,1} & d_{n1} & 0 & 0 \\ & & & & & \\ d_{1l} & d_{kl} & d_{k+1,l} & d_{nl} & 0 & 0 \\ 0 & 0 & d_{k+1,l+1} & d_{n,l+1} & 0 & 0 \\ & & & & & \\ 0 & 0 & d_{k+1,q} & d_{nq} & 0 & 0 \\ 0 & 0 & d_{k+1,q+1} & d_{n,q+1} & d_{n+1,q+1} & d_{s,q+1} \\ & & & & & \\ 0 & 0 & d_{k+1,t} & d_{nt} & d_{n+1,t} & d_{st} \end{bmatrix} \end{array} \qquad (4)$$

On the basis of the industry technology assumption a product *j* can be produced by a certain number of different industries *k*. Each industry *k* needs $b_{ik}$ units of input *i* per unit of industry product *j*, where $b_{ik}$, $i=1 \ldots n$ represents the industry technology of an industry *k*, and each industry *k* produces only a part of the total output of product *j*. This proportion of industry *k* in the total production of product *j* has a notation $d_{kj}$. So all inputs *i* needed to produce one unit of product *j* by different industries can be written as follows:

$$a_{I,ij} = \sum_{k=1}^{n} b_{ik} d_{kj} \qquad \text{(where I refers to the industry technology)} \qquad (5)$$

The above formula shows that input *i* required to produce one unit of product *j* is a weighted average of the input structures of the industries where product *j* is produced: the weights are the proportions $d_{kj}$ of each industry *k* in the total production of product *j*.

To make things clearer, let us use words to develop the above equation for coefficient $a_{ij}$, i.e. the share of the total production of commodity *i* which is used in the total production of commodity *j:*

$$a_{ij} = b_{i1} d_{j1} + b_{i2} d_{j2} + \ldots + b_{ij} d_{jj} + \ldots + b_{it} d_{jt} \qquad (6)$$

It means that $a_{ij}$ is the sum of the share of total production of good *i* used in industry 1, multiplied by the share of industry 1 in the total production of good *j*, plus the share of total production of good *i* used in industry 2, multiplied by the share of industry 2 in the total production of good *j*, … plus the share of total production of good *i* used in industry *j*, multiplied by the share of industry *j* in the total production of good *j*, … plus the share of total production of good *i* used in industry *t*, multiplied by the share of industry *t* in the total production of good *j*.

Thus, through the additional information provided by the shares of various industries in the total production of a particular product (that could be defined as the "market shares" of these industries for this product), we are able to obtain the additional equations which allow for disentangling the initially hidden input structure of particular by-products. These additional equations in the form of equation (6) can be put in matrix format and thus be written as:

$$A_{I(s,s)} = B_{(s,t)} D_{(t,s)} \quad (7)$$

As *B* is a commodity-by-industry rectangular matrix, of dimensions (*s* x *t*). *D* as defined above is an industry-by-commodity rectangular matrix, of dimensions (*t* x *s*). Therefore matrix $A_I$ is a commodity by commodity matrix of dimensions (*s* x *s*). Thus $A_I$ is the input-output coefficient matrix that describes commodities directly required to produce other commodities. It is a square matrix. For UN and US statisticians, in the sources referred to above, as well as for Peter Flaschel in a 1983 article on "Actual Labor Values in a General Model of Production" (Flaschel, 1983, pp. 435-454), which uses their method, this square matrix *A* is invertible. This is a necessity for statisticians in order to calculate the commodity-by-commodity, total requirements matrix (*ITC*), which is derived as:

$$T = [I - A]^{-1} = [I - BD]^{-1} \quad (8)$$

But it is also a necessity for theoreticians, who want to calculate values by using the usual and well known formula:

$V = VA + L$

Which implies that $V = L\,(I - A)^{-1}$

One last difficulty deriving from the existence of joint production and which has to be addressed at this stage is indeed the nature of vector *L*, which appears in the above equation, because if we go back to our initial description of a joint production system in section 6.2. above, we note that the only information that we have regarding the quantities of labor time used in the system are the quantities by industry, which is logical because the nature of by-products does not allow to apportion these quantities to the various joint products of particular industry.

Therefore the row vector of these quantities of labor has for dimensions (*1* x *t*) and is:

$V^I = \left( l_1^I + l_2^I + ... + l_t^I \right)$ where index *I* stands for industry

On this basis, if we wanted to calculate values, using matrix *B*, we would be unable to do it directly, because the unknown values to be determined would be both the values produced by each industry (vector $V^I$), and the values of each commodity, i.e. the elements of vector $V^C$:

$$V^I_{(1,t)} = V^C_{(1,s)} B_{(s,t)} + L^I_{(1,t)} \tag{9}$$

Fortunately, there is a solution to this problem, which comes down to using the same tool as the one devised by statisticians, i.e. matrix *D* which gives us the commodity-output proportions. If indeed we multiply both terms of the above equation by matrix *D*, we obtain:

$$V^I_{(1,t)} D_{(t,s)} = V^C_{(1,s)} B_{(s,t)} D_{(t,s)} + L^I_{(1,t)} D_{(t,s)} \tag{10}$$

As was already demonstrated matrix *D* indeed provides a key for passing from industries to commodities, not only through the transformation of matrix *B*, but also for the transformation of vector $V^I$ and $L^I$ , since it allows us to write:

$V^C = V^I D$ and $L^C = L^I D$, from which we derive :

$$V^C_{(1,s)} = V^C_{(1,s)} A_{(s,s)} + L^C_{(1,s)} \tag{11}$$

Now that we have been able to overcome the difficulties linked to the existence of by-products, with the help of a method developed by statisticians and perfectly suited to the features of exclusive by-products, through the design of a matrix supposed to “well behave”, the calculation of actual labor values seems to be quite simple. From a theoretical point of view, however, things cannot be as simple as they seem, and this for several reasons.

The first reason for which things are not as simple as they might seem has to do with the basic data on which the method is based. Indeed, in its article just cited, Flaschel sticks fully to the SNA method, and since the quantities used by statisticians in their input-output tables are the monetary values of the commodities, he introduces prices in the calculation of the share of each industry in the production of joint products. This explains that the variables which appear in Flaschel’s article are therefore what he calls the “relative sales values”, from which are derived the monetary market shares of these commodities. Prices are therefore introduced in the calculation of values, a position that Flaschel fully endorses when he writes: “for our intents, labor values *v* were made to depend on prices *p* by the use of sales values $C_{kj}$” (Flaschel, 1983, p. 453).

However such a position is wrong for various reasons. First, because to know the market shares implies to know which quantities have been sold on the market, and at which prices. But when we stand at the theoretical level of production and values, there are no prices, because as we already mentioned previously, values are logically and chronologically anterior to the sale of commodities at monetary prices, which appear only after production has taken place, and when this production is being sold on the market.

Another reason has also been emphasized by Toker, and has to do with the distinction between average values and individual values of commodities, which necessarily appear when the same commodity can be produced by several methods. This author indeed writes: “unlike single-production systems, in joint production systems prices at zero rate of profit are still prices *par excellence* but not labor values; in the latter perfect competition can ensure, at any rate of profit (including zero) the uniqueness of prices, but not the uniqueness of labor values” (Toker, 1984, p. 152)..

A last reason why Flaschel should not have based his calculation of labor values on monetary market shares is that values and prices do not have the same dimension: values have the dimension of time, which is the standard of values, and prices are pure dimensionless numbers, or scalars. Therefore they cannot be put at the same level, which means that values are not a special form of prices which would appear when the rate of profit is zero.

Moreover the use of monetary market shares is not needed as a mathematical necessity for the determination of values in the case of joint production, since matrix *D* above can perfectly be defined in terms of ratios between the quantities of a commodity produced in various industries and the total quantity of this same commodity produced by the whole production system: it is a matrix of “commodity-output proportions”. This is indeed what has been done above (p. 158) by writing: $D = V\hat{q}^{-1}$, with the elements of *V* being mere quantities.

There is also another reason for which things are not as simple as they might seem, which has to do with the existence of several distinct categories of commodities. Indeed, as soon as we introduce, as it is the case in this section, a distinction between final commodities and intermediate commodities, which is indispensable in a theory where production is seen as a transformation of the later into the former, things become more complicated than they are in the statistical world where this distinction is not taken into account. Then final goods are not transformed into intermediate goods, and there are two types of these last ones, which - as we already showed in previous chapters, has some consequences on the nature of matrices involved in the representation of production.

As a result matrix *A*, as the product of both matrices *B* and *D* [which themselves are as exposed above in (2) and (4)], presents itself in the following form, with matrices B and D decomposed into sub-matrices:

$$A = \begin{bmatrix} B_{11} & B_{12} & B_{13} \\ 0 & B_{22} & B_{23} \\ 0 & 0 & 0 \end{bmatrix} * \begin{bmatrix} D_{11} & D_{12} & 0 \\ 0 & D_{22} & 0 \\ 0 & D_{32} & D_{33} \end{bmatrix} = \begin{bmatrix} B_{11}D_{11} & B_{11}D_{12}+B_{12}D_{22}+B_{13}D_{32} & B_{13}D_{33} \\ 0 & B_{22}D_{22}+B_{23}D_{32} & B_{23}D_{33} \\ 0 & 0 & 0 \end{bmatrix} \quad (12)$$

It is obvious that the resulting matrix *A* is singular and therefore is not invertible, because matrix *B* is itself a singular matrix: its *s* - *n* last rows are zeros, which simply reflects the fact that final goods by definition do not enter into the production of any good. By getting rid of these last rows, we can nevertheless rewrite matrix A as a new matrix $A^*$, which can be written as:

$$A^* = \begin{bmatrix} A_{11} & A_{12} & A_{13} \\ 0 & A_{22} & A_{23} \end{bmatrix} \qquad (13)$$

It is clear that $A^*$ is a rectangular matrix of dimensions ($n$ x $s$), and therefore is not invertible.

As for the dimensions of the six submatrices contained in matrix A*, they are the following:
$A_{11}$ is a square matrix with $k$ rows and $k$ columns, of dimensions ($k$ x $k$)
$A_{12}$ is a rectangular matrix with $k$ rows and $n$ - $k$ columns, of dimensions ($k$ x $n$-$k$)
$A_{13}$ is a rectangular matrix with $k$ rows and $s$ - $n$ columns, of dimensions ($k$ x $s$-$n$)
$A_{21}$ is a zero rectangular matrix, with $n$ – $k$ rows and $k$ columns, of dimensions ($n$-$k$ x $k$)
$A_{22}$ is a square matrix with $n$ - $k$ rows and $n$ - $k$ columns, of dimensions ($n$-$k$ x $n$-$k$)
$A_{33}$ a rectangular matrix with $n$ - $k$ rows and $s$ - $n$ columns, of dimensions ($n$-$k$ x $s$-$n$)

An observation which can be made from this matrix $A^*$ is that we can isolate from it another submatrix $A^{\#}$, which is composed of the first two rows and first two columns of submatrices of matrix $A^*$, such as:

$$A^{\#} = \begin{bmatrix} A_{11} & A_{12} \\ 0 & A_{22} \end{bmatrix} \qquad (14)$$

In fact submatrix $A_{11}$ gives us the technical coefficients that correspond to the share of intermediate goods of the first type which enter in their own production, submatrix $A_{12}$ those which correspond to the share of these same intermediate goods of the first type which enter in the production of intermediate goods of the second type, and matrix $A_{22}$ those which correspond to the share of these last goods which enter only in their own production.

To be sure $A^{\#}$ is a square matrix of dimensions $n$ x $n$, but it columns are not linearly independent, which implies that $A^{\#}$ is a singular matrix and therefore is not invertible. Like we already did for the calculation of values in the absence of joint production (see sub-section 2.2. of chapter 6), we must therefore solve the problem of value determination by using a step by step approach. Indeed, once the system is split into three subsystems, it becomes possible to represent it in the following matrix format, where we have three matrix equations:

$$V_1 = V_1 A_{11} + L_1 \qquad (15)$$

$$V_2 = V_1 A_{12} + V_2 A_{22} + L_2 \qquad (16)$$

$$V_3 = V_1 A_{13} + V_2 A_{23} + L_3 \qquad (17)$$

Only the first one is an independent matrix equation, in which $A_{11}$ is a $k$ by $k$ matrix with entries $a_{ij}$. Such a matrix corresponds to what is usually called an open Leontief model, i.e. to an economy where there is some external source of demand for each industry: here this demand corresponds to the share of the production of intermediate commodities of the first type which is needed for the production of other intermediate and final commodities.

Matrix $A_{11}$ is positive ($A_{11} > 0$), because $a_{ij} \geq 0$ for all $i$ and $j$ and $a_{ij} > 0$ for some $i$ and $j$. We know also that $A_{11}$ has row sums not exceeding 1, meaning that:

$$\sum_j a_{ij} \leq 1 \text{ for all } i$$

Such a matrix is called sub-stochastic (it would be stochastic if each row sum were 1). In such a case the demonstration has been made by Peterson and Olinick that $(I - A_{11})^{-1} > 0$. To be sure, they make the demonstration for matrices where column sums (and not row sums) are not exceeding 1, because they use the transposed matrix of matrix $A_{11}$, but this does not change either the demonstration or its outcome.

Let us first recall that by definition, matrix $A$ is productive if there is a nonnegative vector $X$ such that $X \geq AX$. Their demonstration is then based on their theorem 5.1. (Peterson and Olinick, 1982, pp. 221-239):

Theorem 5.1.: A sub-stochastic matrix $A$ is productive if and only if $I - A$ is non-singular.

From theorem 5.1., matrix $I - A$ is invertible and necessarily then $(I - A_{11})^{-1} > 0$. The solution vector is $X = (I - A_{11})^{-1} B$.

This ensures that values are always positive, which allows us to write:

$$V_1 = (I - A_{11})^{-1} L_1 \qquad (18)$$

Then, knowing $V_1$, the vector of values of intermediate commodities of the first type, from equation (4) we get $V_2$: the vector of values of intermediate commodities of the second type, and since $A_{22}$ is a square matrix of dimensions $n - k$ by $n - k$, we can solve the second equation:

$V_2 = V_1 A_{12} + V_2 A_{22} + L_2$, by writing: (19)

$$V_2 = (I - A_{22})^{-1} (V_1 A_{12} + L_2) \qquad (20)$$

Finally, knowing $V_1$ and $V_2$, from equations (18) and (20) we obtain $V_3$, the vector of values of final commodities, without any other additional calculation than displayed by equation (17):

$$V_3 = V_1 A_{13} + V_2 A_{23} + L_3 \qquad (21)$$

We thus demonstrated that values can be calculated in a coherent way in case of joint production. Thus we have arrived now at the end of this chapter on joint production, which has shown that contrarily to Sraffa's and Morishima's assertions it is perfectly possible to introduce joint production in a theoretical system without having negative labor values. In fact such peculiarities appear in Sraffa's theory because of the need for him to build a Standard commodity which appears to be quite strange in his system, and in Morishima's theory because of the introduction of strange assumptions like the existence of non-productive sectors that use some inputs in larger quantities than the produced quantities of these same inputs.

This conclusion according to which there is no such thing as negative quantities of labor has the merit to be in line with intuition, because in the real world, where indeed joint production does exist, albeit in small proportions compared to overall production, this particularity is never associated to negative values, being understood that, as Flaschel mentions it rightly: "the labor value of a jointly produced free good should be zero" (Flaschel, 1983, p. 449)..

Although we disagree with Flaschel on a particular point, i.e. the need to use prices in the calculation of values, we cannot but adhere to his conclusion which confirms the relevance of the labor theory of value and the existence of "actual labor values in a general model of production". What this chapter has shown indeed is that as long as it stays confined to the theoretical sphere of production and is not unduly extended to that of exchange, Marx's labor theory of value is perfectly compatible with joint production. On the basis of this finding we can state a new general principle:

**Principle 15**: Joint production, defined as the production of several distinct commodities by a single method of production, is compatible with the existence and calculation of positive labor values, as long as different methods produce these commodities in different proportions.

All this allows us to come back in a concluding chapter of the second part of this book to the concepts of production and value.

## Chapter 9. Conclusion on the concepts of production and value

The purpose of this short chapter is to recapitulate the principal lessons which have been drawn from the second part of this book on production and values. At the end of this part, it seems indeed useful to come back to these concepts, and to give more precisions about their content. We will come back to the principle that production is anterior to and as such independent from exchange, and that it never creates any surplus. We shall discuss the existence of values, as an operational concept, and their field of validity, and show that values are independent from the social structure of the productive system.

### 1. Production is independent from exchange

What this whole second part has shown is that to think the concept of production with as a point of departure the concept of transformation is perfectly possible and feasible, without encountering any contradiction. Moreover, this concept of production does not preclude at all to define values as the common property of commodities obtained through this transformation of other commodities. It is so because such a transformation always involves the intervention of human labor, which is the common point that makes them homogeneous, at a purely theoretical level, and therefore equivalent. And this indeed can be achieved, at least from a purely theoretical point of view, without any reference to what happens to these commodities after they have been produced, i.e. their exchange and distribution.

This approach is made possible because production cannot be defined otherwise than as a transformation, which ends up with the apparition of new material entities, thus called products or commodities. In other words, production is not an *ex nihilo* creation but a creation through the transformation of other and therefore pre-existing material entities.

As such, production is a much more general reality and concept than exchange. Production as transformation is first a natural and widespread phenomenon, which means that every existing thing in nature is the result of some prior transformation of other elements: this is the sense of Lavoisier's law, which implies that in the physical world there is no such thing as spontaneous generation. Production is an activity which takes place in the physical world and abides by Lavoisier's law, which means therefore that prior elements are themselves the result of a past transformation, and so on and so forth back to the primeval elements, i.e. the elements that appear in Mendeleev's table. We know now that these primeval elements were produced in the Big Bang and later in the heart of stars. Physical production has existed in nature for billions of years well before the existence of humanity.

To be sure economic production most often does not go so far as to use directly these primeval elements as the building blocks of its activity, but it uses raw materials, or primary commodities. Some of them are considered as "non-produced", but in fact this only means that they have not been produced by man in a process that could be called "human production", because they have been produced previously by nature itself. This is obvious in the case of all mineral primary commodities, like iron ore or oil. When we say that iron ore or oil are produced, it means more precisely that they are extracted from their deposit and separated from other mineral constituents. In the case of iron ore mining or oil extraction, both activities must be seen as a physical transformation. In the case of oil, for instance, it took millions of years for nature to produce it through physical and chemical processes, meaning that human production by itself does not create crude oil, which already exists deep in underground deposits. Production of crude oil consists in multiple activities which start with exploration, continue with drilling and end up in pumping crude oil up (if needed) to surface.

In the case of agricultural products, things seem a little less obvious, because at first sight some of these products do not seem to preexist to human intervention. It should be clear however that the vast majority of agricultural product existed in nature before and without any human activity whatsoever. Grains of wheat or rice were produced by the corresponding grasses and could be gathered by men before these cereals started being cultivated by them, around 10 000 years ago. Game was hunted before animals were domesticated, and fruits were picked before the cultivation of fruit trees.

That this natural production seems to add something to the seeds which are sowed is also an illusion, because like minerals the products of the land are also the result of the transformation of pre-existing natural elements, through biological processes like photo-synthesis. Using solar energy transported by photons, such a process dissociates carbon dioxide from the air to obtain carbon atoms used by cellular machinery to crate complex organic elements. The soil provides water brought by rain as well as nutrients like nitrogen, potassium and phosphorus. From a physical point of view, land is therefore nothing else than a substratum which stores all of these elements, including many organic compounds, as well as bacteria and fungi that often play a role in the production process of many plants. What human production does in the case of agricultural production is to organize, rationalize and accelerate these natural processes, thus increasing the quantities of commodities which are its outputs.

All these examples show well that production can be analyzed and understood as a concept without the existence of man or society, and therefore of any kind of exchange whatsoever between human beings.

## 2. Production never creates any surplus

These explanations make it easier to understand why human production, no more than natural production, never creates any surplus. This fact is certainly more straightforward in the case of mineral production, where the main elements which are going to be produced (for instance iron) already exist and are already present underground – albeit in another form (for instance hematite ore), before any human production takes place. In the case of iron, this human production consists in the extraction of this ore and in the separation of iron, first from the waste rock and second from its oxide, which is indeed a transformation (a change of form) and corresponds to our definition of production. But there is no material or physical difference, and hence no surplus, between the preexisting iron content of iron ore stored underground and the same iron once the iron ore treatment has been carried out.

Similarly to mineral production, it would be an abuse of language to consider that primary commodities coming from agricultural production constitute a surplus. As we already noted previously this is the result of wrongly considering that the seeds which are sowed are exactly identical to the grains which are harvested, which in all rigor is not true. This observation is still easier to understand in the case of fruit production, where the fruits which are picked on the trees are physically distinct from the seedlings which are planted to grow into fruit trees.

In both cases, what creates the illusion of a physical surplus is the existence of a number of elements (like hematite, or water from rain, carbon dioxide from the air, etc.) which are all going to be transformed into mineral or agricultural products (like iron or wheat), but pre-exist to any human activity, and in this case are therefore not the result of any human production. This is the reason why these elements may be considered as non-produced, which to some extent might look like another abuse of language, unless we precise that it means in fact that these elements are non-produced by human activity, i.e. by labor.

This finding allows us to understand that basic commodities in Sraffa's sense[9] are made of raw materials (from mining or agricultural origin) and of all the commodities, mostly from industrial origin, that are used to produce these raw materials. Since it is obvious that non-basic commodities are themselves made of some raw materials that will therefore appear in the equations describing their production, it is clear that the corresponding share of these basic products will not appear simultaneously in the equations describing their own production. This phenomenon will also concern the commodities, other than raw materials, that are used in the production of basic goods. All this implies that, even though the existence of a surplus is impossible at the scale of the whole production system, a surplus necessarily appears in the sub-system describing the production of basic commodities. In turn this explains why the matrix representing this sub-system is productive and sub-stochastic.

As far as this notion of surplus is concerned, a last point must be emphasized, regarding the equations describing the system of production of basic commodities, such as they are written in Sraffa's "Production of commodities…" (Sraffa, 1960, § 2, p. 4). These equations are indeed misleading when they make the same raw material appear simultaneously on both the left and right hand sides of the same equation, because they tend to blur the fact that there is no input of any raw material in its own production. Sraffa writes for instance:

240 qr. wheat + 12 t. iron + 18 pigs → 450 qr. wheat

Whereas mathematically this is equivalent to:

12 t. iron + 18 pigs → 210 qr. wheat

However, in the real world, in view of what economically (and mathematically) actually happens, we should rather write, for instance:

60 qr. *wheat seeds* + 12 t. iron + 18 pigs → 270 qr. wheat

[9] Sraffa's definition is the following: "the criterion is whether a commodity enters (no matter whether directly or indirectly) into the production of *all* commodities. Those that do that we shall call *basic*, and those that do not, *non-basic* products" (Sraffa, 1960, § 6, p. 8).

Obviously, for the whole production system, there must be another separate equation (not represented here), for the production of wheat seeds, a commodity which is produced from wheat but is not the same as the wheat that has just been harvested, because it has to undergo some transformation in order to be transformed into seeds. We can note that there is no surplus of wheat, because all the wheat that is produced is used in the production of other goods, which can be final goods or other intermediate goods, like in the case of wheat seeds especially. There is no surplus of wheat seeds either, because all these seeds are employed in the production of wheat.

Sraffa’s presentation recalled above thus prevents us from seeing the double origin of the apparent surplus in the sub-system of basic commodities, which is firstly the existence of non (humanely) produced goods which are not represented in the equations (here water, carbon dioxide, etc.) and secondly the use of part of these same basic commodities in another sub-system, i.e. the sub-system which corresponds to the production of non-basic commodities.

It remains nevertheless that globally, for the system of production as a whole, there cannot be any surplus. First because all of the final commodities, taken together, that appear on the right hand side of the sub-system of equations representing their production, are nothing more than the transformation of intermediate commodities, also produced, which appear on the left hand side of the same sub-system of equations. The apparent surplus of intermediate commodities which appear in the sub-systems of equations corresponding to their own production (either for the first type or for the second type) is fully absorbed in the production of final commodities. Therefore there cannot be any surplus of intermediate commodities.

As for these intermediate commodities, since they are not themselves produced ex nihilo, because production is a transformation, they are produced from the transformation of pre-existing elements which do not appear in the equations describing the system of production, because they are not the result of human production. To be sure nothing forbids to introduce these primary elements on the left-hand side of the equations, and therefore it would be theoretically possible to make them appear as such, if for instance we wanted to show the material balance of production. But then their value would have to be zero, because they have not been produced by man: no human labor has been spent in their production. This is the reason why they are generally not taken into account.

## 3. The existence of values and their domain of validity

What the whole second part has shown is that values are not only an intellectual concept, defined in the field of economic theory, but also an operational one, anchored in the real world, because it is perfectly possible in any circumstance, including the existence of joint products, to calculate values empirically. All that is needed to this end is to dispose of data external to economic theory regarding the techniques or methods of production, and among these techniques about the labor time, in physical units, spent when using these various techniques. Values are therefore not only a concept, but a reality.

Values were defined above in chapter 5 on the basis of Marx’s works, and through the concept of abstract labor, which we defined rather simply as labor reduced to its only dimension of time, which can be done in the context of the capitalist mode of production. Tis because wage labor is indeed generalized in the capitalist mode of production. However, in the real world this particular feature of labor which implies that it has to be measured as a function of time appears as soon as labor is paid in money, because this particular type of payment in most circumstances implies the specification of a monetary amount paid for a given quantity of time, which can be an hour, a week, a month or even a year.

But this feature is not limited per se to the capitalist mode of production. The monetary payment of labor against a wage is a phenomenon which took place in the real world in other modes of production, anterior to the capitalist mode of production: even in the Middle-Age wage labor already existed in cities, were the corporations of craftsmen used to enact rules for the payment of wages to companions: cathedrals builders did not work for free and were indeed paid not in kind but in the form of monetary wages. Workers received an agreed sum, paid in cash, at the end of the performance of their work. For instance on the building sites the organization of this work implied the measure of time spent to carry out the prescribed tasks: the first clocks appeared late on the sites, but the measurement of the time thanks to hour glasses has always been a concern of the employers.

What this means is that the concept of abstract labor can be generalized to a broader area than only the capitalist mode of production and therefore that the concept of value itself can be considered as still relevant for other types of modes of production, to the extent that wage labor exists in the corresponding societies. This presupposes the existence of money, because wage labor is paid in money, but money is also an institution which has existed for millennia and well before the development of the capitalist mode of production.

On this question we agree therefore with Frederick Engels, who in his supplement to Capital Volume 3, which he edited and published in 1894, wrote: “Starting with this determination of value by labor-time,...The most important and most incisive advance was the transition to metallic money… In a word: the Marxian law of value holds generally, as far as economic laws are valid at all, for the whole period of simple commodity production — that is, up to the time when the latter suffers a modification through the appearance of the capitalist form of production” (Engels, 1894).

This deserves to be emphasized, because using the concept of value can be useful for the theoretical analysis of the economic functioning of other modes of production. A good example of this type of analysis can be found in the works of Maurice Dobb, and in particular in his well-known “Studies in the development of capitalism”, an important book that he published in 1946.

This also shows the tight link which exists between money and value, and which makes that one cannot exist without the other, in the real world as well as conceptually. This will become even more obvious in the next part, devoted to distribution and prices. Before examining this question, a last point must however be touched upon, which is the independence of values from the structure of the productive system.

## 4. Values are independent from the structure of the productive system

Let us first underline that by structure of the productive system we do not mean the technical structure of production, since - as we have shown so far - values are completely dependent for their calculation on the techniques which are used in the production process, and which are given as a set of data external to the economic theory as such.

What we mean by "structure of the productive system" is the structure of property over the entities, i.e. the enterprises, which are using and implementing these techniques in the course of the production process. For instance a supply chain for a particular type of commodities can often be fragmented among many enterprises: indeed a technical process can be cut in successive technical stages, each of them producing an intermediate commodity to be further transformed downstream in the process. Then each of these technical stages can generally be taken care of by a different enterprise. At the other end of the spectrum, all of these technical stages can be fully taken care of within a single enterprise, which is fully integrated for the production of this particular commodity. This does not necessarily mean that such an enterprise is necessarily in a situation of monopoly, because there can be several such integrated enterprises competing on the same market.

But this does not change the way of calculating values, because this more or less important integration of the production process and the productive system has no bearing on the system of equations giving the values of commodities, which as we have seen only reflects the average production conditions for a particular commodity, whatever the number of enterprises that are involved in its production and their location on the supply chain.

If this observation seems self-evident, it will take on its full meaning in part III which now follows, devoted to distribution and prices.

# PART III - DISTRIBUTION AND PRICES

So far the analysis of commodities and of their values remained confined to the area of production, since it was possible indeed to demonstrate that value, as a property defining the equivalence of these commodities, which is nothing else than their common dimension, could be well defined within the theoretical boundaries of production. The moment has come however to go to the next stage, that of distribution, which can only physically take place once commodities have been produced. Any society or mode of production is indeed characterized by the particular form in which the product is distributed among its members. As we shall see, the first step toward this distribution occurs at the very stage of production, through the payment of wages. This is the case because in the real world wages are not evenly distributed to workers according only to the physical time during which they have been working. The influence of this phenomenon will be dealt with in a first chapter, chapter 10, devoted to the definition of social and monetary values. All the other chapters of this third part, from chapter 11 through chapter 16, will address the distribution of the product between workers and capitalists, and its influence on prices.

# Chapter 10. From physical values to social and monetary values

This chapter will allow us to go further in the definition of production as a social process, which involves various sorts of labor. This will lead to the discussion of the concepts of simple and complex labor, closely related to the question of the differentiation of wages. We will show that there cannot be any economic law of wage differentiation, which is a multi-factorial phenomenon. Finally this will allow us to explore the consequences of these findings on the theory of value and prices.

## 1. Production as a social process

So far we have focused and insisted on production as a technical and even a physical-chemical process. In this connection, labor has been considered essentially as an activity that can be quantified by using a physical variable, which is time, precisely because labor takes time and has therefore a time dimension. This is the reason why values cannot but have the dimension of time, and why we called them physical values. This is true whatever the society and historical time, as long as labor is accomplished in order to produce goods necessary for sustaining individual and social life, and to the extent that these goods are produced for exchange. Then labor itself can be considered as a commodity, because it is exchanged on a regular basis against a monetary wage, paid itself on a time basis.

But production is obviously much more than a physical and technical process: it is also a social process, socially organized in ways which change in accordance with the various modes of production. This explains that in the real world each country, society or culture has its own way to organize production socially and to value various kinds of labor differently. Moreover this organization and valuation usually change over time.

It is for instance possible to imagine, as a kind of thought experiment, two identical physical systems of production, with the same output of the same commodities, the same machines, the same workers, the same labor time spent for each production, and the same composition of the product, but with different monetary values for the commodities produced by these otherwise identical systems, because the differences in the social organization of production result in a different monetary valuation of different types of labor.

To go further in the understanding of production and value, it is therefore necessary to take into account these differences, and specify in more details the system under review. In the context of this work, we confine ourselves to the capitalist mode of production, where means of production are the property of individuals, and where labor is wage labor, with wages paid in money for given amounts of time. Moreover, although the amount of money wages and the period of time for which they are paid out are generally specified in contracts which exist before any work is carried out, and therefore before production starts, wages are generally paid once the various kinds of work corresponding to these contracts have been performed, i.e. at the end of a specified working period, during which commodities have been produced. It is also a reason why value and money are consubstantial.

These various features allowed the definition of abstract labor and the calculation of physical values, i.e. values expressed in physical time, whatever the differences in the nature of the different types of labor involved in the production of different commodities. This can be achieved without needing to know the precise amount of money paid for various kinds of labor. But as soon as a cycle of production is over and wages are paid at the end of this same cycle, then physical values are not enough to account for a thorough understanding of the system.

Indeed there is still one question which must be addressed, and this is the precise amount of money paid for the quantity of physical time spent in this production cycle. In the real world this amount is not purely proportional to the quantity of physical time, and therefore to the physical value of the corresponding product. This phenomenon results from the difference in wages paid for identical quantities of physical time, a discrepancy which in turn comes mainly from the existence of differences in the qualification of workers: it corresponds to the problematic that Marx called the reduction of complex labor to simple labor.

## 2. Simple labor and complex labor

To begin with, let us point out that for Marx simple labor does not mean unskilled labor, because any labor requires some kinds of basic skills, like for instance being able to read. Simple labor corresponds to this average, basic level of skills. Complex labor is therefore labor with additional skills, usually acquired by education and/or training, which allows to perform more complicated tasks than those performed by workers with this average basic skills.

Marx was well aware that this difference in the qualifications of labor had a direct repercussion on values. In Book I, Chapter 1, Section 2 (The Twofold Character of the Labor Embodied in Commodities) he wrote indeed: “A commodity may be the product of the most skilled labor, but its value, by equating it to the product of simple unskilled labor, represents a definite quantity of the latter labor alone[15]. The different proportions in which different sorts of labor are reduced to unskilled labor as their standard, are established by a social process that goes on behind the backs of the producers, and, consequently, appear to be fixed by custom. For simplicity’s sake we shall henceforth account every kind of labor to be unskilled, simple labor; by this we do no more than save ourselves the trouble of making the reduction” (Marx, 1867, p. 32).

However and at the same time, Marx was excluding from the outset a potential way of making such a reduction, by making clear that he did not want to base his definition of values - in terms of complex labor reduced to simple labor, on the differentiation of wages: in note 15 of chapter 1 of Capital, appearing within the quote in the preceding paragraph, he considered indeed that he was speaking of “the value of the commodity in which that labor time is materialized", and that "wages is a category that, as yet, has no existence at the present stage of our investigation" (Marx, 1867, p. 54).

The consequence of this position is that there is not in Marx's writings either a law or even a rule of transformation from complex labor to simple labor. This is well emphasized by J.-L. Cayatte in an article dating back to 1984, where he examines the question of simple labor and complex labor in Marx's works. He finally summarizes Marx's attitude as follows: "Between 1847 and the French version of the Capital, there is no doubt that Marx avoids the issue… However the result of it all is that a systematic comparison of the passages in which he tackles the difficulty leads to two different definitions of simple labor and five different definitions of complex labor. Should one be willing to establish a law of reduction of complex labor to simple labor, Marx's text should not be followed too closely and the inner logic of labor theory of value seems to be the one and only reliable guideline" (Cayatte, 1984, p. 220).

This is this inner logic which has been guiding us in our analysis. Indeed the whole approach followed so far in this book differs from this conception of Marx, since it has shown that wages and abstract labor, i.e. labor reduced to its dimension of physical time, and therefore wages and values, are consubstantial: labor value could not exist theoretically as a concept if wages did not exist simultaneously as the monetary payment of labor time. In opposition to Marx, this is the reason why we think that the differentiation of wages provides a proper mechanism, which is moreover a perfectly coherent one from a theoretical point of view, for the reduction of complex to simple labor.

If we try to understand why Marx rejected such an idea of starting conceptually from the actual differentiation of wages, it seems that it was for fear of being accused of empiricism, i.e. of developing a conception of knowledge based on empirical evidence, and because he thought that this kind of position would lead him to circular reasoning, since it would come down to starting from a reality that the theory is precisely aiming at explaining.

We think on the contrary that on this particular point Marx was wrong, not only because wages and values are conceptually linked, but also because the circular reasoning can occur only if we want to establish the laws of wages differentiation within the boundaries of economic theory, or more precisely of the theory of value itself. The circle is broken, though, if we acknowledge that wages differentiation is such a complex phenomenon that it can only be dealt with outside the realm of economic theory at such, by using a whole range of human sciences. Then, by becoming a set of variables that cannot be determined by economic theory alone, it becomes external to this theory, and can be considered as a given data.

## 3. The differentiation of wages as a first stage of distribution

This does not mean that the matter cannot be discussed. Understanding the difference between simple and complex labor remains in any case an important problem, because behind it the question raised is the difference in wages between workers with different levels of skills, which means that we are here at the heart of a problem which is a problem of distribution.

As we already indicated, distribution of the social product is usually considered as taking place once production itself has taken place and once there are consequently commodities to be distributed, in particular between workers and capitalists. But the differences which in the real world characterize the level of wages in accordance with the qualification of labor lead us to emphasize again what we already pointed out: that even before production takes place a first distribution has already been organized through the existence of labor contracts established in money amounts: it is a distribution between workers themselves, due to the existence of a hierarchy in wages associated to different skills or positions within the production process. This is the reason why this question was not touched upon in the second part, but is dealt with only now in this third part of the book devoted to distribution and prices, all the more so that wage is the monetary price of labor.

Since wages are paid in money, the payment of wages is the first theoretical step which occurs in the transformation of physical values into monetary prices paid for commodities on the market. At this stage however this first transformation only involves labor and corresponds to the transformation of physical values measured in physical labor time into monetary values corresponding to the payment of this physical labor time on the basis of predefined amounts of money paid for predefined quantities of time. The payment of wages on the basis of a labor contract takes the form of a payment of a certain number of monetary units per a predefined unit of time, multiplied by a given quantity of these same time-units.

Wage has therefore a double dimension: as a money amount it is a pure dimensionless number, i.e. a scalar, but since this amount cannot exist without the specification of a given amount of physical time, always associated to this amount of money, it is by its very nature a scalar per unit of time. Therefore, once all the payments for wages have been made, which implies that wages are multiplied by quantities of time, physical values disappear to be replaced by monetary values, i.e. the monetary value of the commodities which have been produced with the intervention of these different kinds of labor. Since all types of labor are not paid the same amount of wages per time unit, this necessarily introduces a distortion between physical values and monetary values.

The question which remains to be asked, and if possible answered, is whether this distortion can be theoretically understood and these monetary values theoretically determined. In other words, we must ask ourselves whether there is a purely economic explanation to the differentiation of wages. In Marxist language such a question aiming at understanding from a theoretical point of view the range or hierarchy of wages can be considered as derived from the question of the reduction of complex labor to simple labor.

## 4. An attempt by Marx to find a law of wage differentiation

After Marx expressly indicated, as we saw above in section 2 from a first quote from chapter 1 of “Capital”, that he wanted to “save (himself) from the trouble of making the reduction” from complex to simple labor (Marx, 1867, p. 32), one would think that Marx would no more touch upon this question. But he was not afraid of being confronted with contradictions, because he decided nevertheless to go back to the problem, once it had been reformulated into the question of the varying cost of labor-power, which is equivalent to the varying amount of wages, in chapter 6 of book 1, entitled "the Buying and Selling of Labor-Power".

Marx considers there that the key for understanding the reduction from complex to simple labor is education, and so he writes in Chapter 6 of Capital: "In order to modify the human organism, so that it may acquire skill and handiness in a given branch of industry, and become labor-power of a special kind, a special education or training is requisite, and this, on its part, costs an equivalent in commodities of a greater or less amount. This amount varies according to the more or less complicated character of the labor-power. The expenses of this education (excessively small in the case of ordinary labor-power), enter *pro tanto* into the total value spent in its production" (Marx, 1867, p. 121).

Marx thus considers that education and training costs constitute an element of the value of labor-power, of which wage is the monetary equivalent. In the following chapter he thus explains that labor-power which has benefited from such training has not only a greater value, but also creates proportionately greater values than unskilled labor-power. He writes: "All labor of a higher or more complicated character than average labor is expenditure of labor-power of a more costly kind, labor-power whose production has cost more time and labor, and which therefore has a higher value, than unskilled or simple labor-power. This power being higher-value, its consumption is labor of a higher class, labor that creates in equal times proportionally higher values than unskilled labor does"..."In every process of creating value, the reduction of skilled labor to average social labor, *e.g.*, one day of skilled to six days of unskilled labor, is unavoidable. We therefore save ourselves a superfluous operation, and simplify our analysis, by the assumption, that the labor of the workman employed by the capitalist is unskilled average labor" (Marx, 1867, p. 137-138).

The solution thus proposed by Marx is not devoid of any logic, since it is based in a way on a kind of law of conservation of labor time or value. The labor carried out in order to obtain a qualification is indeed placed in reserve, and manifests itself as a source of value once it is used in a production process. If we remove innate talents from the causes of qualifications, because they appear mainly in the creation of non-reproducible goods like artistic performances or art works, for which the theory of value is irrelevant, there is no objection to assume that in the real world differences in qualifications come essentially from differences in prior education.

Following Ronald Meek in his "Studies in the labor theory of value" published in 1973, it is therefore possible to imagine an elementary calculation allowing the conversion of a quantity of complex labor into a quantity of simple labor. If $T$ is the total productive period of a worker at the basic level of qualification, and if $t$ is the duration of additional education and training needed to reach an upper level of qualification, then one hour of complex labor will be equivalent to $1+\frac{t}{T-t}$ hours of simple labor, if the productive period is the same. It might even be possible to refine the calculation, by considering that training time itself could be considered as a more and more complex labor as the training continues.

The problem with this approach is not that it is wrong *per se*, it is that it has a limited explaining capacity. Indeed if we try to apply the above formula in a developed country like France, there the minimum education and training time is approximately 10 years for a worker at the bottom of the salary scale, starting at the age of 16 a professional life of 44 years. These can be considered as representative conditions for simple labor. At the other end of the salary scale, an executive typically starts working around the age of 26 after a 20 year education and training time, leading to an MBA or a PhD. Even if we assume a shorter professional life of for instance 40 years for the latter, then the increased labor time content for an executive should be: $1+\frac{10}{40-10}$ = 1,33. This would justify a wage higher by 33% for an executive, whereas in France in 2013 monthly net salaries in the private sector ranged from 1200 euros for the first decile of workers to 8061 euros for the 99$^{th}$ centile, i.e. a spread of 572%. These data come from a study about wages in the private sector and public enterprises published in 2015 by the French National Institute of Statistics and Economic Studies (INSEE Première N° 1565, September 2015).

Similar demonstrations could be done in a number of other developed countries, which would show that salary scales are much wider and inequalities much higher among wage-earners than can be explained by simple differences in education. On top of the above finding it must be added that each country has its specific salary range: there are some countries like the Scandinavian ones with a narrower and more egalitarian range, and some countries like the US with a much wider range and a much more unequal distribution. It is thus clear from this demonstration that the range of wages has far more complicated determinants than the influence of education and training time. We will try in the next section to list a few of them.

## 5. Wage differentiation as a multi-factorial phenomenon

In the real world the question of wage differentiation is addressed by a number of institutions which play a role in implementing employment or unemployment compensation policies. By accumulating over the years huge amounts of data about wages, they have generally been able to construct some good models of wage predictability, linking the level of wages for any individual to a number of economic, social and institutional parameters. From that we can infer some coherent explanations about the main factors responsible for the hierarchy of wages at a given point in time and in a given country.

One such model has been developed in France by APEC (the French Association for the Employment of Executives). Its data are based on APEC's Survey of Professional Status and Compensation in 2014. The purpose of this survey is to collect information on the salaries of current managers. It was conducted with private sector executives. As many as 14 000 people responded on their professional status, training, career, position as private sector executive, and on the company in which they held this position. The salary questions covered both the fixed portion and the variable portion of compensation (variable salary, variable bonus, incentives and participation), as well as benefits and in-kind benefits. The results obtained enabled APEC to calculate and produce a set of data for annual gross salaries expressed in euros.

The model was developed using the method of variance analysis to determine the criteria that influence the salaries of executives and to measure this influence. These criteria are regrouped in three categories corresponding to the individual characteristics of a worker, his/her company, and his/her position. The wage estimation model takes into account all the coefficients of variation associated with these criteria and returns an average wage for each "profile". The model identifies 11 parameters, belonging to these categories, to explain the level of wages which can statistically be obtained by an average worker, and therefore his/her positioning on the salary scale.

The first of these three categories has thus to do with individual characteristics (work experience, education, etc.), and is therefore the training and profile of a worker: the nature (University or specialized graduate school) and level of his/her main degree, the number of years of his/her professional experience. A second category of parameters concerns the company where he/she works in: its sector of activity (for instance does it belong to the financial sector or the industrial one?), its size (number of its employees, importance of its turn over). A third category has to do with the position held by a worker within a company: the nature of the function (sales, marketing, accounting, etc.…); the location of the company (in a big city or remote rural area); the nature of the employment contract (which can be more or less protective); the existence and importance of a budget to manage; the existence and importance of his/her hierarchical responsibility (particularly in terms of the number of workforce and managers that he/she supervises); whether the position is involved in international activities (with relations with subsidiaries, suppliers, customers abroad); and lastly whether the worker has or not a variable share in his/her salary.

What this empirical work shows unambiguously is that what Marx wrote in book I of “Capital” about education as the main determinant of wage differentiation, provides us with only a very limited view of the determinants of wages in the real world. Marx was thus wrong on this particular question. This survey gives a much more complex view of the determinants of wages. To be sure models such as the one that we just described can shed some light on the reasons for the differentiation that can be observed in salary scales in the real world. But there still remains factors that cannot be quantifiable, especially when we realize that salary scales that can be explained by this type of models differ from one country to another, for cultural, historical or political reasons.

Furthermore one important parameter is not taken into account by the APEC model, and it is gender. Indeed, in a country like France women’s wages were 24% lower than that of men in 2014, according to the National Institute of Statistics and Economic Studies (INSEE). While inequalities decline slightly (the spread was 27% in 1995), they remain very marked among managers and high-income earners. Part of the wage gap is explained by the more frequent use of part-time and less-valued jobs: 44.8% of female jobs are concentrated in a few low-paying sectors such as public administration, health, education or social action, according to INSEE. Yet when comparing wages for equivalent conditions (sector, full-time, age, etc.), there remains a gap of 9.9% which constitutes "pure discrimination". These inequalities are all the more curious because in schools, girls perform better than boys. Among the younger generations, they are more educated: 31.3% of women aged 25 to 34 have a diploma corresponding to a bachelor’s degree and above, compared with only 26.4% of men. For the first time in 2013, 49% of managers entering the labor market are women. However, wage differences are noticeable upon graduation: for instance Sciences Po graduate females are paid 28% less than their male counterparts.

Such a gender pay gap is quite common in all countries, but it is stunning to note that even in the European Union, where some cultural homogeneity is supposed to exist among member countries, this gap varies significantly across EU Member States. For the economy as a whole, in 2015 women's gross hourly earnings were on average 16.3 % below those of men in the European Union (EU-28) and 16.8% in the euro area (EA-19). Across Member States, the gender pay gap varied by 21 percentage points, ranging from 5.5 % in Italy and Luxembourg to 26.9 % in Estonia! These data come from the Gender pay gap statistics of EUROSTAT (March 2017). More generally these empirical data point out to the fact that the salary scale can be quite different from one country to another: to take but one last example it is for instance a well-known fact that engineers are better paid in France than in Germany, compared to ordinary workers.

The conclusion to be drawn from all these diverse sets of facts is that it is illusory to imagine that anybody may be able to construct a complete and coherent economic theory of wages differentiation. It is so because this phenomenon is the outcome of very complex social and political mechanisms. At a given point in time and in a given country, the differentiation of wages can only be explained by a multiplicity of factors, which have to do with ideology, institutions and the legal system – in particular the type of labor law, politics, culture, history. Obviously one such factor may also be the particular bargaining power of a given profession. The weight of trade unions can certainly play a role, but this may itself depend on the location of a particular group of workers within a supply chain, which may give a particular profession a capacity of nuisance by blocking the functioning of the economy as a whole.

It must thus be realized that two workers with exactly the same level of qualification, the same job and the same position in the hierarchy of two different companies may have two distinct levels of wage, due to a number of factors, like for instance gender or the location of each company. An important consequence of this recognition is that there is no unicity of the price of labor: exactly the same labor can have multiple prices. This is an empirical finding, which can also be explained theoretically, by statistical models such as the APEC model which was referred to, but should incite economists to modesty, and prevent them to think that an economic theory is able to explain everything which seem to belong to its field.

From all the above reflections we can now understand also why it remained impossible for Marx himself to establish a satisfactory definition (not to mention a measure) of the degree of complexity of labor. It is simply because like wage differentiation, to which it is closely related, labor complexity is such a complicated economic and social phenomenon that it cannot be explained by any universal law.

All this means finally that production at a theoretical level cannot be fully understood without realizing that it is underpinned by a first stage of distribution, i.e. the distribution of wages among workers, which has multiple determinants. Owing to the ideological, cultural, political, institutional and historical nature of these determinants, which go well beyond the framework of economic theory, this phenomenon has to stay among the primary data of the theory, and as a largely exogenous data.

# 6. The consequences on the theory of value and prices

## 6.1. From physical to social values

In the theoretical description of the economic system we have to start from the fact that although production is a continuous process, it can be divided into cycles of a finite duration. Whatever such a cycle may be, it necessarily takes place, to begin with, on the basis of a particular set of techniques, which include the quantities and various types of labor involved in the production process. For production to take place producers need also to have hired workers to perform these different types of labor and entered into contracts establishing the precise money amounts of wages that have to be paid for these different types of labor. This presupposes the knowledge of the wage scale, but from the last section we can now fully understand that it is impossible to provide a purely economic theory of the differentiation of wages corresponding to this scale. This implies that the wage scale must therefore be taken as a given set of social data, imported from outside economic theory, as well as all of the technical data.

Establishing the level of wages in money terms implies that money is also necessarily present in the system, since it has to be used for the hiring of workers and the payment of wages. This also means that we must now go from the physical value of newly produced commodities to their monetary price, at which they are sold against money on the market. But there is an intermediate step, which comes from the fact that before posting the price of their commodities producers have first to account for the monetary amount of the wages that they have to pay, and which represents their monetized labor cost. Since these money payments are made in exchange for specified quantities of physical labor time corresponding ultimately to the physical values of commodities, it seems appropriate to call the corresponding amounts of money the monetary values of these commodities, so as to distinguish them from their monetary prices. Empirically and logically, these monetary prices will come into play one step further than these monetary values, if only because in the real world it takes some time after production before products are brought to the market to be sold. These monetary prices are thus conceptually derived from this monetary values, on which they need to be based.

The inevitable conclusion to be drawn from the recognition of these monetary values as factual elements of the real world is that the use of money wages is conversely the only method available for the conversion of complex to simple labor. Money wages provide indeed the weights that allow for the translation of monetary values to what should be called social values, i.e. physical values weighted by the various levels of wage corresponding to each different kind of labor. In order to keep the overall amount of labor time identical when going from physical values to these social values, the notion of simple labor should be replaced as we shall see now by that of average labor. The average labor time can indeed be defined unambiguously as the labor time corresponding to the average wage per unit of time, i.e. the ratio between the total wage bill and the total quantity of labor time expressed in physical units.

Before showing it rigorously, and because it is instructive, let us start by briefly describing Morishima's attempt to address the joint problems of heterogeneity of labor and wage differentiation in Chapter 14 "The labor theory of value revisited" of his already mentioned book about Marx's Economics (Morishima, 1973, pp. 190-196), and by recalling the difficulties that he faces.

### 6.2. Morishima and his difficulties with the heterogeneity of labor

It is only at the end and in the very last pages of his book that Morishima indicates that: "we have so far assumed that all labor is homogeneous" (p. 190). He then recalls that Marx however recognized the heterogeneity of labor and decides to tackle the question himself. For that he does not try to criticize Marx (like we did earlier), but on the contrary wants to follow him closely by considering that skilled or complex labor is produced by education, which costs an equivalent in commodities. On this basis, Morishima intends to convert any kind of complex labor into simple, unskilled labor, and for doing so he uses a model where complex labor is itself produced by simple labor and commodities!

We will not expand on the model itself, but suffice it to say that Morishima discovers that the result of this model "is in conflict with his (Marx's) theory of exploitation, unless the conversion ratios are determined to be proportional to the wage rates of the different kinds of labor" (Morishima, 1973, p. 192). Indeed on the basis of our author's primary assumptions, and as he indicates: "we may have several groups of workers exploited at different rates, in contradiction to Marx's two-class view of the capitalist economy" (Morishima, 1973, p. 193). At first we could thus think that the conclusion that Morishima draws from this acknowledgment comforts our own position, since he continues as follows: "To avoid this difficulty, we have to abandon the scientific determination of the conversion rates by the formulas (21) and (22) and simply convert different kinds of labor into unskilled labor in proportion to their wages" (Ibidem).

Morishima then gives a formula which would allow this conversion of complex to simple labor by using money wages, but it is to immediately reject its relevance: "In this case, although the rate of exploitation is equalized throughout all kinds of labor, the values do not satisfy the postulate of independence from market conditions and may easily fluctuate from period to period as relative wages change".

It is thus on the basis of such an argument, combined with his prior criticism regarding joint production (which was discussed previously), that Morishima then concludes his whole book on Marx's Economics by rejecting the labor theory of value! But he is doubly wrong.

He is wrong first because labor is not produced. To be sure, labor can be considered as a commodity because like other commodities it exchanges on the labor market against money. But it is the only similarity it shares with other commodities. Indeed, and unlike other commodities, it does not result from a transformation: labor is not a substance which would itself be the end-product of the incorporation of other substances, thus transformed into labor! Neither is there reciprocally any such thing as labor incorporated into commodities, because the only scientific dimension of labor is physical time, and time is not a dimension that can be incorporated in anything, which should remind us that a metaphor does not in itself constitute a scientific statement. This means that what Morishima names in the above-mentioned quote "the scientific determination of the conversion rates" (by the "production cost" of different kinds of skilled labor), in spite of the fact that it is inspired by some writings from Marx, has nothing scientific. So it is no surprise that this attempt was doomed to failure.

Morishima is also wrong on his second point, according to which wages are determined by market conditions, implying therefore, if money wages are used for the conversion of labor times of different kinds into standard (simple) labor, that values will depend on market conditions, whereas values are supposed ultimately to determine market prices. But as we already showed in the previous section of this chapter, although it is true that there is a labor market, in the real world and unfortunately for economists, the labor price on this particular market cannot be economically determined, i.e. determined only within the field and with the categories of economic theory.

Once this fact is recognized, it simply means that economic theory should not try to determine some variables that are simply impossible to determine rigorously in its own field. Such a lucid recognition has in itself nothing that should invalidate economic theory, as long as it uses these thus external or exogenous variables as a basis for the determination of other variables within its own theoretical field.  The fact that, as Morishima writes it "values may easily fluctuate from period to period as relative wages change" (p. 193) is not a cause for abandoning a theory of value which produces such an outcome, but for analyzing the consequences of these fluctuations, otherwise not economically explained.

To take only one example concerning France, the minimum wage was increased suddenly by 35 % and all the other wages by 7% on June 1st 1968, and 3% more on October 1st of the same year. At the same time a reduction of the working time (number of hours worked in a standard labor week) was initiated. This was not the result of any particular economic mechanism, but of a general strike which triggered a tough negotiation ended by the "Grenelle Agreements". This occurred in the context of a major political event: the May 1968 movement. Obviously no economic theory will ever explain such wage fluctuations… On the contrary economic theory has to integrate these fluctuations and their consequences in its theoretical architecture.

## 6.3. A solution based on money wages and average social labor

There is a solution to the integration of labor heterogeneity and wage differentiation (as well as wage fluctuations) in the theory of value and prices. This solution requires to start from wages as they exist in the real world, which must be taken as a given set of data, and to use them as a basis for the definition of an average social labor. To show it, let us consider a simplified system of values, inspired - up to a certain extent, by the system which we just discussed (exposed in Morishima, 1973, pp. 191-193). Although we disagree with him, we borrow some of his notations, which will make it easier to show where we differ from him. We assume that there are only intermediate commodities of the first type - with $h$ such commodities, and $n$ kinds of labor $l_k \ (k=1,...,n)$. The total quantities of the different types of labor used in the system, measured in units of physical time, are represented by the column vector $L = \begin{bmatrix} l_1 \\ .. \\ l_n \end{bmatrix}$

The total labor time spent in the system is thus: $L_{total} = \sum_{k=1}^{n} l_k = \left(l_1 + \cdots + l_n\right)$

Let us take $w_i$ as the money wage-rate for one equivalent unit of physical time corresponding to these $n$ different kinds of labor (thus with also $i=1,...,n)$. The $n$ wage-rates are the elements of the row vector:

$$W = \left(w_1,...,w_n\right)$$

The global amount of the wage bill spent in the whole system is therefore $W_{total} = WL$, from which we get the average wage-rate of the system $\bar{w}$, which is a weighted mean, the weights being the various quantities in physical time of the different kinds of labor, $l_1$ to $l_n$:

$$\bar{w} = \frac{WL}{\sum l_k} = \frac{w_1 l_1 + ... + w_n l_n}{l_1 + ... + l_n} = \frac{WL}{L_{total}} \qquad (1)$$

From this equation it is clear that that we can define one unit (for instance one hour, or one month, or one year) of average social time, corresponding exactly to one identical unit (hour or month or year) of physical time, because the average wage-rate $\bar{w}$ is precisely calculated per unit of physical time (over the overall amount of this time). Knowing this average wage-rate, we can now express all the various wage-rates in terms of this average wage-rate, allowing us to define a new row vector, called $\bar{W}$, such as:

$$\bar{W} = \left(\frac{w_1}{\bar{w}}, \cdots, \frac{w_n}{\bar{w}}\right) \qquad (2)$$

Morishima does the same kind of conversion, but to convert $n$ types of labor into what he calls "labor *n +1* as the standard labor, i.e. the unskilled, simple labor or human labor in the abstract" (Morishima, 1973, p. 193). Although his conversion is mathematically correct, it makes it more difficult to define and carry out what we can call a transformation of physical values into social values, determined on the basis of the average social labor associated to this average wage-rate, since he confuses simple labor and abstract labor. In any case this is not the crux of the matter, which is – as we have explained, what Morishima calls wrongly the "market determination" of the various wage rates.

In fact $\bar{W}$ gives us the conversion ratios of all the various kinds of labor into an average social labor, because it is obvious from equation (1) above that we have:

$$L_{total} = \frac{WL}{\bar{w}} = \frac{w_1 l_1 + ... + w_n l_n}{\bar{w}} = \frac{w_1 l_1}{\bar{w}} + \cdots + \frac{w_n l_n}{\bar{w}} \tag{3}$$

$$L_{total} = \frac{w_1}{\bar{w}} l_1 + \cdots + \frac{w_n}{\bar{w}} l_n = \bar{l}_1 + \cdots + \bar{l}_n \tag{4}$$

To produce a unit of good *j* (with *j =1,..., h*), $a_{ij}$ units of good *i* and $l_{kj}$ units of direct labor *k* (with *k = 1, ..., n*) are required. Then the conditions of production can be represented by two matrices:

$$A = \begin{bmatrix} a_{11} & ... & a_{1h} \\ .. & & .. \\ a_{h1} & ... & a_{hh} \end{bmatrix} \text{and}\, R = \begin{bmatrix} l_{11} & ... & l_{1h} \\ .. & & .. \\ l_{n1} & ... & l_{nh} \end{bmatrix}$$

The quantities of direct labor of the different kinds necessary for the production of the *h* commodities can be converted in average social labor time, and then added so that we can define a row vector of these homogenized quantities of direct labor. Its *h* elements are:

$$\bar{l}_j = \frac{w_1}{\bar{w}} \cdot l_{1j} + \cdots + \frac{w_j}{\bar{w}} \cdot l_{ij} + \cdots + \frac{w_n}{\bar{w}} \cdot l_{nj} \text{ (with } i = 1, ..., n, \text{ and } j = 1,..., h) \tag{5}$$

It is easy to see that in the row vector below the $\bar{l}j$ elements (with *j =1,...,h*), are obviously not the same as the $\bar{l}_i$ (with *i =1,...,n*), in equation (4) above, since the index *j* no more represents a particular kind of labor (converted into average social labor), but the global quantity of average social labor (homogenized labor), corresponding to various kinds of labor used directly in the production of a particular commodity *j*:

$$\bar{l} = \left(l_1, ..., l_j, ..., l_h\right) \tag{6}$$

Then we can also use the conversion vector $\bar{W}$ to calculate the amounts of average social labor time used directly and indirectly in the production of commodities:

$\bar{L} = \bar{W}R + \bar{l}$ where $\bar{W}$ is the conversion vector as defined above by equation (2), and where $\bar{W}R$ gives us the quantities of average social labor time used in the production of intermediate commodities *i* entering in the production of commodity *j*.

If $\bar{V} = \left(\bar{v}_1, \cdots, \bar{v}_h\right)$ is the vector giving the social values of commodities, now expressed in average social labor time, they are then determined by:

$$\bar{V} = \bar{V}A + \bar{L} \tag{7}$$

We have thus demonstrated that the introduction of the heterogeneity of labor and of the associated wage differentiation does not prevent this system from determining values, which are still expressed in units of physical time but must now be called social values, because at the level of individual commodities they will necessarily differ from physical values as we had calculated them so far, i.e. on the only basis of physical labor time spent in their production, whatever its kind. Indeed these last ones did not take into account the heterogeneity of labor, i.e. the existence of complex labor of various kinds, in proportions which obviously differ from one commodity to another.

Moreover we can also very easily go from social values to money values, simply by multiplying social values expressed in average social labor time by the average wage rate!

In section 5 of chapter 6 we demonstrated the equivalence of total direct labor time and total product value. It follows that another important consequence of this transformation of physical values into social values and money values is that at the global level we can now identify the total physical value of the product to its total social value (which is nothing else than what we called $L_{total}$) and translate it into the total money value of the product, i.e. the overall wage bill, since by definition $W_{total} = L_{total}\bar{w}$!

This shows us that for the production process as such to be carried out, assuming obviously the existence of the capitalist mode of production, i.e. of workers and capitalists - these last ones being by definition the owners of existing fixed capital, all that is needed is the hiring of workers, whose labor alone (with the assistance of fixed capital), will be enough to produce intermediate commodities and final commodities: consumption goods as well as new fixed capital.

On the basis of the preceding demonstration we can now state two new principles, each of them with its own corollary:

**Principle 16**: In the real world labor is heterogeneous: there are many kinds of labor, which differ from one branch of production to another and according to the degree of qualification of labor. This is the reason why there is a differentiation of wages. But the precise range of this differentiation cannot be determined within the ambit of economic theory, since it depends on many factors external to this theory, and which are of a political, ideological, cultural and institutional nature.

**Corollary to principle 16**: the differentiation of wages, as it exists in the real world, must be considered as a given set of data for economic theory, which is one reason why this theory should rather be named political economy.

**Principle 17**: Taking as given the existing wage scale (which provides the level of wages for every kind of labor) allows for the conversion of different kinds of labor expressed in physical time into average social labor, or labor which is paid the average money wage, while keeping the same unit of physical time. This in turn makes it possible to calculate the average social values of all commodities, expressed in average social labor time.

**Corollary to principle 17:** Multiplying the social values of commodities by the average wage (in money units per unit of average social time) gives their money value at the stage of production.

### 6.4. The nullity of monetary profits at the stage of production

What we just highlighted, i.e. the correspondence between values as average social labor time and wages as the money payment of this labor time looks strongly like a kind of rehabilitation of Adam Smith's view of values as commanded labor. But everybody also knows that this theory has been rejected, and to some extent rightly so, because Smith did not clearly distinguish conceptually between the stage of production of commodities and the stage of their circulation (i.e. their exchange on the market), and therefore between their value and their price, as their exchange ratio against money on the market. It was thus easy to show that the monetary price of commodities included a profit on top of wages spent for the production of these commodities, a profit which moreover was not proportional to wages. It ensued that the price of a commodity could not be equal to the quantity of labor it commanded.

However, if we restore the distinction between a concept of value which is coherent at the sole stage of production and a concept of monetary price which is relevant at the stage of circulation, then the original idea of Smith regains its coherence, provided that we can show that the existence of profits as an income is itself not relevant at the stage of production, whatever the appearances may be.

In fact this is not so difficult to demonstrate, at macroeconomic level. First let us consider that capitalists start from scratch a new production process and that this production is carried out during a period corresponding to the normal interval for the payment of wages. For this process to be carried out they need money to hire workers, and have to borrow it from the banks. If the economy is fully integrated, at least for the production of each final commodity, with only one firm producing both such a final commodity and simultaneously all of the intermediate commodities needed to produce it, this firm will only have to borrow the amount of wages to be paid at the end of the period specified for such a payment. It is immediate that no profit will appear or be realized at the stage of production: it is only at the stage of circulation, when they have sold their final commodity on the market, that capitalists will be able to realize a monetary profit, by selling their product at a price higher than the wages paid for its production. From these proceeds they will then be able to repay their debt to the banks.

If now the economy is disaggregated, firms producing final commodities will have to buy intermediate commodities that they need for this purpose from other firms producing these commodities. And they will have to buy them at a monetary price including a monetary profit. Thus, on top of borrowing to pay for their own wages, these first firms will have to borrow in order to pay for the purchase of these intermediate goods, at a price including not only the amount corresponding to the wages paid for the production of these intermediate goods, but also the amount corresponding to a profit for their producers, and will thus have a corresponding and additional debt of an equal amount.

The share of the proceeds from the debt of these first firms utilized to pay for the sales of the second ones, by definition (since they are spent) cannot be used for the repayment of the corresponding debt and therefore cannot cancel globally. The phenomenon is not the same at the next level, because the second tier of firms that are paid for selling intermediate goods to final producers do not need to borrow for paying their workers' wages and their own suppliers: this payment is included in the price that they have charged to the first firms, equal to the amount that they have received. But the amount that on turn they have to pay to their own suppliers offsets the amount received by these suppliers, and so on and so forth at each stage of the production process: to each realization of sales we necessarily have a corresponding payment to the next level of suppliers, which therefore reduces the global profits by an identical amount. What remains nevertheless at each production level is a net monetary profit.

Ultimately, when all these exchanges have taken place, the amount of net monetary profits realized by all the suppliers of intermediate goods is necessarily equivalent to and cancelled by the corresponding amount of debt that has to be repaid by the initiators of the process: the producers of final commodities, because precisely they had to borrow for paying an amount which included all of these net profits. As long as they have not sold their commodities as final goods in the sphere of circulation, this debt is exactly offsetting the sum of net profits realized by the suppliers of intermediate goods, for an identical amount. And it is only outside of the sphere of production, with the sale of final commodities on the market, that the repayment of their debt will be made possible and will allow for the disappearance of these negative profits.

In fact one can summarize this whole process by saying that the monetary profits realized by the producers of intermediate commodities are realized at the expense and correspond to negative profits of the producers of final commodities, who are their only buyers. Although it may seem somewhat as a caricature it is a logical and mathematical truth.

It follows from this demonstration that no net profits are created and can exist at global or macroeconomic level at the stage of production: taken globally, capitalists as a whole cannot realize any profits at this stage, through the exchange of intermediate commodities between themselves. This means that as an income, profits cannot be a category of production, and that the money value of the product is made only of wages.

The situation is quite different as regards wages: unlike profits, wages have a purchasing power at the very moment when they are paid to workers, and this for two reasons: first because no worker would accept to work if the wage that he is going to be paid by contract had no purchasing power, and second, because workers do not have to borrow the money that they receive as wages for the payment of their labor, and thus have no corresponding debt to be repaid once they receive the money amount of their wage. Since the main characteristic of an income is to have a purchasing power, wages are therefore an income at the stage of production, and they are the only income, which is completely coherent with the result achieved above at the end of sub-section 6.3., and according to which at the stage of production the global wage bill is equivalent to the money value of the product.

### 6.5. A solution which is reminiscent of Keynes

The reader who is familiar with the "General Theory" and remembers its chapter 4 "the Choice of Units" will note that the point to which we have arrived is not so far from the position adopted by Keynes, who writes: "In dealing with the theory of employment I propose, therefore, to make use of only two fundamental units of quantity, namely, quantities of money-value and quantities of employment. The first of these is strictly homogeneous, and the second can be made so. For, in so far as different grades and kinds of labor and salaried assistance enjoy a more or less fixed relative remuneration, the quantity of employment can be sufficiently defined for our purpose by taking an hour's employment of ordinary labor as our unit and weighting an hour's employment of special labor in proportion to its remuneration; i.e. an hour of special labor remunerated at double ordinary rates will count as two units. We shall call the unit in which the quantity of employment is measured the labor-unit; and the money-wage of a labor-unit we shall call the wage-unit [5]. Thus, if $E$ is the wages (and salaries) bill, $W$ the wage-unit, and $N$ the quantity of employment, $E = N \times W$ (Keynes, 1936, p. 41).

One needs to realize that to become homogeneous and thus assume the status of true measurement units these quantities of employment must necessarily be expressed in time units (of physical time), and therefore be measured in weeks, months, or years of labor time. Then there is an obvious connection between this particular choice of unit and the labor theory of value, although Keynes was careful not to use this expression, which would have made him appear as a disciple of Marx, and be even more misunderstood! To be sure, Keynes does not come to this conclusion on the basis of the same approach as that developed so far in this work. Indeed, against what he calls "the postulates of the classical economics", he wants to follow a more pragmatic and empirical approach, and develop a theory which accounts for the real world. However it seems very clear from this quote that on the basis of somewhat different assumptions he arrives at a similar result.

What deserves also to be noted is that to address the question of the heterogeneity of various "grades and kinds of labor" Keynes does not enter into theoretical considerations, but simply weights the different kinds of labor "in proportion to their remuneration", again almost exactly like we did in subsection 6.3. above. A first difference lies in the fact that we provided all due justification for doing so. A second difference comes from the fact that Keynes uses what he calls ordinary labor, equivalent to Marx's "simple labor", as his unit of labor time - and this is an empirical notion, whereas we use average labor, i.e. labor paid the average wage, which is an abstract concept, as our standard of labor time.

As regards the second "unit of quantity", as Keynes calls it, this is for him the quantity of "money-value", and the use of this expression shows in passing that there is some confusion in Keynes's mind about the distinction between the terms "value" and "price". Indeed Keynes seems most of the time to consider them as equivalent. This might be due to the fact that the General Theory is not a theory of production as such, but a monetary theory of production (Keynes, 1933). For him there is no production without money, because wages are paid in money, and obviously this money is not a commodity and even less a share of the product:

Keynes's theory is fully incompatible with Sraffa's theory. As soon as wages are paid in money, a quantity of money can be multiplied by the quantity of employment, and everything can be measured in money. Prices are thus monetary prices, and the concept of value as distinct from the concept of price does not seem to be needed by Keynes: it does not come to his mind that its quantity of employment could define the value of the product, in terms of physical time, independently of its monetary value, measured by what he names the wage bill, which differs necessarily from the monetary price of this product, because of the existence of profit.

This absence of a clear distinction between monetary value and monetary price certainly explains the very complicated developments devoted in the General Theory to the supply price of the output, at macroeconomic (aggregate) level, and to the user cost, which requires a whole appendix to chapter 6 on "the Definition of Income, Saving and Investment". In fact Keynes introduces this question in chapter 3, where it is clear that for him the only factor of production is labor, with the aggregate supply function defined only as a function of employment: $Z = \varphi(N)$. Then what he names "supply price" corresponds to what we have called the monetary value of the product: we have shown that it is nothing else than the overall wage bill. As for the monetary price of the product, as distinct from its monetary value, it corresponds to what Keynes calls *D, i.e. "*the proceeds which entrepreneurs expect to receive from the employment of *N* men, the relationship between *D* and *N* being written $D = f(N)$, which can be called the *aggregate demand function"* (Keynes, 1936, p. 25).

It must also be understood that both terms used by Keynes, i.e. supply and demand, correspond to the supply and demand of *final* commodities only, to the exclusion of any intermediate commodities. It means that for the producers of final goods (be they consumption goods or capital goods), if they are not integrated enterprises, the "supply price" in terms of employment does not correspond only to their own employment, but also to the employment of the producers of the intermediate goods that these producers of final goods have used in the production of these final goods.

This is something that Keynes had understood, even though his explanations are somewhat confused, as it appears from footnote 3 to this same chapter 3: "The reader will observe that I am deducting the user cost both from the proceeds and from the aggregate supply price of a given volume of output, so that both these terms are to be interpreted net of user cost; whereas the aggregate sums paid by the purchasers are, of course, gross of user cost. The reasons why this is convenient will be given in Chapter 6. The essential point is that the aggregate proceeds and aggregate supply price net of user cost can be defined uniquely and unambiguously; whereas, since user cost is obviously dependent both on the degree of integration of industry and on the extent to which entrepreneurs buy from one another, there can be no definition of the aggregate sums paid by purchasers, inclusive of user cost, which is independent of these factors. There is a similar difficulty even in defining supply price in the ordinary sense for an individual producer; and in the case of the aggregate supply price of output as a whole serious difficulties of duplication are involved, which have not always been faced. If the term is to be interpreted gross of user cost, they can only be overcome by making special assumptions relating to the integration of entrepreneurs in groups according as they produce consumption-goods or capital-goods which are obscure and complicated in themselves and do not correspond to the facts. If, however, aggregate supply price is defined as above net of user cost, the difficulties do not arise. The reader is advised, however, to await the fuller discussion in Chapter 6 and its appendix" (Keynes, 1936, p. 33).

To better understand what Keynes intends to say, even though he does not express it very clearly, let us make a thought experiment, and imagine a capitalist economy that is fully integrated, with therefore only one firm producing everything, i.e. both consumption goods and capital goods, as well as all of the intermediate goods needed to produce these final goods. Then we cannot but realize that the factor cost is only made of wages, which constitute what Keynes calls the aggregate supply price and what we call the monetary value of the product, i.e. the wage bill. Moreover it is obvious that in a capitalist mode of production by definition workers do not buy capital goods. If we add the assumption that capitalists do not buy consumption goods, knowing that capital goods will not have to be bought on the market by the integrated firm (since it has produced them and already owns them), then the only goods sold on the market are consumption goods.

The situation is such that, if all these goods are sold, and the market is cleared, the overall price of these consumption goods must necessarily be equivalent to the wage bill W, which constitutes for Keynes "the proceeds which entrepreneurs expect to receive from the employment of *N* men" or "the aggregate demand" D, which is thus equal to this supply price. However the factor cost of these consumption goods alone does not correspond to the employment of *N* workers, but only to the employment of workers for the production of consumption goods. If we name it *Nc*, and the corresponding wage-bill for these workers *Wc*, we can see that these proceeds include a notional profit, which amounts to $Wp = W – Wc$.

It results from this demonstration that the disaggregation of the production system which is the norm in a capitalist economy hides the fact, that Keynes had recognized at least implicitly, that profits are a transfer income. We can see also why for Keynes fixed capital does not transfer any value to the product, which derives from the deduction of the user cost from the price of output, being understood that the price of output includes nevertheless all the new fixed capital produced in each period[10], because this fixed capital consists of final goods. All this will be made even clearer in the next chapter, devoted to the realization of the product.

We have now arrived at a point where we can express another principle:

**Principle 18**: The wage bill is the measurement of the monetary value of the product.

From what we have discovered so far we can now understand the fundamental importance of wages as an economic variable: it is indeed the determination of wages and of their differentiation, by a number of factors which are mainly out of the reach of economic theory, which allows this theory to determine simultaneously the monetary values and the social values of commodities and of the product.

[10] It also includes the consumption of intermediate goods, which contributes to obscuring the problem.

# Chapter 11. Distribution, surplus-value and profit

In the previous chapter we saw that the first step of distribution, i.e. distribution between workers themselves, takes place at the stage of production, and in fact even before production starts, when wages are negotiated and their level established. We also explained why this particular distribution depends on a number of factors which do not pertain to economic theory as such. We intend now to engage in an attempt to understand a new phase of distribution, i.e. distribution between workers and capitalists. We will show that it is a distribution of commodities, but also of surplus-value, a concept which will be explained through a particular mechanism, i.e. the transformation of values into monetary prices. This process itself is at the origin of the creation of monetary profits, which will be demonstrated through a simple model, based on simple reproduction.

## 1. Distribution as distribution of commodities and surplus value

Whatever the mode of production, and in any society, distribution is not only the distribution of monetary income, but ultimately the distribution of commodities which have been previously produced and (in a capitalist economic system) are thus available to be bought on the market, between the members of this society. What Marx highlighted is that it is the distribution of two quite different categories of commodities: consumption goods, on the one hand, and means of production or capital goods, on the other hand. Due to the specific nature of means of production, and to their appropriation by those who are by definition named capitalists, which allows them to produce goods to be sold on the market and to hire workers, distribution is also the distribution of men into social classes exercising distinct functions in the production process. This point has been emphasized by Althusser in his book written with Balibar "Reading Capital", published in English in 1970 (Althusser, 1970, p. 167).

Because of these different social functions, it is an obvious empirical fact that workers cannot collectively obtain on the market the share of the product made of means of production, or fixed capital (if they did so, they would by definition no longer be workers, but would have become capitalists), and therefore that they only buy consumption goods. In fact in the real world they only buy a part of these goods, because capitalists, as normal human beings, also need to buy consumption goods). This is the case because the wages that they earn never allow them to buy back the whole quantity of commodities that they have produced, but only a share of these consumption goods sold on the market by capitalists.

Such a result is achieved through a very simple mechanism: the monetary price of the consumption goods obtained by workers, which is equivalent to the total wage bill (assuming at this stage that it is wholly spent) is greater than the monetary value of these same consumption goods as defined in the preceding chapter, i.e. the total sum of direct and indirect wages needed to produce these consumption goods. In monetary terms this difference is called profit. If we stay at this theoretical level, i.e. at the level of circulation of money and commodities, we will remain blocked in the classical economic world of Adam Smith and Ricardo.

But the novelty of Marx is to go back to the world of production and values, which allows him to realize that the social value of these consumption goods obtained by workers, in terms of labor time, is lower than the total social value of all the goods that they have produced: this difference, between the overall social value of the product and the social value of the consumption goods obtained by the workers, is nothing less than what Marx names "surplus-value". In his book "For Marx", a collection of previously published articles, published in English in 1969, Althusser considered this theoretical finding as part of an "epistemological breakthrough", not only with classical economists, but "in the history of Marx's thought, a basic difference between the ideological 'problematic' of the Early Works and the scientific 'problematic' of *Capital"* (Althusser 1969, p. 13). As this same author writes also in his previously cited book "Reading Capital": "the sensitive point in this revolution concerns precisely *surplus-value.* Their failure to think it in a word which was the concept of its object kept the classical economists in the dark, imprisoning them in words which were merely the ideological or empirical concepts of economic practice" (Althusser, 1970, p. 148).

If money prices were equal to money values, then workers would be able to buy back the totality of the product, i.e. both consumption goods and fixed capital goods: the wage bill would be equivalent to the money price of the whole product. The condition for the theoretical existence of surplus value is therefore the distinction between values and prices, and more precisely values and money prices. It is because they were never able to make this clear distinction that Smith and Ricardo failed to come to the same conclusion as Marx's. To be sure this difference is easy to understand once we realize that values and prices are not defined at the same level of the economic system: values are defined in the sphere of production, whereas prices are relevant in the sphere of circulation, where distribution takes place. This explains why they do not pertain to the same measurement system, and have distinct dimensions: time for values and no dimension for prices, which are pure scalars. But this difference still exists between money values (i.e. labor-time converted into money through the payment of wages) and money prices: the money value of the product differs from its money price, precisely because of the existence of profits.

Morishima is therefore perfectly right when he says in chapter 5 "Surplus-value and exploitation" of his previously cited book that: "in the capitalist economy (unlike the society with simple commodity production) values and prices, in general, no longer coincide; they should be distinct. For this reason Marxian economics, unlike orthodox economics, has dual accounting systems: one system in terms of value and the other in terms of price. But many people from both camps (such as Sweezy, Joan Robinson and Samuelson) have confused the two, since Marx himself sometimes confused them." (Morishima, 1973, p.46). Morishima provides indeed a striking example of this confusion with this citation of Marx, who writes in Volume 3 of Capital: "if this surplus value is related to the total capital instead of the variable capital, it is called profit, $p$, and the ratio of the surplus value $s$ to the total capital $C$, or s/C is called the rate of profit, $p'$. Now substituting for $s$ its equivalent $s'v$, we find $p' = s'v/C =$ $s'v/c+v$, which equation may be expressed by the proportion $p': s' = v/C$; the rate of profit is related to the rate of surplus-value as the variable capital is to the total capital" (Marx, 1894, chapter 3, § 3).

Marx's mistake comes from the fact that surplus-value is clearly defined in terms of values and not of prices, and that therefore the rate of surplus values $s/v$ belongs to the domain of values, as well as the rate $s/C$, because $C$, like $s$, has to be expressed in value terms (it is the value of total capital) and not in terms of prices: if C were the price - and not the value, of constant capital it would itself incorporate a profit, and the ratio would be a ratio between quantities expressed in different dimensions: labor time for s and scalars (prices) for C.

To keep Marx's terminology, this explains why the ratio $s/C_v$ (the ratio of surplus value to the value of total capital in value terms, which we thus write $C_v$) is necessarily different from what is usually called the rate of profit in economic theory, i.e. the ratio of monetary profit over the monetary price of capital $p/C_p$. In the rest of this book, we will stick to the usual sense of the concept of rate of profit, called $r$, as a ratio of quantities expressed in money. And since the concept corresponding to the ratio $s/C_v$ expressed in value terms does not seem to exist, but is nevertheless closely related to the previous concept, we will call it "rate of profit in value" and call it $r_v$.

These remarks inevitably raise the problem of the transformation of values into prices, which has provoked an endless debate among economists.

## 2. From values to prices, or the transformation problem

We will not expand too much on this question, which Marx addresses in chapter IX of book 3 of Capital, entitled "Formation of a general rate of profit (average rate of profit) and transformation of the values of commodities into prices of production". In trying to resolve it, he made a number of mistakes.

Indeed he starts from the postulate that there is an identity between the total sum of profits and the total surplus value and assimilates the production cost (in terms of prices) to the sum of variable capital (or value of wages) plus the value of constant capital (the value of intermediate goods plus the value transferred by fixed capital). Then it is clear that he can only get the sum of values equal to the sum of prices.

This postulate implies that Marx's solution comes down to allocate the same overall surplus-value differently in the price system as it is in the value system: instead of being proportional to the value of labor-power (what he calls variable capital, or the value of wages), surplus-value becomes proportional in the price system to the value of the total capital (variable plus constant capital). We know that all this cannot but be erroneous, if only because we are in two systems of different dimensions, social (or physical) time for values and scalar for prices, and that these two systems are therefore wrongly mixed.

In any case Marx himself was aware that his method was wrong, or to say the least approximate, when he wrote in chapter IX of Book 3 of "Capital": "This statement (that the sum of the prices of production of all commodities produced in society - the totality of all branches of production - is equal to the sum of their values, author's note) seems to conflict with the fact that under capitalist production the elements of productive capital are, as a rule, bought on the market, and that for this reason their prices include profit which has already been realized, hence, include the price of production of the respective branch of industry together with the profit contained in it, so that the profit of one branch of industry goes into the cost-price of another" (Marx, 1894, p. 122).

The first economist to propose a solution to this problem was Bortkiewicz, a Russian economist, in an article published in 1907: "On the Correction of Marx's Fundamental Theoretical Construction in the Third Volume of *Capital*". This article was translated by Sweezy and published in English as an appendix to Böhm-Bawerk's book. In this article Bortkiewicz rightly indicated that "This solution of the problem cannot be accepted because it excludes the constant and variable capitals from the transformation process, whereas the principle of the equal profit rate, when it takes the place of the law of value in Marx's sense, must involve these elements" (Bortkiewicz, 1907, p. 201).

We will not develop in details Bortkiewiz's solution. Suffice it to say that he exposes a system with three branches or sections (the same as Marx in book 2 of Capital), producing respectively the means of production (constant capital), wage-goods (variable capital) and luxury goods. He chooses the price of luxury goods as the unit of measurement. The system is in a state of simple reproduction, which makes it in fact quite similar to Sraffa's system of production prices.

This system allows Bortkiewicz to calculate the rate of profit and the prices, and to show that it does not give Marx's result according to which the sum of values is equal to the sum of prices. Obviously he finds like Sraffa that the rate of profit is determined only in the first two sections, that produce constant capital and variable capital, which is contradictory to Marx's formula, according to which it is the totality of constant capital (including capital in the luxury goods section) which plays a role in the determination of the rate of profit. It is therefore no surprise that most of Bortkiewicz's article is devoted to the criticism of Marx's formula to calculate the profit rate: $r = \frac{s}{C} = \frac{s}{c+v}$.

The ultimate avatar and logical outcome of Bortkiewicz's approach is Sraffa's system of production prices, in which production prices and the rate of profit can be determined directly on the basis of technical data, in terms of physical quantities, including the quantities of wage-goods consumed by workers. And this can be done even before the calculation of values.

This conception, which has been designated as the inverse transformation theory, was expressed in the most complete manner by Samuelson, in an often cited article: "Understanding the Marxian Notion of Exploitation: A Summary of the So-Called Transformation Problem Between Marxian Values and Competitive Prices". In this article, published in 1971, he wrote: “just as Marx turned Hegel on his head, we turn Marx upside down, and can say in a dozen repetitive ways that this total of profit is not allocated by the value system according to where it was ‘really produced’ but rather falls to the share of each aliquot part of the total social [variable] capital out of the total social profit” (Samuelson, 1971, p. 417). And also, one page further: “a theory of exploitation wage can be based solely upon the analysis of profits and prices (with all the Leontief-Sraffa algebraic complications that are involved in these real-world relations)” (Samuelson, 1971, p. 418).

Since we have criticized at length the system of production prices in chapters 3 and 4 of this book, we will not repeat this criticism here. Let us just state again that in any case values, on the one hand, and production prices, on the other hand, cannot be compared, because they do not have the same dimension: values, determined at the stage of production, have the dimension of physical time, and production prices have the dimension of the standard commodity, i.e. an heterogeneous and permanently changing bunch of commodities.

Values and production prices belong indeed to two completely different universes. Moreover the system of production prices suffers anyway from a number of internal flaws that we have fully exposed. Our criticism is quite sufficient to sweep away all this problematic of inverse transformation, and all the more so that the theory of production prices is of no help to understand how profit is formed and realized, a question that we will now address.

## 3. The realization of profits, in a simplified model

To correctly address the question of the formation and realization of profits, it is better to adopt a step by step approach. Since the realization of profits implies the sale of commodities against money, which is the first step of the reproduction of the system, we will start with a simple model of reproduction, which we will first define in the next subsection.

### 3.1. A model based on simple reproduction

This model is based on the following assumptions, which correspond to the mode of exposure of Marx:

- The system of production is the pure capitalist mode of production, in the sense that there are only workers and capitalists.

- Capitalists are the owners of the means of production, and they employ salaried workers. As far as production is concerned, they are represented by firms, corresponding to what national accounts call corporations or quasi-corporations, non-financial as well as financial, and household firms, provided that the latter employ at least one salaried worker. But the capitalists are also individuals who are the ultimate owners of the firms in question, and as such they are part of households in the sense of national accounts.

- Workers by definition do not own any means of production, and sell their labor power to the capitalists, against payment of a money wage: they are not paid in consumer goods. This point is important, and in particular introduces an important difference between the capitalist mode of production and other modes like the slave mode of production or the feudal one. The question of the nature of money has already been discussed, and we just need to point out once more that money is by no means a commodity.

- The economy consists of two sections, section I, which produces fixed capital goods, and section II, which produces consumer goods. The question of whether there are different production methods is not considered relevant, because the choice of techniques has already been made and the system self-reproduces itself from period to period. We are in the theoretical world of commodities and abstract labor and as a result, physical goods, which Marx and others call use-values, do not appear as such in the model. We must emphasize the fact that these two sections have nothing in common with the sectors that appear in the theory of production prices: the sector of basic goods (of whatever type) and the sectors of luxury goods. Indeed both fixed capital and consumer goods are luxury goods in the sense of the theory of production prices, which means that, once they have been produced, they never re-enter the production process to be transformed through it.

- Surplus-value is the difference between the total value of the whole social product and the value of the consumer goods which workers can obtain by exchanging their money-wages for commodities, called conventionally the value of their labor-power. The rate of surplus-value is thus the ratio between the surplus-value at global level and the value of labor-power.

- The rate of surplus-value as defined above, and sometimes called the exploitation rate, is uniform in the two sections: surplus value is indeed considered as a global social phenomenon, which concerns all the workers taken collectively, and not as an individual phenomenon. This rate is, however, uniform only within the framework of a given social system, corresponding to a separate and autonomous political organization and therefore usually to a State.

- It is assumed for the moment that capitalists do not consume and that workers do not save, which simplifies the reasoning, but as will be seen is not without consequence on the demonstration. This assumption will then be abandoned when the consumption of capitalists is introduced into a second, more complex model.

- All profits are saved and invested in the section where they are formed, which implies that there is no transfer of savings between the two sections.

- On the other hand, the assumption according to which workers do not save will be preserved, for it is consubstantial to the very definition of workers as salaried workers employed by capitalists who own the means of production. Since saving is by definition identical to income less consumption, it follows that if workers saved, this saving would by the same token be used to purchase goods other than consumer goods and hence for buying means of production. But in this case workers would ipso facto turn into capitalists, which would be contradictory!

- This raises the question of the status of real estate, whose value is considered in the national accounts as the "gross fixed capital formation of households", which is supposed to legitimize the treatment of their counterpart in income as constituting “households savings". In fact, dwellings are durable goods, and even if they can obviously be leased, they cannot be considered as capital in the sense of Marx: capital is in fact such because it is not only a means of production, but the support of the employment of salaried workers who make it produce (which is obviously not the case of housing, even when rented). Real estate status, and consequently that of household savings, should therefore be modified in national accounts. This question of the conceptual status of real estate is discussed more comprehensively in an article that I published on Piketty’s best-selling book (Capital in the 21st Century): “Again on Piketty’s *Capital in the Twenty-First Century* or Why National Accounts and Three Simple Laws Should not Be a Substitute for Economic Theory” (Flamant, 2015a, pp. 24-25).

- Moreover, means of production correspond only to fixed capital. This implies that there is no section specifically producing circulating capital, which is produced within each of the two sections I and II. In each section production is thus integrated, in the sense that each section produces, with the aid of means of production and labor, the totality of its circulating capital, i.e. intermediate goods necessary for the production of its final goods, and that disappear as such (in their original form) in the course of the production process (upstream of it), in order to reappear downstream, after transformation, by being incorporated into these final goods. In his book published in 1981: “Structural change and economic growth: a theoretical essay on the dynamics of the wealth of nations” Pasinetti indicates that: “in a vertically integrated model, the criterion is the process of production of a final commodity, and the problem is to build conceptually behind each final commodity a vertically integrated sector which, by passing through all the intermediate commodities, goes right back to the original inputs” (Pasinetti, 1981, p. 113). He shows that for every final good one can obtain the vector of the vertically integrated circulating capital stock of the corresponding branch, which can be regarded as a kind of composite good, called vertically integrated production capacity unit. The same procedure applies to labor coefficients, so that for each final good it is possible to obtain the integrated labor quantity and the corresponding unit of vertically integrated production capacity. On this point Pasinetti was relying on Sraffa, who in appendix A on sub-systems of “Production of commodities…” indicates: “Although only a fraction of the labor of a sub-system is employed in the industry which directly produces the commodity forming the net product, yet, since all other industries merely provide replacements for the means of production used up, the whole of the labor employed can be regarded as directly or indirectly going to produce that commodity” (Sraffa, 1960, p. 105).

- It results from the preceding assumption that, unlike the position adopted by Marx, constant capital (the means of production) does not include circulating capital, and is therefore constituted only of fixed capital. This makes it unnecessary to take into account a third section producing circulating capital. In addition to the resulting theoretical simplification, this is much more in line with the real world where final producers do not advance circulating capital, but instead usually buy it with a credit from their suppliers of intermediate goods.

- The economy is in a state of simple reproduction, from period to period, which in Marxist terms means that there is no accumulation of capital, whether it is constant or variable, from period to period, but a mere maintenance of the value of the existing stock of capital.

- To simplify the argument, it is assumed that the part of constant capital constituted by circulating capital is produced and incorporated into final goods in the same period in which it is produced. There is therefore no transmission of value from one period to the next by the part of constant capital constituted by raw materials or intermediate goods, in the absence of stocks of circulating capital.

- Finally the value produced in each period, which is equal to the quantity of living labor expended (directly or indirectly) during this same period to produce all of the final commodities, breaks down into $L_I$, the value produced in section I, and $L_{II}$, the value produced in section II.

Having reached the present stage, the reader who considers that the analysis of the simple reproduction carried out by Marx in Book II of Capital is correct, will certainly pretend that the model does not allow the analysis of the rate of profit, within the framework of these assumptions. In the case of simple reproduction, the value of the newly produced means of production is indeed not considered by Marx as surplus-value, since it merely replaces the value transmitted to the produced commodities by the existing stock of constant capital, for an identical amount. Surplus-value then exists only to the extent of the unproductive consumption of capitalists, which is impossible here, since it has been assumed that the capitalists do not consume. Surplus value and the rate of profit should therefore be zero.

In fact, we have already proved in previous chapters that fixed capital cannot transmit its value to the final product for many reasons. Let us just recall that in a theory of production as a transformation, the assumption of the transmission of value of constant capital is not legitimate if it applies to fixed capital. And constant capital is here constituted only of fixed capital, for the reasons already just explained above (cf. the second and third paragraphs of the previous page). It follows that Marx's reproduction schemes are erroneous. For the reader who would be interested in a full demonstration, based on the internal incoherence of these schemes, such a demonstration is provided in Appendix 2 at the end of the present book.

It will now be shown that the rejection of the assumption of transmission of value of fixed capital is perfectly compatible with the realization[11] of the product, whatever the type of reproduction envisaged, and this without falling into the type of contradictions that can be encountered in Marx's reproduction schemes. This approach will be explained, first for the realization of the product in Section II and then for the realization of the product in Section I.

[11] The word "realization" has the meaning of "sale", but refers also to the realization of profits through the sale of the product, which explains that we prefer to use it rather than "sale".

### 3.2. The realization of the product and profits in Section II

In order for consumer goods to be realized, it is sufficient to assume that in exchange for the amount of monetary wages corresponding to the total newly produced value $L_I + L_{II}$, workers obtain consumer goods for a $L_{II}$ money value only, which implies that an $L_I$ value is thereby transferred without compensation to the capitalists of Section II who sell them these goods. This is achieved by assuming only that consumer goods are sold at an $L_I + L_{II}$ price higher than their $L_{II}$ monetary value, which allows capitalists of section II to levy an $L_I$ value equal to the total surplus value, but which corresponds now, after realization of the product in this consumer goods section, to the profits of this section alone. Let us call S this surplus value, and k the rate of surplus-value, by definition it is $k = \frac{S}{L_{II}}$. (1)

As in this very simple model $S = L_I = kL_{II}$ (by definition of k) the price of consumer goods, named $P_{II}$ must therefore be $P_{II} = L_{II} + kL_{II} = (1+ k) L_{II}$ (2)

In this scheme, what is generally referred to as the margin rate in the macroeconomic sense in Section II, i.e. the ratio of profits named $\Pi_{II}$ over the price of consumption goods produced by section II, is thus equal to $\frac{\Pi_{II}}{P_{II}} = \frac{L_I}{L_I + L_{II}} = \frac{kL_{II}}{L_{II}(1+k)} = \frac{k}{1+k}$ (3)

This very simple mechanism just described is nothing else than the mechanism of surplus-value extraction, which is thus implemented through a necessary mechanism of transformation of money values into prices: in this simplified model it is sufficient that in exchange for their total monetary wages, corresponding to the production in money value of all the goods produced, workers are able to obtain only an $L_{II}$ money value, that of consumer goods only. This is so because, unlike values, which reflect the conditions of production, prices also reflect, beyond that, the conditions of distribution of the product between workers and capitalists, since they are the instrument of this distribution.

It should be noted that *w* is the average monetary wage per unit of labor time. It has therefore a double dimension: that of money divided by the dimension of time. For simplicity, we can therefore assume that $w = 1$, but only if one takes into account the fact that $w = 1$ is a two-dimensional ratio (monetary unit/unit of labor time). In this case, as we have already explained at length, we may write $W = L_I + L_{II}$, but care must be taken that the quantities $L_I$ and $L_{II}$ are no longer average social values expressed in terms of labor time but are money values $wL_I$ and $wL_{II}$ expressed in monetary units that are pure scalars.

In any event, this makes it theoretically necessary to introduce the concept of value of labor-power as a concept that must be distinct from that of wages itself, since we need to express the difference between wages as the price of labor and the value of the goods obtained on the market in exchange for that monetary wage, which is the value of labor-power. In this simplified model, the overall value of labor-power (let us call it *V*) is thus equal to $L_{II}$.

The value of the labor-power in section II, let us call it $V_{II}$ corresponds to the share of the total value $L_{II}$ of consumer goods that the workers in section II obtain in exchange for their money wages (which amount to $W_{II} = L_{II}$). It is proportional to the share $\frac{L_{II}}{L_I + L_{II}}$ which is obtained by all workers in both sections in the total value of the whole product. This value $V_{II}$ is therefore equal to:

$$V_{II} = \frac{L_{II}}{L_I + L_{II}} \cdot L_{II} = \frac{L_{II}}{(1+k)L_{II}} \cdot L_{II} = \frac{L_{II}}{1+k} \tag{4}$$

Since by definition $S = kV$ (we recall that $V$ is the value of the total labor-power), the surplus-value in this same section II, which we can call $S_{II}$ is thus $S_{II} = k\frac{L_{II}}{1+k}$. We verify that the ratio between the surplus-value and the value of labor-power $V_{II}$ in section II is equal to k. If we name $\Pi_{II}$ the monetary profits realized in section II, then the relation between these profits, which are equal to $L_I$ (in money value), and the labor price (money wages) in section II, which we have called $W_{II}$ (setting w = 1), is therefore equal to $\frac{L_I}{L_{II}} = k$ : it is thus equivalent in section II to the rate of surplus-value.

### 3.3. The realization of the product and profits in Section I

What about section I, which produces fixed capital? As regards the value of labor-power in section I (let us name it $V_I$), it is for similar reasons equal to $\frac{L_I}{1+k}$. As it is known that by definition $L_I = k\,L_{II}$, it is easily verified that the sum of the values of labor-power of the two sections, which is $\frac{L_{II}}{1+k} + \frac{L_I}{1+k}$, is equal to $L_{II}$, the total value of labor-power. The surplus-value in section I is therefore $S_I = \frac{k}{1+k} L_I$, and since by definition we have $L_{II} = \frac{L_I}{k}$, the sum of the surplus-values of the two sections is:

$$S_I + S_{II} = \frac{k}{1+k} L_{II} + \frac{k}{1+k} L_I = \frac{k}{1+k} \cdot \frac{L_I}{k} + \frac{kL_I}{1+k} = \frac{L_I}{1+k} + \frac{kL_I}{1+k} = L_I \tag{5}$$

With regard to the realization of the product in section I, we decide to call $a$ the fraction of the value of the total fixed capital that is accumulated (but not transferred to the product) in section I, and to call $b = 1 - a$ the fraction that is accumulated (but not transferred to the product) in section II.

Capitalists from section II will necessarily buy the fixed capital that they need to invest for maintaining their stock of capital from the capitalist of section I, who precisely produce this fixed capital. In order for these capitalists of section II to obtain in exchange for the money value $L_I$ thus levied on the workers only a fraction $bL_I$ of the money value $L_I$ of the newly produced means of production, it is necessary and sufficient that the capitalists producing in section I the fixed capital used in the production of consumer goods, withdraw themselves, on the occasion of the sale of this fixed capital to capitalists of section II, the remaining fraction $aL_I$ of the value $L_I$ .

Since it is trivial that $L_I = (1-a)L_I + aL_I$ , this is what happens when these means of production to be used in section II, and of a value $bL_I = (1-a)L_I$ are sold at a price $\mathrm{L}_\mathrm{I}$ , equal to $\frac{1}{b}bL_I$ , or $\frac{1}{1-a}\cdot(1-a)L_I$ . This price $\mathrm{L}_\mathrm{I}$ corresponds to the expense of the total profits of Section II, and allows the capitalists of section I who have insured this sale to levy in turn through this sale a profit equal to $aL_I$ .

What remains to be done is to show how this remaining money value of the means of production needed for the production of fixed capital by section I is disposed of and realized. These means of production will be the object of successive purchases within Section I, going back upstream in the production of fixed capital. At this stage of our demonstration, it is appropriate to make an assumption about the share of the value of the available fixed capital used and purchased at each of these successive levels.

To be sure in the real world these shares are different from one level to the other. But taking that into account would introduce in our model a number of new variables, which would complicate unduly this attempt to shed some light on the realization of profits. In any case we know that these shares are necessarily always lower than one. Moreover the initial assumption has been made that capitalists in section I use a fraction $a$ of the total value of fixed capital produced globally by the productive system.

Therefore the simplest solution, to avoid the difficulty that we just mentioned, is to continue to assume that at each successive stage capitalists of section I always purchase the same fraction $a$ of the total money value of fixed capital still available after the previous purchases have been made. Let us consider it as a first assumption.

1) Capitalists in section I always purchase a fraction $a$ of the fixed capital available

It is on this basis that we shall now draw up a table, numbered 11.1. This table gives the prices, the money values realized and the money values levied as profits, on the occasion of each successive sale of fixed capital produced by section I. These variables are provided first for the sale to capitalists of section II, which will constitute the first row of our table, and then for the successive sales between capitalists of section I themselves, which constitute the following rows of this table 11.1.

As regards the first row, we have already seen above that by definition the value of the fixed capital obtained by the capitalists of section II and therefore the value of the fixed capital realized by those of section I by the corresponding sale, is a fraction $bL_I = (1-a)L_I$ of the total money value $L_I$ of this fixed capital, such realization being made at a price corresponding to $L_I$ and allowing the capitalists of section I who make this sale to make a profit. It is these three elements which will appear in the table for this first row, knowing that the realized values appear there as a difference between the monetary prices and the profits, and as indicated above, express in fact the price of the corresponding labor, i.e. the amount of monetary wages.

In order to fill in the second row of the table, the rule should be used that, since we are here in a state of simple reproduction, all profits made at one stage on the sale of fixed capital are spent at the next stage for the purchase of fixed capital. It follows that the value levied as profits at one stage is identical to the price paid in the next stage. The other rule derives from the assumption made above that the purchase made at that price only permits to obtain a value equal to a fraction $a$ of the value $(1-b)L_I = aL_I$ of the fixed capital still available for section I. The value which is thus obtained is $a(aL_I) = a^2L_I$. Conversely, the price/value ratio is equal to $\frac{aL_I}{a^2L_I} = \frac{1}{a}$. The realized money value, in the form of wages, is therefore $a^2L_I$ and the value levied or profit is the difference between the price and the realized value, i.e.:

$$aL_I - a^2L_I = a(1-a)L_I \qquad (6)$$

At the next level, for the third row of the table, the price is therefore necessarily equal to $a(1-a)L_I$ (i.e. the profits from the previous stage), and the corresponding realized money value is therefore a fraction $a$ of this price, i.e. $a^2(1-a)L_I$. Since the remaining fixed capital has a value equal to $aL_I - a^2L_I = a(1-a)L_I$, we can verify that the realized money value is equal to a fraction $a$ of the value of this fixed capital still available at the end of the preceding stage. The profits levied at this stage amount to the difference between the price and the realized value, namely: $a(1-a)L_I - a^2(1-a)L_I = a(1-a)(1-a)L_I = a(1-a)^2 L_I$.

The other rows of the table are obtained according to the same scheme, and therefore by recurrence. It is therefore unnecessary to explain them one by one. Let us note only that for row *n* we get a price equal to $a(1-a)^n L_I$, which is the sum of the realized money value, i.e. $a^2(1-a)^n L_I$ and of profits equal to $a(1-a)^{n+1}$. The complete table can now be given below:

**Table 11.1. Prices, money values and profits in section I (first case)**

| Prices | Realized money values = labor price = wages | Levied values = profits |
|---|---|---|
| $L_I$ | $bL_I = (1-a)L_I$ | $aL_I$ |
| $aL_I$ | $a^2L_I$ | $a(1-a)L_I$ |
| $a(1-a)L_I$ | $a^2(1-a)L_I = a^2L_I - a^3L_I$ | $a(1-a)^2 L_I$ |
| $a(1-a)^2 L_I$ | $a^2(1-a)^2 L_I$ | $a(1-a)^3$ |
| … | … | … |
| $a(1-a)^n L_I$ | $a^2(1-a)^n L_I$ | $a(1-a)^{n+1}$ |

We can verify that for each row of the above table we get the following identity:

Price = realized money value + levied value (profit).

The total price of the product for all the fixed capital sold by section I, named $P_I$, is the sum of the prices in the first column, namely:

$$\Sigma \text{ prices} = P_I = L_I + aL_I + a(1-a)L_I + a(1-a)^2 L_I + ... + a(1-a)^n L_I$$

$P_I = L_I + aL_I\left(1+(1-a)+(1-a)^2+...+(1-a)^n\right)$, which gives us when $n \to \infty$:

$$P_I = L_I\left(1+a\frac{1}{1-(1-a)}\right) = L_I\left(1+\frac{a}{a}\right) = L_I + L_I = 2L_I \qquad (7)$$

It can be seen that the price of total fixed capital is twice the fixed capital price $L_I$ paid by section II, so that the fixed capital for section I (from the first row of the table) has the same price $L_I$ ! But this for a different realized money value: $(1-a)L_I$ for section II and $aL_I$ for section I.

The sum of the realized values for all the fixed capital produced from section I is in fact:

$$\Sigma \text{ realized money values} = (1-a)L_I + a^2L_I + a^2(1-a)L_I + a^2(1-a)^2 L_I + ... + a^2(1-a)^n L_I$$
$$= (1-a)L_I + a^2L_I\left(1+(1-a)+(1-a)^2+...+(1-a)^n\right)$$

When $n \to \infty$, it is easy to check that this amount of $\Sigma$ realized money values is equal to:

$$\Sigma \text{ money values} = (1-a)L_I + a^2L_I\left(\frac{1}{1-(1-a)}\right) = (1-a)L_I + a^2L_I\left(\frac{1}{a}\right) = (1-a)L_I + aL_I = L_I \qquad (8)$$

It can be seen that the total amount $L_I$ of the money values realized by section I is equal to the sum of the value of the fixed capital sold to section II, that is $(1-a)L_I$, and of the value of the fixed capital sold within section I, or $aL_I$.

As for the sum of the levied values, i.e. the overall profit of section I, named $\Pi_I$, it is equal to:

$$\Sigma \text{ profits} = \Pi_I = aL_I + a(1-a)L_I + a(1-a)^2 L_I + ... + a(1-a)^n L_I$$

$\Pi_I = aL_I + a(1-a)L_I\left(1+(1-a)+(1-a)^2+...+(1-a)^n\right)$,

When $n \to \infty$, this gives us:

$$\Pi_I = aL_I + a(1-a)L_I\left(\frac{1}{1-(1-a)}\right) = aL_I + a(1-a)L_I\frac{1}{a} = aL_I + (1-a)L_I = L_I \qquad (9)$$

It can also easily be seen that the total profits of section I, $L_I$, is equal to the sum of profits realized in the sale of fixed capital to section II, i.e. $aL_I$, and profits realized in the sale of fixed capital within section I itself, i.e. $(1-a)L_I$.

It follows of course that the sum of prices in section I, i.e. 2 $L_I$, is equal to the sum of the realized money values $L_I$, the money value of the product of Section I, on the one hand, to which is added the sum of profits (the levied money values), or again $L_I$, on the other hand. To be sure these are very specific values, and we can legitimately question the reason for these particular figures. Here is the explanation: given that the price of the product in section I, i.e. 2 $L_I$, is equal by construction to $\frac{1}{1-\frac{1}{2}}L_I$, it follows that for section I as a whole value $\frac{1}{2}$ is the fraction of the value of the total available capital obtained on average through the overall spending of the profits of the two sections, which is 2 $L_I$. This is so because the sum of the parameters $a$ and $b$ is equal to 1, and therefore because their mean is equal to $\frac{1}{2}$. We may, therefore, ask ourselves what would be the case if another hypothesis had been made on the fraction of the available capital obtained at each stage by capitalists of section I. Suppose now, for example, that these capitalists of section I obtain at each stage not a fraction $a$, but as capitalists of section II, a fraction $b$ of the fixed capital still available. Let us consider it as a second assumption.

2) Capitalists in section I always purchase a fraction $b$ of the fixed capital available

This implies that the means of production are always sold at a price equal to $\frac{1}{b}$, i.e. $\frac{1}{1-a}$ times their value. This makes it possible to draw up a new table giving prices, realized money values and money values levied from each successive exchange of fixed capital between capitalists of section I. For the sale of fixed capital to capitalists of section II, which corresponds to the first row of the new table, there is obviously nothing changed compared to the previous table.

It remains to be seen how the remaining value $aL_I$ of the means of production utilized for the production of fixed capital is disposed of and realized. These means of production are procured in section I by going successively upstream in the production system. To make these successive purchases appear, we continue to assume that the means of production are sold at each stage at a price equal with this new assumption to $\frac{1}{b}$, or $\frac{1}{1-a}$ times their value. This makes it possible to draw up table 11.2, giving prices, realized values and values levied during each successive exchange of fixed capital between capitalists of section I.

**Table 11.2. Prices, money values and profits in section I (second case)**

| Prices | Realized money values = labor price = wages | Levied values = profits |
|---|---|---|
| $L_I$ | $bL_I = (1-a)L_I$ | $aL_I$ |
| $aL_I$ | $abL_I = a(1-a)L_I$ | $aL_I(1-b) = a^2L_I$ |
| $a^2L_I$ | $a^2bL_I = a^2(1-a)L_I$ | $a^2L_I(1-b) = a^3L_I$ |
| $a^3L_I$ | $a^3bL_I = a^3(1-a)L_I$ | $a^3L_I(1-b) = a^4L_I$ |
| … | … | … |
| $a^nL_I$ | $a^nbL_I = a^n(1-a)L_I$ | $a^nL_I(1-b) = a^{n+1}L_I$ |

We can verify that for each line of table 11.2 we have: price = realized value + levied value (or profit).

The sum of prices, i.e. the price of the overall product of section I, named $P_I$, is equal to:

$$\Sigma \text{prices} = P_I = L_I\left(1+a+a^2+...+a^n\right) = L_I\left(\frac{1}{1-a}\right) \text{ when n} \to \infty \qquad (10)$$

The sum of values obtained in section I is equal to:

$$\Sigma \text{realized values} = bL_I + abL_I + a^2bL_I + ... + a^nbL_I = bL_I\left(1+a+a^2+...+a^n\right)$$

When n $\to \infty$ this sum is equal to:

$$\Sigma \text{realized values} = \text{b. } \frac{1}{1-a} L_I\text{, equivalent to } \frac{1-a}{1-a} L_I = L_I \qquad (11)$$

As for the total of levied money values, i.e. the overall profits of section I, or $\Pi_I$, it is equal to:

$$\Sigma \text{profits} = \Pi_I = L_I\left(a+a^2+...+a^n\right)$$

$$\Pi_I = aL_I\left(1+a+a^2+...+a^n\right) = \text{L}_\text{I} \frac{a}{1-a} \text{ when n} \to \infty \qquad (12)$$

It follows, of course, that the sum of prices in section I is equal to the sum of realized values $L_I$, i.e. the money value of the product of section I, on the one hand, to which is added the sum of profits (levied values), on the other hand, since:

$$P_I = L_I \frac{1}{1-a} = L_I + L_I \frac{a}{1-a} = L_I\left[1+\frac{a}{1-a}\right] = L_I \frac{1}{1-a} \qquad (13)$$

However this second case that we just examined may also be regarded as a special case with regard to the fraction of the available fixed capital purchased at each successive stage by capitalists of section I. It seems therefore necessary to examine a more general case, where this fraction no longer takes values $a$ (our first assumption) or $b$ (our second assumption), but a different but constant value. Let us consider it as a third assumption:

3) Capitalists in section I purchase a specified fraction $u$ of the available fixed capital

Let us call $u$ this value, with $0 < u < 1$, and draw up again the table of successive purchases of fixed capital, showing prices, realized values and levied values from each sale at each of the successive stages:

**Table 11.3. Prices, money values and profits in section I (third case)**

| Prices | Realized money values = labor price = wages | Value of available fixed capital | Levied values = profits |
|---|---|---|---|
| $L_I$ | $(1-a)L_I = bL_I$ | $L_I$ | $aL_I$ |
| $aL_I$ | $auL_I$ | $aL_I$ | $a(1-u)L_I$ |
| $a(1-u)L_I$ | $au(1-u)L_I$ | $a(1-u)L_I$ | $a(1-u)^2 L_I$ |
| $a(1-u)^2 L_I$ | $au(1-u)^2 LI$ | $a(1-u)^2 L_I$ | $a(1-u)^3 L_I$ |
| … | … | | … |
| $a(1-u)^n L_I$ | $au(1-u)^n L_I$ | $a(1-u)^n L_I$ | $a(1-u)^{n+1} L_I$ |

It appears that the gross profit margin for sales of fixed capital to capitalists of Section I, i.e. figures appearing from the second row of the table, is variable and equal to $1-u$.

It can be seen from the table that the sum of prices is equal to:

$$\Sigma \text{prices} = P_I = L_I + aL_I + L_I\left[a(1-u)L_I + a(1-u)^2 L_I + ... + a(1-u)^n\right]$$

$$P_I = L_I(1+a) + L_I a(1-u)\left[1+(1-u)+(1-u)^2 ... + (1-u)^n\right]$$, i.e. when $n \to \infty$:

$$P_I = L_I\left[(1+a) + a(1-u)\frac{1}{u}\right] = L_I\left[(1+a) + \frac{a}{u} - a\right] = L_I\left(1+\frac{a}{u}\right) \quad (14)$$

The sum of realized money values in section I is equal to:

$$\Sigma \text{realized money values} = bL_I + auL_I + au(1-u)L_I + au(1-u)^2 L_I ... + au(1-u)^n L_I$$
$$= L_I\left[(1-a+au) + au(1-u)\left[1+(1-u)+(1-u)^2 + ... + (1-u)^n\right]\right]$$

Which gives us when $n \to \infty$:

$$\Sigma \text{realized money values} = L_I\left[(1-a+au) + au(1-u)\frac{1}{u}\right] = L_I\left[(1-a+au) + a - au\right] = L_I \quad (15)$$

Finally the sum of profits in section I, named $\Pi_I$, is equal to:

$$\Sigma \text{profits} = L_I\left[\left(a+a(1-u)+a(1-u)^2+...+a(1-u)^n\right)\right]$$

$$= aL_I\left[\left(1+(1-u)+(1-u)^2+...+(1-u)^n\right)\right]\text{, i.e. when n} \to \infty :$$

$$\Pi_I = aL_I\left(\frac{1}{1-(1-u)}\right) = aL_I\frac{1}{u} = L_I\frac{a}{u} \tag{16}$$

Here again, the sum of prices in section I is equal, on the one hand, to the sum of realized values, $L_I$, i.e. the value of the product of section I (which, with $w = 1$, corresponds to the price of labor in this section), to which is added the sum of profits (or levied values) of this section, on the other hand, since:

$$L_I\left(1+\frac{a}{u}\right) = L_I + L_I\frac{a}{u} = \Pi_{II} + \Pi_I = \Pi \tag{17}$$

We can also verify that the two cases examined above in 1) and 2) are indeed particular cases of this more general case, since the first case corresponds to a situation in which $u = a$ and the second case to a situation where $u = b = 1-a$. To be sure, an even more general case would be the situation prevailing in the real world, where at each stage of successive purchases of fixed capital corresponds a distinct and different coefficient. Nevertheless, since each of the coefficients $u$ at each stage has a value such that $0 < u < 1$, it is clear that the corresponding numerical series is convergent. The economic interpretation of these results is that the mechanism highlighted here is such that the whole value of the product of section I can always be sold through this mechanism.

After seeing above in subsection 3.2. how the value produced in section II and the profits of this section are realized, it has in any event been demonstrated in section 3.3. that all the value produced in section I and the corresponding profits can be realized through a very simple mechanism of successive purchases of fixed capital, and that the reproducibility of the system is therefore ensured. This demonstration was conducted without having recourse to the hypothesis of the transmission of value of fixed capital. It is thus verified at this level that this value transmission is therefore not a necessary condition for the realization of the whole product and for the reproduction of the productive system.

Moreover, it must be noted that the demonstration which has just been made is valid both for simple reproduction and for extended reproduction, since no assumption has been made on the value of the fixed capital used (in a final way) in each of the two sections of production: this value indeed could have been equal to, or less than, or greater than the value of fixed capital accumulated in the preceding period.

But this model, however simplified it may seem, and maybe because of its simple nature, has a potential to teach us a few interesting lessons, as we shall try to demonstrate in the next section.

# 4. Some characteristic quantities within the framework of the simplified model

At this stage, it is possible to calculate the values taken by some characteristic quantities in the framework of this simplified model. This will be done for the price of the global product, the ratio between wages and profits, and the investment rate, knowing that the calculation of the profit rate itself will only be addressed in the next chapter.

## 4.1. The price of the overall product

Taking up the three cases discussed in the previous chapter for the realization of the product in section I, we obtain the following prices for the overall product of the two sections (let us call Y this price):

1) In the first case, where $u = a$, we have:

$$Y = (L_I + L_{II}) + 2L_I = 3L_I + L_{II} = 3kL_{II} + L_{II} = L_{II}(1+3k) \tag{18}$$

$$Y = 3L_I + \frac{L_I}{k} = L_I\left(3 + \frac{1}{k}\right) \tag{19}$$

2) In the second case, where $u = b = 1-a$, we have :

$$Y = (L_I + L_{II}) + L_I\left(\frac{1}{1-a}\right) = L_I\left(1 + \frac{1}{1-a}\right) + \frac{L_I}{k} = L_I\left(\frac{2-a}{1-a} + \frac{1}{k}\right) \tag{20}$$

$$Y = L_{II}(1+k) + kL_{II}\left(\frac{1}{1-a}\right) = L_{II}\left(1 + k + \frac{k}{1-a}\right) \tag{21}$$

3) In the third case, where u can be any number, we have:

$$Y = f(L_I) = (L_I + L_{II}) + L_I\left(1 + \frac{a}{u}\right) = L_I\left(2 + \frac{a}{u}\right) + \frac{L_I}{k} = L_I\left(2 + \frac{1}{k} + \frac{a}{u}\right) \tag{22}$$

$$Y = f(L_{II}) = L_{II}(1+k) + kL_{II}\left(1 + \frac{a}{u}\right) = L_{II}\left(1 + 2k + k\frac{a}{u}\right) \tag{23}$$

## 4.2. The profit/price of labor-power ratio

1) The profit/wages ratio in section II

In fact, it has already been pointed out above (see section 3.1 of this chapter) that in section II the ratio between profits, which may be called $\Pi_{II}$ and are equal to $L_I$, and the price of labor (wages), which may be called $W_{II}$, is equal to $\frac{\Pi_{II}}{W_{II}}$ : it is therefore equivalent in section II to the rate of surplus-value. It follows in passing, which is not without interest, that from a statistical and empirical point of view the ratio between wages and the overall price of consumer goods (a different ratio) is a good indicator of the rate of surplus-value existing in an economy, whatever it is.

2) The profit/wages ratio in section I

Using again the three cases examined in the preceding section regarding the realization of the product in section I, we obtain the following values for this ratio, which is $\frac{\Pi_I}{W_I}$ :

- In the first case where $u = a$, we get :

$\frac{\Pi_I}{W_I} = \frac{L_I}{L_I} = 1$, which shows well the peculiar nature of this case. (24)

- In the second case where $u = b = 1 - a$, we get :

$$\frac{\Pi_I}{W_I} = \frac{L_I \frac{a}{1-a}}{L_I} = \frac{a}{1-a} \qquad (25)$$

- In the third case (the general case) where $u$ can take any value, with $0 < u < 1$, we get:

$$\frac{\Pi_I}{W_I} = \frac{L_I \frac{a}{u}}{L_I} = \frac{a}{u} \qquad (26)$$

Is striking to observe that for a given rate of surplus-value, which as it happens is k, there is a priori a notable difference between the profit/wages ratio in section II, producing consumption goods, where this ratio is always equal to *k*, i.e. the rate of surplus-value, and the ratio in section I, producing fixed capital, where it is independent from *k*, et depends in fact only on parameters $a$ and $u$ .

However, if we have a look at the ratios which correspond to each stage of successive purchases of fixed capital in section I, they are of the general following form, for the i[th] stage of fixed capital spending within section I:

$$\frac{\Pi_{I,i}}{W_{I,i}} = \frac{a(1-u)^{i+1} L_I}{au(1-u)^i L_I} = \frac{1-u}{u} \qquad (27)$$

From this last equation we can derive that for $u = a$ we have: $\frac{\Pi_{I,i}}{W_{I,i}} = \frac{1-a}{a}$ (28)

And that for $u = b = 1 - a$, we get: $\frac{\Pi_{I,i}}{W_{I,i}} = \frac{1-(1-a)}{(1-a)} = \frac{a}{1-a}$ (29)

3) The profit/wages ratio for the system as a whole

- In the first case where $u = a$, this ratio is: $\frac{\Pi}{W} = \frac{kL_I(a+u)}{uL_I(1+k)} = \frac{2ak}{a(1+k)} = \frac{2k}{1+k}$ (30)

- In the second case where $u = b = 1-a$, it is: $\frac{\Pi}{W} = \frac{kL_I(a+1-a)}{(1-a)L_I(1+k)} = \frac{k}{(1-a)(1+k)}$ (31)

- In the general case it becomes: $\frac{\Pi}{W} = \frac{L_I + L_I\left(\frac{a}{u}\right)}{L_I + L_{II}} = \frac{L_I\left(1+\frac{a}{u}\right)}{L_I\left(1+\frac{1}{k}\right)} = \frac{k(a+u)}{u(1+k)}$ (32)

### 4.3. The I/Y investment rate and the share of profits in the product $\Pi$/Y

In the situation described by the simplified model under consideration, which is a simple reproduction situation where capitalists do not consume, it goes without saying that all profits are invested. As a result, the I/Y investment rate and the share of profits in the product, $\Pi$/Y, can only be identical.

Note also that if one reasoned purely in terms of values, the overall investment rate of the system, in a situation of simple reproduction, would be nothing other than the ratio between the value of the fixed capital newly produced and that of the global product, or:

$$\frac{I}{Y} = \frac{\Pi}{Y}(values) = \frac{L_I}{L_I + L_{II}} = \frac{kL_{II}}{L_{II}(1+k)} = \frac{k}{1+k} \qquad (33)$$

The investment rate or the share of profits in section I alone would be equal to:

$$\frac{I_I}{Y_I} = \frac{\Pi_I}{Y_I}(values) = \frac{aL_I}{L_I} = a \qquad (34)$$

As for the investment rate or the share of profits in Section II, they would be equal to:

$$\frac{I_{II}}{Y_{II}} = \frac{\Pi_{II}}{Y_{II}}(values) = \frac{bL_I}{L_{II}} = \frac{(1-a)kL_{II}}{L_{II}} = k(1-a) \qquad (35)$$

If we now use the results in terms of prices obtained above, we get obviously slightly different expressions.

For the overall investment rate, or for the share of profits in the overall product, they are written in the form of the following ratio:

$$\frac{I}{Y}=\frac{\Pi}{Y}\text{(prices)} = \frac{L_I\left(1+\frac{a}{u}\right)}{L_I\left(2+\frac{1}{k}+\frac{a}{u}\right)}$$

$$\frac{I}{Y}=\frac{\Pi}{Y}(prices)=\frac{\frac{a+u}{u}}{\frac{k(2u+a)+u}{ku}}=\frac{k(a+u)}{k(a+u)+u(1+k)} \tag{36}$$

In the case where $u=a$, this ratio becomes:

$$\frac{I}{Y}=\frac{\Pi}{Y}(prices)=\frac{k(a+a)}{k(a+a)+a(1+k)}=\frac{2ak}{2ak+a(1+k)}=\frac{2k}{3k+1} \tag{37}$$

In the case where $u=b=1-a$, the ratio changes as follows:

$$\frac{I}{Y}=\frac{\Pi}{Y}(prices)=\frac{k(a+1-a)}{k(a+1-a)+(1-a)(1+k)}=\frac{k}{k+1+k-a-ak}$$

$$\frac{I}{Y}=\frac{\Pi}{Y}(prices)=\frac{k}{k(2-a)+1-a}=\frac{k}{k+bk+b} \tag{38}$$

As regards the investment rate or the profit share in Section I, which is simply expressed, they are equal to:

$$\frac{I_I}{Y_I}(prices)=\frac{L_I\frac{a}{u}}{L_I\left(1+\frac{a}{u}\right)}=\frac{\frac{a}{u}}{1+\frac{a}{u}}=\frac{a}{a+u} \tag{39}$$

Finally, the investment rate or the share of profits in section II, which also have a simple expression, are equal to:

$$\frac{I_{II}}{Y_{II}}\text{(prices)} = \frac{L_I}{L_I+L_{II}}=\frac{kL_{II}}{L_{II}(1+k)}=\frac{k}{1+k} \tag{40}$$

## 5. A few lessons from the simplified model

One could think that such a simplified model cannot teach us anything about the real world because its assumptions are precisely so simple that they do not correspond to this real world. It is true indeed, for instance, that in the real world capitalists do consume, and in fact, on an individual basis, an average capitalist consumes most probably much more than an average worker. The assumption that there is no capitalist consumption in this simplified model seems therefore quite unrealistic. But going back to these assumptions, this model can nevertheless be rich in lessons.

### 5.1. The simplified model interpreted as excluding capitalists

Firstly, since there is no capitalist consumption in this model, one might imagine that we could go one step further and consider that there are no capitalists at all in the model. Then these assumptions would correspond to a world where there would be no private property of fixed capital, meaning that this capital would be the collective property of all workers, who would ipso facto be the owners of all the commodities which are produced, be they consumption goods or fixed capital goods. In such a situation, there would be no surplus-value, and one could think that there would be therefore no need for a difference between the money value and the price of commodities, since it is this difference which allows for the levy of surplus-value under the form of profit. In this interpretation, each worker would obtain a value equivalent to that of the commodities which it has contributed to produce, in proportion to its contribution to production, measured by the level of his wage.

This idea would nevertheless be wrong, since what ultimately production is destined to satisfy is consumption, and since fixed capital can be accumulated (or even contemplated!) but by definition cannot in any case be consumed like a consumption good, which means that workers of section I producing fixed capital cannot be interested in the property of these goods that they produce, or of their value, but quite obviously are rather interested, like workers in section II, in the acquisition of consumer goods. This reminds us that fixed capital, considered as a whole, is produced only to be employed – ultimately, in the production of consumption goods. We go back there to the old conception, dating back to Böhm-Bawerk and the Austrian school, of capital viewed as a production detour, a process also called “roundaboutness”.

This implies that no more than capitalists, workers in section I producing fixed capital goods cannot be satisfied to obtain the very goods that they have produced, for a value corresponding to the average social labor which they have spent to produce these goods. Since - under the particular assumption of this subsection, there is no private property of fixed capital, they might be considered as the collective owners of all this fixed capital. But even though a worker in section I might thus be considered as the owner of a fraction of fixed capital which he has contributed to produce (to the tune of the amount of his wage), this would not provide him with the consumption goods he needs, in order to make a living.

It results from this recognition that all workers of both sections need to buy all of the consumption goods which have been produced by workers of section II only, and that the overall price of these goods must necessarily be equal to the overall amount of wages, which is higher (by an amount $L_I$ or $W_I$) than their value $L_{II}$ or $W_{II}$. This in turn implies that whatever the mode of production, the existence of fixed capital explains and implies the necessity for prices of consumption goods to be higher than their money value.

Secondly, this also implies the existence of money as a general medium of exchange, which is the only way to avoid that workers of section I enter into some kind of weird exchanges of fixed capital goods against consumption goods: one cannot see how an economic system would work with such a kind of barter trade, although such a system constitutes the basis of neoclassical economics! Here we can see that, through its role as a general medium of exchange, money plays a very important role in allowing for the distribution of the product of the system to take place. In the present case this is the distribution of consumption goods among workers: the fact that wages are paid in money allows workers of section I to share without any difficulty the consumption goods produced in section II with workers of this last section. This distribution function of money plays such an essential role that it might be worth considering it as a fourth function of money.

Thirdly, and as a consequence of what precedes, one must realize that, by interpreting the assumptions of this one-step-further model (without private property of fixed capital) as meaning that there are no capitalists in this system, we arrive at a situation where there are also no true profits, because there is no need for them: profits would materialize only in case of private property of fixed capital, i.e. if there were no collective property of this kind of goods.

To continue with the examination of the operation of this model without capitalists, we can imagine a way to materialize the fact that workers of both sections, once having shared all of the consumption goods produced, would also be the collective owners of the fixed capital produced in section I. As we have seen previously, since the price of all consumption goods is equal to their value $L_{II}$ multiplied by $1 + k$, the simplest way would be to deliver to the buyer of any consumption good a certificate representing the ownership of fixed capital, for an amount equivalent to the price of this good multiplied by $\frac{k}{1+k}$ . Even more simply, the receipt delivered for each purchase could mention this amount, for instance together with the amount of value-added tax (VAT) when such a tax exists: this VAT might itself be seen as a kind of collective contribution to the State budget, involving the right to enjoy public services provided by it.

It would nevertheless be understood that a certificate issued on the occasion of the purchase of whatever consumption good would not be exchangeable against money or any other good. Indeed it would just be a symbol of sharing the collective property of the new fixed capital produced during the period during which this consumption good has also been produced. Moreover, because there are no capitalists in this system, as we already pointed out, there would also be no profits, which means that the property of fixed capital would not give the right to sell this type of good and more importantly the right to derive a profit from its ownership. But, apart from bare ownership, it would have obviously kept its even more important feature, i.e. the right to use fixed capital, in the present case for the production of commodities (we shall come back below to these various attributes of property).

### 5.2. The simplified model and money creation

An other important characteristic of this simplified model has to do with money creation. Continuing to assume that there are no capitalists and that at each period production starts from scratch, this production could not be sold if the corresponding wages had not been paid previously, and wages could not be paid if money had not been obtained from the institutions which create it, i.e. banks: in other words the public entity in charge of managing the productive system needs to borrow from them in order to pay for these wages. The amount of money which thus needs to be borrowed, in order to be injected in the circuit through the payment of wages, corresponds precisely to the overall amount of these wages, i.e. $W_I + W_{II} = L_I + L_{II}$ (with $\bar{w}=1$). As for the repayment of this amount, it poses no problem as long as the collective property of fixed capital allows to consider the whole productive system as a unique giant integrated enterprise, owning both sections of production.

However, if each one of the two sections had a kind of legal autonomy and distinct legal personality, meaning that the collective property of commodities would be split between two sets of workers with two separate entities managing each section (in accordance with the structuration of the productive system between these two sections), there would undoubtedly be a problem, because a distinct amount of money $W_I = L_I$ would have been borrowed by section I, but would accrue to section II through the purchase of consumption goods by workers of section I. In order to get this amount back and repay the borrowed amount of money, section I would thus have to sell fixed capital to section II for a similar amount $W_I$. Then $W_I$ would appear as a kind of virtual profit, corresponding exactly to the sale of fixed capital, for the part of it needed by section II, and which would allow for the repayment of the same amount of money initially borrowed in order to pay for the wages of section I.

This shows us that it is the dissociation of property (be it collective or not) which creates the logical need for a transfer of income from wages, which are the original or primary income. It equally implies the differentiation of price and value also in section I, and the need for the price of fixed capital to be higher than its value, like in the case of consumption goods. Indeed the price $W_I = L_I$ of the fixed capital goods bought by section II would necessarily be higher than their value, since these fixed capital goods would represent not the whole production of fixed capital, but only the part $(1-a)L_I$ of fixed capital needed by and used in section II, and not the remaining part $aL_I$ of fixed capital needed by section I itself. The fixed capital corresponding to this last part would not be sold: because it has only one collective owner, it would stay in section I where it has been produced, and therefore neither a particular price nor further virtual profits would be involved in its distribution.

However, if section I itself were divided into several legal entities corresponding to the technical structuration of the production process between the different kinds of fixed capital needed at each of its stages, then transactions to sell and buy this fixed capital between these separate entities would have to take place, and would involve new amounts of virtual profits. It is this process that we have described earlier in this chapter, when we examined the realization of profits in section I. We showed above (see equation (16) p. 213) that in this case the total amount of (virtual) profits in section I would be $\Pi_I = L_I \frac{a}{u}$.

### 5.3. Surplus-value and monetary profits

The simplified model exposed in this chapter shows therefore that whatever the type of property of fixed capital (individual or collective) it is the existence of multiple entities, each being entitled to a property right on its fixed capital and its products - which logically implies the existence of successive transfers of income originating from wages. Moreover these entities can be linked or not to the technical structuration of the productive system, because there can be several entities for a particular technical stage, but this technical aspect is not what only matters. Incomes resulting from transfers occurring between these entities are logically called profits, since their origin is linked to the sale and purchase of commodities through monetary prices higher than their values, even though profits in this simplified model are to some extent somewhat virtual, as long as it is interpreted as a model without capitalists.

Indeed, if we abandon the idea of a collective property, and go back to the case of individual property, implying that we reinstate the existence of capitalists, let us just consider that this property covers the right to perceive money profits from the sale of products: we shall go back in the next subsection on the various attributes of property. Nevertheless, and by construction in this model, profits cannot be used to buy consumption goods, although this is clearly the ultimate goal of any normal individual, even he is a capitalist. As for the amount of these profits, apart from profits made through the sale of consumption goods (which in the simplified model only correspond to purchases made from wages), this amount depends entirely only on the structuration of the productive system in section I: it is in fact the degree of integration of production in this section (measured by the degree of property fragmentation) which determines its precise amount.

In other words, for a given amount of surplus value and profits in section II, realized in this section when workers buy consumption goods, the amount of monetary profits depends on the structure of property in section I, which is overlapping on the technical structure of production in this section. It is indeed easy to understand that even in a capitalist system, if the fixed capital of section I is entirely produced by only one huge capitalist enterprise, then the total amount of monetary profits for the whole productive system will simply be $L_I = kL_{II}$ , i.e. only monetary profits obtained in section II, because in section I the income $L_I = W_I$ earned from the sale of fixed capital to section II at this price $L_I$ will be totally offset by the repayment of the total amount of wages which have been paid in section I.

On the contrary, if capitalist property in section I is split among at least as many private companies as there are stages in the production process in this section, then the amount of profits in this section will be $\Pi_I = L_I \frac{a}{u}$ , as indicated above.

### 5.4. The simplified model and profits as a transfer income

Finally, this demonstration sheds some light on a mechanism which is at work behind the formation of all incomes created through a transfer occurring in the sphere of circulation, and by which the expense of an income creates a new income. We will now expand a little on this interesting mechanism, as far as profits are concerned.

It should first be noted that, as regards the realization of fixed capital produced in section I, fixed capital exchanges are "successive" only for the sake of clarity. This succession is logical and not chronological, because these exchanges can perfectly be simultaneous. There is no need to refer to a particular period or to successive periods, in particular because of the existence of money and credit: during each period, however short it may be, and in fact at any moment, capitalists of section I sell the needed quantity of fixed capital to those of section II, and sell to one another in section I the quantity of fixed capital corresponding to their own needs. It is important to note that it is this mechanism based on the spending of profits that does make it possible to produce fixed capital goods in section I.

In this mechanism of successive spending of profits in section I that was exposed above in section 3, we thus find, but with a different method, a result which had been well brought to light by Bernard Schmitt, in his book “Monnaie, salaires et profits” (“Money, wages and profits”), which we cited previously, on the basis of a theory of money and monetary circuit in which profit is a transfer income. However Schmitt’s theory does not explain the modalities of this transfer through a reproduction scheme, as we did, for it is an axiomatic theory of money, founded on its circuit, and not a complete theory of production and distribution. In the model developed here, on the contrary, we see clearly that profits are the result of a double transfer:

- A first transfer, which takes place on the occasion of the expenditure of wages by workers of the two sections of production, for the purchase of consumer goods from capitalists of section II. It is through this first transfer that surplus-value is levied, once and for all, under the form of monetary profits, and that its amount is fixed, and is no longer liable to change. In this simplified model, this transfer simultaneously sets the amount $L_I$ of monetary profits for capitalists of section II.

- A second transfer, of a very different nature, since it only involves capitalists, and which takes place on the occasion of the purchase by capitalists of section II, from those of section I, of the fixed capital that they need to produce consumer goods. This purchase has no influence on the amount of surplus-value, which has already been determined, but it fixes the distribution of this total surplus-value $L_I$ between capitalists of the two sections. The share accruing to capitalists of section II is $(1-a)L_I = bL_I$, as we saw in the previous section, and that accruing to the capitalists of section I is $aL_I$, which also corresponds to the amount of monetary profits going to the capitalists who ensured this sale. This first transfer to section I will then be followed by (logically) successive transfers within section I, transfers which will ensure the distribution of this surplus-value $aL_I$ and of the corresponding profits between the capitalists of this same section.

These observations provide us with the opportunity to go a little further as regards the difference between primary incomes and secondary or transfer incomes:

As regards primary incomes, they are such because they are paid and earned on the occasion of the production of commodities, which logically precedes their circulation. In our model where all labor is made of wage-labor, wages are for this reason the only primary incomes. It is so because the payment of wages is indeed a necessary and sufficient condition for production to take place. A necessary condition, because no production can be carried out if workers are not previously hired and their wages specified: these wages are earned at the same pace as production and paid according to a specified periodicity. Thus, any amount of wages corresponds to a period during which work has been performed and production has been carried out. We can note in passing that this period can sometimes be even shorter than the time needed for the resulting commodities to be brought to the market.

The payment of wages is also a sufficient condition for production to take place because once workers are hired and at work, no distribution of (new) profit is needed for production to be carried out: this might seem difficult to understand in a disaggregated economy, but becomes obvious if we imagine that the productive system is made of a single fully integrated firm, producing both consumption goods and all of the intermediate and fixed capital goods needed for the production of the former. This means that no purchases involving the payment of profits have to be made by this single firm from other firms. This observation confirms and reinforces the conclusion that we had already reached above (see subsection 6.4. p. 191), and which is the nullity of profits at the stage of production.

Finally, wages are linked to production by their very nature: as we have seen they have the dimension of a scalar quantity per unit of time, whereas the value of what is produced has itself the dimension of time: it is the payment of wages which transforms this value in money value and reciprocally, by a process of codetermination.

Conversely, as regards profits, they can only form at the level of distribution of the product, if there is a first money transfer from workers to capitalists of section II, when these workers spend their wages to buy consumption goods, and this at a monetary price which is necessarily higher than the money value of these same goods. As for profits realized by capitalists of section I producing fixed capital goods, they will never be realized if there is not a second money transfer, i.e. if capitalists of section II do not invest in the purchase of fixed capital.

This mechanism of realization of the product and profits in section I brings us back to a principle first highlighted by Keynes, who introduced it in his “Treatise on Money”, published in 1930, six years before his “General Theory…”. It is the principle of the "widow's cruse", according to which it is the expense of profits which creates profits. For Keynes, however, this principle must have not seemed so important because he writes: “There is one peculiarity of profits (or losses) that we may note *in passing*, because it is one of the reasons to segregate them from income proper, as a category apart. If entrepreneurs choose to spend a portion of their profits on consumption [...] the effect is to increase the profit on the sale of liquid consumption goods by an amount exactly equal to the amount of profits which have been thus expended [...] Thus, however, much of profits entrepreneurs spend on consumption, the increment of wealth belonging to the entrepreneurs remains the same as before. Thus, profits, as a source of capital increment for entrepreneurs, are a widow's cruse which remains un-depleted, however much of them may be devoted to riotous living" (Keynes, 1930, chapter 10, p. 139).

However our simplified model shows that the mechanism that Keynes considered as a "peculiarity" is not such a particular device and must on the contrary be generalized to explain the realization of profits, which at the same time sheds lights on their nature as a transfer income. Moreover, when Keynes described this - in fact fundamental - mechanism of the widow's cruse, he was referring to the spending of profits to buy consumption goods, whereas we have demonstrated that this mechanism is also at work even when there is no capitalist consumption. In any case Keynes drew our attention to what we think is a neglected but similarly fundamental element of the system: the consumption of capitalists. This element is in our view of such importance that we will now devote to it the full chapter which follows.

However, before addressing this question, we can go a little further in the understanding of profits, by asking the question of their nature, i.e. the reasons that make it possible for capitalists to realize them through a transfer from primary incomes. The answer is quite simple, because it is clear that this capacity to realize and keep profits is due to the capitalists' ability to fix prices at a level which is higher than wages, i.e. the primary incomes which they have payed to the workers that they employ, such a level that it allows them to levy their profits out of the amount of their sales. This capacity to fix prices derives itself from the fact that capitalists are the owners of the commodities which they sell on the market.

We are here at the heart of the system, this heart being the social institution of property, and more precisely private property. This fundamental institution is defined by law, and in capitalist societies it has kept all the main characteristics that it had under Roman law, according to which property rights are divided in three categories: usus, fructus and abusus.

- Usus contains the rights to use the thing that is owned, such as inhabiting a house or an apartment. In the case of the capitalist system property of fixed capital gives the right to put it at work with hired workers.

- Fructus contains the right regarding the products of a property, such as the fruits of a tree or the crops gathered from a piece of land, or in the capitalist system the commodities produced by any production process involving fixed capital, because this capital, as much as labor, is supposed to be productive in itself.

- Abusus contains the right to dispose of a property, such as selling, modifying, or even destroying it, which allows capitalists to sell obviously fixed capital itself, but also all commodities coming from the production process, since - as a part of fructus, they are the property of the producer.

All incomes are created as money incomes, to begin with wages, and all transfer incomes exist because they result from a transfer of property, i.e. the transfer of the property of some commodity from the seller to the buyer in exchange for money. It is important to understand that such a transfer can occur only because the commodity which is transferred pre-exists - as the property of the seller - to this exchange against a specified amount of money, this amount being the monetary price of the transferred commodity. In fact there would not be any exchange if property did not pre-exist: property is a precondition of exchange. This explains that in the neoclassical fairy tale, there must be an endowment of commodities to individuals, before they go to the market, even though no production has taken place previously, a situation which defies the most basic logic.

This has a very profound consequence, because if an exchanged commodity already exists, it means ipso facto that it has already been produced previously, and that these exchanges by themselves do not create anything or any additional value. This explains why these transferred incomes cannot produce themselves anything or create any value, since value already exists: it has been created earlier as a dimension of commodities resulting from the production process itself. Therefore these exchanges cannot do anything else than shifting the ownership of pre-existing commodities by moving them from the seller to the buyer in opposite directions to pre-existing monetary income. This is the reason why these transferred incomes cannot therefore be anything else than a redistribution of pre-existing primary incomes, i.e. wages in this simplified model.

If we took into account what Engels called “simple commodity production”, then incomes from self-employed producers, corresponding to the sale of commodities just produced from their own labor, should also be considered as primary incomes. We concur indeed with Engels, who writes: “In a word: the Marxian law of value holds generally, as far as economic laws are valid at all, for the whole period of simple commodity production” (Engels, 1894, supplement to Capital, Vol. 3, Law of value and rate of profit, on-line version).

It follows that apart from wages, which are revenues paid for buying workers labor, which allows to use this labor in the production process, or incomes of self-employed producers, all other revenues appear in the sphere of circulation, where all exchanges of commodities take place, and are revenues from property, be they profits from the property of fixed capital, rents from the property of housing, interests from the property of money assets, or amounts of money earned from the sale of existing assets (real ones, like lands, real estate, second-hand goods, or financial ones, like companies shares or bonds).

At the end of this chapter, it is now time to summarize the main lessons through some principles that we can derive from all these observations, firstly on surplus-value, and secondly on incomes.

**Principle 19**: In the capitalist mode of production, fixed capital is by definition the property of capitalist but not of workers, who own neither the means of production nor the commodities which they create through their labor. The simple reproduction of this system as it is implies therefore that workers cannot get on the market, from the spending of their wages, the property of at least the part of the product consisting of means of production. The difference between the whole product of labor and the share obtained by workers is defined as surplus-value.

**Principle 20**: Surplus-value is created in the sphere of production, but extracted in the sphere of circulation, where commodities are distributed among various stakeholders. This results from the fact that wages, as the money price of labor-power representing the money value of the whole product, can buy only a share of this product consisting in consumption goods, whose price is always higher than their money value. Surplus-value thus materializes under its monetary form as profits in the section producing consumption goods. These profits are then distributed among capitalists.

**Principle 21**: The only primary incomes are the revenues created on the occasion of the production process, and are the monetary counterpart or remuneration of labor, i.e. wages in a pure capitalist mode of production.

**Principle 22**: Profits, as any income other than wages or the remuneration of labor, derive from the property of commodities or assets. They are transfer revenues which arise from exchanges against money of previously produced commodities (or assets) taking place in the circulation process.

# Chapter 12. A more realistic model integrating capitalist consumption

In this new model it will be shown that taking into account capitalist consumption not only brings about a much more realistic view of the functioning of the capitalist system, but also sheds light on a very important element of this system as far as production and distribution are concerned. This will be done by exploring first the consequences on some key variables of this introduction of capitalist consumption, as a new and fundamental element of the model. The implications of this important modification on the realization of the product and profits in both sections of the productive system will be discussed as a second step.

## 1. The main changes to the initial simple model

Let us first introduce the consumption of capitalists, that we call $C_c$, in a system which is still in a state of simple reproduction. This consumption is of course and by nature made only by capitalists as individuals, since they benefit in one way or another (dividends, etc.) from the distribution of profits made by firms (because these firms by definition do not buy any consumption goods). We assume that this consumption represents a constant fraction, called *c*, of the $L_{II}$ value of consumer goods produced in each period, with ($0 < c < 1$, and $\bar{w} = 1$).

We thus have $c = \dfrac{C_c}{L_{II}}$ and therefore capitalist consumption is by definition: $C_c = c\, L_{II}$ (1)

What then becomes the system of values? In this context, the value of the product of each section (let us name these products $Y_I$ and $Y_{II}$ respectively) remains unchanged at $L_I$ for section I, $L_{II}$ for section II and $L_I + L_{II}$ for *Y*, i.e. the total product. Moreover, by definition, the amount of the total surplus-value S levied on workers increases correspondingly from $L_I$ to $L_I + cL_{II}$, which is equivalent to $k\, L_{II} + c\, L_{II}$, i.e. $L_{II}\,(k + c)$, and the value of the total labor-power decreases in proportion as it passes from $L_{II}$ to $L_{II}\,(1 - c)$. As for the value of the accumulated capital at each period $I_t$ (simply called *I* in a situation of simple reproduction) it is still equal to $L_I$. Let us also call $S = L_{II}\,(k + c)$ the new surplus-value resulting from these new values, and *V* the new value of the labor-power which corresponds to it. Let us call $k_c$ the new rate of surplus value, which is thus written:

$$k_c = \frac{S}{V} = \frac{L_{II}\left(k+c\right)}{L_{II}(1-c)} = \frac{k+c}{1-c} \tag{2}$$

$$\Rightarrow S = Vk_c = V\,\frac{k+c}{1-c} \tag{3}$$

Knowing that the rate of surplus-value $k_c$ is by definition the same in the two sections, and that the value of the product of each section is by definition equal to the value of labor-power plus the surplus-value, $L_I = V_I$ (1+ $k_c$) and $L_{II} = V_{II}$ (1+ $k_c$). The value of labor-power is immediately derived from that:

- in section I: $V_I = \dfrac{L_I}{1+k_c} = \dfrac{L_I}{1+\dfrac{k+c}{1-c}} = \dfrac{L_I}{\dfrac{1+k}{1-c}} = \dfrac{L_I(1-c)}{1+k}$ (4)
- and from a similar calculation in section II: $V_{II} = \dfrac{L_{II}}{1+k_c} = \dfrac{L_{II}(1-c)}{1+k}$ (5)

It can easily be verified that the sum of these two values is equal to the overall value $V$ of labor-power:

$$V_I + V_{II} = \frac{L_I}{1+k_c} + \frac{L_{II}}{1+k_c} = \frac{kL_{II}}{1+\frac{k+c}{1-c}} + \frac{L_{II}}{1+\frac{k+c}{1-c}} = \frac{L_{II}(1+k)}{\frac{1+k}{1-c}} = L_{II}(1-c) = V \quad (6)$$

Surplus-values $S_I$ and $S_{II}$ in each section are also easily obtained, since it is by definition equal to the value of labor-power multiplied by the rate of surplus-value $k_c$ (see equation 2).

In section I, surplus-value $S_I$ is therefore equal to: $k_c \dfrac{L_I}{1+k_c} = \dfrac{L_I \frac{k+c}{1-c}}{1+\frac{k+c}{1-c}} = L_I \dfrac{k+c}{1+k}$ (7)

In section II, the calculation is similar, with $S_{II}$ equal to: $k_c \dfrac{L_{II}}{1+k_c} = L_{II} \dfrac{k+c}{1+k}$ (8)

It can also be verified that their sum is equal to total surplus-value, as previously defined:

$$S = S_I + S_{II} = L_I \frac{k+c}{1+k} + L_{II} \frac{k+c}{1+k} = kL_{II} \frac{k+c}{1+k} + L_{II} \frac{k+c}{1+k} = (k+1) L_{II} \frac{k+c}{1+k} = L_{II}(k+c) \quad (9)$$

# 2. Realization of the product and profits in a model with capitalists consumption

## 2.1. Realization of consumption goods and profits in section II

As we saw previously, for the realization of the product to take place, the level of prices must be such as to allow this realization. The difference that is now introduced, in comparison to the simple model previously exposed, is that the realization of consumer goods of a money value $L_{II}$ must now take into account the consumption of capitalists, and therefore the fact that workers can no longer buy the totality, but only a fraction (1 - $c$) of this value $L_{II}$.

Let us first recall that in the simplified model without capitalist consumption, workers produce final goods with an $L_I + L_{II}$ value, for which they receive, in monetary units, wages amounting to $W = L_I + L_{II}$ (with $\bar{w} = 1$). Then when they spend these wages they obtain in return consumer goods for only an $L_{II}$ value, corresponding to the value of labor-power, since in a simple reproduction situation the price of these consumption goods is fixed at $L_I + L_{II}$, or $L_{II}$ $(1 + k)$ (see above). This is the simple mechanism by which surplus-value is levied. The profit realized in this section is then $L_I = kL_{II}$. By calling $s_{II}$ the ratio between these profits in section II and the price of labor in the same section, it is, as we have seen, identical to the rate of total surplus-value:

$$s_{II} = \frac{kL_{II}}{L_{II}} = k \qquad (10)$$

In the present and more complex model of simple reproduction with capitalist consumption, the reduction in the value of workers consumption to (1 - $c$) $L_{II}$ does not in any way imply a fall in the price of labor, which remains fixed at $L_I + L_{II} = (1 + k) L_{II}$ (i.e. the amount of wages with $\bar{w} = 1$).

In addition, we know that: $\frac{1}{1-c} = \frac{c}{1-c} + \frac{1-c}{1-c} = \frac{c}{1-c} + 1 \Rightarrow \frac{c}{1-c} = \frac{1}{1-c} - 1$

Moreover the rate of surplus value $k_c$ can be expressed as $k_c = \frac{(1+k)-(1-c)}{1-c} = \frac{1+k}{1-c} - 1$

This gives us conversely: $\frac{1+k}{1-c} = 1 + k_c$ (11)

For these wages which stay at an amount of $(1 + k) L_{II}$ to buy only a fraction (1 - c) of the consumer goods, it is therefore necessary and sufficient that consumer goods of a money value equal to $L_{II}$ be sold at a price now increased in such a way that it passes from:

$W = P_{II} = (1 + k) L_{II}$, as they were in the previous and simple model, to:

$$P_{II} = \frac{1}{1-c} P_{II} = L_{II} \frac{1+k}{1-c}, \text{ or } P_{II} = L_{II}\left(1+k_c\right), \qquad (12)$$

This price $P_{II}$ is such that it gives the possibility to realize in this section II an additional monetary profit equal to this capitalist consumption, i.e.:

$$C_c = cL_{II} \cdot \frac{1+k}{1-c} = cL_{II}\left(1+k_c\right) \qquad (13)$$

It is indeed easy to verify that this price $P_{II}$ makes it possible to realize the total $L_{II}$ value of consumer goods, since by adding the consumption of workers $C_w = W$ and that of capitalists, one obtains the overall price of consumer goods. Indeed, since $W = L_I + L_{II} = L_{II}(1 + k)$, we can see immediately that:

$C_w + C_c = W + Cc = L_{II}(1 + k) + c\, L_{II} \cdot \frac{1+k}{1-c}$, which is equal to:

$$W + Cc = (1\text{- c})\, L_{II} \cdot \frac{1+k}{1-c} + c.L_{II}\frac{1+k}{1-c} \;=\; L_{II} \cdot \frac{1+k}{1-c} = L_{II}\left(\frac{k+c}{1-c}+1\right) = L_{II}\left(1+k_c\right) = P_{II} \qquad (14)$$

From this last equation, we can consequently derive that workers obtain a fraction (1 - *c*) $L_{II}$ of the (money) value $L_{II}$ of consumer goods whose overall price is $P_{II} = L_{II} \cdot \frac{1+k}{1-c}$, with capitalists of section II obtaining the remaining $c\, L_{II}$ fraction of this same value.

The price for these workers' consumption goods is: $\left(1-c\right)L_{II} \cdot \frac{1+k}{1-c} = \left(1-c\right)L_{II}\left(1+k_c\right)$ (15)

While the price paid by capitalists for their own consumption goods is: $cL_{II}\frac{1+k}{1-c}$ (16)

When we pass from values to prices, we may ask what happens then to surplus-value, which we know [from equation (9)] that it is equal in this new model to $L_{II}\ (k + c)$, and to the rate of surplus-value $k_c$, which is known to be given by equation (2), and to be equal to $\frac{k+c}{1-c}$.

Within the framework of this model, when one ceases to reason in money value in order to pass to prices, we pass also by the same token from surplus-value to profits. Let us call $\Pi_{II}$ the profits realized in the consumer goods section. The profits realized here go from $\Pi_{II} = L_I = k\ L_{II}$, as they were in the simple model, to:

$$\Pi_{II} = kL_{II} + cL_{II} \cdot \frac{1+k}{1-c} = L_{II}\left[k + c\left(\frac{1+k}{1-c}\right)\right] = L_{II}\left[\frac{k\left(1-c\right)}{1-c} + \frac{c\left(1+k\right)}{1-c}\right] =$$

$$\Pi_{II} = L_{II}\left(\frac{k+c}{1-c}\right) = L_{II}k_c \qquad (17)$$

It must be noted that it is different from total surplus value $L_{II}\left(k+c\right)$ given by equation (9).

Finally, keeping in mind equation (2) above, if we call $\pi_{II}$ the ratio between the profits made in the consumer goods section and the price of labor in the same section, it goes from $\pi_{II} = k$ to:

$$\pi_{II} = \frac{\Pi_{II}}{W_{II}} = \frac{L_{II}\left(\frac{k+c}{1-c}\right)}{L_{II}} = \frac{k+c}{1-c} = k_c \qquad (18)$$

This shows us that in the consumer goods section, when we reason in terms of prices, the new profit-to-wage ratio remains equivalent to the new rate of overall surplus value $k_C$, exactly as it was in the simple model of chapter 11.

We will need to come back to the realization of consumption goods and profits in section II, but before doing that, assuming temporarily that this realization has taken place, we will examine the realization of fixed capital goods and profits in section I, and make some calculation regarding the amount of a few variables reflecting the distribution of the whole product. These two steps will indeed make it easier to address again the question of realization in section II.

### 2.2. Realization of fixed capital goods and profits in section I

What about section I, which produces fixed capital? Before going further, let us recall that $a$ remains the fraction of the total value of fixed capital accumulated in section I, and b the fraction accumulated in section II, with as a consequence $a+b=1$. First of all, we will deal with the realization of fixed capital sold to capitalists of section II.

#### 2.2.1. Fixed capital and profits realized from the sales to section II

We know that the realization of the product implies that capitalists of section II must obtain, in exchange for their profits, equal to $L_{II}\left(\frac{k+c}{1-c}\right)=L_{II}k_c$ and levied on the workers' wages, first only a fraction equal to $(1-a)L_I=bL_I$ of the money value $L_I$ of the newly produced means of production, and secondly a part of consumption goods. But because of similar conditions (which could result from competition) between capitalists of both sections, it seems logical also that capitalists of section I should obtain in addition a part of the total $cL_{II}$ money value of consumer goods accruing globally to capitalists. There is indeed no particular reason why the consumption of capitalists should be entirely reserved for those of section II, while those of section I would not have access to consumer goods! If we call $c_{II}$ the share of total consumption which will be obtained by capitalists of section II, and $c_I$ the share of total consumption which will be obtained by capitalists of section I, then we have:

$$c=c_I+c_{II} \Leftrightarrow c_I=c-c_{II} \Leftrightarrow c_{II}=c-c_I \tag{19}$$

From this we can derive the absolute price of capitalist consumption in section I and section II, which are respectively:

1) For section I, $C_I=c_IL_{II}(1+k_c)$ or, since $L_{II}=\frac{L_I}{k}$, we have also $C_I=c_I\frac{L_I}{k}(1+k_c)$ (20)

2) For section II, $C_{II}=c_{II}L_{II}(1+k_c)=c_{II}\frac{L_I}{k}(1+k_c)$ (21)

The level of this share of consumption which goes to capitalists of section I is not predetermined, but presupposes that they must realize accordingly a profit allocated to this consumption, and which depends on the price level at which they will themselves sell first to capitalists of section II the fixed capital which the latter need, in order to implement the production of consumer goods. It is only once the price of this fixed capital is determined and its value realized through the sale to capitalists of section II that the share of consumption of capitalists of section I in total consumption can be determined, albeit not immediately.

Firstly, these capitalists of section I must indeed deduct from their sale price the amount of wages that they have to pay to their workers, corresponding to the money value of the fixed capital which they sell to capitalists of section II, i.e. $(1-a)L_I$. The remaining amount is their gross profit, but it is not yet the profit available for the purchase of consumption goods, because they must also deduct from it the amount of their own purchase of fixed capital.

Secondly, after the sale of fixed capital of a money value of $bL_I = (1-a)L_I$, the available quantity of fixed capital has by construction a money value of $aL_I$. With the assumption that they would buy in turn, like the capitalists of section II, a fraction $b = (1-a)$ of this remaining fixed capital of a money value equal to $aL_I$, i.e. a fraction $a(1-a)L_I$, in a situation of simple reproduction, the lowest price that they could pay for this fixed capital, if the sellers did not make a profit, would be $a(1-a)L_I$. Of course this cannot be the case: therefore at first, before the purchase of this fixed capital has been made we can only know what would be the maximum amount of the difference between their gross profit and the price of the fixed capital that they buy, which is equivalent to their maximum potential consumption.

From these observations and findings we can draw an important conclusion, which is that each particular share of capitalist consumption is always a residue, depending on the price paid for the purchase of fixed capital. Firstly, the profits of capitalists of section II are a residue: their first component is the part transferred from the spending of total wages, once capitalists of section II have paid their own section II wages. The amount of this residue is decided by them when they set the price of consumption goods that they sell. Their second component comes from the sale of the remaining consumption goods to capitalists of both sections (since their own wages have already and necessarily been recovered, because it is obvious that $L_I + L_{II} > L_{II}$ ). Globally, however, the difference between the total price of consumption goods and total wages (equivalent to the price of workers consumption) is by construction equivalent to the price of consumption goods bought by capitalist of both sections, whose overall money value is simultaneously determined, as well therefore as the share of all capitalists in total consumption. Thus capitalist consumption as a whole can therefore be considered as a residue, its amount being simply:

$P_{II} - W = L_{II}(1+k_c) - (L_I + L_{II}) = L_{II}(1+k_c) - L_{II}(1+k) = L_{II}(k_c - k)$, which is equal to:

$$P_{II} - W = L_{II}\left(\frac{k+c}{1-c} - k\right) = L_{II}\left(\frac{(k+c) - k(1-c)}{1-c}\right) = L_{II}\left(\frac{c(k+1)}{1-c}\right) \Rightarrow$$

$$P_{II} - W = cL_{II}\frac{k+1}{1-c} = cL_{II}(1+k_c) \qquad (22)$$

It is easy to check that this last equation in fact gives us the exact amount of $C_c$ which we had already derived from equation (13).

Going back to profits realized by capitalists of section II, they have first to use them to buy fixed capital from capitalists of section I, otherwise simple reproduction could not be realized. And it is only the difference between these profits in section II and the price of fixed capital bought from section I which can be spent for the purchase of consumption goods. It follows that this consumption of capitalists of section II cannot but be considered itself as a residue.

Moreover the distribution of these profits of section II as a residual income, first between capitalists of both sections, and secondly among capitalists of section I, depends on the various prices charged to the buyers of fixed capital. Since it is these residual profits, when wages have been paid and fixed capital bought, that are used for buying consumption goods at a pre-determined price, it follows that capitalist consumption at each stage is always a residue.

We are now going to produce a table, given below in **2.2.2.**, displaying all the corresponding prices and values. In order to do that we shall first trace the various transactions of these capitalists of section I who sell fixed capital to those of section II. Capitalists producing in section I the fixed capital $I_{II}$ destined for the production of consumer goods must themselves levy on the occasion of the sale of this fixed capital to those of section II, not only the remaining fraction $aL_I$ of the value $L_I$ (that will enable them to obtain the fraction of the fixed capital $I_I$ which they need to invest), but also a fraction of the $L_{II}$ money value of consumer goods. How can they do that?

In fact this comes down to a common question of margin levying: we know that to take a margin $a$ as a fraction (or percentage) off and within a variable $y$, for instance a sale price, is mathematically equivalent to writing: $y = ay + x$, with variable $x$ being then considered as the corresponding cost price. This identity in turn implies:

$$y = ay + x \Rightarrow y(1-a) = x \Rightarrow y = \frac{x}{1-a} \Rightarrow y = x\left(1+\frac{a}{1-a}\right) \Rightarrow \frac{y}{x} = 1+\frac{a}{1-a}$$

This means simply that to levy a margin $a$ within a sale price it is necessary and sufficient to levy a markup $\frac{a}{1-a}$ over the corresponding cost price.

To apply this finding to capitalist consumption in section I, let us call $c^*$ a parameter corresponding to the ratio of consumption of all capitalists of section I which can be obtained through selling to capitalists of section II at a price $\frac{1}{1-c^*}$. The markup will thus be $\frac{c^*}{1-c^*}$.

Then we can recall that:

$$\frac{1}{1-c^*} = \frac{c^*}{1-c^*} + \frac{1-c^*}{1-c^*} \Leftrightarrow \frac{1}{1-c^*} = \frac{c^*}{1-c^*} + 1$$

Moreover it is trivial that $\frac{\frac{c*}{1-c*}}{\frac{1}{1-c*}} = c*$

All this implies that a fraction $c*$, and thus a quantity $\frac{c*}{1-c*}L_I$ of the price of fixed capital $\frac{1}{1-c*}L_I$ can be obtained when these means of production, which here have a value $bL_I = (1-a)L_I$, by definition, are sold to capitalists of section II at a price $\frac{L_I}{(1-c*)}$ which by construction is equal to $\frac{1}{(1-a)\cdot(1-c*)}$ times their money value $bL_I$, because this value is equal by construction to $(1-a)L_I$. Indeed we know that $\frac{1}{1-a} = \left(1+\frac{a}{1-a}\right) = \frac{1-a}{1-a} + \frac{a}{1-a}$

This means obviously that:

$$\frac{1}{1-a}L_I = \left(1+\frac{a}{1-a}\right)L_I = \frac{1-a}{1-a}L_I + \frac{a}{1-a}L_I = L_I + \frac{a}{1-a}L_I \qquad (23)$$

And multiplying by $(1-a)L_I$, we get: $(1-a)L_I\frac{1}{1-a} = (1-a)L_I\left(1+\frac{a}{1-a}\right) = (1-a)L_I + aL_I$

Since $\frac{(1-a)L_I}{(1-a)\cdot(1-c*)}$ is obviously equal to $\frac{L_I}{1-c*}$, this price is also equal to:

$$\frac{L_I}{1-c*} = \frac{L_I c*}{1-c*} + \frac{L_I(1-c*)}{1-c*} = L_I\left(\frac{c*}{1-c*}+1\right) = L_I + L_I\frac{c*}{1-c*} \qquad (24)$$

In passing, we can develop the general formula giving the ratio of price over value from these last equations (23) and (24), for a case where there are two margins, i.e. $a$ and $c*$ by writing:

$$P = L\frac{1}{(1-a)(1-c*)} = L\left[\left(1+\frac{a}{1-a}\right)\left(1+\frac{c*}{1-c*}\right)\right] =$$

$$L\left(1+\frac{a}{1-a}+\frac{c*}{1-c*}+\frac{ac*}{(1-a)(1-c*)}\right) = \left(1+\frac{a(1-c*)+c*(1-a)+ac*}{(1-a)(1-c*)}\right)$$, which is equal to:

$$P = L\left(1+\frac{a-ac*+c*-ac*+ac*}{(1-a)(1-c*)}\right) = L\left(1+\frac{a(1-c*)+c*}{(1-a)(1-c*)}\right) \qquad (25)$$

The corresponding labor price (wages) is by definition $bL_I = (1-a)L_I$, and the gross profits levied by capitalists of section I who sell to those of section II are therefore, by difference between the price of fixed capital sold to section II, i.e. $\frac{L_I}{(1-c^*)}$, and wages $(1-a)L_I$ :

$$\frac{L_I}{1-c^*} - (1-a)L_I = L_I\left[1+\frac{c^*}{1-c^*}\right] - (1-a)L_I = L_I - (1-a)L_I + L_I\frac{c^*}{1-c^*} =$$

$$aL_I + L_I\frac{c^*}{1-c^*} = L_I\left(a+\frac{c^*}{1-c^*}\right) \qquad (26)$$

It is easy to verify that the sum of these last two terms, i.e. the price of labor $(1-a)L_I$ and the gross profits, i.e. $L_I\left(a+\frac{c^*}{1-c^*}\right)$, is indeed equal to the price of the product, i.e. the fixed capital $I_{II}$ of a money value $(1-a)L_I$ sold by capitalists of section I to those of section II:

$$(1-a)L_I + L_I\left(a+\frac{c^*}{1-c^*}\right) = L_I + \frac{L_I c^*}{1-c^*} = L_I\left(1+\frac{c^*}{1-c^*}\right) = \frac{L_I}{(1-c^*)} \qquad (27)$$

From this last equation we can derive the amount of profits of capitalists of section I who sell to those of section II. Since they come as a deduction of the profits available for consumption in section II, their amount is in fact equivalent to the overall consumption of all capitalists of section I. This allows us first to calculate the value of parameter $c_I$, i.e. the share of capitalists of section I in total consumption. This can be done in terms of $c^*$, which is the gross margin rate, and as such a given parameter, a ratio whose value is decided by capitalists of section I selling fixed capital to those of section II. In terms of prices $c_I$ is such that:

$$L_I\frac{c^*}{1-c^*} = c_I L_{II}(1+k_c) = c_I\frac{L_I}{k}(1+k_c) \Rightarrow$$

$$\frac{c^*}{1-c^*} = \frac{c_I}{k}(1+k_c) \Rightarrow c^* = (1-c^*)c_I\frac{1+k}{k(1-c)}$$

With $\gamma = \frac{c^*}{1-c^*}$, we get from this last equation: $c_I = \frac{c^*}{1-c^*}\cdot\frac{k(1-c)}{(1+k)} = \gamma^*\frac{k(1-c)}{1+k}$ (28)

This gives us reciprocally (with all calculations made): $c^* = \frac{c_I(1+k)}{k(1-c)+c_I(1+k)}$ (29)

It is important to understand that $c_I$ and $c^*$, although they are both related to the consumption of all capitalists from section I, correspond to two different concepts and are thus two different variables. Indeed $c_I$ is the share of this consumption within total consumption *C*, whereas $c^*$ is the gross profit margin (a percentage). Multiplied by a selling price $\frac{1}{1-c^*}$ it gives $\frac{c^*}{1-c^*}$ which is in absolute value the fraction of the total gross profit margin intended for consumption for all capitalists from section I, but as a ratio is a markup, i.e. the ratio of this margin over the money value of capital sold to section II (on this see above, p. 233).

Therefore, if numerators of both fractions are the same, denominators are different: it is total consumption *C* for $c_I$ , and the money value of fixed capital $I_{II}$ sold to capitalists of section II for $c^*$.

These profits or this gross profit margin as an absolute amount are indeed given as an expression of $L_I$, i.e. the whole money value of all fixed capital produced in section I, by equation (26). But they can be expressed in terms of their money value going to section II, which is $(1-a)L_I$, by simply multiplying both the numerator and denominator of equation (26) by $(1-a)$. Then the equation giving the amount of profits becomes:

$$L_I\left(a+\frac{c^*}{1-c^*}\right)=(1-a)L_I\left(\frac{a}{1-a}+\frac{c^*}{(1-a)(1-c^*)}\right)=(1-a)L_I\left(\frac{a(1-c^*)+c^*}{(1-a)(1-c^*)}\right) \qquad (30)$$

This equation allows us to check that equation (25) above giving the price as a function of value is perfectly correct. Indeed, remembering that the general formula giving the sale price of fixed capital sold to section II is $\frac{1}{(1-a)(1-c^*)}$ times its money value, we have therefore:

$$(1-a)L_I+(1-a)L_I\left(\frac{a(1-c^*)+c^*}{(1-a)(1-c^*)}\right)=(1-a)L_I\left(1+\frac{a(1-c^*)+c^*}{(1-a)(1-c^*)}\right) \qquad (31)$$

From this last equation (31) we can deduct that $a(1-c^*)+c^*$ is the ratio of profits over the sale price, or the total gross profit margin (used not only for consumption but also for investment) in percentage.

A last amount which can easily be calculated is that of the profits remaining to these capitalists of section I who sell fixed capital to those of section II, once they have themselves bought the fixed capital that they need from other capitalists in section I. It is indeed only the amount of these residual profits which is ultimately available for consumption. Their amount is thus the difference between their overall profits, as it had been determined by equation (26), and the price of the fixed capital that they must themselves buy, from the other capitalist of section I. It can be written:

Residual profits $= L_I\left(a+\frac{c^*}{1-c^*}\right) - a\frac{L_I}{1-c^*} = L_I\left[\frac{a(1-c^*)+c^*-a}{1-c^*}\right] = L_I\left(\frac{c^*(1-a)}{1-c^*}\right)$

This amount, which corresponds to the profits available for consumption for the capitalists of section I selling fixed capital to section II, gives us therefore their consumption, which is equivalent to these profits – since we are in a system of simple reproduction, and is:

$$(1-a)L_I \frac{c*}{1-c*}. \qquad (32)$$

By difference with the overall consumption of capitalists of section I, i.e. $L_I \frac{c*}{(1-c*)}$, the consumption of capitalists of section I selling to the same section must be $aL_I \frac{c*}{1-c*}$. (33)

### 2.2.2. Fixed capital and profits realized from the sales within section I

Here we maintain the assumption already made within the framework of the simple model (in which parameter $u$ is equal to $b$, i.e. to $1-a$). This assumption corresponds to the second case examined in the previous chapter, according to which capitalists in section I always purchase a fraction $b = (1-a)$ of the fixed capital $(1-b) = a$, still available after the previous purchase already made of a fraction $b$.

Capitalists of section I who have sold to those of section II the fixed capital indispensable for the production of consumer goods therefore buy their own fixed capital (a fraction of $I_I$) at a price equal to $(1-b)\frac{(1-a)L_I}{(1-a)(1-c*)} = a\frac{L_I}{1-c*}$, and so on, and so forth, until all of the fixed capital produced for section I has been realized, knowing that all the profits realized during the successive sales of fixed capital will have been used in subsequent corresponding purchases.

To verify this, it is sufficient to draw again a table such as those in the preceding chapter, providing, for each successive purchase of fixed capital within section I, the price, the realized money value (wages) and the levied value (profits). To do that we must now consider that, instead of fixed capital being sold like in the simplified model at a price equal to $\frac{1}{1-a}$ times its money value, it is now sold at a price corresponding to $\frac{1}{1-a}\cdot\frac{1}{1-c*}$ times its money value, including now a share of profits left available for consumption. All these data are collated in table 7 below.

For each successive level, the realized profits are also obtained, symmetrically to the calculation of equation (26) above, by difference between the price and the realized money value (which corresponds to wages paid). For the second row of the following table, we get:

$$a\frac{L_I}{(1-c*)} - a(1-a)\,\mathrm{L_I} = aL_I\left(\frac{1}{1-c*}-(1-a)\right) = a\left(\frac{c*}{1-c*}+1-(1-a)\right) = aL_I\left(a+\frac{c*}{1-c*}\right) \qquad (34)$$

Similarly, profits available for consumption are obtained at each stage by difference between the overall realized profits and the price of the purchase of fixed capital, which is:

$$aL_I\left(a+\frac{c*}{1-c*}\right) - a^2\frac{L_I}{1-c*} = aL_I\left(\frac{a-ac*+c*-a}{1-c*}\right) = aL_I\frac{(1-a)c*}{1-c*} = a(1-a)L_I\frac{c*}{1-c*} \quad (35)$$

The sequence of these successive purchases appears in table 12.1, where a column has been added to show the amount of profits available for consumption.

**Table 12.1. Prices, money values and profits in section I, with capitalist consumption**

| **Realized money values = labor price = wages (with w=1)** | **Price of the product** | **Levied values = realized profits** | **Profits available for consumption** |
|---|---|---|---|
| $(1-a)\,L_I = b\,L_I$ | $\frac{(1-a)L_I}{1-a}\cdot\frac{1}{1-c*} = \frac{L_I}{1-c*}$ | $L_I\left(a+\frac{c*}{1-c*}\right)$ | $(1-a)L_I\cdot\frac{c*}{1-c*}$ |
| $a(1-a)\,L_I = ab\,L_I$ | $a\frac{L_I}{1-c*}$ | $a\,L_I\left(a+\frac{c*}{1-c*}\right)$ | $a(1-a)L_I\cdot\frac{c*}{1-c*}$ |
| $a^2(1-a)\,L_I = a^2b\,L_I$ | $a^2\frac{L_I}{1-c*}$ | $a^2L_I\left(a+\frac{c*}{1-c*}\right)$ | $a^2(1-a)L_I\cdot\frac{c*}{1-c*}$ |
| … | … | … | … |
| $a^n(1-a)\,L_I = a^nb\,L_I$ | $a^n\frac{L_I}{1-c*}$ | $a^nL_I\left(a+\frac{c*}{1-c*}\right)$ | $a^n(1-a)L_I\cdot\frac{c*}{1-c*}$ |

We recall that all quantities appearing in this table are prices, since we are in the framework of the monetary realization of the product, which is sold against money, and we have already demonstrated why these prices are pure scalars. It is easy to see that the sum of profits available for consumption is:

$a(1-a)\left[1+a+a^2+...+a^n\right]L_I\frac{c*}{1-c*} = aL_I\frac{c*}{1-c*}$, the exact amount given by equation (33).

# 3. The values of a few variables reflecting the distribution of the product

We can deduct from the last table above the values taken by a certain number of characteristic quantities of the model with capitalist consumption, to begin with the price of the total product itself, a product which is the very object of distribution.

## 3.1. The price of total product

It is important to determine the price of total product Y since it corresponds to the sum of the prices of all the final goods that are distributed. To determine this price Y we just have to add the price of consumption goods $Y_{II}$ and the price of fixed capital goods $Y_I$ .

1) The price of all consumption goods $Y_{II}$ is already known from equation (12). It is:

$$Y_{II} = P_{II} = L_{II} \cdot \frac{1+k}{1-c} = L_{II}\left(1+k_c\right)$$

2) The price of fixed capital goods is obtained as follows :

The sum of the prices of fixed capital goods, which corresponds to the price of the total product of section I, i.e. $Y_I$ or $I$, is obtained by summation of the second column of table 7 above, namely:

$$Y_I = I = \frac{L_I}{1-c^*}\ (1+a+a^2+...+a^n) = \frac{L_I}{1-c^*}\cdot\left(\frac{1}{1-a}\right) = \frac{L_I}{(1-a)(1-c^*)} \text{ when n } \to \infty \qquad (36)$$

The price of fixed capital $I_{II}$ for section II producing consumer goods is also:

$$I_{II} = \frac{(1-a)L_I}{(1-a)(1-c^*)} = \frac{L_I}{(1-c^*)}$$, which gives us, knowing that $\frac{c^*}{1-c^*} = \gamma^*$

$$I_{II} = \frac{L_I}{(1-c^*)} = L_I\left(\frac{c^*}{1-c^*}+1\right) = L_I\left(1+\gamma^*\right) \qquad (37)$$

As for the price of fixed capital $I_I$ for section I, it is thus:

$$I_I = \frac{aL_I}{(1-a)(1-c^*)} = \frac{a}{1-a} L_I \frac{1}{1-c^*}$$, which gives us, setting $\frac{a}{1-a} = \alpha$

$$I_I = \alpha L_I\left(1+\gamma^*\right) = L_I\left(\alpha+\alpha\gamma^*\right) \qquad (38)$$

We verify that the sum of these last two prices is equal to the price of the total fixed capital produced by section I, as provided by equation (36) above:

$$I = Y_I = I_{II} + I_I = \frac{L_I}{(1-c^*)} + \frac{aL_I}{(1-a)(1-c^*)} = \frac{(1-a)L_I + aL_I}{(1-a)(1-c^*)} = \frac{L_I}{(1-a)(1-c^*)} \quad (39)$$

So that: $I = Y_I = L_I\left(\frac{1}{1-a}\cdot\frac{1}{1-c^*}\right) = L_I\left[(1+\alpha)(1+\gamma^*)\right] = L_I\left(1+\alpha+\gamma^*+\alpha\gamma^*\right)$ (40)

We can check that: $Y_I = I = I_{II} + I_I = L_I\left(\alpha+\alpha\gamma^*\right) + L_I\left(1+\gamma^*\right) = L_I\left(1+\alpha+\gamma^*+\alpha\gamma^*\right)$ (41)

It is a somewhat complicated expression, but it gives us the price of all fixed capital in terms of its money value $L_I$, and it follows that $\left(\alpha+\gamma^*+\alpha\gamma^*\right)$ is in fact the expression of the gross profit margin over this money value or cost price $L_I$.

We will therefore name this ratio $\pi_I = \alpha+\gamma^*+\alpha\gamma^*$ (42)

As for the price of total product $Y$, it is therefore:

$Y = Y_I + Y_{II} = \frac{L_I}{(1-a)(1-c^*)} + L_{II}\cdot\frac{1+k}{1-c}$, which gives us, setting $\frac{c}{1-c} = \gamma \Leftrightarrow \frac{1}{1-c} = 1+\gamma$

$$Y = L_I\left(1+\alpha+\gamma^*+\alpha\gamma^*\right) + L_{II}\left(1+k\right)\left(1+\gamma\right) = kL_{II}\left(1+\alpha+\gamma^*+\alpha\gamma^*\right) + L_{II}\left(1+k\right)\left(1+\gamma\right) =$$

$$Y = L_{II}\left(1+\gamma\right) + kL_{II}\left(2+\alpha+\gamma+\gamma^*+\alpha\gamma^*\right) = L_{II}\left[\left(1+\gamma\right) + k\left(2+\alpha+\gamma+\gamma^*+\alpha\gamma^*\right)\right] \quad (43)$$

At this stage we can simplify further notations by naming $\psi$ an expression representing a good part of the second member of the above equation, so that $\psi = 1+\alpha+\gamma+\gamma^*+\alpha\gamma^*$. Replacing this value by $\psi$ in equation (43) we get:

$$Y = L_{II}\left[\left(1+\gamma\right) + k\left(1+\psi\right)\right] \quad (44)$$

## 3.2. The share of wages and profits in the product

The shares of wages and profits are indeed very important parameters, since we are here at the very heart of distribution, insofar as it is a fundamental aspect of any economic system. To determine these shares we must obviously start by displaying the amount of wages and profits, knowing that the former has already been determined.

### 3.2.1. The amount and share of wages

1) The amount of wages

We set $W = W_I + W_{II}$, with $W_I$ = wages in section I, and $W_{II}$ = wages in section II

This amount is already known, since it is $W = \bar{w}\left(L_I + L_{II}\right)$, which means that $W = L_I + L_{II}$ with $\bar{w} = 1$. Thus wages are $W_I = L_I$ in section I and $W_{II} = L_{II}$ and in section II.

In passing it seems interesting to verify that in section I the sum of the money values which are realized on the occasion of the sales of fixed capital corresponds to the amount of wages $W_I$ paid in this section. This sum of realized money values in section I appears in the first column of table 7, and is equal to:

Σ realized values = $bL_I + abL_I + a^2bL_I + ... + a^n bL_I$

= b.L$_I$ $(1 + a + a^2 + ... + a^n) = bL_I \frac{1}{1-a}$ when n $\to \infty$

Since we know that by definition $b = 1 - a$, then $bL_I \frac{1}{1-a} = \frac{(1-a)L_I}{(1-a)} = L_I$

We have thus checked that the sum of realized money values in section I is indeed $L_I$, which is the labor price $W_I$ with $\bar{w} = 1$, and corresponds to the money value of the whole fixed capital produced by section I.

2) The share of wages in total product

In section II this share is rather simple, since it is $\frac{W_{II}}{P_{II}} = \frac{L_{II}}{L_{II}\left(1+k_c\right)} = \frac{1}{1+k_c}$, which gives us:

$$\frac{W_2}{P_{II}} = \frac{1}{\frac{1+k}{1-c}} = \frac{1-c}{1+k} \tag{45}$$

In section I the expression is the following:

$\frac{W_I}{P_I} = \frac{L_I}{\frac{L_I}{(1-a)(1-c^*)}} = (1-a)(1-c^*)$, which can also be written as:

$$\frac{W_I}{P_I} = \frac{L_I}{L_I\left(1+\alpha+\gamma^*+\alpha\gamma^*\right)} = \frac{1}{\left(1+\alpha+\gamma^*+\alpha\gamma^*\right)} = \frac{1}{1+\pi_I} \tag{46}$$

As for the share of total wages in total product, taking the price of total product from equation (43), this share is expressed as:

$$\frac{W}{Y} = \frac{L_I + L_{II}}{L_{II}\left[(1+\gamma)+k\left(2+\alpha+\gamma+\gamma^*+\alpha\gamma^*\right)\right]} = \frac{L_{II}\left(1+k\right)}{L_{II}\left[(1+\gamma)+k\left(2+\alpha+\gamma+\gamma^*+\alpha\gamma^*\right)\right]}$$

$$\frac{W}{Y} = \frac{1+k}{1+\gamma+k\left(2+\alpha+\gamma+\gamma^*+\alpha\gamma^*\right)} = \frac{1+k}{1+\gamma+k\left(1+\psi\right)} \tag{47}$$

It is quite a complicated variable, which will be discussed later.

### 3.2.2. The amount and share of profits

1) Profits in section II

We already know from equation (17) above that $\Pi_{II}$, i.e. the amount of profits in section II, is equal to: $\Pi_{II} = L_{II}\left(\frac{k+c}{1-c}\right) = L_{II}k_c$ (48)

2) Profits in section I

From the data in the third column of table 7 above we know that the sum of levied values in section I, i.e. the total profits realized in this section, which we name $\Pi_I$, is equal to:

$\Pi_I = L_I\left(a + \frac{c*}{1-c*}\right)\left(1+a+a^2+...+a^n\right)$, which gives us, when n $\to \infty$, with $\frac{c*}{1-c*} = \gamma *$:

$$\Pi_I = L_I\left(a + \frac{c*}{1-c*}\right)\left(\frac{1}{1-a}\right) = L_I\left(\frac{a}{1-a} + \frac{c*}{(1-a)(1-c*)}\right) = L_I\left(\alpha + \gamma * + \alpha\gamma *\right) = L_I\pi_I \qquad (49)$$

These profits are used for a part to purchase fixed capital, and for the balance to purchase consumption goods. From the data in the fourth column of the same table, we know that the sum of profits available for consumption in section I, that we name $C_I$ is equal to:

$$C_I = \left[(1-a)L_I \cdot \frac{c*}{1-c*}\right]\ (1+a+a^2+...+a^n) =$$

$$C_I = \left[(1-a)L_I \cdot \frac{c*}{1-c*}\right]\left(\frac{1}{1-a}\right) = L_I\frac{c*}{1-c*} = L_I\gamma * \qquad (50)$$

Since we know that $\frac{L_I}{1-c*} = L_I\frac{c*}{1-c*} + L_I$ and that $\frac{L_I\frac{c*}{1-c*}}{L_I\frac{1}{1-c*}} = c*$, by dividing each term of the equality by $\frac{1}{1-c*}$, we obtain the trivial identities:

$L_I = L_I c* + L_I(1-c*)$, or $c*L_I = L_I - (1-c*)L_I$, or $\frac{c*}{1-c*}L_I = \frac{L_I}{1-c*} - L_I$

Conversely, it can easily be verified that the difference between $\Pi_I$, the sum of profits in section I, on the one hand, and the consumption of capitalists of this section, i.e. $C_I$, on the other hand, is equal to $I_I$, i.e. the sum of prices of fixed capital used in section I and bought by capitalists in this section. Indeed we have:

$\Pi_I$ - $C_I$ = $L_I(\alpha + \gamma * + \alpha\gamma *)$ - $L_I\gamma *$ = $L_I(\alpha + \alpha\gamma *)$

And we know from equation (38) above that we have indeed $I_I = L_I(\alpha + \alpha\gamma *)$ (51)

It is thus verified that $I_I$ is equal to $\Pi_I$ - $C_I$, or that $\Pi_I = I_I + C_I$

3) The total amount of profits for both sections is therefore:

$$\Pi = L_I\left(\alpha+\gamma^*+\alpha\gamma^*\right)+L_{II}k_c = kL_{II}\left(\alpha+\gamma^*+\alpha\gamma^*\right)+L_{II}k_c = L_{II}\left[k\left(\alpha+\gamma^*+\alpha\gamma^*\right)+k_c\right] \quad (52)$$

This can be written more simply as $\Pi = L_{II}\left(k\pi_I + k_c\right)$

4) The share of profits in the total product can now be calculated, as:

$$\frac{\Pi}{Y} = \frac{L_{II}\left[k\left(\alpha+\gamma^*+\alpha\gamma^*\right)+k_c\right]}{L_{II}\left[\left(1+\gamma\right)+k\left(2+\alpha+\gamma+\gamma^*+\alpha\gamma^*\right)\right]} = \frac{k\left(\alpha+\gamma^*+\alpha\gamma^*\right)+k_c}{\left(1+\gamma\right)+k\left(2+\alpha+\gamma+\gamma^*+\alpha\gamma^*\right)}$$

Knowing that: $k_c = \frac{k+c}{1-c} = k\frac{1}{1-c}+\frac{c}{1-c} = k\left(1+\gamma\right)+\gamma$, this ratio can be somewhat simplified, to eliminate $k_c$ from the numerator, and it then becomes:

$$\frac{\Pi}{Y} = \frac{\gamma+k\left(1+\alpha+\gamma+\gamma^*+\alpha\gamma^*\right)}{1+\gamma+k\left(2+\alpha+\gamma+\gamma^*+\alpha\gamma^*\right)}$$

This last expression is quite complicated, but it can also be written more simply as:

$$\frac{\Pi}{Y} = \frac{\gamma+k\psi}{1+\gamma+k\left(1+\psi\right)} \quad (53)$$

### 3.2.3. The profits/wages ratio

It is already known that $\pi_{II}$, the ratio of profits to wages in section II, is equal to the rate of surplus-value $k_C$. We can also calculate the ratio of profits over wages in section I, which we call $\pi_I$.

$$\pi_I = \frac{\Pi_I}{W_I} = \frac{L_I\left(\alpha+\gamma^*+\alpha\gamma^*\right)}{L_I} = \alpha+\gamma^*+\alpha\gamma^* \quad (54)$$

We can check that it is exactly the same expression as that which was given by equation (42).

Using the same notations as above, we can now express more simply the ratio of total profits over total wages, which we called $\pi$, and which can be written:

$\pi = \frac{\Pi}{W} = \frac{L_I\left(\alpha+\gamma^*+\alpha\gamma^*\right)+L_{II}k_c}{L_I+L_{II}}$, which in turn can be expressed as:

$$\pi = \frac{kL_{II}\left(\alpha+\gamma^*+\alpha\gamma^*\right)+L_{II}k_c}{L_{II}\left(1+k\right)} = \frac{k\left(\alpha+\gamma^*+\alpha\gamma^*\right)+k_c}{1+k} \quad (55)$$

By replacing $k_c$ by its expression as a function of $k$, i.e. $k_c = k\left(1+\gamma\right)+\gamma$, we get:

$$\pi = \frac{k\left(\alpha+\gamma^*+\alpha\gamma^*\right)+k\left(1+\gamma\right)+\gamma}{1+k} = \frac{k\left(1+\alpha+\gamma+\gamma^*+\alpha\gamma^*\right)+\gamma}{1+k} = \frac{k\psi+\gamma}{1+k} \tag{56}$$

### 3.2.4. The share of capitalists from section I in total consumption

At this stage it is now possible to define the $c_I$ share which capitalists of section I get out of total consumption. By calculating the ratio of capitalist consumption in section I, i.e. $C_I$, given by equation (50), over the total price of consumer goods, i.e. $P_{IIc} = L_{II} \cdot \frac{1+k}{1-c}$, we obtain:

$$c_I = \frac{L_I \frac{c^*}{1-c^*}}{L_{II} \frac{1+k}{1-c}} = \frac{L_I}{L_{II}}\left[\frac{c^*(1-c)}{(1-c^*)(1+k)}\right] = k\left[\frac{c^*(1-c)}{(1-c^*)(1+k)}\right] \tag{57}$$

We can check that this value of $c_I$ is strictly equivalent to the value that we could have calculated on a different basis, from equation (28) above, which is indeed the case, because we had found:

$$c_I = \frac{c^*}{1-c^*} \cdot \frac{k(1-c)}{(1+k)} \tag{58}$$

Since we know that $\frac{1-c}{1+k} = \frac{1}{1+k_c}$, this expression can also be written:

$$c_I = \frac{c^*}{1-c^*} \cdot \frac{k}{1+k_c} \tag{59}$$

Conversely we can derive from this last equation a new way of calculating $c^*$. Indeed from equation (59) we get:

$\frac{c^*}{1-c^*} = \frac{c_I\left(1+k_c\right)}{k}$, which implies:

$c^*k + c^*c_I\left(1+k_c\right) = c_I\left(1+k_c\right)$, from which we finally get:

$$c^* = \frac{c_I\left(1+k_c\right)}{k+c_I\left(1+k_c\right)} \tag{60}$$

The consumption of capitalists of section I cannot exceed a maximum constituted by the total consumption of capitalists, whose amount is determined by a fraction of the price of consumption goods fixed by capitalists of section II. We recall that this price is $P_{II} = \frac{L_I + L_{II}}{1-c} = \frac{L_{II}(1+k)}{1-c} = L_{II}\left(1+k_c\right)$, and that the total consumption of capitalists $Cc$ is thus by definition determined by the total share of capitalists in total consumption, called c, multiplied by this price:

$$Cc = c\frac{L_{II}(1+k)}{1-c}\text{, or also } Cc = L_{II}\gamma(1+k) \qquad (61)$$

The maximum capitalist consumption of Section I, i.e. the maximum of $C_I$, whose amount is given by equation (50), as $C_I = L_I \frac{c^*}{1-c^*}$ is therefore reached when $c_I = c$, which implies that $c_{II} = 0$, and in this case we obtain from equation (59) the following equality:

$$c_I = k\frac{c^*(1-c)}{(1-c^*)(1+k)} = c$$

This implies that: $\frac{c}{1-c} = \frac{k}{1+k}\cdot\frac{c^*}{1-c^*}$, which can also be written:

$$\frac{c^*}{1-c^*} = \frac{c}{1-c}\cdot\frac{1+k}{k}$$

With the notations adopted, a maximum for $c_I$ is therefore reached when we have:

$$\gamma^* = \gamma\frac{1+k}{k} \qquad (62)$$

This means that capitalists of section I could theoretically (even though this should happen only in the theoretical world) appropriate the totality of consumer goods going to capitalists, for any level of *c*, provided that the price of fixed capital sold to capitalists of section II, i.e. $\frac{L_I}{1-c^*}$ is set at such a level of c* that this equality given by equation (62) is verified. In this case all of the profits of capitalists of section II would be used only for the purchase of fixed capital from section I and there would be no income left to them for buying consumption goods.

### 3.3. A quick consistency test

In order to ensure that the formulas in the last three sub-sections are correct, a quick check can be performed. It can for instance be verified that in section I the price of fixed capital, provided by equation (28), i.e. the price of the total product of this section, corresponding to the income of this section when all this fixed capital is realized, is equal to the sum of wages $L_I$ and profits $\Pi_I$ - given by equation (49), distributed in this section. It is indeed the case, since as can be seen in the second equation below, by performing the calculation of the numerator, the different terms of this numerator compensate and are eliminated, with the exception of $L_I$, the only remaining term.

$$Y_I = \frac{L_I}{(1-a)(1-c^*)} = L_I(1+\alpha+\gamma^*+\alpha\gamma^*) = I_I + I_{II} \qquad (63)$$

$$W_I + \Pi_I = L_I + L_I\left(\frac{a}{1-a}+\frac{c^*}{(1-a)(1-c^*)}\right)=\frac{L_I(1-a)(1-c^*)+aL_I(1-c^*)+c^*L_I}{(1-a)(1-c^*)}$$

Although this formula seems at first glance quite complicated, all the figures but one indeed cancel one another, to give in fact a much simpler formula:

$$W_I + \Pi_I = \frac{L_I}{(1-a)(1-c^*)} = L_I\left(1+\alpha+\gamma^*+\alpha\gamma^*\right) = L_I\left(1+\pi_I\right) \qquad (64)$$

Finally and similarly, it can also be verified that the sum of prices of fixed capital bought by section I - given by equation (38), plus the consumption of capitalists of this same section – taken from equation (50), plus also the wages paid to its workers, is equal to the price of the whole fixed capital produced and realized by section I. This sum is:

$$Y_I = I_I + C_I + W_I = \frac{aL_I}{(1-a)(1-c^*)} + L_I\frac{c^*}{1-c^*} + L_I \qquad (65)$$

Knowing that $\frac{c^*}{1-c^*} = \frac{1}{1-c^*} - 1$, this expression becomes:

$$Y_I = L_I\left[\frac{a}{(1-a)(1-c^*)} + \left(\frac{1}{1-c^*}-1\right)+1\right]$$

$$Y_I = L_I\left(\frac{a}{(1-a)(1-c^*)} + \frac{1-a}{(1-a)(1-c^*)}\right) = \frac{L_I}{(1-a)(1-c^*)} = I_I + I_{II} \qquad (66)$$

Having reached this point, all these mathematical formulas appear to be quite abstract and rather complicated, not to say somewhat esoteric. This is the reason why it seems both necessary and useful to provide a numerical example, which will be closer to the real world, and will be exposed in the following chapter. Before going to this next chapter we need nevertheless to go back to the problem of realization of products and profits in this model.

# 4. Again on the realization of the product and profits

## 4.1. A little problem of logics

At first sight, everything had seemed to go well when we touched upon the realization of the product (made of consumption goods) and profits of section II. But the attentive reader will probably have noticed that there had been in our reasoning a hidden assumption, which is that all of the consumption goods are sold, meaning not only those sold to workers, necessarily at a price equivalent to the total amount of their wages, i.e. $W_I + W_{II} = L_I + L_{II} = L_{II}(1+k)$, but also the remaining part of these consumption goods, i.e. the part sold to capitalists of both sections in exchange for an equivalent amount of their monetary profits. The price of these remaining consumption goods is necessarily equal to:

$$L_{II}\left(1+k_c\right)-L_{II}\left(1+k\right)=L_{II}k_c-L_{II}k=L_{II}\left(\frac{k+c}{1-c}-k\right)=L_{II}c\frac{1+k}{1-c}=cL_{II}\left(1+k_c\right) \quad (67)$$

It must be noted in passing that we find the same amount for capitalist consumption as that which had been calculated in another way through equations (12) to (16) above.

To buy these goods capitalists of section II could hypothetically use all the profits realized from the sale of consumption goods at a price $L_I + L_{II}$ to all workers of both sections, whereas they have paid their own workers an amount of wages equal to $W_{II} = L_{II}$ : the profits made from this sale are therefore $L_I$ . But in general there is no reason why these remaining consumption goods should be sold at such a price. It is only in a particular case, if the profits realized in section II from the sale to workers only are exactly equal to the price of these remaining consumption goods, that this could happen. This situation would imply that:

$$L_I = cL_{II}\frac{1+k}{1-c} \Rightarrow kL_{II} = cL_{II}\frac{1+k}{1-c} \Rightarrow k\left(1-c\right)=c\left(1+k\right) \quad (68)$$

The last equation implies that such a situation can only occur if we have either:

$$c=\frac{k}{1+2k} \text{ (if } k \text{ is given), or } k=\frac{c}{1-2c} \text{ (} c \text{ being given)} \quad (69)$$

There is no reason for these parameters *c* and *k* to be linked in such a way, but even if it were the case this would not resolve our problem of realization, because it would imply that:

- Capitalists of section I could not buy any consumption goods (all of them having already been bought by capitalists of section II);

- There would be no profits left to capitalists in section II for them to buy fixed capital from section I.

Outside of this purely hypothetical and as such quite strange situation, the problem of realization of the remaining part of consumption goods produced in section II, on top of those that can be bought from wages and from a part of profits realized in this section (the part which is not spent on fixed capital), thus remains to be solved. We must therefore find a way to solve it, which implies to make first some general remarks on money and exchanges.

### 4.2. Some remarks on money and exchanges

We have shown that money is not solely a means of measuring values and more precisely prices. We know that it is also a store of values and we have highlighted its role as a medium of exchanges. This explains why, even though wages are paid at given intervals in time, they are spent daily, i.e. on a continuous basis, which means that like production, expenditures are made in a continuous process. It is so because at the beginning of any period there is always an existing stock of money, recorded in banks' balance sheets. Moreover money, when it is spent by a buyer and gained by a seller, does not lose its purchasing power. It only loses it when it disappears, i.e. when loans are repaid to banks which created it.

Bearing that in mind, we can infer that even though the mechanism that we described for the realization of the product implied that exchanges be logically successive, the fact that these exchanges in the real world take place over time on a continuous basis makes it possible to consider them as simultaneous, or quasi-simultaneous. As far as profits are concerned, the fact that they are realized at the same pace as the expenditure of wages, and thus equally on a continuous basis, also implies that they can also be spent on a continuous basis.

All these remarks bring us to the recognition that the mechanism highlighted in section 4 of chapter 10, in order to explain the formation of profits in section I, a mechanism first explained by Keynes who named it the widow's cruse, can be generalized and understood as working on a continuous basis. These remarks allow us to go back to the problem of the realization of all of the consumption goods produced in section II.

### 4.3. The realization of profits and product

Let us suppose first that money is borrowed by capitalists and injected in the productive system only for the payment of wages, i.e. for an amount $W_I + W_{II}$, that this whole amount is spent on consumption goods, on a continuous basis, and that the profits that they generate are spent in the same proportions as in the system described in this chapter, to buy both consumption goods and fixed capital goods. This also implies that the various exogenous parameters: $k, a$, $P_{II}$ (or $c$) and $c_I$ keep exactly the same value during this process.

This means that capitalists in section II, when they pay wages for an amount $W_{II}$, produce consumption goods for a value of $L_{II}$ and a price of $L_{II}\left(1+k_c\right)$, whereas they are able to sell to workers only an amount $W_I + W_{II}$ of these same goods, equivalent to $L_{II}\left(1+k\right)$. This implies that they can sell at first only a fraction $\frac{1+k}{1+k_c}$ of all the consumption goods available, which also corresponds to the workers' share of consumption goods. Since we know that:

$$1+k_c = 1+\frac{k+c}{1-c} = \frac{1+k}{1-c}\text{, it follows that } \frac{1+k}{1+k_c} = \frac{1+k}{\frac{1+k}{1-c}} = 1-c \qquad (70)$$

Not surprisingly, the share of workers is thus $1-c$, as could be expected. What about the initial profits that capitalists can make from this sale? To be sure, the value of consumption goods sold at a price $L_I + L_{II} = L_{II}\left(1+k\right)$ is lower than $L_{II}$, because capitalists do not sell to workers all the produced consumption goods but only a fraction, the value of which is equivalent to the overall price of goods sold to workers divided by $1+k_c$ (the ratio of price over value). This fraction is therefore $\frac{L_{II}\left(1+k\right)}{1+k_c} = L_{II}\left(1-c\right)$, as just seen. However these initial profits in section II are not the difference between this price $L_{II}(1+k)$ and this last amount, because the mere fact that capitalists in this section are able to sell their goods at this price $L_{II}(1+k)$ implies that they have already paid previously these whole wages for an amount $L_{II}$, and not $L_{II}(1-c)$.

These initial profits, which they can spend, are therefore the difference between the reduced volume of sales and the amount of wages $W_{II} = L_{II}$ which they have paid actually, and they are also lower than the full profits of section II determined at the beginning of this chapter, in subsection 2.1., from equation (17). Therefore, instead of being $k_c L_{II}$, this initial amount of profits, which for this reason we will name $\Pi_{II}^1$, must be considered as the difference between the amount sold to workers and the whole wages paid in section II:

$$\Pi_{II}^1 = L_{II}(1+k) - L_{II} = kL_{II} \tag{71}$$

This last amount in fact corresponds to the amount of profits in a model without capitalist consumption and represents only a percentage of $k_c L_{II}$, i.e. the amount of profits which would be obtained by the sale of the whole product of section II. This percentage is:

$$\frac{\Pi_{II}^1}{\Pi_{II}} = \frac{kL_{II}}{k_c L_{II}} = \frac{k}{k_c} = \frac{k}{\frac{k+c}{1-c}} = \frac{k(1-c)}{k+c} \tag{72}$$

From this equation it is quite easy to express these initial profits in terms of the final or targeted ones:

$\Pi_{II}^1 = \frac{k(1-c)}{k+c}\Pi_{II} = \frac{k(1-c)}{k+c}k_c L_{II}$. It is easy to check that this is equivalent to:

$\Pi_{II}^1 = \frac{k(1-c)}{k+c}\left[\frac{k+c}{1-c}L_{II}\right] = kL_{II}$, which is not a surprise!

From this we see that the amount of money profits in section II, which by construction is equivalent to surplus-value, cannot initially be increased by the rise in the price of consumption goods from $L_{II}(1+k)$ to $L_{II}(1+k_c)$, compared to a situation without capitalist consumption, through the sale of these goods to workers, as long as the remaining part of consumption goods has not itself been realized by being sold to capitalists of both sections, who have to spend a part of their profits to this end. Even if capitalists of section II used the totality of this realized amount of profits to buy consumption goods, they would obtain a share of these goods equivalent to:

$$\frac{\Pi_{II}^{1}}{P_{II}}=\frac{kL_{II}}{\left(1+k_{c}\right)L_{II}}=\frac{k}{1+\frac{k+c}{1-c}}=\frac{k\left(1-c\right)}{1+k} \tag{73}$$

At the same time, the share of consumption goods available after workers have bought their own share is itself equivalent to:

$$\frac{\left(1+k_{c}\right)L_{II}-\left(1+k\right)L_{II}}{\left(1+k_{c}\right)L_{II}}=\frac{\frac{1+k}{1-c}-1+k}{\frac{1+k}{1-c}}=\frac{1+k-\left(1+k\right)\left(1-c\right)}{1+k}=1-\left(1-c\right)=c \tag{74}$$

It is again not a surprise to find out that this share is $c$, i.e. the share of total capitalist consumption from both sections out of the total production of consumption goods.

Thus if we want to compare the share of initial profits $\Pi_{II}^{1} = kL_{II}$, to $c$apitalist consumption, i.e. $C_{c}=c\left(1+kc\right)L_{II}=c\frac{1+k}{1-c}L_{II}$, there is a value of parameters such that $\Pi_{II}^{1}$ and $C_{c}$ are equal. This is the case when $kL_{II}=c\frac{1+k}{1-c}L_{II}$, which implies that we should have between k and c the following relation: $\frac{k}{1+k}=\frac{c}{1-c}$

But this is indeed quite a particular case, and if we leave it to go back to the most general case, what we have in the system when the initial profits of section II materialize is an amount of monetary profits $\Pi_{II}^{1}=kL_{II}$, which is lower than the targeted amount of profits, i.e. $k_{c}L_{II}=\frac{k+c}{1-c}L_{II}$, by a factor $\frac{k\left(1-c\right)}{k+c}$, and these initial profits are not sufficient to realize the purchase of both consumption goods and fixed capital needed by section II for the amounts which would allow for the realization of the product in both sections.

It seems therefore appropriate to adopt a realistic approach, by supposing that these initial profits are spent exactly in the same proportions which would allow for the realization of the whole product if these profits were of the targeted amount. The problem then consists in understanding what happens when all the parameters which constitute the exogenous variables remain unchanged, and when we start the spending of profits from section II with an amount of profits which is $kL_{II}$ instead of being supposed immediately to be $k_{c}L_{II}$, as we postulated above in section II of the present chapter.

Indeed, if the same proportion of section II profits is invested in the purchase of fixed capital goods from section I, it means that instead of investing from the beginning an amount $I_{II} = \frac{L_I}{1-c^*}$, section II will at first only be able to invest an amount reduced in similar proportions $\frac{k}{k_c}$, i.e. to $I_{II}^1 = \frac{kL_I}{k_c(1-c^*)} = \frac{k(1-c)L_I}{(k+c)(1-c^*)}$ (75)

This in turn will affect in a similar proportion the profits of section I, which will reduce the amount of total investments of section I to:

$I_I^1 = \frac{kaL_I}{k_c(1-a)(1-c^*)} = \frac{k(1-c)aL_I}{(k+c)(1-a)(1-c^*)}$, and the total product of section I to:

$$Y_{II}^1 = I^1 = I_I^1 + I_{II}^1 = \frac{kL_I}{k_c(1-a)(1-c^*)} = \frac{k(1-c)L_I}{(k+c)(1-a)(1-c^*)}.$$

Finally it is the amount of total capitalist consumption for both sections that in turn at this stage, will not be:

$C_c = cL_{II}(1+k_c) = cL_{II}\frac{1+k}{1-c}$, but only a $\frac{k}{k_c}$ proportion of it:

$C_c^1 = \frac{k}{k_c}C_c = \frac{k}{k_c}c(1+k_c)L_{II} = \frac{k(1-c)}{k+c}\left(\frac{1+k}{1-c}\right)cL_{II}$, which is equivalent to:

$$C_c^1 = c\frac{1+k}{k+c}kL_{II}. \qquad (76)$$

To summarize our approach, we introduced capitalist consumption in the model through an increase in surplus-value and profits in section II based on an initial increase in the prices of consumption goods (compared to the situation without capitalist consumption). Then we showed that the expense of wages generates profits in section II for an amount $\Pi_{II}^1 = kL_{II}$. In turn we just showed that these profits generate a capitalist consumption $C_c^1 = c\frac{1+k}{k+c}kL_{II}$.

A key element in this demonstration is now to recognize that this capitalist consumption corresponds to pure additional profits in section II, for the simple reason that the whole amount of wages paid in this section (i.e. $W_{II} = L_{II}$) has already been deducted from the sale price of consumption goods bought by workers, in the calculation of $\Pi_{II}^1$, and can thus already be repaid to the banks from which this amount has initially been borrowed. These profits, which we name $\Pi_{II}^2$, are therefore exactly equal to this $C_c^1$ amount of consumption, and therefore we can write:

$$\Pi_{II}^{2} = C_{c}^{1} = c\frac{1+k}{k+c}kL_{II} \tag{77}$$

In turn these profits will generate an additional consumption by capitalists from both sections, which – with all parameters remaining unchanged, is necessarily proportional to the ratio of these new profits in section II to the total profits of this section, and such that we can write:

$$\frac{C_c^2}{C_c} = \frac{\Pi_{II}^2}{\Pi_{II}} \Rightarrow C_c^2 = C_c \frac{\Pi_{II}^2}{\Pi_{II}} =$$

$$C_c^2 = c\left(1+k_c\right)L_{II}\left(\frac{ck\frac{1+k}{k+c}}{k_c}\right) = c\left(\frac{1+k}{1-c}\right)\left(\frac{ck\frac{1+k}{k+c}}{\frac{k+c}{1-c}}\right)L_{II} = c\left(\frac{1+k}{1-c}\right)\cdot ck\left(\frac{1+k}{k+c}\cdot\frac{1-c}{k+c}\right)L_{II} =$$

$$C_c^2 = c^2k\frac{\left(1+k\right)^2}{\left(k+c\right)^2}L_{II} = kc^2\left(\frac{1+k}{k+c}\right)^2 L_{II} \tag{78}$$

In turn this consumption corresponds to pure profits $\Pi_{II}^3 = C_c^2$, which can themselves be spent again, thus generating new consumption and new profits, and so on, and so forth, ad infinitum, so that the sum of all these profits in section II can be written:

$$\sum_{i=1}^{n}\Pi_{II}^{i} = \Pi_{II}^{1} + \Pi_{II}^{2} + \ldots + \Pi_{II}^{n} =$$

$$\sum_{i=1}^{n}\Pi_{II}^{i} = kL_{II}\left[1+\frac{c\left(1+k\right)}{k+c}+\left(\frac{c\left(1+k\right)}{k+c}\right)^2+\ldots+\left(\frac{c\left(1+k\right)}{k+c}\right)^n\right] \tag{79}$$

We recognize the sum of a geometric series, which converges when $n \to \infty$, because its common ratio is $\frac{c\left(1+k\right)}{k+c}$, which is obviously less than 1 since $c$ is necessarily lower than 1, with that we have always $ck < k$. Using the well-known formula for such a sum, we thus get:

$$\Pi_{II} = \sum_{i=1}^{n}\Pi_{II}^{i} = kL_{II}\left[\frac{1}{1-\frac{c\left(1+k\right)}{k+c}}\right] = kL_{II}\left[\frac{k+c}{k\left(1-c\right)}\right] = L_{II}k_c \qquad \text{(QED)} \tag{80}$$

As for capitalist consumption, since we have $C_{II}^{n} = \Pi_{II}^{n+1}$ it is easy to see from equation (79) that the sum of the successive amounts of capitalist consumption is equal to:

$$\sum_{i=1}^{n} C_c^i = \sum_{i=1}^{n} \Pi_{II}^i - \Pi_{II}^1 = \Pi_{II} - kL_{II} = (k_c - k) L_{II} \Rightarrow$$

$$C_c = \left( \frac{k+c}{1-c} - k \right) L_{II} = \frac{c(1+k)}{1-c} L_{II} = c(1+k_c) L_{II} \qquad \text{(QED)} \qquad (81)$$

We have thus demonstrated that when exogenous parameters stay unchanged, the spending of the initial profits of section II, through the successive spending of section I and section II profits, allows for the full realization of the full profits of section II (equation 79) and for the full realization of capitalist consumption. The conclusion that can be drawn from the above demonstration, is that when money injected in the productive system only corresponds to the amount of wages, it does allow for the realization of the full product in the sphere of circulation, because the expected amount of profits and capitalist consumption in section II and consequently in section I, can be realized, through the successive expense of profits.

In passing, we can also confirm that setting the price of consumption goods to $P_{II} = L_{II}(1+k_c)$ ensures to reach the targeted amount of capitalist consumption, as a share $c$ of indifferently the value or the price of consumption goods. In other words with the formulation of the rate of surplus-value as $k_c = \frac{k+c}{1-c}$ which we used so far, we get indeed:

$$C_c = P_{II} - (L_I + L_{II}) = L_{II}\left(1 + \frac{k+c}{1-c}\right) - L_{II}(1+k) = L_{II}\left(1 + \frac{k+c}{1-c} - (1+k)\right) =$$
$$= L_{II}\left(\frac{k+c-k+kc}{1-c}\right) = L_{II} c \left(\frac{1+k}{1-c}\right)$$

And since we know that $\frac{1+k}{1-c} = 1 + \frac{k+c}{1-c} = 1 + k_c$, then we have $C_c = cL_{II}(1+k_c)$

It is also immediate that the ratio of capitalist consumption $C_c$ over the total price of consumption goods is:

$$\frac{C_c}{P_{II}} = \frac{cL_{II}(1+k_c)}{L_{II}(1+k_c)} = c$$

We have thus demonstrated once more that the formulation of the rate of surplus-value (and share of profits in section II) which we have used so far is perfectly correct, and allows for the full realization of the whole product in both sections of the productive system.

The reason for such an outcome has to do with the nature of money as a store of value. Indeed when money is created at the stage of production to pay for wages, since - as we have demonstrated earlier, money wages are the measure of the money value of the product, this money has a purchasing power which corresponds exactly to this money value. When workers purchase consumption goods they do not get the value corresponding to what they have produced, but only a share $\frac{1}{1+k_c}$ because the price of these goods exceeds their money value by a factor $1+k_c$. And the remaining part $k_c$, which corresponds to surplus-value, has kept its purchasing power, which has been converted in monetary profits.

Consequently it is simply this same mechanism, which reduces by a factor $\frac{1}{1+k_c}$ the consumption of workers, while transferring a share of value $k_c$ to capitalists of section II, which is also at work with capitalists expenditures. We had shown in subsection 2.2. above that, assuming that all profits had been realized in section II, this mechanism ensured the realization of profits and fixed capital goods in section I producing this fixed capital. We have now shown that it also works for the realization of profits and product in section II, i.e. for the consumption of capitalists.

Indeed the fact that to reduce workers consumption (and free a part of consumption goods for capitalist consumption) the price of commodities has to be increased to the necessary extent, compared to their money value, to be sure also reduces the amount of money value – i.e. the purchasing power of money, which can be realized by capitalists in the spending of profits to buy consumption goods, but at the same time it transfers the remaining part, which keeps its purchasing power, to other capitalists! And the same price differential over value works in section I for transferring purchasing power on fixed capital among capitalists.

Indeed, as we have seen so far, any increase in the price of any goods by adding to the initial value a percentage $x$ is equivalent to levying a percentage $\frac{x}{1+x}$ out of the purchase at this increased price $1+x$ or to multiplying the initial value by $1+x$. But then this primary levy also leads to the creation of a secondary levy of the same percentage $x$ on the purchasing power of all secondary incomes derived from these primary incomes, when they are spent on the same goods. Thus these secondary incomes themselves can obtain only a $\frac{1}{1+x}$ fraction of the increased price of these goods (because $1-\frac{x}{1+x}=\frac{1}{1+x}$). This reduction in the purchasing power of money incomes in terms of the money value which they can obtain, transfers however a fraction $\frac{x}{1+x}$ of the purchasing power. This gives us a new principle.

**Principle 23**: A levy imposed on the purchasing power of primary incomes through an increase in the price of consumption goods also constitutes a transfer of purchasing power to secondary money incomes. Spending these incomes creates a new levy on secondary incomes when they buy consumption goods as well as fixed capital, and a new transfer of purchasing power. The same process can continue as long as purchasing power is remaining and ensures henceforth the full realization of the whole product from both sections of the productive system.

It is therefore obvious that the only money which is really needed in the system corresponds to money created as a counterpart for production and labor incomes, since it allows for the full realization of the product from both section II and section I. It is so because this money first allows for an initial realization of profits in section II, which is used to buy consumption goods as well as fixed capital in the appropriate amounts. At the same time these expenditures also transfer purchasing power to capitalists and among them, a process which converges up to the full expense of this purchasing power.

Although Keynes's remarks on the widow's cruse concerned only the consumption of entrepreneurs (as he called capitalists), we had previously generalized this mechanism to account for the formation of profits in section I. It is therefore not surprising to see that his remarks account for section II and capitalist consumption, and are therefore fully validated. Indeed it is worth quoting him again: "There is one peculiarity of profits (or losses) that we may note *in passing*, because it is one of the reasons to segregate them from income proper, as a category apart. If entrepreneurs choose to spend a portion of their profits on consumption [...] the effect is to increase the profit on the sale of liquid consumption goods by an amount exactly equal to the amount of profits which have been thus expended [...] Thus, however much of profits entrepreneurs spend on consumption, the increment of wealth belonging to the entrepreneurs remains the same as before. Thus, profits, as a source of capital increment for entrepreneurs, are a widow's cruse which remains un-depleted, however much of them may be devoted to riotous living" (Keynes, 1930, p. 139).

Keynes remarks were not really supported by a full and detailed explanation, which we just provided, expanding its range of action to the whole sphere of circulation. The mechanism is undeniable, and it goes much further than Keynes himself imagined, because we expanded it to investment in fixed capital within section I. This mechanism is however only true and can only work on the basis of a few conditions which make it possible. Indeed as we have demonstrated, in order to increase their consumption by spending their profits, if we start from a situation without capitalist consumption, capitalists will increase their profits from $kL_{II}$ to $k_cL_{II}$ . What we have shown it that this implies to raise the rate of surplus-value from $k$ to $k_c$ , which has to be achieved through an increase in the price of consumption goods. As for the share of profits devoted - as Keynes writes it, to "riotous living", i.e. to the consumption of luxury goods, it needs a change in the composition of consumption goods towards an increased production of this kind of goods.

One other thing that was not mentioned by Keynes is that this widow's cruse mechanism cannot work indefinitely, and has a limit. To be sure, as we have demonstrated, capitalist consumption translates into profits for an identical amount, just as Keynes had indicated: "the effect is to increase the profit on the sale of liquid consumption goods by an amount exactly equal to the amount of profits which have been thus expended" (ibidem).

This is true because wages have already been paid in the sector of consumption goods, but these profits cannot in turn be used by capitalists of section II only to buy other consumption goods, because a part of them must necessarily be spent on the purchase of fixed capital. Otherwise the whole system would be blocked: if capitalists of section I did not sell fixed capital goods to those of section II, they could not purchase consumption goods at the next stage of spending. This is the reason why these profits will translate into a lower amount of expenditure on consumption goods at the next stage, which will make the system gradually converge towards its built-in levels of profits and capitalist consumption. This is all the more needed that there is in fact a limit on the consumption of capitalists, which is the value of consumption goods available once workers have themselves bought their share of these goods. It must also be understood that, at the completion of the process, all monetary profits have been spent and have thus progressively disappeared, as well as money wages.

# Chapter 13. A numerical example of a model with capitalist consumption

In this chapter we will provide an example, and will first start by recalling what are the exogenous variables that appear in it. This will allow us to then show how it is possible to determine from them a certain number of endogenous variables. This will make it possible finally to calculate the amounts of all the variables relating to the distribution of the product.

## 1. Exogenous variables

The first exogenous variable is the value of the product and its distribution between the two sections, which are essentially technological data:

- The product social value is $Y = L = 150$ (in time units), which is distributed between a value $Y_I = L_I = 50$ for the first section producing fixed capital, and a value $Y_{II} = L_{II} = 100$ for the second section producing consumption goods.

- Taking also the average nominal wage per unit of time as a unit, equal to 1, i.e. $\bar{w} = 1$, then the money value of the product is $W = \bar{w}L = L = 150$, with $W_I = \bar{w}L_I = L_I = 50$ and $W_{II} = \bar{w}L_{II} = L_{II} = 100$.

This implies that the ratio of the product social value in both sections $k = \frac{L_I}{L_{II}}$ is also an exogenous variable. It is essentially a technical parameter, although it reflects the distribution of value between the two sections of production, and from its definition in the first simplified model may be considered as the primary rate of surplus-value.

- In this numerical example it is: $k = \frac{L_I}{L_{II}} = \frac{50}{100} = 0.5$

The second variable, which is also a technological datum, is the share in value of fixed capital going to section I, for the production of fixed capital, over the total quantity of fixed capital. It is named $a$, and since it is the ratio of investment in section I to total investment, we have, for this numerical example:

$$a = \frac{I_I}{I} = 0.4 = 40\ \%$$

Since the share of total fixed capital going to section II, called $b$, is obtained by construction from its difference with $a$, the share of total fixed capital going to section I, to which it is the complement to 1, it must similarly be considered as a complementary exogenous variable. It means that in this example its value is necessarily:

$b = 1 - a = 1 - 0.4 = 0.6 = 60$ %

A third variable is the price of consumption goods produced by section II, which is 200, in money units. So we have $P_{II}$ = 200. But $P_{II}$ is also by definition and equivalently the amount of total consumption: $C$ = 200, for a money value equal to $L_{II}$ = 100.

Although that may seem difficult to grasp at this stage in view of the apparent complication of the model, the only other exogenous variable is $c_I$ , the share of consumption of capitalists of section I over total consumption. Let us consider that $c_I$ = 0.15 = 15 %.

Indeed all the other variables derive from these four variables, as it will now be demonstrated.

## 2. Endogenous variables

The first endogenous variable is the consumption of workers, which is identical by construction to the overall amount of wages for both sections, since in a situation of simple reproduction all wages are spent in the purchase of consumption goods, i.e.:

$C_W = L_I + L_{II} = W_I + W_{II} = 50 + 100 = 150$

We can also obtain directly from this last figure the share of workers in total consumption, the second endogenous variable, which we call $c_w$ , and which is:

$c_w = \frac{C_W}{C} = \frac{150}{200} = 0.75$ = 75 % (it is also equivalent to $c_w = 1 - c = 1 - 0.25 = 0.75$ )

Workers can therefore buy 75 % of the total price of consumption goods, and identically 75 % of their value (i.e. $L_{II}$ = 100), which amount to 75. This gives us at the same time the value of labor-power, since it is equal by definition to the value of the consumption goods which can be purchased by workers. We call $V$ this third variable and thus have $V$ = 75.

From the amount of workers consumption $C_W$ we have immediately by difference a fourth endogenous variable, the consumption of capitalists, which is:

$C_C = C - C_W = 200 – 150 = 50$

From this last amount we immediately get a fifth endogenous variable, which is the share of capitalist consumption in total consumption:

$c = \frac{C_C}{C} = \frac{50}{200} = 0.25 = 25$ %

From $c$ and $c_I$ (an exogenous variable) we can directly obtain the share of consumption of capitalists of section II over total consumption. As we know it is called $c_{II}$ and is our sixth endogenous variable, calculated by difference, because by definition $c = c_I + c_{II}$ , which gives us therefore:

$$c_{II} = c - c_I = 0.25 - 0.15 = 0.10 = 10\ \%$$

Now, the share of capitalist consumption $c$ out of total consumption gives us our certainly most important endogenous variables (the seventh and eight ones), since they are nothing less than:

- First, the amount of surplus-value S, i.e. the value of fixed capital plus the value of capitalist consumption:

$$S = L_I + C_c = 50 + 25 = 75$$

- Second, the actual rate of surplus-value for this model, which we have defined as $k_c$ , and which is by construction:

$$k_c = \frac{k+c}{1-c} = \frac{0.5+0.25}{1-0.25} = 1 = 100\ \%$$

We can easily check that this rate of surplus–value is strictly identical to the rate which can also be calculated from two other different formulas:

- The first one based on the price of consumption goods, knowing that $P_{II} = L_{II}\left(1+k_c\right)$, which gives indeed the rate of surplus-value as:

$$k_c = \frac{P_{II} - L_{II}}{L_{II}} = \frac{200-100}{100} = 1 = 100\ \%$$

- The second one based on the ratio of surplus-value over the value of labor-power, which is:

$$k_c = \frac{S}{V} = \frac{75}{75} = 1 = 100\ \%$$

A last variable which must be determined is c*, which as we have seen is a parameter linked to $c_I$ : $c*$ has been defined in the last chapter as corresponding to the share of consumption of capitalists of section I selling to capitalists of section II contained in a price $\frac{1}{1-c*}$, this share being thus $\frac{c*}{1-c*}$. It must be understood that it is not the full profit margin, since a part of this profit margin is being spent for the purchase of fixed capital, but only the remaining part of this profit margin, which can be realized after fixed capital has been bought. Indeed we have seen that the sale price itself is in fact $\frac{1}{\left(1-a\right)\left(1-c*\right)}$ times the cost price (or money value).

This remaining share is therefore available for consumption and thus spent on consumption goods.

We know from equation (58) that $c^* = \frac{c_I\left(1+k_c\right)}{k+c_I\left(1+k_c\right)}$, so that with $c_I$ = 0,15 this amount is:

$$c^* = \frac{0.15(1+1)}{0.5+0.15(1+1)} = \frac{0.3}{0.8} = 0.375 \text{ or } 37.5\ \%$$

It follows that the profit margin intended for consumption over the rest of the price is:

$$\frac{c^*}{1-c^*} = \frac{0.375}{1-0.375} = \frac{0.375}{0.625} = 0.6 = 60\ \%$$

Now that all the main variables have been defined or calculated, we are able to see easily how the income and product are distributed among all stakeholders.

# 3. The distribution of income and product

## 3.1. In section II

In section II producing consumption goods, production has a money value of 100, equal to the wage bill in this section ($W_{II} = L_{II} = 100$) and a monetary price of 200, which gives an amount of profits of 100. In passing it is noteworthy that, as previously pointed out, the profits/wages ratio in section II, i.e. 100/100 = 1, is exactly equal to the rate of surplus-value $k_c$ of 100 %.

These monetary profits are obtained by successive sequences of spending:

The purchase of consumption goods by workers from their wages $W_I + W_{II} = L_I + L_{II} = 150$ generates an amount of initial profit $\Pi_{II}^1 = kL_{II}$ = 0.5 x 100 = 50

These initial profits are then spent, both on consumption goods and fixed capital, by capitalists of section II, and capitalists of section I spend themselves a part of the profits which they make from this spending for the purchase of consumption goods, for an amount which from equation (75) in the previous chapter is equal to:

$$C_c^1 = c\frac{1+k}{k+c}kL_{II} = \frac{0.25(1+0.5)}{0.5+0.25}0.5\text{x}100 = \frac{0.375}{0.75}50 = 25$$

This last amount is itself equal to $\Pi_{II}^2 = 25$

The same phenomenon is reproduced from this second "wave" of profits $\Pi_{II}^2$, to obtain an amount of consumption $C_c^2$ which from equation (77) in the previous chapter is equal to:

$$C_c^2 = kc^2 \left(\frac{1+k}{k+c}\right)^2 L_{II} = 0.5(0.25)^2 \left(\frac{1+0.5}{0.5+0.25}\right)^2 100 = 0.5(0.0625)(4)100 = 0.125*100 = 12.5$$

This last amount is itself equal to $\Pi_{II}^3 = 12.5$

From the last chapter, we know that this series continues and converges to obtain finally a sum of profits realized and spent in section II, which - as expected - is given by equation (78):

$$\Pi_{II} = kL_{II}\left[\frac{k+c}{k(1-c)}\right] = k_c L_{II} = 1*100 = 100$$

We have shown also that these profits correspond to a total amount for capitalist consumption which is, from equation (79):

$$C_c = \frac{c(1+k)}{1-c} L_{II} = c(1+k_c) L_{II} = 0.25(1+1)*100 = 50$$

With their total profits amounting to $\Pi_{II} = 100$, capitalists of section II must buy the quantity of fixed capital which they need, i.e. by construction $b = 1-a = 1-0.4 = 0.6 = 60$ % of the value of all fixed capital available, which is 60 % of $L_I = 50$, corresponding to an amount of 30, in money value.

The price that they have to pay for this fixed capital is $(1-a)L_I \frac{1}{(1-a)(1-c*)} = \frac{L_I}{1-c*}$

$$\frac{L_I}{1-c*} = L_I\left(1+\frac{c*}{1-c*}\right) = 50\left(1+\frac{0.375}{1-0.375}\right) = 50\left(1+\frac{0.375}{0.625}\right) = 50(1+0.6) = 80$$

As a result, the remaining profits available for consumption by capitalists of section II once this fixed capital is paid for are:

$$C_{II} = L_{II}k_c - L_I \frac{1}{1-c*} = L_{II}k_c - kL_{II}\frac{1}{1-c*} = L_{II}\left(k_c - \frac{k}{1-c*}\right) = 100\left(1-\frac{0.5}{0.625}\right) = 100(1-0.8)$$

$C_{II}$ = 100 – 80 = 20

This implies that $c_{II} = \frac{20}{200} = 0.1 = 10\%$ of total consumption.

### 3.2. In section I

1) Capitalists selling fixed capital to section II sell it as already seen for a price of 80, for a money value (wages) of 30, leaving them an amount of profits of 50.

Fixed capital available for section I after this sale has a money value of $aL_I$ = 0.4 x 50 = 20

In line with the assumption previously made, they buy a fraction $b = 1 - a = 0.6$ of the value of this capital, i.e. 60% of 20 =12

From their profits of 50 we know from table 7 in chapter 12 that for the purchase of this fixed capital they pay a price of $a\frac{L_I}{1-c*}$, which is equal to $0.4\frac{50}{1-0.375} = \frac{20}{0.625} = 32$

The balance of profits available for their consumption is therefore: 50 - 32 = 18, which allows them to buy consumption goods for an identical price of 18, and a value of 9.

2) At the next level, capitalists selling fixed capital to the previous ones in section I sell it for a price of 32, corresponding to a money value of 12 (wages), as we just saw, leaving them with an amount of profits of 20.

Fixed capital available after this sale has a money value of 20 - 12 = 8

In line with the assumption previously made, they buy a fraction $b = 1 - a = 0.6$ of this capital, i.e. 60% of 8 = 4.8

From their profits we know from table 7 in last chapter that to buy this fixed capital they pay a price of $a^2 \frac{L_I}{1-c*}$, which is equal to $0.16\frac{50}{1-375} = \frac{8}{0.625} = 12.8$

The balance of profits available for consumption is therefore: 20 - 12.8 = 7.2 which allows them to buy consumption goods for an identical price of 7.2, and a value of 3.6.

3) At the next level, capitalists selling fixed capital to the previous ones in section I sell it for a price of 12.8, for a money value of 4.8 (wages), leaving them an amount of profits of 8.

Fixed capital available after this sale has a money value of 8 – 4.8 = 3.2

In line with the assumption previously made, they buy a fraction $b = 1 - a = 0.6$ of this capital, i.e. 60% of 3.2 = 1.92

From their profits we know from table 7 in last chapter that to buy this fixed capital they pay a price of $a^3 \frac{L_I}{1-c*}$, which is equal to $0.064\frac{50}{1-375} = \frac{3.2}{0.625} = 5.12$

The balance of profits available for consumption is therefore: 8 - 5.12 = 2.88 which allows them to buy consumption goods for an identical price of 2.88, and a value of 1.44.

4) It does not seem necessary to continue this demonstration, all the more so that we can see that after only three stages in the sales of fixed capital the total amount of these sales in money value has already reached 12 + 4.8 + 1.92, which makes a total of 18.72 out of a value of fixed capital intended for section I of 20 (50 minus 30 sold to section II). Thus 93.6 % of this amount of 20 have already been realized, after only three "rounds"!

5) For section I as a whole we will have the following figures:

i. The total money value of fixed capital produced is as we know $L_I = W_I = 50$

ii. The total price of fixed capital sold to this section and bought within it (which as we know has a value of $aL_I = 0.4 \times 50 = 20$ ), is given by equation (30), as:

$$I_I = \frac{aL_I}{(1-a)(1-c^*)} = \frac{0.4 \times 50}{(1-0.4)(1-0.375)} = \frac{20}{0.6 \times 0.625} = \frac{20}{0.375} = 53.33$$

iii. The total amount of profits realized in this section (including profits from the sale of fixed capital to section II) is given by equation (39) as :

$$\Pi_I = L_I \left( \frac{a}{1-a} + \frac{c^*}{(1-a)(1-c^*)} \right) = 50 \left[ \frac{0.4}{0.6} + \frac{0.375}{0.6 \times 0.625} \right] = 50(0.667+1) = 83.33$$

iv. The total of profits available for consumption in this section is provided for each successive stage by table 7 in the previous chapter as:

$$(1-a) L_I \cdot \frac{c^*}{1-c^*} + a(1-a) L_I \cdot \frac{c^*}{1-c^*} + a^2 (1-a) L_I \cdot \frac{c^*}{1-c^*} + \ldots + a^n (1-a) L_I \cdot \frac{c^*}{1-c^*} =$$

$$(1-a) L_I \frac{c^*}{1-c^*} \left(1 + a + a^2 + \ldots + a^n\right) = (1-a) L_I \frac{c^*}{1-c^*} \left( \frac{1}{1-a} \right) = L_I \frac{c^*}{1-c^*}$$

This gives us, as the total amount of profits available for consumption in section I:

$$\Pi_I - I_I = 83.33 - 53.33 = 30$$

We can check that is exactly the same amount as that derived from the above equation:

$$L_I \frac{c^*}{1-c^*} = 50 \frac{0.375}{0.625} = 50 \times 0.6 = 30$$

We can also check that it is indeed equal to $C_I$, i.e. the consumption of capitalists of section I, whose share $c_I$ in total consumption is exogenous and given as $c_I = 0.15$ . $C_I$ is indeed:

$$c_I\, C = 0.15\, C = 15\% \times 200 = 30$$

We have thus checked that all these figures for all of these different endogenous variables correspond perfectly to all the equations that we had devised in order to calculate them from the exogenous variables. This validates these equations.

# 4. The values of a few variables reflecting the distribution of the product

## 4.1. The price of total product

The price of total product $Y$ is equal to the price of consumption goods $Y_{II}$ = 200 plus the price of fixed capital goods $Y_I$, which is itself the sum of prices of fixed capital sold to section II, i.e. 80, and fixed capital sold to section I, i.e. 53.33, so that $Y_I$ =133,33.

We can check that the total price of fixed capital given by equation (41) is indeed:

$$P_I = L_I\left(1+\alpha+\gamma^*+\alpha\gamma^*\right) = 50\left(1+0.667+0.6+0.4\right) = 50\times 2.667 = 133.33$$

Thus we have $Y = Y_I + Y_{II} = 133.33 + 200 = 333.33$

We can also check that this last amount is equivalent to that given by equation (43) as being:

$$\text{Y} = L_{II}\left[\left(1+\gamma\right)+k\left(2+\alpha+\gamma+\gamma^*+\alpha\gamma^*\right)\right] =$$

$$Y = 100\left[\left(1+0.333\right)+0.5\left(2+0.667+0.333+0.6+04\right)\right] = 100\left(1.333+0.5\times 4\right)$$

Y= 100 (1.33 + 2) = 100 x 3.33 = 333.33

This is indeed the case, which by the same token validates the formula.

## 4.2. The share of wages and profits in the product

The share of wages in section I over the price of fixed capital produced in this section is given by the initial expression of equation (45), and equal to:

$$\frac{W_I}{P_I} = \left(1-a\right)\left(1-c^*\right) = 0.6 \text{ x } 0.625 = 0.375$$

We can check that it is indeed equal to:

$$\frac{W_I}{P_I} = \frac{50}{133{,}33} = 0.375$$

The share of wages in section II over the price of all consumption goods produced in this section is given by equation (44) and equal to:

$$\frac{W_2}{P_{II}} = \frac{1-c}{1+k} = \frac{1-0.25}{1+0.5} = 0.5$$

We can check that it is indeed equal to:

$$\frac{W_2}{P_{II}} = \frac{L_{II}}{L_{II}\left(1+k_c\right)} = \frac{100}{\left(100\right)\left(1+1\right)} = \frac{1}{2} = 0.5 = 50\ \%$$

The share of wages in the product, i.e. over the total price of this product, is given by equation (46) and such as:

$$\frac{W}{Y} = \frac{1+k}{1+\gamma+k\left(2+\alpha+\gamma+\gamma^*+\alpha\gamma^*\right)}$$

This formula may seem rather complicated, but we know that $k$ = 0.5 and that the values of parameters $\alpha$ , $\gamma$ and $\gamma^*$ are the following:

$$\alpha = \frac{a}{1-a} = \frac{0.4}{0.6} = \frac{2}{3} = 0.667\ \ ;\ \gamma = \frac{c}{1-c} = \frac{0.25}{0.75} = \frac{1}{3} = 0.333\ ;\ \gamma^* = \frac{c^*}{1-c^*} = \frac{0.375}{0.625} = 0.6$$

Therefore the expression above can easily be calculated as:

$$\frac{W}{Y} = \frac{1+0.5}{1+0.333+0.5\left(2+0.667+0.333+0.6+0.4\right)} = \frac{1.5}{1.333+0.5\left(4\right)} = \frac{1.5}{3.333} = 0.45$$

Moreover, we can check that this amount is exactly equivalent to the ratio of the overall wages $W$ = 150 over the price of the product $Y$ = 333.33

As for profits, in section II their amount is $\Pi_{II} = L_{II}k_c = 100\times 1 = 100$, corresponding to the amount already found by difference between the price of consumption goods and wages in this section. We already know that their share in the product of Section II is equal to:

$$\frac{\Pi_{II}}{P_{II}} = \frac{L_{II}k_c}{L_{II}\left(1+k_c\right)} = \frac{k_c}{1+k_c} = \frac{1}{2} = 50\%$$

Profits in section I were already calculated in the subsection **3.2.** as being $\Pi_I$ = 83,33 (see above p. 263) and also the price of section I product, as being $P_I = 133.33$ (see above p. 264). Hence their share in the product of section I is the ratio $\frac{\Pi_I}{P_I} = \frac{83.33}{133.33} = 0.625$

We can check that it is their share in the product of section I calculated from equations (48) and (40):

$$\frac{\Pi_I}{P_I} = \frac{L_I\left(\alpha+\gamma^*+\alpha\gamma^*\right)}{L_I\left(1+\alpha+\gamma^*+\alpha\gamma^*\right)} = \frac{\alpha+\gamma^*+\alpha\gamma^*}{1+\alpha+\gamma^*+\alpha\gamma^*} = \frac{1.667}{2.667} = 0.625$$

Total profits for both sections are therefore 100 + 83.33 = 183.33, which corresponds exactly to the amount that can be calculated from the formula of equation (51):

$$\Pi = L_{II}\left[k\left(\alpha+\gamma^{*}+\alpha\gamma^{*}\right)+k_{c}\right]=100\left[0.5\left(0.667+0.6+0.667\times 0.6\right)+1\right]=$$
$$\Pi = 100\left[0.5\left(1.667\right)+1\right]=183.33$$

The share of profits in the total product is thus: $\frac{\Pi}{Y}=\frac{183.33}{333.33}=0.55$ = 55% and we can check that it is equal to the ratio given by the developed formula of equation (52):

$$\frac{\Pi}{Y}=\frac{\gamma+k\left(1+\alpha+\gamma+\gamma^{*}+\alpha\gamma^{*}\right)}{1+\gamma+k\left(2+\alpha+\gamma+\gamma^{*}+\alpha\gamma^{*}\right)}=$$
$$\frac{\Pi}{Y}=\frac{0.33+0.5\left(1+0.667+0.333+0.6+0.4\right)}{1+0.33+0.5\left(2+0.667+0.333+0.6+0.4\right)}=\frac{0.33+0.5\times 3}{1.33+0.5\times 4}=\frac{1.83}{3.33}==0.55$$

### 4.3. The profits/wages ratio

It is already known that $\pi_{II}$ , the ratio of profits to wages in section II, is equal to the rate of surplus-value $k_c$ . Thus we have $\pi_{II}=\frac{\Pi_{II}}{W_{II}}=\frac{100}{100}=1$ =100 %

We can also calculate the ratio of profits over wages in section I, which we call $\pi_I$ .

This share is equal to $\pi_I=\frac{\Pi_I}{W_I}=\frac{83.33}{50}=1.667$ = 167 %

We can check that it is exactly the same as the share given from the more general formula of equation (53):

$$\pi_I=\frac{\Pi_I}{W_I}=\frac{L_I\left(\alpha+\gamma^{*}+\alpha\gamma^{*}\right)}{L_I}=\alpha+\gamma^{*}+\alpha\gamma^{*}$$

$$\pi_I=0.667+0.6+(0.667\times 0.6)=1.667 = 167\ \%$$

Finally, as for the ratio of total profits over total wages, which is called $\pi$ , it is:

$$\pi=\frac{\Pi}{W}=\frac{183.33}{150}=1.222 = 122\ \%$$

We can again check that it corresponds to the general formula of equation (55):

$$\pi=\frac{k\left(1+\alpha+\gamma+\gamma^{*}+\alpha\gamma^{*}\right)+\gamma}{1+k}=$$

$$\pi = \frac{0.5\left(1+0.667+0.333+0.6+0.4\right)+0.333}{1.5} = \frac{0.5\times 3+0.333}{1.5} = 1.222 = 122\ \%$$

We have thus verified that this model, as it was constructed, actually allows us to understand the realization of the totality of the product of the two sections, when we introduce the consumption of capitalists.

However, this product is realized according to specific modalities that differ in each of the two sections. In section II, profits corresponding *inter alia* to this consumption are obtained from the sale of consumption goods to workers of each of the two sections, as well as to capitalists of both sections. In section I, profits are obtained first from the sale of fixed capital to capitalists of section II and then from the sale of the remaining fixed capital within section I itself. These profits are therefore logically formed from the redistribution to section I of the surplus-value initially realized under the form of monetary profits in section II, which are then disseminated within section I.

Having completed the exploration of this model, we can now develop a few teachings that can be derived from it, regarding the significance of distribution and prices. This will be done in the next chapter.

# Chapter 14. Distribution and prices: a close connection

To be sure, the model with capitalist consumption developed in the last chapters has the capacity to shed some light on the main characteristics of distribution and prices in the capitalist mode of production. But it relies on a special case, i.e. a particular assumption regarding the fraction of the fixed capital available and bought at each stage by capitalists in section I, after previous sales of fixed capital have been made: this assumption is the fixity of this parameter, which was called *u* in sub-section 3.3. of chapter 11, even if it was different from *b*. It seems therefore useful to get rid of this assumption, in order to get a more realistic view of distribution. We will first expose this generalization of the model with capitalist consumption, which will then enable us first to better understand how microeconomic prices are set, and second the process of transformation of values into monetary prices.

## 1. A generalization of the model with capitalist consumption

In order to examine a more general situation, as a first step it is necessary to reintroduce this parameter $u$ in a model with capitalist consumption. This makes it possible to establish a new table, named 14.1, with $u \neq b$ and therefore $u \neq 1-a$. This explains why the first row of the table, which shows the sale of fixed capital to capitalists of section II – where the share of fixed capital by definition stays as $u_1 = b = 1 - a$, is not homogeneous to the others.

**Table 14.1 - Prices, money values and profits in section I, with capitalist consumption**

| Realized values = labor price = wages (w = 1) | Price of the product | Levied values = realized profits | Profits available for consumption |
|---|---|---|---|
| $(1-a)\ L_I = b\,L_I$ | $\frac{(1-a)L_I}{1-a}\cdot\frac{1}{1-c*}=$ $\frac{L_I}{1-c*}$ | $L_I\left(a+\frac{c*}{1-c*}\right)$ | $(1-a)L_I\cdot\frac{c*}{1-c*}$ |
| $au\mathrm{L_I}$ | $a\frac{L_I}{1-c*}$ | $aL_I\left(\frac{1}{1-c*}-u\right)$ | $auL_I\cdot\frac{c*}{1-c*}$ |
| $au(1-u)\ L_I$ | $a(1-u)L_I\frac{1}{1-c*}$ | $a(1-u)L_I\left(\frac{1}{1-c*}-u\right)$ | $au(1-u)L_I\cdot\frac{c*}{1-c*}$ |
| $au(1-u)^2\ L_I$ | $a(1-u)^2L_I\frac{1}{1-c*}$ | $a(1-u)^2L_I\left(\frac{1}{1-c*}-u\right)$ | $au(1-u)^2L_I\cdot\frac{c*}{1-c*}$ |
| … | … | … | … |
| $au(1-u)^n\ L_I$ | $a(1-u)^nL_I\frac{1}{1-c*}$ | $a(1-u)^nL_I\left(\frac{1}{1-c*}-u\right)$ | $au(1-u)^nL_I\cdot\frac{c*}{1-c*}$ |

Nevertheless, and to begin with, we introduce *u* as a constant, as the table shows, just as a first step allowing for an easier understanding of this matter. We must also note that the parameter which corresponds at each stage to the ratio of profits over product price, which is usually called the gross profit margin, and is obtained by dividing for each row of the table the profits in the third column by the product price in the second column, is no longer 1 – u (like in table 3 of chapter 11), but becomes now 1 - *u* (1 – c*). The detail of the calculation of each term is not repeated, since this above table 8 is in fact nothing else than a combination of table 6 of chapter 11 and table 7 of chapter 12.

If we want to check the validity of this table, we can make a quick test by calculating the sum of prices for all of the fixed capital produced and sold by section I. It is obtained by summing the quantities in column 2 of the table:

$$P_I = \frac{L_I}{1-c^*}\left[1+a\left\{1+(1-u)+(1-u)^2+\ldots+(1-u)^n\right\}\right]$$

$$P_I = \frac{L_I}{1-c^*}\left[1+a\left(\frac{1}{1-(1-u)}\right)\right]$$

Knowing that $1-u=1-b=a$ , $P_I$ becomes:

$$P_I = \frac{L_I}{1-c^*}\left[1+\frac{a}{u}\right] = \frac{L_I}{1-c^*}\left[1+\frac{a}{1-a}\right] = \frac{L_I}{1-c^*}\left(\frac{1}{1-a}\right) = \frac{L_I}{(1-c^*)(1-a)} \qquad \text{[QED]} \qquad (1)$$

Indeed we have $P_I = I = Y_I$ , i.e. exactly the price of fixed capital given by equation (39) of chapter 12.

Before going further, we must recall that the exogenous parameters that we defined as such in the previous chapters must be considered as invariant at least at a macroeconomic level within the framework of a given reproduction scheme. These parameters are *k*, $a$ (and therefore $b=1-a=u_1$ ), $P_{II}$ from which we derive *c* (the share of capitalist consumption in total consumption), and *c** (the parameter linked to capitalist consumption in section I).

In a system generalizing the scheme provided in table 8 on the previous page, we must now consider that, outside the sale of fixed capital from section I to section II, whose amount is fixed and exogenous, parameter *u* is different for each stage of fixed capital sales within section I: in other words there are as many $u_i$ as there are such stages, with the first one being eliminated from the first row (corresponding to the sale of fixed capital to section II) as being equal to $b=1-a$, when it is divided by $(1-a)$ in the expression of the price. This first *u* (or $u_1$ ) is thus fixed by the technological constraint that section II gets in any case from section I a share $b=1-a$ of the total value of fixed capital.

The multiplicity of the other $u_i$ parameters makes the various terms in the table become more complex. All the more so that there is also no reason why capitalist consumption should be the same at each stage: in the real world there is on the contrary every reason to think that the fraction of the value of consumption goods obtained at each stage by capitalists will be different from one stage to another.

Obviously this multiplicity introduces more complexity in the process of price fixation and product realization, but it is nevertheless the price to pay for a better understanding of the real world. Therefore, instead of having just one fraction c* corresponding to $c_I$, i.e. the share of global consumption obtained by all of the capitalists of section I, we will have as many $c_1^*$, $c_2^*,\ldots,c_n^*$ fractions as there are stages. However the overall consumption of capitalists in section I, as an exogenous parameter, must stay at a level $c_I$ which remains a constant (and thus c* = $c_1^*$) as it appears in the first row of the above table). With these new assumptions regarding *u* and *c** we can thus explore what will happen of prices, taking into account this additional complexity.

This in fact comes down to leaving the universe of macroeconomic prices, where we have been evolving so far, i.e. prices of two broad categories of commodities (consumption goods and fixed capital) linked by global constraints, and going to the area of microeconomic prices, i.e. prices such that one price corresponds theoretically to one single commodity.

## 2. From macroeconomic to microeconomic prices

### 2.1. Prices of fixed capital

To continue with prices of fixed capital, and starting therefore with them, let us explore the consequences of having different values at each stage of exchanges within section I for parameters *u* and *c**. While doing that, we must consider first that the overall price of consumption goods produced by section II obviously does not change.

Similarly, the price of fixed capital sold to this section II (in the first row of the table) will not change because of changes in the value of parameter *u*, since this parameter $u_1$ has been defined as a share of values, and not as a share of prices, and as we have just indicated is constrained at this stage by representing in any case the exogenous share $b=1-a$ of the value of fixed capital going to section II: $u_1=b=1-a$ .

The prices that will change because of differing parameters *u* are the prices for the sales of fixed capital within section I from the second row and onward. Moreover, capitalists selling fixed capital to section II (first row) will now obtain a fraction of the value of total consumption depending on $c_2^*$, capitalists selling fixed capital to them a fraction depending on $c_3^*$, and so on. Therefore all of the prices of fixed capital goods have to adjust because of changes in the fraction of the overall value of consumption goods which capitalists selling fixed capital to other ones are able to get, by adjusting their prices.

Thus, although the price of fixed capital sold to section II, appearing in row 1 stays as $\frac{L_I}{1-c^*}$, the price for fixed capital in the second row, instead of $a\frac{L_I}{1-c_2^*}$ becomes $\left(1-u_2\right)\frac{L_I}{1-c_2^*}$.

For row 3 of the table, the price thus becomes $\left(1-u_2\right)\left(1-u_3\right)L_I\frac{1}{1-c_3^*}$, and for row 4 $\left(1-u_2\right)\left(1-u_3\right)\left(1-u_4\right)L_I\frac{1}{1-c_4^*}$, and so on and so forth. The general formula for the price of fixed capital of the m$^{th}$ stage appearing in row *m* can then be written:

Price of fixed capital (row m) = $\left[\prod_{i=2}^{i=m}\left(1-u_i\right)\right]L_I\frac{1}{1-c_m^*}$ (2)

As for the realized money value of fixed capital, it stays unchanged as $\left(1-a\right)L_I = bL_I$, as far as fixed capital sold to section II is concerned, in the first row of the table. But it becomes $au_2L_I$ for the money value of fixed capital in the second row sold to the previous stage, and $au_3\left(1 - u_2\right)L_I$, in the third row. For the fourth row it becomes $au_4\left(1 - u_2\right)\left(1-u_3\right)L_I$, and so on and so forth. The general formula for the realized value of fixed capital of the m$^{th}$ stage appearing in row *m* can then be written:

Realized value of fixed capital (row m) = $au_m\prod_{i=2}^{i=m}\left(1 - u_i\right)L_I$ (3)

In this more general system, instead of having at each stage $b=1-a$ as a fixed parameter, we have $b=u$ as a variable parameter, meaning in other words that at each stage the share of the remaining value of fixed capital which is obtained is not predetermined. It follows that the various fixed ratios which we had defined in chapter 11 become variable ones:

The ratio of sale price over money value (wages) becomes: $\frac{1}{u_i\left(1-c_i^*\right)}$ (4)

Conversely the ratio of money value (or cost price) over sale price becomes: $u_i\left(1-c_i^*\right)$ (5)

Thus the gross profit margin (or ratio of profits over sale price) becomes: $1-u_i\left(1-c_i^*\right)$ (6)

The ratio of profits over money value (or cost price) becomes: $\pi_i = \dfrac{1}{u_i\left(1-c_i^*\right)} - 1$ (7)

In this more general case one might wonder whether all of these successive exchanges of fixed capital sold with changing margins $1-u_i\left(1-c_i^*\right)$ at each stage *i* will make it possible to clear the whole product of section I.

Indeed the sum of prices of fixed capital sold at each level is:

$$I = \frac{L_I}{1-c_1^*} + \left(1-u_2\right)\frac{L_I}{1-c_2^*} + \left(1-u_2\right)\left(1-u_3\right)\frac{L_I}{1-c_3^*} + \ldots + \prod_{i=2}^{i=n}\left(1-u_i\right)\frac{L_I}{1-c_n^*} \quad (8)$$

Whereas we know from equation (39) that the sale of the whole fixed capital corresponds to a situation where for macroeconomic reasons linked to simple reproduction ($a$ and $c^*$ being macroeconomic parameters) we have:

$$I = Y_I = I_I + I_{II} = \frac{L_I}{\left(1-a\right)\left(1-c^*\right)} \quad (9)$$

To be sure this creates a macroeconomic constraint on the values of all $u_i$ and $c_i^*$, but there is no particular reason to think that there would not be at least one system of values for these variables which would not meet the objective of selling the whole fixed capital. The reason to believe that such a situation may exist is that the system is quite flexible. First, because each stage depends on the previous one, which implies that at each stage there is a possibility to adjust the corresponding variables to the situation already achieved at the previous stage. And second because there is an important element of flexibility embodied in the system: as indicated previously all of $c_i^*$, which define the consumption of capitalists at each stage, correspond to residual levels of consumption. In a situation of simple reproduction corresponding to a self-replacing state the variation of consumption levels would thus play the role of the adjustment parameter which would allow to meet the desired level of investment, and thus to absorb all the production of fixed capital.

Finally there is a last element of flexibility coming from the fact that at each stage we are still in the real world at a macroeconomic level, because in the real world each stage corresponds to a number of different commodities composing fixed capital, each commodity with its own price. This multiplies the number of different parameters $u_i$ and $c_i^*$, but also the possibilities of adjustment.

This brings us to the question of microeconomic prices for consumption goods produced in section II.

## 2.2. Prices of consumption goods

In the real world there are a lot of consumption goods and therefore their total price given by equation (12) in chapter 11 cannot but be anything else than an average.

We recall that this total price is $P_{II} = L_{II}\left(1+k_c\right) = L_{II}\left(1+\frac{k+c}{1-c}\right) = L_{II}\left(\frac{1+k}{1-c}\right)$

To be sure *k* is a technological constant, but it plays this role only at macroeconomic level, and it is the same for *c*. At microeconomic level there is no particular reason why *k* and *c* should be the same for whatever consumption good, as long as their level allows for the desired level of investment in fixed capital and for a residual amount of consumption.

It follows that the various fixed ratios which can be derived from this general formula become variable ones:

The ratio of sale price over money value (wages) becomes: $\frac{1+k_i}{1-c_i}$ (10)

Conversely the ratio of money value (or cost price) over sale price becomes: $\frac{1-c_i}{1+k_i}$ (11)

Thus the gross profit margin (i.e. profits over sale price) becomes: $\frac{k_i+c_i}{1+k_i}$ (12)

The ratio of profits over money value (wages or cost price) becomes: $\frac{k_i+c_i}{1-c_i}$ (13)

Therefore for each price the margin over the money value (wages) which is given by this last equation can be variable from one consumption good to another, knowing that the weighted average for all goods (the weights being the respective values) must stay at $\frac{k+c}{1-c}$, i.e. the predetermined rate of surplus-value, which is a fundamental parameter of the whole economic system, as a key for understanding distribution and the modalities of its reproduction.

The constraint created by this average must not be underestimated, because it means that for the reproduction of the system to take place, then the sum of prices of consumption goods (weighted by their respective values), must be such that $P_{II} = L_{II}\left(\frac{1+k}{1-c}\right)$. For given levels of *k* and *c,* this creates an interdependence between all of the prices of consumption goods.

## 2.3. Prices of intermediate goods

So far, because it was in the logics of a model concerned with problematics of reproduction, we dealt only with final goods, be they consumption goods or fixed capital goods. The price of intermediate goods, comprising both wages and a profit margin included in their sale price, was supposed to be included in the price of final goods: the amount of wages (or money value) in the price of final goods therefore included the wages spent in the production of these intermediate goods as well as the profit margin contained in their sale price to final producers.

But intermediate goods must necessarily be introduced, if only because they can play an important role at macroeconomic level, by influencing both distribution and the price of final goods. Recalling the role played by the price of oil at world level is enough to justify that.

However it seems quite difficult, for the determination of their price, to devise a general formula similar to that of consumption goods or fixed capital goods, if only because a good number of intermediate goods are not specific to only one of both sections of the productive system, but on the contrary can be used simultaneously by both sections, to produce consumption goods as well as capital goods. Their influence on the final price of goods may also vary greatly form one intermediate good to the other, depending *inter alia* on their substitutability.

At any rate this influence depends on a case by case microeconomic analysis, which might imply sometimes to begin with a mesoeconomic one (like for oil products), which is not the object of the present work. It is the reason why we prefer to make the simple assumption that for each particular intermediate good its price is simply the sum of its money value (wages) and of a profit margin that we will call $\pi_i$ , and which is given as an exogenous data for each good. As it will be shown now this will be sufficient to go from values to prices.

# 3. From values to prices

## 3.1. A few calculations of values and prices

1) The value system

Let us first consider that there are *n* commodities, and that their values are the elements of a vector of values $V = [v_1 \quad ... \quad v_n]$. Since it is a row vector these elements are named $v_j$ (with *j = 1,..., n*). We continue to define the value $v_j$ of one unit of commodity *j* as the average social labor time it takes to produce commodity *j*. In order to transform values into prices without unnecessarily complicating the demonstration, we will assume that there are no intermediate commodities of the second type (as defined previously), and that we have *n* commodities, i.e. *k* intermediate commodities of the first type and *n - k* final commodities.

Since commodity *j* has been produced through the transformation of various other intermediate commodities, $v_j$ can be defined first as the sum of the direct labor time $l_j$ needed to produce commodity *j*, plus the indirect time corresponding to the value of the commodities that have been transformed to produce commodity *j*, which is the value of its means of production, i.e. the intermediate commodities *i* used in its production, with all of them supposed to be of the first type, and thus with *i = 1,…k.*

Thus $v_j = a_{1j}v_1 + a_{2j}v_2 + .... + a_{jj}v_j + ... + a_{kj}v_k + l_j$ or $v_j = \sum_i^k a_{ij}v_j + l_j$ (14)

In chapter 6 we had already determined a set of values for commodities, with a second method based on a system of two matrix equations (see section 3 of chapter 6).

Let us name $V_1$ the row vector of values of intermediate goods and $V_2$ the row vector of values of final goods, $A_{11}$ and $A_{12}$ the matrices of intermediate coefficients, $L_1$ and $L_2$ the vectors of quantities of direct labor for the production of intermediate goods and final goods respectively. The matrix equations that we can write for values are:

$$V_1 = V_1 A_{11} + L_1 \tag{15}$$

$$V_2 = V_1 A_{12} + L_2 \tag{16}$$

Which gives us, as solutions to these equations:

$$V_1 = L_1 \left(I - A_{11}\right)^{-1} = L_I \left(I + A + A^2 + ... + A^n\right) \tag{17}$$

$$V_2 = L_1 \left(I - A_{11}\right)^{-1} A_{12} + L_2 \tag{18}$$

2) The price system

To go from this system of average social values to a price system, we first have to transform it into a monetary system of money values, which only implies to set as the unit linking both systems the level of the average monetary wage per physical time unit of labor time, i.e. $\bar{w} = 1$ (knowing that $\bar{w}$ itself is not a scalar but has the dimension of a scalar divided by a time). Multiplying values by $\bar{w} = 1$ gives us the system of money values which is made of pure scalars and can be considered as the mirror of the former, and thus can be written exactly with the same notations. Thus the equations do not change, only the unit changes: it is a unit of physical time for the first system of average social values, and a unit of money for the system of money values, with $\bar{w}$ as the transformer.

The second thing that we need to do in order to determine the price system is to introduce profit margins, which cannot be determined at macroeconomic level, although they are submitted to macroeconomic constraints, but as we just indicated have to be provided by mesoeconomic and microeconomic analysis.

If we were doing that within the framework of standard economic theory, we would have first to use our profit margins to calculate profit rates, by dividing each amount of profit appearing in the production of a commodity by the amount of fixed capital used in its production. But we have not introduced so far the amount of the stock of accumulated fixed capital in our theoretical framework, because in fact this was not needed. As a kind of fallback solution we could consequently use the same formulation as Marx and Sraffa and decide to apply the profit margin to the stock of circulating capital. Indeed Marx considered that circulating capital was a part of what he called constant capital, and that it appeared in the denominator of the expression giving the profit rate. As for Sraffa, in most of “Production of commodities…” he calculates the profit rate only over the cost of circulating capital made of basic commodities, and thus the expression giving the price of a particular intermediate commodity is:

$$p_j = \left(a_{1j} p_1 + a_{2j} p_2 + .... + a_{jj} p_j + ... + a_{kj} p_k\right)\left(1 + r_j\right) + l_j w_j \tag{19}$$

However using this definition of prices would not be in line with the necessities of the reproduction of the system, as they were exposed in the previous chapters, which showed that the amount of profits and thus the profit margin incorporated in the price of each particular good must be such as to allow for each particular producer to buy all the fixed capital needed to carry out the production process of each particular good, on the one hand, and to buy some share of consumption goods available on the market, on the other hand.

These considerations imply that the expression giving the price of a particular intermediate commodity should rather be:

$$p_j = \left(a_{1j} p_1 + a_{2j} p_2 + .... + a_{jj} p_j + ... + a_{kj} p_k\right) + l_j w_j \left(1 + \pi_j\right) \tag{20}$$

Determining the price system implies to know all of the various profit margins $\pi_j$ (we must insist on the fact that they are not profit rates). Each one of these margins corresponds to the production of one particular good. We know that these $\pi_j$ are exogenous data, and as such they can be integrated in a system of matrices defining the price system. We just need first to define two simple matrices, one for intermediate goods and the other for final goods.

The first matrix concerns the *k* intermediate goods: it is a diagonal matrix (and thus a square one) of dimensions $(k,k)$, named $\Pi_1$, in which the elements $\pi_i$ of the diagonal correspond to the profit margins over wages in all the different branches which produce these intermediate goods. This matrix is therefore of the general form:

$$\Pi_1 = \begin{bmatrix} \pi_1 & 0 & .. & 0 \\ 0 & \pi_2 & .. & 0 \\ .. & .. & .. & .. \\ 0 & 0 & .. & \pi_k \end{bmatrix} \tag{21}$$

The second matrix concerns the *n - k* final goods. It is also a diagonal matrix of dimensions $(n-k, n-k)$, named $\Pi_2$, in which the elements of the diagonal are $\pi_i = \dfrac{1}{u_i\left(1 - c_i^*\right)} - 1$ for fixed capital goods - see equation (7), and $\frac{k_i + c_i}{1 - c_i}$ for consumption goods - see equation (13), i.e. the respective profit margins over cost prices in both sections. It is of the general form:

$$\Pi_2 = \begin{bmatrix} \pi_{k+1} & 0 & .. & 0 \\ 0 & \pi_{k+2} & .. & 0 \\ .. & .. & .. & .. \\ 0 & 0 & .. & \pi_n \end{bmatrix} \tag{22}$$

If $l_j w_j$ is expressed in units of average social labor multiplied by $\bar{w}$ (the average nominal wage), then it is possible to replace $L_1$ by $W_1$ and $L_2$ by $W_2$, just to emphasize the fact that we are in the domain of monetary variables.

The price system can therefore be determined without particular difficulties. Let us name $P_1$ the row vector of prices of intermediate goods, and $P_2$ the row vector of prices of final goods. $P_2$ combines the prices of fixed capital goods $P_I$ and of consumption goods $P_{II}$ :

$$P_2 = (P_I, P_{II})$$

Then on the basis of equation (17) we can write:

$$P_1 = A_{11}P_1 + W_1(I + \Pi_1) \tag{23}$$

$$P_1 = W_1(I - A_{11})^{-1}(I + \Pi_1) \tag{24}$$

$$P_1 = W_1(I - A_{11})^{-1} + W_1(I - A_{11})^{-1}\Pi_1 \tag{25}$$

Equation (23) can be written:

$$P_1 = W_1\left(I + A_{11} + A_{11}^2 + ... + A_{11}^n\right) + W_1\left(I + A_{11} + A_{11}^2 + ... + A_{11}^n\right)\Pi_1 \tag{26}$$

Before writing the equation giving the price of final goods we can observe that the first member on the right side of equation (25) provides us with the total amount of wages paid for the production of intermediate goods. For each good, we can get the amounts of direct wages paid for labor directly employed in the production of the various gods (i.e. $W_1 I$ ) and for the production of intermediate goods that are used at the next levels: $W_1 A_{11}$, $W_1 A_{11}^2$, and so on. For a given commodity, it is this total amount which would correspond to the amount of wages appearing in the equations of the reproduction scheme of the previous chapters.

We have also defined profit margins as appearing at each level of the production system, which implies that a profit margin at one level only applies to wages spent at this level, which are in any case the only wages known at this level. Therefore profits at one level cannot be calculated on wages at another level. The implications for writing the equation giving the price of final goods are that wages and profits realized at the last level of their production as final goods are additive to profits and wages which are part of the price of intermediate goods transformed into these final goods. Then on the basis of equation (18) we can write:

$$P_2 = P_1(A_{12}) + W_2(I + \Pi_2) \tag{27}$$

Combining equation (24) with equation (27), we obtain:

$$P_2 = \left[W_1(I - A_{11})^{-1} A_{12}\right](I + \Pi_1) + W_2(I + \Pi_2) = \tag{28}$$

$$P_2 = \left[W_1(I - A_{11})^{-1} A_{12} + W_2\right] + \left[W_1(I - A_{11})^{-1} \Pi_1 A_{12}\right] + W_2 \Pi_2 \tag{29}$$

The last equation (29) seems somewhat complicated, but it means only that the price of final goods can always be reduced to a sum of wages and profits:

1) First a sum of wages, i.e. $W = W_1(I - A_{11})^{-1} A_{12} + W_2$, comprising:
   a. Wages paid indirectly in the production of intermediate goods transformed into final goods : $W_1(I - A_{11})^{-1} A_{12}$,
   b. Wages paid at the last stage of production of these final goods: $W_2$ ;
2) Secondly a sum of profits:
   a. Profits realized in the production of intermediate goods: $W_1(I - A_{11})^{-1} \Pi_1 A_{12}$
   b. Profits realized at the last stage of production of final goods: $W_2 \Pi_2$

Because of the assumptions that we made, these profits are additive, and thus the overall profit margin over wages in the production of a particular final good can be considered as the weighted average of all profit margins over wages at all stages of production of intermediate goods transformed into this final goods as well as at the last stage of transformation into this final good. The weights are obviously the amounts of wages at each stage. This weighted average would correspond to the profit margins as defined in the previous chapters.

### 3.2. Some preliminary findings

What can be immediately seen from the final equations giving prices is the complete interdependence between prices and distribution.

1) First because prices are monetary prices and derive from values that are money values equivalent to wages, which themselves are a fundamental distribution variable, even though we have shown that wages depend on a number of parameters which are not of a purely economic nature. It is indeed impossible to deny that the average level of nominal wage, a fundamental parameter of the economic system, as well as the hierarchy of these nominal wages among various categories of labor and workers, depend on a number of elements of an ideological, political, institutional and sociological nature, as well as others of an economic nature, like the level of labor productivity or the level of unemployment.

2) Second, because prices depend also on the level of profit margins which are to some extent arbitrary, and have important degrees of freedom. However, as soon as prices of consumption goods are set by capitalists producing these goods, the most fundamental variables of distribution are simultaneously determined: the share of the value of the whole product going to workers and to capitalists is fixed as the rate of surplus-value, as well as the share of the value of consumption goods going to capitalists (who in any case get by definition the whole value of fixed capital). Similarly, as soon as capitalists selling fixed capital goods set their price, first the share of the value of consumption goods going to both categories of capitalists (producing either consumption goods or fixed capital goods) is determined - this value of capitalist consumption itself having been predetermined at the previous step. Simultaneously setting the price of fixed capital goods and the profit margins to which these prices correspond also determines the sharing of consumption goods among capitalists producing fixed capital.

The only way to change this distribution of values among the various stakeholders is to change prices, which, for a given level of wages, implies to change the various profit margins, on consumption goods as well as on fixed capital goods.

Thus we are lead to realize that distribution and prices are consubstantial: to change one changes the others and reciprocally. This makes us understand that the only economic variables which are totally invariant to changes in distribution are values, because values belong to another conceptual realm than prices, having been determined at the level of production. On the contrary prices, which cannot be determined as long as production has not taken place, are variables defined at the level of distribution and depending on distribution variables, i.e. the various levels of wages and profit margins, just as much as these distribution variables depend on prices.

It means that for a given system of production, to which corresponds one and only one set of values, there can be an infinity of possible distribution schemes, which distribute values among various stakeholders, each of these specific schemes being linked to a corresponding price system. Reciprocally each price system is linked to a specific distribution scheme. Therefore, without denying that production parameters obviously influence prices, they do that via values, and prices should be considered much more as distribution prices than as production prices. At a theoretical level, there is no such thing as production prices: there only are distribution prices, which is indeed a fundamental finding.

3) Third, it must be emphasized that it is only for intermediate goods that we have an equation corresponding to Sraffa's system of production prices for basic commodities, and which uses the Leontief inverse matrix $\left(I - A_{11}\right)^{-1}$. This shows that this method for calculating prices is only relevant for intermediate goods, to the extent that the level of money wages and the levels of margins over wages are given for each particular commodity. Indeed for Sraffa all final goods like consumption goods are luxury goods. As for fixed capital goods we have shown earlier that for Sraffa they are wrongly considered as not being final goods, but as a particular species of circulating capital. This specificity comes from their transferring their value (in fact their price) to products during more than one production cycle, but as we also showed Sraffa fails to introduce this special feature in his system in a non-contradictory way. We shall come back in more details on this distinction between various categories of commodities.

This use of the Leontief matrix is made possible only because in the equations giving values as well as prices for these intermediate goods we have the same goods and prices on both sides of the equations [see equations (15) and (23)], whereas for final goods their price appears only on the left side of the equations, and depends on other prices on the right side, i.e. precisely those of intermediate goods.

This should call for caution when using Leontief inverse matrix in systems with heterogeneous categories of commodities, as if all commodities could be considered as intermediate goods, which is not at all the case.

4) As a corollary to the previous finding, although it is obvious that there is production of fixed capital in our reproduction schemes, we have been able to devise both a value and a price system for commodities in a scheme of simple reproduction without this fixed capital transferring any value to the product. This did not prevent fixed capital to be sold, first to capitalists of section II producing consumption goods, and secondly to capitalists of section I producing fixed capital.

5) Finally, it is notable that the rate of profit does not appear in the equations defining prices, even for intermediate goods. This might seem weird, but derives from the fact that fixed capital is not an intermediate good and does not transfer its value to the product. This does not imply that we would be unable to define profit rates, and in fact we shall do that later. Indeed when we know profit margins we also know the absolute amounts of profits, and we could calculate the profit rate by dividing this amount by the price of the respective stocks of capital.

This shows nevertheless that the emphasis put by all economic literature (including Marx himself) on the rate of profit is grossly mistaken. We shall also come back to this point, but the fact that we do not need to know the profit rates to calculate prices draws our attention to the fact that we do not need either to know the price of the stock of fixed capital in order to determine the price system. It is true that the system determines the price of the various assets constituting fixed capital and which are produced during the period. It would also enable us to determine the price of the total fixed capital stock in operation, if we knew the lifetime of these various assets, as a technical data, and the price at which they were initially bought. But prior knowledge of this price of the stock of fixed capital is not directly necessary for determining the price system, in particular since fixed capital does not transmit its value to the product.

In fact these finding coincide well with the real world, where most businesses do not know precisely what is their own profit rate and moreover would usually be unable to calculate this profit rate, because they would have difficulties in the valuation of their fixed capital (something that is usually subcontracted to specialized external auditors).

6) If the rate of profit is therefore not an indispensable theoretical element, it follows that the $\pi_i$ s are the important variables. The role of microeconomic theory should be precisely to determine for each particular $\pi_i$ the reasons for its level and the various elements influencing its setting and its evolution, as well as its own influence on other prices. Individual prices, and therefore $\pi_i$ s, depend on multiple factors, many of them being external to economic theory, and of an ideological, political or institutional nature. Indeed their role as a distribution variable sometimes puts them at the heart of power struggles. However this does not preclude that in many cases it might be possible for a given commodity to isolate among these factors a number that would be of a purely economic nature, like:

- its production cost and the level and evolution of labor productivity in its production process ;
- the structure of its market and the degree of competition which it implies: its degree of monopoly or its more or less oligopolistic nature ;
- the main elements influencing the level of demand for it:
    - the level of income of consumers and its evolution;
    - its nature as a product: article of subsistence like a staple, or luxury good ;
    - its degree of substitutability to other products, the influence of fashion, etc.;
- and therefore all the main determinants of its supply and demand.

x x x x x

Before drawing all the lessons from the method of price determination which has been developed so far, if we want to progress towards a better theoretical understanding of microeconomic prices, we have to look at a final and important element of prices that has not been integrated so far in our method of price determination, and which is the rent of the land (or any other non-produced means of production). We touched upon the subject when we criticized the way used by Sraffa to integrate land (or any non-produced means of production) in his theory. But it is time now to go further and propose a coherent theory of rent, which will be done in next chapter.

# Chapter 15. A coherent theory of rent

This chapter elaborates on what has been demonstrated so far to expose a coherent theory of rent. The first section starts by recalling that that there are two kinds of rent: differential rent and absolute rent. To illustrate how rent is fixed, it exposes four different systems of prices for the same commodity (wheat), produced on lands of different "fertilities". The first system is characterized by zero global rent, the second one by zero total rent and equalization of prices by the State, the third one by a zero differential rent on marginal land, and the fourth one by a zero differential rent and a positive absolute rent on marginal land. The second section can thus analyze type II differential rent for Sraffa and Marx, i.e. the case of the production of wheat with different techniques on homogeneous land. All this allows in a third section to understand better the effect of rent on the price system and distribution.

## 1. The nature of rent, differential rent and absolute rent

Let us begin by recalling that rent is supposed to be an income corresponding to the "contribution to production" of the non-produced means of production, such as land, and that this definition includes other unproduced natural resources such as mines, or more precisely mineral deposits. In line with the model developed so far, the essential difference with fixed capital as a means of production is the unproduced character of these resources, which implies that it is impossible to calculate for them a cost of production as it is the case for machines. Another difference is that land is supposed to be immutable, and therefore has an infinite life span. But this difference does not apply to mines, which all end up being exhausted.

In Sraffa's book "Production of commodities…", there is an analogy between these unproduced means of production, such as land, on the one hand, and fixed capital, on the other hand, which consists in their presence on the left side of the equations defining the prices of production. But, unlike fixed capital, their absence on the right side of the system of equations does not, however, permit them to be included in the net standard product or in the composition of the standard commodity. This also explains why their price is calculated differently: fixed capital may appear among final goods on the right side of the equations defining prices, whereas land does not appear there, and its price is determined thereby from the rent it provides. Moreover, it is clear that there is no consumption of land in the production process, and therefore it is out of the question to treat it as an intermediate good, through any depreciation, which is not conceivable.

Finally, as we already saw, the price system is calculated by retaining, for each product for which land is used, the only equation in which this rent is zero, and where therefore land does not appear even on the left side of the equations. As has also been shown, for this reason, rent remains the stumbling-block of Sraffa's price system, with fixed capital - for other reasons, since the "fertility" of land varies with distribution and the system of prices, which makes it impossible to determine which is the land without rent without knowing first the price system! Once again, a theory is invalidated by the circularity of its reasoning.

These different characteristics of land, as well as our model, justify Marx when he points out that, as for fixed capital, the remuneration of land exists only because of its appropriation and more precisely because of the private property of land. This explains why in the real world rent is not zero even on the marginal land, where the cost of production is the highest, but where, nevertheless, there is a positive rent. This is what Marx calls absolute rent. If it did not exist, the landlord owning this marginal land would have no interest in renting it for cultivation. One could argue that this is due to the scarcity of land, and that if there is free land then the marginal rent must be zero. But this objection is based on a confusion over the meaning of the word "free", which means not only "not cultivated" but "not the object of appropriation". It does not take into account that precisely in the capitalist mode of production there is no free land, in the sense of land without owners. One cannot, therefore, cultivate even a "free" land, but this time in the sense of a land "not cultivated so far", and thus such as scarcity cannot manifest itself, unless a rent is paid to its owner.

Another proof of the fact that rents remunerate the ownership of unproduced means of production, and not their "productivity", comes from the fact that land, like fixed capital, does not create value: it transmits no value to the product. If land transmitted value, it is difficult to see why the marginal land would not transmit it, especially since with the extension of cultivated areas, a marginal land at a given moment ceases necessarily to be a marginal one later. In fact, like fixed capital, land is a catalyst, and in this case it is the support, or rather the substratum of the production process for agricultural products. This support can even be dispensed with in the case of so-called landless production, like hydroponic production, realized without any land. It follows that rent, as a remuneration for the ownership of land, is just like profits taken also from surplus-value, and consequently follows identical mechanisms.

In the model which has been described hitherto, and as we shall now show, the introduction of rent no longer runs up against the difficulties encountered in Sraffa’s system: there is no need to know first the land without rent, in order to determine only after that - and simultaneously, distribution and prices. To introduce rent into our system, we just need to start by assuming that there is only one commodity, e.g. wheat, produced with lands of different “fertility”, and with different quantities and proportions of circulating capital, considering at first that there is no rent. In order to introduce rent, we will proceed in stages, which will be done by assuming that there are always *n* kinds of lands of different “fertility”, all producing wheat with different production processes, and by examining successively four systems of equations in which rent exists but manifests itself in different ways.

### 1.1. The price of wheat with a zero total rent

We know that if we want to determine values and prices through systems of equations there can be only one single value and one single price for the same commodity, whatever the conditions of production. There can be only one equation giving the value of wheat, otherwise the system would be overdetermined. But this equation cannot in any case be the equation giving the conditions of production on the marginal land. This is true for the calculation of values, because it is the totality of the labor expended to produce wheat that must be taken into account, and not only that expended on the marginal land, which, moreover, is not known, since its determination depends on the price system. This is true also for the calculation of prices, because otherwise one would fall back into a circular reasoning: the system of equations allowing the calculation of prices assumes that the land without rent is known, but in order to know what is this land without rent, one must know the price system!

Therefore, the equation giving the price of wheat must necessarily take into account all the intermediate commodities and labor inputs used on all the lands simultaneously under cultivation to produce the total quantity of wheat. On this basis, we can calculate the value of wheat, like that of any other commodity, using equation (17) in the previous chapter:

$V_1 = L_1 (I - A_{11})^{-1}$ where the value of wheat is one of the values, e.g. $v_w$ (1)

Similarly, assuming that wheat is an intermediate commodity, its price can be calculated using equation (23) in chapter 14 (see p. 278), where – for the sake of comparison with Marx and Sraffa, and since it does not change the reasoning in any way, $\pi$ as the profit margin is replaced by $r$ as the corresponding profit rate, this profit rate being moreover applied to the price of circulating capital:

$$P_1 = W_1 L_1 + A_{11} P_1 (I + R_1) \tag{2}$$

In which the price of wheat is one of these prices, for example $p_w$:

$$p_w = \bar{w} l_w + a_w p_1 (1 + r_w) \tag{3}$$

$p_1$ and $a_w$ are vectors, $\bar{w}$ is the average nominal wage, taken as the wage unit ($\bar{w}$ = 1), $l_w$ is a given quantity of average social labor and $r_w$ is the average profit rate for the whole wheat producing branch.

This price of wheat will then and necessarily be an average price, like all other prices $p_i$. It should be kept in mind that each equation of the price system does not reflect the conditions of production of a given firm, but those of a whole branch, consequently composed of all enterprises producing together the total quantity of each commodity, each with its own production methods and techniques, which have no reason to be the same as those of all the other firms producing the same commodity.

At the enterprise level, firms therefore have different conditions of production, which means that they do not realize the average rate of profit, but a specific rate of profit, which may be higher or lower than the average rate of profit. In the case of commodities for which land is used as a means of production, the difference in production conditions and therefore in production costs may originate firstly from the heterogeneity of land, which for the same area, with the same amounts of circulating capital and labor, and with a given rate of profit, produce different quantities, and therefore have a different yield and fertility. This is even more true if the amounts of circulating capital and labor differ, as is generally the case.

For a given average price of wheat $p_w$ and average rate of profit $r_w$, the most “fertile” land with the lowest production cost will thus benefit from a rent corresponding to what Marx calls the differential rent of type I. But there is also a situation where production conditions may differ and the cost may be higher or lower per unit of quantity produced due to the use of different production techniques on a homogeneous land. This will make appear - for the firms with the lowest cost of production on the same land, what Marx considers to be also a rent, which he calls differential rent of type II.

This phenomenon can be translated into equations because, once prices are known, including the average unit price of wheat $p_w$ and the average rate of profit $r_w$ in the branch producing wheat, the equation giving the average price of wheat can be disaggregated within the price system, making land appear, and thus we can write as many equations as there are lands of different “fertility”, so that we have, for *n* lands of distinct "fertility":

$$\bar{w}l_1 + a_1 p_1 (1 + r_w) + \rho_1 \Lambda_1 = p_w$$

$$\bar{w}l_2 + a_2 p_1 (1 + r_w) + \rho_2 \Lambda_2 = p_w$$

--------------------------------------------

$$\bar{w}l_g + a_g p_1 (1 + r_w) + \rho_g \Lambda_g = p_w \qquad \text{with } \rho_g = 0 \qquad \text{(System 1)}$$

--------------------------------------------

$$\bar{w}l_{n-1} + a_{n-1} p_1 (1 + r_w) + \rho_{n-1} \Lambda_{n-1} = p_w$$

$$\bar{w}l_n + a_n p_1 (1 + r_w) + \rho_n \Lambda_n = p_w$$

In this system $a_i$ and $p_1$ which are known vectors, $\Lambda_i$ are the quantities of land of different fertility in units of area (hectare for example), and $\rho_i$ is the rent per unit of area. It should be emphasized that the system has been standardized to show only the unit prices, so that the distinct quantities $\Lambda_i$ are the inverse of the physical return of the land concerned: this is the required land area in order to produce a unit quantity of wheat, which therefore varies from one land to another according to its fertility.

This system of equations is interesting, first of all because it makes it possible to calculate without difficulty the amount of the different rents $\rho_i$, all the other variables being known. It is also because the total rent is necessarily nil (it is recalled that the absence of rent was a necessary condition for calculating prices, which are average prices). This implies that the following equation is always satisfied: $\sum \rho_i \Lambda_i = 0$

But it immediately follows that there are lands where rents are necessarily negative, and that this is the *sine qua non* condition for the simultaneous existence of positive rents, since the sum of the two kinds of rents must necessarily cancel out each other. Thus, land can be classified according to the level of its rent, ranging from the one where the positive rent is the highest, noted $\rho_1$, to that where the rent is the most negative, noted $\rho_n$, through a land where the rent is null, for instance for $i = g$ (with consequently $\rho_g = 0$). We have $\sum_{i=1}^{i=g} \rho_i \geq 0$, and we have also necessarily $\sum_{i=g+1}^{i=n} \rho_i < 0$: the sum of the positive rents is equal to the sum of the negative rents. It should be emphasized that the order of land fertility is deduced from the order of rents, but that the land marked $\Lambda_1$ is not necessarily the one where the physical yield (in quantity of wheat per hectare) is the highest. In other words, $\Lambda_1$ is not necessarily the smallest area, and the property Λ1 < Λ2 < .... < Λg < … < Λn-1 < Λn is not necessarily verified. This is because in each equation we find also the expressions $\bar{w}l_i + a_i p_1(1 + r_w)$, which vary with distribution.

One might think that such a system is absurd because it is devoid of all reality: it is hard to see how landlords could accept to pay a negative rent in order to rent their land! But we will show through two additional and different examples that this system is not absurd in all cases, because everything depends on the nature of land ownership.

The first example corresponds to the case where all of the capitalist enterprises of the branch (the "farmers") have redeemed the land they exploit to the landowners and sell all their wheat at the price $p_w$. In this case, everything happens as if the best performing firms in the sector had a rate of profit higher than the average rate of profit, and thus realized an over-profit, and as if the least performing firms, with higher production costs, obtained a rate of profit lower than the average rate of profit, realized by the average “farmer” or firm for which the rent is zero. In fact, we are in a situation where there is no rent, strictly speaking, but because there are no "pure" landowners. The system, however, can work, since all firms nevertheless make a profit (otherwise they would not exist), and because those whose profits are reduced (through negative rents) by the lesser fertility of the lands that they own necessarily bought these lands at prices lower than that of other lands of greater fertility. The reverse is also true: the super-profits offset the higher prices paid for the purchase of these other lands.

### 1.2. The price of wheat with a zero total rent and equalization by the State

This situation corresponds to a second example, in which all lands are nationalized, and consequently owned by the State, which leases them to “farmers”, or capitalist enterprises which use them for the production of wheat. Assuming that the State wishes to maintain the price of wheat at the $p_w$ level corresponding to the average cost of production, it is sufficient for this purpose, first, to forego the benefit of an overall net income for the lands it owns, and then to put in place an equalization scheme. In this case, the State will tax all the companies for which the rent is positive (which make an additional profit over the average profit rate), and pay a subsidy to all the companies for which the rent is negative, taxes $(-\rho_1, \ldots, -\rho_{g-1})$ and subsidies $\left(+\rho_{g+1}, \ldots, +\rho_n\right)$ being equal and opposite to the amount of each rent (the amounts are seen from the point of view of firms, which explains the minus sign). The total amount of taxes is equal to the total amount of subsidies. The net cost to the State is therefore nil, but although it is the owner of the land, it pays nothing to itself. The system of equations then becomes, with $\sum_{i=1}^{g-1}\rho_i > 0$ , $\sum_{i=g+1}^{i=n}\rho_i < 0$ , and $\rho_g = 0$:

$$\bar{w}l_1 + a_1p_1(1 + r_w) + (\rho_1 - \rho_1)\Lambda_1 + \boldsymbol{\rho_1\Lambda_1} = p_w$$
$$\bar{w}l_2 + a_2p_1(1 + r_w) + (\rho_2 - \rho_2)\Lambda_2 + \boldsymbol{\rho_2\Lambda_2} = p_w$$

$$\dots\dots\dots\dots\dots\dots\dots\dots\dots\dots\dots\dots\dots\dots\dots\dots\dots \qquad \text{(System 2)}$$

$$\bar{w}l_g + a_gp_1(1 + r_w) + (\rho_g - \rho_g)\Lambda_g + \boldsymbol{\rho_g\Lambda_g} = p_w$$
$$\dots\dots\dots\dots\dots\dots\dots\dots\dots\dots\dots\dots\dots\dots\dots\dots\dots$$
$$\bar{w}l_{n-1} + a_{n-1}p_1(1 + r_w) + (\rho_{n-1} - \rho_{n-1})\Lambda_{n-1} + \boldsymbol{\rho_{n-1}\Lambda_{n-1}} = p_w$$
$$\bar{w}l_n + a_np_1(1 + r_w) + (\rho_n - \rho_n)\Lambda_n + \boldsymbol{\rho_n\Lambda_n} = p_w$$

It can be seen that for each category of land the tax or subsidy cancels the amount of rent, this amount being positive when $\sum_{i=1}^{g-1}\rho_i > 0$ , or negative when $\sum_{i=g+1}^{i=n}\rho_i < 0$ . Rent is therefore no longer an element of price $p_w$, while these taxes and subsidies create for the State a corresponding positive or negative income, noted in bold characters ($\boldsymbol{\rho_1\Lambda_1}$, $\boldsymbol{\rho_2\Lambda_2}$,..., $\boldsymbol{\rho_g\Lambda_g}$, ..., $\boldsymbol{\rho_{n-1}\Lambda_{n-1}}$, $\boldsymbol{\rho_n\Lambda_n}$), which replaces rent as an element of the price, knowing that both global rent and global State income are zero. Therefore the price of wheat $p_w$ remains unchanged. Equalization has resulted in a mere transfer of positive or negative incomes from enterprises, for which there is no longer any rent, to the State, whose total income is zero, as is total rent. Thus all firms are equal as regards the rate of profit that they can realize.

These two examples where the price of wheat remains the average price $p_w$ and where global rent is nil clearly show that the previous system is not absurd and can exist under certain conditions. Moreover, there are cases where equalization schemes of this type can be managed by firms themselves, without necessarily involving the State through taxes and subsidies.

### 1.3. The price of wheat with a zero differential rent on marginal land

In the real world, if the State is not the owner of any land, then lands are the property of landowners who lease them to capitalist enterprises. It is clear, first of all, that there can be no negative rent for any landowner, because it would mean that they would have to pay for their land to be used! This implies, first of all, that price $p_w$ is fixed at a level $p'_w$ which totally suppresses the negative rents which appeared in the previous price system, instead of letting these rents appear on the least fertile lands before being compensated by the State.

This price $p'_w$ must be such that it cancels the highest negative rent $\rho_n$ on the least fertile land, noted $\Lambda_n$. This implies that the price $p_w$ be increased by an amount $-\rho_n\Lambda_n$, equivalent to its absolute value $|\rho_n|\Lambda_n$, in order to cancel the negative rent on the marginal land. Since the price of wheat is the same regardless of the category of land, this increase must apply to all lands. The price system must therefore become:

$$\bar{w}l_1 + a_1p_1(1 + r_w) + \rho_1\Lambda_1 + |\rho_n|\Lambda_n = p'_w \qquad \text{thus } p'_w = p_w + \rho_i\Lambda_i + |\rho_n|\Lambda_n$$

$$\bar{w}l_2 + a_2p_1(1 + r_w) + \rho_2\Lambda_2 + |\rho_n|\Lambda_n = p'_w \qquad \text{(we recall that } \rho_{i=1}^{g-1} > 0 \text{ and } \rho_{i=g+1}^{i=n} < 0\text{)}$$

…………………………………………….

$$\bar{w}l_g + a_gp_1(1 + r_w) + \rho_g\Lambda_g + |\rho_n|\Lambda_n = p'_w \qquad \text{with } \rho_g = 0$$

……………………………………………. (System 3)

$$\bar{w}l_{n-1} + a_{n-1}p_1(1 + r_w) + \rho_{n-1}\Lambda_{n-1} + |\rho_n|\Lambda_n = p'_w$$

$$\bar{w}l_n + a_np_1(1 + r_w) + (\rho_n - \rho_n)\Lambda_n = p'_w \Rightarrow \bar{w}l_n + a_np_1(1 + r_w) + 0 = p'_w$$

For the marginal land at the price level $p'_w$, with $p'_w = p_w + |\rho_n|\Lambda_n$, the rent is therefore nil. On the other hand, firms using land $\Lambda_{n-1}$ will pay a rent equal to $\rho_{n-1}\Lambda_{n-1} + |\rho_n|\Lambda_n$, and so on: firms using land $\Lambda_1$ will pay a rent $\rho_1\Lambda_1 + |\rho_n|\Lambda_n$.

### 1.4. The price of wheat with a zero differential rent and a positive absolute rent on marginal land

The fact that the rent on the least fertile land is zero in price system 3 shows, however, that this case does not yet correspond to a situation where land is appropriated by landowners, which implies that the owner of the least fertile land refuses to provide it to a farmer for free and therefore without any rent. The landowner in question must therefore receive a rent, which corresponds to what Marx calls the absolute rent. Let us call $\rho_a$ the level of this rent per area unit. This rent can be collected only if it is passed on in price, and since there is only one price of wheat, therefore all the landowners will benefit from this rent, at the same level for all. The price system then becomes: $p''_w = p'_w + \rho_a\Lambda_i$

$$\bar{w}l_1 + a_1p_1(1 + r_w) + (\rho_1 + \rho_a)\Lambda_1 + |\rho_n|\Lambda_n = p''_w \qquad \text{we have } p''_w = p'_w + \rho_a\Lambda_i$$

$$\bar{w}l_2 + a_2p_1(1 + r_w) + (\rho_2 + \rho a)\Lambda_2 + |\rho_n|\Lambda_n = p''_w \qquad \text{with } \rho_a > 0 \text{ and } \rho_n < 0$$

……………………………………………………. (System 4)

$$\bar{w}l_g + a_gp_1(1 + r_w) + (\rho_g + \rho_a)\Lambda_g + |\rho_n|\Lambda_n = p''_w \text{, with } \rho_g = 0$$

……………………………………………………

$$\bar{w}l_{n-1} + a_{n-1}p_1(1 + r_w) + (\rho_{n-1} + \rho_a)\Lambda_{n-1} + |\rho_n|\Lambda_n = p''_w$$

$$\bar{w}l_n + a_n p_1(1 + r_w) + \rho_a \Lambda_n = p''_w$$

What lessons can be learned from this analysis? The first of these lessons is that rent and profit are revenues of the same nature: this is shown by the case in which capitalist enterprises are themselves owners of the land they exploit, in which case the introduction of land into their means of production may leave prices and the average rate of profit unchanged and be interpreted simply as the appearance of additional differences between the actual rates of profit of the different enterprises, other than those resulting from the difference of production techniques used to produce the same good, a situation which must also be addressed.

## 2. Type II differential rent for Sraffa and Marx

This example and this remark lead us to the case of the production of wheat with different techniques on homogeneous land. This case is dealt with by Sraffa in § 87 of "Production of Commodities by Means of Commodities", and leads him to state in § 88: "While the scarcity of land thus provides the background from which rent arises, the only evidence of this scarcity to be found in the process of production is the duality of methods on lands of the same quality: if there were no scarcity, only one method, the cheapest, would be used on the land and there could be no rent" (Sraffa, 1960, p. 91).

This remark from Sraffa is totally erroneous. Indeed, if these homogeneous lands are cultivated by one and the same capitalist enterprise (a single "farmer"), the diversity of the techniques used by this farmer does not matter. This diversity may arise from historical reasons, which can come from differences in the rhythm of introduction of new machines: the average rate of profit of such a firm depends on the average cost of these different techniques, so that there can be only one equation for this quality of land in the price system, which remains unchanged, and the rent remains that already determined for this category of land.

It is only if several different capitalist firms are cultivating these homogeneous lands of the same fertility, and when each of them is using a different technique, that the system of equations for the corresponding lands must then be modified. Let us suppose that these lands have the same fertility, and that this corresponds to a land of rank $\Lambda_f$. Let us also assume that there are three firms rated 1, 2 and 3 which cultivate this land and use different techniques materialized by different vectors $a_{f1}, a_{f2}$ and $a_{f3}$, and different quantities of labor $l_{f1}$, $l_{f2}$ and $l_{f3}$. The equation hitherto unique for land $\Lambda_f$ becomes:

$$\bar{w}l_{f1} + a_{f1}p_1(1 + r_{w1}) + (\rho f + \rho_a)\Lambda_f + |\rho_n|\Lambda_n = p''_w$$

$$\bar{w}l_{f2} + a_{f2}p_1(1 + r_{w2}) + (\rho_f + \rho_a)\Lambda_f + |\rho_n|\Lambda_n = p''_w \qquad \text{(System 5)}$$

$$\bar{w}l_{f3} + a_{f3}p_1(1 + r_{w3}) + (\rho_f + \rho_a)\Lambda_f + |\rho_n|\Lambda_n = p''_w$$

The price $p''_w$ is given and the rent that has already been fixed for the category of homogeneous land $\Lambda_f$ has no reason to change, since overall conditions of production have not changed. The only thing that can vary is the rate of profit: we can see in this system of equations that the average rate of profit existing for the three enterprises considered as a whole has been replaced by three separate profit rates, $r_{w1}$, $r_{w2}$ and $r_{w3}$ for each of the three firms, respectively. We thus verify that rent is not explained by scarcity: besides, the lands of fertility *f* may have any rank in the order of fertility. These homogeneous lands do not necessarily correspond to the marginal land, whereas in the example of Sraffa this additional hypothesis must be made so that its reasoning introducing scarcity can be pursued, or there is only one and unique category of land, which is even less realistic. But our reasoning remains valid even if land $\Lambda_f$ is replaced by marginal land $\Lambda_n$.

Indeed, Sraffa's mistake here is to confuse profit and rent: whether or not in the production process there are non-produced means of production such as land, the difference in production techniques within a branch producing the same product is inevitable, unless we assume that there is only one firm per branch, which would be absurd. Even the fact that a given firm employs only one technique is surely infrequent, since fixed capital is rarely replaced in a block and at one point in time, which at all times leads to the coexistence of techniques of a slightly different nature in the same firm. Consequently, it is the difference in production techniques, and not scarcity, which explains the differentiation of profit rates within the same industry.

This analysis allows us to conclude by validating our initial statement: just as profit does not reward productivity, but the property of means of production, rent as well rewards the property and not the “fertility” or “scarcity” of land or any other non-produced means of production. Profit and rent are therefore revenues of the same nature, since both are levied on surplus-value, and suppose the redistribution of this surplus-value, of which they represent distinct parts. This implies that owners of non-produced or rent-producing means of production are competing, whether they realize it or not, with the other owners of means of production for the redistribution of this surplus value.

This redistribution of surplus-value, however, is effected in different ways. Profit is taken to allow the replacement of fixed capital supposed to have been consumed in the production process, through the depreciation of capital in use, but it has been shown that it is also used to a large extent for the consumption of capitalists. Rent is levied by the owners of unproduced means of production, who may also be capitalists, who are not prohibited from purchasing land or other unproduced means of production. With respect to the use of rent under the model developed so far, rent can be consumed or used for the purchase of fixed capital, being understood that in the latter case landowners also become capitalists. Moreover, and as we have shown, rent takes two different forms, which are of the same nature, but which can nevertheless be analytically distinguished: differential rent and absolute rent, and on this point one can only partially correct Marx.

Differential rent is due to the heterogeneity of land, to their difference in fertility. Knowing that the order of land classification may vary according to the change in the rate of profit and the price system, there is nothing such as an absolute fertility, which is why we must be careful when using this term. But the phenomenon corresponds in any case to what Marx called differential rent of type I. With regard to what Marx called differential rent of type II, resulting from differences of production techniques on homogeneous lands, and which Sraffa considers to be a rent of scarcity (of this homogeneous land), on the other hand, it has been demonstrated that this rent does not exist as such, because it actually resolves in profit rate differentials. Marx and Sraffa are therefore both wrong on this point.

Of course, this position is partly a question of vocabulary, but if we agreed to name rent what is in fact a difference between rates of profit in the production of the same good, then there would be rents everywhere. This is nevertheless the position of a few economists, such as Jean-Marie Huriot, in an article of synthesis, published in 1983: "*Rentes différentielles et rentes absolues: un réexamen"* ("Differential rents and absolute rents: a review"). It is true that for this economist, rent always refers to scarcity, and that scarcity is everywhere, whereas in the analysis which has just been conducted rent has been explained without needing to introduce the concept of scarcity. It is therefore essential to reserve the term of rent not only to the case of unproduced means of production but also to cases where the difference in production costs arises from the heterogeneous nature of these non-produced means of production, and not just from differences in the nature of the techniques employed. Otherwise, even with the existence of unproduced means of production, it is better to use the term quasi-rent, which would therefore correspond to the differential rent of type II for Marx, or to the case of homogeneous land with different techniques for Sraffa.

An example will make it easier to understand, that of the rent of location. If it is simply what Huriot names in his article the distance-differential rent, which is noted by him as $\rho\delta$, this "rent" arises from the greater or lesser distance from a land in relation to the market (the furthest being the rent-less land). However if we go back to our example of wheat production, this "rent" is in fact based on the simple addition of a transport cost proportional to this distance among the inputs of wheat production. The greater the distance, the more distant the land and the higher the transport cost, hence the lower the rent, to the point of canceling out at the maximum economically possible distance, that of the marginal land. But this higher transport cost actually corresponds to nothing else than a change in the production technique, in which transport cost has an increasingly important technical coefficient, as the distance of the cultivated land increases. This reduces the rate of profit in relation to the average rate of profit on all the lands concerned, which may be either homogeneous or heterogeneous: one can have lands otherwise highly fertile at a great distance, and less fertile lands nearby, or homogeneous lands at different distances.

In these different cases, since it is an unproduced means of production - land, which is used, and since it is one of its characteristics (its distance from the market), which is involved, we can admit to use the term quasi-rent for surplus profits resulting from the difference in transport costs with respect to the most distant land. On the other hand, if any manufactured good is produced at different distances from the market, the difference in transport costs is equivalent, from a formal point of view, to a difference in techniques. But if there is no use of unproduced means of production, the differences in the rate of profit resulting from this difference in techniques do not, in our opinion, justify the use of the term quasi-rent in relation to and as a kind of explanation for these differences.

In both cases, however, what also justifies not to use the term "rent" is that the price of the produced commodity remains unchanged, as there is no particular reason for it to change, whereas in the case of rent proper we saw well that its existence, whether differential or absolute, necessarily raised prices. On the other hand, in the case of a pseudo-rent of location, there is no rent strictly speaking because the property of land is not the property of distance. Moreover, it is sufficient for the market to change its location for quasi-rents or profit rate differentials to be modified, without any change in property rights.

The analysis performed so far now allows us to provide two new principles concerning rent.

**Principle 24***:* Rent is the income that comes from the property of non-produced means of production, such as land or mines, and is collected by the owners of these means of production. There are two kinds of rent: differential rent and absolute rent. Differential rent comes only from the heterogeneity of each type of non-produced means of production, i.e. from differences in their intrinsic characteristics, and varies according to this heterogeneity. Because of its origin in property rights, absolute rent is the rent perceived even on the least "productive" (i.e. with no differential rent) of a particular type of non-produced means of production (otherwise it would not be rented). It is also perceived on all the other non-produced means of production of the same type.

**Principles 25***:* There is no rent as such coming from the heterogeneity of the techniques which can be employed by various firms using homogeneous non-produced means of production, which already pay a rent for their use. This heterogeneity only results in profit rate differentials. However, owing to the use of non-produced means of production, the corresponding differences in production costs from the average cost can be called "quasi-rents", to distinguish them from other types of profit differentials.

## 3. The effect of rent on the price system and distribution

It is system of equations (4) above which fully accounts for the effects of rents on the price system. Compared to the price system that would prevail in the absence of rent and would result in an average price of wheat equal to $p_w$, it shows that taking rent into account leads to a price increase, such that the unit price of wheat $p_w$ is increased by the amount of rent, and thus becomes:

$$p''_w = p'_w + \rho_a \Lambda_i \qquad \text{with } \rho_a > 0\,,\ \underset{i=1}{\overset{g-1}{\rho_i}} > 0 \text{ and } \underset{i=g+1}{\overset{i=n}{\rho_i}} < 0 \qquad (4)$$

And since we have: $p'_w = p_w + \rho_1\Lambda_1 + |\rho_n|\Lambda_n$ (see system of equations 3), we get:

$$p''_w = p'_w + \rho_a\Lambda_i = p_w + (\rho_i + \rho_a)\Lambda_i + |\rho_n|\Lambda_n \tag{5}$$

Equation (5) shows that in the real world – apart from the marginal non-produced means of production, on which there is no differential rent, there is no such thing as a "pure" rent, because all rents are a combination of a differential rent and an absolute rent.

We recall that $\rho_a$ is the level of absolute rent, which must be paid even to the owner of the least fertile land to make him decide to rent his land, and that $\rho_n|\Lambda_n|$ is equal in absolute value to the amount of the deficit (which we called negative rent) that would appear on marginal land if the price of wheat were set at the level corresponding to the average rate of profit and the absence of rent. The level of $\rho_n$ is not arbitrary, since it is determined for a given price system and a given rate of profit by the conditions of production on the least fertile land. On the other hand, the level of $\rho_a$ depends on what might be called the balance of power between owners of non-produced means of production and capitalists who pay the rents. In the real world this balance of power is generally arbitrated or regulated by the State, but in the case of some primary commodities like oil, it can also be the subject of negotiations between States and oil companies, and even of conflicts between States, as shown by the way oil fields are exploited.

The effect of rent on the rate of surplus-value and on the redistribution of surplus-value is different according to whether rent appears in the production of consumer goods or in the production of fixed capital, i.e. in Section II or in Section I of the production system.

With respect to a situation where there is no rent, and with the assumption of a constant profit rate, the levy of rent in Section II results in an increase in the overall price of consumer goods, which is known to be equal in the absence of rent to $Y_{II} = L_{II} \cdot \frac{1+k}{1-c}$. Since $L_{II}$ and $k$ do not vary, the introduction of rent is equivalent to an increase in $c$ (the share of the value of consumption going to capitalists) by an amount $c'_{II}$ (with $c + c'_{II} = c'$). This amount $c'_{II}$ is the share that the consumption of rent receivers in Section II represents in total consumption. The overall price of consumer goods thus becomes:

$$Y'_{II} = L_{II} \cdot \frac{1+k}{1-(c+c'_{II})} \tag{6}$$

In Section I, the drawdown of rent results in an increase in the price of fixed capital, which is known to be equal in the absence of a rent to $Y_I = \frac{L_I}{(1-a)(1-c^*)}$, with again a constant profit rate. The introduction of rent in this section can only take the form of an increase in total rent corresponding to $c'_I$, i.e. the additional share of total consumption obtained by rent receivers in section I. The price of fixed capital produced in Section I then becomes:

$$Y_I = \frac{L_I}{(1-a)\left[1-\left(c^{*}+c^{*}{'}\right)\right]} \qquad (7)$$

As for the rate of surplus-value, that is $k_c = \frac{k+c}{1-c}$, only the rent affecting the price of consumer goods modifies it, by increasing it to $k_c' = \frac{k+(c+c')}{1-(c+c')}$, and we know that $k_c'$ is also the profits/wages ratio in Section II. The consumption of rentiers in Section II is therefore levied on the consumption of workers. On the other hand, rent levied in section I does not modify the rate of surplus-value, but affects the distribution of surplus-value to the detriment of capitalists: the consumption of rentiers in Section I is deducted from the total consumption of capitalists, through the rise in the price of fixed capital caused by the collection of the rent.

Now that these clarifications have been made, we shall try to answer briefly the question of what the distinct effects of an increase in differential rent or absolute rent may be.

In the first place, an increase in absolute rent without an increase in the differential rent implies that the use of non-produced means of production remains unchanged. To take the example of wheat, there is no increase in wheat production and therefore in the cultivation of new, less fertile land. We are then brought back to the analysis which has just been made of the introduction of rent, which necessarily includes an element of absolute rent, in a system in which there was no rent. We have just seen that the introduction of rent comes down to levying a share of surplus-value on the sale of any good or service by selling it with an unchanged value but at an increased price which integrates the amount of the rent. As such, this price is therefore situated above the price which would result from the application and the collection of the average pre-existing rate of profit for the production system and the section concerned. This is not linked to an increase in the cost of production itself, but is due to the fact that the producer of that good or service is subject to the market power of the owner of unproduced means of production, with the assumption that he can pass on the amount of this rent in his sale price.

The mechanism for levying this amount is in any case similar to that which makes it possible to realize a surplus-value in the form of profits through the fixing of prices, which clearly shows the common nature of rent and profit, either in the form of average profits or of quasi-rents. Depending on whether the commodity for which the pricing power that will generate a rent or quasi-rent is produced in Section I or Section II, levying an absolute rent or a quasi-rent actually amounts, as we have seen, to an increase in the value of variables $c$ or $c^*$, and thus increases the price of consumer goods or fixed capital at global level.

We also just saw that the increase of $c$ and hence the increase of $\gamma$ leads to an increase in the rate of surplus-value $k_c$. The result on the overall rate of profit will be analyzed later, but the object of this work is not to carry out a detailed analysis of all the consequences of an increase in the amount of absolute rent. It is important to note, however, that the appearance or increase of an absolute rent or a quasi-rent in a production system can only have a disruptive effect on the realization of the product at global level, since such a phenomenon necessarily modifies the conditions for the sharing of surplus-value and hence for the realization of profits, which depend in part, as we must recall, on the successive expenditure (in the logical sense) of previously realized profits.

In the case of workers, the increase in surplus-value necessarily triggers a reduction in their share in total consumption (hence in their demand for consumer goods), and in a decrease in the value of labor-power. As regards now capitalists, there is no particular reason to imagine that the collectors of absolute rents and quasi-rents spend them in conditions which make it possible to keep unchanged the reproduction of the system.

In particular, if the collection of a rent or a quasi-rent concerns the sector producing fixed capital, two cases are possible. If capitalists of section I can pass it on through their price, they will keep their profits at an unchanged level. But if they cannot pass it on through their own price, the result is a corresponding reduction of their profits for capitalists of the same section. To avoid it, they will therefore seek to pass on the corresponding levy by reducing wages.

As for capitalists of section II producing consumer goods, in reaction to the rise in the price of fixed capital resulting from the appearance or increase of a rent in section I, they are able to directly increase the rate of surplus-value by increasing the price of consumer goods. But if they cannot pass on the rent increase through an increase in their price to maintain their own profits, they too will seek to reduce wages. Let us conclude, then, that any increase in a rent or a quasi-rent can only have an inflationary effect and create a strong incentive for a fall in wages, and the cumulative effect of these two phenomena must inevitably lead to a fall in the consumption of workers.

As regards the effects of an increase in differential rent, they are complex, and therefore will be analyzed only very briefly here.

To limit ourselves to a few preliminary considerations, the effects of differential rent are not necessarily those foreseen by Ricardo in his “Essay on Profits”. Indeed, the increase in differential rent linked to the use for production of new non-produced means of production, in this case linked to the cultivation of new lands, can be analyzed as resulting from an increase in the average cost of the commodity concerned.

Indeed, since the price integrating the differential rent is $p'_w = p_w + |\rho_n| \Lambda_n$ this means that on the new marginal land the negative differential between the cost of production and the average price (excluding rent) i.e. $p_w$, has increased. Moreover, even with an unchanged production technique on a new cultivated land, the decrease in the quantity of wheat produced resulting from the lower fertility of the land is equivalent to an increase in the unit value of the commodity concerned, which comes as well from an increase in direct labor (l) than indirect labor (vector $a$ ) used in its production. In both cases, this increase in value reflects an increase in the cost of production. But this does not have the same effect according to the section concerned: it raises $L_I$ if the non-produced means of production concerned is used for the production of fixed capital, or $L_{II}$ if the non-produced means of production concerned is used in the production of consumer goods.

In the first case we have an increase in the rate of primary surplus-value $\frac{L_I}{L_{II}} = k$, and in the second case a decrease of this rate! Since $k_c = \frac{k+c}{1-c}$, we can deduct from this that, in the first case where the increase in the differential rent occurs in Section I, there is an increase in $k_c$. Its influence on the rate of profit will be touched upon later. The parallel increase in variable $c^{*\prime}$ has no effect on the rate of surplus value.

On the other hand, in the second case, where the increase in differential rent occurs in Section II, things are not as simple. Since $k'_c = \frac{k+(c+c')}{1-(c+c')}$, the decrease in $k$ resulting from the increase in $L_{II}$ goes indeed in the direction of a decrease in $k'_c$. But if capitalists of section II can raise their prices up to the increase in differential rent, by passing it on to the prices of consumer goods, then the corresponding fall in the rate of surplus-value $k'_c$ is offset by the increase in c', the share of the consumption of rentiers in total consumption, which is added to variable c whose level remains unchanged. The question of knowing whether the final result will be a fall or an increase in the rate of surplus-value $k_c$ depends on the values taken by $k$ and $c'$ !

It is therefore only in the first case, where differential rent increases in section I, that one is assured to be in the "Ricardian" situation where an increase in differential rent and the rate of surplus-value entails a decline in the rate of profit. In the second case, where the differential rent increases in Section II, a fall in the rate of profit remains possible, but not certain, if the increase in $L_{II}$ and the decrease in $k$ are strong enough to prevail on the rise of c'.

However, and in both cases, an increase in differential rent necessarily entails a redistribution of surplus-value, and therefore a modification of its distribution to the detriment of capitalists and in favor of rentiers. These preliminary considerations as regards the consequences of differential rent will thus be no more developed. However they allow us to state two last principles regarding rent.

**Principle 26**: Actual rents are always a combination of differential and absolute rents. Whatever their combination, the nature of rent is not different from that of profits: rent is also a transfer income, which is levied as a part of total surplus-value. Changes in differential rents can result from an exogenous change in the price system. They also correspond to changes in the scale of production, implying the use of additional non-produced means of production (in the case of an increase). Such changes therefore always entail a change in the value of the product and the price system.

**Principle 27**: Theoretically, for a given and fixed value of the product, if an increase in rent (hence absolute rent) could leave prices unchanged, the amount of surplus-value would not change, and the amount of rent would be deducted from profits. In practice, such an increase will have repercussions on prices, whose magnitude depends on the balance of power between the three involved groups of agents: rentiers, capitalists and workers. The ultimate effect on the amount of surplus-value and its distribution between rentiers and capitalists will depend on the balance of power between these groups, under the arbitration of the State.

Now that we have completed, with the examination of the question of rent, the exploration of the various determinants of prices, it is time to take stock of all the theoretical consequences of the different results which we have achieved in the last two parts of this book, as regards the conceptual status of values and prices. This will be the object of the next chapter.

# Chapter 16. Some theoretical lessons from the models

We shall in this chapter go back to the main lessons learned from the two models presented in previous chapters, beginning with chapter 11. These models allowed for a step by step approach to the process of circulation. They were of a great heuristic value for understanding how it works in the capitalist mode of production, a process which starts with the distribution of primary incomes paid in money at the end of the production process, and continues with the distribution of incomes and commodities produced among the various stakeholders. Among the lessons learned, a first one concerns the nature of fixed capital, and is addressed in a first section. A second one concerns the articulation of values and prices, again examined in a second section. A third one concerns the mode of formation of profits: the third section shows that there is necessarily some kind of conflict between capitalists themselves for the sharing of profits. A particular case concerns the sharing of profits between the productive system and the banking system. In a final and fourth section, all this allows us to propose some ideas about what could be a truly microeconomic price theory.

## 1. The nature of fixed capital

### 1.1. Fixed capital and circulating capital

Fixed capital is defined by most economists in opposition to circulating capital. This is a source of inaccuracy, which is symptomatic of the confusion that prevails when these issues are discussed. A good example of this theoretical position is provided again by Pasinetti, in "The Notion of Vertical Integration in Economic Analysis", an article published eight years before his book previously cited, who merely states on this question that "technology requires both circulating capital goods (which are used up within one year) and fixed capital goods (which last for more than one year)" (Pasinetti, 1973, § 2, p. 3). The distinction between fixed capital and circulating capital is thus not perceived as being a difference of nature, and therefore as a conceptual difference, but only as a difference of degree, the criterion being the period of production, empirically determined, which implies moreover that commodities could pass from one category to the other without any difficulty, depending on the length of the period! As for the adoption of the year as the usual reference period, it seems to come from the fact that agricultural production often (but not always) has an annual production cycle, or that the year is the standard accounting or fiscal period.

Let us therefore reiterate that there is a fundamental difference in nature between the two categories of capital, a difference which has nothing to do with the period of use or with the lifespan of the goods concerned. The criterion of distinction is the disappearance or not, in the production process, of the commodities concerned, in their original form, a disappearance which is generally, but not always, accompanied by their transformation as an incorporation into a final product (or another more developed intermediate product) in this process. This is the case for raw materials, which disappear in one form, but to reappear under another, in the cycle of production. This is not the case, however, for services or goods such as energy (electricity for example) which disappear as such during the process, but without being materially "incorporated" (no more than labor) in the goods produced. Together with the fact that circulating capital can never be considered as a final good, since it does not leave the production process to be sold to end-users, this justifies that the value of circulating capital is found in the value of the final product, unlike the value of fixed capital.

It is therefore this criterion of disappearance-reappearance, but under another form, which defines circulating capital proper. In contrast the characteristic of fixed capital, besides the fact that it requires the prior existence of wage labor to make it perform its function, precisely through an interaction with labor, is that it remains immutable throughout the production process - whatever the duration of this process, during which it does not disappear as such. A machine, even when discarded, always has the same physical form, whatever its degree of wear, which generally affects only some of its parts, and none of its components - even when they are out of order, ever reappears in any form in the commodities produced with it.

There is therefore no justification for the idea that fixed capital would be "consumed productively" (as a mere raw material), and that it should therefore be treated as an intermediate consumption, transferring its value to the product, whereas it is clearly a final commodity, and as such an integral part of the final product.

Once again, at the root of the problem we find a confusion between economic theory and the empirical accounting practice whereby fixed capital is depreciated, usually on an annual basis, in a firm accounts. But this only reflects the social rule that any owner of fixed capital used in production has a vested right to recover over time the value of this fixed capital through the sale of its products. It is this accounting rule, which is purely empirical - and consubstantial with the normal operation of capitalist enterprises, which takes the place of a true logical and economic demonstration of the so-called transmission of value of this fixed capital. However such a demonstration is never produced by theoreticians who adopt this transmission as if it were self-evident, beyond pure and simple affirmations of principle not demonstrated, whereas the opposite is done, in the form of a demonstration by the absurd, in Appendix 2 of this book.

### 1.2. Intermediate goods, final goods and fixed capital

The reasons for which we considered - whatever the model, that fixed capital did not transmit in any way its value to the product, should therefore be perfectly clear. As we just indicated, this was not an assumption as such, but the result of a demonstration, which did not prevent at all these models from giving a coherent explanation of the formation of incomes and the distribution of commodities throughout the circulation process.

In contrast, at the heart of a number of problems which have been faced by economic theory, including Marxist analysis, lies the assumption of the transmission of value of fixed capital, an assumption which is never demonstrated, and acts in fact as a postulate of most economic theories. This explains why it is important to discuss this question a little more, precisely because of its influence on circulation, and since this postulate may explain why there are almost no reproduction schemes in economic theory.

Indeed this postulate actually comes down to treating fixed capital as an intermediate good. But any intermediate good is by definition incorporated into a final good, to become a mere part of it. Consequently, the only value produced is that of final goods, which obviously includes the value of all the intermediate goods incorporated in them, a value that cannot, however, be recognized independently a second time as the value of a final output. It is so because whatever the good it must be either an intermediate good or a final good, but cannot be both at the same time. This is a question of pure logics.

Yet this is exactly the way most economists treat fixed capital, considering it simultaneously as an intermediate good that disappears as such in the production process and whose value is therefore passed on to the goods into which it enters, and as a final good whose value is included as such in the total produced value, independently and on top of its value incorporated into other final goods. This simultaneous treatment of fixed capital as an intermediate good and as a final good is a logical blunder: because either fixed capital is an intermediate good, but in this case the value of the product of the period is equal to the value of the only final goods which remain outside of fixed capital, and cannot be anything else than consumption goods, whose value is therefore identical to the newly produced value of this same period. Or fixed capital is a final good in the same way as consumption goods, but then it cannot, by definition of what is a final good, transmit any value, because only intermediate goods transmit their value. The value of the product is thus again identical to the newly produced value (that of consumer goods as well as that of fixed capital).

Thus the analysis of realization in the circulation process shows that the treatment of fixed capital as an intermediate good is not admissible, since it requires that fixed capital be considered simultaneously as a final good, in the interpretation of Marx's reproduction schemes, or that fixed capital itself is produced without the aid of fixed capital, in a second interpretation. Marx's reproduction schemes are analyzed in Appendix 2: "The erroneous nature of Marx's reproduction schemes in Book II of Capital", at the end of this book.

It must be stressed again, however, that the mechanism of value transmission is fully justified as regards intermediate commodities, since this transmission is accompanied by their disappearance in its original form in the production process, just to reappear as transformed (under a modified form) through a physical incorporation into final goods. And we know since Lavoisier that in the real world nothing can be lost and thus can disappear in a process, without reappearing in the end of it. For its part fixed capital does not disappear in the process of production, and by the same does not emerge physically transformed from this process.

The transmission of value of true intermediate goods thus rests on a material basis, which is verified empirically in the real world, whereas the hypothetical transfer of value of fixed capital would appear metaphysical, and related to some kind of transmigration, if one forgot that it is based on the accounting practice of the capitalist mode of production and the legal ideology that underlies it.

All these remarks explain why traditional economic theory cannot but encounter a number of conceptual problems with the treatment of fixed capital, to the extent that even statisticians also have some difficulties in practice to deal with fixed capital, as we shall now see.

### 1.3. The ambiguous treatment of fixed capital in national accounts

It is no surprise that national accounts are also affected by the contradiction that we highlighted, since they consider that intermediate consumption also includes consumption of fixed capital, i.e. the depreciation of fixed capital during the period considered, as a result of normal wear and predictable obsolescence. Indeed the international System of National Accounts (SNA) indicates: "The distinction between intermediate consumption and gross capital formation depends on whether the goods and services involved are completely used up *in the accounting period* or not. If they are, the use of them is a current transaction recorded as intermediate consumption; if not it is an accumulation transaction" (United Nations, 2009, page 8 - words in italics are from the author). And the same manual defines consumption of fixed capital as: "the decline, *during the course of the accounting period*, in the current value of the stock of fixed assets owned and used by a producer as a result of physical deterioration, normal obsolescence or normal accidental damage" (ibidem, p. 123).

But since this intermediate consumption of fixed capital, unlike that of circulating capital per se, is not deducted from the value of production to give the value added (the balance of the production account), it follows that this value added is gross, that is to say, superior, for an equivalent amount, to the creation of value realized during the period. It is indeed what the SNA tells us: "Value added represents the contribution of labor and capital to the production process… However, capital in the form of fixed capital has a finite life length. Some part of value added should therefore be regarded as the reduction in value of fixed capital due to its use in production. This allowance is called consumption of fixed capital" (ibidem, p. 103).

This theoretical position is an absurdity, but it is never seen as such by national accountants, because they contradict themselves without even realizing it when they consider that the sum of the actual outputs of the branches ... less intermediate consumption - *intermediate consumption of which is excluded depreciation* (Italics characters have been added by the author) - is equal to the value of the goods and services assigned to the end uses, and to the sum of the added values. This last position adopted by national accountants makes it possible to verify the identity between the value of the product of the period (the GDP) and the purchases of the period, which allows the compilation of the national accounts without coming up against the contradictions which were just mentioned. But one must be aware that this position adopted surreptitiously (somehow), in practice is contradictory to the basic theoretical position that, in so far as it is an intermediate consumption, depreciation of capital should not be a distinct part of the value created in the period.

The equation: "value added = gross wages + taxes + interest and dividends + savings retained by companies to finance investment", is indeed correct, because it shows that depreciation, included in retained savings, is used by companies to purchase the newly produced fixed capital, which indeed is a part of the value created during the period, and therefore of the net production of the period. If, in accordance with the theoretical position of the national accountants, assimilating depreciation of capital and intermediate consumption, depreciation was removed from the domestic product, i.e. the creation of value of the period, which does not include intermediate consumption, it would become impossible to understand how the newly produced fixed capital could be realized during that same period.

The only logical theoretical solution, because it is consistent with the equality of value created and value realized (and for national accountants this equality is always verified, since stocks are supposed to be bought by the companies that bear them), is therefore to consider that the depreciation of capital, and the amortization supposed to represent it, is not an intermediate consumption. Therefore, depreciation does not represent, except from an accounting and tax point of view, a fraction of the value of the old fixed capital. At macroeconomic level, on the contrary, it represents a fraction (whose importance certainly depends on the accounting and tax rules) of the newly produced fixed capital.

It follows from this that the purchase of capital goods constituting fixed capital must be considered as a final purchase, like for example the purchase of an automobile, and that what national accounts call gross value added is in fact a net value added.

Finally, it is not uninteresting to recall one more that in his “General Theory” Keynes finds that counting what he calls the “user cost”, i.e. in fact depreciation, into the value of the product, results in making its amount dependent on the degree of integration of industry. This is the reason which Keynes refers to for excluding the user cost from the price of global supply and the value of the product, which is tantamount to rejecting the assumption that fixed capital transmits its value to the product! But he does indicate rightly that the total amount spent by buyers of final goods is gross of user cost. It is striking to note that this heterodox position of Keynes has been obscured, because it was very likely misunderstood by almost all of its commentators.

To understand Keynes's position, however, let us recall what we demonstrated in subsection 5.2 of chapter 11 (see pp. 220-221). To do that, it was sufficient to imagine a totally integrated economy, in which there would be no more than one capitalist enterprise. In this case there would be no need for any depreciation of capital to finance the purchase of fixed capital, since it would be produced by this enterprise itself and therefore would not appear on the market to be sold and bought by other capitalist enterprises. Capital goods constituting fixed capital would remain final goods, but not being subject to any exchange they would not even need to have a price. As for the sale of consumer goods, in the absence of capitalist consumption their price would just allow the recovery of all wages paid to all workers, without neither the possibility nor the necessity of including in their price any depreciation. It would be capitalist consumption alone which would imply a price higher than the level of total wages. It is therefore the existence of a multiplicity of capitalist enterprises that makes depreciation necessary, as a means of obtaining fixed capital through successive exchanges between the enterprises that produce it.

## 1.4. Fixed capital and some contradictions in economic theory

A wrong concept of fixed capital cannot but create contradictions in economic theory. In this connection, when assessing once more the question of the so-called transmission of its value to the product, one needs to recall a demonstration previously made by Bernard Schmitt and Alvaro Cencini, in a book published in 1976: "La pensée de Karl Marx - Critique et Synthèse ("Karl Marx's thought – Criticism and Synthesis"). This demonstration shows that the transmission of value of fixed capital is made possible only because of a faulty duplication. The $L_I$ value deemed to be transmitted by the amortized capital is indeed counted a first time in the value realized at the time of the production of this capital, when it is sold on the market. To count this value a second time in the value of consumer goods produced by the operation of this fixed capital through its interaction with labor, amounts to counting twice the value of a good which is exchanged only once.

The fact of counting twice the value of fixed capital goods that are realized only once, when these goods are bought by capitalists, is illogical. Since the value of fixed capital was realized in the period of its own production, it must not be counted a second time in the product of the current period, since fixed capital, once bought by capitalists and employed in the production process, does not come out to present itself again on the products market. As a result, contrary to what Marx thought, the value of the product of the period is identical to the value produced in the period, as long as there are no stocks of circulating capital.

So Adam Smith was right, when he wrote: "In the price of corn, for example, one part pays the rent of the landlord, another pays the wages or maintenance of the laborers and laboring cattle employed in producing it, and the third pays the profit of the farmer. These three parts seem either immediately or ultimately to make up the whole price of corn. A fourth part, it may perhaps be thought is necessary for replacing the stock of the farmer…and other instruments of husbandry. But it must be considered, that the price of any instrument of husbandry, such as a laboring horse, is itself made up of the same time parts; the rent of the land upon which he is reared, the labor of tending and rearing him, and the profits of the farmer, who advances both the rent of this land, and the wages of this labor. Though the price of the corn, therefore, may pay the price as well as the maintenance of the horse, the whole price still resolves itself, either immediately or ultimately, into the same three parts of rent, labor and profit" (Smith, 1776, p. 84: words in bold characters are from the author).

What Smith had understood was that it is not the old fixed capital which transmits its value or part of it (a value confused by Smith with a price, and thus made for of wages, rent and profit), to the new one (or to any other type of good), but the newly produced capital which adds its value - made also for Smith of wages, rent and profit, to the old one (the stock of existing fixed capital) either by replacing obsolete capital (in a self-replacing state) or by increasing this stock. That neo-classical economists do not understand that is not surprising, that Sraffa does not grasp that either is a little more surprising, but that Marx himself misunderstood this phenomenon is much more surprising, and might be due to the strength of the dominant ideology. As Marx and Engels wrote in a Critique of the German Ideology:

"The ideas of the ruling class are in every epoch the ruling ideas, i.e. the class which is the ruling material force of society, is at the same time its ruling intellectual force" (Marx & Engels, 1845, p. 21).

What must be understood is that once it has been produced and realized in the circulation process corresponding to its production period, old fixed capital becomes an asset. At this stage the incomes corresponding to its production have already been created and spent for purchasing it in the circulation process. Therefore like any other second hand good, if it is sold again on the market, it is not produced again, which will not create neither a new income nor a new value, because the income obtained from this second sale by its seller will be cancelled by the income spent by its buyer. This mere displacement of a commodity with a money transfer in the opposite direction does not create any new income or value for the economy as a whole, because only a new production process can do that.

# 2. Again on the articulation between values and prices

## 2.1. The mode of determination and the significance of prices in both models

The operation of the simplified model as described in chapter 11, with regard to the realization of the product, has shown that the method of determining prices in section I differs significantly from that in section II. It is because prices in section I are those of fixed capital goods which are exchanged between capitalists only, and are thus fixed independently of the distribution of the product between workers and capitalists: in other words the rate of surplus-value, i.e. *k*, does not intervene in their fixing, contrary to what is the case in section II.

Indeed we had already seen that in section II the margin rate is $\frac{k}{1+k}$, on the understanding that under this simplified model the rate of surplus-value: *k*, as the ratio between $L_I$ and $L_{II}$ reflecting the distribution of labor between the two sections, also refers to what can be considered as a particular set of techniques in use, i.e. to a state of technology. As $\frac{k}{1+k}=1-\frac{1}{1+k}$, it follows in passing that in section II the coefficient which plays the same role as coefficient $u$ plays in section I is $\frac{1}{1+k}$, which expresses the share of the product obtained by workers in this section II, and that its inverse 1 + *k* is the multiplier by which $W_{II}$ wages need to be multiplied to obtain the price of the product in this same section.

In contrast, in section I there are in the general case two ratios, both independent of distribution, which intervene in the setting of prices:

- The first one is ratio $a$, because this ratio refers only to the distribution of the value of fixed capital employed in the system between section I and section II;
- and second, in the general case, ratio $u$ or rather a series of numbers that represent it (in the case of a multiplicity of separate ratios $u_i$ at each stage of purchases of fixed capital within section I), and which express both the greater or lesser weight that fixed capital purchases have at each stage in relation to the fixed capital available, but also the greater or lesser power over the fixing of the selling price, because we must remember that 1 - $u$ is a margin rate.

We also note that in the simplified model, for the same invariable value of the product of the two sections, i.e. $L_I + L_{II}$, and the same distribution of this product between workers, who obtain a value $L_{II}$, and capitalists, who obtain a value $L_I$ (equal to $kL_{II}$ ), one can have a multiplicity of prices of the global product, according to the values taken by parameters $a$ and $u$, since according to equation (23), above p. 215, we have: $Y = L_{II}\left(1+2k+k\frac{a}{u}\right)$.

In the model with capitalist consumption, the price of the product is given by equation (43) p. 240, and is clearly more complicated, since it is $Y = L_{II}\left[(1+\gamma)+k(2+\alpha+\gamma+\gamma^*+\alpha\gamma^*)\right]$, which shows even more that for a given value the price of the product can greatly change as a function of several variables which reflect the distribution of the product between the various stakeholders.

On its own, this demonstration of the variability of the price of the global product, as a function of the value of various parameters, shows that we cannot define these parameters as a function of prices, unless we fall into circular reasoning, which is precisely what happens with neo-classical theory. It is thus shown that the passage through values is an indispensable theoretical precondition to be able in a second step to think (at a conceptual level) about the nature and level of prices.

Incidentally, one of these parameters, parameter $u$, in the model with capitalist consumption, plays a special role in section I. It certainly acts as a coefficient expressing the share of available fixed capital obtained by the buyers of the capital in question. But from the sellers' point of view, this coefficient is nothing else than the inverse of coefficient $\frac{1}{u}$, i.e. the multiplier applied to wages in section I to obtain the selling price, as shown in table 6 of chapter 11. Moreover, the complement to 1 of $u$, i.e. 1 – $u$, is none other than the margin rate, i.e. the ratio of profit over the price of the product, which corresponds in national accounts to the ratio EBITDA/value added. Seen from this angle, the magnitude of this $u$ coefficient expresses the balance of power which exists between section I capitalists in fixing the selling prices of the various means of production constituting fixed capital. Depending on this balance of power which governs the capacity of capitalists to influence this margin (the higher is the margin, the lower is u) this is the consumption of capitalists which is at stake, beyond the share of fixed capital obtained at each stage.

### 2.2. System of values and price system

Starting with the simplified model, the first of the lessons which can be derived from it on this matter is that the value system is a system that to some extent reflects the distribution of the product between workers and capitalists because it gives the value of commodities going to both categories, but which before all reflects the structure of the productive system, through the influence on the one hand of the distribution of workers between the two sections producing consumer goods and fixed capital respectively, and on the other hand through the distribution of fixed capital between these two sections of the productive system.

The first type of distribution is expressed through the absolutely fundamental parameter $k$, since it is also the rate of surplus-value, at least within the framework of the simplified model. The second type of distribution is expressed by parameter $a$, which is the proportion of the total value of fixed capital which is used in section I, and refers to the structure of the productive system. From this point of view, this parameter corresponds to a state of technology, but it should be emphasized that in the context of a particular assumption adopted supra (see page 212), when $b$ is replaced by $u$ (or $u$ is equal to $b$), this influence on prices comes from the fact that $a$ is then also the homogeneous margin rate $1-u=1-b=a$ used by capitalists of section I in setting their prices. Finally, in this simplified model, the rate of surplus-value is endogenous and depends solely on the ratio between the amounts of average social labor employed in each of the two sections of the economy: the more average social labor is needed for the production of fixed capital, in proportion to that of consumption goods, and the higher is this rate of surplus-value.

This observation helps us to understand that surplus-value is a social and global phenomenon, which explains why there is only one rate of surplus-value, which must necessarily be the same for all workers, whatever their wage level, or the section of the productive system, or the firm, where they work in. It is so because exploitation in its economic sense is not an individual phenomenon. In other words there cannot be different rates of surplus-value $k_i$ or $k_{ci}$. Indeed if it were the case this would come down to imagining that workers would only buy the commodities produced by the firm which employs them, an assumption which would be completely absurd. For instance trying to calculate a distinct rate of surplus value in section I would imply that workers in this section would buy fixed capital, which would be a pure nonsense!

What both models have shown also is that the price system associated to the system of values validates the latter by ensuring the reproduction of the system at the level of circulation, i.e. through the realization of the product of the two sections, even when there is capitalist consumption. The price system allows this reproduction to take place by ensuring the distribution of the value of commodities produced, on the one hand between workers and capitalists, and on the other hand between capitalists of the two sections. It thus allows the realization and the distribution of fixed capital between both sections, while ensuring that this distribution is maintained, through the role played by parameter $a$. The price system therefore has a dual logic: a logic of distribution of the product at the stage of circulation and a logic that could be called technological, but which over-determines that of distribution. From this point of view, prices as described and discussed here, as monetary prices derived from a transformation of values, reflect the macroeconomic constraints expressed through the action of these different parameters.

Another lesson that emerges from this analysis is that the value system and the price system are closely related, and that both are indispensable to the understanding of the other: all variables that we examined so far, whether expressed in values or in prices, are expressed in terms of the fundamental parameter $k$ , which we already called the system basic rate of surplus-value, and which accounts for the distribution of workers and value produced between the two sections of production.

It is therefore clear that the value system is the underpinning of the price system, and frames the magnitude taken by its variables, even if it is true that there is a feedback from the price system on the value system as soon as the consumption of capitalists comes into play, through the intervention of parameters c and c *, which belong initially to the world of prices. The role played by these two parameters testifies to the fact that in the capitalist mode of production the objective of capitalists is not only to maintain themselves as such, thus perpetuating this mode of production through a fixing of prices which allows them to recover the entire fixed capital produced at each production cycle, but that their ultimate objective, through the perpetuation of their ownership over the means of production, is in fact to appropriate as substantial a share as possible of the production of consumption goods.

This seems simple at first glance for capitalists of section II, through the fixing of the price of consumption goods at a level which allows them first of all to fix the amount of surplus-value, while prefixing the share of the production of consumer goods that capitalists as a whole are going to appropriate. It should be remembered that in the day-to-day running of developed capitalist economies this is easy for producers of mass consumer goods, which in most sectors find themselves in the position of oligopolistic "price-makers" facing a multitude of "atomized" consumers. In fact it is less simple than it could seem for capitalists of section II because this initial global share of consumption goods going to capitalists will have itself to be shared with capitalists of section I, as we shall see in sub-section 3.1. below.

These considerations allow us to formulate a new principle regarding the nature of prices.

**Principle 28***:* Because of their close connection to money values, which themselves reflect the conditions of production, it is not wrong to consider prices as also reflecting these conditions, and thus as production prices. However this is a very partial view of reality. First because money values themselves are depending on the distribution of wages between workers. And second because their fixing is linked to the extraction of surplus-value (for section II prices), and to the distribution of this surplus-value between capitalists (for section I prices). As a result, prices must be further considered as distribution prices, being understood 1) that the understanding of distribution involves many factors, a number of which have nothing to do with economy as such, because they are of a sociological, ideological, institutional and political nature, and 2) that distribution itself must be seen ultimately as a distribution of commodities.

# 3. The conditions of profit formation

## 3.1. The dual regime of setting shares of profit

Through the operation of the simplified model, let us recall that the mechanism explaining the realization of the product and profits in section I was identified as similar to the Keynesian principle of the "widow's cruse", meaning that it is the expenditure of profits that creates profits and makes it possible to realize the product of section I. But this was in fact a generalization of Keynes's ideas, which were formulated initially to apply to capitalist consumption only, i.e. to capitalists spending on consumption goods.

As we already indicated in subsection 5.4. of chapter 11, this simplified model allowed us to understand that profits are the result of a double transfer:

- a first transfer, which takes place on the occasion of the expenditure of wages by workers of the two sections, for the purchase of consumer goods from the capitalists of section II. It is through this first transfer that surplus-value is extracted, once and for all, in an amount which is fixed and no longer liable to change. This transfer simultaneously fixes the amount of monetary profits (i.e. $L_I$) for capitalists of section II.

- a second transfer, of a very different nature because it only involves capitalists themselves, and which is carried out on the occasion of the purchase by capitalists of section II, from those of section I, of the fixed capital which they need to produce final consumption goods. This purchase has no influence on the amount of surplus-value, which has already been determined, but it fixes the distribution of this total surplus-value between capitalists of the two sections. The share going to capitalists of section II is $\left(1-a\right)L_I = bL_I$, as we saw in chapter 11, and that going to capitalists of section I is $aL_I$, which also corresponds to the amount of monetary profits going to the capitalists who in section I ensured this sale. This first transfer to section I will then be followed by successive transfers (logically), within section I, transfers which will ensure the distribution of this surplus-value and corresponding profits between capitalists of this same section.

Now, and we will have the opportunity to come back to that, the first transfer is not in itself problematic, at least as a transfer, because it is a transfer between workers who are price-takers only and capitalists who are price-makers. On the contrary, transfers between capitalists as transfers for the distribution of a prefixed amount of surplus-value, are a zero sum game, and as soon as capitalist consumption is introduced, their results, outside the model and in the real world, can be problematic. Before developing this point, we must emphasize the fact that we have therefore two regimes for the realization and determination of profits.

Indeed, and without returning to all the details of the profit-making process highlighted in previous chapters, it must be stressed that the double system of transfers comes down to recognizing that there are two distinct regimes for the realization of profits.

The first regime concerns the constitution or creation of profits, and corresponds to their initial realization in section II. At the risk of repeating ourselves, it is directly linked to the fact that workers by definition buy consumption goods only, and therefore it is from the fixing of their price that the amount of surplus-value derives and materializes in the form of section II profits. These profits are hence nothing else than surplus-value realized in its monetary form through a transfer resulting from the expenditure of money wages. In passing, this explains why Marx's formula expressing the overall rate of profit in terms of values remains valid for the determination of the rate of profit in terms of prices in this section.

However, when we introduce capitalist consumption in the model, things become more complicated. Indeed this first transfer is a kind of priming of the pump, because it does not allow for the full realization of surplus-value under the form of monetary profits in this section, where only a part of all goods produced have been sold to workers, by definition, since this is a condition for capitalist consumption. This full realization can only be completed when all consumption goods are sold, including those bought by capitalist of both sections. This implies that not only capitalists of section II, but also capitalists of section I must realize their profits, from which they will buy consumption goods. These last profits will start being realized when capitalist of section II use their own profits to buy fixed capital goods, which leads us to the second regime of profit realization.

The second regime concerns the distribution of surplus-value under the form of pre-existing profits, through their investment in the purchase of fixed capital, by capitalists of both sections. These purchases allow for the realization of additional profits in section I through the fixing of the price of fixed capital goods, and for the distribution of these profits and means of production among capitalists themselves. This process starts with a first distribution between capitalists of section II producing consumption goods and those of section I producing fixed capital, when the former buy fixed capital from the latter, and continues with the distribution of profits and fixed capital between capitalists of section I themselves. The logically "successive" exchanges of these means of production, through the "successive" spending of profits, ensure this distribution, while respecting the "technological" constraint represented by parameter $a$ .

However the introduction of capitalist consumption in chapter 12 has shown that these two regimes are linked, because as soon as workers cease to be the only buyers of consumption goods, the interaction between capitalists of both sections ceases to be limited to the purchase of fixed capital by capitalists of section II. Not surprisingly, since it was Keynes’s initial idea, the widows’ cruse mechanism plays its role here: the spending of profits on consumption goods indeed results in new and additional profits in section II. But there is a limit on these profits and on capitalist consumption, which is the amount of surplus-value originally extracted from workers through the pricing of consumption goods, which means that capitalists of both sections have to share a prefixed quantity of these same goods: it is the quantity corresponding to the value of consumption goods not bought by workers. Rather similarly, the share of section II capitalists in consumption goods is also prefixed as soon as they have bought the fixed capital which they need from capitalists of section I.

This explains, as chapter 12 also showed, that the realization of profits in both sections is in fact more complicated than it would seem at first, because it implies a series of interactions between capitalists of both sections. In other words, the fact that the amount of profits of section II capitalists is theoretically prefixed, as soon as workers are paid and prices of consumption goods are fixed, does not mean ipso facto that all of these profits are realized. For that to happen, they need also to use their own profits to buy their own consumption goods, and sell the remaining part of consumption goods to capitalists of section I. In turn capitalists of section I need to buy their share of fixed capital (supposed to be predetermined by technological constraints) and the remaining part of consumption goods.

At the completion of this interactive and iterative process all commodities of both sections have been sold and at the same type all profits have been spent, so that there no longer exists any monetary profits in the system, which constitutes a necessary condition for the realization of the totality of surplus-value, through its distribution to its capitalist beneficiaries. This surplus-value, extracted under the monetary form of profits, ultimately takes the form of consumption goods and fixed capital appropriated by capitalists. Profits indeed disappear under their monetary form, because – if the system works smoothly, through their spending on commodities these profits ultimately go back to the firms which had initially borrowed corresponding amounts of money, and must necessarily use them to repay the banks, an action which destroys the amount of money previously created for the payment of wages at the beginning of the production cycle.

The conclusion that emerges from this analysis is that in any case there is an irreducible heterogeneity in the way the shares of profits and profit margins between the two sections of the productive system are realized and determined. As we shall see latter, this cannot but create a heterogeneity in the rates of profits proper, and in any case has the potential to create conflicts between capitalists. All the more so that in the real world spending on final goods, i.e. fixed capital and consumption goods, are not the only ones made by firms, which also buy intermediate goods, as we shall see in next subsection.

### 3.2. The conflict between capitalists for the distribution of surplus value

What clearly comes out from the analysis resulting from the model with capitalist consumption, introduced in chapter 12, is that the question of the appropriation of consumption goods by capitalists does not depend only on the price of these goods, but lies also behind the fixation of the price of fixed capital goods. Indeed surplus-value extracted in the form of profits in section II becomes immediately the subject of a dispute as to its distribution among capitalists themselves, depending on the level of prices of fixed capital. This conflict aims first at fixing the share of consumption goods accruing to capitalists of section II, compared to that accruing to capitalists of section I. It aims secondly to fix the share of capitalists producing fixed capital for those of section II compared to the other capitalists of section I, and so on at each stage of investment spending within section I.

However, this conflict cannot affect the amount of surplus value. Indeed parameter $c$ *, which appears in the first row of the table in chapter 12, and accounts - as defined, as being the share of consumption of capitalists of section I selling to capitalists of section II, contained in a price $\frac{1}{1-c*}$, only defines the price of fixed capital, and has no influence on the price of consumption goods. It is the relative market power of the first ones over the others that will govern the fixation of c*, which in a situation of fragmentation of this power could moreover rather be conceived as a set of $c_i^*$ elements, each one with its own different magnitude.

Going upstream in the productive system, fixed capital suppliers are more likely to be reduced to the sole role of subcontractors, and it is also easy to understand that small producers of specific types of fixed capital are by nature in this category, in the face of oligopolistic groups of section I firms producing on a larger scale more standardized goods, which can therefore be used interchangeably in section I and section II, and whose market is much bigger. The situation may be the same for producers of intermediate goods.

Up to now indeed we made the assumption that each section of the productive system might be defined as the integration to final producers of all enterprises producing intermediate commodities used in the production of final commodities split between consumption goods, on the one hand, and fixed capital goods, on the other hand. If we abandon this assumption that the productive system is integrated, then in each supply chain, corresponding to the production of either a consumption good or a fixed capital good, we have a number of enterprises selling intermediate goods that are more and more transformed (when we go downstream in the system) to other enterprises, along the chain going from primary commodities to commodities sold to final producers, then to wholesalers and ultimately to retailers. Each one of the exchanges which takes place all along this chain comes as a deduction of the potential profits of the buyer, and simultaneously as an increment to the profits of the seller. This explains why these exchanges are potentially conflictual ones. In the real world these conflicts are usually resolved by negotiations, which most often end up by the signature of a contract between the parties involved.

These situations are not at all theoretical, and to give but one example of such a conflict in the real world, we can briefly mention the case of the food sector in France. As reported in “Le Monde” newspaper, in a 2017 article by Laura Motet, each year there is a kind of price war between mass-market retailers and producers in this food sector. “It takes the form of negotiations between producers, manufacturers and retailers to negotiate a fair sharing of value added. These negotiations between mass-market retailers and their suppliers, whether they are small-scale producers in cooperatives or large international industrial groups, start in October and must normally be concluded at the end of February. They are most often conflictual, and must fix for the following year the tariffs of the products bought by the big mass-market retailers to garnish their shelves” (Motet, 2017, October 11).

“At the outset, suppliers are obliged by law to indicate the "average forecast price" of the main agricultural raw materials they use. However these negotiations are not only about prices, but also about the layout of products on the shelves. In a context of strong competition between manufacturers, being present at eye level or at the shelves head is crucial.

Distributors and suppliers sometimes fail to reach an agreement before the last day of February, which results in products being then removed from the store. But some distributors practice these dereferences even before, during negotiations, with a view to show the supplier the shortfall if the agreement is not signed, and to induce him to lower its price" (ibidem).

"Beyond these illegal practices, the stumbling block of these negotiations remains the determination of the sale price. One week before the end of the 2017 negotiations, only a minority of suppliers had agreed with distributors. "Signing faster does not mean signing for good conditions unfortunately for some companies [...]. Some companies give in or sign because the balance of power with their client is particularly unfavorable", says the National Association of Food Industries. According to suppliers, distributors do not intend to sign these agreements without obtaining a tariff reduction. A decrease which is not always in accordance with the very volatile price of raw food materials" (ibidem).

Recognizing the complexities of the real world, on which we tried to shed some light, gives us the possibility to derive a new principle:

**Principle 29***:* Prices, in so far as they are distribution prices, are the object and the resultant of multiple conflicts, which may or may not be arbitrated by the State. They ensure the continuity of the capitalist mode of production by ensuring the distribution of fixed capital among capitalists. But even more they also ensure the collection and distribution of consumption goods among these same capitalists. They make us realize that the ultimate goal of capitalists is not only the accumulation of fixed capital, but also the accumulation of consumption goods.

Apart from these conflicts localized along supply chains, a particular case of conflict over the sharing of surplus-value has to do with the existence of banks, and will be addressed now.

### 3.3. Sharing surplus-value with the banking system

Within the framework of the analysis conducted so far, banks are a particular category of capitalist enterprises, which primarily sell the services of money creation and financial intermediation. Since these activities are carried out through lending, their main remuneration is obtained by charging an interest rate to their borrowers.

It goes without saying that in this analysis the interest rate by itself has nothing to do with the profit rate, because setting it is quite specific to the monetary and financial sphere, and this is not the object of this book. The global financial crisis of 2008-2009 and the developments which have followed have amply showed that the interest rate is certainly not a price that can be purely determined by supply and demand, if only because there is no exogeneity of the money supply. Central Banks intervention in the wake of this crisis has also showed that the interest rate is before all institutionally determined. It should be clear therefore that the rate of interest and the rate of profit are distinct variables, whose evolution can follow vastly diverging paths.

However, it is striking that this position is far from obvious for many economists. As a good example, we can cite a recent IMF working paper by Goes: "Testing Piketty's Hypothesis on the Drivers of Income Inequality", where all calculations are made by squarely identifying the rate of return on capital with the long term rate of interest. Indeed Goes indicates: " To derive the second variable of interest – namely, the real return on capital net of real GDP growth – I take yearly averages of nominal long-term sovereign bond yields, calculate the post-tax nominal rates by deducting corporate income taxes and subtract from them annual percent changes in GDP deflators and real GDP growth" (Goes, 2016, p. 6)! At least this author recognizes that "the choice of sovereign bond yields as a proxy for returns on capital is not self-evident" (ibidem). When one knows that in the last ten years in most developed economies the gap between both rates has surely exceeded 10 %, with negative rates on sovereign bonds yields adjusted for inflation and GDP growth, all this is nothing but a theoretical aberration!

Going back to interest rates, and from the point of view of individuals, interest rate paid to them on deposits can be considered with Keynes as the reward for parting with liquidity, which has nothing to do with a rate of profit. If we adopt the point of view of banks the interest rate charged to borrowers is the price of a service which to be ensured implies to pay a lower interest rate to lenders or depositors, and both rates are equally independent of the rate of profit. Their difference is usually considered as the price of the services of money intermediation (and creation). The difference between the rate paid to banks by borrowers and the rate actually paid to depositors represents what national accountants call charges for financial intermediation services indirectly measured[12]. We can call it more simply the price of banking services.

This price corresponds to what US banks call operating revenue, British banks operating income and French banks net banking income. It can be considered as a product, in this case the "financial and banking services of intermediation and money creation". The value of this product corresponds, as for the rest of the system, to the amount of money wages in the banking sector, if it is consolidated as an integrated sector. Indeed these services like any others have a cost: banks have to pay wages to their workers, and buy intermediate commodities. Thus the difference between the price of these services and the wages paid to ensure it is like profits a transfer income.

This gross profit of the banks, that is to say the difference between this operating revenue or net banking income and the wages of the sector, is of the same nature as the profits of the rest of the system, and is constituted in the same way through the sharing of surplus-value. It is easily understandable that if the profits of the banking system tend to increase more rapidly than other profits, this increase is necessarily obtained at the expense of the rest of the productive system, and is in no way additive to the rest of profits. So there is also a conflict between capitalists of the banking and financial sector and capitalists of the non-financial sector for the sharing of surplus value.

[12] In practice things in national accounts are more complicated, because of the intervention of a reference rate of interest, but there is no need to enter into such complications within the framework of this book.

It must also be understood that, if they have to be recorded within the framework of our models, and like for other types of commodities which are sold to workers as well as to capitalists, these financial intermediation or banking services must be split between section II and section I. In any case these activities, which are indispensable for production to be carried out, can be considered as productive activities, and as such are creating value.

However, since the beginning of the 80's (in fact since 1978), there has been a constant trend of deregulation in the USA which allowed banks and financial institutions to engage more and more into another type of activities, which is usually called investment banking. On the broad question of deregulation, one can refer to a short but well documented and synthetic article by Sherman, from the Center for Economic and Policy Research: "A Short History of Financial Deregulation in the United States" (Sherman, 2009, pp. 1-15).

As for investment banking, this activity has two main lines of business, called the sell side and the buy side. The "sell side" involves trading securities for cash or for other securities (e.g. facilitating transactions, market-making), or the promotion of securities (e.g. underwriting, research, etc.). The "buy side" involves the provision of advice to institutions that buy investment services. Up to 1999, there was in the USA a separation between investment banking and commercial banks which was repealed in 1999 by the Gramm-Leach-Bliley Act. A similar trend of deregulation happened in other advanced economies. This allowed a huge development of these activities by commercial banks, which involved securitization, the creation of sub-primes, etc. All these new developments ultimately provoked the global financial crisis of 2008.

It is important to understand that most of the huge profits made in this business, before it collapsed, did not come from any production at all, but from the exchange of already existing financial assets., an activity which is indeed pure speculation, i.e. the purchase of assets (commodities, goods, or real estate) with the hope that it will become more valuable at a future date. In finance, speculation is also the practice of engaging in risky financial transactions in an attempt to profit from *short term fluctuations* in the market value of a tradable financial instrument - rather than attempting to profit from the underlying financial attributes embodied in the instrument such as capital gains, dividends, or interest.

If we mention it, this is to state that speculation as such creates absolutely no value. In fact it does not create any net profits either, because the corresponding activities concern existing assets, whose sale against money only symmetrically moves the property of these assets from the seller to the buyer, and moves money in the opposite direction (like for any exchange of second-hand goods). Therefore any profits made from this activity by an economic agent cannot constitute a net income at macroeconomic level, because they correspond to a symmetric reduction of income for another agent. When these profits are spent on real goods, they just reduce the remaining amount of existing surplus-value which has to be shared by all other capitalists.

More generally, surplus-value realized by capitalists in the financial sphere cannot but reduce the amount of surplus value realized in the sphere of real production of goods and services. It is therefore easy to understand that the financialisation of capitalism which started at the beginning of the 80s in most of advanced capitalist economies has resulted over the years in a reduction in the share of profits going to capitalists in the sphere of real production, thus reducing their capacity to invest.

All these elements give us now the possibility to state several new principles concerning the role played by the banking sector (principle 30), the conceptual status or nature of interest rate (principle 31), the nature of money (principle 32) and finally the status of banking services (principle 33).

**Principle 30**: The banking sector plays a fundamental role in the capitalist mode of production, by creating the money which among others is indispensable for the payment of wages. The service that the sector renders, by providing this money as well as the service of financial intermediation, has like all services a value measured by the amount of average social labor time spent in the sector, and a money value, measured by the amount of money wages paid to the sector' workers, who contribute like all workers to the creation of surplus-value. The macroeconomic cost of the services of money creation and intermediation is made of wages and of a cost which is the creditor interest rates paid to depositors of money in the banking sector. To this cost must be added the profits of the banking sector, to obtain the price of the service.

**Principle 31**: Like rent is the price paid to or income received by the owners of non-produced goods for obtaining the right to use them during a defined period, creditor interest rates must be considered as the price paid to or income received by the owners of money assets so that banks can use these assets during a defined period. This is why interest rate can be viewed as the price received for parting with liquidity. Interests paid by borrowers are similarly the price paid by them to banks to be able to use defined amounts of money during a specified period, and which constitutes the banks gross income. It follows that interest as an income remunerates the property of money assets, owned either by depositors or by banks. Therefore, like profits or rents, interests are a transfer income.

**Principle 32**: Formally money should be considered as produced by banks, to some extent ex nihilo, because it is not the result of a transformation. But it must be considered nevertheless as non-produced, because – like land for instance, it does not belongs as such to the production process, where it does not result from a transformation and is not itself transformed. It is so because money is not a good that could be sold against itself, which would be a nonsense. Only the right to use it for a definite period is sold (also like for land), against payment of an interest. As such money cannot be considered neither as a final good, which could be consumed or stored, because at global level it always has to be repaid, which ultimately destroys it, nor as an intermediate good or service, which would disappear in the production process.

**Principle 33**: Apart from various fees, the price of the services of money creation and financial intermediation rendered by banks corresponds to the debtor interest rates charged by banks to borrowers. Their overall amount during a given period of time is part of the total price of the product of this period. It includes wages paid in the banking sector, i.e. the money value of bank services, creditor interests paid to depositors and bank profits. Both creditor interests and bank profits are part of global surplus-value. Bank services can be sold both to firms and individuals, be they capitalists or workers. When they are sold to firms, they play the same role as intermediate commodities, and their price becomes a part of the price of the final product, either of consumption goods or fixed capital goods, depending on the section which the firms are part of. When they are sold to individuals, they are part of their consumption, and thus play the same role as consumption goods, of which they are a particular type.

Competition among capitalists for the appropriation of the largest possible amount of surplus-value, especially in an economic world dominated by a limited number of huge multinational firms, most often oligopolistic and struggling to dominate world markets, leads us to an image of the economic world which is far away from the fairy tale behind the neo-classical paradigm. What kind of micro-economic price theory could correspond to this type of situation is the question which we will briefly address now.

# 4. A truly micro-economic price theory

## 4.1. The circularity of neo-classical price theory

Let us recall that neo-classical theory conceives exchange as a barter trade between economic agents which are endowed with different goods before exchanges take place. Supply and demand of these goods depend on the subjective utility which they attach to these goods. Exchange is conceived as an equivalence relation, and the determination of all prices, considered as exchange ratios, takes place on the market when these prices ensure that equilibrium - defined by the equality of all supplies and demands, is reached simultaneously for all goods.

In this paradigm prices are relative prices, defined in terms of one of the goods considered as the numeraire. The theory is however circular, which invalidates it, because the existence of this numeraire, in the form of one of the goods chosen as a standard or measurement unit, must be postulated before exchanges take place. This is indeed a logical precondition, made necessary to allow an auctioneer to announce a first price system, on the basis of which a trial and error process will take place. This process is supposed to lead to an equilibrium allowing actual exchanges to be carried out. It is thus necessary that a first price system exists, even before prices are determined...This is equivalent to the assumption that all economic agents can express the subjective utility that they attach to all goods in the same measurement unit, before exchanges take place. The circularity of such a reasoning shows that exchange alone is unable by itself to provide the basis for a coherent theory of value and prices.

This implies that the nature of values and prices must be defined independently of exchange, which is precisely the approach which has been followed so far: values are defined at the level of production, and prices at the level of circulation, assuming the existence of money as a non-commodity, and within a set of macroeconomic constraints which are indispensable for the reproduction of the economic system.

As regards now a theory of the determination of micro-economic prices, understood as a one by one determination of the price of each single good (and not as their simultaneous determination, like in neo-classical theory), it must necessarily take into account, as its basis, all of the global constraints which have been examined so far. Such a micro-economic theory is not the object of this book, but we shall nevertheless touch upon a few requirements which it should meet, restraining ourselves to laying just a few milestones, and referring specifically to recent microeconomic theories.

### 4.2. Exchange as the expression of a balance of power

Prices of commodities and distribution of incomes are two sides of the same phenomenon, as we believe it has now been demonstrated, because exchange is not only a transfer of a commodity from a seller to a buyer, but at the same time a transfer of money from the buyer to the seller, which reduces the income of the buyer and increases the income of the seller. It is therefore no wonder, when their income is at stake, that the relationships between buyer and seller may express a balance of power, at least on each occasion where it is possible to discuss or influence the price at which actual exchange is going to take place.

We just showed in the last subsection that this is generally the case for the price of fixed capital, of which the various components are exchanged between capitalists, where there are no so many agents involved. On the contrary, it is most often not the case for consumption goods, which are sold in huge quantities to myriads of consumers at fixed prices. This explains that workers usually cannot discuss the price of the commodities that they buy, which is a fundamental phenomenon for the extraction of surplus-value.

However, workers are usually able individually or through their trade unions to discuss the level of their money wages. This means that workers have the possibility through such a negotiation to influence the amount of surplus-value. It must be understood nevertheless that any gain obtained on this front can easily be reversed by an increase in the price of consumption goods. In passing we may note that we touch here upon a mechanism which might well be at the heart of a theory of inflation, but we shall not expand on this question, since it is not the subject of this book.

Moreover, we must take into account the fact that these activities of buying and selling are decentralized within the economic system, and as soon as we abandon the wrong idea of the existence for all markets of a Walrasian auctioneer, this fact has very important theoretical consequences. Indeed this simple observation means that one good can have several prices at the same time. We already saw in the previous chapter on the theory of rent that one same good could have a number of different production costs, or money values, due to the heterogeneity of the conditions of production. Not only because of the existence of unproduced means of production, but also because of the multiplicity of producers and the diversity of these conditions of production, even in the absence of unproduced goods, these phenomena being at the heart of a coherent theory of rent and quasi-rent, on the one hand, and of a differentiation of profit rates, on the other hand.

We must now acknowledge that, as well as there is a great heterogeneity and diversity in the conditions of production, there is also a great heterogeneity and diversity in the conditions of circulation and exchange. The same good can therefore have different prices at the same time depending on a multiplicity of factors, such as the type of firm or store from which it can be purchased, its location, the quantities purchased, etc.…

This multiplicity of prices, even for strictly identical goods, is a well-known fact that can be observed by anybody in the real world, and what a theory - even economic theory, should explain, is the real world and not a world of fantasies. Let us recall that Wittgenstein rightly said that “the world is everything that is the case” and that “the totality of facts determines both what is the case, and also all that is not the case” (Wittgenstein, 1921, p.1, propositions 1 and 1.12). Of course this conception of prices completely invalidates most of orthodox economic theory, to begin with neo-classical theory, because this theory is based on the contrary on the explicit or implicit postulate of unicity of prices, according to which one good can have only one price. It is a precondition to write the equations of general equilibrium or even Sraffa’s equations, which would otherwise be totally indeterminate.

Besides, this observation of the multiplicity of prices for one good is perfectly coherent with the fact that all prices which have been dealt with so far are not micro-economic prices in the sense of prices of single goods, but are always statistical averages between a multiplicity of prices. This is the case for all the variables which we introduced so far in our models. Especially, the fundamental labor price $\overline{w}$ is itself the average nominal wage rate of millions of workers. The prices of each section of the productive system, i.e. the price of consumption goods $P_{II}$ , produced by section II, as well as the price of fixed capital goods $P_I$ , are both an average of prices of thousands of individual goods. Recognition of this fact leads to a new basic principle:

**Principle 34**: A macroeconomic theory is a theory where all prices are statistical averages, and consequently where the magnitude of all “real” variables, since they are measured in terms of prices of their constitutive commodities, is itself a statistical average. A microeconomic theory of prices should therefore restrain itself to the disaggregation of these statistical averages, with a view to understanding the formation of the price or multiplicity of prices of each particular commodity included in these same averages.

Arrived at this point, there are not many economic theories which can correspond to our theoretical background and qualify to explain the determination of micro-economic prices. However, owing to the fact that once the macro-economic setting is in place each particular price, at micro-economic level, must reflect in one way or another a particular balance of power, and that in the real world prices are often determined through negotiations which usually end up by the signature of a contract, it seems to us that the question of pricing at micro-economic level might be addressed, at least in part, by what is usually known as the theory of contracts, which is often considered as being itself a part of the theory of incentives.

The theory of incentives is based both on the theory of information and on the theory of games. The pioneers of the field, Jean-Jacques Laffont and David Martimort, present the most thorough yet accessible introduction to incentives theory to date, in their book published in 2002: "The Theory of Incentives. The Principal–Agent Model". They show that the mere existence of informational constraints may generally prevent the principal from achieving allocative efficiency. Since in the real world informational constraints are everywhere, this theory is in itself sufficient to ruin the neo-classical paradigm and its so-called theory of prices which – as we already pointed out, is anyway logically inconsistent because of its circularity.

Although Laffont and Martimort do not seem aware of this last problem, they indicate in the introduction of their book that "general equilibrium theory proved apt to powerful generalizations and able to deal with uncertainty, time, externalities, extending the validity of the *invisible-hand* as long as the appropriate competitive markets could be set up. However, at the beginning of the seventies, works by Akerlof (1970), Spence (1974), and Rothschild and Stiglitz (1976) showed in various ways that asymmetric information was posing a much greater challenge, and could not satisfactorily be imbedded in a proper generalization of the Arrow-Debreu theory. The problems encountered were so serious that a whole generation of general equilibrium theorists gave up momentarily the grandiose framework of GE to reconsider the problem of exchange under asymmetric information in its simplest form, i.e., between two traders, and in a sense go back to basics. They joined another group trained in game theory and in the theory of organizations to build the theory of incentives, that we take as encompassing contract theory, principal-agent theory, agency theory and mechanism design" (Laffont and Martimort, 2002, p. 13).

From this quote it is perfectly clear that a theory like the theory of incentives obviously does not aim at determining all prices simultaneously, as neo-classical theory tries to do – without succeeding in this endeavor anyway, and is on the contrary a true microeconomic theory, as we defined it, which seems to be compatible with the analyses developed so far. The theory of incentives is however not alone to qualify as such. For instance "mechanism design", which is the last theory cited above, deals with incentive-compatible mechanisms, which satisfy certain incentive constraints. The Nobel Prize lecture by Roger B. Myerson precisely addresses the perspectives on mechanism design in economic theory (Myerson, 2007, pp. 320-341).

Besides that, another quite interesting microeconomic theory, which was recently developed, is known as "evolutionary microeconomics". This is also the title of a book by Jacques Lesourne, André Orléan and Bernard Walliser, where the authors clearly state that their approach is differentiated from the Walrasian model, when they write: "the idea underlying the evolutionary approach is to abandon the hypothesis of centralization, which lies at the heart of Walrasian "tatonnement", and to replace it by decentralized processes of information, negotiation and exchange. Here we are poles apart from the theory described above: in our simple market agents pair up randomly, and can contract, if they so wish, at the prices they negotiate, even outside a situation of equilibrium. *A priori*, therefore, there is nothing in such a structure to ensure price unicity. Each elementary transaction can take place at its own specific price. Contrary to the previous approach, price unicity, if it appears, must be interpreted as an emergent property, as a pure product of competitive forces, and not as the expression of a *prior* postulate provided by the hypothesis of the auctioneer" (Lesourne, Orléan, Walliser, 2006, pp. 48-49).

Whatever the elements of understanding that both the theory of incentives or evolutionary microeconomics may bring about as regards the determination of individual prices, it remains that to have a full understanding of a particular price and its evolution in the real world may also require to mobilize other fields of social science, because many prices if not all of them cannot be considered as the result of purely economic mechanisms. Indeed the level of a particular price may be also the outcome of sociological, cultural, ideological or even political phenomena.

Finally, aaccepting the multiplicity of prices, both in reality and in theory, makes it easier to understand that similarly there is a multiplicity of profit rates among firms which produce a particular commodity, as well as among the subsectors which produce different commodities, or between section I and section II. This means that the rate of profit, which plays quite an important role in most economic theories (be they neo-classical, Sraffian or Marxist) should be viewed much more as a statistical average and a theoretical concept, which reduces and simplifies the complexity of the real world, like an enlarging zoom, which loses track of details, than as a tangible reality. In other words the average rate of profit, although it is an intellectual reality, something that Althusser would have called "concrete in thought", has no material existence as a tangible element of the real world, which would be an indispensable element of any price, and could be empirically observed. Such a view cannot but have a bearing on what should be the aim of microeconomic theory, which we can translate in a new principle. Neither this average rate of profit should be linked in one way or another, to the rate of interest, which is a very different variable.

**Principle 35**: The aim of microeconomic theory should be to understand and explain the divergence between individual prices and statistical averages. This should be done first for the various prices of a same commodity, which come from differences in the conditions of production and in the rates of profit, as well as in the level of rents. This should also apply to the explanation of the difference between rates of profit affecting the production and distribution of different commodities.

Whatever the way the origin and nature of profit is explained among various economic theories, there seems nevertheless to be a consensus among many economists of various schools of thought, to consider that the question of the level and evolution of the profit rate is closely related to the reproduction of the economic system and the existence of crises. For this reason, these points are worthy of further consideration. However the rate of profit is not a price, but the ratio of an income rewarding the property of fixed capital to the value or price of this same capital. Since the present third part of this book is devoted to distribution and prices, we consider that these points need therefore to be discussed in a new and last part of this book, devoted to the rate of profit and the reproduction of the economic system.

# PART IV - REPRODUCTION AND THE RATE OF PROFIT

We already mentioned that the profit rate is a fundamental variable in neo-classical theory, where it is defined as equivalent to the marginal productivity of capital, itself considered as a factor of production. In this theory the interaction of supplies and demands under the supervision of an auctioneer is supposed to ensure the equalization of profit rates when equilibrium is reached. Neoclassical theory therefore deals with an average profit rate which prevails in the whole economic system. We showed that this is not theoretically justified, if only because capital is not a factor of production. It cannot create any surplus, but only plays the role of a catalyst in the production process, where it boosts labor productivity.

The profit rate is also a prominent variable, and is also an average at the same scale in Sraffa's theory of production prices, even though it must be exogenously given. However it is wrongly defined as a rate applying to intermediate goods, even before the problematic introduction of fixed capital in the theory. Although Sraffa's theory invalidates the idea that the rate of profit could be considered as the measure of the marginal productivity of capital, it is supposed to provide a distribution key for profits to share with labor the surplus created through the production process. We showed however that there is no such thing as a surplus, and that the introduction of fixed capital in this theory is theoretically unsound, which ultimately invalidates the theory of production prices, and means that commodities do not produce commodities.

Marx for his part correctly defines profit as the monetary realization of surplus-value, but he too considers that competition brings about an equalization of profit rates, leading to the existence of an average rate of profit for the whole economic system. Indeed he writes in Chapter 9 "Formation of a general rate of profit (average rate of profit) and transformation of the values of commodities into prices of production", in volume 3 of Capital: "the rates of profit prevailing in the various branches of production are originally very different. These different rates of profit are equalized by competition to a single general rate of profit, which is the average of all these different rates of profit. The profit accruing in accordance with this general rate of profit to any capital of a given magnitude, whatever its organic composition, is called the average profit. The price of a commodity, which is equal to its cost-price plus the share of the annual average profit on the total capital invested (not merely consumed) in its production that falls to it in accordance with the conditions of turnover, is called its price of production" (Marx, 1894, p. 108). Moreover Marx establishes a law of the tendential fall of the profit rate which is supposed to provoke a crisis of the capitalist mode of production, a question which we shall also examine in this last part.

The reader might therefore consider it strange that, apart from the chapter on rent, we did not introduce so far in our models the average rate of profit as such, although they integrate fixed capital and profits. But our average rate of profit in the chapter on rent is not at all an average rate at the scale of the economic system: it is a statistical average rate prevailing in a branch producing only one commodity. In fact if we did not introduce a general average rate concerning the whole economic system until now, it is simply because there was no theoretical need to do so: the average rate of profit is an endogenous variable, which cannot be determined as long as all values and prices have not been determined. In our theory the rate of profit is nothing else than a resultant, a mere statistical ratio between profits as a transferred money income deriving from the property of fixed capital and the corresponding amount of this fixed capital. And we showed that logical constraints governing the realization of the product are incompatible with the theoretical existence of a general profit rate, other than a mere statistical average. This corresponds truly to what can be observed in the real world.

From what was said above, and since it takes a great place in theoretical debates, it remains nevertheless that the question of nature and evolution of the rate(s) of profit certainly presents some theoretical interest and deserves to be dealt with, within the framework of our analysis, as well as the question of its potential influence on the occurrence of crises concerning the capitalist mode of production. This task will be carried out in the following chapters of this last part.

# Chapter 17. Profits, rate of profit and reproduction

This chapter is dedicated first to a more in-depth analysis of profits. It starts in a first section by exploring the complex nature of profit rates, which explains the practical difficulties to measure them in the real world, due in part to their multiplicity. This leads to propose then some alternative indicators, like the amounts of profits or the profit/wages ratios. A second section shows how the reproduction of the system is problematic, which explains the probability of crises coming from inherent difficulties of reproduction, and why Keynes viewed Say's law as essentially not true. A third section shows that there are some remedies to crises. One of them, autonomous investment, leads to the analysis of the Keynesian multiplier, and shows that the value of this multiplier cannot be higher than 1. A fourth and final section discusses briefly the link between crises and evolution of the profit rate, to conclude that is less than obvious, and that the Marxist law of the tendency of the rate of profit to fall needs also similarly to be reassessed.

## 1. The complex nature of profit rates

Apart from the fact already mentioned that there was no need to introduce the rate of profit in our theory so far, there are a few other reasons explaining why we did not do it. One of them is that we did not introduce the denominator of the expression giving the profit rate, i.e. the accumulated stock of fixed capital and its measurement, because a precondition for doing that is first of all to have a coherent theory of value and prices. Now that it is the case, we will address this question in the last part of this book, devoted to the evolution of the profit rate as related to the evolution of other variables. Another reason for not introducing the profit rate so far is obviously, as we just indicated at the end of last chapter, that there is a multiplicity of profit rates. Before addressing this aspect of the question, it is also worth to recall that even from an empirical point of view, there is a genuine difficulty just to measure actual or “real” profit rates.

### 1.1. The practical difficulties to measure profit rates in the real world

In fact, the difficulty of measuring profit rates in the real world stems from the problems that capitalist firms themselves have in practice in measuring the denominator of the expression giving the profit rate, i.e. the stock of capital, particularly in view of their conception of the nature and extent of this capital stock.

Within the framework developed by the International Integrated Reporting Council (IIRC), an international committee working on the issue of business accounting, a company's capital is made up of several asset classes: financial capital (securities) and capital equipment (buildings, factories, equipments); natural capital (natural resources, impact on the environment) and social capital (reputation, ability to deploy in a given context); and lastly, the "intangible" capital formed by the human capital (qualifications, training, organization, etc.) coupled with intellectual capital (knowledge, know-how, patents, ability to innovate).

The measurement of each of these components is difficult because it depends on norms, always subject to interpretation, and judgments which are necessarily subjective. Effective measurement of the capital stock in practice thus depends on multiple conventions. It is therefore extremely hazardous, and thus largely devoid of theoretical relevance. This explains that companies often entrust this task of measuring the value of their capital stock to consultants or chartered accountants. This difficulty can be clearly observed in corporate mergers, since the value of the acquired company is most often the result of a process of negotiation of this value that is not scientific in nature. So much so that in a number of cases the value finally retained turns out to appear a posteriori as erroneous.

It should be added that capitalist enterprises have a strong interest, for tax reasons, in minimizing the amount of their taxable profit, thereby reducing the posted profit rate. As for the tax rules themselves, which come into play when it comes to measuring after-tax profit, they are so complex, particularly in terms of depreciation, that large companies employ many tax professionals for the purpose of tax optimization, which often involves the relocation of profits to lodge them in tax havens, by playing on internal transfer prices between subsidiaries.

The result of all these difficulties and all these manipulations is that most capitalists firms themselves are probably not aware of their own rate of profit! Both the numerator and denominator of the fraction giving the rate of profit are thus in practice tainted with serious uncertainties which forbid to make the profit rate play the role which theoreticians, neoclassical like Marxist ones, would like to assign to it. The fact that the managers of firms do not know for sure the rate of profit does not in any way seem to bother them, since they prefer to concentrate on the rate of return on equity (ROE) which is another concept: in corporate finance, it is a measure of the profitability of a business in relation to the book value of shareholder equity, also known as net assets, or assets minus liabilities.

In any case the real world shows that there is a multiplicity of profit rates, a phenomenon also deriving from strong theoretical reasons, which will now be examined.

## 1.2. The multiplicity of profit rates

### 1.2.1. Some technical or “macroeconomic” reasons

The models used so far were based initially on the assumption of the homogeneity of the margin rates (considering for instance that we had $1-u=a$ ). But in the real world there is no reason why the margin rates should be the same for all sectors and all firms. It was especially the case in the first and simplified model, where they corresponded to the share of available fixed capital obtained at each stage of the production process. In general, as we had indicated when studying this model, there can be as many distinct values for the $u$ coefficient, and therefore for the margin rate $1-u$, as there are sectors or firms.

Admittedly, at each stage of fixed capital purchases within section I there can be a distinct rate of profit, but its formulation is very simple, since in the simple model all profits are invested. Consequently, the amount of fixed capital is nothing more than the sum of profits, identical from period to period, accumulated in the form of investments in fixed capital, as long as this capital is used in the production process. Therefore, assuming that the lifespan of fixed capital corresponds to *t* times the production period, the relation between the amount of profits at each stage and the corresponding fixed capital is expressed in the form:

$$r_{Ii}(price) = \frac{a\left(1-u_i\right)^i L_I}{a\left(1-u_i\right)^i t_i L_I} = \frac{1}{t_i}$$

The formula is effectively simplified to the point that the profit rate appears to be the inverse of the average fixed capital lifespan, the same fixed capital being employed at each particular stage of fixed capital purchases within section I. But there is absolutely no reason for the actual lifespan of fixed capital to be the same at each of these stages. Even if these lifespans are not necessarily very different, profit rates are nonetheless formally different: the longer the service life, the lower the rate of profit, which is logical since the amount of profit needed at every period to renew the fixed capital that has reached the end of its productive life is itself all the more limited as its life-time is long.

### 1.2.2. Some “microeconomic” reasons

As soon as we introduced capitalist consumption in our second model, it became obvious that prices and therefore profit rates do not have only a macro-economic dimension, and that they also have a micro-economic component, because the realization of profit appears then as a way, not only to reproduce the stock of fixed capital, as in the simple model, but also for the owners of this capital to get a share of capitalist consumption which they would like to be as large as possible. This introduces a powerful incentive for capitalists to increase their prices as much as market conditions allow them to do. There is no question of denying that this makes them dependent on these market conditions and thus partly, among others, to the law of supply and demand. For neo-classical economists, competition is in this context supposed to push for the equalization of profit rates. But neo-classical theory has never produced a rigorous and convincing demonstration of how competition leads to such an outcome.

Of course, capital is supposed to flow off from sectors where demand is low, and the rate of profit supposed to be equally low as a result, in order to flow into sectors where demand is high and the rate of profit supposed to be high: these moves are supposed to bring about an equalization of profit rates. But it is the starting point of this reasoning, according to which a low rate of profit necessarily coincides with a weak demand, which is not demonstrated but postulated. Whereas in the real world the rate of profit can be low or high for other reasons than the weakness or strength of demand, including macroeconomic reasons such as those developed so far.

Conversely, we can see that the whole strategy of capitalist enterprises consists precisely in limiting competition in order to obtain as much market power as possible, going as far as practicable to obtain some kind of monopoly power, which can be achieved among others by taking advantage of market imperfections, such as asymmetric information. The natural objective of any firm is to achieve the power to decide freely of its own prices in order to maximize its corresponding margin rate. And there is no reason for this market power to be the same for all sectors or all goods.

Markets are also marked by the existence of multiple externalities, which do not allow prices to play the role assigned to them by neo-classical theory, a role supposed among other things to lead to the equalization of profit rates. Such a functioning of real markets then pushes not to an equalization but rather to a differentiation of profit rates. It is therefore clear that in the capitalist mode of production the preponderant situation corresponds to a multiplicity of profit rates. This is a good reason, and in itself a sufficient one, to desacralize the concept of average rate of profit and reduce its importance as regards its role of a performance indicator supposed to guide business investment.

If such a multiplicity of profit rates seems for all these reasons inevitable in a state of simple reproduction, it is also a consequence of the dynamics of the system, because as soon as we abandon the assumption of simple reproduction there is no reason for each sector to grow at the same rate as the others. In fact the demand for each particular good, which depends on many factors, usually grows at its own rate, which implies that growth is never a homothetic process, a question on which we will elaborate a little.

### 1.2.3. Growth is not a homothetic process

If we leave for a moment the theoretical world of simple reproduction to go to that which Marx called enlarged reproduction, corresponding to a situation of growth of the economic system, it is a fact of experience that growth is never a transformation without a deformation of a number of economic parameters.

The economist who best described and theorized the dynamics of the capitalist system is certainly Joseph Schumpeter, in his well-known book, “Capitalism, Socialism and Democracy”, originally published in 1942, where he coined the expression of “creative destruction” to characterize the usual growth process of the capitalist mode of production. The whole sentence where this expression appears deserves to be quoted: “The opening up of new markets, foreign or domestic, and the organizational development from the craft shop and factory to such concerns as US Steel illustrate the process of industrial mutation that incessantly revolutionizes the economic structure from within, incessantly destroying the old one, incessantly creating a new one ... [The process] must be seen in its role in the perennial gale of creative destruction; it cannot be understood on the hypothesis that there is a perennial lull” (Schumpeter, 1976, p. 83).

Schumpeter continues: “The first thing to go is the traditional conception of the modus operandi of competition. Economists are at long last emerging from the stage in which price competition was all they saw. As soon as quality competition and sales effort are admitted into the sacred precincts of theory, the price variable is ousted from its dominant position. However, it is still competition within a rigid pattern of invariant conditions, methods of production and forms of industrial organization in particular, that practically monopolizes attention. But in capitalist reality as distinguished from its textbook picture, it is not that kind of competition which counts but the competition from the new commodity, the new technology, the new source of supply, the new type of organization (the largest-scale unit of control for instance) - competition which commands a decisive cost or quality advantage and which strikes not at the margins of the profits and the outputs of the existing firms but at their foundations and their very lives" (Schumpeter, 1976, p. 84).

We could not say better, which explains the length of the quote, but such a situation as described by Schumpeter implies that at any point in time there are a number of firms which go bankrupt or at least stop their activity and conversely a number of new firms who start their operations. This necessarily means that at any time there is a wide spectrum of profit rates, from the negative rates of firms which are expelled from their market and closing, or have just been created and are struggling for growth, to the highly positive ones of successful start-ups.

More simply, and even without creative destruction, the fact that each particular sector has its own rate of growth implies that the need for new investments differs from one sector to the other, and that ceteris paribus less profits need to be invested in sectors with a low growth rate than in sectors with a high growth rate. Although there is no direct correspondence between this observation and the rate of profit, this also goes in the direction of the differentiation of profits rates.

### 1.2.4. The differentiation of profit rates is inevitable

From the foregoing we can understand first of all the "technical" dimension of profits and the rate of profit. Indeed their amount or magnitude have to be such as making it possible to recover the fixed capital indispensable for the reproduction of the system, and this in section I as well as in section II. Thus when the rates of growth are heterogeneous from one commodity to the other, the technical components, i.e. the amounts of profits needed for the sole reproduction of the system, become similarly heterogeneous.

But we can also understand the dual nature of profits, because beyond this technical dimension, and as the monetary expression of surplus-value, profits are also the source of capitalist consumption. This gives to profits and the rate of profit a social dimension which, as we pointed out, is expressed differently in each of the two sections of production. We therefore understand why there is no reason a priori for the rates of profit to be identical in both sections. Within each section, the same social dimension plays its role for each particular commodity, meaning that in each section there is no reason for profit rates to be identical in the production of different commodities. Moreover there is neither a particular reason nor a particular mechanism in the economic system to ensure that both dimensions, i.e. the "technical" one and the social one, should necessarily be related.

In other words, the price system which theoretically allows the system to adjust and reproduce itself harmoniously has no reason to be necessarily the price system which derives from conflicts for the actual sharing of surplus-value. This last price system itself has no reason either to be the price system ensuring the equalization of profit rates between the two sections of production, or within each section between the firms belonging to it. From this point of view competition between capitalists, supposed to ensure the equalization of these rates in neo-classical theory, in so far as it is a competition for the distribution of a prefixed and non-extensible amount, actually plays a reverse role of additional differentiation, aimed at ensuring not only the replacement of fixed capital, but also the maximum level of consumption made possible by the balance of power prevailing in each particular supply chain. The ultimate objective of capitalists as individuals is indeed their level of consumption, the ownership of fixed capital being only a necessary mediation to achieve it.

Finally, competition itself is a driving force which does not play its role only through the price system, but also through innovation and a permanent change of structures and organizations, which ends up by a continuous reshuffling of market shares and profit rates. The globalization process, through its relocations, is also conducive to a wide range of profit rates.

Let us conclude on this point that if capitalists really wanted to achieve an average rate of profit, they would need to know at the outset the value of this average rate, or at least have some approximate idea of its value, which is quite impossible for all the reasons which have been exposed. Therefore when making strategic decisions regarding for instance the nature and magnitude of their investments they have to rely on other indicators.

Before touching upon this question, it may be useful to recall that neo-classical theory itself has come to deal with the multiplicity of profit rates.

### 1.2.5. Neo-classical theory and the multiplicity of profit rates

Having reached this stage, it is interesting to note that within the framework of the theory of general intertemporal equilibrium, which constitutes to some extent the culmination of the neo-classical theory, one of the few authors to have published on the subject, Edmond Malinvaud, also comes to the conclusion that there is a multiplicity of "rates of profit", in his "Lectures on Microeconomic Theory" (Malinvaud, 1975, for the third French edition).

We will not develop here a full and detailed criticism of this theory[13]. Suffice it to say that these rates are defined in fact as commodity own-rates of interest. On this background, in order to introduce fixed capital considered as a factor of production, Malinvaud introduces the concept of equilibrium rate of interest, such that "for every commodity and every period, the market own-rate of interest is equal to the subjective own-rate of interest of every consumer and to the technical own-rate of interest of every enterprise". But, as he has the honesty to recognize it himself in the third edition of his "Lectures": "In general the own-rates of interest corresponding to one period and to different commodities do not coincide. For them to coincide, discounted prices should be such that ratios $\frac{p_{q,t+1}}{p_{qt}}$ take the same value for all commodities. But such an eventuality does not have a priori any reason to occur" (Malinvaud, 1975, p. 244).

[13] A detailed criticism can be found in my PhD thesis: Flamant, C. (1977). "*Hypothèses et concepts dans les théories du capital, de la valeur et de la répartition* ». University of Paris I - Pantheon-Sorbonne. pp. 138-157.

The theory of intertemporal equilibrium in practice remains unable to introduce on the basis of its own assumptions the concept of an equilibrium rate of interest. The notion of interest rate, introduced on a purely definitional basis, cannot in any case be linked to the marginal productivity of a "factor of production", i.e. fixed capital. Neo-classical theory, in spite of the diversity of its attempts and refinements, is unable to define fixed capital as a set of produced means of production and at the same time as a factor of production having at equilibrium a uniform remuneration: the profit rate (or the interest rate in this case). When it succeeds in defining fixed capital as a factor of production, it is only by suppressing the first term of its definition, by means of adopting the assumption of an economic world containing in fact only one commodity, which negates all its problematic of value.

### 1.3. Alternative indicators: the amount of profits and the profit/wage ratio

It follows from the above discussion that for capitalists the rate of profit is not a relevant indicator for the management of their enterprises. It is so because neither in theory, which highlights the multiplicity of profit rates, nor in practice, the profit rate cannot be a good indicator for the day to day management of a capitalist enterprise, owing to the difficulty to measure it, i.e. to know precisely its magnitude, and even less where a firm's particular profit rate might stand among a cloud of individual profit rates, mostly unknown. Moreover, as an average between a wide range of individual rates, it cannot be a good instrument to analyze in a simple way the functioning of the capitalist system.

In fact, the profit rate is practically not used as an indicator, either by capitalists themselves, for example to guide their investment decisions, or by investors in the stock market, who prefer to use the ratio of net profit per share to guide their speculative activity, or even by national accountants to measure the performance of firms. Nobody seems to refer to the existing overall rate of profit to assess the overall performance of a national economy. The search for the maximum profit rate is therefore not the central engine behind the operation of the capitalist system.

But this should not imply that the search for the maximum amount of profit is no longer a determining criterion of the functioning of capitalist enterprises. It is indeed a quantity which, not being a ratio between two distinct variables, is much easier to apprehend. Thereby, if capitalists have nevertheless a maximization objective regarding their profits, which is generally the case, this objective indeed remains the actual amount of monetary profits.

This is clearly seen in current practice with the importance given by business accounting to EBITDA (earnings before interest, taxes, depreciation and amortization), which corresponds at macroeconomic level of national accounts to what is called gross operating surplus. It is therefore safe to assume that the true motive behind the behavior of capitalist firms is the maximization of the amount – and not the rate - of profit.

This search for maximum profit takes place in an ideological framework delimited by neo-classical theory, in which this profit is supposed to correspond only to the marginal productivity of capital. From a microeconomic point of view, for a given price level and revenue provided by a given commodity produced by an enterprise, the amount of profits is established as a difference between value added and the amount of wages. The amount of profits is thus perceived rightly, at least at this level, as a competitor to the amount of wages in the sharing of this value added. It follows that the search for maximum profit leads to the search for a reduction in labor costs, both at the level of firms and at global level. With the rise of neo-liberalism, this decrease is supposed to be achieved among others by a removal of rigidities supposed to plague the labor market.

Let us therefore conclude that, in the real economy and the world as it is, what capitalists are looking for is first and foremost a rise in the profit/wage ratio. As we have seen, moreover, that profit is nothing else than monetized and redistributed surplus-value, even to the point that in section II producing consumption goods this profit/wages ratio is precisely equal to the rate of surplus-value (i.e. $k_c$ , or $\frac{k+c}{1-c}$ ). We thus come back to what constitutes the true and actual engine of the capitalist mode of production: the search for the highest possible rate of surplus-value. But as we will now show, the operation of such an engine can have adverse effects for the desired goal, inducing a structural instability of the system.

All these findings allow us to formulate a new principle:

**Principle 36**: The average rate of profit cannot be considered as a meaningful element of economic theory. Indeed, apart from the difficulty to simply measure it in the real world, it is nothing else than a mere statistical average. The real world, as well as a number of theoretical factors deriving from the nature of profit formation and realization, show that there is a multiplicity of profit rates both in theory and practice, and that the nature of economic activity, based on competition among firms, always pushes to a differentiation of these profit rates. The profit/wage ratio plays a more important role for the understanding of economic processes.

## 2. Reproduction of the system is problematic

### 2.1. Reproduction is possible, but not probable

The analysis performed in both the simple model and the model with capitalist consumption lead us to observe that there is a set of magnitudes for the various variables which allows for the full realization of the product, in the sense that all of the produced commodities are sold and all of the incomes are realized and spent.

This makes us consider that both models are “almost” in the universe of Say’s law, according to which all commodities which are produced can be sold, because they can be bought by all the incomes created on the occasion of production. This is the meaning generally accepted for this quote from chapter XV “Of the Demand and Market for products” of his “Treatise on Political Economy", originally published in 1803: "It is worthwhile to remark, that a product is no sooner created, than it, from that instant, affords a market for other products to the full extent of its own value. When the producer has put the finishing hand to his product, he is most anxious to sell it immediately, lest its value should diminish in his hands. Nor is he less anxious to dispose of the money he may get for it; for the value of money is also perishable. But the only way of getting rid of money is in the purchase of some product or other. Thus, the mere circumstance of the creation of one product immediately opens a vent for other products" (Say, last English edition in 2001, p. 57).

However we have demonstrated in this book and from our models that not all incomes, but only wages (or more generally the remuneration of labor) are created on the occasion of production, since incomes deriving from property are created in the sphere of circulation, as transfer incomes realized through exchanges of commodities against money. This is the reason why we wrote "almost" in the last paragraph. But the outcome is nevertheless the same, since - as we showed, these incomes are indeed clearly sufficient to buy the whole product.

Say argued that because production necessarily creates demand, a "general glut" of unsold goods of all kinds was impossible. But he considered nonetheless that if there were an excess supply of one good, there should be a shortage of others: "the glut of a particular commodity arises from its having outrun the total demand for it in one or two ways; either because it has been produced in excessive abundance, or because the production of other commodities has fallen short" (Say, 2001, p. 57). It is clear from this quote that Say at least accepted the possibility of disajustments between supply and demand on individual markets.

Coming back to our models, the introduction of capitalist consumption did not call into question the theoretical possibility of a complete realization of the product, nor consequently of a harmonious reproduction of the system. But this could only take place because on paper, i.e. in the world of theory, all of the different parameters involved in this reproduction, and in particular prices, adjusted to levels that effectively allowed this harmonious reproduction. However, in practice and in the real world, there is no reason to believe that these parameters will be precisely set at such levels for three reasons.

First of all, and this is the first one, there is no reason to think that monetary prices, as they are set independently by each capitalist enterprise, will allow for each commodity, be it a consumption good or a fixed capital good, to sell all the quantities produced, and thus to achieve exactly the expected profits, on the basis of which production was launched. This would imply that each entrepreneur, even before equilibrium is reached, would have a perfect information on all the prices and quantities which can be known only at this point. Therefore if prices are not equilibrium ones – and nothing can guarantee that they are such in a world of imperfect information, there will be excess supply on some markets, and at the same time excess demand on others. Assuming now that prices are more "viscous" than quantities produced, this can only result in an adjustment first by quantities. This hypothesis is not devoid of any empirical basis, because in the real world it is easier to change the daily production of a production unit (workshop or plant), something which can be done almost immediately, than to change a price catalog, which cannot be done every day.

On the other hand, and this is a second reason for maladjustments, once profits have been realized, and even assuming that they were realized in accordance with the original expectations, there is nothing to suggest that they will be spent autonomously by each capitalist in a way that achieves exactly the closure of the circuit and the reproduction of the system, whether for the purchase of consumer goods or for investment in fixed capital.

This comes from the fact that investment is a bet on the future, that the future is uncertain, and that this uncertainty is not probabilizable. Here one cannot but mention once more Keynes, who in chapter XII of the General Theory on "The state of long-term expectations", develops the idea that investment, insofar as it results from individual decisions of entrepreneurs, is influenced by their “animal spirits”: “Most, probably, of our decisions to do something positive, the full consequences of which will be drawn out over many days to come, can only be taken as a result of *animal spirits* - of a spontaneous urge to action rather than inaction, and not as the outcome of a weighted average of quantitative benefits multiplied by quantitative probabilities” (Keynes, 1936, chapter 12, p. 103).

It is no wonder therefore that Keynes viewed Say's law as essentially not true, and considered that effective demand, in the meaning of expected demand, rather than supply, is the key variable that determines the overall level of economic activity. Consequently there is no reason to expect enough aggregate demand to produce full employment.

Moreover, there is a third reason for maladjustments: while the dominant economic theory considers capital as a factor of production which - again when equilibrium is reached, must ensure a rate of profit identical to all its owners, the beginning of the present chapter has demonstrated on the contrary that the reproduction of the system implies a differentiation of these rates of profit, which is incompatible with this neo-classical representation of reality. But this differentiation of profit rates is a factor of increased rivalry between capitalists, by logically pushing those who realize profit rates deemed lower than those of their competitors to do everything possible to increase the profits levied on their own workers. As will be seen below, this creates a constant downward pressure on wages and upward pressure on profit shares and thus profit rates, which pushes primarily to a rise in the rate of surplus-value. Such a rise results from an action aiming at bringing down the level of wages, since this is indeed the most immediately accessible parameter. It is this same pressure that is at the origin either of productivity investments in the country where capitalist firms are located, or of investments abroad, in the form of relocations, which have exploded in advanced economies over the last 40 years, with the development of globalization.

To overcome these causes for maladjustment, capitalism has an absolute need for coordination between agents, in the form of a collective regulation, and it makes sense that this regulation should be the task of the State, as representative of the community and general interest. Indeed there has been in the last century an important regulation by the States, which prevailed in the aftermath of the second world war and continued for a good thirty years in most developed capitalist countries, up to the end of the seventies. This is certainly, together with the needs linked to Europe's reconstruction, the main reason for the high rates of economic growth and low levels of unemployment registered in the main occidental capitalist economies up to the beginning of the eighties.

However, a neo-liberal ideology based on questioning the role of the State in all areas has developed in the opposite direction from the beginning of the 1980s. This ideology flourished in the name of an individualism proclaiming the benefits of individual freedom in all areas, and conversely the harmfulness of the State and its regulation role. The progressive implementation of this ideology's recommendations all along the following decades has resulted in the constitution of what could be called a neo-liberal mode of regulation of capitalism, if this expression was not a magnificent oxymoron, because the objective of the neo-liberals is on the contrary the most extensive deregulation of the economy. The resulting long term crisis, which began early in the 1980s, has been marked by a generalized decline of growth rates and real wages in practically all of capitalist economies which adopted this ideology. It is a sign of the failure of this ideology and of the absence or failure of economic policies that should have ensured a smooth functioning of the capitalist mode of production, at the national level as well as internationally.

These last developments provide us with the logical bases to express a new principle:

**Principle 37**: It might exist a set of variables which make it theoretically possible to guarantee a balanced reproduction of a capitalist economy, in the neo-classical sense of ensuring a simultaneous equilibrium on all markets. However such a reproduction is practically impossible to obtain, for a number of equally theoretical reasons. First, because of the decentralized nature of such an economic system, which implies that decisions by capitalists, who fix prices and make investment decisions, are made independently, moreover in situations of imperfect information, and not coordinated. Second, because these decisions are made within time frames and based on expectations about the future which cannot be rational, because of the natural uncertainty which characterizes what happens in the future. Third because the dynamics of the system, if they are left unregulated by State authorities, go in the direction of a permanent increase in the rate of surplus-value, which in itself has destabilizing effects.

## 2.2. Crisis as coming from inherent difficulties of reproduction

It can be shown that it is the constant search for the maximization of profits, although it is the engine of the capitalist mode of production, which is also at the same time its worst enemy. And this because of its harmful consequences on the reproduction of the system, which are at the roots of its crises. This will be shown first by using essentially the simple model, since it is good enough for such a demonstration, without unduly complicating it.

### 2.2.1. Reproduction in section II

Let us suppose that after a production cycle which has been carried out without any difficulty, capitalists of section II, for whatever reason (for example, the perception that the amount of their profits is insufficient), want to increase these profits during the next cycle. For any capitalist who has a microeconomic perception of his business, the easiest way to do this is to reduce the amount of wages. This can be achieved either by reducing the amount of money wages paid to workers, something which can be hard to obtain, or by reducing the number of workers, which is generally simpler to achieve, while betting that increased productivity of remaining workers will maintain physical output at its previous level.

In the context of the simplified model, this results in a reduction in $L_{II}$, which falls from $L_{II}$ to $L'_{II}$, with $L'_{II} < L_{II}$, and hence in $W_{II}$, which decreases to $W'_{II}$ (with an unchanged money wage). If the price of the product of section II, i.e. the supply of consumer goods, is kept unchanged at $L_I + L_{II} = W_I + W_{II}$ (with $\bar{w} = 1$), ipso facto the demand for these goods emanating from workers will be reduced to the equivalent of $L_I + L'_{II} < L_I + L_{II}$. The value of the product of section II will also go down from $L_{II}$ to $L'_{II}$, and the price of the sold product will therefore be lowered to $L_I + L'_{II}$. Depending on whether the rise in labor productivity has offset or not the decline in the number of workers employed, we will be in a situation either of a maintained level of the quantities which are produced, or a fall in these quantities (a recession). The realized profit will not have increased, although its rise was the initial objective, but it will nonetheless remain fixed to the same amount $L_I$. The profit/wage ratio in section II, which in this section is, as we know, identical to the rate of surplus-value, i.e. $\frac{L_I}{L_{II}} = k$, will have increased, becoming $\frac{L_I}{L'_{II}} = k'$, with $k' > k$. But the number of workers employed will have stayed at its reduced level $L'_{II}$: unemployment will have increased.

### 2.2.2. Reproduction in section I

It has just been observed that, as a result of a fall in the cost of labor initiated by capitalists of section II, they record a fall in the overall amount of their sales and the failure of their attempt to increase the amount of their profits, which stays at $L_I$.

Let us suppose also that, since they maintained the amount of their profits at their previous level, and in order to ensure the growth of the physical productivity of labor in section II, capitalists in this section wish nevertheless to maintain the level of their investments, by continuing to buy fixed capital from section I for a price $L_I$, equivalent to the amount of their profits. If this is the case, then there is no reason at first sight that the initial decision of capitalists in section II has repercussions on the sales, employment and profits of section I. But can it be so?

To answer this question, let us recall that this analysis is based on a more general assumption of the uniqueness of the rate of surplus-value, considered as the result of a global social phenomenon, a rate which is the same for all workers, whatever the section in which they work. As unemployment has increased as a result of decisions taken in section II, the bargaining power of all workers has inevitably decreased, which points to a subsequent fall in wages, first in section II, but also in section I. It is indeed recalled that the objective pursued by section I capitalists, like those in section II, is indeed to increase the amount of their profits, which at the micro-economic level of each enterprise inevitably means to lower the amount of wages, which is made easier by the rise in unemployment.

If this reduction in wages materializes, then the nominal wage level in section I will drop from $L_I$ to $L'_I$. Paradoxically, this will result in a relative drop in the rate of surplus value, compared to the rate $\frac{L_I}{L'_{II}} = k'$, since a fall in the amount of wages in section I will have repercussions on the sales of consumer goods in section II, which will not reach a level $L_I + L'_{II}$, but rather a level $L'_I + L'_{II} < L_I + L'_{II}$. For this same reason, capitalists of section II will see their profits go down from $L_I$ to $L'_I$ (with $L'_I < L_I$), and therefore will not be able to maintain their purchases of fixed capital at its previous $L_I$ level. They will have to reduce their investment to an $L'_I$ level, which will reduce the amount of sales and profits of section I by the same amount. With a coefficient $a$ unchanged, the profit of section I actually becomes $L'_I \frac{a}{1-a} < L_I \frac{a}{1-a}$. This can only result in a fall in output and employment in section I, coming after the fall recorded in section II, ceteris paribus (the question of rising labor productivity is here left out).

To avoid burdening the demonstration, we will not discuss in detail what would happen if the initial decline in sales of consumption goods in section II, due to the fall in wages in this section, translated into a direct and proportional decline in section II purchases of fixed capital from section I, but it is not difficult to imagine that the effect on sales and therefore on output and employment in section I would be just as negative as in the case that has just been developed.

The new rate of surplus-value would then become $k'' = \frac{L'_I}{L'_{II}}$ a rate which might possibly be lower than the initial rate of surplus-value. The question of how this evolution can be reflected in the evolution of the average rate of profit will be examined in the following chapters, which deal precisely with the evolution of this last rate.

It can be concluded nevertheless that if the capitalist mode of production is characterized not only by an inherent antagonism between workers and capitalists, but also by competition among capitalists themselves, the latter pushes each one of them to realize the maximum amount of profits, and therefore the maximum amount of surplus-value, since profits are only surplus-value realized and redistributed among the capitalists under a monetary form.

And it has just been shown, albeit in a simplified framework, that the simultaneous pursuit of maximum profit by all capitalists, through the fall in the amount of wages which is its logical corollary at micro-economic level, leads at macroeconomic level to the reverse of the desired objective, and brings the system into a deflationary spiral of declining values, employment, and ultimately profits. These findings highlighted by the above demonstration allow us to formulate a new principle:

**Principle 38**: It is the very engine of the capitalist system, i.e. the pursuit of maximum profit at micro-economic level, which at macro-economic level constitutes a powerful factor of structural instability. Starting with a reduction in wages, it leads to cascading maladjustments that do not allow a full realization of the product and therefore hamper the reproduction of the system.

To be sure, the capitalist mode of production has not been functioning well over the last 40 years, at least in the most advanced economies, corresponding roughly to the OECD member countries. However, apart from the world financial crisis of 2008, followed by a deep recession in almost all of these countries, there has not been simultaneous true or prolonged recessions in these countries during this period, but rather a general reduction in the rate of GDP growth, accompanied by a significant decrease in the rate of productivity growth, a stagnation of real wages and a rise in income and wealth inequalities. This evolution might be the sign that the deflationary spiral evoked in the last paragraph has been counterbalanced by some kind of remedies against the fall of the system into an outright crisis. We will try to identify them in the next section.

## 3. Some remedies against crises

Quite paradoxically, capitalist consumption can be a remedy to such a crisis situation as described in the last section, together with another remedy which is autonomous investment.

### 3.1. Capitalist consumption as a remedy for crises

To understand how profits can be at the same time a scourge and a solution to the difficulties of reproduction which we just pointed out, we have to refer again to this fundamental mechanism highlighted by Keynes, who might not have realized its full scope: the widow's cruse.

To show that, we have obviously to get capitalist consumption back into the picture, which precisely corresponds to our second model with capitalist consumption. Then we can go back to subsection 2.2.1. above, where the insufficient demand in section II was due to a reduction in the amount of wages paid in this section. Supposing now that this fall in wages comes from a reduction, not in the number of workers, but in the level of nominal wages, then the value of commodities produced in section II in terms of average social labor is unchanged, as well as their price, even though their money value has decreased, from $W_{II}$ to $W'_{II}$, with $W'_{II} < W_{II}$.

But let us now consider that capitalists from section II can borrow an amount $W_{II} - W'_{II}$, corresponding to the reduction in demand for consumption goods from their own workers. Then, if they spend this borrowed money to buy the consumption goods left available for sale in section II (because they were not sold to workers due to the fall in their purchasing power), we are back to the mechanism of the widow's cruse highlighted in chapter 12. Capitalists of section II are able to buy all the consumption goods left over, and for which all wages amounting to $W'_{II}$ have already been paid, which means that the purchase of these goods will generate pure profits and allow to sell all of the commodities produced in section II.

This will however take place at a price which is globally unchanged but cannot correspond to the same rate of surplus-value. Indeed, instead of being $P_{II} = W_{II}\left(1+k_c\right) = W_{II}\left(1+\dfrac{k+c}{1-c}\right)$ this price is now $P_{II} = W'_{II}\left(1+k'_c\right) = W'_{II}\left(1+\dfrac{k+c'}{1-c'}\right)$.

In other words the rate of surplus-value has increased, from $k_c$ to $k'_c$, because of the increase in capitalist consumption (limited to capitalists of section II where wages have been cut) from a $c$ share to a $c'$ share of the value of consumption goods, with $c' > c$.

However, for this to happen, several conditions have to be met:

- First, for each capitalist of section II the borrowed amount must be related to his share in the market for consumption goods, since it must correspond to the amount of additional profits which he will get by selling his whole production ;
- Preferences of capitalists must be the same as those of workers if they had kept their purchasing power, precisely for the same reason.

If these conditions are fulfilled, then the consumption goods market will be cleared, capitalists of section II will have realized and spent a higher amount of profits, and they will be able to repay all the borrowed money to the banks. In other words money creation can be used by capitalists, not only for the payment of wages, but also to increase their consumption and their amount and share of profits.

This leads to the important conclusion that there are theoretically no limits to the amount of surplus-value and profits which can be obtained by capitalists, by having recourse to money creation, as long as they are able to repay the corresponding additional borrowings through increased profits. For a given quantity of fixed capital and investment, the increase in consumption is matched by additional profits: all that capitalists need to do after lowering the nominal wage rate and/or increasing the price of consumption goods is to absorb the consumption goods which workers are made unable to buy. In other words additional capitalist consumption substitutes to the missing workers' consumption.

Initially, the corresponding additional profits go to capitalist of section II, but undoubtedly those of section I will be able also to get an increased share of consumption goods, simply by increasing the price at which they sell fixed capital to capitalists of section II: they will thus increase their share of capitalist consumption. Once more we can see how prices play their distribution role, by distributing consumption goods between capitalists and workers and among capitalists themselves.

All this is nevertheless theoretical, because it implies that all of these additional profits must be spent on consumption goods, whereas in the real world it is obvious that the mechanism which we just described cannot work as smoothly and must necessarily face some implementation problems.

First because workers, especially in democratic countries where trade unions are strong, can oppose the reduction in their wages, be they nominal or real. This resistance explains that capitalists seek to relocate their activities from rich countries with high wages to poorer countries where wages are much lower. This process, known as globalization, has been and still is a powerful tool to overcome the resistance opposed by workers. The progressive reduction in the share of wages in the national income of all developed countries since the beginning of the eighties indeed testifies that this instrument has succeeded, and has correspondingly raised the share of profits in global income. This is the mechanism behind the rise of income inequalities in all of these countries. Over the years it is no surprise that these inequalities of incomes have translated to even higher inequalities of wealth. And since to benefit from lower wages implies for capitalists to invest in low-wage countries, the transfers of fixed capital involved have lowered the rate of growth in advanced economies and lead to their de-industrialization.

A second cause for difficulties comes from the fact that capitalists do not have the same structure of consumption as that of workers taken collectively: the richer they are and the higher is the share of luxury goods in capitalist consumption. As they grow richer, this implies therefore a reorientation of the productive system towards the production of luxury goods, which can take some time to materialize. Indeed all the available figures about the structure of consumption show that the luxury goods industry has grown faster after the world financial crisis of 2008 than the mass consumption industry.

But there is also a limit on the quantity of luxury goods which can be appropriated by an individual or his family: once somebody owns two or three luxury houses he usually does not need more, and the same can be said as far as beautiful yachts, private jets or luxury cars are concerned. As an exception is the case of ultra-rich people who become collectors, owning for instance dozens of houses in which they never live or dozens of luxury cars which they never drive.

This relative saturation of the luxury goods market means that with the rise in income and wealth inequalities a higher and higher share of profits no longer needs to be spent on commodities. Therefore an ever greater share of profits is diverted from the purchase of newly produced commodities to the purchase of assets such as works of art, ancient jewels, master paintings of past centuries, existing houses or palaces, or unproduced goods such as lands or even tropical islands. All of these are material assets, and one important category of assets is that of existing stocks, made of shares which represent the fractional ownership of corporations. These shares are ultimately equivalent to a property on a company’s net assets. They can easily be bought from anywhere in the world once they are listed on one of the world stock markets. The growth of inequalities induced by the rising share of profits causes more investments to be made on stock markets, which pushes these markets upwards.

On top of that and over the last years a new practice has developed, with the same results, i.e. a rise in stock markets indices: this is the increasing recourse of large capitalists companies to the method of buying back their own shares from the market place. There are several reasons for that. First of all, when a company purchases its own shares, since firms cannot be their own shareholder, repurchased shares are absorbed by them: shares are most often cancelled, which reduces the number of shares outstanding. Secondly, a buyback also helps to improve a firm price-earnings ratio (P/E), because when this happens, the relative ownership stake of each investor in a company increases due to the reduction in outstanding shares. There are fewer shares, or claims, on the earnings of the company: net earnings per share (EPS) increase, which mechanically makes share prices to increase. Moreover, buybacks are paid out from cash, which is an asset, and this reduces the assets value on the balance sheet. As a result, return on assets (ROA) actually increases. Return on equity (ROE) also increases because there is less outstanding equity. In general, the market views higher ROA and ROE as positives. All this explains that buybacks are generally considered as increasing shareholder value.

This is certainly true at microeconomic level for a particular firm. At macroeconomic level, buybacks are just a transfer of profits from firms to individual capitalists who sell their shares, so that the global effect on incomes is neutral. However, by reducing the outstanding amount of shares, it corresponds to an equivalent reduction in global net wealth (at least if the reduction in the quantity of shares is not fully compensated by an increase in their price).

As far as the dynamics of an economy are concerned, there is however one problem with this reorientation of profits towards the acquisition of all sorts of assets. Indeed the price of all these assets is soaring, but profits employed this way are not creating any additional output since they are not spent on the market for newly produced commodities and in particular are not invested in new fixed capital, which should be the normal use of profits that are not distributed as dividends to shareholders. This is anyway the condition of future growth. This would not be a big cause for concern if the amounts involved were not huge. However they are enormous. For instance, in the US and for companies belonging to Standard & Poor's 500 alone, the amount of buybacks was 572 billion USD in 2015 and 536 billion USD in 2016, i.e. 3.2 and 2.9 % of GDP, respectively, in each of these years. This is to be compared with the fact that insufficient investment as a proportion of GDP might well be an explanation for the poor growth in productivity in the US economy over the last decades.

Moreover spending profits on assets rather than on real goods only corresponds to a reshuffling in the property of pre-existing assets. Bearing in mind that speculation is the purchase of an asset on a market with the hope that it will become more valuable at a future date, this practice fuels speculation on the assets markets, thus creating the risk of igniting a new financial crisis. All the more so that the last 20 years have seen the development of new and "virtual" assets, such as derivatives, which are highly speculative and potentially more dangerous than real estate or even normal stocks. Subprimes could be considered as such a derivative, linked to subprime loan mortgages. The last world financial crisis of 2008 showed that Minsky, whom we already cited in the introduction of this book, was certainly right when in 1992 he explained in a short article ("the financial instability hypothesis", op. cit.) how speculation was at the root of the financial instability of capitalism, a reason why the last crisis has sometimes been called a "Minsky moment". But he considered that speculation was fueled by the building up of huge amounts of debts. What we just showed is that, if debts certainly play a role to facilitate speculation, they are not the only fuel of this activity, which can be financed also from existing profits, with the same potentially destabilizing results.

To conclude on this diversion of profit spending toward assets, its harmfulness comes not only from its capacity - through the rise in speculation, to trigger bubble bursts and financial crises, but also, as Keynes had indicated, from the deficit in demand which it generates on the market for commodities. However what Keynes was pointing out was the reduction in the spending of profits on consumption goods, whereas it should be clear from our models that a reduction in profit spending on the market for fixed capital has similar adverse consequences, and cannot but induce a shortfall in the realization of the product also for fixed capital goods.

This leads us to understand that the role played by profit spending in a crisis situation can have two antagonistic aspects, which need to be underlined in the following principle:

**Principle 39**: An autonomous increase in capitalist consumption financed through money creation can be a remedy to crises deriving from insufficient demand from workers, since it offsets the slump in workers consumption, and results in a higher share of profits. However higher profits may end up being at least partly spent for the purchase of existing assets instead of commodities. This cannot but perpetuate the insufficiency in demand for commodities and trigger speculation, which has been shown to be at the origin of financial crises.

### 3.2. Another remedy for crises: autonomous investment and the Keynesian multiplier

Chapter 10 "The marginal propensity to consume and the multiplier" of Keynes "General Theory" has popularized the idea that it is possible to define an investment multiplier *k*, such that an autonomous increase in investment $\Delta I$ will create an increase in output $\Delta Y$, so that $\Delta Y = k\, \Delta I$, with *k* linked to the marginal propensity to consume *c* by the following relation: $c = 1 - \frac{1}{k}$, which is equivalent to $k = \frac{1}{1-c}$ (Keynes, 1936, p. 78).

Therefore an obvious remedy to a situation of insufficient demand is for the State or a monetary authority to finance from money creation an additional investment, which is autonomous in the sense that it is not induced by the expected demand of capitalists, a demand supposed anyway to be insufficient to create full employment. This autonomous investment will increase both employment and income.

This idea may appear quite sound, but it is easy to show that the multiplier is not $k = \frac{1}{1-c}$, because in any case *k* is always equal to 1. This demonstration has already been made by Bernard Schmitt, in several of his books. It appears in particular in "Macroeconomic Theory -

A Fundamental Revision", which he published in 1972. But the corresponding demonstration is quite a lengthy one, since it takes more than 30 pages and involves the concepts of "nominal money" and "real money". "Nominal Money and Real Money" is also the title of chapter III of this book, where in section A - Positive Analysis, the author discusses the multiplier (Schmitt, 1972, pp. 123-156). But these particular concepts are not necessary from our point of view to achieve the desired result. We will briefly show that the demonstration is much simpler on the basis of our models, using the mechanism of the widow's cruse.

These models showed indeed that money created by banks at the beginning of a production cycle corresponds initially to the money value of the product, when it is paid to workers in the form of wages, which are a primary income. However we have seen that when incomes are spent on whatever type of commodities, the purchasing power of money in terms of the value of the purchased products, or realized value, is reduced by the levy of profits, which are a transfer income: this is the mechanism behind first the creation of surplus-value, and second its subsequent distribution among capitalists.

Therefore when an autonomous investment, as an additional spending on fixed capital (or in fact on any type of goods, be they fixed capital or consumption goods) takes place, the value which is realized through the purchase is lower than the amount of money spent in exchange for the commodities which are bought. If we call $I_a$ this autonomous investment and if $u_1$ is the ratio of wages $w_1$ to the price of the products $p_1$ bought in this initial purchase, we have $I_a = p_1$. Then this price is $\frac{1}{u_1}$ times the corresponding value $v_1 = w_1$, and the transferred income, i.e. the profit realized from this initial transaction, is $(1-u_1)\,p_1$. Once this amount of money $I_a = p_1$ is spent, its purchasing power has thus been used for an amount equivalent to the realized value of the purchased commodity, i.e. $w_1 = u_1 p_1$, and it cannot be used again: it is so because the corresponding amount of money has to be repaid to the banks from which it had been borrowed by the firms, at the beginning of the production process, in order precisely to pay the $w_1$ amount of wages of the workers employed by these firms.

As a second step, these profits corresponding to an amount $(1-u_1)\,p_1$ are used to buy other products for an identical amount, but only a fraction $u_2$ of their purchase corresponds to a realized value, and so on and so forth. In order to simplify the demonstration, let us consider temporarily that the ratio of the value realized on the occasion of each of these successive transactions is constant and equal to $u_1$. Then the series of realized values is:

Σ realized values = $u_1 p_1 + u_1(1-u_1)\,p_1 + u_1(1-u_1)^2\,p_1 + ... + u_1(1-u_1)^n\,p_1$, which gives:

$$\Sigma \text{ realized values} = u_1 p_1\left[1+(1-u_1)+(1-u_1)^2+...+(1-u_1)^n\right] \quad (1)$$

This is the sum of a geometric series of first term 1 and of constant ratio $(1-u_1)$. We know that the <ins>sum</ins> of a geometric series is finite as long as the absolute value of the constant ratio is less than 1 (which is the case), and *n* goes to infinity. This sum is therefore:

$$\Sigma \text{ realized values } = u_1 p_1 \left( \frac{1}{1-(1-u_1)} \right) = u_1 p_1 \frac{1}{u_1} = p_1 \qquad \text{(QED)} \qquad (2)$$

Thus the total amount of the sequence of realized values is $p_1$: this newly created value corresponds exactly to the initial spending of $I_a = p_1$, which demonstrates that the multiplier is equal to 1. What is nevertheless true is that this sum of realized values is higher than the amount $w_1 = u_1 p_1$ of the initial wages corresponding to the first purchase made from this spending, because equation (1) can also be written:

$$\Sigma \text{ realized values } = w_1 \left[ 1 + (1-u_1) + (1-u_1)^2 + ... + (1-u_1)^n \right] \qquad (3)$$

This in turn gives us:

$$\Sigma \text{ realized values } = w_1 + w_2 + w_3 + ... + w_n \text{, corresponding to:} \qquad (4)$$

$$\Sigma \text{ realized values } = \frac{w_1}{u_1} = p_1 \qquad (5)$$

Obviously in the real world, as previously indicated, there is absolutely no chance to have parameter $u = u_1$ as a constant. There are on the contrary as many $u_i$ parameters as there are firms and successive stages of spending for the amount of money put in circulation by the initial autonomous investment. But what is important is that the values of these $u_i$ are always lower than 1, which is a sufficient condition for the series to converge to a limit. This limit in any case cannot be higher than the money value of the additional product, i.e. the amount of wages paid out for its production, which themselves are equivalent to the initial injection of money corresponding to the autonomous investment.

Why then are so many economists thinking that the multiplier is higher than 1? For instance if the marginal propensity to consume $c$ is 0.75, they consider that the multiplier is $k$ = 4, because we are supposed to have $\Delta Y = \frac{1}{1-c} I_a = kI_a$. The reason behind such a mistake is that they do not understand that the purchasing power of income cannot be kept and transmitted indefinitely, but on the contrary disappears through the spending of income on real goods or services produced during the same period. This period corresponds to that of the payment of primary incomes for this production, and the levy of transfer incomes. These economists do not understand either that the incomes which are transferred in the form of profits on the occasion of each transaction also lose a part of their purchasing power at each subsequent transaction, exactly like primary incomes, and cannot realize a value equivalent to the price of the goods and services which they obtain. These repeated leakages explain indeed that a series of purchases cannot go one indefinitely, but on the contrary converges toward a maximum amount that cannot be greater than 1 compared to the initial purchase.

If we stay in the real world it must also be understood that this multiplier equal to unity is in fact an upper limit, which in order to be exactly reached would imply that there would be absolutely no leakages in the course of the successive spending of profits, meaning that all profits would always be fully spent on real goods and services. But in the real world there are many reasons, as we saw in the last subsection, for some part of the profits at each stage of their realization to be diverted to other uses than buying real goods or services, and this is going undoubtedly to lower below 1 the real effect of the multiplier.

Whatever the precise amount of the multiplier, and even if it is somewhat lower than 1, it remains that its existence provides a theoretical basis explaining why autonomous investment may play a useful role as a remedy to insufficient demand. This all the more so that its corollary is an increase in employment and thus in the creation of value, as long as productivity remains constant. Indeed investment spending as a spending on fixed capital implies that additional workers need to be recruited in order to produce this fixed capital, which creates an additional value. Moreover, once it has been produced and is used in the production process, a new investment implies the recruitment of additional workers to make it work, which also leads to value creation, although this will generally happen in a following period.

More generally autonomous spending can lead to the creation of additional value, if it takes place independently of any reduction in wages or in employment. Then it does not substitute to a reduction in the consumption of workers, because it triggers the production of additional goods and services, implying also the recruitment of additional workers.

The understanding of these mechanisms provides us with possibility to express a new principle.

**Principle 40**: As Keynes had understood, autonomous investment financed by money creation is the most effective tool to help an economic system out of a crisis. However Keynes, like his followers, was wrong in thinking that it would trigger a net multiplier effect such that the outcome in terms of product would be higher than 1. Because of leakages due to the existence of transfer incomes at each stage of spending, the overall amount of the multiplier between autonomous investment or spending and the corresponding product cannot be greater than 1.

## 4. Crises and evolution of the rate of profit

We only briefly evoked so far the mechanism at the origin of financial crises, on which we are not going to elaborate, because it is not the object of this book. We also highlighted the main mechanism at the origin of "real" crises, i.e. to summarize it briefly the contradiction between the effect of pursuing increased profits at micro-economic level and its outcome at macroeconomic level, where it results in insufficient demand. Having identified these mechanisms, and some potential remedies, we could stop here, if there was not an important economic literature regarding the Marxist explanation of crises linked to "the law of the tendency of the rate of profit to fall", which is exposed, under this very title, in part III of volume III of "Capital" (Capital, 1894, pp. 153-186).

The first chapter of this part of "Capital" (chapter 13) exposes "the law as such", the following chapter addresses "counteracting influences", and the third and last one is devoted to the "exposition of the exposition of the internal contradictions of the law".

There are a number of controversies on this question, not only between Marxists and anti-Marxist economists, but even between Marxist authors. Some of this literature has been devoted to confronting theory with the empirical measurement of the rate of profit and its evolution. Since this book is not primarily an empirical analysis, we will not touch upon this aspect of the question. Moreover, we showed that the rate of profit in our theoretical framework is, as a ratio, nothing more than a resultant, and anyway a statistical average of a multiplicity of profit rates. As such one could easily infer that it should not play such an important role in the functioning of the capitalist mode of production.

However, the question has been the object of so much debate among economists that it has undoubtedly some theoretical interest. All this justifies our view that it should be dealt with theoretically before trying to connect its evolution to crises and the reproduction of the mode of production. A theoretical approach is also needed before any empirical attempt at measuring the profit rate and its evolution, as a prerequisite for the understanding of the results which these attempts may deliver. This is the reason why we chose to devote the next and last chapters of this work to a purely theoretical point of view, based on the results already achieved regarding the nature and measurement of the rate of profit and its evolution.

# Chapter 18. The Marxist definition of the rate of profit in the simple model

Before Marx, the question of the long run evolution of the rate of profit and its influence on accumulation was not so controversial, since, as recalled by Tsoulfidis "Many of the great economists, including Adam Smith, David Ricardo, John S. Mill and Joseph Schumpeter, accepted the tendency of the rate of profit to fall and associated with this fall economic crisis as one of basic stylized facts of the evolution of capitalist economies" (Tsoulfidis, 2006, p. 65). But, as we indicated previously, the origin of present controversies is the Marxist law of the tendency of the rate of profit to fall. In order to examine it, we will proceed on a step by step basis, which justifies first to dedicate this chapter to the definition and calculation of the rate of profit, before dealing with its evolution in the following chapter. We will start here by recalling briefly the main thrust of Marx's argumentation, using the simple model which was developed previously.

## 1. Some reminders on the tendency of the rate of profit to fall

On a preliminary basis, since we will return more precisely to these different notions, let us recall first of all that the rate of profit $r$ is defined in the Marxist analysis purely in terms of values, and not in terms of prices. Marx deals with this question in chapter 3 "The relation of the rate of profit to the rate of surplus-value" of volume III of "Capital". There he indeed defines the rate of profit as the ratio of surplus-value $S$ over the sum of constant capital K and what Marx calls variable capital V, i.e. the value of labor-power (because he considers the latter as being also advanced):

$r = \frac{S}{K+V}$, which can be written: (1)

$$r = \frac{\frac{S}{V}}{\frac{K}{V}+1} \qquad (2)$$

It should be noted that we prefer to use letter K, rather than C, as the symbol representing constant capital, since C has already been used to represent consumption, and will continue to be used as such.

Let us recall also that within the framework of Marxist analysis, constant capital is composed of fixed capital - buildings, machines, beasts of burden - which serves for several periods of production, but which, in each, is incorporated into the value of the product only for a part of its own value (in relation to its own wear), and of circulating capital - raw materials, fuel, lighting - which, in each period of production, disappears to be incorporated in full, as well as its value, in the new product. Marx calls them constant because the value of equipment and materials being used in production is conserved and transferred to the new product.

For Marx, the rate of profit is therefore calculated not only over the value of fixed capital, but also over the value of intermediate goods, because this last value is supposed also to be advanced, like in the theory of production prices, and finally – as we already mentioned it, on the value of labor-power, for the same reason.

However, in all the following analysis, we will not use the Marxist concept of constant capital, and will only keep fixed capital at the denominator of the expression giving $r$, for two reasons (a third reason will be added in section 2 of this chapter):

1) First, for an empirical reason: because intermediate goods cannot be considered as advanced for an equivalent and constant period for all of these goods. This period is indeed different from one good to another, depending on various technical factors: it means that the stock circulation rate differs. Certainly for this straightforward and quite obvious reason the current accounting practice does not put intermediate goods at the denominator for the calculation of a rate of return.

2) Second, for a theoretical reason: when all of the intermediate commodities are nevertheless considered as advanced for a similar period of time, like in the theory of production prices, we have demonstrated at length in the first part of this book that this theory was not a coherent one, because it is based on the postulate that there is a surplus of intermediate goods (a condition for profits to appear), whereas as we have shown it is theoretically impossible to demonstrate that there is such thing as a surplus of these goods.

As for variable capital, it corresponds for Marx to the value of living labor employed in the production process, a value considered by him as advanced in the form of wages, which corresponds to what he calls the value of labor-power. Here too one could reject the idea that wages are advanced, and anyway if they are, it is only for quite a short period (one week or at most one month). However the full law of the tendency of the rate of profit to fall is based in Marxist theory on the increase in the organic composition of capital, and this last concept cannot be defined without keeping $V$ at the denominator of the expression giving the rate of profit. If we want to test the Marxist law, even if we do not agree on this Marxist way of calculating this rate, we are therefore obliged, at least temporarily, to stick to Marx's definition, which we will do in the present chapter and the two next ones.

Now that these points have been specified, and on the basis of the above definition, two hypotheses are then necessary for Marx to demonstrate the law of the tendency of the rate of profit to fall:

- the first one is that there exists in the capitalist mode of production a tendency to accumulate, linked to the competition between individual capitals, which results in the increase of the organic composition of capital, which is by definition equal to $q = \frac{K}{V}$ ;

- the second one is the constancy of the rate of surplus value $k = \frac{S}{V}$ or at least the idea that a rise in this rate would come up against limits such that $\frac{S}{V}$ increases less rapidly than $\frac{K}{V} + 1$ ;

The important point in the demonstration is anyway the fact that the evolution of $\frac{S}{V}$ and the evolution of $\frac{K}{V}$ are supposed to be relatively independent, even if some feedbacks are admitted between the two rates (for example the rise of $\frac{K}{V}$ can lead to a fall of V and a higher rate of surplus-value $\frac{S}{V}$), which explains why the decline in $r$ can be countered, and why the law is for this reason only a "tendency".

As we have indicated, this definition of the profit rate by Marx - with *V* at the denominator of the ratio giving the profit rate, is not the only possible one. Another one will prove to be more relevant anyway. However, and whatever the definition, we will always have the stock of fixed capital at the denominator of the ratio. The measurement of this stock is therefore of the essence, and this question will be addressed first, in the next section of this chapter.

## 2. Measuring the stock of fixed capital

In order for all of the terms in ratio $r = \frac{S}{K+V}$ giving Marx's rate of profit to be determined, the $K$ value of the total fixed capital stock in use over the production period has yet to be determined precisely, which we did not do so far. This capital is indeed the result of an accumulation made in previous periods, which raises the question of whether it is necessary to take into account depreciation, and of the nature and extent of this depreciation.

We recall that in our theoretical framework, and as a dimension of fixed capital, value cannot be "lost" to be transmitted to the value of the product, whereas fixed capital itself is not materially transformed into the product.

Taking thus into consideration the lack of transmission of the value of fixed capital to the product, depreciation can be defined not in theoretical and economic terms but more empirically as the accounting practice by which capitalists record what they deem to be the loss of value sustained by fixed capital when part of this capital comes out of the production process - after having been used there. It is thus the practice by which capitalists transfer to workers or more generally to buyers of their products the charge of the replacement of this capital, this depreciation being included in the price of the product, as a part of profits. It is indeed supposed to be equivalent to the difference between gross and net profits, or gross and net operating income, in national accounts.

If we call $t$ the duration of use of fixed capital, which can be considered as its average lifespan in a situation of simple reproduction, the capital stock $K$ in function necessarily loses each year a fraction of its value which corresponds to the capital put in service $t$ years ago, which has therefore reached the end of its lifespan $t$, and which therefore definitively leaves the production process by ceasing to be used, and therefore is discarded (or scrapped), to be replaced by newly produced fixed capital.

This makes it easy to calculate the value of the whole fixed capital stock, called $K$, which after being accumulated during previous periods, is in operation during each period. Its newly accumulated value in simple reproduction is indeed identical from period to period and equal to $I_i$, by calling $i$ any current period. If we suppose that the new fixed capital $I_i$ is put into operation at the beginning of each period, and during $t$ periods, $t$ being the average lifespan of fixed capital, we obtain the following equation:

$$K = \sum_{i=1}^{i=t} I_i = I_1 + I_2 + \ldots + I_t = L_i t \tag{3}$$

Since we are in a regime of simple reproduction, and in a situation without capitalist consumption, we have $I_i = S = L_I$, then:

$$K = St = L_I t \tag{4}$$

On the other hand, if one imagined that each generation of the capital stock $K_i$ in operation during a period $i$ loses a fraction of its value inversely proportional to its lifetime, this would have a consequence similar to the introduction of a linear depreciation of this capital. Fixed capital would then lose value continuously, but this would imply that its effective life is known in advance, and implicitly imply the transmission of lost value, something which we have rejected. Nevertheless, and by way of illustration, in this case capital $K_i$ would be accounted for at its residual value, net of amortization, as soon as the end of its first period of use, and we would obtain:

$$K = \left( I_1 - \frac{1}{t} I \right) + \left( I_2 - \frac{2}{t} I \right) + \ldots + \left( I_t - \frac{t-1}{t} I \right) \tag{5}$$

When performing the calculation the expression becomes:

$$K = I_i \frac{t-1}{2} \text{, that is, if we take } T = \frac{t-1}{2} \text{:} \tag{6}$$

$$K = I_i T \tag{7}$$

That would give in this context $K = S\frac{t-1}{2} = L_I\frac{t-1}{2}$, or $K = L_I T$ (8)

Instead of $K = L_I t$ (see equation 4)

Having reached this point, and since we are studying Marx's conception of the profit rate, the reader could raise an objection concerning what happened of circulating capital (i.e. the value of intermediate commodities) which is supposed also to be a part of *K*, as a component of constant capital. In fact, apart from the two reasons already mentioned above in section 1 (p. 348) there is a third and obvious reason not to include circulating capital in constant capital: it is that in our models and at macroeconomic level where we stand, each section is integrated, in the sense that each section produces, with the aid of fixed capital and labor, the totality of circulating capital, i.e. all of the intermediate goods necessary for the production of final goods (see sub-section 3.1. of chapter 11, p. 203). This means necessarily that values which appear in the models are only the values of final commodities, because the value of circulating capital is already included into the value of these final goods, which are $L_I$ for fixed capital and $L_{II}$ for consumption goods. Therefore it would not be logically coherent to count it again a second time in the value of *K*.

Going back now to the question of the transfer of value, there is clearly a significant divergence in the way capital stock $K$ is measured, between an analysis based on values, in which the transfer of value of fixed capital has been abandoned because it is illogical, and a conception based on the accounting practice of depreciation, which is implicitly based on the notion of transmission of value of fixed capital, assumed to correspond to a continuous loss of this value over time.

Although this latter conception is consistent with Marx's analysis of Capital, it is the source of insurmountable theoretical problems, as we already showed, since depreciation in its accounting sense is necessarily different from depreciation in its economic sense. To be convinced of that let us only ask ourselves how to take into account the transfer of value of a fixed equipment already amortized in the accounting sense, but which continues to be used in the production process, because its actual economic lifespan is greater than its accounting and purely conventional lifespan (a situation which corresponds moreover to the most frequent case). These intractable difficulties explain why this conception of the value of fixed capital stock is not retained in the model presented in this analysis.

Let us note that what is known at a given moment, for example at the beginning of a new production cycle, is the original value of the different elements of fixed capital in use during this cycle, and their date of entry into operation, from which one can derive the average lifespan of this fixed capital up to that moment, since it is an historical and given data. On the other hand, the total lifespan of the various elements of this fixed capital is unknown because it will depend on many factors, such as:

• the subsequent maintenance of these various elements, which is likely to prolong or shorten their total life, depending on the quality and intensity of this maintenance ;

• the pace of technical progress, which can render obsolescent and therefore lead to the decommissioning of certain elements of this fixed capital, even if their maximum possible lifetime is still far from being reached.

The only way to avoid these insurmountable problems is therefore to measure the stock of fixed capital excluding depreciation, and thus to its original value, as long as it is actually used in the production process. This is what is usually called valuation at historical values or prices.

In any case, there is one last reason for doing so, and this is the criterion of practice. Indeed, the practice of capitalist enterprises when it comes to estimating the profitability of an anticipated investment is as everyone knows to use always for this purpose as the cost of capital (of the investment in question) the initial actual cost, in order to calculate what in the jargon of financiers is called ROI ("return on investment"), by equalizing this cost with all the future incomes expected from this investment. Similarly, when it comes to comparing this initial forecast, which led to the realization of a particular investment, to the actual results obtained from the same investment, and therefore to an ex-post evaluation of the merits of this investment, it is obvious that we must reuse the same original value as that used for the initial calculations, that is to say this historical value. This is the method used by financiers and management controllers. Otherwise the comparison between expectations and achievements would no longer make sense.

However, we must be aware that doing so, i.e. using historical values and prices, leads to profit rate levels that will necessarily - and definitely, be lower than using the method taking into account the depreciation of fixed capital. This amounts to increasing the value of capital in proportion to the increase in its lifespan, i.e. by a factor $\frac{t}{\frac{t-1}{2}} = \frac{2t}{t-1}$ (see equations 4 and 6).

We can derive the consequence on the rate of profit, using Marx's formulation, which adds variable capital at the denominator. As a simplified example, let us assume that the rate of surplus-value $k$ is 100 %, meaning that $k = \frac{S}{V} = 1$ or $S = V$ .

Then, for an amount of surplus-value that is unchanged at the numerator of the expression of this rate, in the simple model without capitalist consumption, this implies that $L_I = L_{II}$ , and that the rate of profit calculated with the assumption of depreciation of capital will be:

$r_I = \frac{L_I}{L_I T + L_I} = \frac{L_I}{L_I \frac{t-1}{2} + L_I} = \frac{2}{t+1}$ instead of:

$$r_{II} = \frac{L_I}{L_I t + L_I} = \frac{1}{t+1}\ .$$

For the rate of profit calculated with the value of capital based on the assumption of its depreciation, this method corresponds to a multiplication by a factor 2: the profit rate is doubled.

If, on the other hand, the current definition of the rate of profit were used, considering it as the simple ratio of profit to fixed capital only, the resulting rate of profit, calculated with the assumption of depreciation, would be multiplied by a factor $\dfrac{\frac{1}{t-1}}{\frac{2}{\frac{1}{t}}} = \dfrac{2t}{t-1}$ .

In order to give a numerical example, let us assume that the average lifespan of fixed capital is 10 years. Then with Marx's definition the profit rate would go up from 9.1 % without depreciation of fixed capital to 18.2 % with depreciation. Using the current definition without variable capital in the denominator, the profit rate would go from 10 % to 22.2 %.

We had voluntarily left aside the question of the rate of profit when we had explored simple reproduction in the simplified model of chapter 11, where that was no capitalist consumption. Having dealt with the measurement of the stock of capital, we can now go further in analyzing the rate of profit and its evolution in connection with the organic composition of capital, since we have all the elements to determine this rate and its evolution. Like in the present section, this will be done first in the context of this simple model.

## 3. The average rate of profit in the simple model

In order to simplify the presentation, which is the least that we can expect from a simplified model, we will deal with only one case as regards parameter $u$ , i.e. the share of the value of available fixed capital obtained at each stage of exchanges taking place within section I, among capitalists of this section. To make sure that this case is not artificial, and as close as possible to reality, it seems better to stick to the assumption of case n° 2 examined above in chapter 11, where parameter $u$ is equal to $b$ , i.e. to $1-a$. This amounts to considering that capitalists of section I (apart from those selling fixed capital to section II capitalists) have a homogeneous behavior, and fix their price with the same margin rate $1-u=1-b=a$ .

The reason for this choice comes from the fact that in developed economies coefficient $a$ , as the share of the value of fixed capital employed in the sector producing this same fixed capital (therefore quite close to the share of employment in this sector), seems to have a magnitude ranging between 0.2 and 0.3 (or between 20 and 30%). Moreover with this assumption we also arrive at a margin rate (the same coefficient $a$ , by construction) which is consistent with the values found in these economies, where they are in the same area. If, on the contrary, we had adopted the hypothesis corresponding to case n° 1, with $u$ equal to $a$, the margin rate $1-a$ would have been between 70 and 80%, which would have been rather unrealistic.

Having adopted this assumption, we now have almost all of the elements that make it possible to determine the evolution of the rate of profit in the framework of the model, as depending on the variation of the organic composition of the capital, designated by symbol $q$, whose determination remains to be clarified. This is the reason why we will begin by this determination.

The value of the capital stock has been determined in the last section, with the assumption that the average lifespan of fixed capital, which is a given technical parameter, is noted $t$. Adding the simplifying assumption that $t$ is the same in both sections, we can deduce the organic composition of capital, both in terms of values and in terms of prices.

## 3.1. The organic composition of capital

### 3.1.1. The organic composition of capital in terms of values

1) At global level

Let us recall that $S = L_I$, $V = L_{II}$ and that according to equation (4) we have $K = L_I t$

This gives us:

$$q = \frac{K}{V} = \frac{L_I t}{L_{II}} \tag{9}$$

We know that $k = \frac{L_I}{L_{II}}$ is the rate of surplus-value, and that consequently $L_I = kL_{II}$

This implies that in terms of values $q = \frac{kL_{II}}{L_{II}} t = kt$, or conversely that $k = \frac{q}{t}$ (10)

2) In section II (for the sake of simplicity $t$ has the same value here as in section I)

The organic composition of capital in section II is expressed by ratio $q_{II} = \frac{K_{II}}{V_{II}}$ (11)

We know (see the first row of table 5 in chapter 11) that the value of fixed capital invested annually in section II is:

$$(1-a) L_I = (1-a) kL_{II} \tag{12}$$

It follows that the value of the total accumulated fixed capital is:

$$(1-a) L_I t = (1-a) ktL_{II} \tag{13}$$

Knowing moreover (see equation (4) in chapter 11) that the value of labor power in section II is:

$V_{II} = \frac{L_{II}}{1+k}$, we obtain:

$$q_{II}(values) = \frac{(1-a)ktL_{II}}{\frac{L_{II}}{1+k}} = (1-a)(1+k)kt \qquad (14)$$

3) In section I

The organic composition of capital in section I is expressed by ratio $q_I = \frac{K_I}{V_I}$

The annual investment in values in section I is $q_I = \frac{K_I}{V_I}$ and the capital stock is therefore $aL_I t$ (assuming that the average lifespan of fixed capital is the same in both sections).

The value of labor power in section I is $\frac{L_I}{1+k}$ (see also p. 214), so that we have:

$$q_I(valeurs) = \frac{aL_I t}{\frac{L_I}{1+k}} = at(1+k) \qquad (15)$$

### 3.1.2. The organic composition of capital in terms of prices

1) At global level

We know that the annual investment in terms of prices is $L_I\left(\frac{1}{1-a}\right)$. At global level, the price of the capital stock is therefore $K = \frac{tL_I}{1-a}$ and the price of labor-power (or total wages) is W, which is equal to $L_I + L_{II}$ (with $\bar{w} = 1$).

The organic composition of capital in terms of prices, or $q$ is therefore:

$$q(prices) = \frac{K}{W} = \frac{\frac{tL_I}{1-a}}{L_I + L_{II}} = \frac{\frac{tL_I}{1-a}}{L_I + \frac{L_I}{k}} = \frac{\frac{t}{1-a}}{\frac{(1+k)}{k}} = \frac{kt}{(1-a)(1+k)} \qquad (16)$$

2) In section II

We know that the price of investments during the period in section II is $L_I$ . The price of the stock of capital for section II is therefore $K_{II} = L_I t$ and that of labor-power (section II wages) is $W_{II} = L_{II}$ (with $\bar{w} = 1$ ).

It follows that the organic composition of capital in terms of price is:

$$q_{II}(prices) = \frac{K_{II}}{W_{II}} = \frac{L_I t}{L_{II}} = \frac{L_I t}{\frac{L_I}{k}} = kt \tag{17}$$

It has therefore the same value as the organic composition in terms of values at global level.

3) In section I

We know that the price of investments during the period in section I is $L_I \frac{a}{1-a}$ . The price of the stock of capital for section I is therefore $K_I = L_I t \frac{a}{1-a}$ and that of labor-power (section I wages) is $W_I = L_I$ (with $\bar{w} = 1$ ).

It follows that the organic composition of capital in section I, in terms of prices is:

$$q_I(prices) = \frac{K_I}{W_I} = \frac{\frac{aL_I t}{1-a}}{L_I} = \frac{at}{1-a} \tag{18}$$

We have now all the parameters that we need to define the average rate of profit.

### 3.2. The average rate of profit

At first sight, the rate of profit is normally supposed to belong to the theoretical universe of prices: in the neo-classical theory it is supposed to be the rate of return of capital as a factor of production, whose components are measured at their price. In the theory of production prices, it is defined as a ratio between a share of the surplus and the means of production made of basic goods, both being measured in prices. However in Marx's formulation it is defined as a ratio of values, which has the interest to make it invariable to prices and fluctuations in prices.

This is the reason why we will start by calculating the overall profit rate of the system keeping for the moment the same formula as that used by Marx, which we recalled in the introduction to this chapter. This implies to make calculations from data in value, at least for the time being, and first at global level. Furthermore, simple reproduction implies that there are different rates in both sections.

### 3.2.1. The average rate of profit at global level in terms of values

On this basis, we therefore have $r = \frac{S}{K+V} = \frac{\frac{S}{V}}{\frac{K}{V}+1} = \frac{\frac{L_I}{L_{II}}}{\frac{L_I}{L_{II}}t+1}$

Or, with the adopted notations, $r = \frac{k}{kt+1}$ (19)

and since $k = \frac{q}{t}$ (see equation (10) above), we get $r = \frac{q}{(q+1)t}$ (20)

In these two last equations, $r$ is a function of only two variables: either the rate of surplus-value $k$ or the organic composition of capital $q$, and the average lifespan of capital $t$. Now let us look at the profit rates in section I and section II, because we showed that in our theoretical framework there is no equalization of the rate of profit: let us state again that the difference between these rates is even a necessary condition for realization of the product and thus the reproduction of the system.

### 3.2.2. The average rate of profit in section II

Starting with section II producing consumption goods, we know that the value of labor-power $V_{II}$ is equal to $\frac{L_{II}}{1+k}$, that surplus-value amounts there to $\frac{k}{1+k}L_{II}$, and that accumulation during the period is equal in value to $L_I\left(1-a\right)$. In a simple reproduction regime, the capital stock in value is therefore simply $L_I\left(1-a\right)t$.

As a result, the profit rate in Section II expressed in terms of values is:

$$r_{II}\left(values\right) = \frac{\frac{k}{1+k}L_{II}}{L_I(1-a)t + \frac{L_{II}}{1+k}} = \frac{kL_{II}}{kL_{II}\left(1-a\right)\left(1+k\right)t + L_{II}} = \frac{k}{kt\left(1-a\right)\left(1+k\right)+1} \quad (21)$$

The profit rate in the same section, expressed now in terms of prices, has a much simpler expression, since its formulation leads to the following result:

$$r_{II}\left(prices\right) = \frac{L_I}{L_I t + L_{II}} = \frac{\frac{L_I}{L_{II}}}{\frac{L_I}{L_{II}}t+1} = \frac{k}{kt+1} \quad (22)$$

Thus, *the profit rate in the consumption goods section, expressed in the price system, is absolutely identical to the overall rate of profit, or the average rate of profit at global level,* *expressed in the value system*, as shown in equation (19). We can also observe that the rates of profit calculated in terms of values and in terms of prices coincide in this same section II only when:

$(1-a)(1+k)=1$, from equations (19) and (21)

Which implies that $(1-a)=\frac{1}{1+k}$ or also $1+k=\frac{1}{1-a}$ (23)

This gives us $a=1-\frac{1}{1+k}=\frac{k}{1+k}$ or also $k=\frac{1}{1-a}-1=\frac{a}{1-a}$ (24)

### 3.2.3. The average rate of profit in section I

Continuing with section I producing fixed capital, we know that the value of labor-power in this section is $\frac{L_I}{1+k}$, and that surplus-value there is therefore $\frac{k}{1+k}L_I$. We know also that fixed capital accumulated during one period in this section represents $a$ times the total value $L_I$ of the fixed capital it produces, and that the total accumulated capital represents $t$ times this value, namely:

$$K_I = aL_I t \quad (25$$

So the rate of profit in the section producing fixed capital (let us call it $r_I$), expressed in terms of values, is:

$$r_I(values)=\frac{\frac{k}{1+k}L_I}{aL_I t+\frac{L_I}{1+k}}=\frac{kL_I}{(1+k)aL_I t+L_I}=\frac{k}{a(1+k)t+1}=\frac{k}{akt+at+1} \quad (26)$$

As for the rate of profit of the same section I producing fixed capital, but this time in the price system, it is equal to:

$$r_I(prix)=\frac{L_I\frac{a}{1-a}}{L_I t\frac{a}{1-a}+L_I}=\frac{aL_I}{atLI+(1-a)L_I}=\frac{a}{at+1-a} \quad (27)$$

We can observe that in section I the rate of profit calculated in terms of values and that calculated in terms of prices coincide only when:

$\frac{k}{akt+at+1}=\frac{a}{at+1-a}$, which implies that:

$$a^2tk+a^2t+a=atk-ak+k \quad (28)$$

By putting $k$ in factor, it comes: $k(-a^2t + at - a + 1) = a^2t + a$

That gives then by rearranging: $k\left[\left(-a^2t - a\right) + \left(at + 1\right)\right] = a\left(at + 1\right)$

From which we get: $k\left(at + 1\right)\left(1 - a\right) = a\left(at + 1\right)$

This gives us $k = \frac{a}{1-a}$, which is equivalent, as we already saw in (24) to $a = \frac{k}{1+k}$ (29)

This shows that the condition of equality of rates of profit expressed in value and in price in section I is exactly the same as that already determined for section II (see equation 24).

#### 3.2.4. Equalization of average profit rates in value and price

To conclude on the average profit rate in the simple model, it has already been determined in the value system (see equation (19)), where it is equal to:

$$r\left(values\right) = \frac{k}{kt+1}$$

We still have to determine the average rate of profit in the price system, where it is equal to:

$$r\left(prices\right) = \frac{L_I \frac{1}{1-a}}{L_I \frac{1}{1-a} t + \left(L_I + L_{II}\right)} = \frac{L_I \frac{1}{1-a}}{\frac{L_I kt}{\left(1-a\right)k} + \frac{\left(1-a\right)L_I\left(1+k\right)}{\left(1-a\right)k}} = \frac{k}{kt + \left(1-a\right)\left(1+k\right)} \quad (30)$$

It can be deduced that the overall rate of profit expressed in terms of values and that expressed in terms of prices can be identical provided that we have:

$\frac{k}{kt+1} = \frac{k}{kt + (1-a)(1+k)}$, which simply implies that the following relationship be verified:

$$\left(1-a\right)\left(1+k\right) = 1 \Leftrightarrow \left(1-a\right) = \frac{1}{1+k} \Leftrightarrow 1 + k = \frac{1}{1-a}$$

This expression is equivalent to: $a = 1 - \frac{1}{1+k} = \frac{k}{1+k}$, and also to $k = \frac{1}{1-a} - 1 = \frac{a}{1-a}$

Therefore we may write : $a = \frac{k}{1+k} \Leftrightarrow k = \frac{a}{1-a}$ (31)

This is exactly the same condition as that required for the rate of profit calculated in terms of values and that calculated in terms of prices to coincide in section II producing consumption goods, which resulted from the equation (24), as well as in section I producing fixed capital (see equation 29). For this reason, from now on we will call these relations given by equations

(30) between the magnitudes of $a$ and $k$, and which ensure an identical rate of profit in values and in prices *the proportion of Marx***.**

At the end of all these calculations, it seemed useful to gather in a summary table the different variables and parameters of the simplified model, in terms of values as well as in terms of prices. They all appear in Appendix 3 at the end of the book, where they are provided in table 1 of this appendix. In this table the rate of profit $r$ is expressed as a function of $k$, the rate of surplus-value, rather than as a function of $q$, the organic composition of capital, since most of the variables are expressed in terms of $k$. But we could easily switch from one variable to the other, given that according to equation (10) we have $q = kt$, or conversely that $k = \frac{q}{t}$.

All these preliminary definitions were necessary before exploring what they mean in terms of the evolution of the profit rate within the framework of the simplified model without capitalist consumption. This evolution is the object of the next chapter.

# Chapter 19. The evolution of the rate of profit as defined by Marx in the simple model

We will continue in this new chapter to work on the basis of the Marxist definition of the rate of profit which puts variable capital, or the value of labor power, at the denominator of the expression giving this rate, since it is a necessary condition if we want to express it as a function of both the rate of surplus-value, i.e. $k=\dfrac{S}{V}$, and the organic composition of capital, defined as $q=\dfrac{K}{V}$. It is on this basis that its evolution will be analyzed and graphically represented, both in terms of values and in terms of prices. A numerical example corresponding to the simplified model will also be provided.

## 1. The evolution of the rate of profit and the proportion of Marx

At this stage, and as we just saw that the overall rate of profit expressed on the basis of values and that expressed on the basis of prices differ in the general case, one must inevitably ask what is the right formulation of this rate to be used with a view to discussing its evolution.

Certainly the rate that corresponds to the empirical reality is the rate expressed on the basis of prices. But as the point of departure of the last chapter was to check the validity of the Marxist law of the tendency of the rate of profit to fall, it nevertheless seems preferable to start, at least temporarily, with the Marxist definition of the rate of profit, expressed consequently on the basis of values. Let us recall that at global level and in terms of values we have:

$r=\dfrac{k}{kt+1}$, equivalent to $r=\dfrac{q}{(q+1)t}$

The Marxist definition is indeed that which must be retained if we wish to check the validity (or absence of validity) of Marx's law on the basis of his own hypotheses, knowing that this formula remains valid for the rate of profit expressed in terms of prices if the proportion of Marx is respected (see equations (30) at the end of last chapter).

Moreover, we have seen that the overall rate of profit expressed on the basis of values is in any case equivalent to the rate expressed on the basis of prices in section II producing consumption goods. But it will be shown anyway below, in the course of this chapter, that the results obtained on the basis of values do not differ fundamentally from those obtained by using the general formula of the rate of profit expressed in terms of prices.

The formula defining the rate of profit on the basis of values implies first of all that it is obviously impossible to consider the organic composition of capital $q=kt$ as independent of the rate of surplus-value $k$. On the contrary $q$ depends directly on this rate, since it is defined and deduced from it by the application of a $t$ coefficient that could be called the coefficient of immobilization of fixed capital, itself reflecting the lifespan of this capital. If this lifespan is assumed to be constant, this observation refers to the fact that the organic composition of capital cannot vary without a prior variation in the ratio between the value of fixed capital accumulated in each period and the value of labor power. This supposes a concomitant variation of the rate of surplus-value, and consequently of the proportion according to which workers are distributed between the two sections of production – with $\bar{w}$ being the same in both sections (taking into account the remark made on this subject above).

But the formula expressing the rate of profit allows in a second step to analyze the way in which the evolution of the organic composition of capital acts on the evolution of this rate, and this is where the interest of such a formula turns out to be the greatest.

To analyze the evolution of the rate of profit, we will continue to assume that the average lifespan of fixed capital is constant, which corresponds well to a situation of simple reproduction, and anyway does not seem to be an assumption too remote from reality, at least in the short-term or even the medium-term, given the inertia normally displayed by a variable like the value of the accumulated capital stock.

On this basis, it appears that function $r = f(k) = g(q)$ is a homographic function, i.e. a well-known category of functions, that can be represented as a quotient of two affine functions, of the form $f(x) = \frac{ax+b}{cx+d}$. The derivative of such a function is $f'(x) = \frac{\begin{vmatrix} a & b \\ c & d \end{vmatrix}}{(cx+d)^2}$.

This function $r = g(q)$ can easily be rewritten in such a way that we get:

$$r = \frac{q}{(q+1)t} = \frac{1}{t}\left(1 - \frac{1}{q+1}\right) = \frac{1}{t}\left[1 - f(q)\right] \qquad (1)$$

Where $\frac{1}{t}$ with $t$ as a constant is mathematically a scaling factor, which therefore does not affect the direction of variation of the function.

Moreover, it is trivial that function $f(q) = \frac{1}{q+1}$ is decreasing on $[0; +\infty[$ (2)

From that we deduce that the initial function $r = f(k) = g(q)$ is increasing on $[0; +\infty[$.

This means that any increase in the organic composition of capital corresponds to an increase in the rate of profit, and conversely that any decrease in the organic composition of capital corresponds to a fall in the rate of profit. This leads us to a result that is strictly opposite to that of Marx and the Marxist tradition! This means that at least within the framework of this simplified model, which is in any case consistent with Marx's approach - apart from the transfer of value of fixed capital, the law of the tendency of the rate of profit to fall cannot be sustained on mathematical grounds, and is therefore not logically valid. On the contrary, it must be replaced by the law of the tendency of the rate of profit to rise ! To be sure, this is a fundamental finding, even though this phenomenon is based on quite restrictive assumptions, and must therefore be relativized.

As stated at the beginning of this section (on the previous page), this observation remains true if we replace the formula on the basis of values used above to express $r$ by the formula in terms of prices (see equation (30) at the end of last chapter).

We have indeed: $r\left(prices\right)=\dfrac{k}{kt+\left(1-a\right)\left(1+k\right)}=\dfrac{q}{\left[q+\left(1-a\right)\left(1+k\right)\right]t}==\dfrac{1}{t}\left(\dfrac{q}{q+\alpha}\right)$ (3)

Considering temporarily that $\alpha=\left(1-a\right)\left(1+k\right)$

Then, since $t$ can again be considered as a scaling factor, we can write:

$r=\dfrac{1}{t}f(q)=\dfrac{1}{t}\left(\dfrac{q}{q+\alpha}\right)$ which gives us $f'(q)=\dfrac{\alpha}{\left(q+\alpha\right)^2}$ (4)

We deduce that $f'\left(q\right)$ has the same sign as $\alpha$ at the numerator, and therefore that:

$\alpha=\left(1-a\right)\left(1+k\right)>0\Leftrightarrow\ f\left(q\right)=\dfrac{1}{q+\alpha}$ is strictly increasing (5)

$\alpha=\left(1-a\right)\left(1+k\right)<0\Leftrightarrow\ f\left(q\right)=\dfrac{1}{q+\alpha}$ is strictly decreasing (6)

Now, we know that $k$ is by definition always positive, and that *a fortiori* $k>-1$. So the variations of $f(q)$ depend only on the sign of $\left(1-a\right)$. And since we also know that we always have $a<1$ (otherwise this would imply that the whole fixed capital of the productive system is used in section I), it follows that the rate of profit $r=f(q)$ expressed in terms of prices is always a strictly increasing function of the organic composition of capital $q$ and the rate of surplus-value $k$ since $q=kt$ (and $q$ is deduced from $k$).

It has thus been demonstrated, to be sure within the framework of the simplified model, but on the basis of Marx's own hypotheses - apart from the transmission of value of fixed capital - that in this particular case the law of the tendency of the rate of profit to fall must be replaced by the law of the tendency of the rate of profit to rise, and this even in the context of a definition of the rate of profit in terms of prices, which is absent in Marx.

This tendency to rise, however, has a limit, since the analysis of function $r=f\left(k\right)=g\left(q\right)$ shows that the rate of profit tends asymptotically towards a maximum which is equal to $r=\dfrac{1}{t}$, when the organic composition of capital tends to infinity: the maximum rate of profit is then equal to the inverse of the average lifespan of fixed capital. This is an interesting result, since conversely there does not seem to be a limit to the fall in the profit rate in the Marxist analysis, when the organic composition increases while the rate of surplus-value is supposed to remain constant (these last two evolutions being incompatible simultaneously within the framework of this model). But this new law of the tendency of the rate of profit to rise will perhaps be more meaningful to the reader by means of a graphical representation, which is given below.

## 2. A graphical representation of function *r = f (q)* in terms of values

Indeed, we observe in figure 19.1 that function $r = f\left(q\right)$ is increasing according to the organic composition of capital, and that we therefore have, for a constant fixed capital lifespan, an increase in the rate of profit, linked to the increase in the organic composition of capital, and which tends towards an asymptote equal to the inverse of the lifespan of fixed capital. But we see also that the whole curve representing function $r = f\left(q\right)$ moves downwards when the lifespan of fixed capital gradually increases, which is another factor of increase of the organic composition of capital, but which thus tends in the case of this variable to provoke a decline in the rate of profit.

In passing, this shows that the rate of profit is a complex variable, which, even at the level of simplification where we are here, depends, like the organic composition of capital, of two variables, the rate of surplus-value and the lifespan of fixed capital. Admittedly, in a situation of simple reproduction where the system is supposed to reproduce identically from one period to the next, the lifespan of fixed capital is not supposed to change. It is thus the variation of the rate of surplus-value only which provokes that of the organic composition of capital.

**Figure 19.1 - Representation of function *r (values) = f (q) = q / (q + 1) t*, for different values of t**

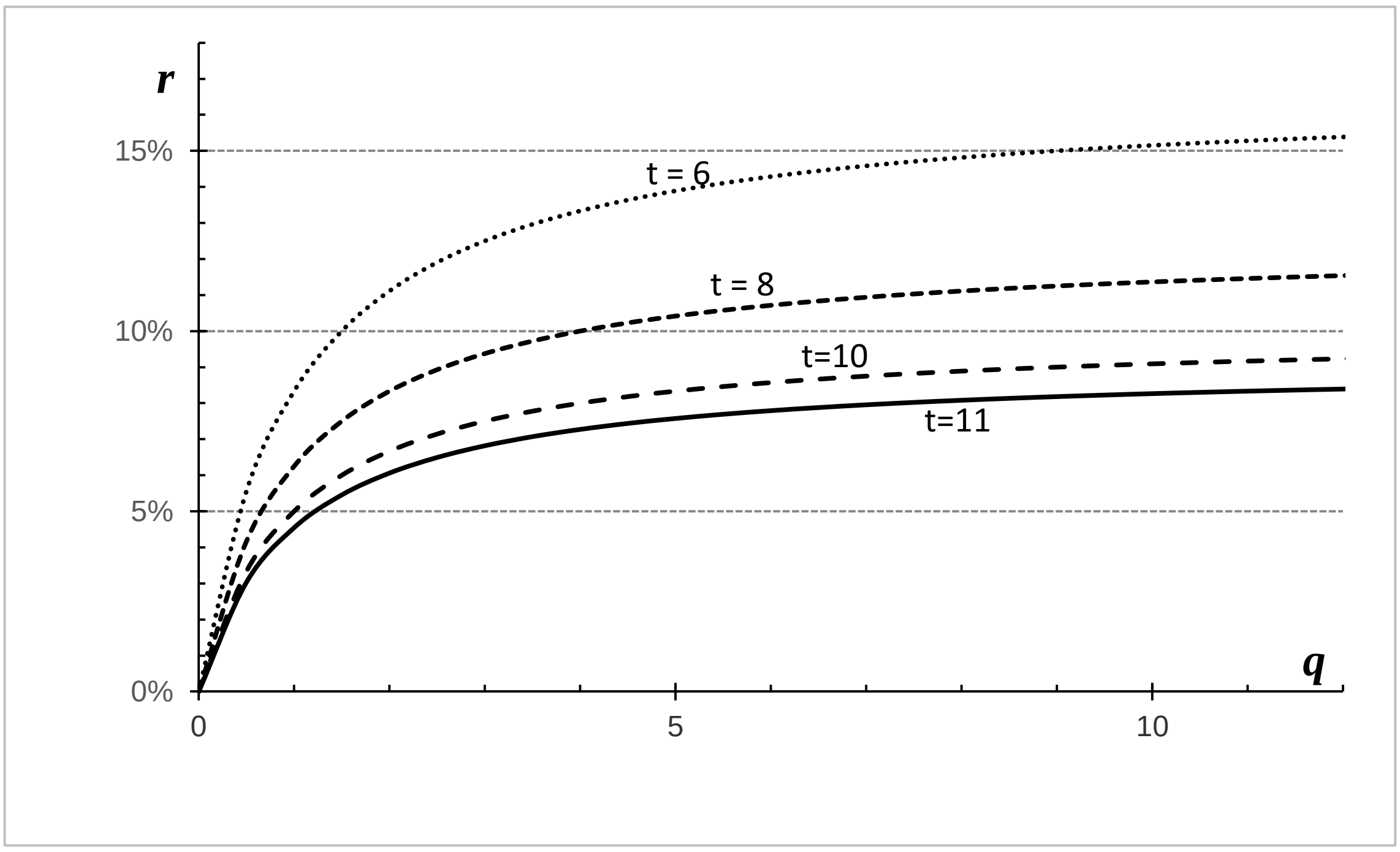

We will see in the next chapter that even in a situation of simple reproduction, other variables, related to the introduction of the consumption of capitalists, will come into play. Moreover, the real world is not in a situation of simple reproduction. Taking expanded reproduction into account could only lead to the introduction of an even greater number of variables, and this is why it will be tackled only briefly in this book, and by means of numerical examples.

The fact remains that the rate of profit in the real world is thus the result of a large number of parameters (among which tax rules regarding capital depreciation have already been mentioned). Their interaction can only lead to permanent fluctuations in the average rate of profit, upwards or downwards, just as the diversity of production conditions from one sector to another and one firm to the other can only play in the direction of a differentiation of profit rates between sectors and firms.

This is the reason why, although the present part started from a question about the relevance of a law that concerns the rate of profit, the reflection carried out here leads once more to put into perspective the importance of this law, especially with regard to the rate of profit, or rather the rates of profit, which are recorded ex-post.

On the other hand, at microeconomic level and in the context of the preparation of investment decisions, it is clear that the expected rate of profit, or what is often referred to as the projected internal rate of return (IRR), certainly continues to play an important and useful role. Here we find the notion of marginal efficiency of capital dear to Keynes, which corresponds to the expected rate of profit as a key variable in the decision to invest. But then we are at microeconomic level, because each firm has its own appreciation of what the level of marginal efficiency of capital should be, whether it leads to a given investment or not. Again, this observation goes into the direction of a multiplicity of profit rates which itself leads to relativizing very strongly the importance of such a parameter as the average rate of profit, both as an expected rate and as a realized rate.

It seemed useful nevertheless to make more concrete this new law of the tendency of the rate of profit to rise with the help of a numerical example, which is thus developed hereafter.

## 3. A numerical example corresponding to the simplified model

This example uses the static comparative method. Two systems A and B are considered in a state of simple reproduction for which organic composition differs, and consequently the distribution of the produced value between each of the two sections of the productive system. In addition, it is assumed that the average lifespan of fixed capital in both cases is 7 years, meaning that with our notations $t=7$. The values for the two systems A and B are:

Section I Section II

A $L_I = 100$ $L_{II} = 400$ , so that the organic composition is $q_A = \frac{K_A}{V_A} = \frac{100 \times 7}{400} = 1,75$

B $L_I = 200$ $L_{II} = 400$ , so that the an organic composition is $q_B = \frac{K_B}{V_B} = \frac{200 \times 7}{400} = 3,5$

We can see that the rate of surplus-value increases sharply from A to B and goes from $k_A = 0.25 = 25$ % to $k_B = 0.50 = 50$ %. The value of the total accumulated capital increases from $K_A = 700$ for system A to $K_B = 1400$ for system B. We thus obtain:

$$r_A = \frac{S_A}{K_A + V_A} = \frac{100}{700 + 400} = 9{,}1\% \text{, or } r_A = \frac{k_A}{k_A t + 1} = \frac{0.25}{0.25 * 7 + 1} = \frac{0.25}{2.75} = 0.091$$

$$r_B = \frac{S_B}{K_B + V_B} = \frac{200}{1400 + 400} = 11{,}1\% \text{ or } r_B = \frac{k_B}{k_B t + 1} = \frac{0.50}{0.50 * 7 + 1} = \frac{0.50}{4.5} = 0.111$$

We can observe that the rise in the organic composition of capital, which doubles (like the rate of surplus-value) from 1.75 to 3.5, is accompanied not by a fall, but by a rise in the profit rate from 9.1 % to 11.1 %.

It is also interesting to note in passing that this increase in the rate of profit is parallel to a decline in the ratio $\frac{Y}{K}$, which to some extent can be considered as representative of the "return" on capital, and which goes from:

$$\frac{100 + 400}{700} = 0,71 \quad \text{to} \quad \frac{200 + 400}{1400} = 0,43$$

The reader may, however, wonder whether the seemingly paradoxical result thus achieved is not due to the restrictive nature of the model assumptions, namely the lack of capitalist consumption and a situation of simple reproduction. If we actually introduce into the model the consumption of capitalists, we will see that the situation will indeed get more complicated and that the results will change. This will be shown in the chapters to come.

## 4. The evolution of the rate of profit in terms of prices

We saw in section 2 of chapter 17, starting from a situation where the rate of surplus-value was $k = \frac{L_I}{L_{II}}$, that an attempt to increase profits in section II had ultimately resulted in a decline in employment and value produced, and this in each of the two sections of the productive system, from $L_I$ to $L'_I$ and from $L_{II}$ to $L'_{II}$, and thus to a new rate of surplus value $k'' = \frac{L'_I}{L'_{II}}$. At the same time, it was found that the amount of profits had decreased in each of the two sections: from $L_I$ to $L'_I$ in section II, and from $L_I \frac{a}{1-a}$ to $L'_I \frac{a}{1-a}$ in section I.

What about the evolution of the rate of profit, globally and in each of the two sections, using now its expression in terms of prices ?

To answer this question, let us begin by emphasizing that the new rate of surplus-value $k''$ may be either greater than or lower than the initial rate of surplus-value $k$, depending on whether the rate of decline of $L_I$ is lower or higher than the rate of decline of $L_{II}$.

To appreciate the evolution of the overall rate of profit in terms of prices as a function of the evolution of the rate of surplus-value $k$, let us recall that its expression is:

$$r\left(prices\right)=\frac{k}{kt+\left(1-a\right)\left(1+k\right)}.$$

We know that in this simple model without capitalist consumption it is an increasing function of $q$. It is likewise an increasing function of $k$, since we know that $q=\frac{kt}{\left(1-a\right)\left(1+k\right)}$, in terms of prices (see equation 16 in chapter 18). Figure 19.2 corresponding to $r\left(prices\right)=f\left(k\right)$ is given for different values of $t$ and for $a=0.25$.

**Figure 19.2 - Representation of function *r (prices) = f (k)* = $\frac{k}{kt+\left(1-a\right)\left(1+k\right)}$ with *a* = 0.25**

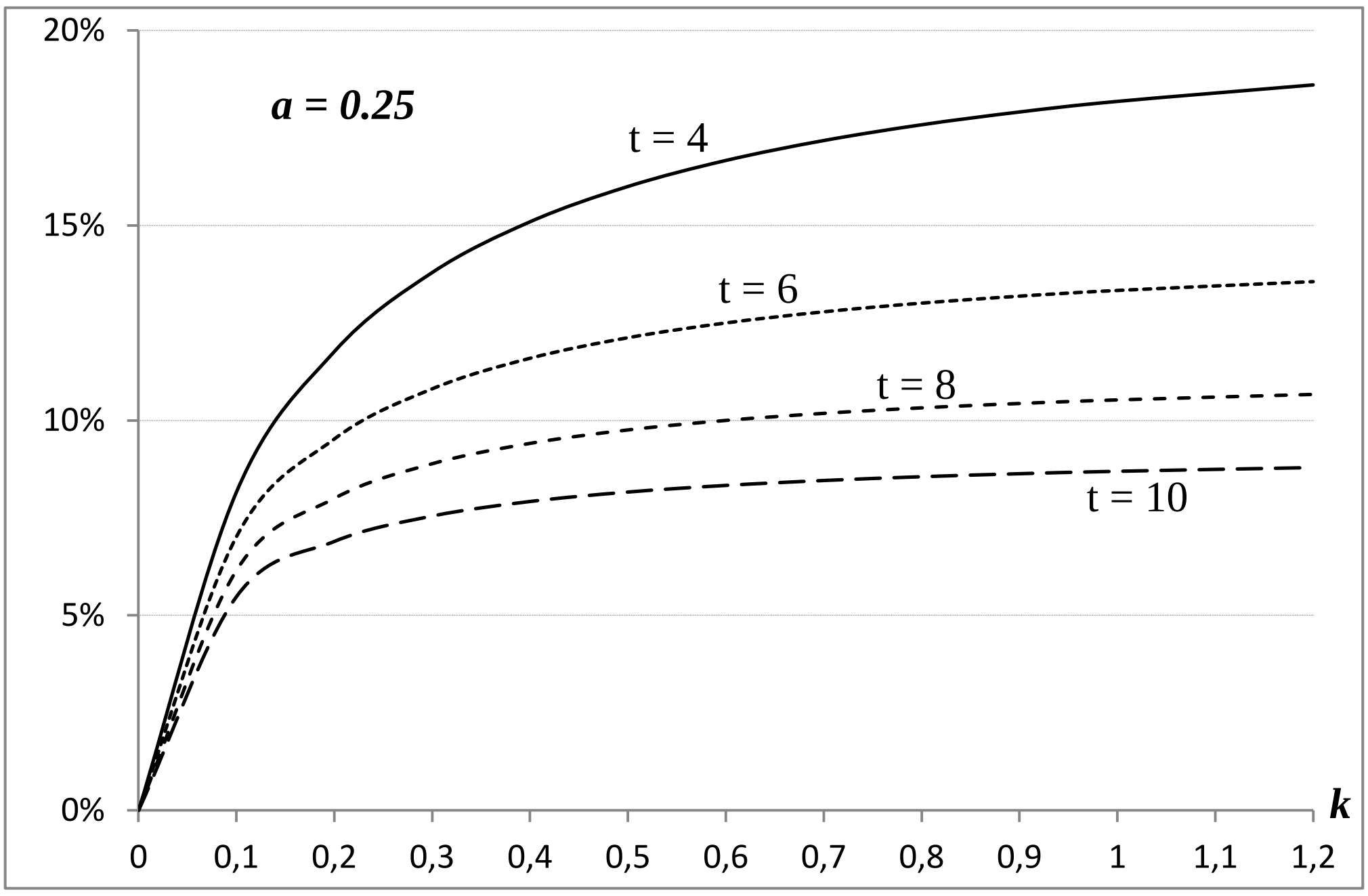

What happens in the case of an upward trend in the rate of surplus-value? In this simple model it implies that in section I $L_I$, corresponding to value produced (and by extension to employment) increases at a higher rate or decreases at a slower rate than $L_{II}$, the value produced and by extension the employment in section II. The graph above shows that an increase in the rate of surplus-value $k$ on the x-axis leads to an increase in the overall rate of profit on the y-axis.

This can happen even in the case of a decrease in the absolute amount of profits, if this decrease takes place at a lower rate than a simultaneous decrease in the value of labor power. But this rise of the global rate of profit can correspond to a divergent evolution for the rate of profit of each section:

- The rate of profit will increase at the same time in section II, where we have also $r_{II} = \frac{k}{kt+1}$, since it is similarly an increasing function of $k$ (its derivative is $\frac{1}{(kt+1)^2}$ ).

- Whereas in section I, where we have $r_I = \frac{a}{at+1-a}$, the rate of profit does not depend on the rate of surplus value k, but depends on the evolution of parameter $a$, i.e. the share of the value of fixed capital goods going to section I.

Figure 19.3 reflects this evolution.

**Figure 19.3 - Representation of function $r_I$ *(prices)* = *f* ( $a$ ) =** $\frac{a}{at+1-a}$

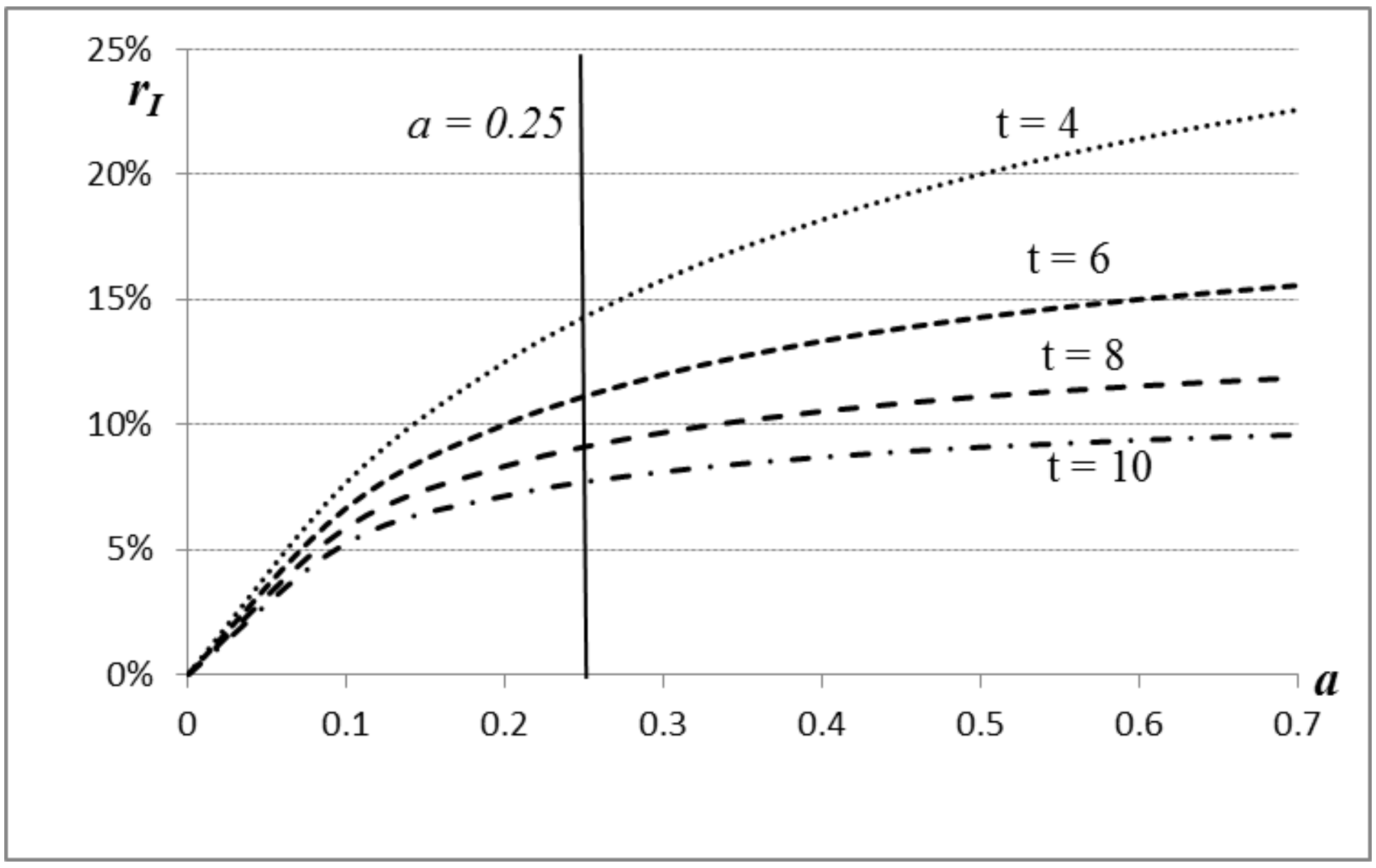

On the x-axis in figure 19.3 we have no more the rate of surplus-value but parameter $a$ instead. It can thus be seen that an increase in the rate of surplus-value $k$ can be accompanied by both an increase in the overall rate of profit and an increase in the profit rate in section II, but with a change in the rate of profit in section I which remains independent of this evolution of the rate of surplus value. Indeed it can take place in any direction - because it depends only on the evolution of parameter $a$ : if $a$ decreases then the rate of profit $r_I$ will itself fall. Its evolution can therefore diverge from that of $r$ and $r_{II}$. But all this can be accompanied by a simultaneous decline in the amount of profits in each of the two sections. This kind of possible evolution of the system goes clearly in the direction of its structural instability.

The least that can be concluded from these contrasting developments is that competition between capitalists, insofar as it seems to act first in the direction of an equalization of the monetary wage rate and/or the level of employment, does not act in the direction of equalizing profit rates. This equalization thus appears to be nothing else than a myth of political economy, to which Marx himself succumbed in Capital, but which has no serious theoretical justification or empirical consistency.

## 5. Rate of profit, rate of accumulation and intensity of capital

To conclude this chapter, it may be interesting briefly to review the relationship between the rate of profit, the rate of accumulation and the intensity of capital in this simplified model.

To this end, we will define here the rate accumulation as the ratio of on the one hand the value of the gross increase in the stock of capital in a period (corresponding to the surplus-value which is reinvested rather than consumed), over the overall gross product of the period, on the other hand. This is different from the usual Marxist definition, which at the numerator takes the net increase in the stock of capital in a period, since we must take into account the fact that in a situation of simple reproduction this net increase is obviously zero. In fact this rate of accumulation is similar *ceteris paribus* to what is usually called by national accountants the investment rate (appearing also in business statistics), i.e. the ratio of gross tangible investment to value added, in terms of prices.

As for capital intensity, it is defined here as the amount of fixed capital present in relation to labor. At the level of the production process, it may be estimated by the capital to labor ratio.

We know from equation (33) in chapter 11, concerning the simple model where the whole amount of surplus-value is invested, that the investment ratio in terms of values is:

$$\frac{I}{Y} = \frac{L_I}{L_I + L_{II}} = \frac{kL_{II}}{L_{II}\left(k+1\right)} = \frac{k}{k+1} \qquad (7)$$

It is clear that this investment rate or rate of accumulation is an increasing function of $k$, the rate of surplus-value.

As for the ratio of capital intensity, it is by definition, also in terms of values:

$$\frac{K}{L_I + L_{II}} = \frac{L_I t}{L_I + L_{II}} = \frac{kt}{k+1} \qquad (8)$$

Capital intensity is also an increasing function of $k$.

Not surprisingly, we can observe that in this model the ratio of capital intensity is equal to $t$ times the rate of investment, $t$ being the average lifespan of fixed capital.

We know also that in this simple model the global rate of profit $r = \frac{k}{kt+1}$ is an increasing function of the rate of surplus-value *k*, since the derivative of its function is positive. In fact it is immediate that it is the same, for an identical reason, for both the rate of accumulation and the rate of capital intensity: both rates increase with the rate of surplus-value.

Going one step further it is interesting to compare the evolution of the profit rate and that of these two last rates, which can be done quite easily by simply writing the expressions giving the ratios of the rate of profit over these two rates.

Indeed the ratio of the rate of profit over the rate of investment is :

$$\frac{r}{\frac{I}{Y}} = \frac{\frac{k}{kt+1}}{\frac{k}{k+1}} = \frac{k+1}{kt+1} \tag{9}$$

And the ratio of the rate of profit over the rate of capital intensity is:

$$\frac{r}{\frac{K}{L_I + L_{II}}} = \frac{\frac{k}{kt+1}}{\frac{kt}{k+1}} = \frac{k+1}{(kt+1)t} \tag{10}$$

The derivatives of both functions are:

$$f'\left(\frac{r}{\frac{I}{Y}}\right) = \frac{1-t}{(kt+1)^2} \text{, and :} \tag{11}$$

$$f'\left(\frac{r}{\frac{K}{L_I + L_{II}}}\right) = \frac{t-t^2}{\left[(kt+1)t\right]^2} \tag{12}$$

Since the average lifespan of fixed capital t is always > 1, it is obvious that both derivatives are negatives, which means that both ratios are a decreasing function of the rate of surplus-value *k*.

What all these equations tell us in economic terms is simply that in a situation of simple reproduction where all surplus-value is invested and transformed into fixed capital:

1) an increase in the rate of surplus-value *k* will necessarily induce an increase in the rate of profit, as well as an increase in both the rate of investment and the rate of capital intensity ;

2) the increase in the rate of profit will take place at a slower pace than the increase in the rates of investment and of capital intensity.

At the end of this chapter, and referring to the interrogation that we had at its beginning concerning the validity of the Marxian law of the tendency of the rate of profit to fall in response to an increase in the organic composition of capital, we can now state a principle which goes against this law:

**Principle 41**: In a situation of simple reproduction where there is no capitalist consumption and all surplus-value is invested, the rate of profit rises with the increase in organic composition of capital, both being the result of an increase in the rate of surplus-value.

# Chapter 20. The rate of profit as defined by Marx and its evolution in the model with capitalist consumption

When the model with capitalist consumption was initially defined, the question of the rate of profit in this model was left aside, and we have now to define it. Beforehand, we need first to define the stock of fixed capital and the organic composition of capital in this model. This last variable is indeed needed if we continue to define the rate of profit as Marx does, i.e. precisely as a function of the organic composition of capital. We will then be able to understand its evolution, precisely as a function of this organic composition. This will be analyzed both in terms of values and in terms of prices. We will show finally that the validity of the Marxist law is dependent on the share of capitalist consumption in total consumption.

## 1. Capital stock and the organic composition of capital

### 1.1. Capital stock

#### 1.1.1. Capital stock in terms of values

In this model, as we saw previously, surplus-value is increased by an amount corresponding to capitalist consumption, and is now equivalent to $S_c = L_I + cL_{II}$ . However in terms of values the fact that only a part of this surplus-value is now invested does not change the fact that this part is still equivalent to the value of fixed capital produced in a period, which remains $L_I$ . Therefore there is no reason for the measurement of the stock of capital to change as compared to the simple model.

Thus at global level this stock stays as:

$$K = L_I t \tag{1}$$

In section I and section II this stock is by definition:

$$K_I = aL_I t \tag{2}$$

$$K_{II} = (1-a) L_I t \tag{3}$$

#### 1.1.2. Capital stock in terms of prices

Although in the Marxist definition the analysis is carried out in term of values, it seems worth it to use this opportunity to give also the expression of the capital stock in terms of prices. At global level we saw in chapter 12 that from equation (39) and (40), and with $\alpha = \frac{a}{1-a}$, the price of fixed capital, the product of section I was:

$$I = \frac{L_I}{(1-a)(1-c*)} \text{ or also } I = L_I\left(1+\alpha+\gamma^*+\alpha\gamma^*\right)$$

(4)

This implies that with an average lifespan of $t$ years the price of the stock of capital is:

$$K = \frac{L_I t}{(1-a)(1-c*)} \text{ or } K = L_I t\left(1+\alpha+\gamma*+\alpha\gamma*\right) \tag{5}$$

In section II producing consumption goods, we can deduce similarly the price of the stock of fixed capital, which we call $K_{II}$, from the price of annual investment $I_{II}$, given by equation (37) in chapter 12:

$K_{II} = \frac{(1-a)L_I t}{(1-a)(1-c*)} = \frac{L_I t}{(1-c*)}$, which gives us, setting $\frac{c*}{1-c*} = \gamma*$:

$$K_{II} = \frac{L_I t}{(1-c*)} = L_I\left(\frac{c*}{1-c*}+1\right)t = L_I t\left(1+\gamma*\right) \tag{6}$$

Following the same procedure for section I producing fixed capital we get the price of the stock of fixed capital from that of its annual investment given by equation (38), which gives:

$K_I = \frac{aL_I t}{(1-a)(1-c*)} = \frac{a}{1-a}L_I t\frac{1}{1-c*}$, which gives us, again with $\frac{a}{1-a} = \alpha$

$$K_I = \alpha L_I\left(1+\gamma*\right)t = L_I t\left(\alpha+\alpha\gamma*\right) \tag{7}$$

From equations (6) and (7) we can check that the total stock of capital can also be written, as the sum of $K_I$ and $K_{II}$, and that we find the same result as that given by equation (4) :

$$K = K_I + K_{II} = L_I t\left(\alpha+\alpha\gamma*\right) + L_I t\left(1+\gamma*\right) = L_I t\left(1+\alpha+\gamma*+\alpha\gamma*\right)$$

## 1.2. The organic composition of capital

### 1.2.1. The organic composition of capital in terms of values

At global level the organic composition of capital is obviously modified, compared to the simple model, because the value of labor power $V$ is reduced by capitalist consumption and becomes $V_c$, so that $q$ becomes $q_c$ (we recall that $q = \frac{K}{V} = \frac{L_I t}{L_{II}} = kt$ )

$$q_c(values) = \frac{K}{V_c} = \frac{L_I t}{L_{II}(1-c)} = \frac{kt}{1-c} = \frac{q}{1-c} \tag{8}$$

From which one can obtain similarly: $\frac{q_c}{t} = \frac{k}{1-c} \Rightarrow q_c = \frac{kt}{1-c}$ (9)

Since we know that by definition $k_c = \frac{k+c}{1-c}$ , we can deduct from this equation that:

$$k_c = \frac{k+c}{1-c} = \frac{k}{1-c} + \frac{c}{1-c} \Rightarrow k_c = \frac{q_c}{t} + \frac{c}{1-c} \quad (10)$$

Remembering that we had $\frac{c}{1-c} = \gamma$ , we can also write: $k_c = \frac{q_c}{t} + \gamma$ (11)

And since equation (10) above implies that: $\frac{k}{1-c} = k_c - \frac{c}{1-c} = k_c - \gamma$ (12)

From equation (11) we check that we have again $q_c = (k_c - \gamma)t = \left[k_c - \frac{c}{1-c}\right]t = \frac{kt}{1-c}$ (13)

One can also calculate the organic composition of capital in the system of values for each section. In section I, the organic composition of capital becomes:

$$q_{Ic}(values) = \frac{atL_I}{\frac{L_I}{1+k_c}} = at(1+k_c) = at\frac{1+k}{1-c} \quad (14)$$

The same calculation for the organic composition of capital in section II results in the following expression:

$$q_{IIc}(values) = \frac{(1-a)tL_I}{\frac{L_{II}}{1+k_c}} = \frac{(1-a)tL_I}{\frac{L_I}{k(1+k_c)}} = kt(1-a)(1+k_c) = kt(1-a)\frac{1+k}{1-c} = \frac{kt(1-a)(1+k)}{1-c}$$

From equation (9) we know that $q_c = \frac{kt}{1-c}$. We can thus replace $\frac{kt}{1-c}$ by $q_c$ in this last equation, which gives us:

$$q_{IIc}(values) = kt(1-a)\frac{1+k}{1-c} = q_c(1-a)(1+k) \quad (15)$$

### 1.2.2. The organic composition of capital in terms of prices

Equation (4) in subsection 1.1.2. above gave us the price of fixed capital produced in section I. As regards the organic composition of capital in terms of prices, and taking as usual $\bar{w} = 1$, it is then:

- For the overall organic composition, i.e. $q_c(prices) = \frac{K_c(prices)}{V_c(prices)} = \frac{K_c(prices)}{W(prices)}$

$$q_c\left(prices\right)=\frac{L_I t\left(1+\alpha+\gamma^*+\alpha\gamma^*\right)}{L_I+L_{II}}=\frac{L_I t\left(1+\alpha+\gamma^*+\alpha\gamma^*\right)}{L_I\left(1+\frac{1}{k}\right)}=\frac{kt\left(1+\alpha+\gamma^*+\alpha\gamma^*\right)}{\left(1+k\right)} \tag{16}$$

But we know from equation (4) that $\left(1+\alpha+\gamma^*+\alpha\gamma^*\right)=\left(\frac{1}{1-a}\cdot\frac{1}{1-c^*}\right)$

So that we can write also: $q_c\left(prices\right)=\frac{kt}{\left(1+k\right)\left(1-a\right)\left(1-c^*\right)}$ (17)

- For the organic composition in terms of prices in section I, we know $K_{Ic}$ from equation (7) above in subsection 1.1.2.:

$$q_{Ic}\left(prices\right)=\frac{K_{Ic}\left(prices\right)}{W_{Ic}\left(prices\right)}=\frac{\frac{atL_I}{\left(1-a\right)\left(1-c^*\right)}}{L_I}=\frac{at}{\left(1-a\right)\left(1-c^*\right)}=at\left(1+\alpha+\gamma^*+\alpha\gamma^*\right) \tag{18}$$

- For the organic composition in terms of prices in section II, we also know $K_{IIc}$ from equation (6) above in subsection 1.1.2.:

$$q_{IIc}\left(prices\right)=\frac{K_{IIc}\left(prices\right)}{W_{IIc}\left(prices\right)}=\frac{\frac{tL_I}{1-c^*}}{L_{II}}=\frac{\frac{tL_I}{1-c^*}}{\frac{L_I}{k}}=\frac{kt}{1-c^*}=kt\left(1+\gamma^*\right) \tag{19}$$

We have defined now all the variables that are needed in order to express the rate of profit in this model of simple reproduction with capitalist consumption.

# 2. The average rate of profit as defined by Marx in the model with capitalist consumption

As we already pointed out, the rate of profit is defined by Marx in terms of values, but can be expressed as a function of various parameters. In this section we will calculate its expression first as a function of the rate of surplus-value, and second as a function of the organic composition of capital.

## 2.1. The average rate of profit expressed as a function of the rate of surplus-value $k_c$

By expressing $r_c$ as a function of the rate of surplus-value $k_c$, the formula giving the average rate of profit $r_c$ at global level in the model with capitalist consumption thus becomes:

$$r_c\,(values) = \frac{\frac{S_c}{V_c}}{\frac{K_c}{V_c}+1} = \frac{\frac{L_I + cL_{II}}{L_{II}(1-c)}}{\frac{L_I}{L_{II}(1-c)}t+1} = \frac{\frac{L_I}{L_{II}(1-c)}+\frac{c}{1-c}}{\frac{L_I}{L_{II}(1-c)}t+1} = \frac{\frac{k+c}{1-c}}{\frac{kt}{1-c}+1} = \frac{k_c}{q_c+1} \tag{20}$$

Using equation (10) above in subsection 1.2.1., from which we can derive that:

$\frac{q_c}{t} = k_c - \frac{c}{1-c} \Rightarrow q_c = \left[k_c - \frac{c}{1-c}\right]t$ , we can therefore write:

$$r_c = \frac{k_c}{q_c+1} = \frac{k_c}{\left[k_c - \frac{c}{1-c}\right]t+1} \tag{21}$$

And since $\frac{c}{1-c} = \gamma$ , then we can define the rate of profit at global level $r_c$ as a function of the rate of surplus-value $k_c$ only, which was the objective of this sub-section. Indeed it can now be written as:

$$r_c = \frac{k_c}{(k_c - \gamma)t+1} \qquad \text{(with } k_c > \gamma \text{ )} \tag{22}$$

### 2.2. The average rate of profit expressed as a function of the organic composition of capital $q_c$

#### 2.2.1. The average rate of profit at global level

In order to express $r_c$ as a function of the organic composition of capital $q_c$ , we can use for this purpose equation (11): $k_c = \frac{q_c}{t} + \gamma$ . Then the formula giving the average rate of profit in the model with capitalist consumption becomes:

$$r_c\left(values\right) = \frac{\frac{q_c}{t}+\frac{c}{1-c}}{q_c+1} = \frac{q_c+\frac{c}{1-c}t}{(q_c+1)t} = \frac{q_c+\gamma t}{(q_c+1)t} \tag{23}$$

#### 2.2.2. The average rate of profit in section I

As regards the expression in the value system of the rate of profit in section I, in order to obtain it we just need to remember first that the rate of surplus-value is the same in both sections and in the whole economic system (we always have $k_c = k_{Ic} = k_{IIc}$ ), and to go back to the formula defining it, namely:

$$r_{Ic}\left(values\right)=\frac{\frac{S_{Ic}}{V_{Ic}}}{\frac{K_I}{V_{Ic}}+1}=\frac{k_{Ic}}{q_{Ic}+1}=\frac{k_c}{q_{Ic}+1}=\frac{\frac{q_c}{t}+\gamma}{q_{Ic}+1}=\frac{q_c+\gamma t}{\left(q_{Ic}+1\right)t} \tag{24}$$

This expression can also be written in terms of $k$, i.e. the primary rate of surplus-value (the rate defined in the simple model), since all these variables are already known:

$$r_{Ic}\left(values\right)=\frac{k_c}{at\left(1+k_c\right)+1}=\frac{\frac{k+c}{1-c}}{at\frac{1+k}{1-c}+1}=\frac{k+c}{at\left(1+k\right)+1-c} \tag{25}$$

### 2.2.3. The average rate of profit in section II

Lastly, as regards the expression of the profit rate in Section II, in order to obtain it in the system of values, it suffices to go first to the formula defining it, namely:

$$r_{IIc}\left(values\right)=\frac{k_{IIc}}{q_{IIc}+1}=\frac{k_c}{q_{IIc}+1}=\frac{\frac{q_c}{t}+\gamma}{q_{IIc}+1}=\frac{q_c+\gamma t}{\left(q_{IIc}+1\right)t} \tag{26}$$

Then we can use equation (15) giving the organic composition of capital in section II, which, like for the rate of profit in section I, allows that this expression be written also in terms of $k$, the primary rate of surplus-value, which gives us:

$$r_{IIc}\left(values\right)=\frac{k_c}{kt\left(1-a\right)\left(1+k_c\right)+1}=\frac{\frac{k+c}{1-c}}{kt\left(1-a\right)\frac{1+k}{1-c}+1}=\frac{k+c}{kt\left(1-a\right)\left(1+k\right)+1-c} \tag{27}$$

We note that this average rate of profit in section II can thus be defined in terms of the variables of both the simple model: $k$, and the model with capitalist consumption: $k_c$ and $c$.

### 2.2.4. The condition for the equalization of profit rates in both sections

Incidentally, we can ask ourselves what are the conditions for the profit rates $r_{Ic}$ and $r_{IIc}$, defined in terms of values, to be identical in each of the two sections. Since the two numerators in the two equations giving $r_{Ic}$ and $r_{IIc}$ are identical and equal to $k+c$ (see equations (25) and (27), to obtain $r_{Ic}=r_{IIc}$ we just need that the denominators of the two ratios expressing these profit rates in each of the two sections be equal, namely that:

$$at\left(1+k_c\right)+1-c=kt\left(1-a\right)\left(1+k_c\right)+1-c$$

This identity can be very easily simplified, just by removing $(1-c)$ and dividing each term on both sides by $t(1+k_c)$, which leaves us with the following identity:

$$a = k(1-a) \text{ or } k = \frac{a}{1-a} \text{ which is equivalent to } a = \frac{k}{1+k} \tag{28}$$

We thus note that despite the modified definition of the rate of profit when the consumption of capitalists is taken into account, the verification of this equality between the rates of profit in section I and section II, in terms of values, implies always the same condition as that which was needed for the equalization of the rates of profits in terms of values and in terms of prices in the simple model, meaning that the structure of the system must be consistent with what we called the proportion of Marx.

### 2.3. The average rate of profit expressed as a function of the variables of the simple model

Indeed it is also possible to express $r_c$ as a function of the organic composition of capital and the rate of surplus-value which were defined in the initial simple model (in the absence of capitalists consumption), thus keeping the notations already used in this simple model, with $k$ and $q$, respectively, for the rate of surplus-value and the organic composition of capital.

Let us recall first that by definition $k_c = \frac{k}{1-c} + \frac{c}{1-c} = \frac{k}{1-c} + \gamma$

And since $k = \frac{q}{t}$ then $k_c = \frac{q}{(1-c)t} + \gamma$ (29)

Let us recall also that on the basis of equation (8): $q_c = \frac{q}{1-c} = \frac{kt}{1-c}$ (30)

To express $r_c$ as a function of $k$, we can to go back to equation (20), in which we replace in the denominator $q_c$ by its value given by equation (29), and the number 1 by $\frac{1-c}{1-c}$, before simplifying the whole expression by $(1-c)$. We then obtain:

$$r_c(values) = \frac{k_c}{q_c+1} = \frac{\frac{k+c}{1-c}}{\frac{kt}{1-c} + \frac{1-c}{1-c}} = \frac{k+c}{kt+1-c} = \frac{k+c}{q+1-c} \tag{31}$$

To express $r_c$ only as a function of $q$, the organic composition of capital, we just need to start from the same equation (20), and knowing that $q = \frac{L_I t}{L_{II}}$, this makes it possible to obtain:

$$r_c\left(values\right)=\frac{\frac{L_I}{L_{II}}+c}{\frac{L_I}{L_{II}}t+(1-c)}=\frac{\frac{L_I}{L_{II}}t+ct}{qt+(1-c)t}=\frac{q+ct}{(q+1-c)t} \tag{32}$$

# 3. The evolution of the rate of profit with the Marxist definition

Now that the rate of profit has been fully defined within the framework of Marx's formulation, i.e. in terms of values, what can we say about its evolution as a function of the organic composition of capital, again in this same model with capitalist consumption?

## 3.1. The evolution of the rate of profits at global level in terms of $q_c$

Let us first start from equation (23) above, that is from the expression of the rate of profit $r_c$ as a function of $q_c$, i.e. $r_c = g\left(q_c\right)=\frac{q_c+\gamma t}{\left(q_c+1\right)t}$, where $\frac{1}{t}$ can be considered mathematically as a scale factor, which therefore does not affect the variation of this function.

So $g\left(q_c\right)$ can be replaced by the homographic function $\tilde{g}\left(q_c\right)= t.r_c=\frac{q_c+\gamma t}{q_c+1}$ (33)

From which we obtain the derivative $\tilde{g}'\left(q_c\right)=\frac{1-\gamma t}{\left(q_c+1\right)^2}$ (34)

We deduce from this last equation that $\tilde{g}'\left(q_c\right)$ has the sign of $1-\gamma t$ and therefore, with regard to function $g\left(q_c\right)$ which differs only by a scale factor, that we have:

$g\left(q_c\right)$ strictly increasing $\Leftrightarrow$ $\gamma t<1$ (35)

$g\left(q_c\right)$ strictly decreasing $\Leftrightarrow$ $\gamma t>1$ (36)

From which we deduce, by expressing $\gamma$ as a function of $c$ (with $0<c<1$), that:

$g\left(q_c\right)$ is strictly increasing if and only if $\frac{c}{1-c}t<1$,

Which implies, since $1-c>0$, that $ct<1-c$, or $c\left(1+t\right)<1$

So that the rate of profit $r_c = g\left(q_c\right)$ is strictly increasing if and only if $c<\frac{1}{1+t}$ (37)

And $r_c = g\left(q_c\right)$ is strictly decreasing if and only if $c>\frac{1}{1+t}$ (38)

### 3.2. The evolution of the rate of profit at global level in terms of *q*

Let us now start from equation (32), i.e. from the expression of the profit rate as a function of $q$, with $r_c(values) = \frac{q+ct}{(q+1-c)t}$, where $\frac{1}{t}$ can again be considered as a scale factor, which as such does not affect the variation of function $r_c = g(q)$.

In order to examine these variations, let us write: $tr_c = \tilde{g}(q) = \frac{q+ct}{q+1-c}$ (39)

It is easy to get the derivative of $\tilde{g}(q)$, which is $\tilde{g}'(q) = \frac{1-c-ct}{(q+1-c)^2}$. (40)

We deduce from this equation that $\tilde{g}'(q)$ has the sign of $1-c-ct$ and therefore that:

$\tilde{g}'(q) > 0$ implies that $1-c-ct > 0 \Leftrightarrow 1-c(1+t) > 0 \Leftrightarrow c(1+t) < 1$

$\tilde{g}'(q) < 0$ implies that $1-c-ct < 0 \Leftrightarrow 1-c(1+t) < 0 \Leftrightarrow c(1+t) > 1$

Since $g(q)$ differs from $\tilde{g}'(q)$ only by a scale factor, we can infer that:

$r_c = g(q)$ is strictly increasing if and only if $c < \frac{1}{1+t}$ (41)

$r_c = g(q)$ is strictly decreasing if and only if $c > \frac{1}{1+t}$ (42)

It is not surprising to find exactly the conditions set by inequalities (37) and (38) above.

This implies that, unlike what has been demonstrated in the simplified model, the Marxist law of the tendency of the rate of profit to fall in the event of an increase in the organic composition of capital is no longer systematically false when one introduces in the system under review the consumption of capitalists.

It is false only when $c < \frac{1}{1+t}$. On the other hand, it is still valid when $c > \frac{1}{1+t}$.

### 3.3. The graphical representation of the evolution of the profit rate in terms of values

We can first represent the evolution of $r_c$, the rate of profit in terms of values, as a function of the evolution of $q_c$, the organic composition of the capital , on the basis of a given value of the average lifespan of fixed capital, i.e. $t$, which in figure 20.1 is assumed to be 5 years.

**Figure 20.1 - Representation of function $r_C$ (values) = $g(q_C)$ = $\frac{q_C + \gamma t}{(q_C + 1)t}$**

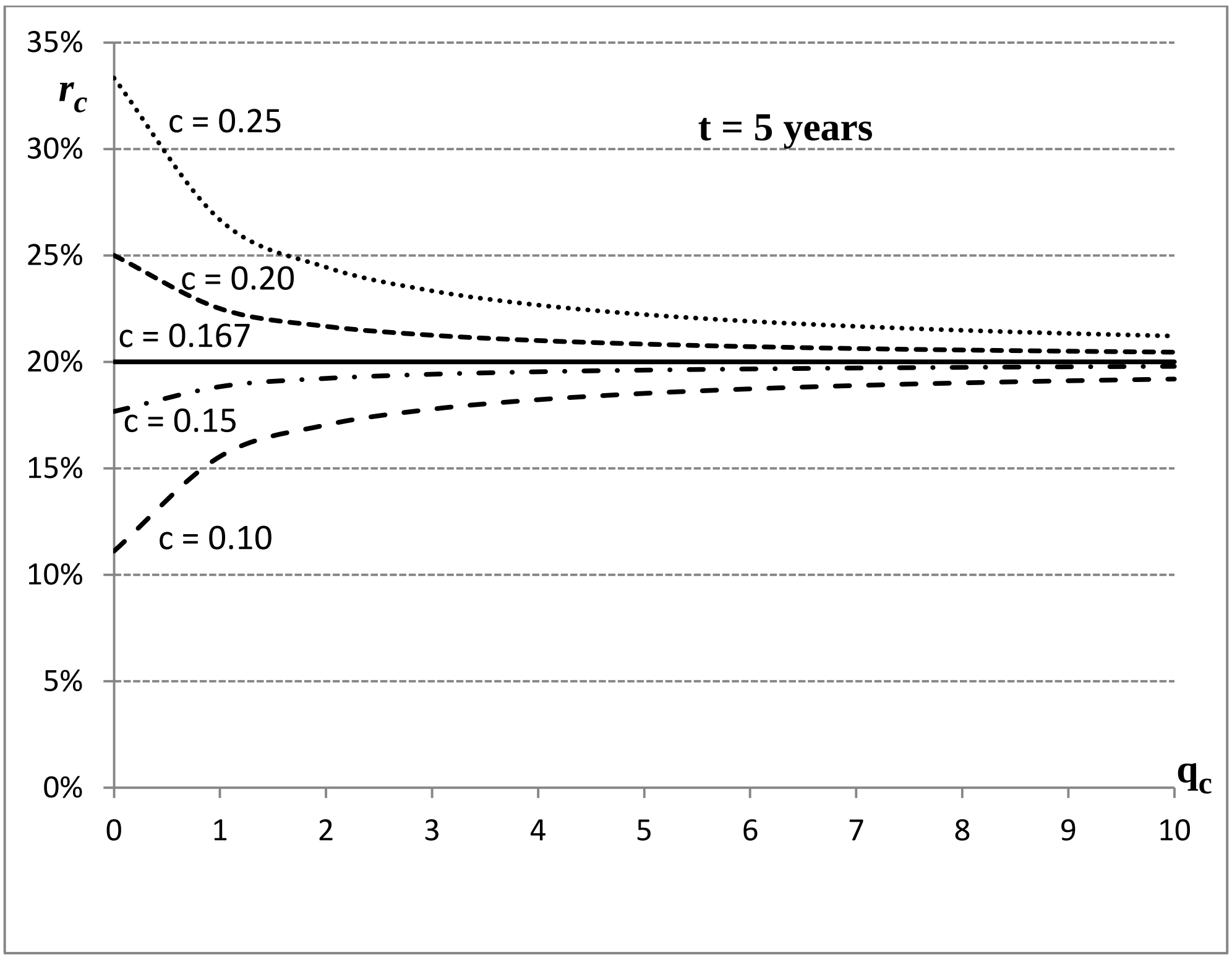

From this figure 20.1 we can see that there is indeed an increase in the rate of profit corresponding to an increase in the organic composition of capital, for the two values of *c* (c = 0.10 and c = 0.15) which are lower than *c* = 0.167 (i.e. $c = \frac{1}{1+t}$ with t = 5 years), as in the simple model without capitalist consumption, which again contradicts, under the condition that $c < \frac{1}{1+t}$, the Marxist law of the tendency of the rate of profit to fall. On the other hand, and contrary to this simple model in which capitalist consumption did not appear, when c exceeds this limit value, which is in fact rather low, the Marxist law of the tendency of the rate of profit to fall becomes true.

It is also quite interesting to note that for this limit value of *c*, i.e. $c = \frac{1}{1+t}$ the rate of profit remains constant, whatever the value taken by the organic composition of capital. Here the value of this constant is $r_C = 0.20$, or 20%, which is maintained whatever the level of $q_C$ when $c = 0.167$. It clearly results from this demonstration that the rate of profit is a function of - among others, the level of capitalist consumption (as a share of total consumption), and as such directly depends on it. This variable thus appears to be a key variable in the functioning of the economic system.

This constitutes indeed quite an interesting finding, since it makes the Marxist law of the tendency of the rate of profit to fall dependent on the level of capitalist consumption, as a share of total consumption, and also indirectly on the average lifespan of fixed capital, the reason being that this last variable is linked to the level of the total stock of fixed capital,and is in fact one of its determinants.

On the basis of equation (22), we can also provide a graphic representation of the evolution of $r_C$, the rate of profit, as a function of the evolution of $k_C$, the rate of surplus-value which takes into account the consumption of capitalists, always with the same average lifespan of fixed capital, i.e. with t = 5 years. This representation appears in figure 20.2, for five different values of $c$ (0.10, 0.15, 0.167, 0.20, 0.25).

**Figure 20.2 - Representation of function $r_C$ (values) = f ($k_C$)** $= \dfrac{k_C}{(k_C - \gamma)t + 1}$ **with** $k_C > \gamma$

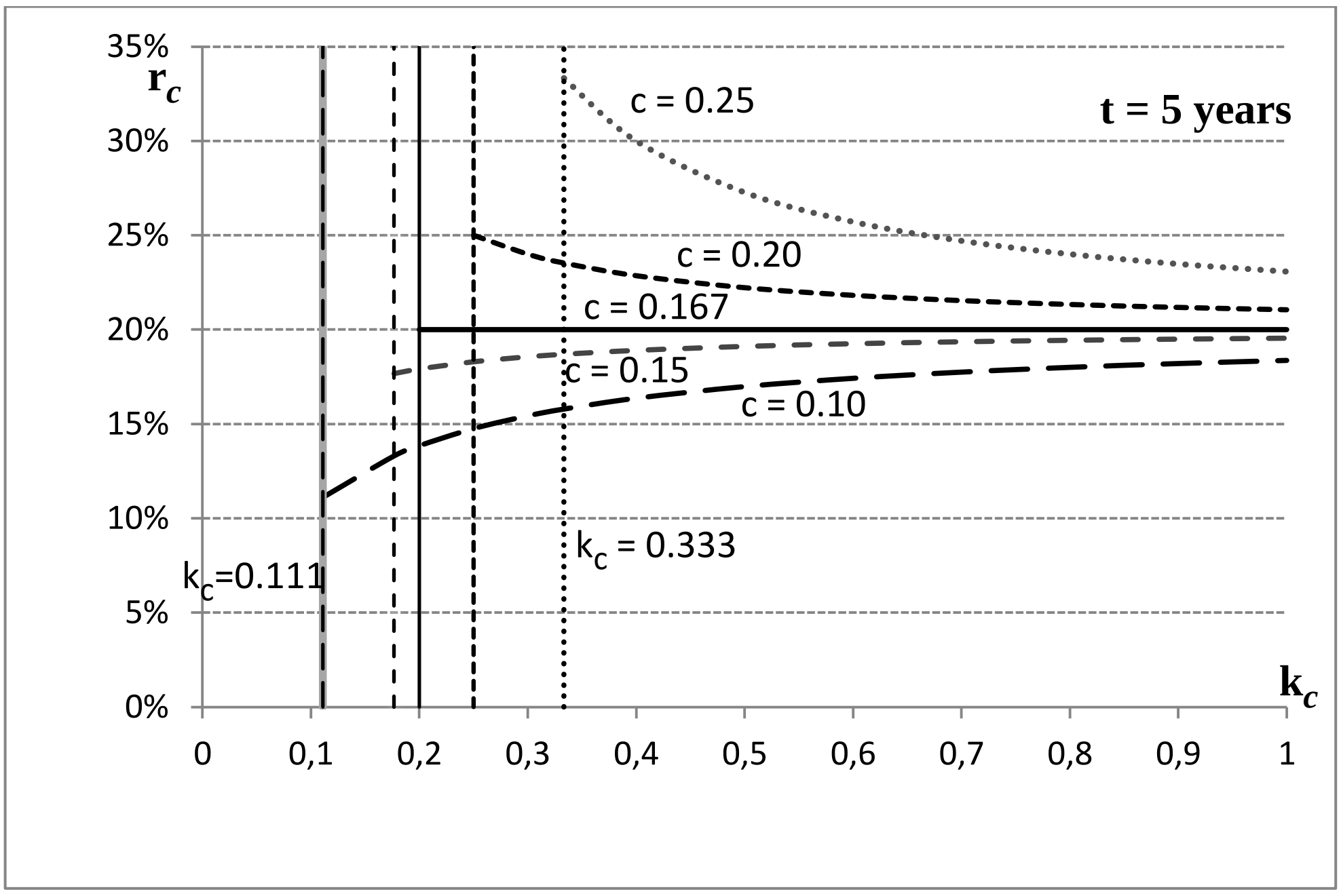

Here also it is easy to see that $c = \dfrac{1}{1+t}$ constitutes a limit value for which the rate of profit becomes a constant, whatever the level of the rate of surplus-value $k_C$. Above this limit $r_C$ is a decreasing function of $k_C$. We can also observe that when this rate of surplus-value increases, whatever the consumption of capitalists, the rate of profit converges asymptotically towards this same limit, which in the chart above stays at 20 %. It is for instance noticeable that, for a given value of *t*, when the rate of surplus-value reaches 100 %, the differences in the rates of profit deriving from the differences in the values of c are reduced.

### 3.4. A numerical example for the Marxist model with capitalist consumption

A numerical example will help to better understand how the model works, in two possible cases, named A and B. Let us take the previous numerical example (see section 3 of chapter 19, p.365-366):

| | Section I | Section II |
|---|---|---|
| A | $L_I = 100$ | $L_{II} = 200$ |
| B | $L_I = 200$ | $L_{II} = 200$ |

We now consider two distinct cases:

**1st case**: $c < \frac{1}{1+t}$ and we consider for example that ***c* = 0,10** and that ***t* = 7**

In situation A the organic composition of capital is:

$$q_c = \frac{K}{V_c} = \frac{L_I \cdot t}{L_{II}(1-c)} = \frac{100 \times 7}{200 \cdot (1-0,10)} = \frac{700}{180} = 3,89$$

And we also have $\gamma = \frac{0,10}{1-0,10} = \frac{1}{9} = 0.11$

The rate of profit is then $r_c = \frac{q_c + \gamma t}{(q_c+1)t} = \frac{3,89 + (0,11 \times 10)}{(3,89+1) \times 10} = \frac{5,0}{48,9} = 10,2\%$

In situation B the organic composition of capital has become:

$$q_c = \frac{K}{V_c} = \frac{L_I \cdot t}{L_{II}(1-c)} = \frac{200 \times 7}{200 \cdot (1-0,10)} = \frac{1400}{180} = 7,78 \text{ (so it has doubled)}$$

And we still have $\gamma = \; = \frac{1}{9} = 0.11$

The rate of profit is then $r_c = \frac{q_c + \gamma t}{(q_c+1)t} = \frac{7,78 + (0,11 \times 7)}{(7,78+1) \times 7} = \frac{8,56}{61,46} = 13,9\%$

There is indeed a rise in the average rate of profit resulting from the increase in the organic composition of capital, which in this example has doubled. We are clearly in a case where the Marxist law of the tendency of the rate of profit to fall is not valid.

**2nd case**: $c > \frac{1}{1+t}$ and we consider for example that ***c* = 0.25** and that ***t* = 10**

In situation A the organic composition of capital is:

$$q_c = \frac{K}{V_c} = \frac{L_I \cdot t}{L_{II}(1-c)} = \frac{100 \times 10}{200 \cdot (1-0,25)} = \frac{1000}{150} = 6,67$$

And we also have $\gamma = \frac{0,25}{1-0,25} = \frac{1}{3} = 0.33$

The profit rate is then $r_c = \frac{q_c + \gamma t}{(q_c + 1)t} = \frac{6,67 + (0,33 \times 10)}{(6,67+1) \times 10} = \frac{10}{76,7} = 13,0\%$

In situation B the organic composition of capital has become:

$$q_c = \frac{K}{V_c} = \frac{L_I \cdot t}{L_{II}(1-c)} = \frac{200 \times 10}{200 \cdot (1-0,25)} = \frac{2000}{150} = 13,33 \text{ (so it has doubled)}$$

And we always have γ = 0.33

The rate of profit is then $r_c = \frac{q_c + \gamma t}{(q_c + 1)t} = \frac{13,33 + (0,33 \times 10)}{(13,33+1) \times 10} = \frac{16,67}{143,3} = 11,6\%$

In this case, we can see conversely a decrease in the average rate of profit resulting from an increase (doubling) in the organic composition of capital. We are thus here in a situation where the Marxist law of the tendency of the rate of profit to fall is perfectly valid. If we had started this demonstration from the alternative formula, i.e. $r_c = \frac{q + ct}{(q+1-c)t}$ (in terms of $q$ instead of $q_c$), the results that we would have obtained of course would have been strictly identical, in both cases.

## 4. Some preliminary findings about the evolution of the rate of profit defined in terms of values as a function of the organic composition of capital

What the last section has clearly established is that the consumption of capitalists as a share of total consumption, and therefore variable *c* which measures this share, plays a decisive role in the way other variables, like the organic composition of capital $q_c$ or the rate of surplus-value $k_c$, influence changes in the rate of profit.

The lessons that we can learn from this finding emerge more clearly from an examination of function $f(c) = \frac{1}{1+t}$, of which the representative curve, as we have demonstrated, constitutes a boundary between the two regimes of evolution of the average rate of profit, and which can thus be designated as the invariance frontier of the rate of profit. This is why this curve is shown in figure 20.3.

We note that when the lifespan of fixed capital increases, the share of consumption of capitalists that corresponds to the boundary where the behavior of the rate of profit changes decreases: for $t$ = 4 years we have $c$ = 0.20, but for $t$ = 7 years, we have $c$ = 0,125.

The area above the boundary, where we have $c > \frac{1}{1+t}$ corresponds to the situation where the function $r_c = \frac{q_c + \gamma t}{(q_c + 1)t}$ is strictly decreasing and where the rate of profit evolves therefore in the opposite direction of the organic composition of capital. This is the situation which corresponds to a fall in the average rate of profit in the event of an increase in the organic composition of capital, and which therefore validates the Marxist law of the tendency of the rate of profit to fall. It is also the situation where, conversely, the decline in the organic composition of capital is accompanied by an increase in the average rate of profit.

The area below the boundary, where we have $c < \frac{1}{1+t}$, corresponds to the situation where function $r_c = \frac{q_c + \gamma t}{(q_c + 1)t}$ is strictly increasing and where the rate of profit evolves in the same direction as the organic composition of capital. This is a situation opposite to that described by Marx, but which nevertheless already resulted from the operation of the simple model, and which corresponds to an increase in the average rate of profit in the event of an increase in the organic composition of capital. This is also the situation where, conversely, the decline in the organic composition of capital is accompanied by a fall in the average rate of profit.

**Figure 20.3 - The invariance frontier of the average rate of profit** $f(c) = \frac{1}{1+t}$

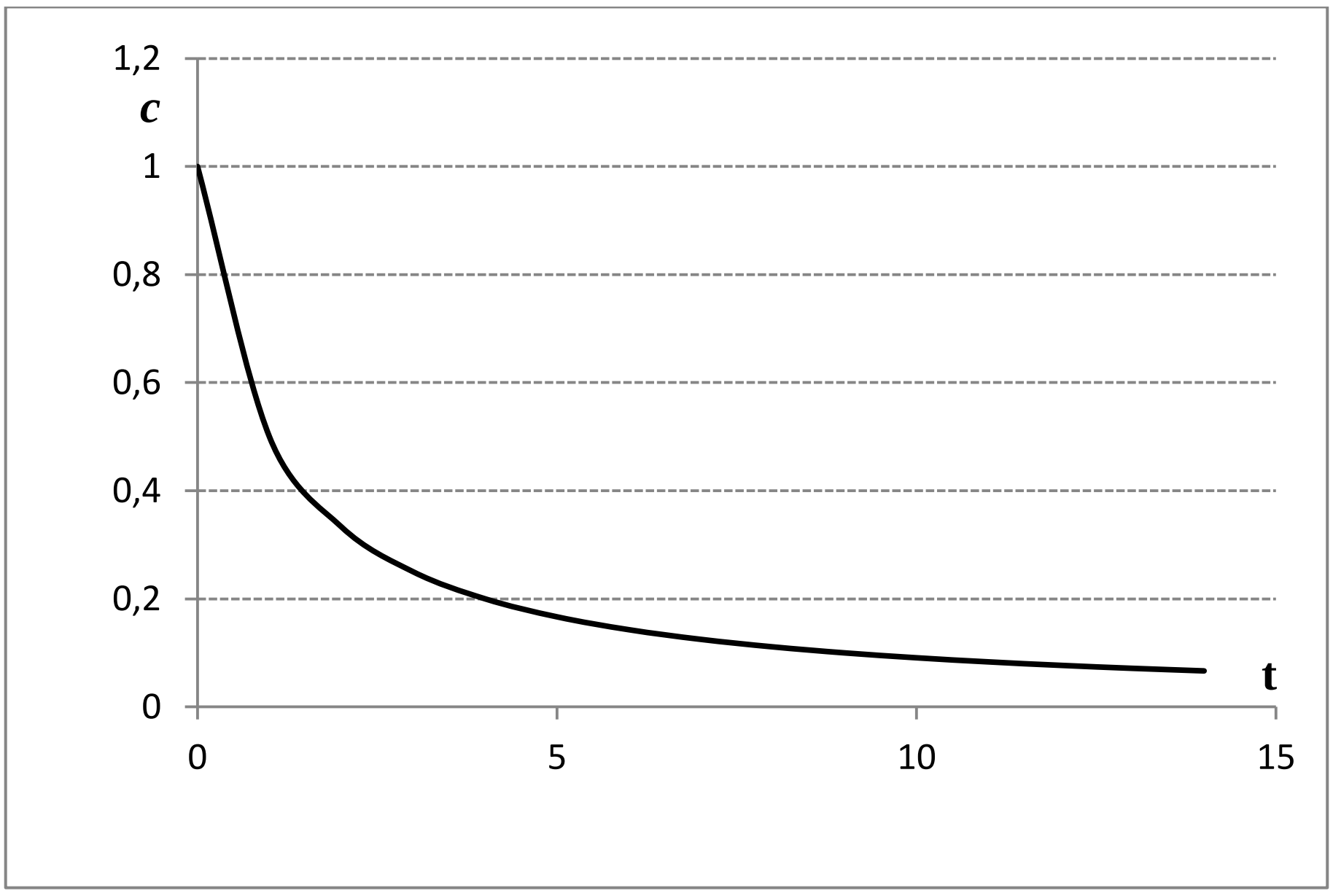

At this stage, it is obviously necessary to ask what is the economic interpretation of this somewhat unexpected mathematical result, or in other words what is the reason for the influence that can be exercised by parameter $c$ on the rate of profit, i.e. by the share of capitalist consumption in total consumption, but also by parameter $t$, the average lifespan of fixed capital. To understand it we must simply return to the basic definition of the rate of profit in Marxist theory.

This rate is defined in the simple model either as: $r = \frac{S}{K+V}$ or as $r = \frac{\frac{S}{V}}{\frac{K}{V}+1}$,

Which corresponds in the model with capitalist consumption to: $r_c = \frac{\frac{S_c}{V_c}}{\frac{K}{V_c}+1}$

It can indeed easily be deduced from this last formula that the influence of $c$ is felt through parameter $V_c$, i.e. the value of labor power, which goes from $V = L_{II}$ in the simple model to $V_c = L_{II}(1-c)$ in the model with capitalist consumption. Indeed the decrease in $V_c$, resulting from a rise of $c$, makes the numerator increase together with the denominator of the above fraction. But the rate of surplus-value in the numerator necessarily increases less rapidly than the organic composition of capital at the denominator, because the growth of the organic composition of capital is amplified by the intervention of parameter $t$, since $K = L_I$ t and $q_c = kt$. For low values of c, this mechanism explains that the increase in c corresponds to a decrease in the rate of profit.

However when $c$ continues to rise, there comes a threshold where the increase in surplus-value which also results from the rise of $c$ (since we go from $S = kL_{II}$ to $S_c = (k+c)L_{II}$), which is - in addition the decline in $V$ - another factor that increases the rate of surplus-value at the numerator of the fraction, can no longer offset the rise in the organic composition of capital at the denominator of the fraction. And this threshold is obviously all the lower as the influence of $t$ becomes stronger, which happens with the increase in the lifespan of fixed capital. When this threshold is reached, and if the consumption of capitalists remains at the level corresponding to this threshold, the rate of profit remains constant, whatever the value of the organic composition of capital. Once this threshold is crossed, and it is different for each value of $t$, the direction of variation of the rate of profit is reversed, which explains why the Marxist law of the tendency of the rate of profit to fall becomes valid again.

However, all these findings have been obtained with a definition of the rate of profit as a ratio of values. To fully analyze the Marxist law, it is therefore interesting to see what happens to this law if we define the rate of profits in terms of prices, because in the real world the rate of profit is calculated as a ratio of prices, and not as a ratio of values. Since the Marxist law links the rate of profit to the organic composition of capital we need nevertheless to keep the price (and not the value) of labor-power at the denominator of the fraction, which is also at the core of this Marxist definition. This is what will be done in the next section.

# 5. The evolution of the rate of profit in terms of prices

In order to define the rate of profit in terms of prices, we need first to determine the price of the capital stock and to define the organic composition of capital in terms of prices, which has already been done in section 1 of this chapter, and therefore makes it easy to define this rate of profits.

## 5.1. The expression of the rate of profit in terms of prices

So far and again in this chapter we started by defining first the rate of profit at global level, before defining it for section I and section II. However to express the rate of profit at global level is quite complicated in terms of prices, as we shall see, which explains that in this subsection we will do the opposite, and go from the simplest formulation to the most complicated one. We will thus start by defining the rate of profit in section I.

### 5.1.1. The rate of profit in section I

To calculate the rate of profit in terms of prices in section I, let us recall that capitalists in this section pay to their workers' wages $W_I$ for an amount $L_I$ (by posing $\bar{w}=1$). They make a profit $\Pi_I$ for an amount $L_I\left(\alpha+\gamma^*+\alpha\gamma^*\right)=L_I\pi_I$ which is provided by equation (48) in chapter 12. They use fixed capital resulting from yearly investments whose annual price, given by equation (38), is $I_I=L_I\left(\alpha+\alpha\gamma^*\right)$. This corresponds to a total quantity of fixed capital employed equal to $tI_I$. The profit rate in terms of prices in section I follows:

$$r_{cI}\ (prices) = \frac{L_I\left(\alpha+\gamma^*+\alpha\gamma^*\right)}{tL_I\left(\alpha+\alpha\gamma^*\right)+L_I} = \frac{\alpha+\gamma^*+\alpha\gamma^*}{t\left(\alpha+\alpha\gamma^*\right)+1} = = \frac{\left(1+\alpha\right)\left(1+\gamma^*\right)-1}{t\alpha\left(1+\gamma^*\right)+1} = \frac{1+\alpha-\dfrac{1}{1+\gamma^*}}{t\alpha+\dfrac{1}{1+\gamma^*}}$$

Since we know that we have also: $\dfrac{1}{1+\gamma^*}=\dfrac{1}{1+\dfrac{c^*}{1-c^*}}=1-c^*$, then we can write:

$$r_{cI}\ (prices) = \frac{1+\alpha-\left(1-c^*\right)}{t\alpha+\left(1-c^*\right)} = \frac{\alpha+c^*}{t\alpha+1-c^*} \qquad (43)$$

It is recalled that parameter $\alpha$ gives a synthetic account of the distribution of fixed capital in value between the two sections: when $\alpha$ increases, this means that the share of fixed capital going to section I increases, and vice versa. Moreover we remember that we have already defined $\gamma=\dfrac{c}{1-c}$, which accounts for the distribution of consumption, expressed in terms of values, between workers and capitalists, $c$ being the capitalists' share. It is also recalled that, as defined in section 2 of chapter 12, $c^*$ is a parameter corresponding to the share of consumption of capitalists of section I selling to capitalists of section II contained in a price $\frac{1}{1-c*}$. This share will thus be $\gamma* = \frac{c*}{1-c*}$. Thus $c*$ is different from $c_I$, the share of capitalists of section I in total consumption. We have shown in chapter 12 that $\frac{c*}{1-c*}$ corresponds to the fraction of the gross profit margin intended for consumption of all capitalists from section I over the value of fixed capital that they sell to section II. This explains that $c*$ and $c_I$ are linked by a rather complicated relationship, given by equation (28) of chapter 12, and which is: $c_I = \gamma* \frac{k(1-c)}{(1+k)}$.

### 5.1.2. The rate of profit in section II

To calculate the profit rate in Section II, it is recalled that capitalists of this section pay wages $W_{II}$ for an amount equivalent to $L_{II}$ (with $\bar{w}=1$), make a profit $\Pi_{II} = k_c L_{II}$, and use it partially to buy from those of section I the quantity of fixed capital that they need, of a value $L_I(1-a)$, for a price equal to $I_{II} = \frac{L_I}{1-c*}$. The price of the total fixed capital in function in section II is $tI_{II}$. The rate of profit in terms of prices in section II then follows (it is recalled that $L_I = kL_{II}$):

$$r_{cII} = \frac{\Pi_{II}}{t.I_{II} + L_{II}} = \frac{k_c L_{II}}{\frac{tkL_{II}}{1-c*} + L_{II}} = \frac{k_c(1-c*)}{kt+1-c*}$$

Since we know that $k_c = \frac{k+c}{1-c}$ and $\frac{1}{1-c*} = 1 + \frac{c*}{1-c*} = 1+\gamma*$, this can be written:

$$r_{cII} = \frac{k+c}{(1-c)\left[kt(1+\gamma*)+1\right]} \qquad (44)$$

### 5.1.3. The overall rate of profit

Lastly, with regard to the overall rate of profit in the price system, its calculation is complicated by the fact that it is the sum of two types of profit derived from the two sections of the productive system. These two types of profit, which follow two different logics, are aggregated in the numerator of the fraction which defines the rate of profit: in section II, profits are constituted in a logic that reflects the levy of surplus-value, through the purchase of consumer goods by workers; in section I, profits are obtained on the occasion of the sales of fixed capital, in a logic of redistribution of this same amount of surplus-value between capitalists of the two sections. The calculation is therefore as follows, with the price of fixed capital given by equation (5) above in this chapter:

$$r_c\ (prices) = \frac{\Pi_I + \Pi_{II}}{K+V} = \frac{L_I\left(\frac{a}{1-a} + \frac{c^*}{(1-a)(1-c^*)}\right) + k_c L_{II}}{t\frac{L_I}{(1-a)(1-c^*)} + L_I + L_{II}} \Rightarrow$$

$$r_c\ (prices) = \frac{L_I\left(\frac{a}{1-a} + \frac{c^*}{(1-a)(1-c^*)} + \frac{k_c}{k}\right)}{tL_I\left[\frac{1}{1-c^*}\cdot\frac{1}{1-a}\right] + L_I\frac{1+k}{k}}$$

This gives us:

$$r_c\ (prices) = \frac{\alpha + (1+\alpha)\gamma^* + \frac{k_c}{k}}{t\left[(1+\alpha)(1+\gamma^*) + \frac{1+k}{kt}\right]} = \frac{\alpha + \gamma^* + \alpha\gamma^* + \frac{k_c}{k}}{t\left[1+\alpha+\gamma^*+\alpha\gamma^* + \frac{1+k}{kt}\right]}$$

We know also that $k_c = \frac{k}{1-c} + \frac{c}{1-c} = k\frac{1}{1-c} + \frac{c}{1-c} = k(1+\gamma) + \gamma$, from which we obtain:

$$r_c\ (prices) = \frac{\alpha + \gamma^* + \alpha\gamma^* + \frac{k(1+\gamma)+\gamma}{k}}{t\left[1+\alpha+\gamma^*+\alpha\gamma^* + \frac{1+k}{kt}\right]} = \frac{k(\alpha+\gamma^*+\alpha\gamma^*) + k(1+\gamma) + \gamma}{kt(1+\alpha+\gamma^*+\alpha\gamma^*) + 1 + k}$$

$$rc(prices) = \frac{k(1+\alpha+\gamma+\gamma^*+\alpha\gamma^*)+\gamma}{kt(1+\alpha+\gamma^*+\alpha\gamma^*)+1+k} \tag{45}$$

In the above equation, the rate of profit in terms of prices is expressed as a function of $k$, i.e. the primary rate of surplus-value. But at this point, since we want still to explore the validity of the Marxist definition even when it is expressed in terms of prices, it is now possible and equally interesting to define the rate of profit as a function of the organic composition of capital. We know indeed, as we saw it above (see equations 10 to 12 in this chapter) that:

$q = kt$, $q_c = \frac{q}{1-c} = \frac{kt}{1-c}$, and that $k_c = \frac{q_c}{t} + \frac{c}{1-c} = \frac{q_c}{t} + \gamma$

We can thus rewrite $r_c(prices)$ as a function of the initial organic composition of capital $q$ :

$r_c(prices) = \frac{\frac{q}{t}(1+\alpha+\gamma+\gamma^*+\alpha\gamma^*)+\gamma}{q\left(\frac{1}{t}+1+\alpha+\gamma^*+\alpha\gamma^*\right)+1}$, and since $\frac{1}{1-c} = 1+\gamma$, we obtain:

$$r_c\left(prices\right)=\frac{q\left(1+\alpha+\gamma+\gamma^*+\alpha\gamma^*\right)+\gamma t}{qt\left(\frac{1}{t}+1+\alpha+\gamma^*+\alpha\gamma^*\right)+t}=\frac{1}{t}\cdot\frac{q\left(1+\alpha+\gamma+\gamma^*+\alpha\gamma^*\right)+\gamma t}{q\left(\frac{1}{t}+1+\alpha+\gamma^*+\alpha\gamma^*\right)+1} \quad (46)$$

These expressions of the rate of profit, in terms of prices, first as a function of the rate of surplus-value, and second as a function of the organic composition of capital, allow us now to see what is its evolution as a function of both variables.

## 5.2. The evolution of the average rate of profit in terms of prices

It can be seen that the expression above in equation (46), which expresses the rate of profit (in terms of prices) as a function of the organic composition of capital, still remains a homographic function. To discuss the evolution of the rate of profit, it is therefore sufficient to analyze this function. To simplify the discussion, let us temporarily assume that:

$$1+\alpha+\gamma+\gamma^*+\alpha\gamma^*=a \quad (47)$$

$$\frac{1}{t}+1+\alpha+\gamma^*+\alpha\gamma^*=b \quad (48)$$

We also assume that $t$ is constant.

Then function *r* (*prices*) becomes with these notations:

$$r_c\,(prices)=f\left(q\right)=\frac{1}{t}\cdot\frac{aq+\gamma t}{bq+1} \quad (49)$$

And since $\frac{1}{t}$ can be considered as a constant scale factor, we can rewrite this function as:

$$g\left(q\right)=\frac{aq+\gamma t}{bq+1} \quad (50)$$

The increasing or decreasing nature of *r* (*prices*) = $f\left(q\right)$ depends on the positive or negative nature of its derivative, to which we can substitute the derivative of $g\left(q\right)$, i.e.:

$$g'\left(q\right)=\frac{a-b\gamma t}{\left(bq+1\right)^2} \quad (51)$$

Since its denominator is always positive, the derivative has the sign of its numerator. It follows that:

$a-b\gamma t>0\Leftrightarrow r_c\left(prices\right)=f\left(q\right)$ is strictly increasing

$a-b\gamma t<0\Leftrightarrow r_c\left(prices\right)=f\left(q\right)$ is strictly decreasing

The condition of positivity of the derivative $g'(q)$ is thus that $a - b\gamma t > 0$, which gives us, developing from the definitions of equations (46) and (47) above:

$$1+\alpha+\gamma+\gamma^*+\alpha\gamma^*-(\frac{1}{t}+1+\alpha+\gamma^*+\alpha\gamma^*)\gamma t > 0$$

Knowing that $\frac{1}{t}+1+\alpha+\gamma^*+\alpha\gamma^* > 0$, because all of its terms are positive, this expression can be rewritten under the form:

$\gamma t < \dfrac{1+\alpha+\gamma+\gamma^*+\alpha\gamma^*}{1+\frac{1}{t}+\alpha+\gamma^*+\alpha\gamma^*}$, which gives $\gamma < \dfrac{1+\alpha+\gamma+\gamma^*+\alpha\gamma^*}{1+t(1+\alpha+\gamma^*+\alpha\gamma^*)}$, from which we can deduce:

$$\gamma+\gamma t(1+\alpha+\gamma^*+\alpha\gamma^*) < 1+\alpha+\gamma+\gamma^*+\alpha\gamma^* \Rightarrow \gamma t(1+\alpha+\gamma^*+\alpha\gamma^*) < 1+\alpha+\gamma^*+\alpha\gamma^*$$

From which we draw that $g'(q) > 0$ if and only if $\gamma t < 1$ (52)

Conversely $g'(q) < 0$ if and only if $\gamma t > 1$ (53)

It is recalled that $\gamma = \dfrac{c}{1-c}$, and that $c$ is the share of capitalist consumption within total consumption (with $0 < c < 1$). By replacing γ by its expression as a function of $c$ in inequation (52), we obtain:

$$\gamma t < 1 \Rightarrow \frac{c}{1-c}t < 1 \Rightarrow ct-(1-c) < 0 \Rightarrow ct+c < 1 \Rightarrow c(1+t) < 1 \Rightarrow c < \frac{1}{1+t}$$

It follows that $f'(q)$ is positive and that function *r* (*prices*) = $f(q)$, which determines the rate of profit as a function of the organic composition of the capital, is increasing when coefficient $c$ is lower than $\dfrac{1}{1+t}$, that is when the share of capitalist consumption in total consumption is less than the inverse of the average lifetime of fixed capital plus one. Conversely, function decreases when $c$ is greater than $\dfrac{1}{1+t}$.

To be sure, the calculations that had to be made to achieve this result were rather cumbersome, if not very complex. In spite of that, we should not be surprised to find again exactly the same law that was highlighted for the evolution of the rate of profit on the basis of its initial definition in terms of values, which corresponded to Marx's formulation. The identity of both results in fact validates these calculations! It is indeed the opposite that would have been amazing! Profits are indeed the form taken by surplus-value transformed by the intervention of the price system when it is levied and distributed among capitalists. Thus there was no reason to imagine that the passage through the price system that makes possible their actual realization would have resulted in an evolution of the rate of profit that would have been distinct and therefore a priori different in the value system as compared to the price system.

We had highlighted in the simplified model which was the subject of chapters 18 and 19 a law of evolution of the profit rate opposite to and therefore radically different from that of Marx. But now we see again, this time in terms of prices, that the introduction of capitalist consumption allows to verify a law inverse to the Marxist one only in the hypothesis where the share of capitalist consumption in total consumption is lower than a particular threshold, defined on the basis of the average lifespan of fixed capital.

However, the assumptions that can be made on this average lifespan in the real world show that this threshold is certainly quite low, and decreases with the extension of this lifespan (which is a cause of increase in the organic composition of capital). The average lifespan of fixed capital in capitalist firms in developed countries seems indeed to be in the range of 5 years to 10 years, which corresponds for the share of capitalists consumption in total consumption, with $c = \frac{1}{1+t}$, to thresholds of 16.7% and 9.1%, respectively. We know also that the rate of profit decreases when this average lifespan of fixed capital increases, as figures 1 and 2 have shown (see page 364and 367).

In passing, it should be recalled that the value of fixed assets in operation is reduced when linear depreciation is introduced, which is equivalent to a shortening of the notional average lifespan that can be obtained by relating the fixed capital stock at a given time (measured after depreciation) to the amount of fixed capital invested each year. We have shown above that $T = \frac{t-1}{2}$. As a result, statistics taking into account the depreciation of fixed capital to measure its stock as net of depreciation underestimate its amount and therefore its actual lifespan(for t = 5 years, we thus have T = 2 years). The end-result is that statistical data relying on such a basis necessarily overestimate the rate of profit.

In any case, as soon as this threshold equal to $c = \frac{1}{1+t}$ is crossed up, that is to say as soon as the share of capitalist consumption in total consumption becomes is greater than $\frac{1}{1+t}$, for any given value of t, then the Marxist law of the tendency of the rate of profit to fall due to an increase in the organic composition of the capital becomes valid. Moreover this increase is not a mere tendency, but is on the contrary systematic, within the framework of the assumptions retained. Although his demonstration left much to be desired, Marx's intuition and conclusions on the subject were therefore largely well founded.

It must be noted, however, that this law is reversible, meaning that in the situation likely to be the most realistic, or the most plausible, where $c > \frac{1}{1+t}$, in case of a decline - and no longer a rise, in the organic composition of capital, there is necessarily a rise in the rate of profit. In the real world this can be the case for economic systems where there is a decline in heavy industry, and a development of services requiring less fixed capital. This is certainly a circumstance that has not been foreseen by Marx to characterize the evolution of capitalism, but which must not be ruled out.

To conclude temporarily on the question of the evolution of the rate of profit, it is necessary to underline again that all the demonstration carried out so far was on the basis of the Marxist definition of the rate of profit, which constituted the only way to test the validity of Marx's reasoning. But this definition of the rate of profit puts variable capital (the value or price of labor-power) at the denominator of the expression giving this rate of profit. This is not unrelated as we have just seen with the nature of the results obtained, and in fact is indispensable to introduce the organic composition of capital in the picture. But this leads to a definition of the rate of profit that is not the most common. The current definition is indeed simpler and limits itself to put fixed capital, and fixed capital alone, at the denominator of the expression giving the rate of profit.

At the end of this chapter we are now able to recap the system of values and prices in a summary table, similar to table 4 in chapter 11 for the simplified model, but corresponding now to the model with capitalist consumption. We must stress once again that all the content in this table 2, which appears in appendix 3, is based on the Marxist definition of the rate of profit, which has been the only one examined so far, but is not the current definition of the rate of profit, such as that generally used by economic theory. We will examine below in the next chapter what happens if we leave the Marxist definition of the rate of profit to use the current definition only.

Before that the demonstrations that have been made in this chapter allow us nevertheless to formulate a new principle.

**Principle 42**: In a situation of simple reproduction where capitalist consumption is introduced, then the rate of profit defined as Marx does becomes dependent on the share of this capitalist consumption in total consumption. The validity of the Marxist law of the tendency of the rate of profit to fall when the organic composition of capital increases depends therefore on this share. If this share exceeds a threshold linked to the lifespan of fixed hospital and defined by $c = \frac{1}{1+t}$, the Marxist law is valid. If this share is lower than this threshold the Marxist law is not verified and the rate of profit increases on the contrary as a function of the organic composition of capital.

# Chapter 21. The evolution of the rate of profit - based on its current definition - and the share of profits

In this chapter we will see what happens to the rate of profit when it is defined, not on the basis of the Marxist definition, but of the current definition, i.e. without variable capital at the denominator of the fraction giving its value. We will highlight the interest of this current definition, and will show what is the evolution of the rate of profit both in the simple model exposed earlier and in the model with capitalist consumption. This will also be done both in terms of values and in terms of prices. Graphical representations will be provided for all these cases. But we will not confine ourselves only to analyzing the evolution of the rate of profit, and will also address in the second part of the chapter the question of the evolution of the share of profits. We will conclude this chapter by exploring briefly the link between the rate of profit and the share of profits.

## 1. The interest of the current definition of the rate of profit

The ultimate goal of this work is not to develop a pure theory for just the beauty of it, but to provide theoretical tools to begin to shed some light, through the lessons of the models developed so far, on the way the capitalist system actually functions in the real world. This could also help us to better understand the reasons for the long term crisis into which the capitalist system has progressively entered since the late 1970s or early 1980s, in practically all of the most developed countries. This long term crisis is characterized among others by a dramatic surge in inequalities, which has been well described - if not well explained, in Piketty's much acclaimed book "Capital in the 21st Century". I have criticized Piketty's explanations - but not his data and figures - in my article previously cited: "Again on Piketty's *Capital in the Twenty-First Century"*.

If one wants to leave the realm of theory to try to understand the evolution of the real world, it is a limitation from which it is advisable to free oneself. In so far as one has to come to grips with the reality of crises, a good point of entry for the analyses conducted so far has indeed been the Marxist theory of crises based on the tendency of the rate of profit to fall, and this was a good reason to base these analyzes on the Marxist definition of the rate of profit. But as we have already pointed out, this definition, by including wages at the denominator of the expression defining the rate of profit, does not correspond to the usual way of defining this variable, which considers it rather as a ratio between profits and the value of fixed capital alone (which includes the value of the intermediate commodities used for its production). Indeed in the real world wages are not advanced one year before the realization of the product.

For this reason it seems preferable to avoid that the reasoning which follows be criticized on the basis of a rejection of the Marxist definition of the rate of profit. This is the reason why this chapter will now be devoted to the determination of the rate of profit, still using the models developed so far, but this time on the basis of the current definition of this rate, which is also going to simplify its expression. This will allow us to demonstrate that the conclusions drawn so far remain valid within the modified framework of this current definition.

Moreover, and to remain consistent with our questioning about the determining character of the rate of profit, if only because of the demonstration that a multiplicity of profit rates was necessary for the reproduction of the system, this chapter will also examine how a variable that ultimately appears more fundamental is determined in the model: this variable is the share of profits in global income (or output), which corresponds to what is often called the margin rate in its macroeconomic sense.

## 2. The current definition of the rate of profit

Let us recall that this definition only relates surplus-value (in terms of values) or profit (in terms of prices) to fixed capital alone (in value or price).

### 2.1. Within the framework of the simplified model

In terms of values, we have = $r(values) = \frac{S}{K} = \frac{L_I}{L_I t} = \frac{1}{t}$ (1)

In terms of prices, we know from equation (17) in chapter 11 that total profit is:

$$\Pi = L_I\left(1 + \frac{a}{u}\right) = L_I + L_I \frac{a}{u}$$

Simplifying again by taking $u = b = 1 - a$ (as we already did previously), we have:

$$r_c\,(prices) = \frac{\Pi}{K} = \frac{L_I\left(1 + \frac{a}{1-a}\right)}{L_I \frac{1}{1-a} t} = \frac{L_I \frac{1}{1-a}}{L_I \frac{1}{1-a} t} = \frac{1}{t} \qquad (2)$$

We thus find that the rate of profit is the same in its two expressions, both in terms of values and in terms of prices. Moreover with this definition the organic composition of capital does not intervene in the definition of the rate of profit. The latter is therefore independent from the organic composition of capital, and even from the rate of surplus-value $k$, equal to $L_I / L_{II}$. In fact the rate of profit in the simplified model is only the inverse of the average lifespan of fixed capital. It means that the rate of profit is here purely and simply a technological parameter, knowing however that we are in a situation of simple reproduction, without any capitalist consumption, that the whole surplus-value is invested in fixed capital, and that the rise or fall in the rate of surplus-value, when it occurs, is supposed to have passed on to the totality of the stock of fixed capital (which cannot happen instantaneously, but after a time $t$).

### 2.2. Within the framework of the model with capitalist consumption

#### 2.2.1. The rate of profit in terms of values

Let us start first by writing the most simple expression of the rate of profit in terms of values, which is:

$$r_c(values) = \frac{S_c}{K} = \frac{L_{II}\left(k+c\right)}{L_I t} = \frac{L_{II}\left(k+c\right)}{kL_{II} t} = \frac{k+c}{kt} \tag{3}$$

From this expression, since we want to check the validity of the Marxist law of the evolution of the rate of profit with its current definition, we can derive a new one expressing $r_c(values)$ in terms of $q_c$, the organic composition of capital.

1) The evolution of the rate of profit in terms of $q_c$, the organic composition of capital

If we want to verify whether the Marxist law of the tendency of the rate of profit to fall with a rise in the organic composition of capital is still valid with this current definition of the rate of profit, it is better to do this checking in terms of values, i.e. in the same way as it was formulated by Marx. First, because it is more simple mathematically, and second because in any case we will see below in subsection 3.5. that the rate of profit defined in terms of prices always evolves like the rate of profit defined in terms of values.

To do that implies to redefine the rate of profit $r_c$ by reintroducing the organic composition of capital $q_c$ in the above equation giving $r_c$. This can be done very simply by just dividing both the numerator and denominator of equation (3) above, i.e. $r_c(values) = \frac{S_c}{K}$ by the value of labor power, i.e. $V = L_{II}\left(1-c\right)$, so that we obtain:

$$r_c(values) = \frac{S_c}{K} = \frac{\frac{S_c}{V}}{\frac{K}{V}} = \frac{\frac{L_{II}\left(k+c\right)}{L_{II}\left(1-c\right)}}{\frac{L_I t}{L_{II}\left(1-c\right)}} = \frac{\frac{k+c}{1-c}}{\frac{kt}{1-c}} = \frac{k_c}{q_c} \tag{4}$$

Obviously we find that the rate of profit $r_c$ is now the ratio of the rate of surplus-value with capitalist consumption, i.e. $k_c$, over the organic composition of capital with capitalist consumption, i.e. $q_c$. This rate $r_c$ is thus clearly a function of two variables: $r_c = f\left(k_c, q_c\right)$.

As such this function has no derivative, but a gradient, which we call $G\left[f\left(k_c, q_c\right)\right]$. The formula giving this gradient is:

$$G\left[f\left(k_c,q_c\right)\right]=\begin{pmatrix}\dfrac{\delta f\left(k_c,q_c\right)}{\delta k_c}\\ \dfrac{\delta f\left(k_c,q_c\right)}{\delta q_c}\end{pmatrix}=\begin{pmatrix}\dfrac{1}{q_c}\\ -\dfrac{k_c}{q_c^2}\end{pmatrix}, \text{ which is a field of vectors} \quad (5)$$

This gradient never cancels (which would mean that $q_c$ becomes infinite and that $k_c$ equals zero). Therefore it has no extremum. On the other hand we can see that $\frac{\delta f\left(k_c,q_c\right)}{\delta k_c}=\frac{1}{q_c}$ is always positive and therefore that $r_c\left(values\right)=f\left(k_c,q_c\right)$ is increasing in terms of $k_c$, for instance on the $x$ axis, whereas $\frac{\delta f\left(k_c,q_c\right)}{\delta q_c}=-\frac{k_c}{q_c^2}$ is always negative and therefore is decreasing in terms of $q_c$, for instance on the $y$ axis, this decrease being $-\frac{k_c}{q_c}$ times the increase in terms of $k_c$.

This shows that the Marxist law of the tendency of the rate of profit to fall when the organic composition of capital grows is still verified with the current definition of the rate of profit, and that its validity no longer depends on the value of $c$ like it was the case with the pure Marxist definition, where $c$ had to be higher than $\frac{1}{1+t}$ for the law to be verified (see above subsection 3.2. of chapter 20).

In fact this law can also be highlighted by expressing $r_c=f\left(k_c,q_c\right)$ only in terms of $q_c$ and $c$ (or $\gamma$), which is easy to do, remembering that we can write:

$$k_c=\frac{k+c}{1-c}=\frac{k}{1-c}+\frac{c}{1-c}=\frac{kt}{\left(1-c\right)t}+\gamma=\frac{q_c}{t}+\gamma$$, which was equation (11) in chapter 20.

$$\text{Therefore } r_c=g\left(q_c\right)=\frac{k_c}{q_c}=\frac{\frac{q_c}{t}+\gamma}{q_c}=\frac{q_c+\gamma t}{q_ct}=\frac{1}{t}.\frac{q_c+\gamma t}{q_c} \quad (6)$$

We can see that $g\left(q_c\right)$ is a homographic function of $q_c$, where $\frac{1}{t}$ is a scale factor. From this function we can draw the derivative of $t.r_c=\tilde{g}\left(q_c\right)$, at a scale factor close of $g\left(q_c\right)$. It is:

$$\tilde{g}'\left(q_c\right)=-\frac{\gamma t}{q_c^2} \quad (7)$$

$\tilde{g}'(q_c)$ has the sign of $-\gamma t$, and since both $\gamma$ and $t$ are always positive, $-\gamma t$ is always negative. It follows that, with the current definition of the rate of profit, $r_c = g(q_c)$ is always a decreasing function: an increase in $q_c$ always causes a decrease in the rate of profit $r_c$, for any given value of $\gamma$, and therefore of $c$. However, when $c$ (and thus $\gamma$) are increasing for a given value of $q_c$, then $r_c = g(q_c)$ is going to increase.

2) The evolution of the rate of profit in terms of $k_c$, the rate of surplus-value

We already indicated in the previous page that $\frac{\delta f(k_c, q_c)}{\delta k_c} = \frac{1}{q_c}$ is always positive and therefore that $r_c(values) = f(k_c, q_c)$ is increasing in terms of $k_c$. In other words, the rate of surplus-value $k_c$ may increase at a rhythm which can be higher than the decrease generated by the increase in $q_c$, the organic composition of capital. From what we saw above, we can deduce that an increase in $r_c$ resulting from an increase in $k_c$ will prevail over a decrease resulting from an increase in $q_c$ if this decrease is less than $-\frac{k_c}{q_c}$ times the increase of $k_c$.

This is the reason why the law of the tendency of the rate of profit to fall when the organic composition of capital rises can only be a tendency, because it can thus be countered by a rise in the rate of surplus-value $k_c$. This rise itself can only come from the rise in its own components, i.e. $k$ and $c$. Let us examine below what results from the evolution of $k$ and $c$.

3) The evolution of the rate of profit in terms of $k$, the primary rate of surplus-value and $c$, the share of capitalist consumption.

We saw above that the most simple expression of the rate of profit, given by equation (3) is:

$$r_c(values) = \frac{S_c}{K} = \frac{L_{II}(k+c)}{L_I t} = \frac{L_{II}(k+c)}{kL_{II}t} = \frac{k+c}{kt}$$

To begin with, it is clear that $r_c = \frac{k+c}{kt}$ is again a homographic function of $k$, for a given value of $c$ considered as a constant. This makes it possible to describe the evolution of the rate of profit, no longer as a function of the organic composition of capital $q_c$, which does not appear in this expression, but as a function of the rate of "primary" surplus-value $k$ (and not of $k_c$, the rate of surplus-value with capitalist consumption), being noted that $c$ itself appears at the numerator of the expression. This is a simple function, since its derivative is equal to:

$$f'(k) = \frac{-ct}{(kt)^2} \qquad (8)$$

Since we know that $c > 0$ and $t > 0$, it follows that the derivative $f'(k)$ is always negative, and therefore that function $r_c = f(k)$ is always decreasing, which implies that the increase in the primary rate of surplus-value always results in a lower rate of profit, *ceteris paribus*.

But it is easy to show that an increase in $k$ alone does not necessarily translate into a rise in $k_c$ because by definition $k_c = \frac{k+c}{1-c}$, which means that such an increase in $k$ can be countered by a simultaneous decrease in $c$. To show that it can be useful to establish a relation between $k$ and $c$ which will leave $k_c$ as a constant. Let us thus consider that $k_c = h$ with $h$ = constant.

This simply means that the variation of $k$ and $c$ necessarily cancel each other's, in such a way that:

$$\frac{k+c}{1-c} = h \Rightarrow k+c = h(1-c) \Rightarrow k = -c(1+h)+h \qquad (9)$$

Or alternatively: $c = -\frac{k-h}{1+h}$

In other words if $k$ grows, $c$ must decline in such a proportion that equation (9) remains valid, or that for any given value of $k$ there can be a value of $c$ which cancels its influence on $k_c$, and therefore on the rate of profit as a function of $k_c$. To give an example, with $k$ = 0.25 and $c$ = 0.30 (with $k < 1$ and $c < 1$), we have :

$$k_c = \frac{k+c}{1-c} = 0.7857 \Rightarrow k = -1.7857c + 0.7857$$

Let us take $k_c = h = 0.7857$

Then if $k$ rises to 0.40, then $c$ must be reduced to $c = \frac{k - 0.7857}{-1.7857} = \frac{0.4 - 0.7857}{-1.7857} = 0.216$ for $k_c$ to remain constant. This can easily be checked.

But the fact that the opposite evolution of $k$ and $c$ cancel each other in the expression of $k_c$ does not imply that the same happens in the evolution of $q_c$, the organic composition of capital. Indeed $q_c = \frac{kt}{1-c}$. Thus, taking $t = 10$, when $k$ and $c$ go from 0.25 to 0.40 and from 0.30 to 0.216, respectively, then $q_c$ increases from 3.57 to 5.10, which obviously reduces the rate of profit $r_c = f(q_c)$, from 22% to 15.40 %. Another way for the organic composition of capital to grow "autonomously" would be for the average lifespan of fixed capital to increase, which would reduce the rate of profit in due proportions, since $r_c = \frac{k+c}{kt}$ (see equation (3)).

It is clear nevertheless that the organic composition of capital $q_c = \frac{kt}{1-c}$ is itself an increasing function of three variables: i.e. $k$ the primary rate of surplus-value, $c$ the share of capitalist consumption in total consumption (in value), and $t$ the average lifespan of fixed capital. And it is also obvious that the first two variables, $k$ and $c$, if they are growing simultaneously, increase the rate of surplus-value $k_c = \frac{k+c}{1-c}$ through both the numerator and denominator of the equation giving $r_c$, and therefore counterbalance the increase in the organic composition of capital at the denominator of this same equation.

This will maybe appear more easily by providing a second numerical example:

Let us take $t$ = 10, $k$ = 0.25, and $c$ = 0.2

Thus we have:

$$k_c = \frac{0.25+0.20}{0.8} = 0.5625\text{; } q_c = \frac{0.25*10}{1-0.2} = 3.125 \text{ and } r_c = \frac{k_c}{q_c} = \frac{0.5625}{3.125} = 0.18 = 18\%$$

Then let us imagine that both variables increasing the rate of surplus-value are rising, so that we have now $t$ = 10, $k$ = 0.30, and $c$ = 0.3

Therefore $k_c = \frac{0.3+0.3}{0.7} = 0.8571$; $q_c = \frac{0.30*10}{1-0.3} = 4.2857$ and $r_c = \frac{k_c}{q_c} = \frac{0.8571}{4.2857} = 0.20 = 20\%$

This example shows that an increase in the organic composition of capital $q_c$ resulting from an increase in the primary rate of surplus-value $k$, which alone would induce a decrease in the rate of profit $r_c$, when combined with an increase in the share of capitalist consumption c, causes both the rate of surplus-value $k_c$ and the rate of profit $r_c$ to increase, despite the increase in the organic composition of capital which should normally result in the decrease of this rate of profit. This sheds light on the prominent role played by capitalist consumption in the evolution of other key variables of the system, since both the rate of surplus-value and the rate of profit are always an increasing function of this share of capitalist consumption in total consumption.

### 2.2.2. The rate of profit in terms of prices

In terms of prices, the rate of profit can be calculated by referring to its definitions already given for the different sections of the productive system in the previous chapter, with the suppression of the price of labor-power which was at the denominator.

For capitalists of section II, starting from the first expression of equation 44 in chapter 20, and removing the price of labor-power $L_{II}$ (with $\bar{w}=1$) from the denominator, the rate of profit (let us call it $r_{cII}$ ) is now defined as:

$$r_{cII} = \frac{k_c L_{II}}{\frac{ktL_{II}}{1-c*}} = \frac{k_c\left(1-c*\right)}{kt} = \frac{\left(k+c\right)\left(1-c*\right)}{\left(1-c\right)kt} \tag{10}$$

For capitalists of section I, it seems interesting to make a distinction, already introduced previously, between those who produce fixed capital sold to capitalists in section II and those who produce fixed capital sold to capitalists in section I, also going upstream to all other members of this section I (selling intermediate goods to the former).

This separation of capitalists from section I into two sub-categories will provide a possibility to check the coherence of the expression of these rates of profit.

The former make a profit equal to $L_I\left(a+\gamma *\right) = L_I\left(a+\frac{c*}{1-c*}\right)$, according to equation (26) in chapter 12, and use a quantity of fixed capital whose price (and not value) is equal to $\frac{L_I at}{1-c*}$.

Their rate of profit (let us call it $r'_{cI}$ ) is therefore defined as:

$$r'_{cI} = \frac{L_I\left(a+\gamma *\right)}{L_I at\left(1+\gamma *\right)} = \frac{a+\gamma *}{at\left(1+\gamma *\right)} = \frac{1}{t}\left(1-c*+\frac{c*}{a}\right) \tag{11}$$

The second realize a profit equal to $\alpha L_I\left(a+\gamma *\right)$, taken from table 7 in chapter 12, and use an amount of fixed capital whose overall price is $a\left(\alpha+\alpha\gamma *\right)tL_I$ . Their rate of profit (let us call it $r''_{cI}$ ) is therefore:

$$r''_{cI} = \frac{\alpha L_I\left(a+\gamma *\right)}{a\left(\alpha+\alpha\gamma *\right)tL_I} = \frac{a+\gamma *}{\left(1-a\right)\left(\alpha+\alpha\gamma *\right)t} = \frac{1}{t}\left[\frac{a+\gamma *}{\alpha\left(1-a\right)\left(1+\gamma *\right)}\right] \Rightarrow$$

$$r''_{cI} = \frac{1}{t}\left(\frac{a+\gamma *}{a\left(1+\gamma *\right)}\right) = \frac{1}{t}\left(1-c*+\frac{c*}{a}\right) \tag{12}$$

This allows to verify that this rate of profit is in fact identical to the previous one, which is logical, under the assumption that parameters $a$ (and therefore $\alpha$ ) as well as $c*$ are the same at all levels of the supply chain in section I.

Noting that by definition $a\alpha+a=\alpha$ , the average profit rate for section I as a whole is thus:

$$r_{cI} = \frac{L_I\left(\alpha+\gamma^*+\alpha\gamma^*\right)}{tL_I\left(\alpha+\alpha\gamma^*\right)} == \frac{\left(1+\alpha\right)\left(1+\gamma^*\right)-1}{\alpha t\left(1+\gamma^*\right)} = \frac{1+\alpha-\dfrac{1}{1+\gamma^*}}{\alpha t} = \frac{\alpha+c^*}{\alpha t} \tag{13}$$

Similarly, by developing this last equation, we get:

$$r_{cI} = \frac{\alpha+c^*}{\alpha t} = \frac{1}{t}\left(\frac{a+c^*-ac^*}{a}\right) = \frac{1}{t}\left(1-c^*+\frac{c^*}{a}\right) \tag{14}$$

For the overall rate of profit, referring to table 2 in appendix 3, we have:

$$r_c\ (prices) = \frac{\Pi}{K} = \frac{L_I\left(\dfrac{a}{1-a}+\dfrac{c^*}{(1-a)(1-c^*)}\right)+k_cL_{II}}{t\dfrac{L_I}{(1-a)(1-c^*)}} = \frac{L_I\left(\dfrac{a}{1-a}+\dfrac{c^*}{(1-a)(1-c^*)}+\dfrac{k_c}{k}\right)}{tL_I\left[\dfrac{1}{1-c^*}\cdot\dfrac{1}{1-a}\right]}$$

We know also that: $k_c = \frac{k+c}{1-c} = \frac{k}{1-c}+\frac{c}{1-c} = k\frac{1}{1-c}+\gamma = k\left(1+\gamma\right)+\gamma$

Thus if we replace $k_c$ by the above value $k\left(1+\gamma\right)+\gamma$ , we get:

$$r_c\ (prices) = \frac{\alpha+\left(1+\alpha\right)\gamma^*+\dfrac{k\left(1+\gamma\right)+\gamma}{k}}{t\left[\left(1+\alpha\right)\left(1+\gamma^*\right)\right]} = \frac{k\left(\alpha+\gamma^*+\alpha\gamma^*\right)+k\left(1+\gamma\right)+\gamma}{kt\left(1+\alpha+\gamma^*+\alpha\gamma^*\right)}$$

$$r_c\ (prices) = \frac{k\left(1+\alpha+\gamma+\gamma^*+\alpha\gamma^*\right)+\gamma}{kt\left(1+\alpha+\gamma^*+\alpha\gamma^*\right)} \tag{15}$$

It is obviously different from the equation given in table 2 in Appendix 3, since the definition in this table corresponds to the Marxist definition, which is no longer the case here. In order to simplify this rather complicated expression, let us temporarily pose:

$$1+\alpha+\gamma+\gamma^*+\alpha\gamma^* = a \tag{16}$$

$$1+\alpha+\gamma^*+\alpha\gamma^* = b \tag{17}$$

We can now see that function $r_c = f\left(k\right)$ is still, as one might expect, a homographic function, because it is of the form:

$$r_c\ (prices) = f\left(k\right) = \frac{1}{t}\cdot\frac{ak+\gamma}{bk} \tag{18}$$

Its derivative is therefore $f'(k) = \dfrac{-b\gamma}{(tbk)^2}$ (19)

Since $b > 0$ and $t > 0$, it follows that the sign of this derivative is always negative, and that this function is therefore always decreasing, which implies that when we define the rate of profit in terms of prices, an increase in the primary rate of surplus-value is always reflected in a fall in the rate of profit, like it was the case for the rate of profit defined in terms of values.

To go further in studying function $r_c(prices)$ would imply to go back to equation (15), which is quite complicated, and would therefore involve complex mathematical calculations. This explains that, rather than embarking on cumbersome reasoning, it seemed preferable to use a graphical representation of this function, giving ourselves as examples the value of some of the variables in its price expression involved in the evolution of $r_c$.

# 3. A graphical representation of the evolution of the rate of profit

## 3.1. The setting of parameters

To graphically represent function $r$ and function $r_c$, we need first to give values to the various parameters involved in the expression of these functions.

For the average lifespan of fixed capital $t$, we assume that it is 10 years, i.e. $t = 10$.
In addition, we consider that one-third of total fixed capital is used in section I, i.e. $a = \dfrac{1}{3}$ (going back to our usual definition of parameter $a$).

It follows that $\alpha = \dfrac{a}{1-a} = \dfrac{1/3}{2/3} = \dfrac{1}{2} = 0.5$

If we want to construct function $r_c(values) = f(k)$ for different values of $c$, already used in figures 20.2, we will also have different values for $\gamma = \dfrac{c}{1-c}$. At this point it must be understood that to be sure capitalists of section II can decide the price of the consumption goods that they sell and thus the amount of surplus-value and the overall consumption of capitalists, by setting parameter $c$ as a share of global consumption. But they cannot decide the value of $c_{II}$, i.e. their own share of this global consumption, because it depends on the price at which they buy fixed capital from capitalists of section I.

On the other hand, capitalists of section I can decide the price of fixed capital that they sell to those of section II by setting parameter $c^*$, which has been defined as a gross margin rate, i.e. the ratio of the gross margin allocated to consumption over the selling price of fixed capital. Through the setting of $c^*$ they can influence the value of $c_I$, but they cannot directly decide its precise value, i.e. their own share in global consumption, because it depends on $c$ and the price of consumption goods. Similarly, and this clearly derives from equation (28) in chapter 12, $c_I$ also depends on $k$, the primary rate of surplus-value, and varies with it.

In fact, once the value of $c$, i.e. the share of all capitalists in global consumption, and thus the price of consumption goods, are set by capitalists in section II (which at the same time also set the rate of surplus-value $k_c$), capitalists of section I can obtain indirectly the share $c_I$ that they get in total consumption, by setting $c^*$ and thus the price of fixed capital sold to section II. Therefore the share $c_{II}$ obtained by capitalist of section II is logically determined afterwards, as the difference between $c$ and $c_I$, and must be considered as a residue. This being noted, and in order not to unduly complicate this example, let us make the assumption that $c^*$ is equal to 0.25 and does not change when $c$ changes.

We thus get for parameters $\gamma$ et $\gamma^*$ the following values:

For $c = 0.10$, we have $\gamma = 0.1111$      For $c^* = 0.25$, we have $\gamma^* = 0.333$
For $c = 0.15$, " " $\gamma = 0.1765$
For $c = 0.20$, " " $\gamma = 0.25$
For $c = 0.25$, " " $\gamma = 0.3333$
For $c = 0.30$ " " $\gamma = 0.4286$

For each of the five cases corresponding to the above parameters, the corresponding functions are defined, first in terms of values and secondly in terms of prices.

### 3.2. A graphical representation of function $r_c$ (values).

This function has the very simple general form $r_c = \frac{k+c}{kt}$, since it is not influenced by the distribution of surplus-value in the form of monetary profits among capitalists of both sections, which is mediated by rate $c^*$. It only depends on parameters $c$ and $t$, and obviously on $k$, the primary rate of surplus-value. Therefore it can be written for the five cases above:

Case 1: $r_c$ *(values)* = $\frac{k+0.10}{10k}$ for $c = 0.10$

Case 2: $r_c$ *(values)* = $\frac{k+0.15}{10k}$ for $c = 0.15$

Case 3: $r_c$ *(values)* = $\frac{k+0.20}{10k}$ for $c = 0.20$

Case 4: $r_c$ *(values)* = $\frac{k+0.25}{10k}$ for $c = 0.25$

Case 5: $r_c$ *(values)* = $\frac{k+0.30}{10k}$ for $c = 0.30$

These five functions, with $r_c$ defined in terms of values, correspond to the five values of parameter $c$ shown above, and are represented in figure 21.1, as well as a horizontal line, $r_c$ = 0.10, which represents the constant rate of profit (equal to $\frac{1}{t}$, with $t = 10$) in the absence of capitalist consumption ($c$ = 0).

This line is also the asymptote towards which the different curves tend, when the rate of primary surplus-value $k = L_I / L_{II}$ tends towards infinity.

Three vertical lines have also been drawn, which correspond to rates of primary surplus-value of 20%, 25% and 30% respectively. Therefore the figure is as follows:

**Figure 21.1 - Evolution of the rate of profit $r_c$ with its current definition and in terms of values as a function of the primary rate of surplus-value k, for different values of $c$.**

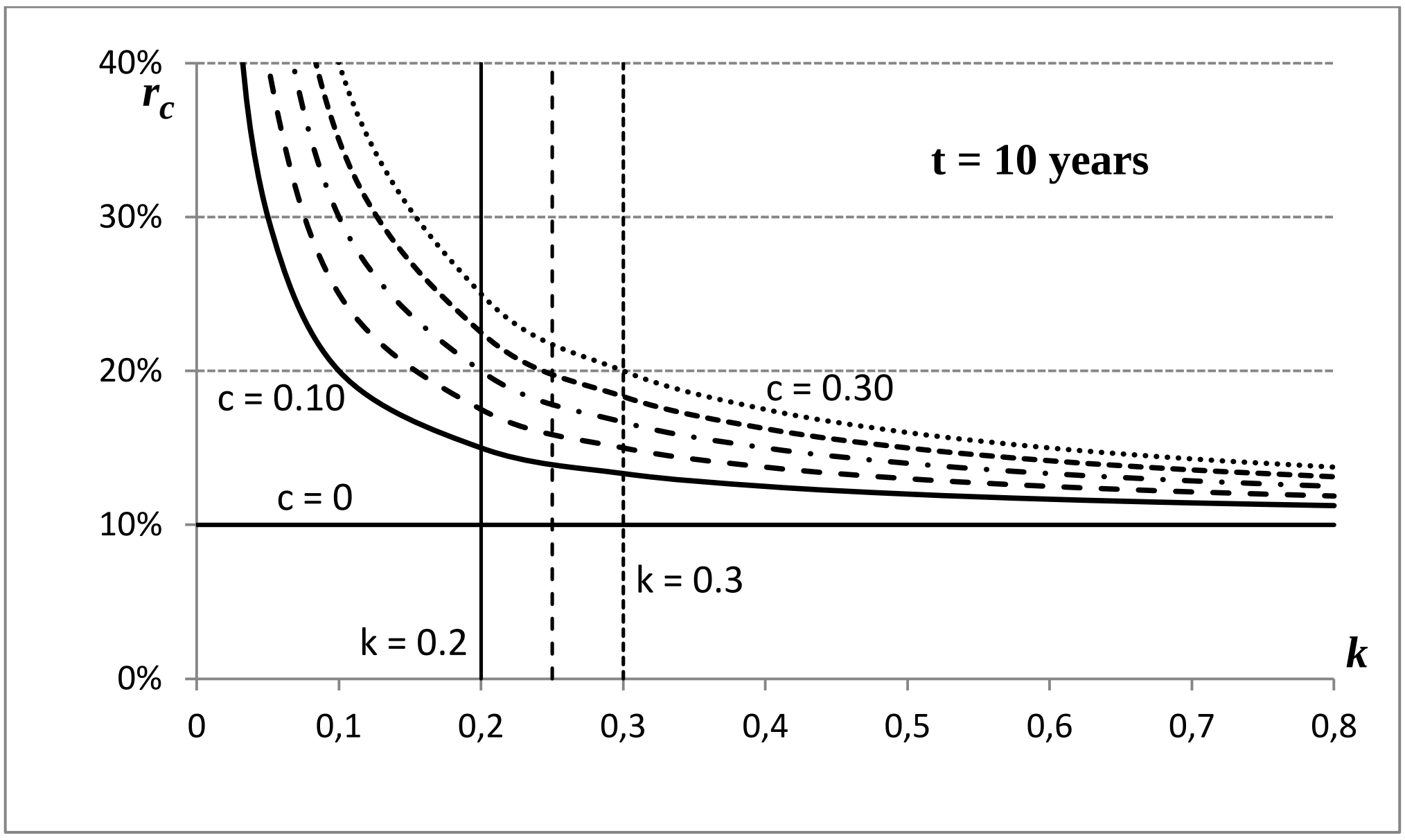

We can easily observe from figure 21.1 that the rate of profit decreases with the increase in the rate of surplus-value, for any given value of parameter $c$. However, this last parameter exerts a strong influence on the rate of profit: the higher is $c$ and the higher is the profit rate. For instance, with a rate of primary surplus-value $k$ equal to 20 %, we have a profit rate of 15% with $c$ equal to 10 %, and of 25 %, with $c$ equal to 30 %. For $k$ equal to 30 %, the profit rate goes from 13.33 % for $c$ equal to 10 %, to 20 % for $c$ equal to 30 %.

One could nevertheless wonder whether a rate of profit expressed in terms of value has any theoretical interest, since in the real world this rate is always calculated in terms of prices. In fact it is an interesting variable, because its amount depends only on the distribution of surplus-value between workers and capitalists as a whole, and whatever the rate of profit in terms of prices, once $k$ and $c$ are given, this rate will not change. On the contrary, the rate of profit defined in terms of prices also depends on $c^*$, i.e. on the price of fixed capital and therefore on the distribution of surplus-value and of consumption goods among capitalists.

### 3.3. A graphical representation for function $r_c$ (prices)

This function has the following general form:

$$r_c(prices) = f(k) = \frac{\Pi_c}{K} = \frac{k(1+\alpha+\gamma+\gamma^*+\alpha\gamma^*)+\gamma}{kt(1+\alpha+\gamma^*+\alpha\gamma^*)}$$ (see equation 15 above)

For the five cases already defined, each one of them corresponding to a different value of parameter $c$, and with again $t=10$, $a=\frac{1}{3}$ (hence $\alpha$ = 0.5), and $c^*$ = 0.25 (hence $\gamma^*$ = 0.333), function $r_c(prices)$ becomes therefore, with all calculations made:

Case 1: $r_c$ *(prices)* $= \frac{2.111k+0.111}{20k}$ for $c$ = 0.10

Case 2: $r_c$ *(prices)* $= \frac{2.176k+0.176}{20k}$ for $c$ = 0.15

Case 3: $r_c$ *(prices)* $= \frac{2.25k+0,25}{20k}$ for $c$ = 0.20

Case 4: $r_c$ *(prices)* $= \frac{2.333k+0.333}{20k}$ for $c$ = 0.25

Case 5: $r_c$ *(prices)* $= \frac{2.429k+0.429}{20k}$ for $c=0.30$

We must stress the fact that $\gamma$, and therefore $c$ (from which $\gamma$ is defined) do not appear in the denominator of the above expression defining $r_c(prices) = f(k)$, because at the denominator of equation (15) we only find $\gamma^*$, and therefore indirectly $c^*$ (from which $\gamma^*$ is defined). Combined with the assumption made above that a does not change and that $c^*$ is fixed and equal to 0.25, this explains that the denominator is always the same, and equal to 20$k$.

The five curves corresponding to the five functions which were just defined appear in figure 21.2, and differ only from the differences in the magnitude of parameter $c$. They allow us to observe that an increase in the primary rate of surplus-value $k$ results in a decrease of the rate of profit: we have the same evolution in terms of prices as we had in terms of values. We can see also that this decline is slowing down once we exceed 0.20 for this rate $k$, and that all curves converge towards an asymptote which is represented by the straight line:

$r_c$ = 0.1 = $\frac{1}{t}$ when $k$ tends towards infinity.

**Figure 21.2 - Evolution of the rate of profit $r_c$ with its current definition and in terms of prices as a function of the primary rate of surplus-value *k*, for different values of $c$.**

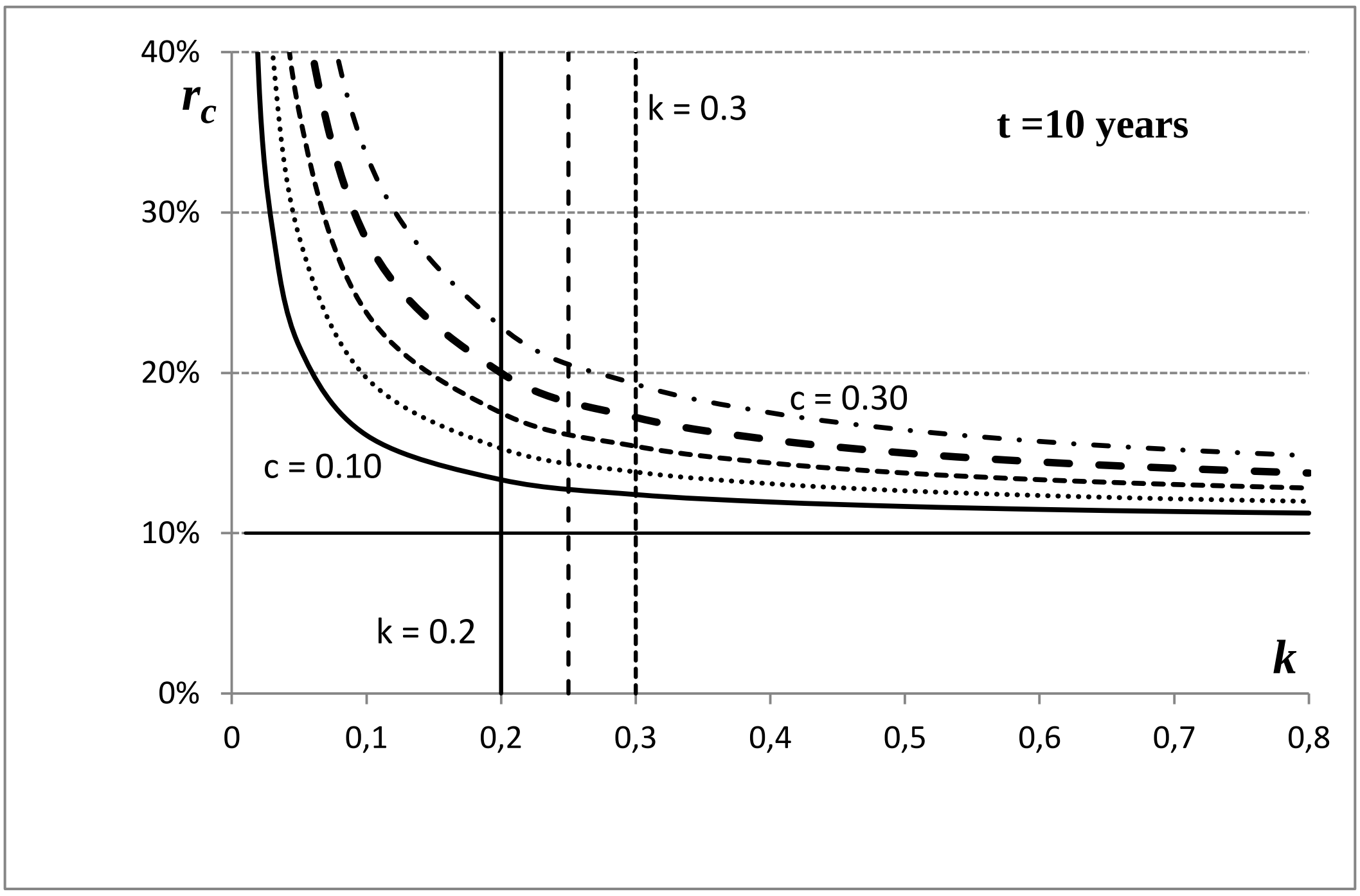

One could say that this evolution to some extent confirms the validity of Marx's law, although within the framework of this graphical representation, we did no longer link the variation in the rate of profit to the organic composition of capital, but rather to the evolution of the rate of surplus-value, for the sake of simplicity. With this provision, the law interpreted in terms of surplus-value would be moreover better verified with this usual definition of the rate of profit, rather than with the Marxist definition including variable capital at the denominator of the fraction giving the rate of profit. On the basis of the usual definition of the rate of profit, its decrease is indeed verified for any value of $c$, and not only for the situations where $c > \frac{1}{1+t}$.

We can also observe that for a given rate of surplus-value, the rate of profit increases with the increase in $c$, which is denoted by the displacement of the curves upward and to the right, which accompanies this increase. Once more we can notice that the rate of profit is actually very sensitive to a rise in $c$: for instance with $k$ = 0.20, the rate of profit almost doubles, going from 13.33 % when c is equal to 0.10, to 22.87 % when $c$ is equal to 0.30. Capitalist consumption thus appears again to be of fundamental importance in explaining the evolution of the rate of profit.

Nevertheless, parameter $k$ maintains a strong influence, which for instance can be registered if we try to know what happens to $c_I$ and $c_{II}$, for a same value of $c$, for instance $c$ = 0.20, when $k$ increases from 0.20 to 0.30. We know from equation (28) in chapter 12 that:

$$c_I = \gamma * \frac{k(1-c)}{(1+k)}$$

Therefore with $c = 0.20$, $c^*$ = 0.25, and thus $\gamma^* = 0.333$, we get:

1) For $k = 0.2$ , which means that $k_c = \frac{k+c}{1-c} = 0.5 = 50$ % :

$$c_I = \gamma * \frac{k(1-c)}{(1+k)} = 0.333 \frac{0.2(1-0.2)}{(1+0.2)} = 0.333 \frac{0.16}{1.2} = 0.0444 = 4.44\% \text{ of total consumption}$$

Which implies that $c_{II} = c - c_I$ = 0.2 - 0.0444 = 0.1556 = 15.56 % of total consumption.

2) For $k = 0.30$, which means that $k_c = 0.625$ = 62.5 %, we get :

$$c_I = 0.333 \frac{0.3(1-0.2)}{(1+0.3)} = 0.333 \frac{0.24}{1.3} = 0.0615 = 6.15\%$$

Which implies that $c_{II} = c - c_I$ = 0.2 – 0.0615 = 0.1385 = 13.85 %

This shows us that an increase in the primary rate of surplus-value, even with a constant share of consumption for all capitalists, results in an increase in the share of consumption for capitalists of section I, at the expense of the share of capitalists of section II.

This gives us a justification to examine more particularly function $k_c = f(c)$, which we will do in next subsection.

### 3.4. The relation between the actual rate of surplus value $k_c$ and the consumption of capitalists

The importance of capitalist consumption as a determinant of the rate of profit is due to the great elasticity of the actual rate of surplus-value $k_c$ with respect to $c$, which derives from the nature of the function defining $k_c$ in terms of $c$. This function is indeed nothing else than the definition of $k_c$ :

$k_c = f(c) = \frac{k+c}{1-c}$, which as a function of $c$ should rather be rewritten now as:

$$k_c = f(c) = \frac{c+k}{-c+1} \tag{16}$$

This is again a homographic function, whose derivative is $f'(c) = \frac{1+k}{(1-c)^2}$ (17)

This derivative is always positive, which implies that the rate of surplus-value $k_c$ is an increasing function of $c$, the share of capitalist consumption in total consumption.

Since $k_c = f(c)$ is a positive and differentiable function of a positive variable $c$, its elasticity at point $c$ is defined as:

$$Ef(c) = \frac{\frac{\delta k_c}{k_c}}{\frac{\delta c}{c}} = \frac{c}{f(c)} f'(c) = \frac{c(1-c)}{k+c} * \frac{1+k}{(1-c)^2} = \frac{c(1+k)}{(k+c)(1-c)} \qquad (18)$$

We can calculate this elasticity for a given value of parameter $k$, taking for instance $k$ = 0.3, and for two values of $c$, for instance for $c$ = 0.10 and $c$ = 0.30 :

For $c$ = 0.10, this elasticity is $Ef(c) = \frac{0.10(1+0.3)}{(0.3+0.1)0.9} = \frac{0.13}{0.36} = 0.36$

For $c$ = 0.30, the elasticity is $Ef(c) = \frac{0.30(1+0.3)}{(0.3+0.3)0.7} = \frac{0.39}{0.42} = 0.93$

Through this simple example, we can see that the elasticity of $k_c = f(c)$ with respect to a relative change of $c$ increases very strongly when $c$ increases: this means that the higher is the share of capitalist consumption in total consumption, the stronger is the influence of its variation on the rate of growth of the rate of surplus-value $k_c$. This is reflected in the shape of the representative curves in the chart below, where they are plotted for four values of $k$, namely:

$k$ = 0.2, $k$ = 0.3, $k$ = 0.4 and $k$ = 0.5, respectively

This strong influence of $c$, which can be observed in the four curves drawn in figure 21.3, appears to be unsurprising, if we have a look at the second derivative of $k_c = f(c)$, which is:

$$f''(c) = \frac{2(1+k)}{(1-c)^3} \qquad (19)$$

**Figure 21.3 - Evolution of the rate of surplus-value $k_c$ with $k_c = f(c) = \dfrac{c+k}{-c+1}$**

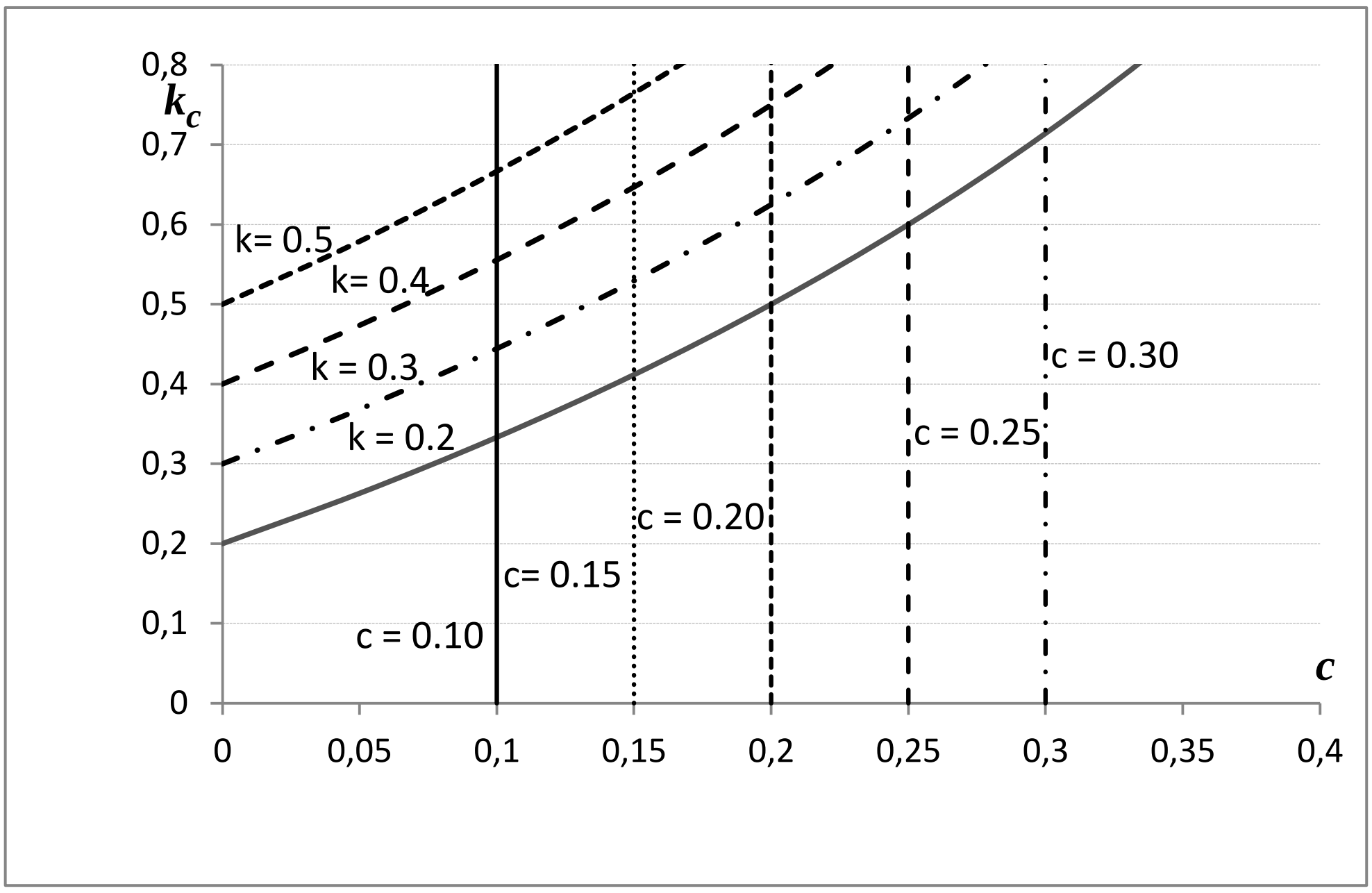

Since we know that $c < 1$, by definition, we also have always $0 < (1-c)^3 < 1$. Since we have also $k > 0$, it follows that we always have $f''(c) > 2$.

The fact that $f''(c) =$ is greater than 2 essentially implies that $f''(c)$ is always positive. In turn this means that $f'(c) = \dfrac{dk_c}{dc}$ is an increasing function of $c$. It follows that when $c$ increases, the variations of $k_c$, i.e. $\delta k_c$, are larger than the variations of $c$, i.e. $\delta c$. Relative variations of $k_c$, i.e. $\dfrac{\delta k_c}{k_c}$, therefore become larger than relative variations of $c$, i.e. $\dfrac{\delta c}{c}$. This means that the elasticity of $k_c$ with respect to $c$ rises when $c$ increases. This appears in the above example, where the elasticity is indeed more than twice as strong (going from 0.36 to 0.93) when c goes from 0.10 to 0.30. This corresponds to the form taken by the representative curves of function $k_c = f(c)$ represented in figure 9 above, together with five vertical lines corresponding to different values of $c$ ($c$ = 0.10, $c$ = 0.15, $c$ = 0.20, $c$ = 0.25, $c$ = 0.30).

Keeping again $k$ = 0.3, we have for these different values of $c$:

1) $k_c = f(c)$ = 0.44 for $c$ = 0.10

2) $k_c = f(c)$ = 0.53 for $c$ = 0.15 (from 1) to 2) $\delta k_c$ = 0,090 and $\delta c$ = 0.05)

3) $k_c = f(c) = 0.625$ for $c = 0.20$ (from 2) to 3) $\delta k_c = 0{,}095$ and $\delta c = 0.05$)

4) $k_c = f(c) = 0.73$ for $c = 0.25$ (from 3) to 4) $\delta k_c = 0{,}105$ and $\delta c = 0.05$)

5) $k_c = f(c) = 0.86$ for $c = 0.30$ (from 4) to 5) $\delta k_c = 0{,}130$ and $\delta c = 0.05$)

This evolution shows the rise in the values of the elasticity of $k_c$ compared to $c$, which becomes greater than one: indeed it reaches 1 for $0.324 < c < 0.325$.

### 3.5. A comparison between two expressions of the profit rate: in terms of values and in terms of prices

A final point worth mentioning is the great similarity of the two curves representing function $r_c$ as a function of $k$, one with the rate of profit defined on the basis of values, and the other where it is defined on the basis of prices. This similarity is apparent although their equations are significantly different, since for two values of parameter $c$, for example $c = 0.10$ and $c = 0.25$, these equations are as follows:

1) For $c = 0.10$, and keeping $\alpha = 0.5$, $t = 10$, and $\gamma^* = 0.333$ :

$$r_c \ (values) = \frac{k+c}{kt} = \frac{k+0{,}10}{10k} \ ;$$

$$r_c \ (prices) = \frac{k(1+\alpha+\gamma+\gamma^*+\alpha\gamma^*)+\gamma}{kt(1+\alpha+\gamma^*+\alpha\gamma^*)} = \frac{2.111k+0{,}111}{20k}$$

2) For $c = 0.25$:

$$r_c \ (values) = \frac{k+c}{kt} = \frac{k+0{,}25}{10k} \ ;$$

$$r_c \ (prices) = \frac{k(1+\alpha+\gamma+\gamma^*+\alpha\gamma^*)+\gamma}{kt(1+\alpha+\gamma^*+\alpha\gamma^*)} = \frac{2.333k+0{,}333}{20k}$$

This similarity appears in Figure 21.4, where the curves of the two functions, in terms of values and in terms of price, have been plotted for each of the two above values of c: $c = 0.10$ and $c = 0.25$. In both cases the proximity of the two curves is also striking.

**Figure 21.4 - Comparison of curves representing $r_c$ in terms of values and prices**

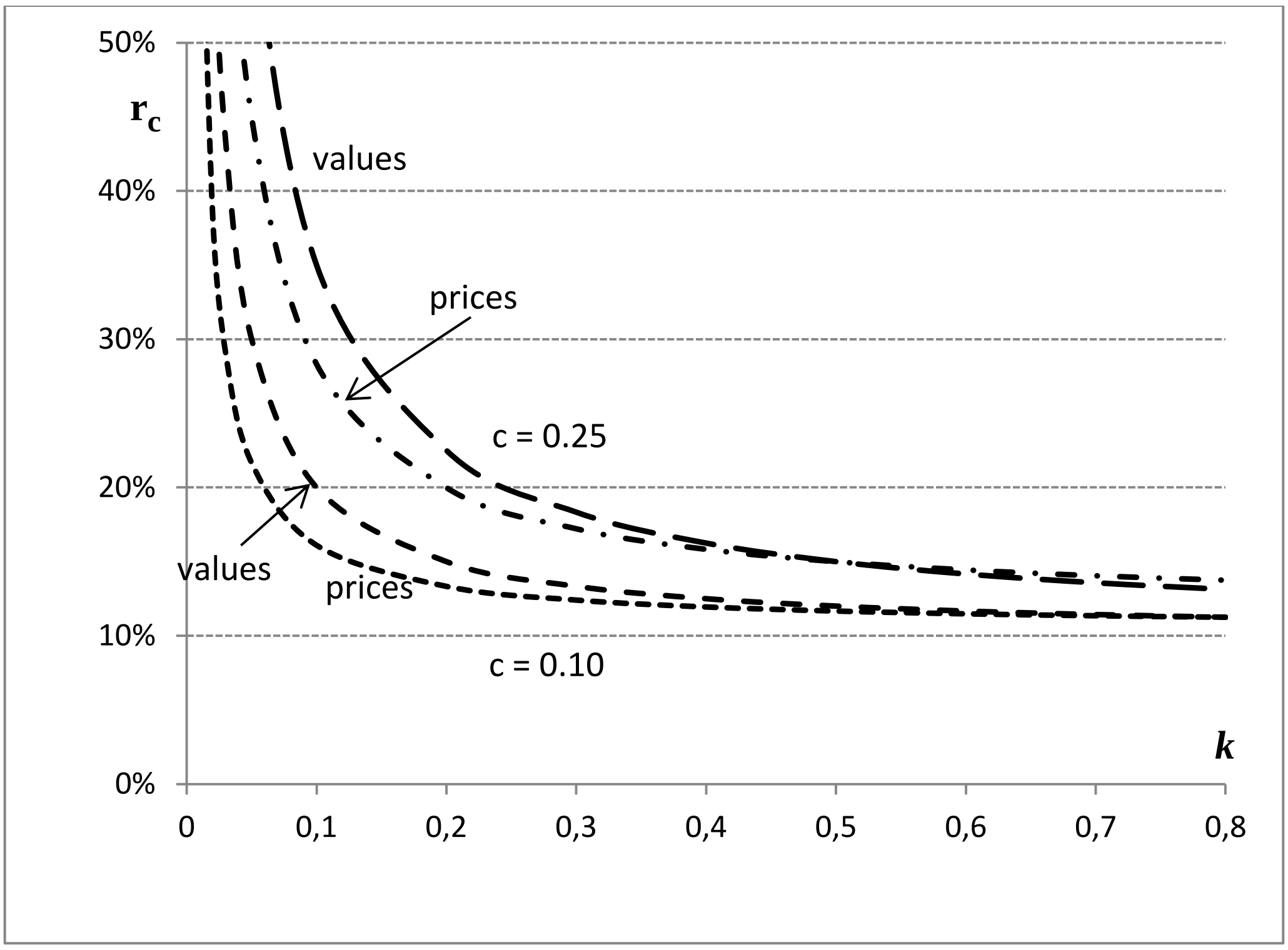

This proximity is not very surprising, since profit is only surplus-value transformed and redistributed through exchanges that take place on the basis of monetary prices. Profit, therefore, is nothing else than the concrete form of surplus-value when it is monetized on the occasion of the realization of the product, which simultaneously ensures the realization of this surplus-value. We can also observe that the curves intersect. As a result, the price curves pass above the value curves when $k$ grows above amounts which may however be too high to correspond to the real world. However, this difference between the curves seems to increase slightly as c increases because the curves then move upwards and to the right, which is related to the correlation between the price level of consumer goods and this rising capitalist consumption.

## 4. The evolution of the share of profits in the product

Paradoxically, all the analysis that has just been conducted on the rate of profit and its evolution, and in particular its dependence on and sensitivity to the evolution of many variables (the rate of surplus-value, certainly, but also the lifespan of fixed capital, as well as the consumption of different types of capitalists) tends to downplay the importance it can have, in the real world of capitalist enterprises, as an effective criterion of decision-making.

This is all the more true as the various rates of profit as they are calculated in practice by capitalist enterprises are highly dependent on tax and accounting rules, particularly those concerning the depreciation of fixed capital. The rate of profit in the real world is therefore a poor performance indicator, and this explains why financial analysts are not mistaken as regards its true significance. As mentioned previously these experts are actually much more interested in the cash flow of a firm, which is referred to in financial jargon as EBITDA ("Earnings before interest, taxes, depreciation and amortization"). This concept is often used, moreover, to evaluate the value of a company in view of an acquisition. It corresponds in national accounts to the gross operating income, but also, in the form of a ratio, to what is called the margin rate.

The concept corresponds in the analysis conducted here to the ratio of total profits over total product, but as we know that this ratio has a fairly complex expression (see equation (52) in chapter 12 and table 2 in appendix 3), and especially that the realization of profits takes place according to a very different logic in each of the two sections, it seemed better to start by calculating the ratio for each of the two sections taken separately, before dealing with the global profit share. Let us call $\pi_I$ and $\pi_{II}$, these two ratios, respectively for section I and section II. We will show that each of these two ratios $\pi_I$ and $\pi_{II}$ is a function of different parameters: the evolution of $\pi_I$ depends on that of $a$ and $c^*$, while that of $\pi_{II}$ depends as already noted above on the evolution of $k$ and $c$.

### 4.1. The share of profits in the product of section I

In section I, to begin with, the profit / product ratio is equal to:

$$\pi_I = \frac{\Pi_I}{Y_I} = \frac{L_I\left(\frac{a}{1-a} + \frac{c^*}{(1-a)(1-c^*)}\right)}{\frac{L_I}{(1-a)(1-c^*)}} = \frac{\frac{a(1-c^*)+c^*}{(1-a)(1-c^*)}}{\frac{1}{(1-a)(1-c^*)}} = a(1-c^*)+c^*$$

If we look at $\pi_I$ as a function of $a$, then we must write $\pi_I$ as above:

$$\pi_I = f(a) = a(1-c^*)+c^* \tag{20}$$

But $\pi_I$ can also be considered as a function of $c^*$ and in this case the function is written:

$$\pi_I = f(c^*) = c^*(1-a)+a \tag{21}$$

We can verify that both functions are linear functions, perfectly symmetrical to each other.

It seemed useful to provide a graphical representation of the corresponding functions, which is often more eloquent than long explanations.

1) Let us start with function $\pi_I = f(a) = a(1-c^*)+c^*$

It can be represented in a figure for different values of $c^*$, i.e. $c^*$ = 0.25, $c^*$ = 0.30, $c^*$ = 0.35, $c^*$ = 0.40 and $c^*$ = 0.45. We recall that $c^*$ is a margin rate which corresponds to capitalist consumption in section I. As for the share of consumption of capitalists of section I in global consumption, it is $c_I$. The corresponding curves appear in figure 21.5, where we also add four straight lines corresponding to $a$ = 0.25, $a$ = 0.30, $a$ = 0.35 and $a$ = 0.40.

**Figure 21.5 - Evolution of $\pi_I$ - the share of profits in section I - as a function of $a$ and for different values of parameter $c*$**

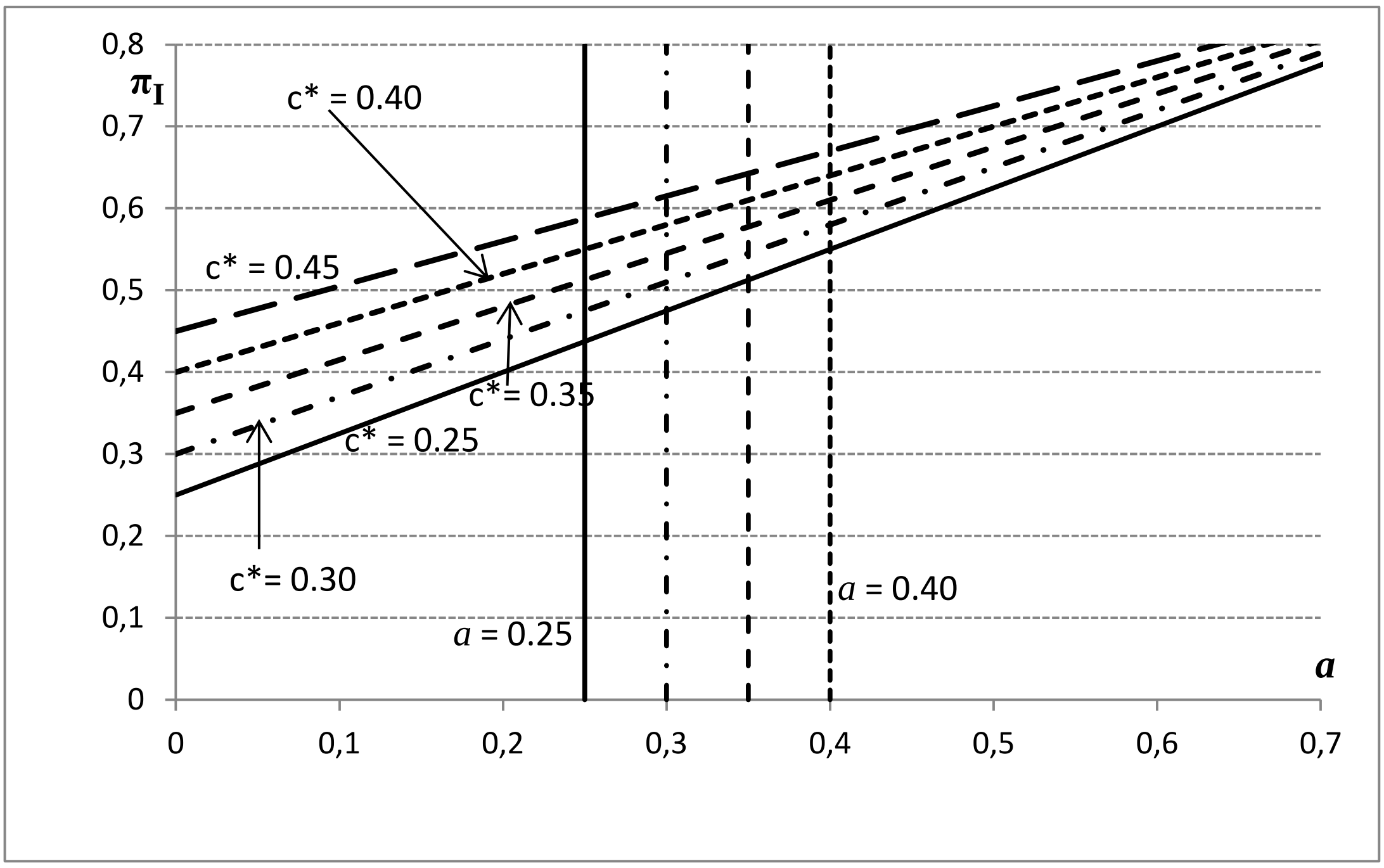

We can calculate that the share of profits is for instance 43.75 % for $c*$ = 0.25 and $a$ = 0.25, and goes up to 67 % for $c*$ = 0.45 and $a$ = 0.40.

Using the formula previously established in chapter 12 (page 235), which links $c*$ and $c_I$, i.e. $c_I = \gamma * \frac{k(1-c)}{1+k}$, we can therefore calculate $c_I$, by taking for instance $k$ = 0.33 and $c$ = 0.3. Then we obtain:

$c_I$ = 5.83 % , for $c*$ = 0.25

$c_I$ = 14.30 %, for $c*$ = 0.45

2) Let us continue by representing now function $\pi_I = f(c*) = c*(1-a)+a$

It can also be represented for different values of $a$, i.e. $a$ = 0.20, $a$ = 0.25, $a$ = 0.30, $a$ = 0.35 and $a$ = 0.40. We recall that $a$ is the share of the total value of produced fixed capital which is used in section I. The corresponding curves appear in figure 21.6, where we also add four straight lines corresponding to $c*$ = 0.25, $c*$ = 0.30, $c*$ = 0.35 and $c*$ = 0.40.

We can see that the share of profits is for instance 40% for $a$ = 0.20 and $c*$ = 0.25 and goes up to 64 % for $a$ = 0.40 and $c*$ = 0.40. We already know that for these last two values of $c*$ (i.e. 0.25 and 0.40) we have $c_I$ = 5.83 % and $c_I$ = 14.30 %.

**Figure 21.6 - Evolution of $\pi_I$ - the share of profits in section I - as a function of $c*$ and for different values of parameter $a$**

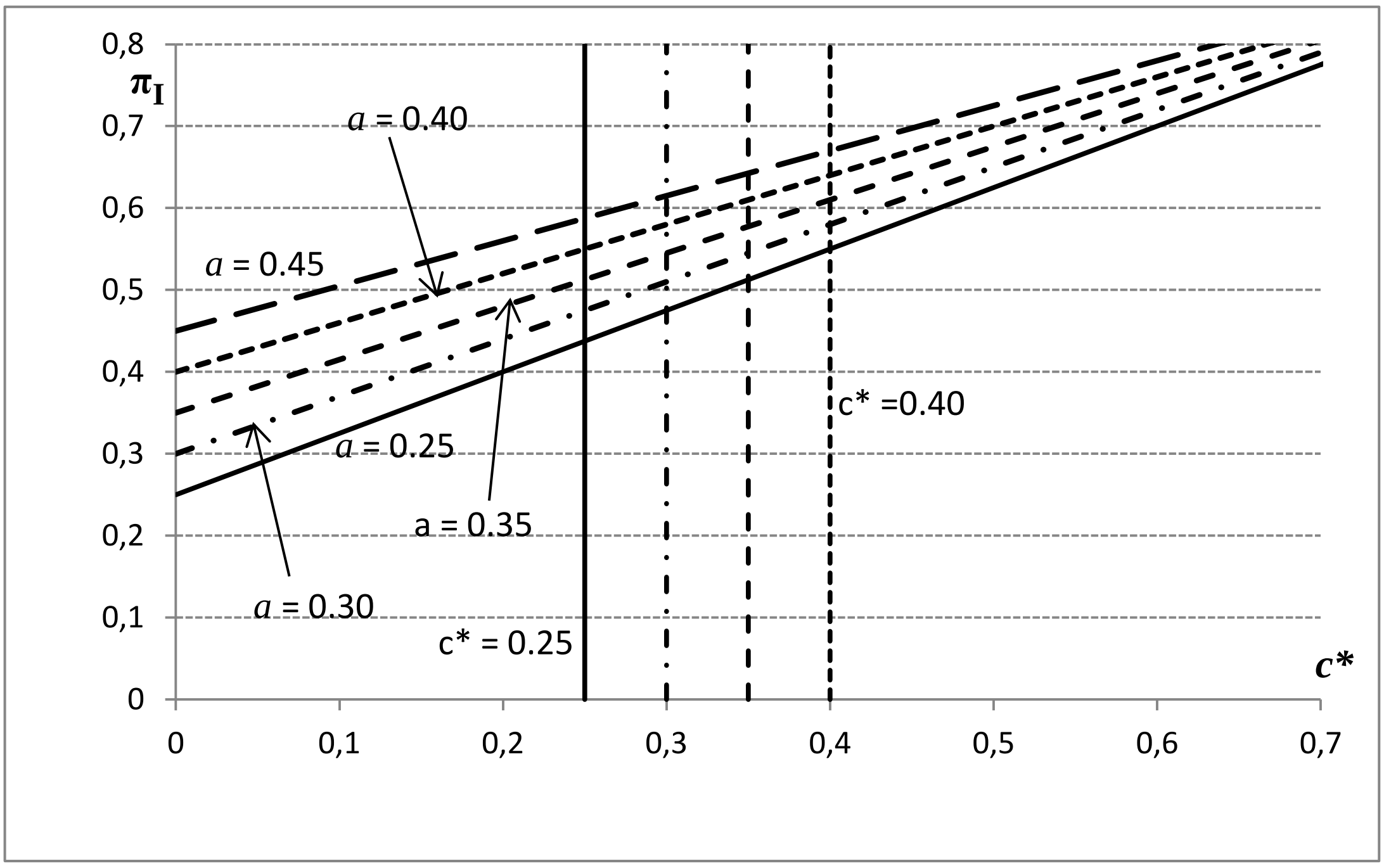

These two figures show us that the share of profits in section I is always increasing as a function of $a$ as well as a function of $c*$ because the derivative of both functions is always positive, since by definition we always have $a<1$ and $c*<1$. Moreover, both functions being of the same form, with a range of values quite similar, $\pi_I$ is quite similarly affected by changes of $c*$ as well as by those of $a$. This shows once again the importance of capitalist consumption, here in section I, in the evolution of the share of profits, or reciprocally the fact that the share of profits plays a prominent role in the determination of capitalist consumption.

### 4.2. The share of profits in the product of Section II

In section II the profit / product ratio is equal to:

$$\pi_{II} = \frac{\dfrac{L_{II}(k+c)}{1-c}}{\dfrac{L_{II}(1+k)}{1-c}} = \frac{k+c}{1+k} \qquad (22)$$

The share of profits in this section is thus a function of only two variables: $k$, the primary rate of surplus-value, and $c$, the share of overall capitalist consumption in global consumption. In order to isolate the influence of each of these two variables, we can study two functions, the first one with $\pi_{II}$ as a function of $k$ and $c$ as a parameter having given values, and the second one with $\pi_{II}$ as a function of $c$ and $k$ as a parameter having given values.

1) Let us start first with $\pi_{II}$ as a function of $k$, which as such is:

$$\pi_{II} = f(k) = \frac{k+c}{1+k} \tag{23}$$

This is a homographic function. Its derivative is $f'(k) = \frac{1-c}{(1+k)^2}$, which is always positive, so that it is an increasing function of $k$, the primary rate of surplus-value.

We can provide a graphical representation of function $\pi_{II} = f(k)$ in figure 21.7. We do that by drawing five curves, each one corresponding to a different value of $c$:

$c$ = 0.10, $c$ = 0.15, $c$ = 0.20, $c$ = 0.25 and $c$ = 0.30.

We represent as well the straight vertical lines corresponding to $k$ = 0.25, $k$ = 0.30, $k$ = 0.35, $k$ = 0.40 and $k$ = 0.45.

We can verify that the share of profits $\pi_{II}$ is for instance 28 % for parameter $c$ = 0.10 and variable $k$ = 0.25, and goes up to 51.7 % for $c$ = 0.30 and $k$ = 0.45.

**Figure 21.7 - Evolution of $\pi_{II}$ - the share of profits in section II - as a function of $k$ and for different values of parameter $c$**

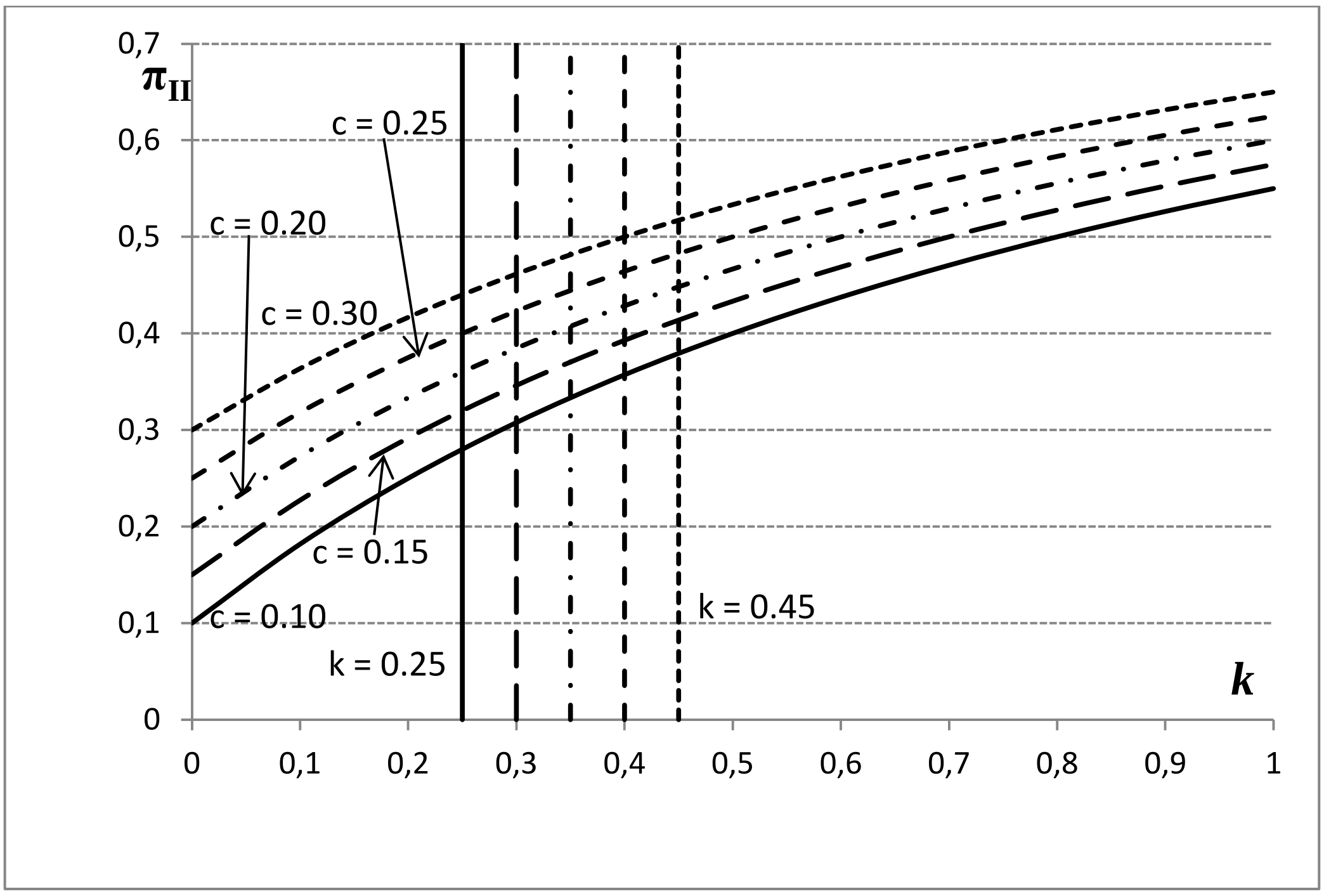

2) Let us continue now with $\pi_{II}$ as a function of $c$, which as such is :

$$\pi_{II} = f(c) = \frac{k+c}{1+k} = \frac{1}{1+k}c + \frac{k}{1+k} \tag{24}$$

It is a linear function whose derivative is $f'(c) = \frac{1}{1+k}$, a positive and constant number for a given value of $k$, so that it is an increasing function with a positive slope equal to $\frac{1}{1+k}$.

We can provide a graphical representation of function $\pi_{II} = f(c)$, in figure 21.8. We do that by drawing five curves, each one corresponding to a different value of $k$:

$k$ = 0.25, $k$ = 0.30, $k$ = 0.35, $k$ = 0.40 and $k$ = 0.45.

We represent as well the straight vertical lines corresponding to:

$c$ = 0.10, $c$ = 0.15, $c$ = 0.20, $c$ = 0.25 and $c$ = 0.30.

We can verify that the share of profits $\pi_{II}$ is obviously exactly the same for these different functions and for the same values of parameter $k$ and variable $c$: $\pi_{II}$ = 28 % for $k$ = 0.25 and $c$ = 0.10 and goes up to $\pi_{II}$ = 51.7 % for $k$ = 0.45 and $c$ = 0.30.

**Figure 21.8 - Evolution of $\pi_{II}$ - the share of profits in section II - as a function of $c$ and for different values of parameter $k$**

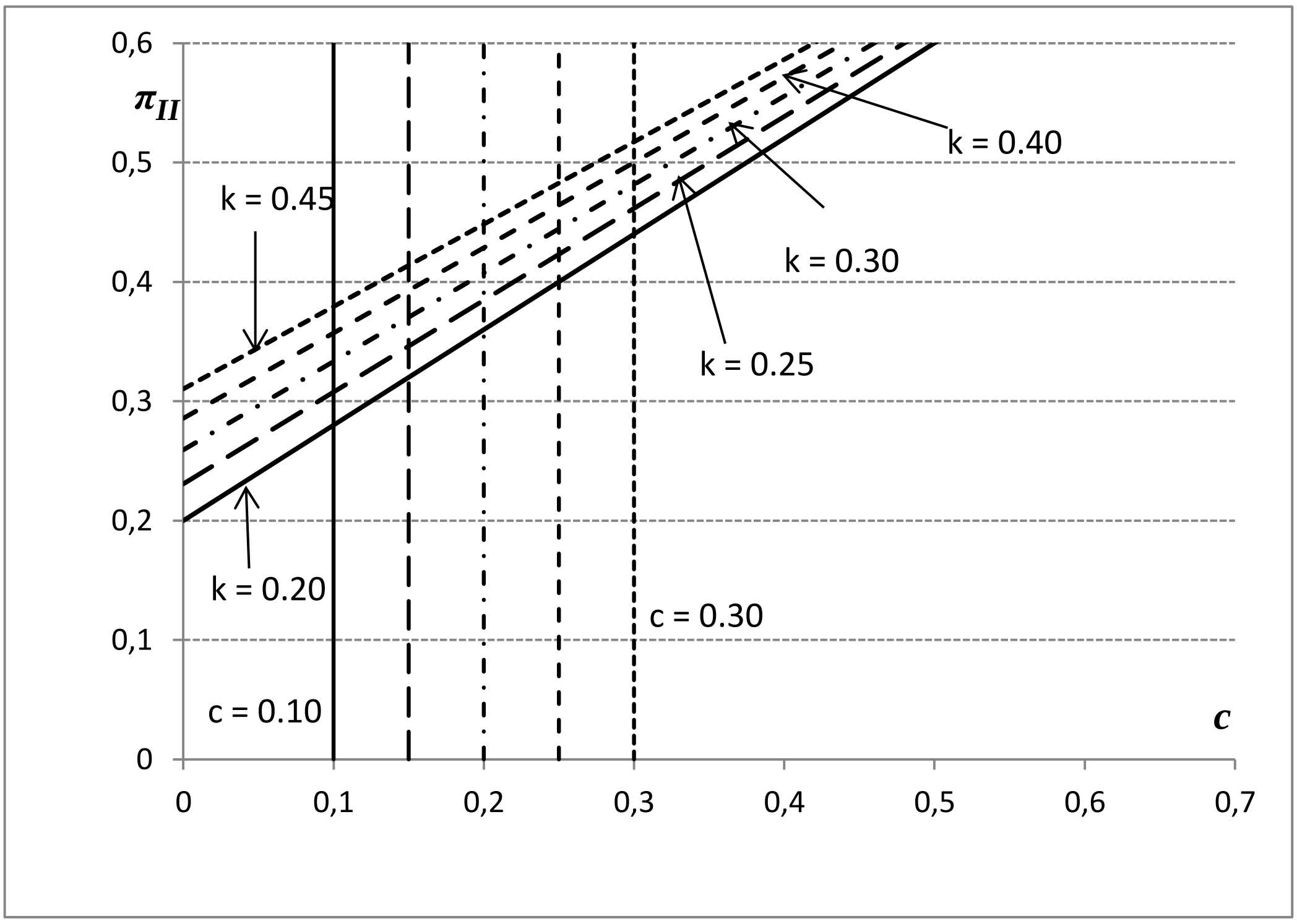

All in all it is notable, if we compare these two last figures, that $\pi_{II}$, the share of profits in section II producing consumption goods, seems more affected by changes in $c$ than by those in $k$, which shows once again the importance of capitalist consumption as a whole as a key element in the functioning of the system.

### 4.3. The share of profits in the overall product

The expression giving the share of profits $\pi$ in the overall product is complicated, because of the two distinct modes of realization of profits that have been highlighted previously. It can nevertheless be calculated.

$$\pi = \frac{\Pi}{Y} = \frac{L_I\left(\frac{a}{1-a}+\frac{c*}{(1-a)(1-c*)}\right)+L_{II}\frac{k+c}{1-c}}{\frac{L_I}{(1-a)(1-c*)}+L_{II}\cdot\frac{1+k}{1-c}} = \frac{(\alpha+\gamma*+\alpha\gamma*)+\frac{1}{k}\left(\frac{k}{1-c}+\frac{c}{1-c}\right)}{1+\alpha+\gamma*+\alpha\gamma*+\frac{1}{k}\left(\frac{1}{1-c}+\frac{k}{1-c}\right)}$$

$$\pi = \frac{\Pi}{Y} = \frac{k(\alpha+\gamma*+\alpha\gamma*)+k(1+\gamma)+\gamma}{k(1+\alpha+\gamma*+\alpha\gamma*)+1+\gamma+k(1+\gamma)} = \frac{k(1+\alpha+\gamma+\gamma*+\alpha\gamma*)+\gamma}{k(2+\alpha+\gamma+\gamma*+\alpha\gamma*)+1+\gamma} \qquad (25)$$

By putting $\Psi = 1+\alpha+\gamma+\gamma*+\alpha\gamma*$, the expression becomes quite simplified, and we get:

$$\pi = \frac{\Pi}{Y} = \frac{k\Psi+\gamma}{k(1+\Psi)+1+\gamma} \qquad (26)$$

If we now consider that $\pi = f(k)$ then its derivative is $f'(k) = \frac{\Psi(1+\gamma)-\gamma(1+\Psi)}{\left[k(1+\Psi)+1+\gamma\right]^2}$ (27)

The condition of positivity of the derivative is $\Psi(1+\gamma)-\gamma(1+\Psi)>0$ (28)

When performing the calculation, it is immediate that this implies $\Psi-\gamma>0$, which is always verified, given the definition of $\Psi$, whose value is always positive and includes $\gamma$. So the share of profits $\pi$ in the overall product $Y$ is always an increasing function of $k$, the primary rate of surplus-value.

The phenomenon is more meaningful by providing a graphical representation of function $\pi = f(k)$, which implies to give a value to its different parameters.

Let us first take $a$ = 0.25, which gives us $\alpha = \frac{a}{1-a} = \frac{0,25}{1-0,25} = 0,333$.

We can then represent our function for different values of $c$, which we arbitrarily take as:

$c$ = 0.10, $c$ = 0.15, $c$ = 0.20 , $c$ = 0.25, and $c$ = 0.30,

These values of $c$ correspond to the following corresponding values for $\gamma = \frac{c}{1-c}$:

$\gamma$ = 0.111, $\gamma$ = 0.177, $\gamma$ = 0.250, $\gamma$ = 0.333, and $\gamma$ = 0.429

We can also assume that there are different values of $c*$:

$c^*$ = 0.25, $c^*$ = 0.30, $c^*$ = 0.35, $c^*$ = 0.40 and $c^*$ = 0.45.

We recall again that $c^*$ is a margin rate which corresponds to capitalist consumption in section I, and is linked to $c_I$, i.e. precisely the share of capitalists of section I in total consumption, by a formula previously established in chapter 12 (page 235), which is $c_I = \gamma^* \frac{k(1-c)}{1+k}$. Since $c_I$ depends not only on $c^*$ (and thus $\gamma^*$), but also on $k$, its value will change with the level of $\pi = f(k)$ and cannot be considered as a parameter with a given value.

The values of $c^*$ correspond to the following corresponding values for $\gamma^* = \frac{c^*}{1-c^*}$:

$\gamma^* = 0.333$, $\gamma^* = 0.429$, $\gamma^* = 0.538$, $\gamma^* = 0.667$, and $\gamma^* = 0.811$

If we want now to give an example of a few numerical values for $\pi$, since $\pi$ is a function of *k* we need to adopt a numerical value for *k*. Let us decide arbitrarily that we have *k* = 20. Then the corresponding values for all these parameters are given in table 21.1.

**Table 21.1. Values of $\pi$ corresponding to different values of the other parameters**

| $\alpha$ | $c$ | $\gamma$ | c* | $\gamma^*$ | $\alpha\gamma^*$ | $\Psi$ | $k$ | $\boldsymbol{\pi}$ |
|---|---|---|---|---|---|---|---|---|
| 0.333 | 0.10 | 0.111 | 0.25 | 0.333 | 0.111 | 1.888 | 0.20 | **0.290** |
| 0.333 | 0.15 | 0.177 | 0.30 | 0.429 | 0.143 | 2.082 | 0.20 | **0.331** |
| 0.333 | 0.20 | 0.250 | 0.35 | 0.538 | 0.179 | 2.300 | 0.20 | **0.372** |
| 0.333 | 0.25 | 0.333 | 0.40 | 0.667 | 0.222 | 2.555 | 0.20 | **0.413** |
| 0.333 | 0.30 | 0.429 | 0.45 | 0.811 | 0.270 | 2.843 | 0.20 | **0.454** |

We get for $\pi$ a range of values going from 29 % to 45.4 %, which are quite plausible.

If we now want to see what happens to $\pi$ when $k$ is no longer set at $k = 0.20$, but becomes a variable, an appropriate way is to give a graphical representation of $\pi = f(k)$. If we keep constant all the other variables, i.e. $\alpha, c, \gamma, c^*, \gamma^*$ and $\Psi$ which appear on each line of the above table we are able to define five distinct functions $\pi = f(k)$.

These five functions are the following ones:

1) π = (1,888 k + 0,111)/(2,888 k + 1,111)

2) π = (2,082 k + 0,177)/(3,082 k + 1,177)

3) π = (2,30 k + 0,250)/(3,3 k + 1,25)

4) π = (2,555 k + 0,333)/(3,555 k + 1,333)

5) π = (2,843 k + 0,429)/(3,843 k + 1,429)

We can now represent these five functions under the form of five curves, each of them corresponding to a line in table 1 above, save obviously for the values of $k$ and $\pi$. They appear in figure 21.9.

**Figure 21.9 – Evolution of the share of profits in the product $\pi$ as a function of $k$**

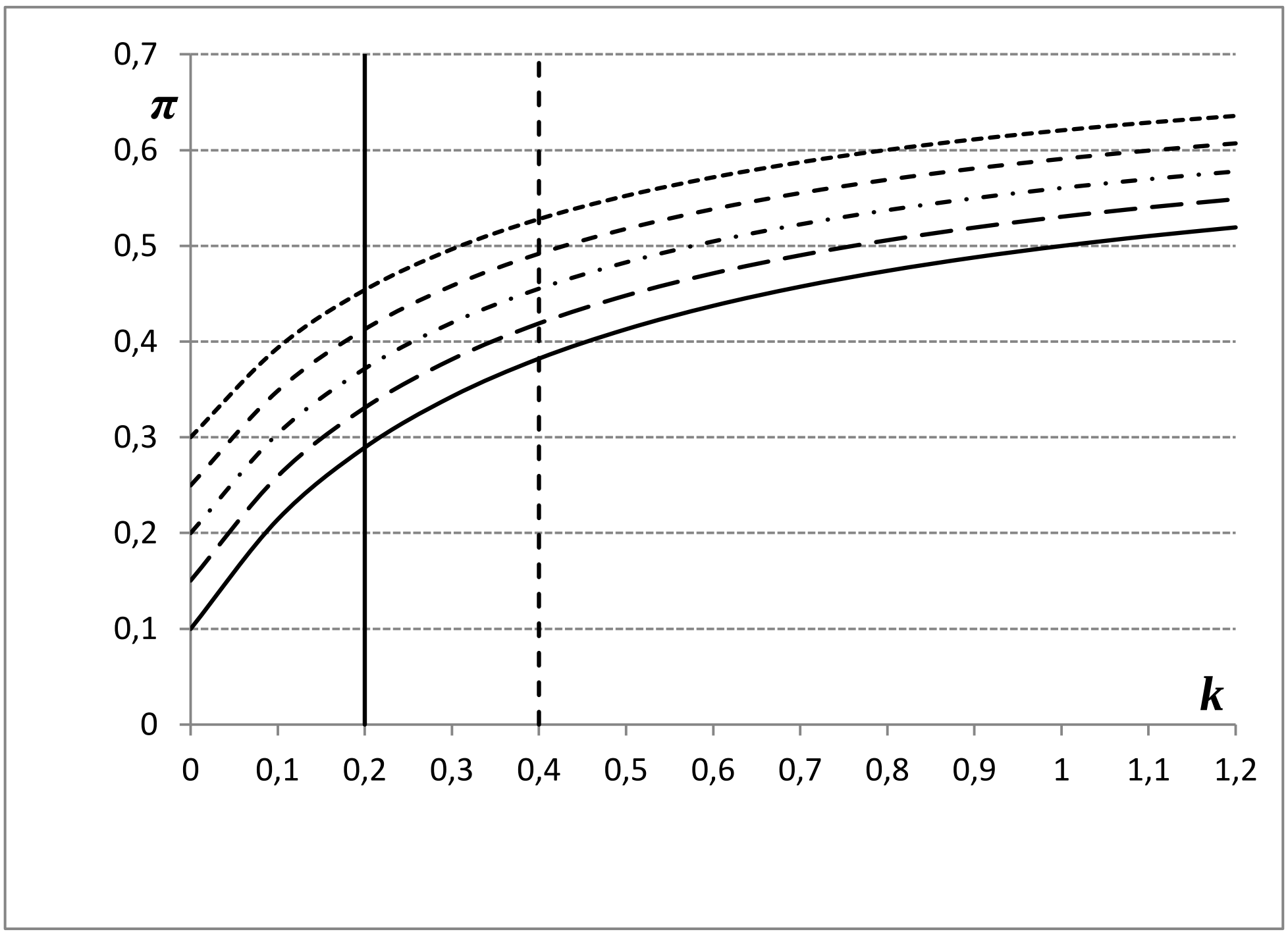

The five curves shown above are representing function $\pi = f(k)$ for five different values of parameters $c$ and $c*$. This figure shows clearly that the share of profits in the product, i.e. the overall gross margin rate in a macroeconomic sense, is an increasing function of the primary rate of surplus-value, and therefore varies inversely with the rate of profit, which is itself a decreasing function of this primary rate of surplus-value. Indeed chart 8 in subsection 3.3 of this chapter showed that in a graphical way. This last result might seem therefore somewhat counterintuitive, if we did not remember that an increase in $k$ causes an increase in the organic composition of capital which, as we also showed at length in the preceding sections of this same chapter, always results in a reduction of the rate of profit.

Another reason for this dissimilar evolution comes from the fact that the rise in the share of profits results ipso facto from a reduction in the share of wages which is at the origin of a rise both in the rate of surplus-value $k_c$, through an increase in $c$, the share of capitalist consumption in total consumption, and in the organic composition of capital $q_c$, itself inducing also, as we just recalled, a reduction in the rate of profit.

To close this chapter, we may now briefly devote a last sub-section to show that the share of profits is obviously linked to the rate of profit, the latter being - as we can easily imagine, a growing function of the former.

### 4.4. The share of profits and the rate of profit

The relationship between these two variables it quite easy to establish, since we know that by definition:

$$r_c = \frac{\Pi}{K} = \frac{\Pi}{Y} \cdot \frac{Y}{K} = \pi \frac{Y}{K} \tag{29}$$

We will not repeat the calculation of $Y$ and $\Pi$, since it has already been carried out in previous sections, and therefore will go directly to its result, which gives us:

$$r_c = \pi \left[ \frac{k\left(2+\alpha+\gamma+\gamma^*+\alpha\gamma^*\right)+1+\gamma}{kt\left(1+\alpha+\gamma^*+\alpha\gamma^*\right)} \right] \tag{30}$$

In order to simplify this expression, we take again $\Psi = 1+\alpha+\gamma+\gamma^*+\alpha\gamma^*$

This ultimately gives us the following expression for $r_c = f\left(\pi\right)$ :

$$r_c = \pi \left[ \frac{k\left(1+\Psi\right)+1+\gamma}{kt\left(\Psi-\gamma\right)} \right] \tag{31}$$

It is immediate to observe that the rate of profit $r_c = f\left(\pi\right)$ is an increasing function of $\pi$ . But we can observe one more time that it remains also a growing function of $a$ , $c$ and $c^*$, and a decreasing function of $k$ , the primary rate of surplus-value. To give a few examples showing the evolution of the rate of profit in parallel with the share of profits, we can use the values of the various parameters appearing in table 21.1 of previous sub-section 4.3., with all of them corresponding again to a unique value of $k$ , given as $k$ = 0.20.

We can see that the increase in the share of profits which is displayed in this table is necessarily due to the increase in $k_c$ (since $k$ is a constant, as well as $a$ ). It has thus been deemed useful to also indicate the effective rate of surplus-value $k_c = \frac{k+c}{1-c}$ . This allows us to provide table 21.2, which gives all the values of the parameters for each line, and thus for each value of $\pi$ that had been calculated.

**Table 21.2 Values of $r_c = f(\pi)$ corresponding to different values of the parameters**

| $\alpha$ | c | $\gamma$ | c* | $\gamma*$ | $\alpha\gamma*$ | $\Psi$ | $k$ | $k_c$ | **$\pi$** | $r_c$ |
|---|---|---|---|---|---|---|---|---|---|---|
| 0.333 | 0.10 | 0.111 | 0.25 | 0.333 | 0.111 | 1.888 | 0.20 | 0.333 | **0.290** | **0.138** |
| 0.333 | 0.15 | 0.177 | 0.30 | 0.429 | 0.143 | 2.082 | 0.20 | 0.412 | **0.331** | **0.156** |
| 0.333 | 0.20 | 0.250 | 0.35 | 0.538 | 0.179 | 2.300 | 0.20 | 0.50 | **0.372** | **0.181** |
| 0.333 | 0.25 | 0.333 | 0.40 | 0.667 | 0.222 | 2.555 | 0.20 | 0.60 | **0.413** | **0.190** |
| 0.333 | 0.30 | 0.429 | 0.45 | 0.811 | 0.270 | 2.843 | 0.20 | 0.714 | **0.454** | **0.207** |

For $\pi$ = 0.290, we have $r_c = f(\pi)$ = 0.138 = 13.8 %

For $\pi$ = 0.331, we have $r_c = f(\pi)$ = 0.156 = 15.6 %

For $\pi$ = 0.372, we have $r_c = f(\pi)$ = 0.181 = 18.1 %

For $\pi$ = 0.413, we have $r_c = f(\pi)$ = 0.190 = 19.0 %

For $\pi$ = 0.454, we have $r_c = f(\pi)$ = 0.207 = 20.7 %

It is easy to see that $r_c = f(\pi)$ is an increasing function of $\pi$, for a given value of $k$, the primary rate of surplus-value, which here is $k$ = 0.20. Moreover these values of $r_c$, ranging from 13.8 % to 20.7 %, are themselves quite plausible. Obviously these values of $r_c$ would be lower with higher values of $k$ and of $q_c$, the organic composition of capital. In any case, what these calculations and examples show clearly is once more that the main driving force behind the increase in the share of profits as well as in the rate of profit is the increase in the rate of surplus-value resulting from the increase in capitalist consumption, which depends on the level of prices in section II producing consumption goods.

The confirmation of the prominent role played by this parameter justifies again the importance given to a model taking into account capitalist consumption. The conclusion that we can draw from the thorough examination of these models, both with the Marxist and current definition of the rate of profit, tends finally to minimize the role played by this variable as such, because it has been shown first that it was before all a statistical average of multiple rates, and second that even its level as an average was influenced by a number of various parameters. Some of these parameters are of a technical and exogenous nature, like $k$, $t$ and $a$. But it is clearly another one, $c$, as the driving force of the rate of surplus-value $k_c$, which is the pivot of the functioning of the whole economic system. Its determination is at the heart and the outcome of a social conflict for the distribution of consumption goods between workers and capitalists, which goes far beyond the realm of economic theory alone.

This examination of the evolution of the rate of profit in relation to various variables - including the share of profits in the product, is now complete, we can now formulate two new principles.

**Principle 43**: When then the rate of profit defined according to its current definition, (i.e. without applying this rate to wages - as if they were advanced), and in a situation of simple reproduction with capitalist consumption, this rate is an increasing function of $k$, the primary rate of surplus-value, $k_c$, the actual rate of surplus-value with capitalist consumption, and $c$, the share of capitalist consumption in total consumption. On the contrary it is decreasing in terms of $q_c$, the organic composition of capital. This confirms the validity of the Marxist law of the tendency of the rate of profit to fall when the organic composition of capital increases. But this law remains a mere tendency, since it can be thwarted by an increase in capitalist consumption, which shows the paramount importance of this variable.

**Principle 44**: The share of profits in the product (named $\pi$) is also an increasing function of the primary rate of surplus-value and of the share of capitalist consumption, and thus necessarily of $k_c$, the actual rate of surplus-value. The rate of profit itself is an increasing function of $\pi$.

However, it must be remembered that the models that we used are valid only for the analysis of simple reproduction. We will therefore look in the following and last chapter of this book at some of the elements that make it possible to analyze extended reproduction. The complexity of the phenomenon, however, will lead us not to engage in the construction of a full-fledged new model and rather to restrain ourselves to a numerical example and some preliminary considerations.

# Chapter 22. Some reflections and first elements of discussion to introduce extended reproduction

Extended reproduction is typically a Marxist concept, which is dealt with in volume II (or book II) of Capital, published by Engels in 1885, after Marx's death. In this chapter we will recall briefly in a first section how this concept is introduced by Marx, before trying to clarify some questions of definitions. Then, since developing a complete model of extended reproduction would be quite a complicated task, we will confine ourselves to the presentation of two numerical examples, the first one without capitalist consumption and the second one with capitalist consumption. Due to this Marxist point of departure, we will use in our examples the Marxist definition of the rate of profit, with variable capital – i.e. the value of labor-power, at the denominator of the expression defining this variable.

## 1. Extended reproduction in Marx's Capital

Marx touches upon the matter on three occasions in volume II of Capital. First in part II "Accumulation and reproduction on an extended scale" of chapter 2, where he indicates: "Accumulation, or production on an extended scale, which appears as a means for constantly more expanded production of surplus-value - hence for the enrichment of the capitalist, as his personal aim - and is comprised in the general tendency of capitalist production, becomes later, however, as was shown in Book I, by virtue of its development, a necessity for every individual capitalist" (Marx, 1885, p. 45).

It must be understood that accumulation here means an increase in the stock of constant capital, which for Marx includes both fixed capital and circulating capital.

Marx deals again with the same question in chapter 17 "The Circulation of Surplus-Value", where the second part has the same title as above and deals essentially with the circulation of money associated with the accumulation of additional constant capital (on top of what is needed in the case of simple reproduction). This explains that he starts by writing: "Accumulation takes place in the form of extended reproduction, it is evident that it does not offer any new problem with regard to money-circulation" (Marx, 1885, p. 209).

Finally, it is in chapter 21, the very last of volume II, again under the same heading, that Marx really addresses the matter, through reproduction schemes, and in a way which is however quite complicated, because he continues to mix it with the problem of the origin of the additional money required to finance additional investments in constant capital, on top of what is needed in simple reproduction.

Contrarily to Marx, we do not think that there is any specific problem of realization in a situation of extended reproduction, mostly because we are not reasoning within the framework of his mainly wrong theory of money, which created a lot of problems in his theory. Moreover, as we showed in chapter 11, Marx's reproduction schemes are erroneous, and there is a detailed demonstration on this question in Appendix II.

## 2. Some questions of definitions

The models for the realization of the product that we exposed so far were based on an assumption of simple reproduction. We will now see what happens to these models if we abandon the simple reproduction hypothesis, in order to integrate extended reproduction into them, as an increase in the rate of surplus-value leading to additional investments and an increase in the stock of fixed capital, which can be accompanied (or not) by a growth in capitalist consumption. It should be understood, however, that rejecting the assumption that fixed capital transmits its value to the product's must lead to redefine what Marx names simple reproduction and extended reproduction.

Indeed for Marx, in the case of simple reproduction, there could not be any surplus-value and therefore any accumulation of capital, as long as there would not be any consumption of capitalists, since in such a situation it is the value transmitted by the fixed capital that is already in use, and not an equivalent amount of surplus-value, which replaces the value lost by this same fixed capital. There would be no growth either, since Marx does not seem to imagine that with a fixed capital with unchanged characteristics there can be any increase in the (physical) productivity of labor, for instance thanks to a better use of existing fixed capital, or by a better organization of the production process with this unchanged fixed capital.

However we have already pointed out that, as soon as this hypothesis of transmission of the value of fixed capital is abandoned, there is necessarily a surplus-value, even without any consumption of capitalists, because it is this surplus-value, and not the value supposed to be "lost" by the fixed capital in operation, which – once monetized in the form of profits, is used to acquire the newly produced fixed capital due to replace the portion which leaves the production process at the end of its lifespan.

Similarly for Marx, extended reproduction corresponds to the accumulation of capital by accumulation of surplus-value, beyond the simple replacement of existing fixed capital, and only in this case can there be for him economic growth in the trivial sense of the term, that is to say growth of the quantities that are produced. Expanded reproduction is therefore in fact the growth of the stock of constant capital by accumulation of surplus-value, and this accumulation – excluding capitalists consumption by definition - has for counterpart only an investment which is net of constant capital replacement, and for fixed capital an investment which is net of the replacement of amortized capital.

In the present work, the rejection of the assumption of the transfer of value of fixed capital to the product implies that extended reproduction is defined by the accumulation of additional surplus-value, beyond the surplus-value already present in simple reproduction and corresponding to a mere replacement or maintenance of the value of fixed capital. This means that extended reproduction is therefore defined by an increase in the value of fixed capital produced by section I, which by itself does not necessarily imply a parallel increase in the overall value of consumer goods produced by section II. Indeed, if the increase in the stock of fixed capital is shared between section I and section II, the latter may be able to increase the physical quantities that it produces, without necessarily having to hire additional workers - i.e. without a rise in the value produced in section II, particularly if the newly invested additional fixed capital saves labor time by increasing its physical productivity. Such a phenomenon also ensures that this accumulation of additional fixed capital is accompanied simultaneously by an increase in the organic composition of capital. A similar result might nonetheless be obtained with a faster rate of increase of the value produced in section I compared with that produced in section II.

As already indicated, we will not try to provide a complicated model for this situation of extended reproduction, and will rather present in the next sections two simple numerical examples, which will be enough to shed some light on what is of interest for us, and which is not the realization of the product, but the evolution of the main variables, like the rate of surplus-value and capitalist consumption, and their influence on the profit rate.

## 3. A first numerical example, without capitalist consumption

In order to figure out what can be the evolution of the average rate of profit in a situation of extended reproduction, we will use a numerical example already exposed above in section 3 of chapter 19 (see pp. 365-366) by looking now at the dynamic sequence going from a first situation of simple reproduction (A), to a second one (B) characterized by an organic composition of capital which is double that of the first one, and a rising rate of profit, which goes from 9.09% to 11.1 %.

Let us assume, for example, that we gradually move from an annual fixed capital investment with a value of 100, in this situation A of a simple reproduction regime allowing only for the replacement of fixed capital at the end of its lifespan (supposed to be 7 years), to an annual capital accumulation of 200, and this in five years, the newly accumulated capital during these successive years having a value of 120, 140, 160, 180 and 200. This means that during the first year the value of the means of production produced with previously accumulated capital increases by 20 %, a priori because of the employment of 20 % of additional workers, all other things being equal. During the second period the increase is 16.7 % and so on with decreasing rates until the value of the newly produced means of production reaches the level 200 at the end of the fifth year.

Taking into account the convention adopted for the decommissioning of fixed capital, and always assuming a seven-year lifespan for this capital, it is necessary to wait until an additional period of seven years has elapsed before we reach a new stable situation, where the totality of the accumulated capital corresponds entirely to the new level attained by yearly investments, and to end up in a situation B of simple reproduction, which constitutes the new cruise regime. The transition from the first simple reproduction regime to the second one, marked by an organic composition of capital which has doubled, thus takes place in a twelve year period. The evolution of the rate of profit and of the different variables which determine it appear in table 22.1.

| **Years** | **0** | **1** | **2** | **3** | **4** | **5** | **6** | **7** | **8** | **9** | **10** | **11** | **12** |
|---|---|---|---|---|---|---|---|---|---|---|---|---|---|
| $V = L_{II}$ Value of labor force | 400 | 400 | 400 | 400 | 400 | 400 | 400 | 400 | 400 | 400 | 400 | 400 | 400 |
| $L_I = S$ value of capital accumulated during each period | 100 | 120 | 140 | 160 | 180 | 200 | 200 | 200 | 200 | 200 | 200 | 200 | 200 |
| $k = \frac{S}{V}$ Rate of surplus-value | 25 % | 30 % | 35 % | 40 % | 45 % | 50 % | 50 % | 50 % | 50 % | 50 % | 50 % | 50 % | 50 % |
| K Value of accumulated Fixed capital | 700 | 700 | 720 | 760 | 820 | 900 | 1000 | 1100 | 1200 | 1280 | 1340 | 1380 | 1400 |
| $q = \frac{K}{V}$ Organic composition of capital | 1.75 | 1.75 | 1.8 | 1.9 | 2.05 | 2.25 | 2.5 | 2.75 | 3.0 | 3.2 | 3.35 | 3.45 | 3.5 |
| K + V capital advanced (Marxist definition) | 1100 | 1100 | 1120 | 1160 | 1220 | 1300 | 1400 | 1500 | 1600 | 1680 | 1740 | 1780 | 1400 |
| $r = \frac{S}{K+V}$ Average rate of profit | 9.1 % | 10.9 % | 12.5 % | 13.8 % | 14.8 % | 15.4 % | 14.3 % | 13.3 % | 12.5 % | 11.9 % | 11.5 % | 11.2 % | 11.1 % |

**Table 22.1 - A numerical example tracing the evolution of the rate of profit with extended reproduction**

At first glance one might think that function $r = \frac{q}{(q+1)\cdot t}$ remains valid to describe the evolution of the rate of profit during this twelve year period, since it goes from 9.1% to 11.1%, which confirms that within the framework of the simplified model without capitalist consumption the Marxist law does not hold, because the rise in the organic composition of capital is accompanied by a rise in the rate of profit. But the numerical values provided by this example lead to note also that the rate of profit does not rise from 9.1 % to 11.1 % following a steady progression. On the contrary this rate goes through a maximum of 15.4 % in the fifth year, then goes down, first quickly and then more slowly, to arrive to the value of 11.1% corresponding to the new situation of simple reproduction.

This evolution of $r$ is thus no longer monotonous and strictly increasing and therefore cannot be explained by function $r = \frac{q}{(q+1)\cdot t}$, which as such does not account for the evolution of the rate of profit between the two simple reproduction regimes observed during year 0 and year 12. But this should not be surprising, if we remember that the definition of this function was developed on the basis of the explicit assumption of a very simple linear relation between the quantities of fixed capital $K$ and of surplus-value $L_I$, since it has been stated that $K = \sum_{i=0}^{i=t} K_i = L_I t$. Such a relationship therefore also exists between the rate of surplus value $k$ and the organic composition of capital, since $q = kt$. If $t$ is a constant, as is the case in this example, this relation becomes even a pure and simple identity ($q = 7k$). It can easily be checked for instance for years 0 and 12.

Now in the present numerical example, this relation $q = kt$ is no longer satisfied, and $t$ is no longer a simple scale factor that cannot influence the variation of function $r = f(q)$, since we do not pass instantly from situation A to situation B. In the meantime, and as it is easy to observe by examining the rate of surplus-value and organic composition in years 1 to 11, and even if $t$ remains constant ($t$ = 7 years), this relation is not verified, and it is observed on the contrary that $q \neq kt$. This is so because the value of the accumulated fixed capital cannot move instantly from one level reflecting a stable regime of the rate of surplus-value to another stable regime, but does so over a period which is all the more extensive (12 years in this example), that the average lifespan of fixed capital is long. There is thus an inertia in the value of the stock of capital, which, because of the seven-year lifespan of each generation of accumulated capital, evolves much more slowly than surplus-value and the rate of surplus-value.

By expressing the evolution of the value of the capital stock $K$ according to the value of additional capital $I_i$ accumulated during the preceding periods (and equal in this example to the surplus-value of each period), it is of course theoretically possible to write a mathematical formula expressing this evolution within this framework: $K = f(I_i) = f(k)$.

This formula would take the form of a linear equation of recurrence, and one could introduce it into the function giving the rate of profit, which would become greatly complicated. But within the framework of this book, this type of mathematical refinement does not seem useful. This is the reason why we kept to a simple numerical example, which appears sufficient for the lessons that one wishes to obtain from it.

What other lessons can we draw from this example? First of all, that the process of extended accumulation allows the profit rate to exceed the limit previously highlighted of $r = \frac{1}{t}$, which for a fixed asset life of 7 years is $r = \frac{1}{t} = 14.29$ %. This limit cannot be crossed in a situation of simple reproduction, no matter how great the organic composition of capital. The transition from a simple reproduction situation to another one marked by a higher organic composition is thus accompanied by a fluctuation in the rate of profit during which it exceeds the level at which it will eventually stabilize, a level which is nevertheless higher than the initial one.

A second lesson is that the rate of profit, if not volatile, is at least more subject to fluctuations than the organic composition of capital. However, these fluctuations are much more moderate than the change in the rate of surplus-value, because of the buffer effect played by the stock of fixed capital. The numerical example shows us that a doubling of the rate of surplus-value results only in a limited increase in the rate of profit, which goes from 9.1 % to only 11.1 %, because of the concomitant doubling of the organic composition of capital.

## 4. A second example, with capitalist consumption

In this example, and as can be seen from Table 22.2, the economy is supposed to be in year 0 in a simple reproduction regime, with a capitalist consumption of 60, corresponding to a ratio $c$ of 60/260 or 23.1 % of the value of consumer goods, which is 260. Since the amount of fixed capital accumulated in each period is supposed to have a value of 50, surplus-value is therefore 60 + 50, i.e. 110, and as the value of the labor force is $V_c = L_{II} - C_c = L_{II}(1-c)$ = 260 (1 – 0.231) = 200, the rate of surplus-value is at the outset 110/200 or 55 %. Assuming now a lifespan of 10 years for fixed capital (3 years more than in the previous example), the amount of accumulated capital is therefore $50*10 = 500$. The profit rate is thus 15.7%.

Then the system enters in year 1 and up to year 14 in a process of gradual and simultaneous increase in the consumption of both workers (+ 10 each year) and capitalists (+ 5 each year), on the one hand, and of investment $I_i$ (+ 5 each year), on the other hand. Thus after 14 years, during which the amount of capitalist consumption has thus reached 130 and that of workers consumption 340 – corresponding to the value of the labor force $V_c$, the system stabilizes at this level. Capitalist consumption has thus risen from 23.1% to 130/470, or 27.7 % of the value of consumer goods. Annual investment has increased from 50 to 120. Surplus-value has thus reached 250, and the rate of surplus-value 250/340, or 74 %, instead of 55 % initially.

| Years | 0 | 1 | 2 | 3 | 4 | 5 | 6 | 7 | 8 | 9 | 10 | 11 | 12 | 13 | 14 | … | 23 |
|---|---|---|---|---|---|---|---|---|---|---|---|---|---|---|---|---|---|
| Value of labor force $L_{II\,(1-c)} = Vc$ | 200 | 210 | 220 | 230 | 240 | 250 | 260 | 270 | 280 | 290 | 300 | 310 | 320 | 330 | 340 | … | 340 |
| $L_I = Ii$ Value of capital accumulated in each period | 50 | 55 | 60 | 65 | 70 | 75 | 80 | 85 | 90 | 95 | 100 | 105 | 110 | 115 | 120 | … | 120 |
| Consumption of capitalists | 60 | 65 | 70 | 75 | 80 | 85 | 90 | 95 | 100 | 105 | 110 | 115 | 120 | 125 | 130 | … | 130 |
| $S_c$ Surplus-value | 110 | 120 | 130 | 140 | 150 | 160 | 170 | 180 | 190 | 200 | 210 | 220 | 230 | 240 | 250 | … | 250 |
| $k_c=S_c/V_c$ rate of surplus-value | 55% | 57% | 59% | 61% | 63% | 64% | 65% | 67% | 68% | 69% | 70% | 71% | 72% | 73% | 74% | … | 74% |
| K Value of total fixed capital | 500 | 505 | 515 | 530 | 550 | 575 | 605 | 640 | 680 | 725 | 775 | 825 | 875 | 925 | 975 | … | 1200 |
| $q_c = K/V_c$ Organ. composition Of capital | 2.50 | 2.40 | 2.34 | 2.30 | 2.29 | 2.30 | 2.33 | 2.37 | 2.43 | 2.50 | 2.58 | 2.66 | 2.73 | 2.80 | 2.87 | … | 3.53 |
| $K + V_c$ Capital advanced | 700 | 715 | 735 | 760 | 790 | 825 | 865 | 910 | 960 | 1015 | 1075 | 1135 | 1195 | 1255 | 1315 | … | 1540 |
| $r_c = S_c/K+V_c$ Average rate of profit | 15.7% | 16.8% | 17.7% | 18.4% | 19.0% | 19.4% | 19.7% | 19.8% | 19.8% | 19.7% | 19.5% | 19.4% | 19.2% | 19.1% | 19.0% | … | 16.2% |

**Table 22.2 - A numerical example tracing the evolution of the rate profit with extended reproduction and capitalist consumption**

These values, achieved in year 14, are maintained for 10 years, i.e. the average lifespan $t$ of fixed capital, until year 23, that is to say until the value of total accumulated fixed capital corresponds to $I_i t$. The rate of profit, after having gone through a maximum of 19.8 % in years 7 and 8, then stabilizes at 16.2 %, or 0.5% more than its initial level.

This new example is instructive, as it shows firstly that a significant increase in the rate of surplus-value, which goes from 55% to 74%, if indeed it translates in the long-term into an equally significant rise in the organic composition of capital $q_c$, which goes from 2.50 to 3.53, does not translate as one would expect at first, on the basis of the model developed in a situation of simple reproduction with no capitalist consumption, into a fall in the average rate of profit.

This takes place even though, as shown in the table on the previous page, we find ourselves in a situation where the hypothesis according to which $t$ = 10 has been maintained, from which it follows that $c > \frac{1}{1+t}$, as well in year 0, where c = 60/320 = 0.188 as in year 23, where $c$ = 130/600 = 0.277. If we were in a situation of simple reproduction, then function $r_c = f\left(q_c\right)$ should be decreasing, as we showed in sub-section 3.1. of chapter 20, and one would expect to record a decrease in the rate of profit. On the contrary, there is a slight increase in the average profit rate at the end of the period, from 15.7% to 16.2%. This can be explained for reasons related to the nature of function $r_c = f\left(q_c\right)$.

The first reason is very simple, and comes from the fact that in table 2 in reference coefficient $c$ is not a constant, which implies that parameter γ in function $r_c = f\left(q_c\right)$ is not a constant either. From then on we have thus different functions giving the rate of profit, one for each value of $c$ and therefore of $\gamma$. This can be verified for example for year 0 and year 23.

In year 0 we have: $\gamma = \frac{c}{1-c} = \frac{0,231}{1-0,231} = 0,30$, which gives us:

$$r_c = \frac{q_c + \gamma t}{(q_c+1)t} = \frac{q_c + 0,30.10}{(q_c+1).10} = \frac{q_c + 3,0}{10q_c + 10,0}$$

Whereas at the end of the period we have: $\gamma = \frac{c}{1-c} = \frac{0,277}{1-0,277} = 0,383$, which gives us:

$$r_c = \frac{q_c + \gamma t}{(q_c+1)t} = \frac{q_c + 0,383.10}{(q_c+1).10} = \frac{q_c + 3,83}{10q_c + 10}$$

From this it follows that in fact the values of $c$ and γ change each year between years 1 to 14, and the corresponding function with this γ coefficient changes therefore similarly. It can thus be seen that $r_c = f\left(q_c\right)$ is not linked by a function of a single variable to the organic composition of capital, but also depends on the evolution of parameter γ, which itself expresses the variations in capitalist consumption relative to that of workers. As an example, we have drawn the two curves corresponding to years 0 and 23, which appear in figure 22.1.

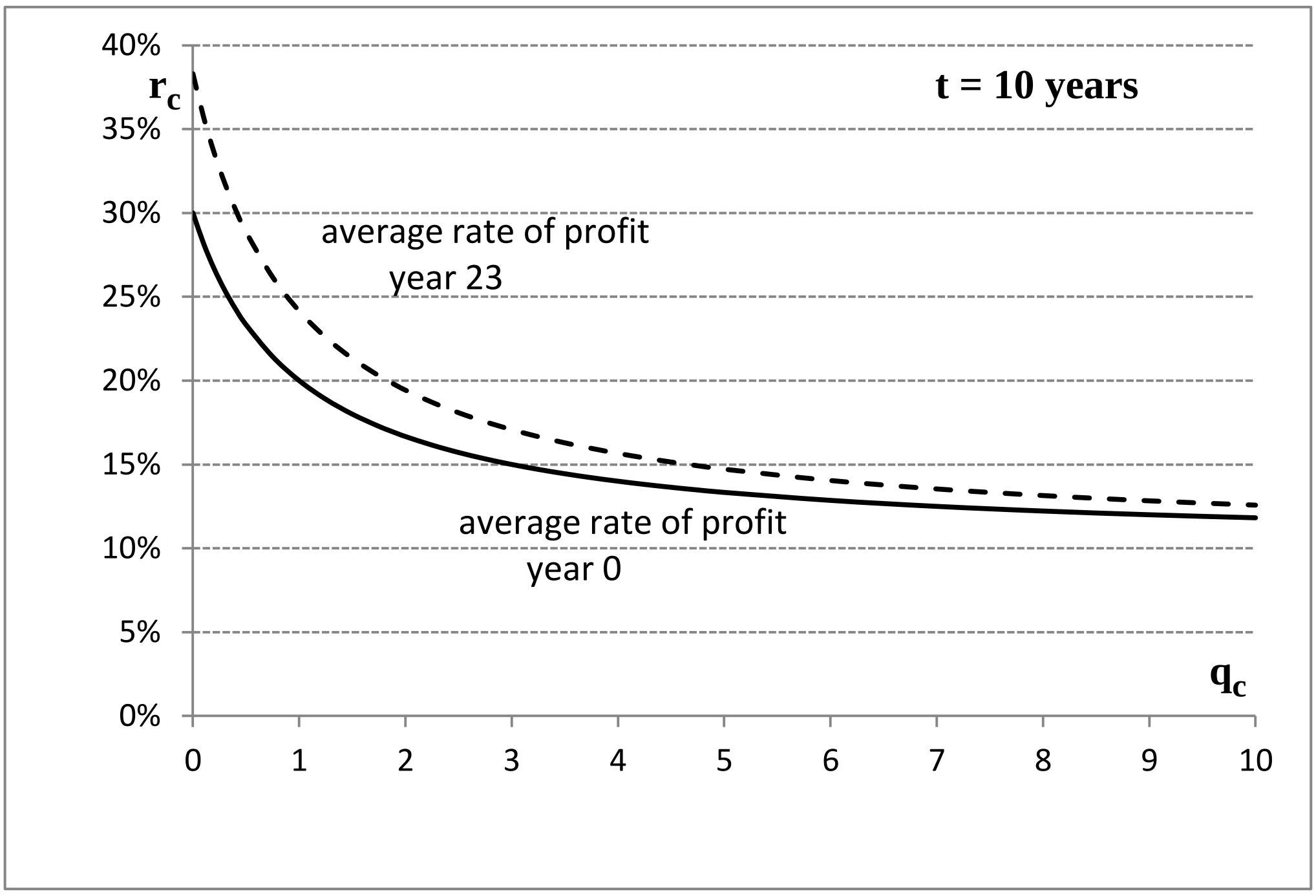

**Figure 22.1 - Curves representing function $r_c = \dfrac{q_c + \gamma t}{(q_c + 1)t}$ in year 0 and year 23**

Moreover, as in the more simple case discussed in section 3 above, where the consumption of capitalists did not appear, the function linking the rate of profit and the organic composition of capital has been defined on the basis of a relationship between the rate of surplus value, which is now $k_c$, and the organic composition of capital, which is now $q_c$. It is relatively simple, as given by equation (13) in chapter 20, that is $q_c = (k_c - \gamma)t$.

However, a second reason explaining that function $r_c = \dfrac{q_c + \gamma t}{(q_c + 1)t}$ cannot any more explain the evolution of the rate of profit from one year to another is similarly that $t$ is no longer a simple scale factor that does not affect the variation of the function, since one does not instantly switch from the situation of year 0 to the situation of year 23. The above relation between $q_c$ and $k_c$ is thus verified in year 0 and in year 23, but in the interval, and even if $t$ remains constant, which is the case in our example where $t$ remains permanently equal to 10, this relation is not verified, and $q_c \neq (k_c - \gamma)t$. This is so for the same reason as before: in a situation of expanded reproduction, the stock of capital at a given instant necessarily depends on all the previous evolution of the accumulation during the $t-1$ preceding years.

The form of function $r_c = f(q_c)$ shows us nevertheless that it is an increasing function of variable γ, which is itself an increasing function of variable $c$. The relative increase in capitalist consumption goes therefore in the same direction as an increase in the rate of profit, which corresponds to intuition. In this numerical example, this increase in capitalist consumption is enough to counterbalance the decreasing effect on the rate of profit of the significant increase in the organic composition of capital. We can also verify from this numerical example that if at the end of the period parameter $c$ remained at its initial level during year 0, i.e. $c$ = 0.231 corresponding to γ = 0.30, but with the same increase in the organic composition, from 2.50 to 3.83, the rate of profit would be 14.1 %, instead of 16.2 %, and would therefore be significantly reduced, also from its initial value of 15.7%. The growth of $c$ thus has the effect of moving $r_c$ curves up and to the right, and this is what explains ultimately the rise in the rate of profit, in spite of the increase in the organic composition of capital.

This shows one more, if it was necessary, the paramount importance of the consumption of capitalists as a driving variable of the economic system.

Incidentally, it is also interesting to note that the rise in the profit rate during the whole 23 year period, even though it is quite limited: from 15.7% to 16.2%, is accompanied by a drop in what we might call the efficiency of capital, rather than the productivity of capital, i.e. ratio $Y / K$ . In year 0 we have indeed $Y_0 / K_0$ = 310/500 = 0.62, while at the end of the period we have $Y_{23} / K_{23}$= 590/1200 = 0.49, which represents a 21% decrease in the ratio between the product and the stock of fixed capital. This divergent evolution shows once more that the rate of profit has nothing to do with any productivity of capital, since, as this example shows, the evolution of the efficiency of capital can very well go in the opposite direction from that of the rate of profit.

## 4. The influence of the lifespan of fixed capital and k$_c$

Before concluding this chapter, another question may be worth analyzing, which is the influence that changes in the lifespan of fixed capital may have on the evolution of the rate of profit. In order to do that, we need to go back to equation (22) in chapter 20, namely:

$$r_c = \frac{k_c}{(k_c - \gamma)t + 1}$$

But now we must interest ourselves to the nature of this function as $r_c = f\left(t\right)$. Its derivative is: $\frac{dr}{dt} = \frac{-k_c\left(k_c - \gamma\right)}{\left[\left(k_c - \gamma\right)t + 1\right]^2}$ , which has the same sign as its numerator.

Since we know that by definition $k_c = \frac{k + c}{1 - c} = \frac{k}{1 - c} + \frac{c}{1 - c} = \frac{k}{1 - c} + \gamma$ , with $k > 0$ and $0 < c < 1$, it follows that we always have $k_c > \gamma$ , which implies that $\forall k_c, \forall \gamma$ , $k_c - \gamma > 0$ .

Therefore the sign of the derivative $\frac{dr}{dt}$ is always negative, which shows clearly that the rate of profit $r_C = f(t)$ is necessarily a decreasing function of the lifespan of fixed capital, any increase in its duration leading to a fall in the rate of profit, because of the rise that it implies in the organic composition of capital $q_C$, obviously for a given value of $k_C$ and $\gamma$.

As for the effect on the rate of profit of a simultaneous increase in the rate of surplus-value, $k_C$, and the lifespan of fixed capital, assuming now that $r_C = f(k_C, t)$, it is given by the following formula:

$$dr_C = \frac{\delta r_C}{\delta k_C} \cdot dk_C + \frac{\delta r_C}{\delta t} \cdot dt \tag{1}$$

As a result, assuming that γ = constant, we get:

$$dr_c = \frac{1}{\left[(k_c - \gamma)t + 1\right]^2} \cdot dk_c - \frac{k_c\left(k_c - \gamma\right)}{\left[(k_c - \gamma)t + 1\right]^2} \cdot dt \tag{2}$$

From this last equation we can infer that to have a positive evolution of the profit rate, such that $dr_c > 0$ implies that:

$$\frac{1}{\left[(k_c - \gamma)t + 1\right]^2} \cdot dk_c > \frac{-k_c\left(k_c - \gamma\right)}{\left[(k_c - \gamma)t + 1\right]^2} \cdot dt \Rightarrow \frac{dk_C}{k_C} > -\left(k_C - \gamma\right)dt \tag{3}$$

or $\frac{dk_C}{k_C} > \left(\gamma - k_C\right)dt$ (4)

From sub-section 3.3. in chapter 20, we know that with the Marxist definition of the profit rate, this rate is a decreasing function of the rate of surplus-value $k_C$ if $c > \frac{1}{1+t}$. Assuming that we are in such a situation (with for instance $t = 10$ and $c > 0.091$), then a decrease in $k_C$, which must cause a rise in the rate of profit, can be compensated by a reduction in the lifespan of fixed capital, so that we have:

$$\frac{dk_C}{k_C} = \left(\gamma - k_C\right)dt \Rightarrow dt = \frac{dkc}{kc\left(\gamma - k_C\right)} \tag{5}$$

For instance an increase in $k_C$ of 10 % (from 0.50 to 0.55), coming from an increase in $k$ (the primary rate of surplus-value), since $c$ and $\gamma$ remain constant, would normally imply a reduction in the rate of profit. But this reduction can be compensated by a reduction of the lifespan of fixed capital, given by this last formula, resulting in an offsetting increase in the rate of profit,

Indeed taking $c = 0.20$, $k_c = 0.50$ and $t = 10$, we have $r_c = \frac{0.5}{(0.5-0.25)10+1} = 0.1429$

Whereas with $\frac{dk_c}{k_c} = 10\% = 0.10 \Rightarrow k_c = 0.55$, we have :

$$\frac{dk_c}{k_c} = 0.10 = (\gamma - k_c)\,dt$$

$$\Rightarrow dt = \frac{0.10}{\gamma - k_c} = \frac{0.10}{0.25-0.50} = \frac{0.10}{-0.25} = -0.40$$

This means that a 10 % increase in $k_c$ can be compensated by a reduction in $t$ from 10 years to 9.6 years. If it is the case, we have indeed:

$$r_c = \frac{k_c}{(k_c - \gamma)t+1} = \frac{0.55}{(0.55-0.25)9.6+1} = \frac{0.55}{2.88+1} = 0.1418$$

The slight difference in both values of $r_c$ comes from the fact that $\frac{dk_c}{k_c}$ is not an infinitesimal amount.

# 5. The realization of the product in the context of extended reproduction with capitalist consumption

## 5.1. Marx's extended reproduction schemes in volume II of Capital

In part III of chapter 21 of volume II of Capital, Marx presents a number of reproduction schemes, that are all of the same form as what he calls an "Initial Scheme for Reproduction on an Extended Scale", and which is as follows:

I. 4,000c + 1,000v + 1,000s = 6,000

II. 1,500c + 750v + 750s = 3,000

Total or annual social product = 9,000

Then he explains that after one year of capital accumulation we can have a second scheme, where capital accumulation has taken place, and such as:

I. 4,400c + 1,100v + 1,100s = 6,600

II. 1,600c + 800v + 800v = 3,200

Total or annual social product = 9,800

What Marx intends to do is to show that the realization (sale) of what he calls the annual social product is possible even in a situation of extended reproduction, when this social product grows from one period to the other due to the accumulation of additional capital (for an amount of 500 in his example). To some extent he succeeds in his endeavor, by bringing into the picture an appropriate amount of money supply, coming from exports, gold production, or previous hoarding, not to mention credit money. Nevertheless all his schemes are logically flawed, for several reasons, which we will mention briefly:

1) The schemes make no distinction between values and prices

With the reproduction schemes we are at the level of circulation, or realization (sale) of the commodities which have been produced. This is the condition for the economic system to be able to reproduce itself. We are thus in a sphere where the corresponding exchanges (purchases/sales) take place through a transfer of money against commodities, for amounts which are monetary prices. Marx is well aware of that, since as soon as the first chapter of volume 2 and in several other chapters he gives numerous examples in terms of £ (British pounds). But all his reasoning is made in terms of values, as if exchanges took place directly at levels corresponding to the values of commodities. This should not be a surprise, because at this stage Marx has not yet elucidated the question of transformation of values into prices, a question that he addresses only in volume III, published by Engels nine years later. And we know that the methodology that he uses to try and solve this problem is itself a wrong one.

In any case, this confusion between values and prices forbids him to understand that money prices of commodities are always necessarily higher than their money values, which is a condition for surplus-value to be extracted under the form of profits in section II (even with no consumption of capitalists) and for surplus-value to be redistributed among capitalists in section I.

2) Constant capital as Marx defines it, i.e., the total value of raw materials and fixed means of production used up in an accounting period, is a flawed concept

Indeed raw materials are by definition intermediate commodities which are always produced in order to be used in the production process, where they are transformed to be incorporated into other intermediate goods, and ultimately into final goods. As such they never appear on the market for final goods, where they are to be bought, in order to be consumed by workers or capitalists - for consumption goods, or to be invested by capitalists - in the case of fixed capital. Therefore they should not be put in the same category (constant capital) as fixed capital, which is made of final goods.

However this is not what Marx does, because he considers that fixed capital, like intermediate goods, transfers a part of its value to the value of the final product. This is why he puts this part into the same category as raw materials, and treats it as a particular kind of intermediate good.

In any case the values of intermediate goods must be calculated first, but within Marx's conceptual framework constant capital produced by section I would include both intermediate goods which enter into the production of other intermediate goods, on the one hand, and constant capital which enters only into the production and value of final goods, on the other hand.

This would lead to a problem for Marx's schemes, because the annual social product would not be composed only of final goods, as it should be, but also of intermediate goods which do not enter into the production of final goods: i.e. those which enter only in the production of other intermediate goods, and as such do not go to the market. This might be more easily understood if we thought of the process of production as fully integrated, with only one huge firm in each section, producing not only the final goods but also all of the intermediate goods necessary for its final production. Then it would be impossible to wrongly count intermediate goods entering only in the production of other intermediate goods as a part of constant capital, and as a part of the annual social product, which would come down to counting them twice. Only intermediate goods entering in the production of final goods would be accounted for. This leads us to a third and final reason for which the extended reproduction schemes in fact cannot teach us anything about the true nature of the reproduction process.

3) The wrong assumption that fixed capital transmits its value to the product.

Indeed with Marx's assumption above, as we already showed, a part of fixed capital would also wrongly be accounted for twice in his reproduction schemes: first as a final good sold on the market of final products, and second for the amount supposed to transmit its value to the final goods into which this lost value would be "incorporated". In fact once whatever good, including fixed capital, has been sold on the market for final products against an income which is transferred from the buyer to the seller, this good cannot be sold a second time as a final good. If it is sold again, it will necessarily, for logical reasons, be sold as a second-hand good, and as such will not create any new income: what is spent by the buyer is obtained by the seller, which means that from a macroeconomic point of view no additional income is created by these transactions concerning goods previously produced and exchanged.

Because of these various flaws there was no chance that Marx's reproduction schemes could be correct, which leads us to propose now what we deem is a right approach to this question.

## 5.2. Another way to deal with reproduction schemes

At the outset, it should be emphasized that the realization of the product is a precondition for its reproduction. This explains that our schemes for realization are the basis for understanding reproduction problems. In the case of simple reproduction the product and its structure are identical from one period to the other, which means that the value of the product does not change, both globally and in each of the two sections of the productive system, and therefore that $L_I$, $L_{II}$ and by definition $k$, which connects them (since $k = \frac{L_I}{L_{II}}$) are constant.

If we want to compare our schemes with Marx's schemes, since they involve circulating capital, as a part of constant capital, the first thing to do is to define correctly the values of both intermediate commodities and final commodities, which has been done already in chapter 6 of the present study. Indeed we showed in this chapter that values could be calculated with a first method as given by the following simplified system of equations:

$V_1 = V_1 A_{11} + L_1$ , for intermediate commodities (6)

$V_2 = V_1 A_{12} + L_2$ , for final commodities (7)

This system could be solved as:

$$V_1 = L_1 \left( I - A_{11} \right)^{-1} \quad (8)$$

$$V_2 = L_1 \left( I - A_{11} \right)^{-1} A_{12} + L_2 \quad (9)$$

We showed also that values could be calculated with a second method as given by another system of equations:

$$V_1 = L_1 X_{11} \quad (10)$$

$$V_2 = L_1 X_{12} + L_2 \quad (11)$$

Finally, we pointed out that both systems were equivalent, because we had:

$$X_{11} = \left( I - A_{11} \right)^{-1} \quad (12)$$

$$X_{12} = \left( I - A_{11} \right)^{-1} A_{12} \quad (13)$$

So that we could replace $X_{11}$ and $X_{12}$ in equations (10) and (11) by their values given by equations (12) and (13) to obtain equations (8) and (9).

This allowed us to show finally that the value of the product is equivalent to total direct labor time. Indeed, calling $y_1$ and $y_2$ the column vectors of the output of intermediate commodities and final commodities, respectively, from equations (8) and (9) we could write that the value of the total product is equal to:

$$V_1 y_1 + V_2 y_2 = L_1 \left( I - A_{11} \right)^{-1} y_1 + L_1 \left( I - A_{11} \right)^{-1} A_{12} y_2 + L_2 y_2 \quad (14)$$

This, from equations (10) and (11), is equivalent to:

$$V_1 y_1 + V_2 y_2 = L_1 X_{11} y_1 + L_1 X_{12} y_2 + L_2 y_2 \quad (15)$$

This last equation gives us indeed the overall quantity of direct labor time used in the production process, first for the production of intermediate commodities, i.e. $L_1 X_{11} y_1 + L_1 X_{12} y_2$ , and second for the production of final commodities, i.e. $L_2 y_2$ .

Going back now to reproduction schemes, it is clear that to establish them we need to make a change in the system of references, because we are dealing here with final incomes, which are spent only to buy final commodities. At this stage, money wages have been paid to workers, and we are in the theoretical realm of money values. Equation (15) shows that all the money value of the product can be considered as equivalent to the total wages paid to all workers. Moreover, since wages and the transfer incomes deriving from them are spent on the markets for final goods, all the transactions to buy and sell intermediate goods are supposed to have taken place previously, during the production process (otherwise production could not have been carried out up to its final stage).

Moreover, the quantity of direct labor time must be distributed no more between the production of intermediate goods and the production of final goods, but between the production of consumption goods and the production of fixed capital goods, this production being integrated for each of them, with section I producing both its own intermediate goods and fixed capital and section II producing its own intermediate goods and consumption goods.

This implies first that the first two terms of equation (15) corresponding to intermediate goods must be split between these two sections, in such a way that the total labor of the system is spent indirectly and directly in each of these two sections. Thus the overall quantity of direct labor time $L_1X_{11}y_1 + L_1X_{12}y_2 + L_2y_2$ is split between the labor time used indirectly and directly in the production of consumption goods, or $L_{II}$, and the labor time used indirectly and directly in the production of fixed capital, or $L_I$. This implies as a second step that this equation be monetized, which is done by multiplying the quantities of labor by the average money wage per unit of time $\overline{w}$. By taking it also as unit so that $\overline{w}$ = 1, we get:

$L_{II} = \overline{w}\ L_{II} = L_{II}$, equivalent to the money value of consumption goods.

$L_I = \overline{w}\ L_I = L_I$ equivalent to the money value of fixed capital goods.

Reproduction implies also that exchanges take place at monetary prices which must allow first for the creation or extraction of surplus-value $S$. This process, as we have seen, is carried out in section II through the realization of consumption goods and implies that their overall price $P_{II}$ must differ from their overall money value $L_{II}$, and be higher than this value in such a way that this overall price is:

$$Y_{II} = P_{II} = L_{II} + S = L_{II} + L_I + cL_{II}\left(1+k_c\right) = L_{II}\left(1+k_c\right) \qquad (16)$$

This equation shows that this price is the sum of wages in section II ($L_{II} = W_{II}$) plus surplus-value $S$. The realization (or sale) of the product is thus carried out at a price which is the sum of wages in both sections ($L_I + L_{II} = W_I + W_{II}$) spent on consumption goods, plus the consumption of capitalists, as a share $c$ of total consumption, measured by the total price of these goods. This share is the same if measured in terms of values, since in this section prices are globally proportional to values.

From the two last members of this equation, we can check that the rate of surplus-value $k_C$ has exactly its expected value. Indeed since $L_I = kL_{II}$, we can rewrite equation (16) as follows:

$$L_{II} + kL_{II} + cL_{II} + ck_C L_{II} = L_{II} + k_C L_{II} \Rightarrow (k+c)L_{II} = k_C L_{II} - ck_C L_{II} \Rightarrow k_C = \frac{k+c}{1-c}$$

As for the overall price of fixed capital goods $Y_I$, we know that it is :

$$P_I = Y_I = \frac{L_I}{(1-a)(1-c^*)} = L_I\left(1+\alpha+\gamma^*+\alpha\gamma^*\right) \qquad (17)$$

It is immediate that the price of the product of section I is the sum of:

Wages in section I, i.e. $L_I$, and profits in section I, i.e. $L_I\left(\alpha+\gamma^*+\alpha\gamma^*\right)$

Here too we can see that the realization of the product of section I is possible, and therefore accordingly the reproduction of the whole economic system. The fact that reproduction is extended does not create any other additional problem, as opposed to simple reproduction, than the necessary adjustment of prices to reflect the new and supposedly higher amounts of surplus-value, investments or capital accumulation, and wages. But contrarily to Marx's schemes, these schemes above show that a number of parameters, like $k$, $c$, $a$ and $c^*$ (and thus $\alpha$, $\gamma$ and $\gamma^*$), have to adjust simultaneously to the new amounts of surplus-value, investments or capital accumulation and wages, which in the real world is not self-evident.

### 5.3. Reproduction cannot go smoothly in the real world

In the context of this analysis, using a model which itself relies on only the most basic principles of the functioning of the capitalist system, it seemed preferable not to embark on a necessarily complex theoretical analysis of all the parameters concerned by the evolution of this system. But no more than the introduction of the consumption of capitalists, there is no reason a priori that extending this analysis to extended reproduction should call into question the theoretical possibility of the realization of the product.

As long as wages paid to workers are advanced as monetary wages and money is a credit money and an endogenous variable, the system embeds mechanisms that theoretically allow for the product to be realized and for reproduction to take place in each of the two sections of production, provided that prices are set at appropriate levels. To increase the product and create additional value, it is easy for capitalists to borrow additional money in order to finance additional investments and pay additional workers. But for this to happen it would be indispensable for capitalists to have the right expectations, and that these expectations be exactly compatible with each other's.

That prices may be established precisely at the appropriate levels in the reality of the functioning of markets, allowing reproduction to take place smoothly at any time, is indeed quite a serious problem, because it has already been shown that prices reflect distribution, among other things. And what is at stake is not only distribution between workers and capitalists, where the shares of the product going to each category are achieved through the level of wages and the pricing of consumer goods, but also distribution between capitalists themselves, because the distribution of surplus-value among them is achieved through fixing the prices of the various means of production constituting fixed capital. This fixation depends for each particular market on the power relations which specifically characterize their functioning, something that the theory of contracts tries to explain to a certain extent. In such a context, that neither simple reproduction nor extended reproduction can be considered as taken for granted can be understood with two simple examples.

As a first example, although simple reproduction implies an unchanged level of parameter $k$, it is not incompatible with an increase in the rate of surplus-value $k_c$, which means that capitalists of section II increase the price of consumption goods $P_{II}$. This should normally raise their profits and parameter $c$, i.e. their share of consumption goods. It is possible however that capitalists do not increase their consumption, but instead decide to use the additional expected profits - deriving from the rise in prices in section II, to buy some already existing assets. These assets may be art works, or real estate, or shares, but in any case buying them instead of consumption goods will cause the realized sale of consumption goods to be lower than their production, which would prevent the expected rise in profits from materializing.

As a second example with extended reproduction, let us imagine that capitalists in section I decide to invest more, which means that they must produce and accumulate more fixed capital and hire additional workers ($\Delta L_I$) to meet this objective. Then $L_I$ and $k$ are going to increase, and therefore the rate of surplus-value $k_c$ should also increase to $k_c'$, as a result of this increase in $k$. If however capitalists in section II do not increase the price of consumption goods, i.e. $P_{II} = L_{II}\left(1+k_c\right)$, the rise in $k_c$ will not happen, and there will be a shortage of consumption goods which in an open economy would be offset either by additional imports or by inflation.

In a capitalist system where investment and pricing decisions are decentralized there is indeed no reason to think that at any time a coordination is established between capitalists of section I and those of section II. Such a price-setting logic therefore has no particular reason to lead to equilibrium.

All the more so that the main problem which can hamper extended reproduction remains the inherent tendency of the capitalist system to reduce the amount of wages. As we already pointed out it is indeed a normal behavior of capitalists at the level of the firm to consider wages as a cost of production which should always be minimized in order to increase profits. This is based on the standard accounting practice which considers wages as an accounting charge. Business managers seek to achieve this objective by reducing their workforce, i.e. $L_I$ and $L_{II}$ and/or the amount of wages per unit of labor time.

With unchanged prices of consumption goods this microeconomic actions create a permanent trend towards an increase in $k_c$, the rate of surplus-value, but at the same time towards a reduction in the demand for these goods from workers. This reduction could hypothetically be offset by an increase in $c$, the share of capitalist consumption, if the consumption patterns of workers and capitalist were the same, which however is unlikely to be the case. The end result of such a process has every chance to be a chronic shortage of demand for consumption goods, which cannot but affect the level of investment in both sections.

This difficulty of the productive system to reach an equilibrium can also be highlighted on other grounds. First at the level of production relations and the realization of the product, which has been analyzed so far, because the necessary differentiation of profit rates, which it has been possible to demonstrate, can only become more complex in a situation of extended reproduction. Indeed extended reproduction implies a priori a growth of the different branches of the productive system at a specific and distinct rhythm for each of them. This corresponds to investments needs which are quite different from one branch to another: higher in the new and fast developing branches and lower in the old and declining ones. In turn this leads to a need for higher profits and profit rates in the fast growing branches, which is a powerful factor for the differentiation of these rates.

The fact that competition is supposed to lead to the equalization of profit rates, to the extent that it could be demonstrated, would in any case only play against this differentiation, thus making the reproduction of the system more difficult. Indeed it has been shown that the rates of profit must be different, both between section I and section II, as well as within section I, so that the realization of the product and thus the reproduction of the system can take place without any disturbance.

Consequently, if competition really pushed for the equalization of profit rates, then this would simultaneously create the conditions for some misalignments, i.e. the non-realization of the product in at least some branches. All this can only oppose the harmonious reproduction of the system! More generally, there can be no guarantee that profits accrued in various sectors will be reinvested so as to allow for this harmonious reproduction of the system, and in particular its extended reproduction. All this implies that at each production cycle the architecture of the productive system is modified, with a variation of the relative weight of the different branches which compose it, reflecting the variation of the growth rate of each one of them. It is this kind of phenomenon that we designated as constituting a factor of structural instability within the system. In other words, the realization of the entire product is theoretically always possible, but in practice quite unlikely.

It must also be pointed out that this source of instability can only underlie another one, which lies at a second level and comes out of another problem and another type of analysis, because it concerns the specific monetary nature of the capitalist economy. This other source of instability comes from the nature of money and the conditions of its creation, circulation, and destruction. It involves monetary and financial variables, including debt and interest rates. This is however not the object of this work, being understood that the author, as indicated previously, adheres to the analyzes which in the extension of Keynes' theory of money were developed inter alia by Minsky, Schmitt, Keen, Godley, Lavoie and others, and are constitutive of what is often labelled as modern monetary theory (MMT) or circuit theory.

To conclude this chapter by coming back to the evolution of the rate of profit, the numerical examples given in this chapter have shown that taking into account extended accumulation and reproduction does not call into question, in the generality of cases (for the most probable values of the parameters) the fact that an increase in the organic composition of capital is in itself associated to a decline in this rate, which validates Marx's analyzes on this point, however with the important nuance that this trend can be countered by an increase in capitalist consumption, as a share of global consumption. This allows us to formulate a last principle.

**Principle 45**: In a situation of extended reproduction and accumulation of capital, the Marxist law of the tendency of the rate of profit to fall when the organic composition of capital increases remains valid, again as a mere tendency. Such a fall can also be caused by an increase in the rate of surplus-value, as well as by an increase in the average lifespan of fixed capital. However, this evolution 1) is not monotonous over time, because the rate of profit initially tends to rise in response to an increase in accumulation and 2) is always countered by an increase in capitalist consumption.

# Chapter 23 - Conclusion

In this concluding chapter all the lessons learned during the course of this book will allow us to come back first to the object of economic theory (section 1). We will then expose the final view that we can have on the concepts of production, value and money (section 2), and those of distribution, exchange and prices (section 3). All this will allows us to elaborate on the incompleteness of economic theory (section 4), before a few final words (section 5).

However, having reached the end of this work, it may be useful to start its conclusion by pointing out its limitations. Indeed our objective was not to produce a full-fledged economic theory. More simply, since a satisfying, encompassing and broadly accepted economic theory still does not exist at the beginning of the 21st century, our aim was to ask again, two and a half centuries later, the basic questions that were at the heart of the approach of Adam Smith, the founding father of what we would like to be a truly scientific economic theory: "what are the nature and causes of the wealth of nations?" And if this wealth is made of commodities, where do they come from? This immediately brings us to the question of the nature of commodities, and of their production and distribution.

The object of this book was thus to try and logically define these most basic concepts of commodities, production and distribution. Since production and distribution occur in the context of a social and political system, which establishes the rights of its members on commodities, in particular through the social institution of property, we had to take as a basic assumption that the analysis was set in the broader context of the capitalist mode of production, as it is by far the most dominant system on earth. And since in the real world distribution is carried out through exchanges between commodities and money, which take place on the basis of prices, we had also to be very clear about what money is and what it is not, and what is a monetary price. Finally, since prices are usually defined through the concept of value, it was similarly indispensable to go back to this concept.

Without any false modesty, we think that we have succeeded, since all of the main concepts and the basic principles that underlie them have been defined on the way, in a coherent and logical manner, which should give the possibility for other fellow economists to go further with a view to rebuilding little by little the entire economic theory. But from the outset we need to underline that these principles were not put together from scratch. On the contrary they were defined by going back to the old classical economists, who had already asked the right questions, even though they had not been able to answer all of them. We are thus greatly indebted to Smith and Ricardo, as well as to Marx and Keynes.

However we did not answer all questions which are raised by the economic functioning of the real world. Indeed the theoretical bricks that we have laid under the form of basic principles do not constitute a theory of money, which is understandable, because in our view - and in the wake of Keynes, modern monetary theory (MMT) already exists and is precisely a scientific theory of money, which only lacks to be theoretically linked to a theory of value and prices. As such MMT plays the role of a kind of background for our own theory. The views developed here and summarized in our principles do not constitute either a theory of the firm, of investment, growth and development, or a theory of crises, etc. But they open some tracks which should help to move forward in all of these other fields of economic theory, and we are hopeful that this will happen, if indeed we have properly defined the right concepts as regards production and values, distribution and prices, reproduction and profits.

For instance, since our principles show that prices are closely connected to distribution, it should be possible to combine monetary theory and our theory of value and prices to devise a coherent theory of inflation. Finally, using some of our concepts should allow to devise some modified national accounts, with the aim of - among others, providing a better empirical understanding of the functioning of modern economies, as well as of the building up of inequalities which has plagued them over the last 40 years.

Such a task would highlight two additional limitations: first the fact that even though there is money in our analysis, there are no banks as the institutions that create money. Indeed we have voluntarily restricted this approach to two categories of economic agents: workers and capitalists. This allowed us not to unduly complicate the circulation schemes, and to focus on the two main actors who drive a capitalist economic system: workers and capitalists. Second, the State is also missing, for similar reasons, despite the fact that it is the regulating authority of the system and an important actor, which connects to the circuit of incomes to both collect and inject resources into it.

These limitations being acknowledged, this conclusion will aim at synthesizing and summarizing our main findings, under the headings of production, value and money, distribution, exchange and prices. We will do that while briefly recalling how they relate to the economic literature. Moreover, claiming as we do that our principles have a scientific status implies that we start first with a quick reminder on epistemology, as regards the object of a scientific theory.

## 1. The object of economic theory

Since science aims at producing "true" descriptions of the world, its object is supposed to be the real world or rather - since scientific knowledge covers a number of different areas of the world, one of these areas, on which science is supposed to product such a description, taking the form of a theory.

Such areas, in the case of economic theory as for all other theories, have to be built by the theories themselves. This means that a theory must begin by delimiting its perimeter, i.e. the part or aspect of the real world that it intends to represent, and which distinguishes it from other theories. This presupposes that this corresponding part of the real world exists within it, regardless of the theory, and therefore has a material existence, made of material objects, accessible by the sensible, empirical experience that we can have of this real world, i.e. by a first level of knowledge acquired through the senses, especially through observation and experimentation. Building the concepts which define its object, i.e. the basic elements of a theory and the essential interrelations between them, constitutes what we have called the basic principles of a theory, which are nothing else than an appropriate set of propositions.

It seems obvious that such an object of a theory must exist in the real world, and we think that it is clearly the case as regards the theoretical approach which we have followed in this research, because what we wanted to describe is what in the real world has to do with commodities and the way in which they are produced and distributed among the members of a human society, and more precisely a capitalist economy. Indeed the material existence of commodities, in the broad sense of the word, encompassing goods and also services, like transport, teaching etc., can be ascertained and observed by all of us in the real world. We know also by empirical knowledge that these commodities do not fall from the sky and are produced within the economic system. In contrast, the most famous neo-classical definition of Economics, given by Lionel Robbins in his well-known book "An Essay on the Nature and Significance of Economic Science", is that "Economics is the science which studies human behavior as a relationship between ends and scarce means which have alternative uses." (Robbins, 1932, p. 16).

From the start, we can suspect that this definition is made to be quite problematic. First because its main object is not made of material elements but of a particular human behavior, which is an activity corresponding in fact to a methodology. To be sure, this activity can have a material component, to the extent that it concerns indeed "scarce means". However, a serious problem arises here, because one would normally expect that the theory itself should normally tell us what are these means and where they come from, in other words how they have been created, and if they do not appear simply by teleportation, where and how they were produced.

At this stage, however, we are simply told that at the beginning of the neo-classical theoretical story, all individuals have previously received (from whom and from where?) an endowment of various quantities of all these (scarce) means, or goods. Then these people go to a market where they exchange their goods among themselves in order to allow for the program exposed in the above definition to be carried out. This will materialize, once an equilibrium of supplies and demands is reached on the market for each good, in the form of exchange ratios between all of these goods, which will define their value or price (both words being viewed as equivalent), and allow finally for exchanges to take place, at prices or values which are necessarily relative prices or values.

This neoclassical view suffers from several problems. First, there is no production as such in this description. Nobody knows where the goods endowed to each individual come from. Second, exchanges take the form of barter trade, although it does not exist in the real world, or only in small and residual markets. Third, there must be an auctioneer, to register all proposals of supply and demand, and make the market evolve towards equilibrium, through the announcement of successive systems of exchange ratios, or prices, but there is no auctioneer on most markets in the real world. Fourth, to avoid having a myriad of exchange ratios announced at each stage of the process (n! for n commodities), there must be a general equivalent, taking the form of money, which cannot be other than a particular commodity chosen as such, allowing to have only n - 1 exchange ratios if there are n goods. Again in the real world there is no such thing as a commodity-money.

Finally and to go one step further in the working of the mechanism supposed to lead to equilibrium, the demands and supplies must be formulated on the basis of each participant's own utility function, taking into account the particular utility which each of them attaches to each quantity of each commodity. Although it is not mentioned in Robbins's definition, utility is indeed an indispensable ingredient to define the functions of supply and demand.

The problem is that the utility attached to a particular commodity is a subjective feeling which varies from one individual to the other and from one commodity to the other. Therefore, and despite huge quantities of economic literature on the question, nobody has been able to prove that utility is a dimension of commodities in the scientific sense of the word, i.e. a measurable characteristic of commodities. Therefore utility functions cannot be written without postulating that they are made homogeneous by an ad hoc and arbitrary common unit of measurement. This has been called the postulate of homogeneity, or postulate of the existence of a numeraire. Needless to say that such a numeraire as a measurement unit for utilities does not exist in the real world, and it should in any case be conceptually built.

Indeed, even without going so far as to say that there is science only if there is measurement, it is clear that all sciences which deal with quantities, and consequently with numbers, always need to have a measurement of these quantities, and it is obviously the case in Economics. The problem is that this question is almost never addressed, or is left implicit, by the vast majority of economists, which gives the impression that they are always dealing with pure numbers, or scalars (for instance units of money). This leaves obviously unanswered the question of the dimension of the objects that are measured, as a property which needs to be measurable, all the more so that the question is most often not even asked. Such a dimension is characterized by the nature of the unit of measurement, or the way it is defined. Physical science has defined all the fundamental dimensions that material objects can take in the real world, and there are only seven of them. All the other dimensions are necessarily derived dimensions, i.e. a combination of these seven fundamental ones.

Going back to neo-classical theory, when neo-classical economists are confronted to the question of the measurability of their objects, i.e. commodities whose common property is their utility, since they cannot link this property to any known dimension of the real world, they simply avoid the problem by postulating a unit of measurement that is not defined: it should normally be a unit of utility, if only utility was a measurable property! The end result of all this is that neo-classical theory is unable to build its object in a coherent way, and therefore has no object, in the scientific sense of the term. To say it otherwise, Robbins's definition, although it seems prima facie to be based on common sense, is ultimately devoid of any scientific significance, it is a nonsense on which no scientific theory can be built.

This failure of a theory which tries to build its object starting directly from exchanges between commodities, without explaining in the first place how these commodities were produced, validates the approaches of economic theories which want to start building their object from the production of commodities, or what brings them into existence. This is the case with classical economists, Smith and Ricardo, as well as Marx, their most famous heir, and with the approach that we have followed. However, if one wants to build the object of economic theory on the concept of production, one has to be extremely careful in defining this concept, which these economists unfortunately failed to do in a totally coherent way.

## 2. Production, value and money

Unlike the Physiocrats, for whom land was the essential source of any production, Smith, Ricardo and Marx rightly point out that production depends on the productive power of labor, but from this observation Ricardo and Marx conclude that during the activity of production labor can be considered as some kind of a common substance which is passed on to the commodities that it produces. This substance is thus embodied or incorporated into all commodities. Besides they consider that exchange is the exchange of commodities against commodities, if only because money itself - against which commodities are primarily exchanged, is considered by them as a commodity. For them exchange is consequently an equivalence relation between commodities, which defines their value. Values are thus relative values. It results from these premises that the equivalence of commodities on the occasion of their exchange results from the equivalence of the quantities of labor incorporated in them, directly or indirectly, on the occasion of their production. We are here at the heart of the labor theory of value.

The definition of value is quite different for Adam Smith, since in the Wealth of Nations (op. cit.) he writes in chapter 5 “Of the Real and Nominal Part of the Price of Commodities, or of their Price in Labor, and their Price in Money”: “The value of any commodity, therefore, to the person who possesses it, … is equal to the quantity of labor which it enables him to purchase or command” (Smith, 1776, p. 49). Since wages are paid in money, Smith from the start situates himself at the level of monetary prices, and thus confuses price and value. In other words there is an equivalence in exchange between a quantity of money and the monetary price of a commodity, but not between commodities at the level of production.

Going back to Ricardo and Marx’s approach, a first problem that this approach encounters is that it is wrong to consider labor as a substance, because even in the real world labor has no material content, or does not exist as a material or tangible physical object. Labor is an activity, and what we have shown is that as such the only link that can be established between labor and the dimensions of the real world is that this activity, whatever the form it takes, is carried out during time, time being precisely one of the seven existing dimensions defined by physical science. It follows that if labor can allow for the establishment of an equivalence relation between commodities, which gives them their value, it is because this activity has this common dimension of time, which itself has several measurement units, like an hour, a week, a month or a year, all these units being related. Thus values, based on labor time, have the dimension of time.

This explains that seeing values as made of a substance which would be labor is wrong, and has a bearing on the definition of the value of a commodity as incorporating part of the value of fixed capital, a theoretical position which is developed by both Ricardo and Marx. Indeed fixed capital is supposed to transfer to the product, in each production cycle, a part of its value supposed to be lost through its use. This view comes down to seeing fixed capital as quite similar to circulating capital, i.e. intermediate goods or raw materials which are used and in fact transformed in the production process.

Such a view suffers from logical flaws. First, because production is indeed a transformation process, which implies that quantities produced as an output by any productive system - be it an economical or chemical one, through the transformation of other quantities, cannot be more (or less) than these last quantities used as inputs in this process. If they were more, this would mean that spontaneous generation exists, and if they were less, this would mean that spontaneous disappearance or annihilation exists, and none of these proposals can be scientifically supported. It is moreover a fact that fixed capital, unlike intermediate goods or raw materials, does not transfer any substance, i.e. any material component or physical part to the product during a production process: it does not enter into the production process to be transformed by it. Second, because transferring a part of its value would imply that this value, as a dimension attached to fixed capital, could be detached from it and independently transferred to the product. But such an assumption has no logical foundation, because the dimension of an object is an intrinsic characteristic which cannot be detached from it.

In passing we have shown that defining production as a transformation process implies for the same reason that there can be no physical surplus. Indeed, leaving aside the existence of precautionary stocks of some goods, which by definition, if stocked, are not transformed, the quantities of intermediate goods which are produced during a production cycle are necessarily exactly the same as the quantities which are needed to be then transformed into final goods.

In order to balance inputs and outputs in equations supposed to represent the production process, a precise and complete description of this process (usually not performed by economists), should make explicit the existence on the input side of free goods, like water from rain, carbon dioxide from the air, photons from the sun, or nitrogen from the soil. Since they are free goods their price would obviously be zero. This treatment should apply similarly to minerals like iron ore or crude oil stocked underground: the corresponding quantities pre-exist to their transformation, which takes the form of mining or extraction, being understood 1) that minerals in underground deposits and mineral brought to surface should be considered as two distinct commodities and 2) that all quantities produced as intermediate goods are transformed in the production process. This leaves no place to any surplus, since final goods, made from the transformation of all intermediate goods, have no reason to be considered as such: they are not a surplus, but the end-result of this process. This would avoid the nonsense of Sraffa equations where any intermediate good can be produced in higher quantities than those in which it is used in the production process.

In any case we have shown that the conception of production as transformation allows for the definition of values and for their calculation in a coherent way, and that the dimension of these values is time, understood as the time of physics, which is the reason why we proposed to call them physical values. Moreover we have demonstrated that this calculation is possible even in the case of joint production. Thus values are logically defined as a measurable dimension of commodities which makes possible to think of them as equivalent at the stage of production. This explains that on the contrary any approaches defining values at the stage of exchange are doomed to failure.

What we showed also is that values cannot exist without money. Indeed in the real world labor time as such is made of very different and heterogeneous types of labor, what Marx called concrete labor, which can hardly be compared. But 1) wages are paid in money, which as a general measurement for values and according to the mathematical theory of measurement is a pure system of numbers, or scalars, and 2) wages are always defined, even implicitly (in the case of task payment) in terms of a quantity of money paid for a quantity of physical time (be it one hour, one week, one month, or even one year). This means that to any quantity of concrete labor, as long as it is paid for a given time, can always be associated a pure quantity of money, which as a scalar or a pure number without any dimension is totally homogeneous. It follows that once all types of concrete labor are paid, they all become homogeneous since to any kind of labor corresponds a quantity of money, itself defining a quantity of abstract labor corresponding to a quantity of average social labor time. This average social labor time is the ratio of total wages by total labor time.

This draws our attention on the paramount importance of wages. First, because they are the only income created in the sphere of production, and as such a primary income. Second, because their payment in money brings money into the productive system, which will allow for circulation to take place. And third, because of their dual dimension, as a scalar or real number without dimension per unit of time. As such wages are not only a derived unit of measurement of the global standard, the International System of Units, but an operator in the mathematical sense of the term. Indeed using the average wage per unit of time allows to go from the values of commodities defined in physical time (or physical values) to their monetary values, and reciprocally. In turn this allows to calculate social values, a defined as quantities of average social time. This means that to the system of physical values of commodities defined as a number of time units (with the dimension of time) corresponds a system of social values and a system of money values defined in money units, with a tangible existence in the real world. It follows also from this observation that the money value of the global product is equivalent the overall wage-bill of the productive system.

## 3. Distribution, exchange and prices

The problem with the labor theory of value as exposed by Ricardo and Marx is that it defines value at the level of exchange, conceived as a relation of equivalence between commodities, but simultaneously founds this equivalence at the level of production. In so doing it misses the fact that in the real world exchanges do not take place between commodities, but between commodities and money, which itself is not a commodity: in other words money exchanges against commodities and commodities exchanges against money, but commodities do not exchange against commodities. This implies that the equivalence at the level of exchange is an equivalence between money and the monetary price of a commodity, a price which must logically pre-exist to its exchange against money. To say it otherwise exchanges do not define relative prices, but monetary prices. Thus prices are not defined by an equivalence relation between commodities in the sphere of circulation, whereas it is in the sphere of production only that values can be defined as such, i.e. on the basis of such an equivalence relation.

This conceptual divergence between prices and values would not be a problem if commodities were exchanged at prices reflecting solely their content in labor, which once converted in money takes the form of wages. The value of a commodity could thus be initially assimilated to its price, itself being measured by its content in terms of wages. To some extent, it is the conceptual path followed by Adam Smith.

However the big conceptual problem that arises at this stage is that exchanges also allow for completing the distribution of incomes with the distribution of profits, i.e. the income of capitalists. Indeed the only income which is distributed at the level of production is wages, because the determination and payment of wages is a condition for production to take place. As for profits, they cannot appear before previously produced commodities are sold on the market. Moreover these profits are a monetary income, and they are not allocated proportionately to the quantity of labor or wages, but to the quantity of fixed capital.

Ricardo and Marx thus realized that exchanges in the real world did not take place according to the value of commodities, measured as they had defined it in terms of quantities of labor embodied into them in the production process. However they continued to stick to the idea that exchange is defined as an equivalence relation between commodities.

As for Smith he had understood that the quantity of labor that can be purchased by the monetary price of a commodity is higher than the amount of all wages paid directly or indirectly to produce this commodity, precisely because this price also contents profits on top of these wages. That these profits were for him additional to wages is clear from this quotation, taken from Chapter 6 "On the Component Part of the Price of Commodities" of "the Wealth of Nations": "an additional quantity, it is evident, must be due for the profits of the stock which advanced the wages and furnished the materials of that labor" (Smith, 1776, p. 83).

Smith was even conscious that: "In the price of commodities, therefore, the profits of stock constitute a component part altogether different from the wages of labor, and regulated by quite different principles. In this state of things, the whole produce of labor does not always belong to the laborer. He must in most cases share it with the owner of the stock which employs him. Neither is the quantity of labor commonly employed in acquiring or producing any commodity, the only circumstance which can regulate the quantity which it ought commonly to purchase, command or exchange for" (Smith, 1776, pp. 82-83). As for the provenance of these profits, it seems obvious from the above citation that for Smith they were a share of the product of labor.

Indeed he believed that labor was the only factor of production and that fixed capital in itself was not productive, but increased the productivity of labor, as it appears clearly from this citation from Chapter 2 of Book II of the "Wealth of Nations": "The intention of the fixed capital is to increase the productive powers of labor, or to enable the same number of laborers to perform a much greater quantity of work" (Smith, 1776, p. 471). Nevertheless he just fell short to qualify profits as a levy on the workers' product, which as such could have been interpreted as constitutive of an exploitation of these workers. Such a recognition would have been opposite to what appeared as the common view of his time.

Owing to a quite different definition of values, it is not surprising that the views of Ricardo were rather different from Smith's theoretical position. Indeed the title of Section IV of the first Chapter of Ricardo's Principles precisely states: "the principle that the quantity of labor bestowed on the production of commodities regulates their relative value (is) considerably modified by the employment of machinery and other fixed and durable capital" (Ricardo, 1817, p. 22). In other words he realized that prices at which commodities were exchanged were not proportional to wages and thus to their labor content, and were as such different from their values. But he nevertheless stuck to the idea that these prices were revolving around their value, and that consequently, unlike Adam Smith's view, a rise in wages implied a fall in profits, and reciprocally. Therefore, for given conditions of production, these prices would change with the rise and fall of wages and the corresponding fall and rise of profits. This is what led him to the search for an invariable measure of value, to which he devoted Section VI of Chapter 1 "On Value" of his Principles. This Section remained nevertheless inconclusive.

However it is this same kind of reasoning which was followed by Sraffa, with the introduction of profits, not at the level of circulation, but at the level of production, as a share of a surplus in the quantities of commodities produced compared to the quantities used in the production process. This is the reason why Sraffa's theory is known as a theory of production prices, or as a neo-Ricardian theory of prices. We demonstrated in Chapter 4 that this theory is wrong, because it suffers from a number of logical flaws, as regards in particular the nature of commodities and the existence of a surplus of commodities. This should not be surprising, because from a conceptual point of view, prices belong to the sphere of circulation, and not to the sphere of production.

As for Marx, he realized like Ricardo and for the same reasons that prices were necessarily different from values, which explains why he considered that it was necessary to transform values into prices. He devoted considerable efforts to this end in book III of Capital, but the result did not satisfy him. It was the reason why he did not published this third book, which was edited by Frederick Engels and completed by him in 1894, i.e. eleven years after Marx's death. Had he continued in the way in which he had started, with the proper mathematical apparatus, Marx might possibly have arrived to the same point as Sraffa.

In fact Marx's great contribution to economic theory was to combine the Ricardian definition of values, implying that fixed capital is not a factor of production, with Smith's implicit view of profits as a share of the product of labor, and therefore as a share of the value of this product. Starting from this basis, he just had to go to the end of its logic by introducing the concept of surplus-value, which is precisely the share of the value of the product which is levied by capitalists, as the owners of both fixed capital and the product of labor. This logically led him to define the remaining share going to workers, not as the value of their labor, but as the value of their labor-power, a new concept which he needed to introduce for the coherence of his demonstration. This indeed allowed him to introduce the concept of surplus-value, which he defined quite simply as the difference between the value produced in the production process and the value of labor-power, itself being considered as the value of the commodities that workers can obtain in exchange of their wages.

Marx himself considered - and we share this view - that the discovery of surplus-value was one of his main findings. In chapter 2 "Marx and his discoveries" of his book "Reading Capital" (written with E. Balibar) Althusser indeed recalls that in a letter to Engels on 24 August 1867, Marx writes: "The best points in my book are: (1) the two-fold character of labor, according to whether it is expressed in use-value or exchange value. (All understanding of the facts depends on this). It is emphasized immediately, in the first chapter; (2) the treatment of surplus-value independently of its particular forms as profit, interest, ground rent, etc." (Althusser, 1970, p. 79). A logical consequence of the introduction of this concept is that profits do not reward the productivity of capital but its property.

This citation also shows clearly that for Marx profits were a monetary form of surplus-value, and as such a component of monetary prices, but as we just indicated his theory of transformation of values into prices was wrong. As we also showed this mistake came from his theory of money as a commodity, and from his reproduction schemes, which did not give a correct view of circulation and of the formation of profits in a monetary economy, where money is not a commodity and prices are not relative prices, but monetary prices.

In fact, what we elucidated in chapters 11 and 12 of this book is precisely the mode of formation of profits at the level of the realization of the product, i.e. in the sphere of circulation. We succeeded in this endeavor by taking as a point of departure a remark made by Keynes in his "Treatise on Money", where he explains that profits spent on consumption goods are a source of new profits for the entrepreneurs whom theses goods are bought from, which means in other words that the spending of profits generate new profits. Keynes went as far as considering that this source could never dry up, writing indeed: "If entrepreneurs choose to spend a portion of their profits on consumption (and there is, of course, nothing to prevent them from doing this), the effect is to *increase* the profit on the sale of liquid consumption goods by an amount exactly equal to the amount of profits which have been thus expended."…"Thus profits, as a source of capital increment for entrepreneurs, are a widow's cruse which remains undepleted however much of them may be devoted to riotous living" (Keynes, 1930, p. 139).

Reflecting on this remark, our contribution has consisted firstly in generalizing this mechanism to all types of goods, including fixed capital goods, and secondly to show that at each stage of profit spending, and contrary to what Keynes indicates, there is a partial loss in the purchasing power of these profits. This depletion comes from the fact that wages have to be deducted from the amount of the proceeds, because they represent the money value of the product. Thus the purchasing power of these profits is exhausted in this purchase, and for the amount of this monetary value: first because it represents the realization of a money value which has already been produced and cannot be realized twice; and second because initially, at the start of a new production cycle, a corresponding amount of money has to be borrowed from the banks, which create new money on this occasion to pay for wages. The repayment of this borrowed money destroys the money previously created and closes its circuit. As a result, and because of this “leakage”, the widow’s cruse cannot continue forever undepleted as Keynes wrongly thought.

Having in mind this mechanism, we had also to get rid of the misconception that workers (or households) should be considered as capitalists to the extent that they own their homes, an idea which comes from the treatment of housing as fixed capital in national accounting. Since such an ownership does not involve the hiring of any workers, there is indeed no reason to consider real estate owned by individuals as fixed capital, even if it is rented. Housing must be considered instead as a consumption good, to be sure with a particularly long life. Indeed renting a dwelling does not appear as different from selling a temporal share of its use, an act which transfers income form the buyer to the seller, without adding any value to the existing global product.

On this basis, we have used two models, the first one without, and the second one with capitalist consumption, to show that the realization of the whole product is always possible, provided that the price system and the values of some underlying variables fit well to each other.

In the simplified system without capitalist consumption, since the whole surplus-value is restricted to fixed capital, which has to be recovered by capitalists, prices in section II producing consumption goods must be such as they allow 1) for the workers to buy all of the consumption goods available, and 2) for capitalists to recover the full value of fixed capital. This is the case when the global price of consumption goods is equal to the amount of wages in both sections of the productive system, whereas their money value corresponds to wages in the section producing these goods. Then the amount of profits in the section producing consumption goods corresponds exactly to the money value of fixed capital, i.e. wages in the section producing it. It also corresponds to surplus-value measured in money, with the consequence that the ratio of profits to wages in section II is exactly equal to the rate of surplus-value. These profits are then redistributed among capitalists through successive purchases of fixed capital, a process inspired by the widow’s cruse, but more reminiscent of the nesting of Russian dolls. What emerges from this process is that all profits have to be spent for them to be realized.

To be sure, the introduction of consumption in our second model complicates somewhat the description of the realization of the product, as well as the calculation of the values of the main variables, but does not question its main conclusions. Indeed profits in section II still correspond to surplus-value in its money form, a surplus-value whose rate is increased because it now depends not only on the relative weight of fixed capital in the overall product, but also on the value of consumption goods obtained by capitalists.

In fact the main lesson from this last model is maybe the paramount importance of this capitalist consumption as a key variable of the economic system. Indeed, once the appropriate amount of surplus-value has been levied to match the need for investment in fixed capital, all the remaining surplus-value has by construction to be devoted to acquiring consumption goods. This last part can well represent a majority share of surplus-value, higher than the share of investment, and it is certainly its permanent increase that lies behind the rise in inequalities observed over the last decades in most capitalist economies all around the world.

It is not surprising therefore to find again this variable as an important determinant of the evolution of the rate of profit. Our models have indeed allowed us to shed some light on the old controversies among economists about the rise or fall of this rate in parallel with capital accumulation supposed to cause a rise in the organic composition of capital. What we have demonstrated is that Marx’s law of the tendency of the rate of profit to fall is wrong when there is no capitalist consumption. However, with his own definition of the profit rate (with variable capital at the denominator of the expression giving this rate), the law is true as soon as the share in value of capitalist consumption over total consumption exceeds a threshold which is in fact rather low ($c > \frac{1}{1+t}$, where $t$ is the average lifespan of fixed capital). Furthermore, with the current definition of the profit rate, an increase in the organic composition of capital always causes a decrease in the profit rate, *ceteris paribus*. These last two words are important, because we also showed that this decrease can be counterbalanced by an increase in the share of capitalist consumption, causing the rate of profit to rise, despite the increase in the organic composition of capital.

Going back to the divergence between Smith and Ricardo as regards the effect on wages of a rise in profits, we can say now that both were right, but each one in his own reference system. Indeed Smith’s reasoning concerns monetary prices, and thus a rise in profits of course increases these prices, but has no reason to decrease money wages, once they have been established and paid to workers. As for Ricardo, who reasons not in terms of prices, but of labor values, an increase in monetary prices and thus in profits corresponds to a rise in the share of capitalists in the value of the product, and thus to a fall in the share of workers. It is because Ricardo did not have the concept of surplus-value that he was unable to conclude, as Marx would have, that this situation is equivalent to an increase in the rate of surplus-value, if the increase in prices concerns consumption goods. However if the increase in prices were restricted to fixed capital alone, without influencing the price of consumption goods, we have shown that it would only affect the redistribution of surplus-value between capitalists, without materializing in a reduction in the workers’ share of the product.

In any case an important by-product of our demonstration is to make us understand the nature of monetary incomes. Indeed the only primary incomes are wages, because they are the only income whose distribution is needed before production takes place. And this for a quite straightforward reason, which is that nothing would come out of the production process without the labor done by workers, and that workers would not work if they were not paid their wages. Although workers are generally paid after having worked for some time (one week, or at most generally one month), most of the time the commodities that they have produced have not yet reached the market and been sold at the time when they receive their salary. This explains that for classical economists wages were usually considered as advanced. On the contrary the payment of profits is not a condition for the production process to start, and anyway profits could not be paid if commodities were not sold beforehand. Thus it is the realization of the product in the sphere of circulation which allows for the formation of profits.

This elementary finding explains why all incomes other than wages are necessarily transfer revenues derived from wages. Unlike wages, they cannot be linked directly to production, and even more contribute to it, quite simply because they are realized and formed after production has already taken place, i.e. in the sphere of circulation, where previously produced commodities are exchanged against money, which flows in an opposite direction.

The only reason which can be found for the existence of these incomes derived from wages is that they are the counterpart of property, as a social institution giving a right to the use and/or to what is perceived as the fruit of some assets. Profits thus come from the property of fixed capital and rents form the property of non-produced resources like land. All of these transfer incomes are generated through exchanges which can take place only after the production process has been completed and commodities are therefore available to be exchanged.

Profits made in the banking sector also obey to the same logics, and represent a share of the overall surplus-value, whatever the specificities of this sector, where debtor interests correspond to the price of the service of money creation and/or financial intermediation, and creditor interests to the price of parting with liquidity. The evolution of interest rates since the great financial crisis of 2008 has showed definitively that both these rates were heavily influenced by the policy decisions of central banks. Like a number of other prices, they are thus mainly institutionally determined. This finding is important, since it shows that contrarily to neo-classical theory and even to what Sraffa hinted to, the rate of profit and the rate of interest (be it creditor or debtor), are two different variables, which are quite independent because their determination obeys to very different mechanisms.

In passing, as regards Piketty's views on this question, let us recall first that he prefers to talk, not of the rate of profit, but of the rate of return on wealth (a term for him synonymous with capital), which is clearly a rate of return on the ownership of property, knowing that in an interview given on 7 January 2015 he admitted that the marginal productivity explanations of income are wrong, even indicating: "I do not believe in the basic neoclassical model" (Piketty, 2015, Potemkin Review interview). In Piketty's theory the long-term interest rate is part of this global rate of return, it is the rate of return on financial bonds, whereas the rate of profit is the rate of return on more risky investments in real fixed capital. This means that implicitly for him the rate of interest is determined by the rate of profit minus a risk premium. Obviously this is theoretically wrong, and the extremely low level of long term interest rates in developed economies since the world financial crisis has demonstrated that it is also empirically wrong.

In any case all of these incomes other than wages thus appear through the establishment of prices, and none of them can correspond to a primary income corresponding to a creation of value, which by definition can occur only in the sphere of production. Exchanges by themselves cannot generate values ex-nihilo, but only transfer the property of previously produced commodities, by changing their previous distribution. Indeed in the sphere of circulation, any income obtained from a sale corresponds to an income simultaneously spent in a purchase, and which thereby loses part of its purchasing power, corresponding to the money value realized through this purchase, and transfers the remaining part in the form of the seller's profit. One function of these prices is therefore to allow for the creation of these transfer incomes and simultaneously for the distribution of the product. Whereas values can be considered as "production values", this is the reason why these prices, although they derive from values, should not be considered primarily as production prices, but should rather be named "distribution prices".

Acknowledging this particular dimension and role of prices as a determinant of the distribution of commodities has important theoretical implications, which deserve some more elaboration.

## 4. Distribution and the incompleteness of economic theory

Distribution of the available goods is one of the most important activities of any society, from a tribe of hunter-gatherers in the Neolithic period to the highly developed capitalist economies of today. It must have become a powerful factor of inequalities within societies as soon as agriculture started to develop, around 10 000 years ago, which gave birth to the social institution of hereditary private property, at the basis of inequalities of wealth, translating to inequalities of income, as recently recalled in a fascinating book by Walter Scheidel, published in 2017 under the title "The great Leveler. Violence and the History of Inequality from the Stone Age to the Twenty-First Century".

In capitalist economies with fiat money, what we have showed is that the main instrument for the primary distribution of the social product appears to be the price system. It is a fact that distribution of incomes among the members of a given society starts at the stage of production with a first distribution among workers themselves, through wage setting and the pricing of various types of labor.

We have demonstrated that, although this pricing is certainly influenced by what might be called "pure" economic factors, like the situation of supply and demand for each category of labor, the salary scale cannot be reduced to these factors alone. Indeed, once they have been taken into account, it remains an irreducible element, or residue, which can be explained only by non-economic factors. These other factors can be ideological, cultural, political, institutional, and they may have an historical component. In some cases, like the gender bias which works against women, it is easy to understand that labor price can also reflect the balance of power between men and women within human societies. In any case it has nothing to do with the so-called productivity of various types of labor.

It is easy to see the importance of these non-economic factors by comparing two countries with similar levels of development and standard of living. There can be a discrepancy between the relative level of wages for manual and intellectual work, as for instance between France and Germany: in this last country, manual work is comparatively better paid than in France. Or taking self-employed into account, there is a big difference between the incomes of doctors in the USA and France, especially as regards general practitioners.

This is one of the reasons why the theory which has been developed all along this book did not deal with individual prices of particular goods, but only with the overall price of some main categories of goods: final goods, including consumption goods and fixed capital, and intermediate goods. It is thus truly a macroeconomic theory. The global quantities corresponding to each of these categories must be considered as statistical averages, which result from an aggregation approach, and mask the dispersion of individual variables. But it is precisely what made it possible to highlight some general relations between these different variables and to establish basic principles governing the functioning of a capitalist economic system. This gives us the framework within which individual prices are set.

It is our conviction that a microeconomic theory of prices is possible only by taking this framework as a background allowing for setting limits on the magnitude of individual prices, and working then on each price, on a case by case basis, to disentangle the multiple factors which can explain it. In other words microeconomics can only be applied economics[14], with some of the explanatory factors being purely economic ones, like obviously, to begin with, the underlying money value of a particular commodity, which links it to its conditions of production (wages being given), or the situation of supply and demand for it. But many of these factors will be outside the usual economic scope, since they will have to do with cultural traditions, fashion, the institutional setting or even political interference. An important factor is the market power of the producers of a particular good, which depends in particular on the market structure and on the degree of competition among these producers.

This is reflected when we examine among others the price of intermediate goods, which so far we have tended somewhat to put aside, because we have been reasoning on integrated sectors for consumption goods and fixed capital. If we take for instance the price of crude oil, a good example of an intermediate commodity, it is undeniable that it is not determined primarily by its production cost or even by the balance between supply and demand, to the extent that oil supply itself is decided by complex negotiations between sovereign States, some of them being members of the Organization of the Petroleum Exporting Countries (OPEC). As for the price of oil at consumption level, once it has been transformed into refined products, it varies from one country to another with among others the level of taxes.

The case of supply chains is also worth emphasizing, because at each step of these chains there is some kind of negotiations or arm twisting among stakeholders along the chain. It is in particular the case in the agricultural sector, where a myriad of farmers have to face a few big industrial transformers, which themselves must confront with just a few giant supermarket chains. At the end of these chains millions of individual consumers cannot be anything else than disarmed price-takers. Understanding the final price of food products implies therefore a detailed analysis of each case, or each particular supply chain, which cannot be reduced to the neo-classical equations of supply and demand.

[14] An interesting approach in this regard has been coined as « Experimental Economics », but its interest remains limited by its narrow scope. As an example see Duflo, E., (2006) “Field Experiments in Development Economics”. Massachusetts Institute of Technology. Cambridge. USA.

This leads us to emphasize the fact that the whole existing economic theory is based on the fundamental assumption that one good or commodity has only one price. In the case of neo-classical theory, because it is a microeconomic theory, supposed to provide the theoretical foundations of macroeconomic theory, based on the determination of prices in a state of general equilibrium on all markets, and because this task would be totally insurmountable if one same good could have several different prices. In the case of other theories, because they have been contaminated by neo-classical reasoning and the use of supply and demand curves, supposed to be an abstract representation of real markets. They just forget that real markets do not function like that, and it is therefore more than doubtful that supply and demand curves provide a good image of the real world…

On the contrary, the assumption of uniqueness of prices was not formulated in the case of this book, for several epistemological reasons: first that the whole is more than the sum of its parts, secondly that if there are general economic laws, they must be defined at a general level which is the level of an economy as a whole, and thirdly that defining these laws implies to work on statistical averages. Therefore the fact that any particular good can have several different prices simultaneously has absolutely no importance, since values were defined as average social values, and since it is always possible to define an average price and work on this basis.

Recognizing at theoretical level the existence of a multiplicity of prices for the same good helps to understand that there is similarly a multiplicity of profit rates, because it is a necessary corollary. Indeed we never made the assumption of a single profit rate, which would be brought about by competition. On the contrary the models that we used to describe the realization of the product showed already that this realization implied different rates for both sections of the economic system, and also within section I producing fixed capital, even without capitalist consumption.

But the introduction of the consumption of capitalists as codetermined with the level of the prices that they can control is also a powerful motive for each individual firm to increase its profit above the average level of its competitors. Finally, if we reason in terms of dynamics, an economic system is permanently changing due to technical progress, the invention of new products and the spur of competition. Old companies whose products become obsolete die and new firms are created every day to exploit new opportunities. This implies that rates of profits are declining and become negative for the first ones and on the contrary start at negative level for the last ones, then grow and can become sometimes very high for members of the second group, made of successful start-ups. And we should not forget that among start-ups, only a minority is usually successful!

What all this tells us as regards economic theory is that it can be compared to other sciences like for instance human physiology, which is a true science in its own right as long as it deals with an average human being, for whom it has defined first the average value of many parameters, like temperature, blood pressure, heart rate, etc., and various ratios of physiological constants, like cholesterol, uric acid, and so on. Obviously human physiology has also established their interrelationships. It is on the basis of this established and recognized science that doctors intervene to diagnose and possibly cure the disease(s) of their patients. But although they can and indeed do rely on an enormous amount of scientific knowledge, validated by numerous experiments and clinical observations, they know that each patient is different from the others and that there are multiple factors that must be taken into account in each case, like genetic factors, the way of living, the influence of environment, patient and family history, and so on and so forth. This is precisely the reason why physicians themselves consider that clinical medicine is not a science, but an art!

This comparison is illuminating, since we can infer from it that microeconomics, for the same obvious reasons, to be sure can and must rely on macroeconomics and on economic laws, in order to perform its own analyses concerning one particular product, price, or profit rate. But there are many factors which then must be taken into account to understand the individual variables which constitute its object, and these factors come under other disciplines than economic theory, like history, anthropology, sociology, psychology, political science, etc. The contributions of these sciences are indispensable to complete its own analyzes which in themselves cannot be comprehensive. This situation, based on scientific methodology, reflects the complexity of the real world and explains that economic laws cannot be fully defined at microeconomic level. Therefore microeconomics in itself is condemned to always remain incomplete and as such to be more an art than a science!

This incompleteness of microeconomic theory does not mean that it cannot produce scientific knowledge, but that do so it must resort to these other social sciences mentioned above, because in the real world pure economic phenomena do not exist. Economic phenomena are closely related to multiple other fields of reality. Since microeconomics is an integral part of economic theory, and since macroeconomics itself - as we have seen with the case of the salary scale as a determinant of money values, is embedded within other aspects of reality, the inescapable conclusion is the incompleteness of economic theory as a whole. As such economic theory is not in any way different from all the other social sciences.

## 5. A few final words

“Pure logical thinking cannot yield us any knowledge of the empirical world; all knowledge of reality starts from experience and ends in it. Propositions arrived at by purely logical means are completely empty as regards reality” (Albert Einstein, in Ideas and Opinions. 1933).

We have now reached the end of this intellectual journey. Along the way, we have gone from production to values, from values to prices, from prices to distribution and profits. In short we tried to redefine the bases of an economic theory which has been in dire straits for decades, and in fact since neoclassical theory has become hegemonic in universities as well as in publications, and even more in the mind of most rulers all over the world. To get back at this stage to Wittgenstein and some epistemology, but not being myself a philosopher or a specialist of Wittgenstein's epistemology beyond the Tractatus, I relied greatly for the four following paragraphs on an excellent article published in 2005 by Christiane Chauviré: "Wittgenstein, Sciences and Epistemology Today". Quotes are from this article.

She emphasizes the fact that "the direct cognitive or representational content of a scientific theory is necessarily limited", which after all is quite normal since language can only provide a limited image of the real world. As such, a scientific theory gives but an image of reality with a certain number of stipulated elements, or assumptions, connected by logical constraints. Although a theory has a descriptive claim, "it is not intended to capture the inner and hidden essence of phenomena". In other words, proposals made by scientific theories cannot be tightly coupled with the real world, since "the referential and representational link is only indirect" (Chauviré, 2005, pp. 157-179).

Thus, proposals of a theory do not simply reproduce the real world, much less the immediate experience, even supposing that language can do it. These proposals "reproduce only the form of the facts, and cannot account for their full content", because a whole scientific system is an "image" less close to the real world than the basic principles that represent "the state of affairs". But "science is not limited to recording or photographing a reality that verifies it. A scientific theory cannot be a copy of the phenomena that would verify it, because it is quite a complicated and tiered construction" (ibidem).

Indeed economic theory must be viewed as staged on several levels: on the level immediately descriptive of the elementary logical images or basic principles which already retain only the form or the structure of the facts, it must superimpose less basic principles and laws which "move scientific modeling away from a simple factual depiction" (ibidem). The present work must thus be viewed as an attempt to provide essentially the first stage, i.e. basic principles and some immediately derived and less basic principles, which could also be viewed as first laws. They constitute what we chose to call the main principles of political economy.

The theoretical foundations laid here could also be seen as providing the beginning of a grammar of economic theory. Within this grammar basic principles constitute what might be called its semantics, giving a precise meaning to basic concepts like value, price, production, capital, profit etc. As for less basic principles, they provide the first elements of its syntax, providing the relationships between these concepts which structure them.

Basic principles reproduce only formally or schematically the real world, even though their objective is nevertheless to lay the foundations of a scientific system which itself can only be indirectly connected to this real world. This does not mean that basic principles cannot be submitted to verification. On the contrary, as "complete logical images" formally reproducing "states of affairs" they can be directly verified by comparison with the real world, in particular by using all kinds of experiments or statistical data.

Indeed everybody knows from direct experience, for instance from examining his/her bank account, paying with a credit card or making a wire transfer, that money is not a commodity with material existence, but a set of figures which are pure numbers without any dimension, i.e. scalars in the mathematical sense of the word. Everybody also knows empirically that nothing can be produced whatsoever without spending some quantity of labor, or that wages received in payment for labor under an employment contract are defined and paid in money units per unit of physical time, and thereby that all possible types of labor share a common dimension, which is precisely time.

Similarly all of us know that the vast majority of exchanges take the form, not of a barter trade between buyers and sellers of commodities, but of simultaneous purchases and sales where some amounts of money from buyers are given in exchange for commodities provided by sellers. In another vein, watching any machine working as fixed capital in a production process provides a convincing evidence that no material component of this machine is transferred to its final products, in which it would be incorporated. We could continue such a list, but it would be useless.

Although this would certainly be more complicated, by taking advantage of progress made in national accounts statistics it should also be possible to verify the evolution of the average rate of profit as a function of various parameters like capitalist consumption, the stock of productive fixed capital and its average lifespan, etc., and to check our laws of evolution concerning the behavior of profit rates over time. The theoretical principles and proposals established in this book are therefore entirely refutable, and thus certainly meet the criteria to be falsifiable in the sense of Popper.

Hoping nevertheless that they are not going to be refuted in a near future, we are looking forward to the work that could be done by our fellow economists to extend and improve these theoretical foundations. This should be instrumental in finally completely wiping out the imposture and the false pretenses of neoclassical theory.

Because it succeeded in persuading the political elite in Western countries that it had a true scientific status, this bunch of theoretical nonsense without any connection to the real world has indeed been the main support of neo-liberal economic policies that have little by little after 1980 completely reversed the mode of regulation of capitalist economies which had prevailed for more than thirty years after the Second World War. State intervention and Keynesian policies were abandoned in favor of these neo-liberal policies based on deregulation, globalization and financialisation.

These neo-liberal policies have led to a regression in the rates of growth, de-industrialization, a fall in labor productivity, a rise in unemployment rates and a significant increase in inequalities for nearly forty years in almost all the major capitalist countries of the Western world. I provided some elements for an explanation of this evolution, showing how a reversal would be possible, in section 3 of my article previously cited about Piketty's book "Capital in the 21$^{st}$ Century" (Flamant, 2015a, pp. 32-38).

My sincere hope would therefore be that this book could constitute a contribution, in a not so distant future and in its field, i.e. the area of scientific practice and intellectual confrontation, to providing an alternative theoretical background to the neo-classical narrative. This might be of some help for policymakers eager to start a reversal of these harmful neo-liberal policies.

Paris, September 2018

# APPENDIX 1 – FUNDAMENTAL PRINCIPLES OF POLITICAL ECONOMY

**Principle 1**: An elementary commodity is a multidimensional object.

**Corollary to Principle 1**: Commodities are heterogeneous.

**Principle 2**: Commodities must have been produced before being exchanged.

**Principle 3**: Money is not a commodity.

**Principle 4**: Commodities exchange against money and money exchanges against commodities: commodities do not exchange against commodities.

**Principle 5**: Commodities have a monetary price before exchanges take place.

**Principle 6**: As the measurement of a dimension of commodities which appears when it is exchanged against them and called their price, money is a pure set of numbers belonging to the set of positive real numbers $R^+$. As such money has itself no dimension, and is composed of scalar numbers.

**Corollary to Principle 6**: Monetary prices are therefore also dimensionless, scalar numbers.

**Principle 7**: Money is created ex nihilo by banks through credits that they provide to the other economic agents, taking the form of a double asset and a double debt: money is cancelled by the repayment of these credits.

**Principle 8**: Wages are paid in money and have the dimension of dimensionless numbers per unit of time.

**Corollary to Principle 8**: Wages are not paid to workers in wage-goods.

**Principle 9**: Workers buy consumption goods out of their wages, previously paid in money.

**Corollary to Principle 9**: There are no specific consumption goods bought by workers that could be isolated as pure wage-goods.

**Principle 10**: Intermediate goods are bought, not by individuals, but by producers, to enter the production process, where they are physically transformed into other intermediate goods or final goods.

**Corollary to Principle 10**: Goods bought by workers as consumption goods, as well as all the other consumption goods, are not intermediate goods, but final goods.

**Principle 11**: Production, or the activity which creates goods, is a transformation process, which creates new goods through the transformation of pre-existing or already produced goods. As such production does not create any physical surplus.

**Principle 12**: Labor is the human activity which produces goods through the transformation of other goods, and as such is the only factor of production.

**Corollary to Principle 12**: Fixed capital, as well as energy, is not an activity, but a produced good, and as such is not a factor of production. As a good its specificity is that it is not transformed in the production process, but intervenes into it through increasing the productivity of labor. Fixed capital goods are therefore final goods.

**Principle 13**: The only way to think of commodities as homogeneous through a common property, which is by definition their value, is to consider that they are produced by abstract labor, i.e. labor reduced to its only dimension of physical time. Therefore values, and commodities through this common property, have the dimension of time.

**Corollary to Principle 13**: Since time in which values are expressed is the time of physics, measured in physical units like hours, days, weeks, months or years, values at the level of production of commodities must be considered as "physical values".

**Principle 14**: Since fixed capital is not itself transformed in the production process, its value as an intrinsic characteristic of commodities cannot be transmitted to the products in the production of which fixed capital participates only as a catalyst.

**Corollary to Principle 14**: The value of intermediate commodities is transferred to a product because these commodities themselves are transferred to this product through their transformation as a part or an element of it, in the course of its production process. In some cases (like energy) this transformation can be a disappearance.

**Principle 15**: Joint production, defined as the production of several distinct commodities by a single method of production, is compatible with the existence and calculation of positive labor values, as long as different methods produce these commodities in different proportions.

**Principle 16**: In the real world, labor is heterogeneous: there are many kinds of labor, which differ from one branch of production to another and according to the degree of qualification of labor. This is the reason why there is a differentiation of wages. But the precise range of this differentiation cannot be determined within the ambit of economic theory, since it depends on many factors external to this theory, and which are also of a political, ideological, cultural and institutional nature.

**Corollary to principle 16**: the differentiation of wages, as it exists in the real world, as well as the average wage level, must be considered as a given set of data for economic theory, which is one reason why this theory should rather be named political economy.

**Principle 17**: Taking as given the existing wage scale (which provides the level of wages for every kind of labor) allows for the conversion of different kinds of labor expressed in physical time into average social labor, or labor which is paid the average money wage, while keeping the same unit of physical time. This in turn makes it possible to calculate the average social values of all commodities, expressed in average social labor time.

**Corollary to principle 17**: Multiplying the social values of commodities by the average wage expressed in money units per unit of average social time, gives their money value at the stage of production.

**Principle 18**: The wage bill is the measurement of the monetary value of the product.

**Principle 19**: In the capitalist mode of production, fixed capital is by definition the property of capitalists but not of workers, who own neither the means of production nor the commodities which they create through their labor. The simple reproduction of this system as it is implies therefore that workers cannot get on the market, from the spending of their wages the property of at least this part of the product consisting of means of production. The difference between the whole product of labor and the share obtained by workers is defined as surplus-value.

**Principle 20**: Surplus-value is created in the sphere of production, but extracted in the sphere of circulation, where commodities are distributed among various stakeholders. This results from the fact that wages, as the money price of labor-power representing the money value of the whole product, can buy only a share of this product consisting in consumption goods, whose price is always higher than their money value. Surplus-value thus materializes under its monetary form as profits in the section producing consumption goods. These profits are then distributed among capitalists.

**Principle 21**: The only primary incomes are the revenues created on the occasion of the production process, and are the monetary counterpart or remuneration of labor, i.e. wages in a pure capitalist mode of production.

**Principle 22**: Profits, as any income other than wages or the remuneration of labor, derive from the property of commodities or assets. They are transfer revenues which arise from exchanges against money of previously produced commodities (or assets) taking place in the circulation process.

**Principle 23**: A levy imposed on the purchasing power of primary incomes through an increase in the price of consumption goods also constitutes a transfer of purchasing power to secondary money incomes. Spending these incomes creates a new levy on secondary incomes when they buy consumption goods as well as fixed capital, and a new transfer of purchasing power. The same process can continue as long as purchasing power is remaining and ensures henceforth the full realization of the whole product from both sections of the productive system.

**Principle 24**: Rent is the income that comes from the property of non-produced means of production, such as land or mines, and is collected by the owners of these means of production. There are two kinds of rent: differential rent and absolute rent. Differential rent comes only from the heterogeneity of each type of non-produced means of production, i.e. from differences in their intrinsic characteristics, and varies according to this heterogeneity. Because of its origin in property rights, absolute rent is the rent perceived even on the least

"productive" (i.e. with no differential rent) of a particular type of non-produced means of production (otherwise it would not be rented). It is also perceived on all the other non-produced means of production of the same type.

**Principles 25**: There is no rent as such coming from the heterogeneity of the techniques which can be employed by various firms using homogeneous non-produced means of production, which already pay a rent for this use. This heterogeneity only results in profit rate differentials. However, owing to the use of non-produced means of production, the corresponding differences in production costs from the average cost can be called "quasi-rents", to distinguish them from other types of profit differentials.

**Principle 26**: Actual rents are always a combination of differential and absolute rents. Whatever their combination, the nature of rent is not different from that of profits: rent is also a transfer income, which is levied as a part of total surplus-value. Changes in differential rents can result from an exogenous change in the price system. They also correspond to changes in the scale of production, implying the use of additional non-produced means of production (in the case of an increase). Such changes therefore always entail a change in the value of the product and the price system.

**Principle 27**: Theoretically, for a given and fixed value of the product, if an increase in rent (hence absolute rent) could leave prices unchanged, the amount of surplus-value would not change, and the amount of rent would be deducted from profits. In practice, such an increase will have repercussions on prices, whose magnitude depends on the balance of power between the three involved groups of agents: rentiers, capitalists and workers. The ultimate effect on the amount of surplus-value and its distribution between rentiers and capitalists will depend on the balance of power between these groups, under the arbitration of the State.

**Principle 28**: Because of their close connection to money values, which themselves reflect the conditions of production, it is not wrong to consider prices as also reflecting these conditions, and thus as production prices. However this is a very partial view of reality. First because money values themselves are depending on the distribution of wages between workers. And second because their fixing is linked to the extraction of surplus-value (for section II prices), and to the distribution of this surplus-value between capitalists (for section I prices). As a result, prices must be further considered as distribution prices, being understood 1) that the understanding of distribution involves many factors, a number of which have nothing to do with economy as such, because they are of a sociological, ideological, institutional and political nature, and 2) that distribution itself must be seen ultimately as a distribution of commodities.

**Principle 29**: Prices, in so far as they are distribution prices, are the object and the resultant of multiple conflicts, which may or may not be arbitrated by the State. They ensure the continuity of the capitalist mode of production by ensuring the distribution of fixed capital among capitalists. But even more they also ensure the collection and distribution of consumption goods among these same capitalists. They make us realize that the ultimate goal of capitalists is not only the accumulation of fixed capital, but also the accumulation of consumption goods.

**Principle 30**: The banking sector plays a fundamental role in the capitalist mode of production, by creating the money which among others is indispensable for the payment of wages. The service that the sector renders, by providing this money as well as the service of financial intermediation, has like all services a value measured by the amount of average social labor time spent in the sector, and a money value, measured by the amount of money wages paid to the sector' workers, who contribute like all workers to the creation of surplus-value. The macroeconomic cost of the services of money creation and intermediation is made of wages and of a cost which is the creditor interest rates paid to depositors of money in the banking sector. To this cost must be added the profits of the banking sector, to obtain the price of the service.

**Principle 31**: Like rent is the price paid to or income received by the owners of non-produced goods for obtaining the right to use them during a defined period, creditor interest rates must be considered as the price paid to or income received by the owners of money assets so that banks can use these assets during a defined period. This is why interest rate can be viewed as the price received for parting with liquidity. Interests paid by borrowers are similarly the price paid by them to banks to be able to use defined amounts of money during a specified period, and which constitutes the banks gross income. It follows that interest as an income remunerates the property of money assets, owned either by depositors or by banks. Therefore, like profits or rents, interests are a transfer income.

**Principle 32**: Formally money should be considered as produced by banks, to some extent ex nihilo, because it is not the result of a transformation. But it must be considered nevertheless as non-produced, because – like land for instance, it does not belongs as such to the production process, where it does not results from a transformation and is not itself transformed. It is so because money is not a good that could be sold against itself, which would be a nonsense. Only the right to use it for a definite period is sold (also like for land), against payment of an interest. As such money cannot be considered neither as a final good, which could be consumed or stored, because at global level it always has to be repaid, which ultimately destroys it, nor as an intermediate good or service, which would disappear in the production process.

**Principle 33**: Apart from various fees, the price of the services of money creation and financial intermediation rendered by banks corresponds to the debtor interest rates charged by banks to borrowers. Their overall amount during a given period of time is part of the total price of the product of this period. It includes wages paid in the banking sector, i.e. the money value of bank services, creditor interests paid to depositors and bank profits. Both creditor interests and bank profits are part of global surplus-value. Bank services can be sold both to firms and individuals, be they capitalists or workers. When they are sold to firms, they play the same role as intermediate commodities, and their price becomes a part of the price of the final product, either of consumption goods or fixed capital goods, depending on the section which the firms are part of. When they are sold to individuals, these are part of their consumption, and thus play the same role as consumption goods, of which they are a particular type.

**Principle 34**: A macroeconomic theory is a theory where all prices are statistical averages, and consequently where the magnitude of all “real” variables, since they are measured in terms of prices of their constitutive commodities, is itself a statistical average. A microeconomic theory of prices should therefore restrain itself to the disaggregation of these statistical averages, with a view to understanding the formation of the price or multiplicity of prices of each particular commodity included in these same averages.

**Principle 35**: The aim of microeconomic theory should be to understand and explain the divergence between individual prices and statistical averages. This should be done first for the various prices of a same commodity, which come from differences in the conditions of production and in the rates of profit, as well as in the level of rents. This should also apply to the explanation of the difference between the rates of profit affecting the production and distribution of different commodities.

**Principle 36**: The average rate of profit cannot be considered as a meaningful element of economic theory. Indeed, apart from the difficulty to simply measure it in the real world, it is nothing else than a mere statistical average. The real world, as well as a number of theoretical factors deriving from the nature of profit formation and realization, show that there is a multiplicity of profit rates both in theory and practice, and that the nature of economic activity, based on competition among firms, always pushes to a differentiation of these profit rates. The profit/wage ratio plays a more important role for the understanding of economic processes.

**Principle 37**: It might exist a set of variables which make it theoretically possible to guarantee a balanced reproduction of a capitalist economy, in the neo-classical sense of ensuring a simultaneous equilibrium on all markets. However such a reproduction is practically impossible to obtain, for a number of equally theoretical reasons. First, because of the decentralized nature of such an economic system, which implies that decisions by capitalists, who fix prices and make investment decisions, are made independently, moreover in situations of imperfect information, and not coordinated. Second, because these decisions are made within time frames and based on expectations about the future which cannot be rational, because of the natural uncertainty which characterizes what happens in the future. Third, because the dynamics of the system, if they are left unregulated by State authorities, go in the direction of a permanent increase in the rate of surplus-value, which in itself has destabilizing effects.

**Principle 38**: It is the very engine of the capitalist system, i.e. the pursuit of maximum profit at micro-economic level, which at macro-economic level constitutes a powerful factor of structural instability. Starting with a reduction in wages, it leads to cascading maladjustments that do not allow a full realization of the product and therefore hamper the reproduction of the system.

**Principle 39**: An autonomous increase in capitalist consumption financed through money creation can be a remedy to crises deriving from insufficient demand from workers, since it offsets the slump in workers consumption, and results in a higher share of profits. However higher profits may end up being at least partly spent for the purchase of existing assets instead of commodities. This cannot but perpetuate the insufficiency in demand for commodities and trigger speculation, which has been shown to be at the origin of financial crises.

**Principle 40**: As Keynes had well understood, autonomous investment financed by money creation is the most effective tool to help an economic system out of a crisis. However Keynes, like his followers, was wrong in thinking that it would trigger a net multiplier effect such that the outcome in terms of product would be higher than 1. Because of leakages due to the existence of transfer incomes at each stage of spending, the overall amount of the multiplier between autonomous investment or spending and the corresponding product cannot be greater than 1.

**Principle 41**: In a situation of simple reproduction where there is no capitalist consumption and all surplus-value is invested, the rate of profit rises with the increase in organic composition of capital, both being the result of an increase in the rate of surplus-value.

**Principle 42**: In a situation of simple reproduction where capitalist consumption is introduced, then the rate of profit defined as Marx does becomes dependent on the share of this capitalist consumption in total consumption. The validity of the Marxist law of the tendency of the rate of profit to fall when the organic composition of capital increases depends therefore on this share. If this share exceeds a threshold linked to the lifespan of fixed hospital and defined by $c = \frac{1}{1+t}$, the Marxist law is valid. If this share is lower than this threshold the Marxist law is not verified and the rate of profit increases on the contrary as a function of the organic composition of capital.

**Principle 43**: When then the rate of profit defined according to its current definition (i.e. without applying this rate to wages - as if they were advanced), and in a situation of simple reproduction with capitalist consumption, this rate is an increasing function of $k$, the primary rate of surplus-value, $k_c$, the actual rate of surplus-value with capitalist consumption, and $c$, the share of capitalist consumption in total consumption. On the contrary it is decreasing in terms of $q_c$, the organic composition of capital. This confirms the validity of the Marxist law of the tendency of the rate of profit to fall when the organic composition of capital increases. But this law remains a mere tendency, since it can be thwarted by an increase in capitalist consumption, which shows the paramount importance of this variable.

**Principle 44**: The share of profits in the product (named $\pi$) is also an increasing function of the primary rate of surplus-value and of the share of capitalist consumption, and thus necessarily of $k_c$, the actual rate of surplus-value. The rate of profit itself is an increasing function of $\pi$.

**Principle 45**: In a situation of extended reproduction and accumulation of capital, the Marxist law of the tendency of the rate of profit to fall when the organic composition of capital increases remains valid, again as a mere tendency. Such a fall can also be caused by an increase in the rate of surplus-value, as well as by an increase in the average lifespan of fixed capital. However, this evolution 1) is not monotonous over time, because the rate of profit initially tends to rise in response to an increase in accumulation and 2) is always countered by an increase in capitalist consumption.

# APPENDIX 2 - THE ERRONEOUS NATURE OF MARX'S REPRODUCTION SCHEMES IN BOOK II OF CAPITAL

On the basis of the distinction made by MARX, the value produced in each period is $L_I$ for section I and $L_{II}$ for section II. The assumption that fixed capital transfers its value to the product can then be incorporated into the reproduction scheme on the basis of two different interpretations, of which it will be demonstrated that neither of them adequately solves the problem of realizing the product.

## 1. A first interpretation of the transfer of value of fixed capital

This first interpretation consists in considering that in simple reproduction the value lost and transmitted by fixed capital in the two sections of production is identical with the value of the product of Section I, which produces fixed capital.

Let us consider that $a$ is the fraction of the value of fixed capital that is consumed and transferred to the product in Section I, and b the fraction consumed and transferred to the product in Section II.

We thus have $a$ + b = 1 and b = 1 - $a$ with 0 < $a$ < 1 and 0 < b <1 (1)

In order for the reproduction scheme to be in accordance with the adopted interpretation, it is first necessary to determine the value of the product of section I, which represents a certain factor of the value newly produced in this section (the difference being the transfer of value from fixed capital). Let us call x this factor, with $x > 1$ and such that:

$$\frac{\text{value of the section I product}}{\text{Value newly produced in section I : } L_I} = x \qquad (2)$$

The following scheme is thus obtained for the value produced in each section:

Section I: $a(xL_I) + L_I = xL_I$ (3)

Section II: $b(xL_I) + L_{II} = (1-a)(xL_I) + L_{II}$ (4)

From the conditions of production of value in Section I we derive:

$$L_I = (1-a)xL_I \qquad (5)$$

Hence it follows that $x = \dfrac{1}{1-a}$ (6)

It is thus possible to write the reproduction schemes of equations (3) and (4) as a function only of $L_I$, $L_{II}$ and $a$, which gives the following identities, for each of the two sections:

Section I: $a\frac{1}{1-a}L_I + L_I = \frac{1}{1-a}L_I = K$ (7)

Section II: $(1-a)\frac{1}{1-a}L_I + L_{II} = \frac{(1-a)}{1-a}L_I + L_{II}$ (8)

Thus for section II: $L_I + L_{II} = L_I + L_{II}$ (9)

Let us name Y the value of the total output of the two sectors, it is thus:

$$Y = \frac{1}{1-a}L_I + (L_I + L_{II}) \Rightarrow Y = \left(\frac{2-a}{1-a}\right)L_I + L_{II} : \quad (10)$$

Does this scheme allow for a correct analysis of the realization of the product of both sections? Apparently yes, since equations (7) and (8) show first that the value of the variable capital advanced in the two sections, $L_I + L_{II}$, is identical to the value of the product of Section II, which can thus be sold to and bought by workers. Similarly, the value K of the product of Section I, the fixed capital produced during the period, i.e. $\frac{1}{1-a}L_I$ (see equation7), is equal to $\frac{a}{1-a}L_I + L_I$, the sum of the value of fixed capital consumed productively and transmitted to the product in Section I , i.e. $\frac{a}{1-a}L_I$ , and in Section II, i.e. $(1-a)\frac{1}{1-a}L_I = L_I$ respectively.

But this presentation, although arithmetically correct, is nevertheless possible only at the cost of an error of logic which will now be explained and will consequently lead to reject the initial assumption (that the value lost and transmitted by fixed capital in the two sections of production is identical with the value of the product of Section I).

The reason why the above scheme is not logically coherent can be summarized as follows: the interpretation on which the above scheme is based is contradictory with an obligatory corollary of the assumption of the transmission of the value of fixed capital, according to which the value transmitted by fixed capital must by definition be necessarily identical to the value lost by this fixed capital because of its consumption, itself necessarily identical, in fine, to the initial value of this fixed capital. This contradiction results from the fact that the interpretation adopted implies that fixed capital transfers its value several times to the product.

To show it, it suffices to note that the value transmitted to the value of the newly produced fixed capital during one period is then necessarily retransmitted to the following period by the same fixed capital when in turn it becomes to operate, and so on and so forth. The value transmitted by the fixed capital in section I is thus carried out in an iterative manner during the successive production periods.

From the above reproduction scheme, it is in fact immediate, according to equation (7), that at each period a fraction $a$ of the value of the fixed capital produced and realized in section I is constituted of value transmitted by the consumption during that same period of the fixed capital in operation. This fraction $a$ is then in turn transmitted and realized during the periods when this capital is in function, and so on and so forth. The same fixed capital in this way transmits its value over an infinity of periods, in such a way that, if the sum of these successive transmissions is made, one obtains for a fixed capital of an initial value K, equal to $\frac{1}{1-a}L_I$ :

Value transmitted over n periods $= K + aK + a^2K + ... + a^nK = \frac{1}{1-a}K = \frac{1}{(1-a)^2}L_I$ (11)

For example, for $a = \frac{1}{2}$, the value transmitted by the fixed capital K during the successive periods is thus from equation (11) equal to 4 $L_I$, i.e. twice its initial value 2 $L_I$, and for $a = \frac{3}{4}$ the transmitted value is equal to 16 $L_I$, i.e. four times the value 4 $L_I$ of the fixed capital. In passing, one notes the importance of the value of the $a$ coefficient!

Since the value transmitted varies as a function of $a$, and since its compulsory logical corollary: transmitted value = lost value, is not verified, we must conclude that the Marxist assumption of the transmission of value of fixed capital is not consistent with the interpretation of simple reproduction which implies the identity between the value of the fixed capital which is produced and the value which is transmitted during one period.

This explains in passing another strange phenomenon, which may not have escaped the reader, although it may seem anecdotal: the reasoning which was developed on the basis of the hypothesis of transmission of the value of fixed capital to the product implies that the value of the fixed capital K produced in Section I and that of the total product Y depend on the $a$ coefficient, which reflects the share of the total value of fixed capital used and accumulated in Section I, and consequently the distribution of fixed capital between the two Sections of production. One will admit that this is not logically coherent, which confirms that the assumption which has been tested in this subsection leads to absurd reasoning.

## 2. A second interpretation of the transfer of value of fixed capital

It is possible to show that the phenomenon of transmission by fixed capital of a value greater than its initial value does not take place if one considers that the transmission of the value of fixed capital results solely from the fraction of the value of fixed capital newly produced by living labor. We have indeed in section I, again on the basis of equation (7):

Newly produced value $= L_I = (1-a)K$ (12)

Value transmitted over n periods: $= (1-a)\frac{K}{1-a} = K = \frac{1}{1-a}L_I$ (13)

It can be seen that the value K transmitted by the newly produced part $(1-a)K$ of the value of fixed capital is identical to the total value of this capital. It can be inferred that for fixed capital to transfer to the product a value identical to that which it can lose, and which cannot be different from its initial value, it is necessary and sufficient that this transmission concerns only the fraction of its value resulting from living labor.

One can thus be tempted to adopt a second interpretation allowing the integration to simple reproduction of the assumption of transmission of the value of fixed capital. In this interpretation it is therefore considered that the value transmitted by fixed capital during each period is identical to the value of the fixed capital newly produced by living labor implemented in section I.

In this case the value of the product of each of the two sections becomes:

Section I: $aL_I + L_I = (1+a)L_I$ (14)

Section II: $bL_I + L_{II}$ (15)

Is the realization of the product possible in this framework? To answer this question, let us recall first of all that if we follow Marx in his reasoning in Volume II of Capital, there is no surplus value in our model, and consequently no profit, from which it follows that the commodities produced are exchanged at their value. To address the question of realization comes down therefore to ask oneself what are the condition for the commodities produced in both sections to be disposed of at their value. With regard to section II above, the realization of the consumer goods produced there implies that the value of the variable capital advanced in the two sections, i.e. $L_I + L_{II}$ (in the absence of surplus value), is identical to the value of consumer goods. One must therefore have:

$$L_I + L_{II} = bL_I + L_{II}$$

But this identity is verified only if b = 1, which implies ipso facto $a = 0$. This means that the realization of consumer goods requires that the totality of fixed capital be used for the production of consumer goods, And that no fixed capital is used for the production of goods of production!

This very special hypothesis is necessary in order to ensure that simple reproduction takes place in the absence of exploitation of workers, who receive in exchange for the value of their labor power ($L_I + L_{II}$), on the one hand the value newly produced of consumption goods ($L_{II}$) and on the other hand the totality of the value transmitted by the fixed capital, which is by hypothesis (in this second interpretation) only $L_I$.

But such a hypothesis is not logically permissible, since the production of fixed capital without the aid of means of production, besides being totally artificial and therefore unrealistic, means that the capitalists producing the fixed capital, owning in fact no means of production, are not actually capitalists!

# APPENDIX 3 - SUMMARY TABLES OF VALUES AND PRICES OF THE MAIN VARIABLES OF THE MODELS

**Table 1 – TABLE OF VALUES AND PRICES IN THE SIMPLE MODEL**

| | **Output** | | | **Labor power** | | | **Surplus-value/Profit** | | |
|---|---|---|---|---|---|---|---|---|---|
| | Output of section I | Output of section II | Global output | Section I | Section II | Total of two sections | Section I | Section II | Total |
| **Values** | $L_I = kL_{II}$ | $L_{II}$ | $L_I + L_{II}$ | $V_I = \frac{L_I}{1+k}$ | $V_{II} = \frac{L_{II}}{1+k}$ | $L_{II}$ | $\frac{k}{1+k} L_I$ | $\frac{k}{1+k} L_{II}$ | $L_I = k\, L_{II}$ |
| **Prices** | $L_I \frac{1}{1-a}$ | $L_I + L_{II} = (1+k)L_{II}$ | $L_I \left(\frac{2-a}{1-a} + \frac{1}{k}\right)$ | $W_I = L_I$ | $W_{II} = L_{II}$ | $W = L_I + L_{II}$ | $L_I \frac{a}{1-a}$ | $L_I = kL_{II}$ | $L_I \frac{1}{1-a}$ |
| | **Ratio Surplus-value/Value of labor power And Profits/Wages** | | | **Organic composition of capital** | | | **Rate of profit** | | |
| | Section I | Section II | Total | Section I | Section II | Total | Section I | Section II | Total |
| **Values** | $k = \frac{L_I}{L_{II}}$ | $k = \frac{L_I}{L_{II}}$ | $k$ | $at(1+k)$ | $\frac{(1-a)}{(1+k)kt}$ | $q_{=k.t}$ | $\frac{k}{akt + at + 1}$ | $\frac{k}{kt(1-a)(1+k)+1}$ | $\frac{k}{kt+1}$ |
| **Prices** | $\frac{a}{1-a}$ | $k = \frac{L_I}{L_{II}}$ | $\frac{k}{(1-a)(1+k)}$ | $q_I = \frac{at}{1-a}$ | $q_{II=k.t}$ | $\frac{k.t}{(1-a)(1+k)}$ | $\frac{a}{at+1-a}$ | $\frac{k}{kt+1}$ | $\frac{k}{kt+(1-a)(1+k)}$ |

| | Table 2 - TABLE OF VALUES AND PRICES IN THE MODEL WITH CAPITALIST CONSUMPTION | | | | | | | | |
|---|---|---|---|---|---|---|---|---|---|
| | **Output** | | | **Labor power** | | | **Surplus-value and Profit** | | |
| | Output of section I | Output of section II | Global Output | Section I | Section II | Total of 2 sections | Section I | Section II | Total |
| **Values** | $L_I = kL_{II}$ | $L_{II}$ | $L_I + L_{II}$ | $V_I = \frac{L_I}{1+k_c}$ | $V_{II} = \frac{L_{II}}{1+k_c}$ | $Vc = L_{II}(1-c)$ | $\frac{L_I k_c}{1+k_c} = L_I \frac{k+c}{1+k}$ | $\frac{L_{II} k_c}{1+k_c} = L_{II} \frac{k+c}{1+k}$ | $Pl = L_{II}\,(k+c)$ |
| **Prices** | $\frac{L_I}{(1-a)(1-c^*)}$ | $L_{II}\frac{1+k}{1-c}$ | $\sum(P_I + P_{II})$ Cf. équat. 43 in chapter 12 | $W_I = L_I$ | $W_{II} = L_{II}$ | $W = L_I + L_{II}$ | $L_I(\alpha + \gamma^* + \alpha\gamma^*)$ | $L_{II}.kc = L_{II}\frac{k+c}{1-c}$ | $L_I(\alpha + \gamma^* + \alpha\gamma^*) + L_{II}k_c$ |

| | **Ratio Surplus-value/Value of labor power and Profits/Wages** | | | **Organic Composition of capital** | | | **Rate of profit We put Ψ=1+α+γ*+αγ*** | | |
|---|---|---|---|---|---|---|---|---|---|
| | Section I | Section II | Total | Section I | Section II | Total | Section I | Section II | Total |
| **Values** | $k_c = \frac{k+c}{1-c}$ | $k_c = \frac{k+c}{1-c}$ | $k_c = \frac{k+c}{1-c}$ | $at(1+k_c)$ | $\frac{kt(1-a)}{(1+k_c)}$ | $\frac{kt}{1-c}$ | $\frac{k_c}{at(1+kc)+1}$ | $\frac{k_c}{kt(1-a)(1+k_c)+1}$ | $\frac{k+c}{kt+1-c} = \frac{k_c}{(k_c-\gamma)t+1}$ |
| **Prices** | $\alpha + y^* + \alpha y^*$ | $k_c = \frac{k+c}{1-c}$ | Cf. equat. 54 In chapter 12 | Cf. equat. 18 in chapter 20 | $k_c(1+\gamma^*)$ | Cf. equat. 17 In chpater 20 | $\frac{\alpha + c^*}{t\alpha + 1 - c^*}$ | $\frac{k+c}{(1-c)[kt(1+\gamma^*)+1]}$ | $\frac{k(\Psi+\gamma)+\gamma}{kt\left(\frac{1}{t}+\Psi\right)+1}$ |

# DETAILED TABLE OF CONTENTS